\documentclass[10pt,a4paper,twoside]{report}

\usepackage[numbers,sort&compress]{natbib} % <<<<< References in numbered list [1],[2],...
\usepackage{notoccite}

\usepackage[utf8]{inputenc}   % <<<<< Linux
\usepackage[english]{babel} % <<<<< English
\usepackage{nomencl}
\makenomenclature
\RequirePackage{ifthen} 
\ifthenelse{\equal{\languagename}{english}}%
    { % English
    \renewcommand{\nomgroup}[1]{%
      \ifthenelse{\equal{#1}{R}}{%
        \item[\textbf{Roman symbols}]}{%
        \ifthenelse{\equal{#1}{G}}{%
          \item[\textbf{Greek symbols}]}{%
          \ifthenelse{\equal{#1}{S}}{%
            \item[\textbf{Subscripts}]}{%
            \ifthenelse{\equal{#1}{T}}{%
              \item[\textbf{Superscripts}]}{}}}}}%
    }{%
    \ifthenelse{\equal{\languagename}{french}}%
    { % French
    \renewcommand{\nomgroup}[1]{%
      \ifthenelse{\equal{#1}{R}}{%
        \item[\textbf{Symbole romains}]}{%
        \ifthenelse{\equal{#1}{G}}{%
          \item[\textbf{Symboles grecs}]}{%
          \ifthenelse{\equal{#1}{S}}{%
            \item[\textbf{Indices}]}{% lettre inférieure
            \ifthenelse{\equal{#1}{T}}{%
              \item[\textbf{Exposants}]}{}}}}}% lettre supérieure
    }{ % Portuguese
    \renewcommand{\nomgroup}[1]{%
      \ifthenelse{\equal{#1}{R}}{%
        \item[\textbf{Simbolos romanos}]}{%
        \ifthenelse{\equal{#1}{G}}{%
          \item[\textbf{Simbolos gregos}]}{%
          \ifthenelse{\equal{#1}{S}}{%
            \item[\textbf{Subscritos}]}{%
            \ifthenelse{\equal{#1}{T}}{%
              \item[\textbf{Sobrescritos}]}{}}}}}%
    }}%

\usepackage[acronym,nonumberlist,nopostdot,nogroupskip]{glossaries}
\setglossarystyle{long}
\makeglossaries

\newcommand{\acknowledgments}{@undefined} % new LaTeX variable name
\addto\captionsenglish{\renewcommand{\acknowledgments}{Acknowledgments}}
\addto\captionsenglish{\renewcommand{\acronymname}{Glossary}} % Acronyms
\addto\captionsfrench{\renewcommand{\acknowledgments}{Remerciements}}
\addto\captionsfrench{\renewcommand{\listfigurename}{Liste des figures}} % Table des figures
\addto\captionsportuguese{\renewcommand{\acknowledgments}{Agradecimentos}}
\addto\captionsportuguese{} % Nomenclatura
\newcommand{\coverThesis}{@undefined} % new LaTeX variable name
\addto\captionsenglish{\renewcommand{\coverThesis}{Thesis approved in public session to obtain the PhD Degree in}}
\addto\captionsenglish{}
\addto\captionsenglish{}
\addto\captionsenglish{}
\addto\captionsenglish{}
\addto\captionsenglish{}
\addto\captionsfrench{\renewcommand{\coverThesis}{Th\`ese pour l'obtention du Maîtrise des Sciences en}}
\addto\captionsfrench{}
\addto\captionsfrench{}
\addto\captionsfrench{}
\addto\captionsfrench{}
\addto\captionsfrench{}
\addto\captionsportuguese{\renewcommand{\coverThesis}{Disserta\c{c}\~{a}o para obten\c{c}\~{a}o do Grau de Mestre em}}
\addto\captionsportuguese{}
\addto\captionsportuguese{}
\addto\captionsportuguese{}
\addto\captionsportuguese{}
\addto\captionsportuguese{}

\newcommand{\declarationTitle}{@undefined} % new LaTeX variable name
\newcommand{\declarationText}{@undefined}  % new LaTeX variable name
\addto\captionsenglish{\renewcommand{\declarationTitle}{Declaration}}
\addto\captionsenglish{\renewcommand{\declarationText}{I declare that this document is an original work of my own authorship and that it fulfills all the requirements of the Code of Conduct and Good Practices of the Universidade de Lisboa.}}
\addto\captionsportuguese{\renewcommand{\declarationTitle}{Declara\c{c}\~{a}o}}
\addto\captionsportuguese{\renewcommand{\declarationText}{Declaro que o presente documento \'{e} um trabalho original da minha autoria e que cumpre todos os requisitos do C\'{o}digo de Conduta e Boas Pr\'{a}ticas da Universidade de Lisboa.}}

\def\FontLb{% 16 pt bold
  \usefont{T1}{phv}{b}{n}\fontsize{16pt}{16pt}\selectfont}

\def\FontMb{% 14 pt bold
  \usefont{T1}{phv}{b}{n}\fontsize{14pt}{14pt}\selectfont}
\def\FontSn{% 12 pt normal
  \usefont{T1}{phv}{m}{n}\fontsize{12pt}{12pt}\selectfont}

\usepackage{geometry}	
\usepackage{setspace}

\usepackage{indentfirst}

\usepackage{graphicx}

\usepackage{amsmath}  % AMS mathematical facilities for LaTeX.
\usepackage{amsthm}   % Typesetting theorems (AMS style).
\usepackage{amsfonts} % TeX fonts from the AMS

\usepackage{subfig}

\usepackage{ar}

\usepackage{dcolumn}
\newcolumntype{d}{D{.}{.}{-1}} % column aligned by the point separator '.'
\newcolumntype{e}{D{E}{E}{-1}} % column aligned by the exponent 'E'

\usepackage{rotating}

\usepackage[pdftex]{hyperref} % enhance documents that are to be
\hypersetup{colorlinks,       % color text of links and anchors,
            linkcolor=black,  % color for normal internal links
            anchorcolor=black,% color for anchor text
            citecolor=black,  % color for bibliographical citations
            filecolor=black,  % color for URLs which open local files
            menucolor=black,  % color for Acrobat menu items
            urlcolor=black,   % color for linked URLs
	          bookmarks=true,         % create PDF bookmarks
	          bookmarksopen=false,    % don't expand bookmarks
	          bookmarksnumbered=true, % number bookmarks
	          pdftitle={Thesis},
            pdfauthor={Andre C. Marta},
            pdfsubject={Thesis Title},
            pdfkeywords={Thesis Keywords},
            pdfstartview=FitV,
            pdfdisplaydoctitle=true}

\usepackage[figure,table]{hypcap}

\usepackage{multirow}

\usepackage{booktabs}
\usepackage{pdfpages}

\usepackage{algorithm}
\usepackage{algpseudocode}

\usepackage{pdfpages}
\usepackage{xcolor}
\usepackage{bm}
\usepackage{slashed}
\usepackage{tikz}
\usetikzlibrary{decorations.pathmorphing}
\usetikzlibrary{decorations.markings}
\usetikzlibrary{decorations.pathmorphing,decorations.markings,calc}
\usepackage{amssymb}

\usetikzlibrary{decorations.pathmorphing,arrows.meta}

\newcommand{\nn}{\nonumber\\}

\tikzset{
  midarrow/.style={
    postaction={
      decorate,
      decoration={
        markings,
        mark=at position 0.55 with {\arrow{>}}
      }
    }
  }
}

\def\cG{{\cal G}}
\def\cA{{\cal A}}

\def\cP{{\cal P}}
\def\cK{{\cal K}}
\def\cM{{\cal M}}
\def\cC{{\cal C}}
\def\cD{{\cal D}}
\def\cB{{\cal B}}
\def\cT{{\cal T}}

\def\cH{{\cal H}}

\def\cO{{\cal O}}
\def\cV{{\cal V}}
\def\cS{{\cal S}}

\def\cS{{\cal S}}
\def\cQ{{\cal Q}}

\def\beps{{\boldsymbol \epsilon}}

\def\q{{\boldsymbol q}}
\def\R{{\boldsymbol R}}
\def\0{{\boldsymbol 0}}
\def\1{{\boldsymbol 1}}
\def\p{{\boldsymbol p}}
\def\l{{\boldsymbol l}}
\def\k{{\boldsymbol k}}
\def\x{{\boldsymbol x}}
\def\y{{\boldsymbol y}}
\def\n{{\boldsymbol n}}

\def\r{{\boldsymbol r}}

\def\u{{\boldsymbol u}}
\def\v{{\boldsymbol v}}

\def\b{{\boldsymbol b}}
\def\0{{\boldsymbol 0}}

\def\P{{\boldsymbol P}}

\def\bkappa{{\boldsymbol \kappa}}

\def\bn{{\boldsymbol n}}

\def\Tr{{\rm Tr}}
\def\Re{{\rm Re}}
\def\Im{{\rm Im}}

\newcommand{\jewel}{\textsc{Jewel}}

\newacronym{HIC}{HIC}{Heavy-ion collision}
\newacronym{EEC}{EEC}{Energy-energy correlator}
\newacronym{ENC}{ENC}{$N$-\,point energy-correlator}
\newacronym{PENC}{PENC}{Projected $N$-\,point energy-correlator}
\newacronym{BDMPS-Z}{BDMPS-Z}{Baier-Dokshitzer-Mueller-Peigné-Schiff-Zakharov}
\newacronym{QCD}{QCD}{Quantum chromodynamics}
\newacronym{LPM}{LPM}{Landau-Pomeranchuk-Migdal}
\newacronym{DIS}{DIS}{Deep inelastic scattering}
\newacronym{DGLAP}{DGLAP}{Dokshitzer-Gribov-Lipatov-Altarelli-Parisi}
\newacronym{LL}{LL}{Leading logarithm}
\newacronym{QGP}{QGP}{Quark-gluon plasma}
\newacronym{LHC}{LHC}{Large Hadron Collider}
\newacronym{RHIC}{RHIC}{Relativistic Heavy Ion Collider}
\newacronym{PDF}{PDF}{Parton distribution function}
\newacronym{nPDF}{nPDF}{Nuclear parton distribution function}
\newacronym{AA}{AA}{Nucleus-nucleus}
\newacronym{pA}{pA}{Proton-nucleus}
\newacronym{pp}{pp}{Proton-proton}
\newacronym{PbPb}{PbPb}{Lead-lead}
\newacronym{CGC}{CGC}{Colour glass condensate}
\newacronym{IRC}{IRC}{Infrared- and collinear-safe}
\newacronym{RG}{RG}{Renormalization group}
\newacronym{HO}{HO}{Harmonic oscillator}
\newacronym{HTL}{HTL}{Hard thermal loop}
\newacronym{LSZ}{LSZ}{Lehmann-Symanzik-Zimmerman}
\newacronym{OPE}{OPE}{Operator product expansion}
\newacronym{ELF}{ELF}{Energy loss function}
\newacronym{LO}{LO}{Leading order}

\glsaddall

\begin{document}

% Set plain page style (no headers, footer with centered page number)
\pagestyle{plain}

% Set roman numbering (i,ii,...) before the start of chapters

\pagenumbering{roman}
% ----------------------------------------------------------------------
% Cover page
% ----------------------------------------------------------------------
%%%%%%%%%%%%%%%%%%%%%%%%%%%%%%%%%%%%%%%%%%%%%%%%%%%%%%%%%%%%%%%%%%%%%%%%
%                                                                      %
%     File: Thesis_FrontCover.tex                                      %
%     Tex Master: Thesis.tex                                           %
%                                                                      %
%     Author: Andre C. Marta                                           %
%     Last modified : 27 Feb 2024                                      %
%                                                                      %
%%%%%%%%%%%%%%%%%%%%%%%%%%%%%%%%%%%%%%%%%%%%%%%%%%%%%%%%%%%%%%%%%%%%%%%%

%\thispagestyle {empty}

% IST Logo - Signature A
% parameters: bb=llx lly urx ury (bounding box), width=h_length, height=v_length, angle=angle, scale=factor, clip=true/false, draft=true/false. 
%\includegraphics[bb=9.5cm 11cm 0cm 0cm,scale=0.29]{figs/IST_A_CMYK_POS_new.pdf}
%\includegraphics[bb=9.5cm 11cm 0cm 0cm,scale=0.2]{figs/IST_A_CMYK_POS_new.pdf}
\noindent
\includegraphics[width=0.45\textwidth]{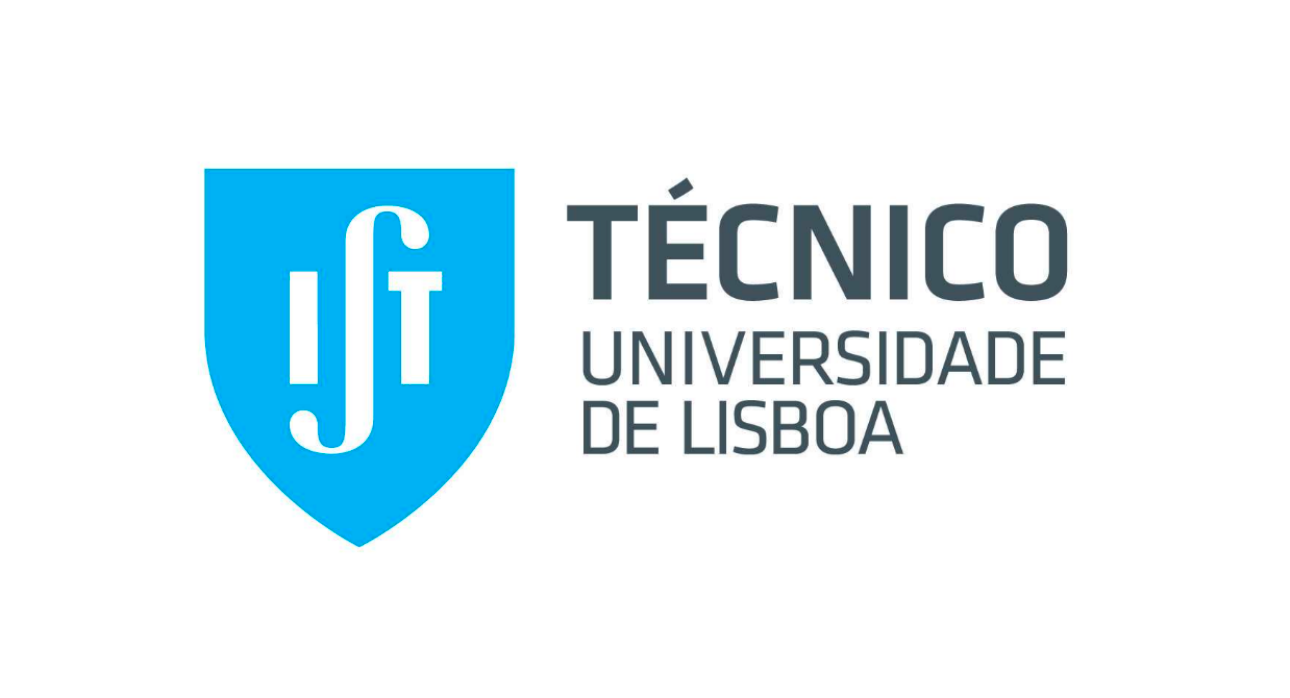}
\par
\thispagestyle{empty}

\begin{center}
%
% Figure (Image or plot)
%\vspace{3cm}
{\FontLb UNIVERSIDADE DE LISBOA} \\
{\FontLb INSTITUTO SUPERIOR TÉCNICO} \\
% height = 50 mm

% Title, author and degree
\vspace{3cm}
{\FontLb Jets as differential probes of hot QCD matter} \\ % <<<<< EDIT TITLE
%\vspace{0.2cm}
%{\FontMn Subtitle (optional)} \\
%\vspace{1.9cm}
\vspace{2cm}
{\FontMb João Miguel Martins da Silva} \\ % <<<<< EDIT NAME
\vspace{3.0cm}
{\FontSn %
\begin{tabular}{ll}
\textbf{Supervisor:} & \textbf{Doctor} José Guilherme Teixeira de Almeida Milhano\\ % <<<<< EDIT NAME
\textbf{Co-Supervisors:} & \textbf{Doctor} Liliana Marisa Cunha Apolinário\\    % <<<<< EDIT NAME
& \textbf{Doctor} Carlos Alberto Salgado López
\end{tabular} } \\
\vspace{2cm}
{\FontSn \coverThesis} \\
\vspace{0.3cm}
{\FontMb Physics} \\ % <<<<< EDIT COURSE
\vspace{1.0cm}

%{\FontLb Draft} \\
%{\FontSn \coverThesis} \\
{\FontSn \textbf{Jury final classification: Pass with Distinction and Honour}}\\
\vspace{1cm}
{\FontMb 2026} \\ % <<<<< EDIT DATE (corresponds to date of oral examination)
\end{center}
 % file "Thesis_FrontCover_new.tex"
\cleardoublepage

%%%%%%%%%%%%%%%%%%%%%%%%%%%%%%%%%%%%%%%%%%%%%%%%%%%%%%%%%%%%%%%%%%%%%%%%
%                                                                      %
%     File: Thesis_FrontCover.tex                                      %
%     Tex Master: Thesis.tex                                           %
%                                                                      %
%     Author: Andre C. Marta                                           %
%     Last modified : 27 Feb 2024                                      %
%                                                                      %
%%%%%%%%%%%%%%%%%%%%%%%%%%%%%%%%%%%%%%%%%%%%%%%%%%%%%%%%%%%%%%%%%%%%%%%%

%\thispagestyle {empty}

% IST Logo - Signature A
% parameters: bb=llx lly urx ury (bounding box), width=h_length, height=v_length, angle=angle, scale=factor, clip=true/false, draft=true/false. 
%\includegraphics[bb=9.5cm 11cm 0cm 0cm,scale=0.29]{figs/IST_A_CMYK_POS_new.pdf}

\noindent
\includegraphics[width=0.45\textwidth]{figs/IST_A_CMYK_POS_new.pdf}
\par
\thispagestyle{empty}

\begin{center}
%
% Figure (Image or plot)
\vspace{-1.1cm}
{\FontLb UNIVERSIDADE DE LISBOA} \\
{\FontLb INSTITUTO SUPERIOR TÉCNICO} \\
% height = 50 mm

% Title, author and degree
\vspace{1cm}
{\FontLb Jets as differential probes of hot QCD matter} \\ % <<<<< EDIT TITLE
%\vspace{0.2cm}
%{\FontMn Subtitle (optional)} \\
%\vspace{1.9cm}
\vspace{0.8cm}
{\FontMb João Miguel Martins da Silva} \\ % <<<<< EDIT NAME
\vspace{0.75cm}
{\FontSn %
\begin{tabular}{ll}
\textbf{Supervisor:} & \textbf{Doctor} José Guilherme Teixeira de Almeida Milhano\\ % <<<<< EDIT NAME
\textbf{Co-Supervisors:} & \textbf{Doctor} Liliana Marisa Cunha Apolinário\\    % <<<<< EDIT NAME
& \textbf{Doctor} Carlos Alberto Salgado López
\end{tabular} } \\
\vspace{0.55cm}
{\FontSn \coverThesis} \\
\vspace{0.15cm}
{\FontMb Physics} \\ % <<<<< EDIT COURSE
\vspace{0.5cm}

%{\FontLb Draft} \\
%{\FontSn \coverThesis} \\
{\FontSn \textbf{Jury final classification: Pass with Distinction and Honour}}\\

\vspace{0.5cm}
{\FontSn \textbf{Jury}} \\[0.2cm]

{\FontSn %
\begin{tabular}{@{}p{0.06\textwidth}@{}p{0.82\textwidth}@{}}

\multicolumn{2}{@{}p{0.88\textwidth}@{}}{
\textbf{Chairperson:} \textbf{Doctor} João Paulo Ferreira da Silva,
Instituto Superior Técnico, Universidade de Lisboa
} \\

\multicolumn{2}{@{}p{0.88\textwidth}@{}}{
\textbf{Members of the Committee:}
} \\

&
\textbf{Doctor} Krishna Rajagopal,
Department of Physics, Massachusetts Institute of Technology, USA \\

&
\textbf{Doctor} Korinna Zapp,
Faculty of Science, Lund University, Sweden \\

&
\textbf{Doctor} José Guilherme Teixeira de Almeida Milhano,
Instituto Superior Técnico, Universidade de Lisboa \\

&
\textbf{Doctor} Patrícia Conde Muíño,
Instituto Superior Técnico, Universidade de Lisboa \\

&
\textbf{Doctor} Ruben Maurício da Silva Conceição,
Instituto Superior Técnico, Universidade de Lisboa

\end{tabular} } \\

{\FontSn %
\begin{tabular}{c}
\textbf{Funding Institution:} \\
FCT: Fundação para a Ciência e a Tecnologia
\end{tabular} } \\

\vspace{0.35cm}
{\FontMb 2026} \\ % <<<<< EDIT DATE (corresponds to date of oral examination)

\end{center}
 % file "Thesis_1st_page.tex"
\cleardoublepage

% ----------------------------------------------------------------------
%  Cover page
% ----------------------------------------------------------------------
%\input{Thesis_FrontCover_new} % file "Thesis_FrontCover_new.tex"
%\clearpage
%
%\null
%\thispagestyle{empty}
%\newpage
%
%
%\input{Thesis_1st_page} % file "Thesis_1st_page.tex"
%\clearpage
%
%\pagenumbering{roman}
%\null
%\newpage
%
%\null
%\newpage
%\clearpage

% ----------------------------------------------------------------------
% Dedication page (optional)
% ----------------------------------------------------------------------
%%%%%%%%%%%%%%%%%%%%%%%%%%%%%%%%%%%%%%%%%%%%%%%%%%%%%%%%%%%%%%%%%%%%%%%%
%                                                                      %
%     File: Thesis_Dedication.tex                                      %
%     Tex Master: Thesis.tex                                           %
%                                                                      %
%     Author: Andre C. Marta                                           %
%     Last modified :  2 Jul 2015                                      %
%                                                                      %
%%%%%%%%%%%%%%%%%%%%%%%%%%%%%%%%%%%%%%%%%%%%%%%%%%%%%%%%%%%%%%%%%%%%%%%%

\null\vskip5cm%
\begin{flushright}
     Aos meus amores, Carolina e Sushi.
\end{flushright}
\vfill\newpage

 % file "Thesis_Dedication.tex"
\cleardoublepage

% ----------------------------------------------------------------------
% Declaration page (mandatory)
% ----------------------------------------------------------------------
%%%%%%%%%%%%%%%%%%%%%%%%%%%%%%%%%%%%%%%%%%%%%%%%%%%%%%%%%%%%%%%%%%%%%%%%
%                                                                      %
%     File: Thesis_Declaration.tex                                     %
%     Tex Master: Thesis.tex                                           %
%                                                                      %
%     Author: Andre C. Marta                                           %
%     Last modified :  29 Jun 2022                                     %
%                                                                      %
%%%%%%%%%%%%%%%%%%%%%%%%%%%%%%%%%%%%%%%%%%%%%%%%%%%%%%%%%%%%%%%%%%%%%%%%

\null\vskip5cm%
\begin{flushleft}
	\declarationTitle \\
	\declarationText
\end{flushleft}
\vfill\newpage

 % file "Thesis_Declaration.tex"
\cleardoublepage

% ----------------------------------------------------------------------
%  Acknowledgments (optional)
% ----------------------------------------------------------------------
%%%%%%%%%%%%%%%%%%%%%%%%%%%%%%%%%%%%%%%%%%%%%%%%%%%%%%%%%%%%%%%%%%%%%%%%
%                                                                      %
%     File: Thesis_Acknowledgments.tex                                 %
%     Tex Master: Thesis.tex                                           %
%                                                                      %
%     Author: Andre C. Marta                                           %
%     Last modified :  2 Jul 2015                                      %
%                                                                      %
%%%%%%%%%%%%%%%%%%%%%%%%%%%%%%%%%%%%%%%%%%%%%%%%%%%%%%%%%%%%%%%%%%%%%%%%

\section*{\acknowledgments}

% Add entry in the table of contents as section
\addcontentsline{toc}{section}{\acknowledgments}

Let me start by thanking my supervisors, Guilherme Milhano, Liliana Apolinário and Carlos Salgado. Guilherme and Liliana introduced me to heavy-ion physics, and my interest in this rich field has only grown since then. They were always available whenever I needed to discuss our projects, and they taught me a great deal, in particular to think carefully about the phenomenological purpose of a given theoretical calculation. Being mentored by Carlos was an incredible opportunity to learn how to think about physics, since in our discussions he was always able to extract the simplest possible conceptual insight from a problem. I am also grateful to Alba Soto-Ontoso, for giving me the opportunity to be mentored by her during a one-year stay at the University of Granada. This stay was indispensable to my PhD path.

I would also like to thank my colleagues at LIP-Lisbon – André Cordeiro and Nuno Olavo – with whom I shared the experience of attending my first conferences, as well as the people I have collaborated with over these five years: Andrey Sadofyev, Ian Moult, Jasmine Brewer, João Barata, Kyle Lee, Matvey Kuzmin, Meijian Li and Wenyang Qian. I am especially grateful to Andrey and João, my unofficial mentors during the second half of my PhD. Andrey taught me how to find the right scientific questions to ask and to dwell on. The discussions I had with him regarding the career paths of an academic researcher were very useful. João taught me a lot over a short period, mainly due to his remarkable awareness of the important problems to solve and to how pragmatic he is in formulating solutions to them. He put me in contact with many of my current collaborators, giving me the opportunity to be involved in many projects.

I am indebted to all my friends. To the Condóminos, for helping me experimentally determine my wine-consumption threshold. To Carol, Duarte, Fé, Hugo, Inês, Paulo, Raquel and Rita, for the friendship and for the mind-clearing habit of running that I developed with some of them, although I suspect we started it because none of us wants to admit that we are pushing 30. To my closest friends – Adriana, Ana, Inês, Inês, Joana, Madalena and Pinto – who kept me sane through never-ending philosophical conversations and dinners.

I could not have completed this PhD without the love and support of my family. I am thankful to my father, José, for his fantastic cooking and for the good, humorous conversations that came with the resulting meals. To my sister, Rita, for her constant good mood and her contagious optimism about my career. To my mother, Graça, who knows right away when I am overworked or bothered by some work-related issue, and who is always there for me when I need to talk. She is a role model for persistence and resilience, and completing this thesis is proof that I am beginning to follow in her footsteps. \textit{Obrigado, mãe}.

Finally, I am deeply thankful to my wife, Carolina. Her constant support is otherworldly, and if I second-guess myself, she is there to make sure there is not a third guess. Once in a while, she decides to say a few random words from my field just to let me know she ``knows’’ what I do for work. It is adorable. She is the one I talk to about every decision I make, and she always gives me genuine and thoughtful input, without judging my overthinking way of processing things. For this, and for much more, \textit{obrigado, mô}. I cannot forget to thank the cutest, most loving dog in the world. I do not really like that you wake me up very early when I just want to sleep a bit more before work, but I also never want you to stop doing that. \textit{Obrigado, minha Sushi}.

~\\

The original work presented in this thesis was supported by the FCT Ph.D. scholarship PRT/BD/\ 152262/2021; by MCIN/AEI (10.13039/501100011033) and ERDF through grant PID2022-139466NB-C21; by Consejería de Universidad, Investigación e Innovación, Gobierno de España and Unión Europea -- NextGenerationEU under grant AST22\_6.5; by the European Research Council project ERC-2018-ADG-835105 YoctoLHC; by ERC-PT A-Projects ``Unveiling'', financed by PRR, NextGenerationEU; and by the FCT projects CERN/FIS-PAR/0024/2019 and CERN/FIS-PAR/0032/2021, funded by CERN, and EXPL/FIS\-PAR/0905/2021, funded by the Portuguese state budget.
 % file "Thesis_Acknowledgements.tex"
\cleardoublepage

% ----------------------------------------------------------------------
%  Abstract (both in English and Portuguese)
% ----------------------------------------------------------------------
%%%%%%%%%%%%%%%%%%%%%%%%%%%%%%%%%%%%%%%%%%%%%%%%%%%%%%%%%%%%%%%%%%%%%%%%
%                                                                      %
%     File: Thesis_Resumo.tex                                          %
%     Tex Master: Thesis.tex                                           %
%                                                                      %
%     Author: Andre C. Marta                                           %
%     Last modified :  4 Mar 2024                                      %
%                                                                      %
%%%%%%%%%%%%%%%%%%%%%%%%%%%%%%%%%%%%%%%%%%%%%%%%%%%%%%%%%%%%%%%%%%%%%%%%

\section*{Resumo}

% Add entry in the table of contents as section
\addcontentsline{toc}{section}{Resumo}

As colisões ultra-relativistas de iões pesados produzem matéria de QCD quente e desconfinada, cuja evolução liga a dinâmica inicial fora de equilíbrio ao plasma de quarks e gluões em regime hidrodinâ\-mico. Os jatos de alta energia, produzidos nos processos \textit{hard} iniciais e modificados ao atravessarem este meio, constituem sondas multiescala das suas propriedades microscópicas. Nesta tese, desenvolvemos ferramentas de QCD perturbativa para usar observáveis diferenciais de subestrutura de jatos na extração de informação detalhada sobre esta matéria.

O enquadramento teórico baseia-se na descrição BDMPS-Z da propagação de partões em campos de cor estocásticos. Neste contexto, estudamos matéria de QCD anisotrópica, substituindo o parâmetro escalar de \textit{broadening} por um parâmetro dependente da direção. Isto permite relacionar correções anisotrópicas às funções de \textit{splitting} no meio com informação de spin e padrões azimutais no fluxo de energia final. Os correladores energia-energia (EECs) são usados como observáveis sensíveis a estes efeitos, incluindo EECs de sabor pesado, EECs diferenciais no ângulo azimutal e projeções em trevo que isolam harmónicos induzidos pela anisotropia.

Examinamos também como a perda de energia afeta a interpretação de observáveis de subestrutura de jatos. O \textit{quantile matching} define a perda de energia reconstruída como um deslocamento entre espectros de jatos no vácuo e modificados pelo meio. Introduzimos \textit{energy loss functions}, que codificam a energia retida por hadrões originados pelo jato, e mostramos como os seus momentos entram em correladores de energia projetados de múltiplos pontos. Por fim, vamos além das funções de \textit{splitting} $1\to2$ modificadas pelo meio, estudando correlações de spin entre emissões sucessivas e correladores de energia de ordem superior com resposta do meio. No seu todo, os resultados mostram que a subestrutura diferencial de jatos sonda propriedades direcionais da matéria de QCD e a dinâmica multiponto de jatos no meio.

\vfill

\noindent\textbf{\Large Palavras-chave:} Plasma de quarks e gluões; colisões de iões pesados; supressão de jatos; subestrutura de jatos; correladores de energia.
   % file "Thesis_Resumo.tex"
\cleardoublepage

%%%%%%%%%%%%%%%%%%%%%%%%%%%%%%%%%%%%%%%%%%%%%%%%%%%%%%%%%%%%%%%%%%%%%%%%
%                                                                      %
%     File: Thesis_Abstract.tex                                        %
%     Tex Master: Thesis.tex                                           %
%                                                                      %
%     Author: Andre C. Marta                                           %
%     Last modified :  4 Mar 2024                                      %
%                                                                      %
%%%%%%%%%%%%%%%%%%%%%%%%%%%%%%%%%%%%%%%%%%%%%%%%%%%%%%%%%%%%%%%%%%%%%%%%

\section*{Abstract}

% Add entry in the table of contents as section
\addcontentsline{toc}{section}{Abstract}

Ultra-relativistic heavy-ion collisions create short-lived droplets of deconfined, hot QCD matter whose evolution connects far-from-equilibrium early dynamics to the hydrodynamic quark-gluon plasma. High-energy jets, produced in early hard scatterings and modified as they traverse this evolving medium, provide multi-scale probes of its microscopic properties. In this thesis, we develop perturbative QCD tools for using differential jet substructure observables to extract detailed information about hot QCD matter through the modifications it imprints on jets.

The theoretical framework is based on the BDMPS-Z description of high-energy partons propagating through stochastic colour fields. Within this setup, we first study anisotropic QCD matter by replacing the usual scalar broadening parameter with a direction-dependent description of transverse momentum broadening. This allows aniso\-tropic corrections to medium-induced splittings to be related to spin information and to azimuthal patterns in final-state energy flow. Energy correlators are then used as phenomenologically relevant observables sensitive to these effects, including heavy-flavour EECs, azimuthally differential EECs and clover projections designed to isolate leading anisotropy-induced harmonics.

In this thesis, we also examine how energy loss affects the interpretation of modifications to jet substructure observables. Quantile matching is used to define reconstructed energy loss as a population-level shift between vacuum and medium-modified jet spectra. We further develop energy-loss functions, which encode the distribution of energy retained by jet-originated hadrons, and show how their moments enter multi-point projected energy correlators. Finally, we go beyond single $1\to2$ medium-induced splittings by studying spin-induced correlations between successive emissions and higher-point energy correlators including contributions from medium response. Overall, the results show that differential jet substructure can probe directional properties of QCD matter and multi-point dynamics of jets propagating through it.

\vfill

\noindent\textbf{\Large Key-words:} Quark-gluon plasma; heavy-ion collisions; jet quenching; jet substructure; energy correlators.

 % file "Thesis_Abstract.tex"
\cleardoublepage

% ----------------------------------------------------------------------
%  Table of contents, list of tables, list of figures and nomenclature
% ----------------------------------------------------------------------

% Table of contents
%
\tableofcontents
\cleardoublepage 

% List of tables
%
% Add entry in the table of contents as section
% \phantomsection
% \addcontentsline{toc}{section}{\listtablename}
% % Generate list

% List of figures
%
% Add entry in the table of contents as section
\phantomsection
\addcontentsline{toc}{section}{\listfigurename}
% Generate list
\listoffigures
\cleardoublepage 

% Nomenclature
%
% Add entry in the table of contents as section
% \phantomsection
% \addcontentsline{toc}{section}{\nomname}
% % Insert nomenclature section produced by \makenomenclature
% \printnomenclature
% \cleardoublepage

% Glossary
%
% Add entry in the table of contents as section
\phantomsection
\addcontentsline{toc}{section}{\acronymname}
% Insert glossary section produced by \makeglossaries
\printglossary[type=\acronymtype,title=\acronymname, toctitle=\acronymname]
%\printglossary[type=\acronymtype,style=long,title=\acronymname, toctitle=\acronymname]
\cleardoublepage

\section*{Preface}

% Add entry in the table of contents as section
\addcontentsline{toc}{section}{Preface}

The research presented in this thesis has been carried out at Laboratório de Instrumentação e Física Experimental de Partículas (LIP) in Lisbon, at the Departamento de Física of Instituto Superior Técnico, Universidade de Lisboa and at Instituto Galego de Física de Altas Enerxías, in Universidade de Santiago de Compostela.

To the present date, I have contributed to the following works:
\begin{itemize}
    \item\cite{Apolinario:2024apr} L. Apolinário, L. Luís, J. G. Milhano, and J. M. Silva,
    ``Towards an unbiased jet energy loss measurement'',
    JHEP 01, 164, January 2025 [arXiv:2409.12238].

    \item\cite{Barata:2024bqp} J. Barata, C. A. Salgado, and J. M. Silva,
    ``Gluon to $q\bar q$ antenna in anisotropic QCD matter: spin-polarized and azimuthal jet observables'',
    JHEP 12, 023, December 2024 [arXiv:2407.04774].

    \item\cite{Barata:2024bmx} J. Barata, I. Moult, and J. M. Silva,
    ``Tracking Energy Loss in Heavy Ion Collisions'',
    September 2024 [arXiv:2409.18174].

    \item\cite{Barata:2025fzd} J. Barata, I. Moult, A. V. Sadofyev, and J. M. Silva,
    ``Dissecting Jet Modification in the QGP with Multi-Point Energy Correlators'',
    March 2025 [arXiv:2503.13603].

    \item\cite{Kuzmin:2025fyu} M. V. Kuzmin and J. M. Silva,
    ``QCD antenna radiative spectrum in dense media within the Improved Opacity Expansion'',
    JHEP 12, 022, December 2025 [arXiv:2507.00143].
    
    \item\cite{Barata:2025uxp} J. Barata, J. Brewer, K. Lee, and J. M. Silva,
    ``Heavy Quark Pair Energy Correlators: From Profiling Partonic Splittings to Probing Heavy-Flavor Fragmentation'',
    August 2025 [arXiv:2508.19404].

    \item\cite{Silva:2025dan} J. M. Silva and A. Soto-Ontoso,
    ``Collinear spin correlations of final-state radiation in dense QCD matter'',
    JHEP 05, 026, May 2026 [arXiv:2511.17737].

    \item\cite{Barata:2025zku} J. Barata, J. G. Milhano, A. V. Sadofyev, and J. M. Silva,
    ``Early-Time Dynamics of Heavy-Ion Collisions through Energy Correlators: celestial blocks and the spacetime structure of out-of-equilibrium QCD matter'',
    December 2025 [arXiv:2512.17009].

    \item\cite{Barata:2026icn} J. Barata, M. Li, W. Qian, C. A. Salgado, and J. M. Silva,
    ``Quantum simulating multi-particle processes in high energy nuclear physics: dijet production and color (de)coherence'',
    April 2026 [arXiv:2604.11616].
    
    \item\cite{Barata:2026pgh} J. Barata, M. V. Kuzmin, I. Moult, A. V. Sadofyev, and J. M. Silva,
    ``Hydrodynamics and Energy Correlators'',
    April 2026 [arXiv:2604.21971].
\end{itemize}

Refs.~\cite{Apolinario:2024apr,Barata:2024bqp,Barata:2024bmx,Barata:2025fzd,Barata:2025uxp,Silva:2025dan,Barata:2025zku} contain a significant portion of the original work included in this thesis.

This thesis develops theoretical tools for studying jet substructure in hot and dense QCD matter, with emphasis on differential observables sensitive to the microscopic dynamics of in-medium jets. The introductory chapter presents the physical setting of the thesis, from QCD as the microscopic theory of the strong interaction to the formation of hot nuclear matter in heavy-ion collisions and the use of jets as probes of this medium.

The perturbative description of jet evolution in vacuum and in matter is reviewed in Chapter~\ref{chap:jet_quenching}. This chapter introduces the soft and collinear structure of QCD radiation and the BDMPS-Z framework used throughout the thesis to describe high-energy partons propagating through stochastic colour fields.

The study of anisotropic QCD matter and its imprints on jet observables is presented in Chapter~\ref{chap:anisotropic_early_time}, which contains results from Refs.~\cite{Barata:2024bqp,Barata:2025uxp, Barata:2025zku}. Ref.~\cite{Barata:2024bqp} studies the $g\to q\bar q$ cross section in anisotropic matter, keeping explicit spin information. Ref.~\cite{Barata:2025uxp} introduces heavy-flavour energy correlators as a controlled probe of medium modifications to partonic splittings. Ref.~\cite{Barata:2025zku} studies early-time dynamics through azimuthally differential energy-energy correlators and their harmonic structure.

Jet energy loss and its relation to differential jet substructure measurements are addressed in Chapter~\ref{chap:jet_energy_loss}, which contains results from Refs.~\cite{Apolinario:2024apr,Barata:2024bmx}. Ref.~\cite{Apolinario:2024apr} develops the quantile-matching procedure used to define reconstructed jet energy loss as a shift between vacuum and medium-modified jet spectra. Ref.~\cite{Barata:2024bmx} introduces energy-loss functions and studies how their moments enter multi-point projected energy correlators.

The final chapter, Chapter~\ref{chap:towards_multi_point_structure}, studies jet substructure observables that go beyond the standard single $1\to2$ medium-induced splitting picture and contains results from Refs.~\cite{Silva:2025dan,Barata:2025fzd}. Ref.~\cite{Barata:2025fzd} studies projected and shape-dependent higher-point energy correlators in matter, including perturbative jet modifications and medium response. Ref.~\cite{Silva:2025dan} studies spin-induced azimuthal correlations between successive emissions and their modification by dense QCD matter.

The thesis closes with Chapter~\ref{chap:conclusions}, where the main conclusions are summarized and possible future directions are discussed. % file "Preface.tex"
\cleardoublepage

% Set arabic numbering (1,2,...) after preface
%

\setcounter{page}{1}
\pagenumbering{arabic}
% ----------------------------------------------------------------------
%  Chapters
% ----------------------------------------------------------------------

%%%%%%%%%%%%%%%%%%%%%%%%%%%%%%%%%%%%%%%%%%%%%%%%%%%%%%%%%%%%%%%%%%%%%%%%
%  Thesis_Introduction.tex  --  Chapter 1: Introduction
%  Tex Master: Thesis.tex
%%%%%%%%%%%%%%%%%%%%%%%%%%%%%%%%%%%%%%%%%%%%%%%%%%%%%%%%%%%%%%%%%%%%%%%%

%% ---------------------------------------------------------------
%% Add to Thesis_Preamble.tex if not already present:
%%   \usepackage{xcolor}
%%   \usepackage{bm}
%%   \usepackage{slashed}
%%   \newcommand{\nn}{\nonumber\\}
%%   \newcommand{\leap}[1]{{\color{red}\sffamily\small\textbf{[LEAP: #1]}}}
%%   \newcommand{\flag}[1]{{\color{blue}\sffamily\small\textbf{[FLAG: #1]}}}
%% ---------------------------------------------------------------

\chapter{Introduction}
\label{chap:introduction}

% ==============================================================
\section{The strong interaction: a historical prelude}
\label{sec:strong_phenomenology}
% ==============================================================
The story of the strong interaction begins with a simple observation: atomic nuclei exist. Protons carry positive electric charge and repel one another, with the Coulomb barrier between two protons at nuclear separations ($\sim 1$~fm) being of order several MeV. Yet, protons and neutrons, collectively known as nucleons, are bound into nuclei and account for the majority of the visible mass in the Universe. A force stronger than electromagnetism and acting over a short range is therefore required to explain this observation. 

The first successful attempt at describing this quantitatively was made by Yukawa in 1935, who postulated that this force between nucleons arose from the exchange of a new particle much heavier than the electron but lighter than the proton -- the pion~\cite{Yukawa:1935xg}. The idea was later supported by the experimental discovery of this new particle, first identified in cosmic ray experiments through its characteristic decay products and measured mass~\cite{Lattes:1947mx}. As accelerator experiments reached higher energies in the late 1950s, this picture proved to be incomplete. Beyond protons and neutrons, an increasing number of unstable, strongly interacting particles were observed. These were later termed hadrons, from the Greek \textit{hadrós}, meaning "stout", "thick" or "heavy", and were divided into baryons, e.g., protons and neutrons, and mesons, the lighter hadrons which in early models were the force mediators, e.g., pions and kaons. These experimental efforts revealed a rich and extensive spectrum of hadronic states which could not all be fundamental particles. Moreover, patterns in the masses, electric charges and decay modes of these states were indicative of a deeper underlying theory.

\subsection{The quark model and color charge}

In the early 1960s, these patterns were systematised through symmetry-based classification schemes, most notably the “Eightfold Way”, independently developed by Gell-Mann~\cite{GellMann:1961ky} and Ne'eman~\cite{Neeman:1961jhl}. The key idea was that the hadron spectrum could be organised according to an approximate $SU(3)$ flavour symmetry, which groups particles into well-defined families (multiplets) according to their quantum numbers, such as electric charge, strangeness and isospin, and predicts specific relations among their masses. Gell-Mann and Zweig~\cite{GellMann:1964nj,Zweig:1964jf} then proposed that this symmetry has a physical interpretation: all hadrons are bound states of more fundamental constituents called \emph{quarks}. In this picture, quarks are spin-$1/2$ particles, like the electron, and they come in three species with different fractional electric charges, the $u$, $d$, $s$ quarks with charges $+2/3$, $-1/3$, $-1/3$, respectively. Mesons are described as $q\bar{q}$ pairs, while baryons are $qqq$ composites. This accounted elegantly for the observed multiplet structure and for the empirical $SU(3)$ mass relations, culminating in what is known as the quark model. 

Within this classification, a previously unobserved state, the $\Omega^-$, was predicted with well-defined properties, namely strangeness $S=-3$. Its discovery at the Brookhaven National Laboratory in 1964 provided a striking confirmation of the quark model~\cite{Barnes:1964pd}. Despite its success, an immediate puzzle arose. The $\Delta^{++}(1232)$ baryon has spin $J=3/2$ and is assigned quark content $uuu$. In its ground state, the spatial, spin, and flavour parts of its wavefunction are all symmetric under exchange of any two quarks. This contradicts Fermi-Dirac statistics, i.e., the requirement that the total wavefunction of a system of identical fermions (half-integer spin particles) be antisymmetric under exchange. Greenberg~\cite{Greenberg:1964pe} and Han and Nambu~\cite{Han:1965pf} resolved this by postulating a new internal degree of freedom -- \emph{colour charge} -- of which each quark carries one of three varieties, labeled red, green, blue. In this framework, the colour part of the baryon wavefunction is antisymmetric (proportional to $\epsilon^{ijk}$), restoring consistency with Fermi-Dirac statistics.

\subsection{Deep inelastic scattering and the parton model}

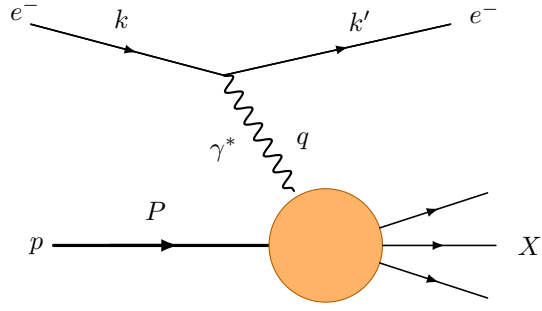
\begin{figure}[t]
\centering
\begin{tikzpicture}[line cap=round,line join=round,>=latex,scale=1.5]

% Key points
%\coordinate (V) at (-0.10,1.05);   % lepton-photon vertex
\coordinate (V) at (0,1.5);   % lepton-photon vertex
%\coordinate (C) at (0.90,0.00);    % centre of hadronic blob
\coordinate (C) at (0.90,0);    % centre of hadronic blob
\def\r{0.50}                       % blob radius

% Incoming proton (shorter, ending at circle boundary)
\draw[very thick] (-1.5,0) -- ({0.90-\r},0);
1.3
%\draw[->,very thick] (-1.55,0) -- (-1.00,0);
\draw[->,very thick] (-1,0) -- (-0.4,0);
\node at (-1.65,0) {$p$};
\node at (-0.6,0.30) {$P$};

% Incoming electron
\draw[semithick] (-1.7,1.95) -- (V);
% \draw[->,semithick] (-1.65,1.95) -- (-1.05,1.38);
\draw[semithick,midarrow] (-1.7,1.95) -- (V);
\node at (-1.75,2.08) {$e^{-}$};
\node at (-0.9,2) {$k$};

% Outgoing electron
% \draw[semithick] (V) -- (1.40,1.95);
\draw[semithick] (V) -- (2,1.95);
\draw[semithick,midarrow] (V) -- (2,1.95);
% \draw[->,semithick] (0.25,1.28) -- (0.85,1.63);
\node at (2.3,2.05) {$e^{-}$};
\node at (1.2,1.98) {$k'$};

% Virtual photon (more vertical, no arrow)
\draw[thick,decorate,decoration={snake,amplitude=2.8pt,segment length=7pt}]
(V) -- (0.62,0.48);
\node at (0,0.85) {$\gamma^\ast$};
\node at (0.7,0.92) {$q$};

% Hadronic blob
\filldraw[fill=orange!60,draw=orange!70!black] (C) circle (\r);

% Final-state lines starting at circle boundary, arrows exactly aligned
% angles and length
\def\ang{18}     % controls spread
\def\L{1}      % controls length (THIS is what you change)

% upper line
\draw[semithick,midarrow] 
({0.90+\r*cos(\ang)},{\r*sin(\ang)}) --
++({\L*cos(\ang)},{\L*sin(\ang)});

% middle line
\draw[semithick,midarrow] 
({0.90+\r},0) --
++(\L,0);

% lower line
\draw[semithick,midarrow] 
({0.90+\r*cos(\ang)},{-\r*sin(\ang)}) --
++({\L*cos(\ang)},{-\L*sin(\ang)});
\node at (2.7,0.00) {$X$};

% DIS invariants
% \node at (-1.10,-1.15) {$Q^2=-q^2$};
% \node at (1.90,-1.15) {$x_B=\dfrac{Q^2}{2P\!\cdot q}$};

\end{tikzpicture}
\caption{Schematic representation of deep inelastic electron--proton scattering, $e^{-}(k)+p(P)\to e^{-}(k')+X$. The exchanged photon carries four-momentum $q=k-k'$, and the Bjorken variable is defined as $x=Q^2/(2P\cdot q)$.}
\label{fig:dis_ep_schematic}
\end{figure}

Despite its success, the quark model initially lacked direct experimental evidence of quarks as physical constituents. This situation changed with the deep inelastic scattering (DIS) experiments at the SLAC National Accelerator Laboratory in the late 1960s~\cite{Bloom:1969kc,Breidenbach:1969kd}. In these experiments, high-energy electrons were fired at protons, allowing the internal structure of the proton to be probed at very short distances (see Fig.~\ref{fig:dis_ep_schematic} for an illustration of the process). The large momentum ($Q^2$) transfers involved, carried by an intermediate photon and reaching values of several tens of GeV$^2$, corresponded to spatial resolutions well below $\sim 1$~fm. By analysing how the electrons were deflected, one could infer how momentum was distributed inside the proton. If the proton were a smooth, extended charge distribution, the scattering cross section would be strongly suppressed at large $Q^2$, reflecting the fall-off of the electromagnetic form factor. Instead, the SLAC experiments found that the structure functions, extracted from the cross section, depend only weakly on $Q^2$ at fixed values of a dimensionless kinematic variable $x$, known as the Bjorken scaling variable~\cite{Bjorken:1969wi}. This behaviour, known as \emph{Bjorken scaling}, suggested that the scattering process probes point-like constituents.

Feynman interpreted these results in terms of incoherent scattering off quasi-free constituents, which he called \emph{partons}, with $x$ identified as the fraction of the proton’s longitudinal momentum carried by the struck constituent~\cite{Feynman:1969ej}. The observed approximate scaling was therefore a direct signature of scattering off nearly free, point-like particles. Furthermore, the structure functions were found to satisfy a relation characteristic of spin-$1/2$ constituents~\cite{Callan:1970yg}, providing support for their identification with quarks. However, the scaling is not exact: the structure functions exhibit a slow, logarithmic dependence on $Q^2$ at fixed $x$, corresponding to \emph{violations of Bjorken scaling}. These deviations show that the simple parton picture is only approximate.
Subsequent neutrino deep inelastic scattering experiments showed that the electrically charged partons, identified with quarks, carry only about half of the proton’s momentum, requiring additional electrically neutral constituents—later identified as gluons—to account for the remainder~\cite{Friedman:1972sy}. Together, these observations provided the first compelling experimental evidence for an internal substructure of nucleons: the charged partons revealed in DIS experiments were naturally identified with the quarks of the quark model.

\subsection{Confinement and asymptotic freedom}

The emerging picture of hadron structure revealed a fundamental tension. DIS experiments indicated that, at sufficiently short distances, nucleons behave as if they are composed of weakly interacting, point-like constituents. Yet, despite extensive experimental searches, no isolated quarks had ever been observed. In fact, all observed hadrons were consistent with being \emph{colour-neutral} combinations of quarks, suggesting that colour-charged particles could not exist in isolation. This phenomenon came to be known as \emph{confinement}. Further insight into confinement was obtained by considering hadrons as bound states of quarks. In particular, mesons were described phenomenologically as quark-antiquark systems, and their observed spectrum suggested that, at large separations $r$, the potential energy between a quark and an antiquark grows approximately linearly,
\begin{align}
  V(r) \simeq \sigma r \,,
\end{align}
where $\sigma$ is the so-called string tension, of order $\sigma \sim 0.9~\mathrm{GeV/fm}$. This behaviour is in stark contrast with electromagnetism, where the potential decreases with distance. It suggests that the force between quarks does not decrease at large distances, but instead remains approximately constant. A useful qualitative picture is that the colour field between a quark and an antiquark is concentrated in a narrow flux tube of approximately constant cross-section. As the separation increases, the energy stored in this tube grows approximately as $\sigma r$, and once it exceeds the energy required to produce a quark-antiquark pair ($\sim 2m_q$), the string breaks, leading to the formation of additional colour-neutral hadrons. Thus, attempts to separate quarks do not produce free quarks, but instead generate new quark-antiquark pairs that form additional colour-neutral hadrons. This process, known as hadronisation, is responsible for the appearance of a multitude of hadrons in the final state of high-energy processes.

Taken together, these observations imply that any consistent theory of the strong interaction must account for both the near-free behaviour of quarks at short distances and their confinement inside hadrons. This requirement is fulfilled in Quantum Chromodynamics (QCD), a quantum field theory in which the interaction strength decreases with decreasing distances or, equivalently, increasing energies. This property, known as \emph{asymptotic freedom}, provides a natural explanation for the observations of DIS: approximate scaling can be understood as reflecting a weak interaction between quarks at high energies, while logarithmic scaling violations arise from a residual, slowly varying interaction. The discovery of this property by Gross and Wilczek~\cite{Gross:1973id} and Politzer~\cite{Politzer:1973fx} provided the theoretical foundation for this description of the strong interaction, which we introduce in the following section.

% ==============================================================
\section{Quantum Chromodynamics}
\label{sec:QCD}
% ==============================================================

\subsection{QCD Lagrangian}
\label{subsec:Lagrangian_coupling}

\begin{figure}[t]
  \centering
  \includegraphics[width=0.65\textwidth]{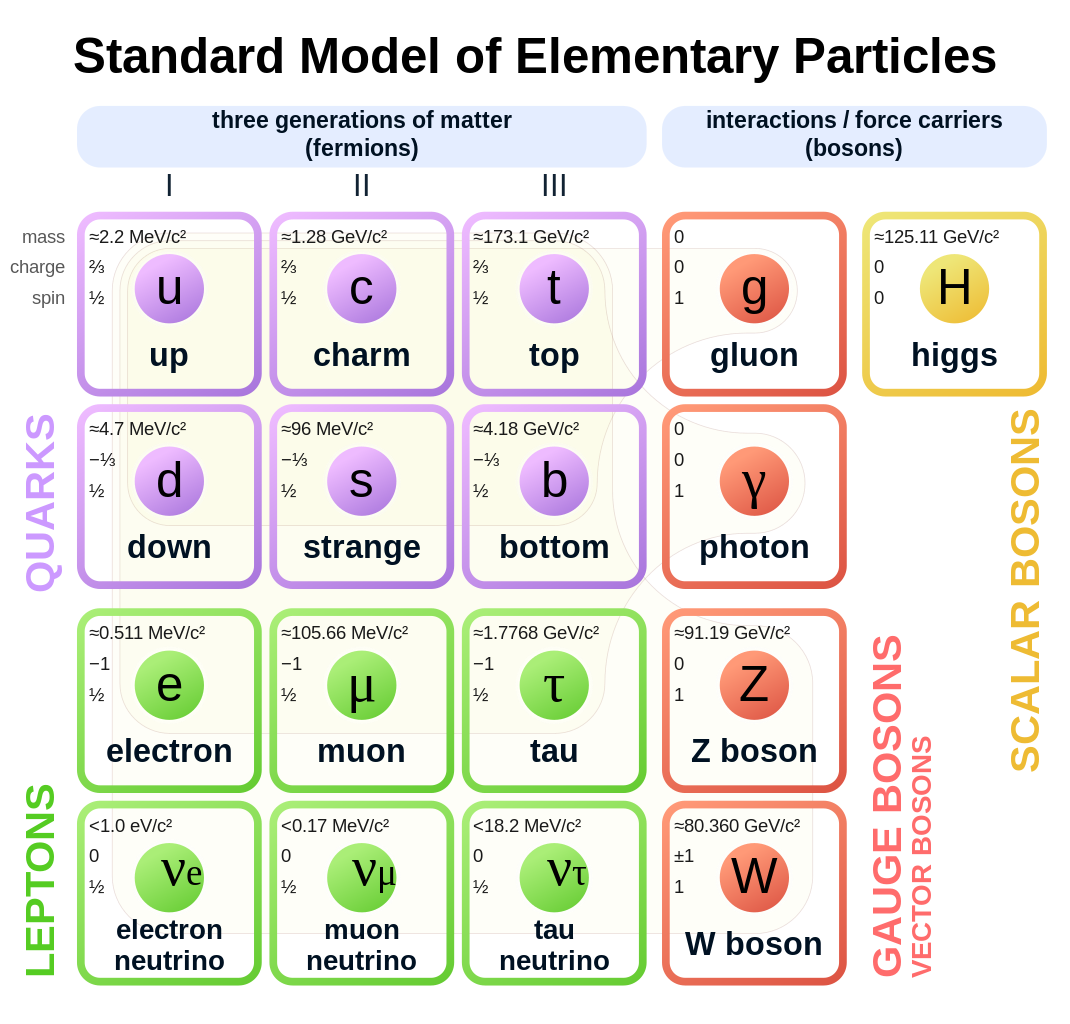}%
  \caption{Particle content of the Standard Model, including the masses, electric charges and spin of all fundamental fields. Adapted from~\cite{ATLAS_SM_chart}.}
  \label{fig:SM_particles}
\end{figure}

In modern particle physics, our theoretical understanding of fundamental interactions is formulated within the framework of quantum field theory (QFT), which provides a consistent description combining the principles of quantum mechanics with special relativity. In this framework, the fundamental degrees of freedom are fields defined over spacetime, and particles arise as quantised excitations of these fields. Over several decades of theoretical and experimental developments, this framework led to the construction of the Standard Model of particle physics, a quantum field theory based on the principle of local gauge invariance that describes the electromagnetic, weak, and strong interactions. It is defined by the symmetry group $SU(3)_c \times SU(2)_L \times U(1)_Y$ and formulated through a Lagrangian that encodes the fields and their interactions. As shown in Fig.~\ref{fig:SM_particles} the field content consists of fermionic matter fields (quarks and leptons), gauge bosons associated with each of the symmetry groups, namely the gluons of $SU(3)_c$ and the electroweak gauge bosons that give rise to the $W^\pm$, $Z^0$, and the photon, and a scalar Higgs field responsible for electroweak symmetry breaking and mass generation.

\begin{figure}[t]
\centering
\begin{tikzpicture}[
    scale=1.25,
    line width=0.7pt,
    fermion/.style={line width=0.7pt},
    gluon/.style={
        decorate,
        decoration={coil, aspect=0.7, segment length=4.5pt, amplitude=2.4pt},
        line width=0.7pt
    }
]
% quark--gluon vertex
\begin{scope}[shift={(0,0)}]
    \coordinate (v) at (0,0);
    \draw[fermion] (-1.25,0) -- (v);
    \draw[fermion] (v) -- (1.15,0.82);
    \draw[gluon]   (v) -- (1.15,-0.82);
\end{scope}
% three-gluon vertex
\begin{scope}[shift={(3.8,0)}]
    \coordinate (v) at (0,0);
    \draw[gluon] (-1.25,0) -- (v);
    \draw[gluon] (v) -- (1.15,0.82);
    \draw[gluon] (v) -- (1.15,-0.82);
\end{scope}
% four-gluon vertex
\begin{scope}[shift={(7.8,0)}]
    \coordinate (v) at (0,0);
    \draw[gluon] (-1.05,0.72) -- (v);
    \draw[gluon] (-1.05,-0.72) -- (v);
    \draw[gluon] (v) -- (1.05,0.72);
    \draw[gluon] (v) -- (1.05,-0.72);
\end{scope}
\end{tikzpicture}
\caption{Elementary interaction vertices of QCD. From left to right: the quark-gluon vertex, the three-gluon vertex, and the four-gluon vertex. Solid and curly lines represent quarks and gluons, respectively.}
\label{fig:QCD_vertices}
\end{figure}
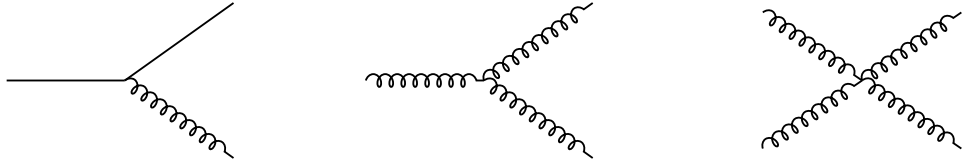

The strong interaction is described by Quantum Chromodynamics (QCD), the sector of the Standard Model associated with the colour gauge group $SU(3)_c$. It is a quantum gauge field theory whose fundamental degrees of freedom are quarks and gluons. Quarks are spin-$\tfrac{1}{2}$ fermions described by Dirac spinor fields $\psi_f$, which come in six flavours, $u, d, s, c, b, t$, extending the lighter quark species introduced in the previous section to include heavier counterparts, which became accessible as accelerator experiments reached progressively higher energies (see Fig.~\ref{fig:SM_particles}). For each flavour, the quark field may be written as a three-component object $\psi_f^i(x)$, where $i=1,2,3$ labels colour, so that quarks transform in the fundamental (triplet) representation of $SU(3)_c$. The introduction of colour, originally motivated by the symmetry properties of baryons in the quark model, here acquires a precise interpretation as the charge associated with a local gauge symmetry. The corresponding interaction fields are the gluons, described by vector fields $A_\mu^a(x)$ with $a=1,\ldots,8$, associated with the eight generators $t^a$ of the $SU(3)_c$ group, which mediate the strong force between coloured quarks and arise from imposing local invariance under $SU(3)_c$ transformations. To see this, one begins with a free Lagrangian for quark fields with masses $m_f$,
\begin{equation}
  \mathcal{L}_{\rm free} = \sum_f \bar\psi_f \left(i\gamma^\mu \partial_\mu - m_f\right)\psi_f,
\end{equation}
where $\gamma^\mu$ are the Dirac gamma matrices. 
This Lagrangian is invariant under global $SU(3)_c$ transformations, i.e., under
\begin{equation}
  \psi_f(x) \to \psi_f'(x) = U \psi_f(x), \qquad U \in SU(3)_c,
\end{equation}
with $U$ independent of spacetime. The gauge principle consists in requiring that this symmetry holds locally, promoting $U$ to $U(x)$. Under such local transformations, the ordinary derivative $\partial_\mu \psi_f$ does not transform covariantly, and the free Lagrangian is no longer invariant. Local gauge invariance can then be restored by introducing gauge fields $A_\mu^a(x)$ and replacing the ordinary derivative by the covariant derivative
\begin{equation}
  D_\mu = \partial_\mu - i g_s A_\mu^a t^a,
\end{equation}
if the gauge fields transform as
\begin{equation}
    A'_\mu(x) = U(x) A_\mu(x) U^\dagger(x) + \frac{i}{g_s} (\partial_\mu U(x)) U^\dagger(x),
\end{equation}
where $A_\mu(x) \equiv A_\mu^a(x) t^a$, $g_s$ is the strong coupling constant and $t^a = \lambda^a/2$ are the generators of $SU(3)$ in the fundamental representation, with $\lambda^a$ the Gell-Mann matrices. Since $SU(3)$ has eight generators, the theory contains eight gauge bosons, identified with the gluons. The full QCD Lagrangian is then given by
\begin{align}\label{eq:L_QCD}
  \mathcal{L}_{\rm QCD} =
  \sum_f \bar\psi_f\!\left(i\slashed{D} - m_f\right)\psi_f
  - \frac{1}{4}F^a_{\mu\nu}F^{a,\mu\nu},
\end{align}
where the field-strength tensor is
\begin{align}
  F^a_{\mu\nu} = \partial_\mu A^a_\nu - \partial_\nu A^a_\mu + g_s f^{abc} A^b_\mu A^c_\nu.
  \label{eq:field_strength}
\end{align}
The first term describes the propagation of quarks and their coupling to the gluon field, while the second term describes the dynamics of the gluons themselves. The resulting vertices of the theory are illustrated in Fig.~\ref{fig:QCD_vertices} The non-Abelian nature of the $SU(3)_c$ gauge group is encoded in the non-commutativity of its generators, $[t^a,t^b] = i f^{abc} t^c$. This structure is reflected in the term proportional to the structure constants $f^{abc}$ in the field-strength tensor, which gives rise to gluon self-interactions -- cubic and quartic. These interactions have no analogue in QED and, being the origin of both asymptotic freedom and confinement, they are a defining feature of QCD. More detailed treatments can be found in standard references such as~\cite{Peskin:1995ev, Kovchegov:2012mbw}.

\subsection{The running coupling}
\label{subsec:running_coupling}

The QCD Lagrangian in Eq.~\eqref{eq:L_QCD} contains a single dimensionless coupling constant $g_s$, which determines the strength of the interaction between the theory's degrees of freedom. In quantum field theory, however, this coupling is not a fixed parameter, as quantum corrections require the parameters and fields of the theory to be redefined through \emph{renormalisation}. This can be understood by considering quantum corrections in perturbation theory, evaluated as a series in \( g_s \) when the coupling is sufficiently small. These corrections are computed using a set of Feynman rules derived from the QCD Lagrangian, which specify the propagators of quarks and gluons and their interaction vertices. This allows physical observables or correlation functions to be expressed as sums of diagrammatic contributions organised order-by-order in the coupling. Beyond leading order, these contributions involve loop diagrams, i.e., diagrams containing at least one closed path of internal propagators, corresponding to integrations over unconstrained momenta. These integrals are generically ultraviolet divergent, i.e., they diverge for large momenta. Renormalisation absorbs these divergences into a redefinition of the parameters at a scale \( \mu \), as to match observable physical quantities, thereby introducing a dependence of the coupling on the renormalisation scale, \( g_s \to g_s(\mu) \). 

Since $\mu$ is an unphysical parameter introduced by this procedure, physical observables must ultimately be independent of it. This constrains the scale dependence of the coupling, resulting in its renormalisation group equation,
\begin{align}\label{eq:beta_function}
  \mu \frac{d g_s}{d \mu} = \beta(g_s),
\end{align}
where to leading order in $g_s$ one finds
\begin{align}\label{eq:beta_function_oneloop}
  \beta(g_s) = -\frac{g_s^3}{48\pi^2}\left(\frac{11}{3}C_A - \frac{2}{3}n_f\right) + \mathcal{O}(g_s^5),
\end{align}
where $n_f$ is the number of active quark flavours and $C_A$ is the quadratic Casimir of the adjoint representation of $SU(3)$ (the number of colors). Notably, the negative sign of the leading coefficient implies that the coupling decreases as the renormalisation scale increases, i.e., weaker interactions at short distances or high energies. This provides a quantitative manifestation of asymptotic freedom, introduced earlier~\cite{Gross:1973id,Politzer:1973fx}, and is a distinctive feature of non-Abelian gauge theories such as QCD. This behaviour can be understood by noting that, in Abelian gauge theories such as quantum electrodynamics, quantum fluctuations tend to screen the electric charge, so that the effective coupling increases at short distances. In contrast, in a non-Abelian gauge theory the self-interaction of the gauge (gluon) field gives rise to an antiscreening contribution, which dominates over the screening effect from fermion loops. As a result, the effective coupling decreases at high energies.

Integrating the renormalisation group equation at leading order, it is convenient to introduce the strong coupling $\alpha_s \equiv g_s^2/(4\pi)$, in terms of which one obtains
\begin{align}\label{eq:running_coupling}
  \alpha_s(\mu^2) = \frac{4\pi}{\beta_0 \ln(\mu^2/\Lambda_{\rm QCD}^2)},\qquad \beta_0 = \frac{11}{3}C_A - \frac{2}{3}n_f\,.
\end{align}
The quantity $\Lambda_{\rm QCD}$ is an integration constant with dimensions of energy, which is invariant under the renormalisation group. Its appearance is a manifestation of \emph{dimensional transmutation}: although the QCD coupling is dimensionless, the solution of the renormalisation group equation introduces a characteristic scale. In contrast to Abelian gauge theories such as QED, where the running coupling remains perturbative over a wide range of energies, $\Lambda_{\rm QCD}$ sets the scale at which the coupling becomes large and non-perturbative effects dominate.

Thus, the solution in Eq.~\eqref{eq:running_coupling} implies a separation of dynamical regimes. At scales $\mu \gg \Lambda_{\rm QCD}$, the coupling becomes small and perturbative methods are applicable. In this regime, processes involving large momentum transfer can be computed systematically as expansions in powers of $\alpha_s$, with radiative corrections organised through the corresponding loop expansion. This short-distance domain underlies perturbative QCD and provides the framework for describing hard scattering processes in lepton--hadron and hadron--hadron collisions. At scales $\mu \sim \Lambda_{\rm QCD}$, by contrast, the coupling becomes large and the perturbative expansion ceases to be reliable. In this regime, QCD exhibits confinement and the spontaneous breaking of chiral symmetry. These phenomena give rise to the observed hadronic spectrum and determine the low-energy
properties of strongly interacting matter. Non-perturbative approaches, such as lattice QCD and Dyson–Schwinger equations, are required to study this regime from first
principles~\cite{Wilson:1974sk,Roberts:1994dr}. Despite the success of techniques of this type, complete analytic understanding of confinement, i.e., why isolated coloured states are not observed at large distances, remains an open problem. In the following section, we make these ideas more concrete in the context of DIS, introduced earlier from a phenomenological perspective. In this setting, the interplay between short-distance dynamics and hadronic structure can be described quantitatively.

\begin{figure}[t]
  \centering
  \includegraphics[width=0.75\textwidth]{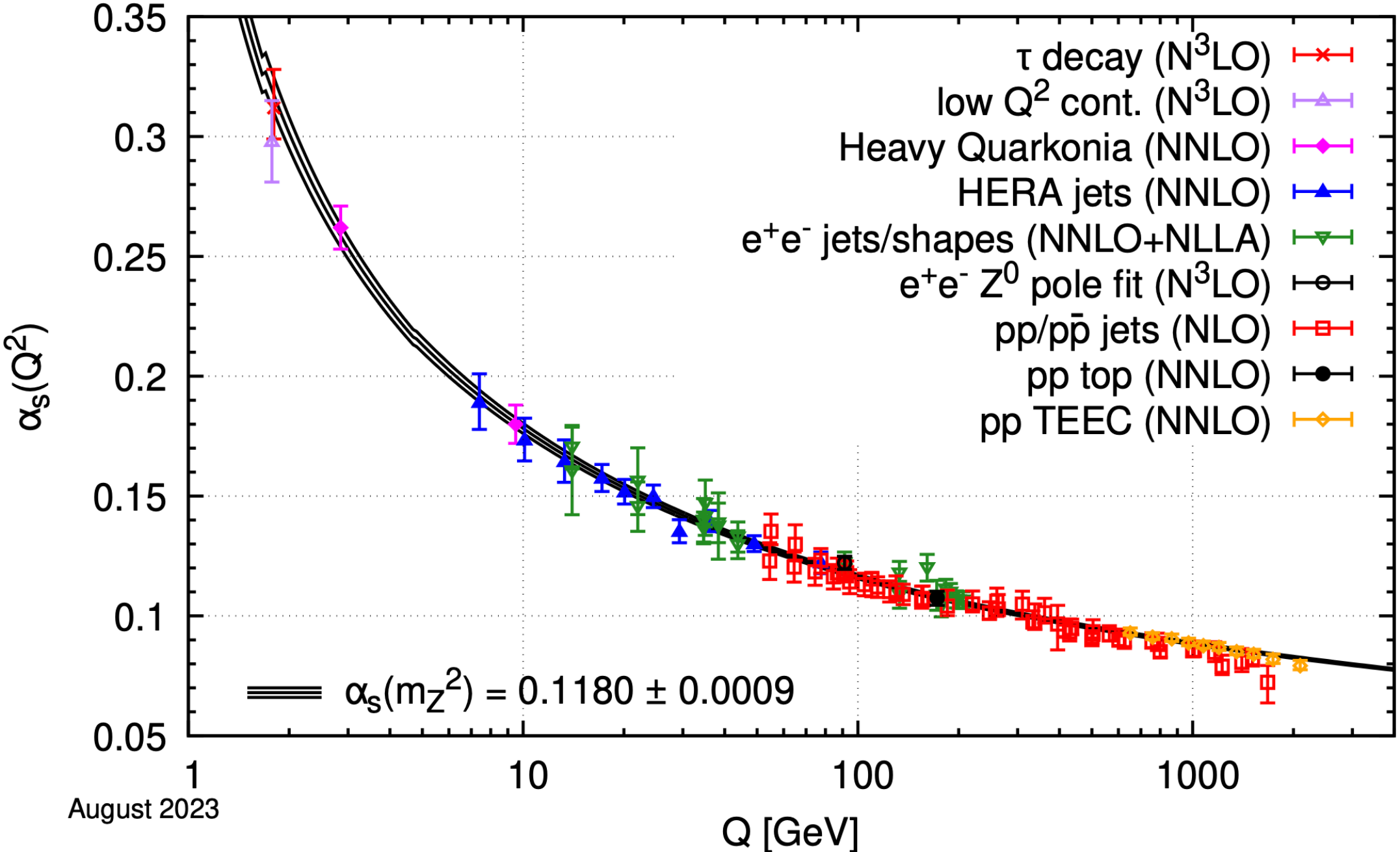}
  \caption{Summary of measurements of the strong coupling $\alpha_s(Q)$ as a function of the energy scale $Q$, illustrating its scale dependence over several orders of magnitude, and the PDG average at the $Z^0$ boson mass scale $Q^2 = m_Z^2$. Adapted from the Particle Data Group review~\cite{ParticleDataGroup:2024cfk}.}
  \label{fig:alphas_running}
\end{figure}
The scale dependence of the strong coupling, shown in Fig.~\ref{fig:alphas_running}, has been tested experimentally over a wide range of energies, from hadronic $\tau$ decays to multi-TeV processes at the LHC, and is found to be in excellent agreement with the logarithmic behaviour predicted by Eq.~\eqref{eq:running_coupling} computed to high perturbative orders. The current world average, evaluated at the $Z^0$ boson mass scale ($\mu^2 = m_Z^2)$, is $\alpha_s(m_Z^2) = 0.1180 \pm 0.0009$~\cite{ParticleDataGroup:2024cfk}, roughly 100 times larger than electromagnetic fine structure constant $\alpha_{\rm em}$ at the same scale, reflecting the intrinsic strength of the strong interactions. 
%The quark masses $m_f$ are further parameters of the theory, ranging from $m_u \approx 2$~MeV for the up quark to $m_t \approx 173$~GeV for the top quark. 

\subsection{DIS, DGLAP, and parton distribution functions}
\label{subsec:DIS_DGLAP}

As discussed in the previous subsection, the running of the strong coupling implies a separation between short- and long-distance regimes, with perturbative methods applicable when the relevant scale satisfies $Q \gg \Lambda_{\rm QCD}$. DIS provides a direct realisation of this situation, as the large momentum transfer $Q^2$ sets the resolution scale of the probe. In this context, the approximate Bjorken scaling observed experimentally can be understood as a consequence of the weak coupling at short distances, while the residual dependence of the structure functions on $Q^2$ arises from perturbative corrections governed by the renormalisation group. The observed scaling violations therefore reflect the scale dependence of the proton’s internal structure, probed at distances $1/Q$ by the photon, and can be described quantitatively within perturbative QCD.

The separation between short- and long-distance dynamics can be made precise within QCD through factorisation theorems~\cite{Collins:2011zzd}, which apply to sufficiently inclusive observables involving a hard scale $Q \gg \Lambda_{\rm QCD}$. In the case of DIS, this leads to the structure functions admitting a decomposition into a short-distance contribution, associated with the interaction of the photon with elementary partons and calculable in perturbation theory, and a long-distance contribution, which encodes the non-perturbative structure of the proton. This decomposition leads to a representation of the DIS structure function $F_2$ as a convolution of perturbative functions $C_a$ and non-perturbative objects $f_{a/p}$ termed \emph{parton distribution functions} (PDFs),
\begin{align}
F_2(x,Q^2) = \sum_a \int_x^1 \frac{dz}{z}\, C_a\!\left(\frac{x}{z},\frac{Q^2}{\mu^2},\alpha_s(\mu^2)\right)\, f_{a/p}(z,\mu^2)\,,
\end{align}
At leading order, this expression reduces to the parton model result, in which the PDFs admit a probabilistic interpretation in terms of the distribution of momentum fraction $z$ of species $a$ (quark or gluon) inside a proton probed at scale $\mu^2$. The scale $\mu$ defines the scale at which long-distance physics is being factorized from short-distance and into the PDFs. In general, one may distinguish between a factorisation scale and a renormalisation scale, although for simplicity a common scale is often employed. Physical observables are independent of this choice up to higher-order corrections.

The dependence of the PDFs on the factorisation scale $\mu$ is not arbitrary, but is governed by perturbative QCD. Requiring physical observables to be independent of $\mu$ leads to evolution equations that determine how PDFs change with $\mu$. These are the Dokshitzer--Gribov--Lipatov--Altarelli--Parisi (DGLAP) equations~\cite{Gribov:1972ri,Altarelli:1977zs,Dokshitzer:1977sg},
\begin{align}
\mu^2 \frac{\partial f_a(x,\mu^2)}{\partial \mu^2}
=
\frac{\alpha_s(\mu^2)}{2\pi}
\sum_b \int_x^1 \frac{dz}{z}\, \mathcal{P}_{a \leftarrow b}\big(z,\alpha_s(\mu^2)\big)\, f_b(x/z,\mu^2)\,,
\end{align}
where the splitting functions $\mathcal{P}_{a \leftarrow b}(z)$ are known perturbatively and encode the structure of parton branchings. These can be interpreted, at leading order, as describing the probability for a parton $b$ to produce a parton $a$ carrying a fraction $z$ of its momentum. Explicit leading-order expressions are given in Appendix~\ref{app:vertex_matrix_elements} and further discussion is given further ahead in the context of jet production. The evolution encoded in these equations can be understood in terms of successive parton branchings as the resolution scale $\mu^2$ increases. Branching processes ($q\to qg$, $g\to gg$, $g\to q\bar q$) redistribute momentum among partons, leading to a scale dependence of the PDFs. In particular, the growth of the gluon distribution at small $x$ and the depletion of quark distributions at large $x$ with increasing $\mu^2$ provide a natural explanation of the observed scaling violations. This behaviour reflects the enhanced probability of collinear parton emission in QCD, a feature that will be discussed in more detail in the context of jet production.

\begin{figure}[t]
    \centering
    \includegraphics[width=\linewidth]{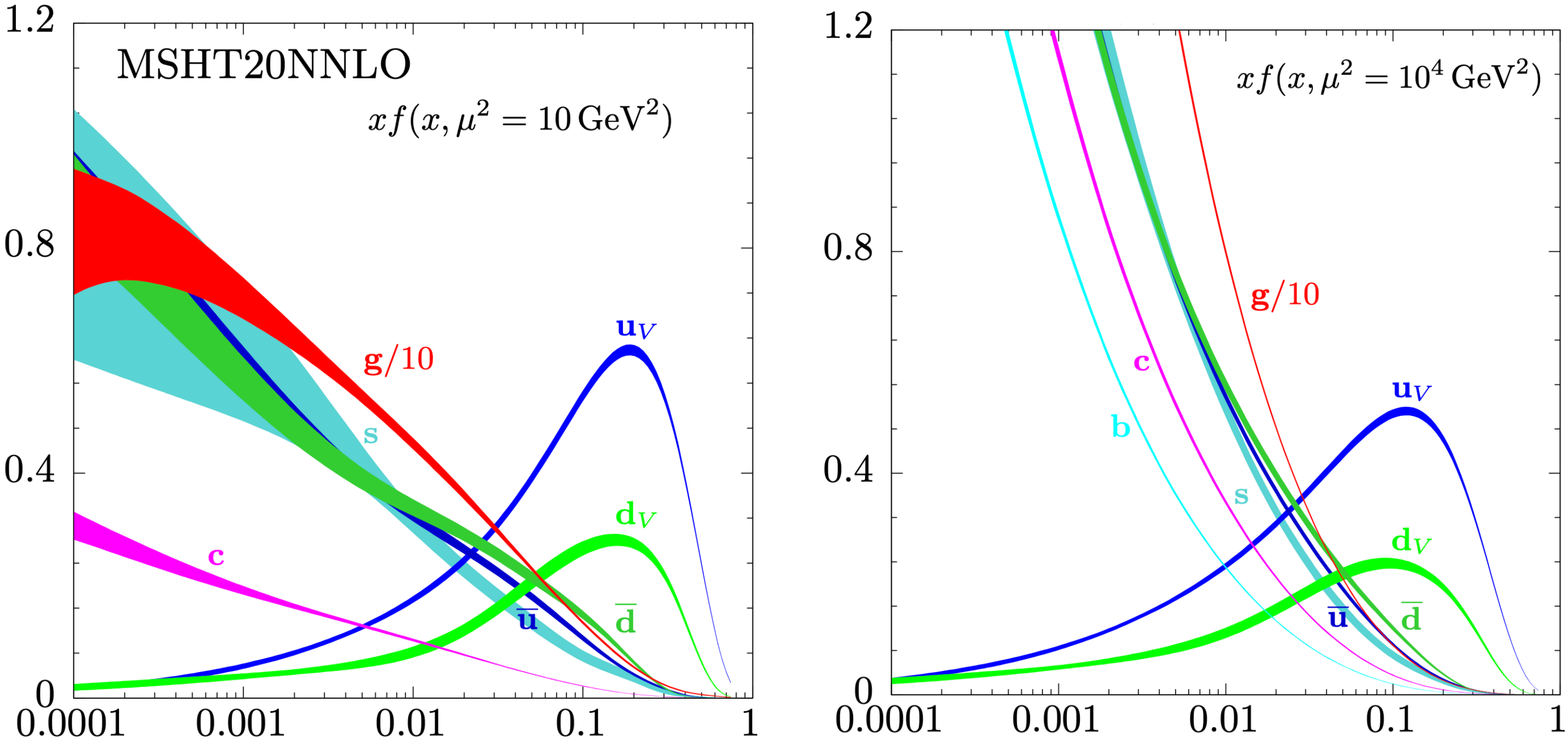}
    \caption{Parton distribution functions of the proton, shown as $x f_i(x,\mu^2)$ for several parton species, obtained in a global QCD analysis at NNLO (next-to-next-to-leading-order), adapted from the Particle Data Group~\cite{ParticleDataGroup:2024cfk}. The left and right panels correspond to $\mu^2 = 10~\mathrm{GeV}^2$ and $\mu^2 = 10^4~\mathrm{GeV}^2$, respectively. The change in shape between the two panels illustrates the scale dependence of the PDFs governed by DGLAP evolution.}
    \label{fig:pdf_evolution}
\end{figure}

In practice, PDFs are extracted from experimental data by parametrising them at an initial scale $\mu_0^2$ and fitting the corresponding parameters to measurements of sufficiently inclusive processes, with DGLAP evolution used in the fit to relate the input distributions to the scales probed by the data. The resulting scale dependence is illustrated in Fig.~\ref{fig:pdf_evolution}, which shows the distributions of quarks and gluons as functions of $x$ at different values of $\mu^2$. As the resolution scale increases, the gluon density grows at small $x$, while the quark distributions are progressively shifted towards lower momentum fractions. These features underlie the observed scaling violations of the DIS structure function $F_2(x,Q^2)$, in particular its rise at small $x$ and suppression at large $x$ with increasing $Q^2$.

%==============================================================
\section{From QCD to QCD matter}
\label{sec:QCD_matter}
% ==============================================================

\subsection{The QCD phase diagram}
\label{subsec:phase_diagram}

QCD as introduced in the previous section defines the dynamics of quarks and gluons at the microscopic level. When considered at finite temperature $T$ and/or baryon chemical potential $\mu_B$, the theory gives rise to a rich set of macroscopic states of strongly interacting matter. This set of states can be organised into a phase diagram, typically represented in the $(T,\mu_B)$ plane. Mapping this diagram amounts to understanding how different forms of complex matter emerge from the fundamental degrees of freedom of QCD, and remains an open problem and active area of both theoretical and experimental research. From the theory side, QCD at finite $T$ and $\mu_B$ is described within thermal field theory~\cite{Kapusta:2006pm}. This underlies, for instance, lattice QCD (LQCD) calculations of thermodynamics quantities, an approach based on the discretisation of spacetime, allowing for non-perturbative numerical evaluation. In practice, however, much of the structure of the phase diagram must be inferred from a combination of LQCD, effective models and experimental constraints. In particular, while the region of vanishing $\mu_B$ is fairly well constrained quantitatively by LQCD, the finite $\mu_B$ part of the diagram remains much less well understood. The interested reader is referred to~\cite{Stephanov:2004wx, Fukushima:2010bq} for a more thorough account of the subject.

\begin{figure}[t]
    \centering
    \includegraphics[width=0.7\linewidth]{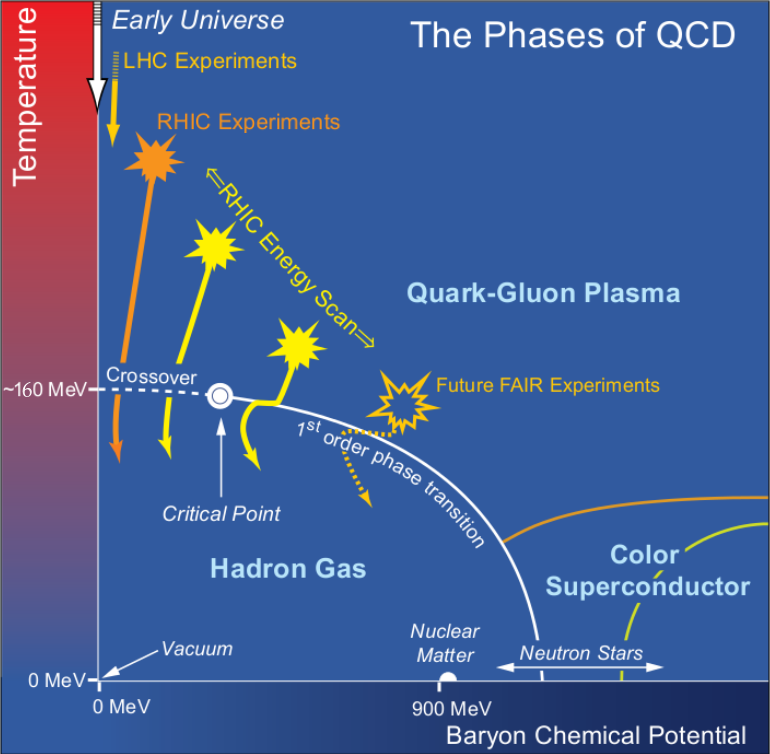}
    \caption{Schematic representation of the QCD phase diagram as a function of temperature and baryon chemical potential. The solid lines represent phase boundaries and the solid circle indicates the conjectured position of the critical endpoint. Illustrative phase trajectories for systems created at different experimental facilities (corresponding to different collision energies) are shown. Adapted from~\cite{796947}.}
    \label{fig:QCD_phase_diagram}
\end{figure}

In Fig.~\ref{fig:QCD_phase_diagram} we present the QCD phase diagram, represented in terms of the temperature $T$, which sets the characteristic energy scale of the system, and the baryon chemical potential $\mu_B$, which determines the net baryon density. The region of low temperature and moderate baryon chemical potential corresponds to ordinary nuclear matter, while increasing temperature or varying density leads to qualitatively different regimes characterised by changes in the relevant degrees of freedom and in the realisation of QCD symmetries, such as chiral symmetry.

\paragraph{Low baryon density.}
At small $\mu_B$, the low-temperature phase of QCD is a confined hadronic phase, well described as a hadronic gas populated mainly by light particles (pions), with heavier states contributing as the temperature increases. In this regime, approximate chiral symmetry associated with light-quarks is spontaneously broken, reflected in a non-vanishing chiral condensate $\langle \bar{q} q \rangle$. As the temperature increases, lattice QCD calculations indicate a rapid but continuous crossover at a temperature $T_c \simeq 155\,\mathrm{MeV}$~\cite{Aoki:2006we,HotQCD:2014kol}. As one further increases the temperature, the system enters a deconfined phase where quarks and gluons constitute the relevant degrees of freedom, restoring approximate chiral symmetry. This state of matter, now known as the quark-gluon plasma (QGP), was phenomenologically developed in the late 1970s, notably by Shuryak~\cite{Shuryak:1978ij}, and later embedded in the space-time picture of ultrarelativistic heavy-ion collisions (HICs) by Bjorken~\cite{Bjorken:1982qr}. Although asymptotic freedom predicts a weakly coupled quark–gluon plasma at sufficiently high temperatures, the conditions achieved in current experiments correspond to a regime of moderate coupling. Combined with the high density of coloured degrees of freedom, this gives rise to a strongly coupled system exhibiting pronounced collective behaviour. Further, note that while this high-temperature phase is expected to extend over finite $\mu_B$, its properties are quantitatively under control primarily in the low $\mu_B$ region. 

\paragraph{High baryon density.}
At large $\mu_B$ the structure of the phase diagram is considerably less well established. The region of low temperature and $\mu_B \approx m_N$, where $m_N$ denotes the nucleon mass, corresponds to ordinary nuclear matter. While keeping $T\approx 0$, as one increases baryon density the system evolves into a degenerate Fermi system, with pressure largely governed by the Pauli exclusion principle. A first-order phase transition between hadronic matter and deconfined quark matter is expected in such region. This transition defines a line which extends to higher temperatures and lower baryon densities and is conjectured to terminate at a critical point. This endpoint is associated with a second-order phase transition, in particular the emergence of long-range fluctuations. Its existence and location are subject to ongoing theoretical and experimental investigation~\cite{Stephanov:2004wx}. At even higher densities, QCD may exhibit additional phases such as colour superconductive matter, where colour gauge symmetry is expected to be broken~\cite{Alford:2007xm}. Finally, the region of simultaneously large $T$ and $\mu_B$ remains largely unknown, lacking a first-principles description.

~\\

Different regions of the phase diagram are realised in distinct physical settings. Compact stellar objects probe cold and dense matter at large baryon density, where the possible presence of deconfined quark matter is expected to affect the equation of state of neutron stars (see e.g.~\cite{Oertel:2016bki}). In the early Universe, approximately \(10\,\mu\mathrm{s}\) after the Big Bang, the system evolved along a trajectory close to \(\mu_B \approx 0\), undergoing the crossover from a high-temperature QGP to a hadronic phase as it cooled~\cite{Aoki:2006we, Busza:2018rrf}. On Earth, ultrarelativistic heavy-ion collisions provide access to this region of the phase diagram, creating a short-lived QGP that reaches temperatures of several times $T_c$ and enabling the study of this phase under controlled conditions. This motivates the experimental heavy-ion collisions programme, currently carried out at facilities such as the Relativistic Heavy Ion Collider (RHIC)\footnote{RHIC completed its final collisions in 2026, after approximately 25 years of operation, and its infrastructure is being repurposed for the Electron-Ion Collider.} and the Large Hadron Collider (LHC), which aims to characterise the properties of the QGP and to constrain the structure of the QCD phase diagram. In the following subsection, we discuss how such collisions are used to create and probe nuclear matter at high temperature, the main object of study of this thesis.

\subsection{Heavy-ion collisions as a laboratory for hot QCD matter}
\label{subsec:HIC}

The experimental study of strongly interacting matter at high temperature and energy density has developed over several decades, progressing through a sequence of accelerator facilities with increasing collision energies. In the 1980s and 1990s, early heavy-ion programmes at the Alternating Gradient Synchrotron (AGS), at the Brookhaven National Laboratory (BNL), and the Super Proton Synchrotron (SPS), at the European Organization for Nuclear Research (CERN), explored nucleus--nucleus collisions, with PbPb systems at the SPS reaching centre-of-mass energies per nucleon-nucleon pair of $\sqrt{s_{NN}} \approx 17$~GeV. In 2000, it was claimed that a new state of matter had been produced in SPS collisions~\cite{Heinz:2000bk}. This was supported by a combination of observations, including the suppression of quarkonium states (in particular the $J/\psi$, a bound state of a charm quark--antiquark pair)~\cite{NA50:2005} and the enhanced production of hadrons containing strange quarks~\cite{WA97:2002}.
%These effects were interpreted as signatures of a medium in which colour screening modifies quark--antiquark binding and in which strange quarks are produced more abundantly than expected in a hadronic gas. 
However, the individual measurements were not conclusive and their interpretation remained subject to uncertainties~\cite{Heinz:2000bk}.

A significant step forward in the experimental situation was achieved with the start of operations of RHIC in 2000, which enabled Au-Au collisions at $\sqrt{s_{NN}} = 130$--$200$ GeV. In contrast to earlier expectations of a weakly coupled quark--gluon gas at high temperature, measurements at RHIC indicated that the produced medium behaved more as a strongly coupled fluid, later interpreted as a strongly coupled QGP. This interpretation was supported by the observation of large elliptic flow coefficients, indicative of strong collective expansion driven by pressure gradients, and by the suppression of high-transverse-momentum hadron yields by about a factor of five relative to expectations from proton--proton collisions. The absence of such suppression in deuteron--nucleus collisions further indicated that these effects arose from final-state interactions in the produced medium~\cite{BRAHMS:2004adc,PHOBOS:2004zne,PHENIX:2004vcz,STAR:2005gfr}. These observations are quantitatively described by relativistic hydrodynamic models, whose successful description of the data requires an early onset of hydrodynamic behaviour, at time scales of order $1$ fm/c~\cite{Bjorken:1983qr,Kolb:2003dz}. The heavy-ion programme was subsequently extended at the LHC, where PbPb collisions at $\sqrt{s_{NN}} = 2.76$ and $5.02$ TeV since 2010 have confirmed and further constrained this picture over a broad kinematic range.

\begin{figure}
    \centering
    \includegraphics[width=0.9\linewidth]{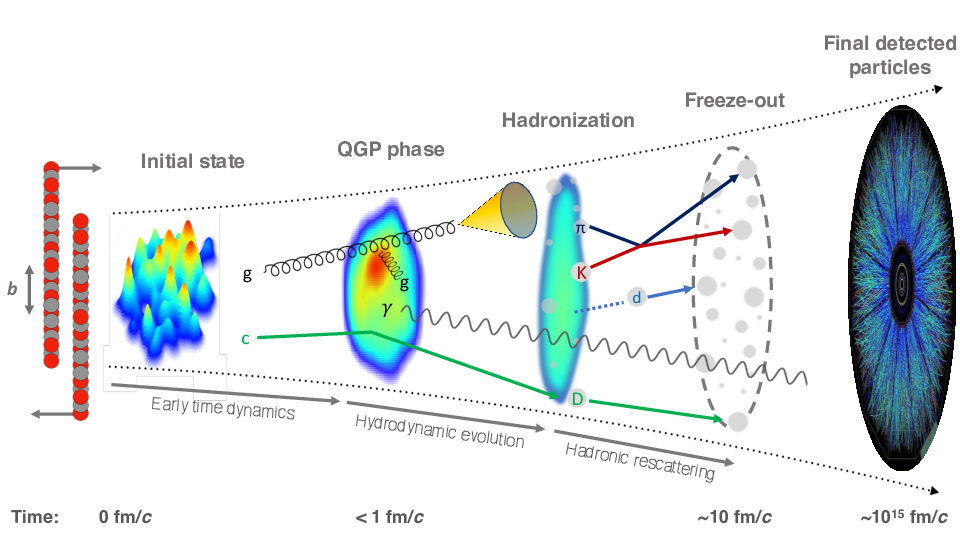}
    \caption{Schematic space–time evolution of an ultrarelativistic heavy-ion collision. Adapted from~\cite{Arslandok:2023utm}.}
    \label{fig:HIC_illustration}
\end{figure}
The experimental programme described above established heavy-ion collisions as a tool to study QCD matter under extreme conditions. Once the produced medium approaches a regime where hydrodynamics is applicable, its evolution can be characterised in terms of temperature and baryon chemical potential, such that the system is probing the QGP phase of the QCD phase diagram described in the previous section. At earlier times, however, the collision proceeds through strongly interacting many-body states that are far from equilibrium and are not represented on the phase diagram. Heavy-ion collisions therefore provide access not only to the equilibrium thermodynamics of QCD matter, but also to its real-time non-equilibrium dynamics. Understanding these phenomena at a quantitative level requires a dynamical description that connects the initial nuclear wavefunctions to the observed hadronic final state. A widely used phenomenological picture is presented in Fig.~\ref{fig:HIC_illustration}, where the collision evolves through several successive stages, characterised by different relevant degrees of freedom and timescales. Currently, this description does not arise from a first-principles solution of QCD in real time, but rather from a combination of effective descriptions of the dominant physics at each stage, e.g., initial state, collective evolution and hadronic freeze-out. Notably, this formulation has proved successful in describing a broad range of heavy-ion collision observables. For detailed reviews on the stage-wise description of HICs and its theoretical foundations, see~\cite{Gelis:2010nm,Heinz:2013th,Gale:2013da,Busza:2018rrf}.

\subsection{Space-time evolution of a heavy-ion collision}\label{subsec:st_evolution_HIC}

In a heavy-ion collision the incoming nuclei are strongly Lorentz-contracted along the longitudinal (beam) direction, effectively interacting as thin discs. For instance, for large nuclei such as Pb and Au, the diameter of these discs is $\sim 10$ fm, while their thickness ranges from $10^{-3}$ to $10^{-1}$ fm depending on the exact beam energy~\cite{Busza:2018rrf}. The geometry of the interaction is characterised by the impact parameter $b$, defined as the transverse distance between the centres of the two nuclei. Central collisions correspond to nearly maximal overlap, while peripheral collisions probe only a small fraction of the nuclear volume. 

Since the impact parameter is not directly measurable, events are first classified experimentally into centrality percentiles (or centrality classes) using observables such as charged-particle multiplicity. To relate these percentiles to the underlying collision geometry, one models the spatial distribution of nucleons inside each nucleus and computes the geometric overlap for a given impact parameter. In practice, this mapping is commonly performed using event-by-event Monte Carlo Glauber simulations, in which nucleon positions are sampled from nuclear density distributions and nucleon--nucleon collisions are assigned probabilistically according to $\sigma_{NN}^{\rm inel}$~\cite{Miller:2007ri,Loizides:2017ack}. This yields average quantities such as the impact parameter, the number of participating nucleons $N_{\rm part}$, and the number of binary nucleon--nucleon collisions $N_{\rm coll}$. The corresponding geometric structure is encoded in Glauber models~\cite{Glauber:1970jm,Miller:2007ri}, where the central quantity in the optical formulation is the nuclear overlap function
\begin{align}
    T_{AB}(\mathbf b)=\int d^2\mathbf s\, T_A(\mathbf s)\,T_B(\mathbf s-\mathbf b),\qquad 
    T_A(\mathbf s)=\int dz\,\rho_A(\mathbf s,z)\, ,
\end{align}
with nuclear density distributions $\rho_A$ typically parametrised by Woods-Saxon distributions~\cite{Woods:1954zz}, with dependence on transverse ($\mathbf s$) and longitudinal ($z$) positions. The overlap function determines, for instance, the optical Glauber expectation value of the number of binary collisions through
\begin{align}
    N_{\rm coll}(\mathbf b)\sim \sigma_{NN}^{\rm inel}\, T_{AB}(\mathbf b),
\end{align}
where $\sigma_{NN}^{\rm inel}$ is the inelastic nucleon--nucleon cross-section. For the most central (0--5\%) PbPb collisions at $\sqrt{s_{\rm NN}} = 5.02$ TeV, one finds Glauber-model estimates of $\langle N_{\rm part} \rangle \approx 384$ and $\langle N_{\rm coll} \rangle \approx 1760$~\cite{Loizides:2017ack}, corresponding to a large nuclear overlap region. This thesis focuses on such semi-central to central collisions, where conditions for QGP formation, i.e. sufficiently large deposited energy density, are most favourable.

Following the nuclear overlap, the produced system undergoes a rapid longitudinal expansion. The natural variable used to describe the subsequent evolution is the proper time $\tau=\sqrt{t^2-z^2}$, measured from the moment of nuclear overlap.

\paragraph{Pre-collision state ($\tau < 0$) and the \emph{Colour Glass Condensate}.}

The Glauber description captures the distribution of matter in the incoming nuclei, providing a framework to compute the geometric overlap and the initial energy deposition. At the partonic level, however, this nucleonic picture is insufficient to describe the small-scale structure of the collision, and a more appropriate description is required.

At ultrarelativistic energies ($\sqrt{s}$), the semi-hard bulk of particle production occurs at momentum transfers $Q \sim$ few GeV, which forces the partons inside each nucleus to carry momentum fractions $x \sim Q/\sqrt{s} \ll 1$. In this regime, the small Bjorken-$x$ enhancement of gluon radiation drives rapid growth of the gluon distribution $xG(x,Q^2)$, as described at leading order by BFKL evolution~\cite{Kuraev:1977fs,Balitsky:1978ic}. This evolution is complementary to the DGLAP evolution discussed in Section~\ref{subsec:DIS_DGLAP}, which describes the enhancement of collinear radiation. The resulting growth implies that each nucleus is effectively a Lorentz-contracted disk with a dense population of small-$x$ gluons. For sufficiently small $x$, recombination processes $gg \to g$ become important and linear evolution must be replaced by non-linear dynamics, described by the BK/JIMWLK equations~\cite{Balitsky:1995ub,Kovchegov:1999yj, Iancu:2003xm}. This leads to the emergence of a saturation scale $Q_s(x,\sqrt{s})$, at which gluon splitting and recombination balance. Since $Q_s \gg \Lambda_{\rm QCD}$, the coupling remains weak, $\alpha_s(Q_s)\ll1$, while the gluon occupation number becomes large, $f\sim\alpha_s^{-1}\gg1$, leading to the phenomenon of \emph{gluon saturation}. In this regime the small-$x$ gluon fields, with amplitudes $A\sim1/g$, are more naturally described as classical $SU(3)$ Yang-Mills fields, with quantum fluctuations suppressed relative to the classical background.

These features are systematically described by the \emph{Colour Glass Condensate} (CGC)~\cite{McLerran:1993ni,McLerran:1993ka,Gelis:2010nm}, an effective field theory built on the separation of scales between the fast, large-$x$ (valence) colour sources and the slow, small-$x$ gluon fields that emerge in the high-energy limit. The large-$x$ partons, frozen by time dilation on the timescale of the collision, act as static stochastic colour sources for the small-$x$ gluon fields, with the dynamics of the latter governed by the Yang--Mills equations sourced by the former. 

Beyond accounting for the bulk of particle production in heavy-ion collisions, gluon saturation is a fundamental prediction of QCD at high density, reflecting the non-Abelian self-interaction of gluons and the unitarisation of scattering amplitudes at small-$x$. Experimentally, HERA deep-inelastic scattering data~\cite{H1ZEUS:2015qlf} are broadly consistent with the expected slowdown in the growth of $xG(x,Q^2)$ at small-$x$~\cite{GolecBiernat:1998js,Stasto:2000er}, although an unambiguous identification of the saturation regime has not yet been established. The experimental search for gluon saturation thus remains active and is one of the central motivations for the Electron-Ion Collider (EIC), whose kinematic reach and precision are designed to provide decisive tests of small-$x$ QCD dynamics~\cite{AbdulKhalek:2021gbh}.

\paragraph{Initial overlap and \emph{Glasma} formation ($\tau \sim 0$ fm/c).}
When the two nuclei pass through each other, their respective colour fields interact and generate a highly out-of-equilibrium state characterised by strong longitudinal chromoelectric and chromomagnetic fields in the forward light cone, i.e. in the region between the two receding Lorentz-contracted nuclei, localised in transverse domains of size $\sim 1/Q_s$~\cite{Kovner:1995ja,Krasnitz:1998ns,Lappi:2006fp,Gelis:2010nm}. This state, often referred to as the \emph{Glasma}, inherits properties from the CGC as initial conditions and is dominated by coherent field configurations rather than particle-like excitations. The deposited energy density is correspondingly very large, estimated to be of order of tens of GeV/fm$^3$ at times $\tau \sim 0.1-1$ fm/c in RHIC and LHC collisions~\cite{Bjorken:1982qr,Busza:2018rrf}. The corresponding stress tensor is strongly anisotropic: the longitudinal pressure, associated with expansion along the beam direction, is initially much smaller than the transverse pressure and in classical Yang-Mills descriptions it is even negative, while the transverse pressure remains positive~\cite{Lappi:2006fp,Berges:2013eia}. In practice, early-time Glasma properties are commonly computed by solving the classical Yang--Mills equations with CGC initial conditions~\cite{Krasnitz:1998ns,Lappi:2003bi,Lappi:2006fp}. Modern implementations such as IP-Glasma additionally incorporate event-by-event geometry fluctuations and impact-parameter dependent small-$x$ structure~\cite{Schenke:2012wb,Schenke:2012hg}. 

It is also at these earliest timescales that hard partonic scatterings -- the source of the high transverse-momentum probes such as jets, central to this thesis -- take place. Their characteristic formation time scales as $\tau \sim 1/Q$, so that for sufficiently hard processes with $Q \gg Q_s$ they occur earlier than most of the early-time classical field dynamics, whereas for more moderate hard scales they can overlap with this stage.

\paragraph{Pre-equilibrium evolution and \emph{hydrodynamisation} ($0 < \tau \lesssim 1$~fm/$c$).}

Following the initial overlap, the Glasma fields decohere and their large occupancies decrease as the system expands on a timescale of $Q_s^{-1}$, leading to a pre-equilibrium stage more naturally described in terms of quark and gluon quasiparticle excitations. The medium remains far from equilibrium and exhibits strongly anisotropic momentum distributions inherited from the rapid longitudinal expansion. At sufficiently weak coupling, the evolution of this stage can be described at leading order in $\alpha_s$ using QCD effective kinetic theory (EKT), introduced by Arnold, Moore and Yaffe~\cite{Arnold:2002zm} and developed further in subsequent studies (see e.g.~\cite{Kurkela:2018vqr}). In this description, quark and gluon quasiparticles undergo elastic scatterings ($2\to2$) and inelastic splittings ($2\to 3$) which are essential for redistributing momentum and driving isotropisation of an expanding background. Modern numerical implementations embed such descriptions into realistic event-by-event simulations with fluctuating initial conditions before matching to subsequent hydrodynamic evolution~\cite{Kurkela:2018wud,Keegan:2016cpi}. These approaches provide a quantitatively controlled description of entropy production and of the gradual decrease of the initial pressure anisotropy.

Modern studies based on EKT and pre-equilibrium matching to hydrodynamics have shown that viscous hydrodynamics can become applicable before full isotropisation or local thermal equilibrium is reached~\cite{Kurkela:2018vqr,Kurkela:2018wud,Keegan:2016cpi}. More precisely, the energy--momentum tensor approaches the constitutive form of viscous hydrodynamics on timescales of order $\tau \sim 1$ fm/c at LHC energies~\cite{Kurkela:2015qoa,Kurkela:2018wud}. Such an early onset of fluid behaviour, referred to as \emph{hydrodynamisation}, is non-trivial, since it can occur while sizeable momentum-space anisotropies persist and before local thermal equilibrium is reached. It is closely connected to the concept of a hydrodynamic attractor, whereby systems starting from different far-from-equilibrium initial conditions evolve toward a universal trajectory that smoothly connects to viscous hydrodynamic behaviour long before complete equilibration is achieved~\cite{Romatschke:2017vte,Berges:2013lsa}.

This regime thus interpolates between the early classical-field dynamics and the later near-equilibrium quark-gluon plasma, providing a framework in which both the emergence of collective behaviour and the approach to thermal equilibrium\footnote{More broadly, understanding thermalisation in non-Abelian gauge theories remains a central problem in real-time many-body QFT. For a pedagogical introduction to this topic, see~\cite{Berges:2004yj}.} can be studied within controlled weak-coupling descriptions of QCD~\cite{Schlichting:2019abc}.

\paragraph{Hydrodynamic QGP phase and collective expansion ($1 \lesssim \tau \lesssim 10$ fm/c).}
After hydrodynamisation, the subsequent evolution of the bulk medium is described well by relativistic viscous hydrodynamics. In this regime, the system is described in terms of locally defined energy density, pressure, and fluid velocity fields, with an equation of state based on lattice QCD calculations at small baryon chemical potential~\cite{HotQCD:2014kol,Borsanyi:2013bia}. Pressure gradients drive rapid longitudinal and transverse expansion, converting the internal energy into collective motion and cooling the medium from temperatures of order a few hundred MeV toward the crossover region near $T_c \simeq 155$ MeV.

\begin{figure}[t]
  \centering
  \includegraphics[width=0.48\textwidth]{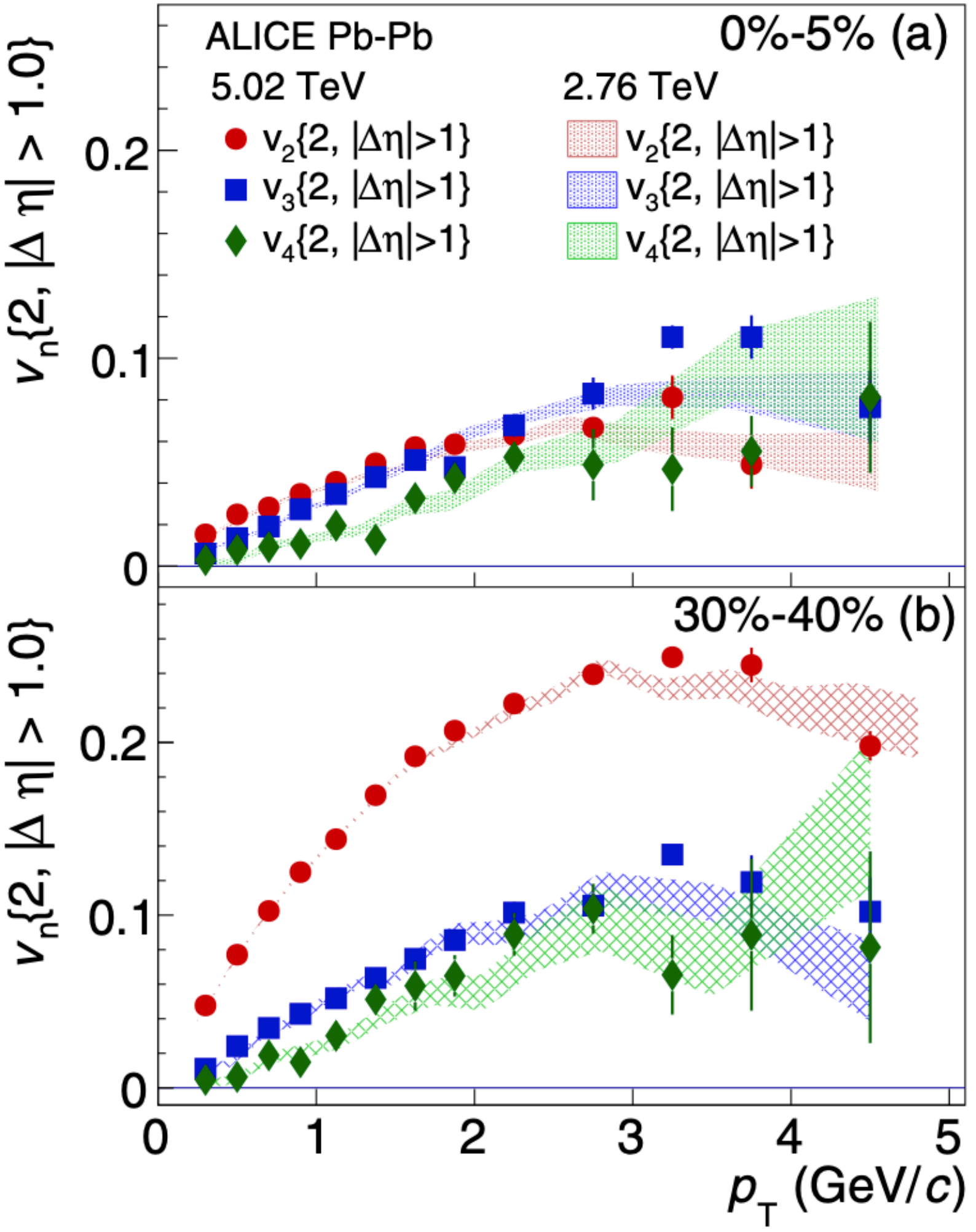}
  %{figs/intro/v2_ATLAS.pdf}
  \includegraphics[width=0.48\textwidth]{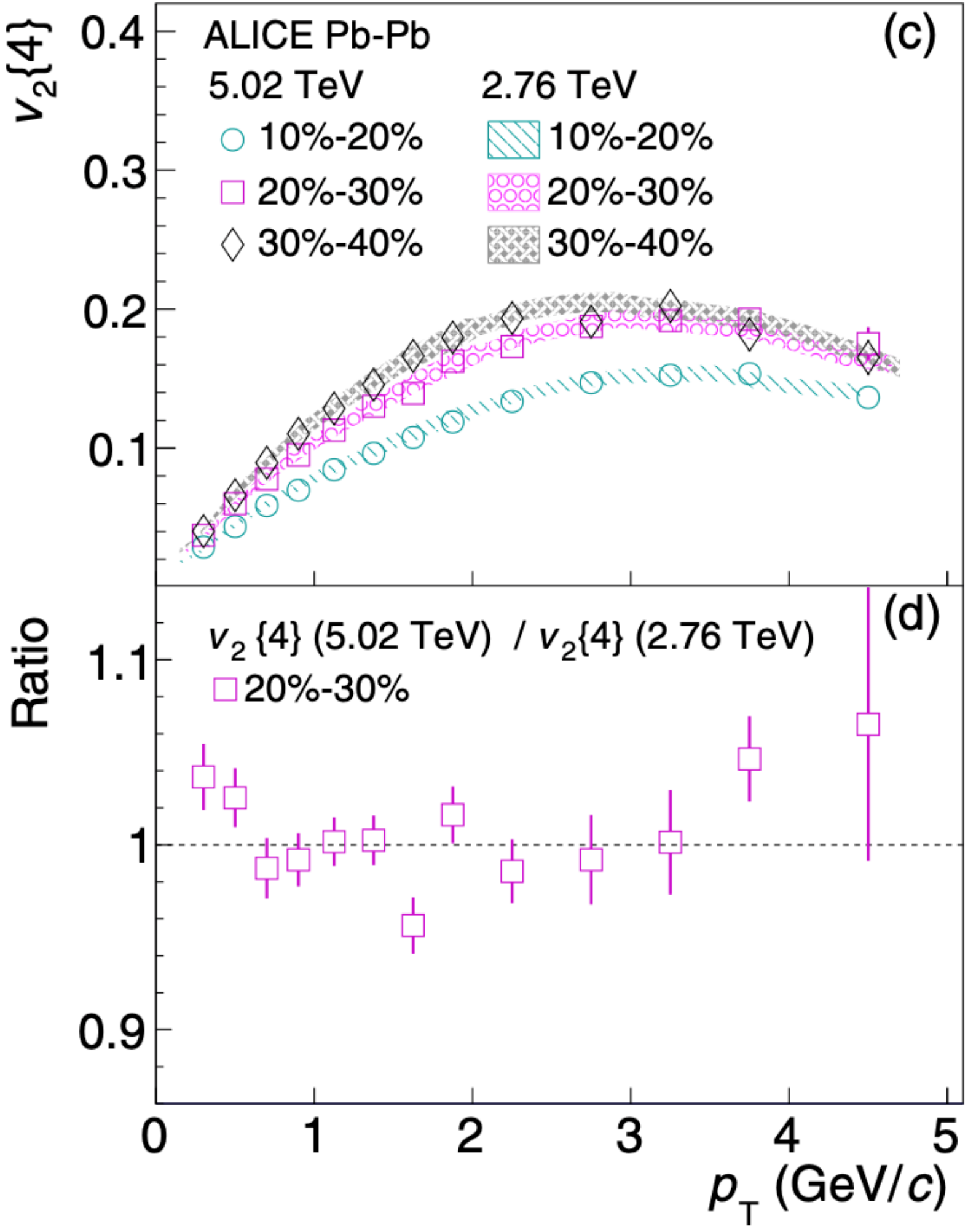}%{figs/intro/v3_ATLAS.pdf}
  \caption{Anisotropic flow coefficients for charged particles, as a function of their transverse momentum, in PbPb collisions at $\sqrt{s_{\rm NN }}=5.02$ TeV, compared with results at 2.76 TeV. The left panels shows $v_2$, $v_3$ and $v_4$ computed using two-particle correlations, for two centrality classes and the right panels shows $v_2$ computed using four-particle correlations. Using correlation between a higher number of particles aims at reducing contributions not related to collective flow. Adapted from the ALICE analysis in ~\cite{ALICE:2016ccg}.}
  \label{fig:v2_v3}
\end{figure}

This hot deconfined phase is identified as a QGP, discussed in Section~\ref{subsec:phase_diagram}. Since the lifetime and size of this  system are both microscopic, it cannot be observed directly and needs to be inferred from several complementary final-state observables sensitive to its properties:

\noindent\emph{\underline{Collective flow:}} A central signature of the QGP is the observation of large azimuthal anisotropies in the momentum distribution of produced hadrons. These are conventionally characterised through the Fourier decomposition
\begin{align}
  \frac{dN}{d\phi} \propto 1 + 2\sum_{n=1}^{\infty} v_n \cos\left[n(\phi-\Psi_n)\right],
\end{align}
where $\phi$ is the azimuthal angle, $\Psi_n$ is the event-by-event orientation angle of the $n$-th harmonic anisotropy (the event plane), and $v_n$ are the corresponding flow coefficients. In the absence of substantial rescattering, the initial spatial anisotropy of the overlap region would not be efficiently converted into final-state momentum anisotropy. In non-central collisions, the dominant contribution is typically the elliptic flow coefficient $v_2$, generated by the almond-shaped geometry of the nuclear overlap, while event-by-event density fluctuations generate higher harmonics such as $v_3$ and $v_4$ (see Fig.~\ref{fig:v2_v3} for a recent measurement of $v_n$). The large flow coefficients, first measured at RHIC in ~\cite{STAR:2000ybh} and at the LHC in~\cite{ALICE:2010suc}, are well described by viscous hydrodynamic simulations and indicate a rapidly equilibrating, strongly collective medium with a small specific shear viscosity $\eta/s$~\cite{Romatschke:2017ejr}.
This led to classification of the QGP as a \emph{near-perfect fluid}. Phenomenological extractions generally favour values of order $\eta/s \sim 0.08$--$0.20$~\cite{Bernhard:2019bmu,Parkkila:2021yha}, not far above the conjectured lower bound $\eta/s = 1/(4\pi)$ obtained in strongly coupled gauge theories with holographic dual descriptions (AdS/CFT)~\cite{Kovtun:2004de}.

\noindent\emph{\underline{Strangeness enhancement:}} Strange hadron production is enhanced~\cite{Koch:1986ud,WA97:2002} in nucleus--nucleus collisions relative to appropriately scaled proton--proton collisions, with the effect increasing with the strange\-ness content of the hadron~\cite{ALICE:2017jyt}. In a hot deconfined medium, strange quarks can be produced efficiently through partonic scattering processes, e.g., $gg\to s\bar s$ since $T> m_s$, so that their abundance approaches near-equilibrium values during the lifetime of the fireball. Enhanced yields of $K$, $\Lambda$, $\Xi$, and $\Omega$ hadrons are consistent with this picture~\cite{ALICE:2017jyt}.

\noindent\emph{\underline{Quarkonium suppression:}} Heavy quark--antiquark bound states provide a sensitive probe of deconfinement~\cite{Matsui:1986dk}. In a colour-screened medium, the binding potential between heavy quarks is reduced, suppressing the survival probability of states such as $J/\psi$ and the $\Upsilon$ family relative to proton--proton collisions scaled by the number of binary nucleon collisions. Because different states have different binding energies, a sequential suppression pattern is expected, with more weakly bound states dissociating at lower temperatures. Measurements at the LHC, particularly in the bottomonium ($b\bar b$) sector, where late-stage recombination is expected to be less important, are broadly consistent with this picture~\cite{CMS:2018zza}.

\begin{figure}[t]
  \centering
  \includegraphics[width=0.46\textwidth]{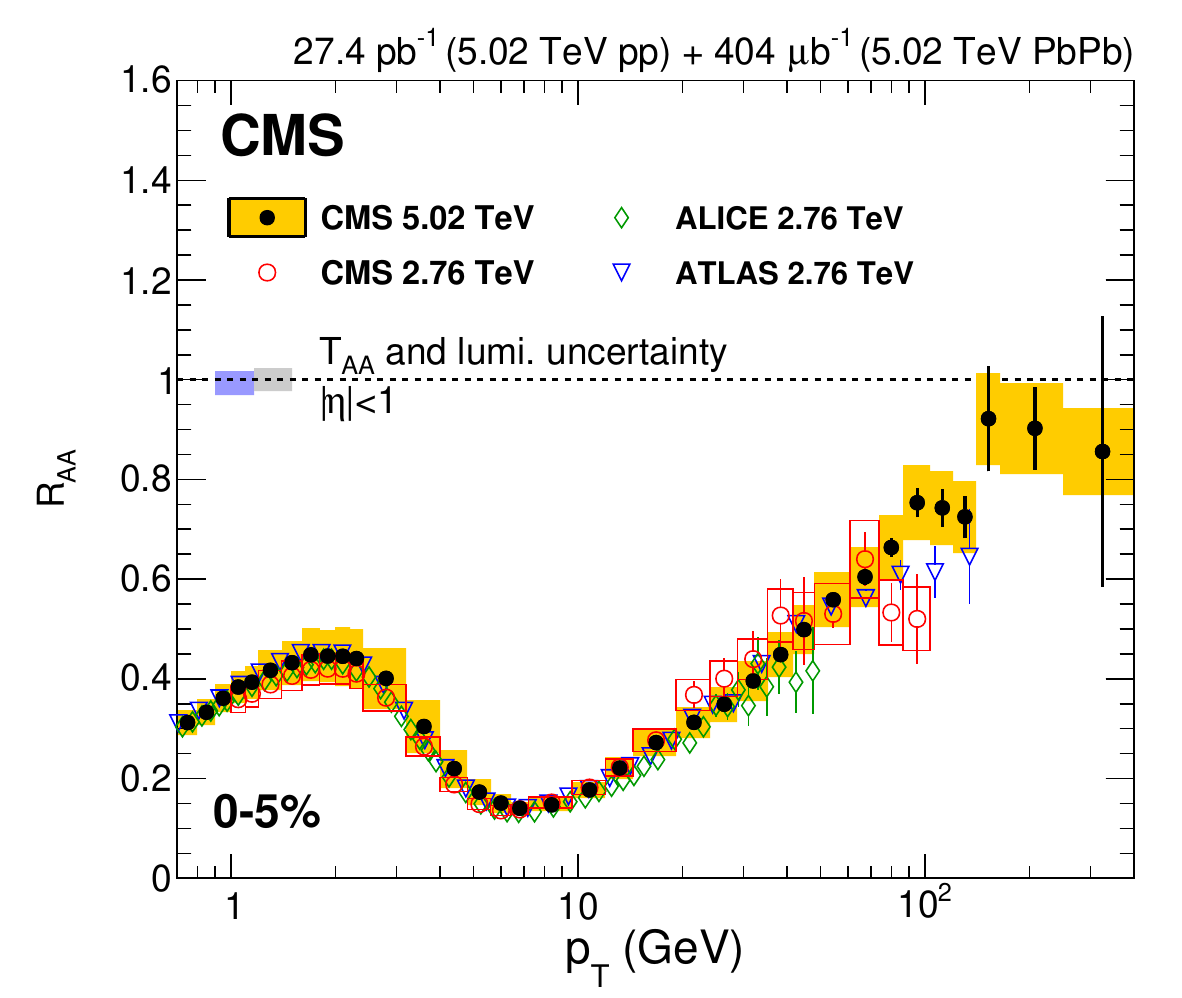}
  \includegraphics[width=0.53\textwidth]{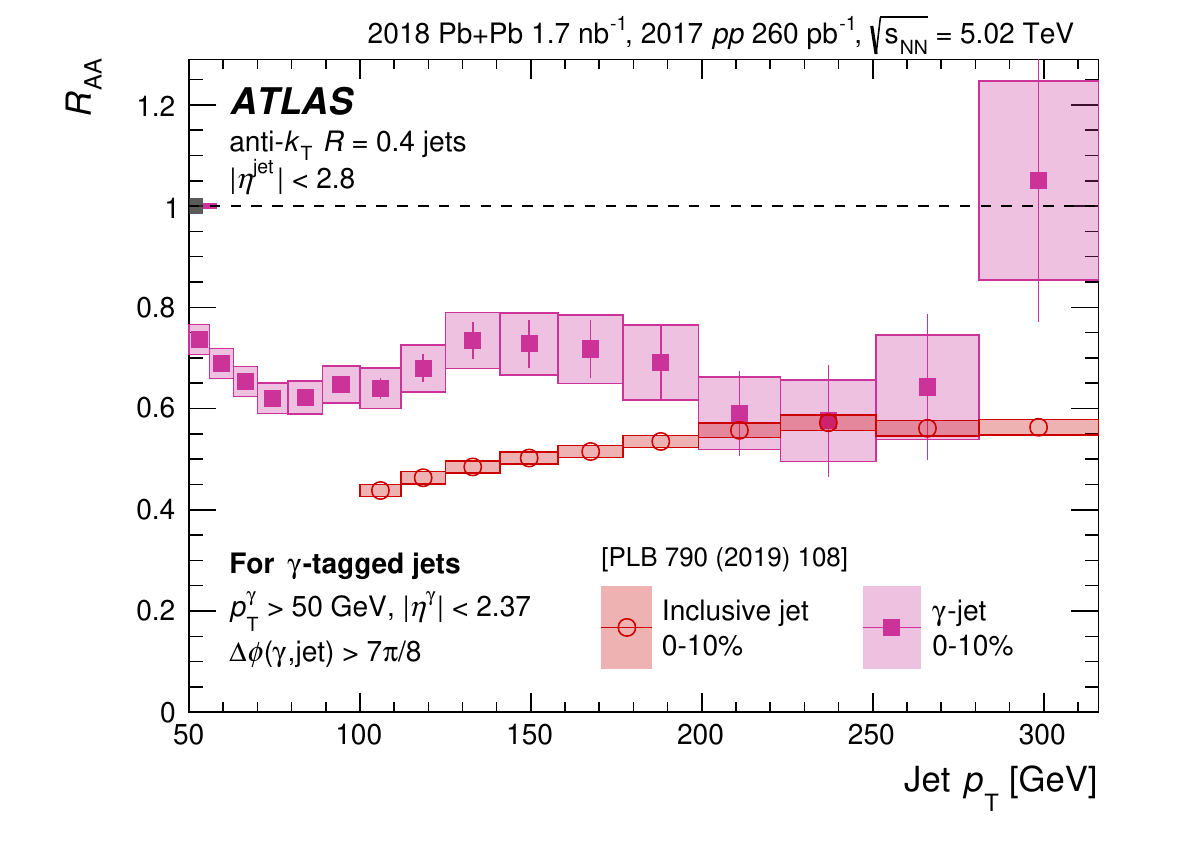}
  \caption{Nuclear modification factor $R_{AA}$ for charged-hadrons (left, adapted from the CMS analysis in~\cite{CMS:2016xef}) and jets (right, adapted from the ATLAS analysis in~\cite{ATLAS:2023iad}) PbPb collisions. The left panel combines results for both $\sqrt{s_{NN}} = 2.76$ TeV and $5.02$ TeV, for various experiments and in the 0-5\% centrality class. The right panel shows results for inclusive jets superimposed with $\gamma+$jet results (events where a jet is produced back-to-back with a photon), in the 0-10\% centrality class and for $\sqrt{s_{NN}} = 5.02$ TeV.}
  \label{fig:exp_RAA}
\end{figure}

\noindent\emph{\underline{High-$p_T$ hadron and jet suppression:}} Energetic partons produced in initial hard scatterings subsequently traverse the hot and dense medium created in the collision. Their interaction with the deconfined plasma leads to a suppression of high-$p_T$ hadron and jet - a collimated bunch of high energy hadrons - yields relative to proton--proton expectations. A standard measure is the nuclear modification factor
\begin{align}\label{eq:intro_RAA}
    R_{AA}(p_T)=\frac{1}{\langle N_{\rm coll}\rangle}
\frac{dN_{AA}/dp_T d\eta}{dN_{pp}/dp_T d\eta},
\end{align}
which compares heavy-ion yields to the proton-proton baseline scaled by the number of binary nucleon-nucleon collisions in the HIC. At RHIC, values $R_{AA}\simeq0.2$ were first observed for neutral pions at $p_T\sim5$~GeV in central AuAu collisions~\cite{PHENIX:2004vcz} (see Fig.~\ref{fig:exp_RAA} for a recent measurement of charged hadron $R_{AA}$). At the LHC, convincing evidence for jet quenching was first observed in~\cite{CMS:2011iwn}, in the form of the transverse momentum imbalance between two jets produced back-to-back in azimuth (\emph{dijet asymmetry}). Subsequent measurements show strong suppression over a wide kinematic range extending to several hundred GeV, with the effect strongest in the most central collisions and gradually reduced in more peripheral events~\cite{CMS:2021vui,ALICE:2023waz, ATLAS:2023iad} (see Fig.~\ref{fig:exp_RAA} for a recent measurement of jet $R_{AA}$). In the right panel of Fig.~\ref{fig:exp_RAA}, the rise of $R_{AA}$ at very high $p_T$ reflects both the energy dependence of in-medium interactions and the steeply falling production spectrum. The modifications imprinted by the produced hot matter on jets, central to this thesis, go beyond suppression. These medium-induced modifications, generally termed \emph{jet quenching}, will be discussed in detail in the following section.

~\\ 

Finally, it is worth stating that the theoretical descriptions of the QGP combine several complementary approaches. Lattice QCD determines equilibrium thermodynamic properties such as the equation of state and fluctuations of conserved charges. Relativistic viscous hydrodynamics provides a description of the collective expansion at macroscopic scales. Perturbative thermal field theory and kinetic approaches are used to estimate quantities associated with sufficiently hard momentum scales or quasiparticle transport, including scattering rates and transport coefficients. Gauge/gravity duality, while not a description of QCD itself, provides useful qualitative insights into strongly coupled QGP dynamics, including the small-viscosity limit discussed above.

\paragraph{Hadronisation and freeze-out ($\tau \sim 10$~fm/$c$).}
As the system expands and cools to $T=T_c$, it undergoes a crossover transition from the QGP to a hadronic phase. Inelastic interactions cease first, fixing the relative abundances of different hadron species at \emph{chemical freeze-out}, while elastic scatterings continue to redistribute momenta. At later times, when the mean free path becomes comparable to the characteristic size of the resulting hadronic gas, elastic interactions also cease, leading to \emph{kinetic freeze-out}. The final-state hadron yields and momentum distributions thus encode information about both the collective expansion of the system and the conditions at the kinetic freeze-out.

% ==============================================================
\section{High-energy probes of QCD matter}
\label{sec:hard_probes}
% ==============================================================

\subsection{Hard processes and factorisation}
\label{subsec:factorisation}

The previous section framed a heavy-ion collision as a time-dependent many-body QCD system, evolving from non-equilibrium early dynamics through a deconfined plasma phase and culminating in hadronic freeze-out. Alongside this bulk evolution, there are rare short-distance scatterings with momentum transfers $Q$ far exceeding characteristic medium scales, such as the temperature $T$, which produce \emph{hard probes}. In this context, a hard probe denotes a final state generated in a process calculable in perturbative QCD, i.e., $Q \gg \Lambda_{\rm QCD}$. This separation of scales implies that the scattering  process is largely insensitive to both long-distance confining physics and to the evolution of the underlying matter. In fact, the associated production time of the probe is parametrically set by the inverse hard scale, $\Delta t \sim 1/Q$, which for sufficiently large $Q$ is much shorter than the characteristic time over which the bulk medium approaches hydrodynamic behaviour, $\tau_{\rm hydro}\sim 1$ fm/c (see Section~\ref{subsec:st_evolution_HIC}). Thus, the resulting energetic probe is created at early times and subsequently propagates through the evolving matter, retaining information both about its perturbative production and the imprints of interactions with the matter. For this reason, hard probes provide valuable constraints on the microscopic properties of evolving HIC matter and their time evolution.

The production of hard probes can be formulated systematically within QCD through factorisation theorems, which we briefly introduced in Section~\ref{subsec:DIS_DGLAP}. The hierarchy $Q \gg \Lambda_{\rm QCD}$ renders the underlying short-distance scattering perturbative and provides the scale separation on which such descriptions rely. For suitable observables, the leading contribution can then be organised into perturbatively calculable short-distance coefficients and functions encoding long-distance QCD dynamics, with residual terms suppressed by powers of $1/Q$. For sufficiently inclusive observables that do not resolve additional soft or collinear scales at leading power, this organisation reduces to the standard collinear factorisation. As a simple example, for the production of a sufficiently inclusive final state $F$ characterised by a hard scale $Q$ in collisions of nuclei $A$ and $B$, the leading-power cross section may be written schematically as~\cite{Collins:1989gx,Campbell:2017}
\begin{align}\label{eq:factorisation_hard_probe}
    d\sigma_{AB\to F}
    =
    \sum_{a,b}\int dx_a\,dx_b\,
    f_{a/A}(x_a,\mu_F^2)\,
    f_{b/B}(x_b,\mu_F^2)\,
    d\hat{\sigma}_{ab\to F}(x_a,x_b,Q,\mu_F,\mu_R)
    +
    \mathcal{O}\!\left(\frac{\Lambda_{\rm QCD}^2}{Q^2}\right)\,,
\end{align}
where the sum runs over incoming parton species $a$ and $b$. The functions $f_{i/A}(x_i,\mu_F^2)$ denote the nuclear parton distribution functions (nPDFs) of the incoming nucleus, evaluated at momentum fraction $x_i$ and factorisation scale $\mu_F$, which we introduced in Section~\ref{subsec:DIS_DGLAP}. The partonic coefficient $d\hat{\sigma}_{ab\to F}$ describes the hard subprocess producing the final state $F$ and is calculable perturbatively, with $\mu_R$ the renormalisation scale. The final term represents contributions beyond leading power.
\begin{figure}[t]
    \centering
    \includegraphics[width=0.7\linewidth]{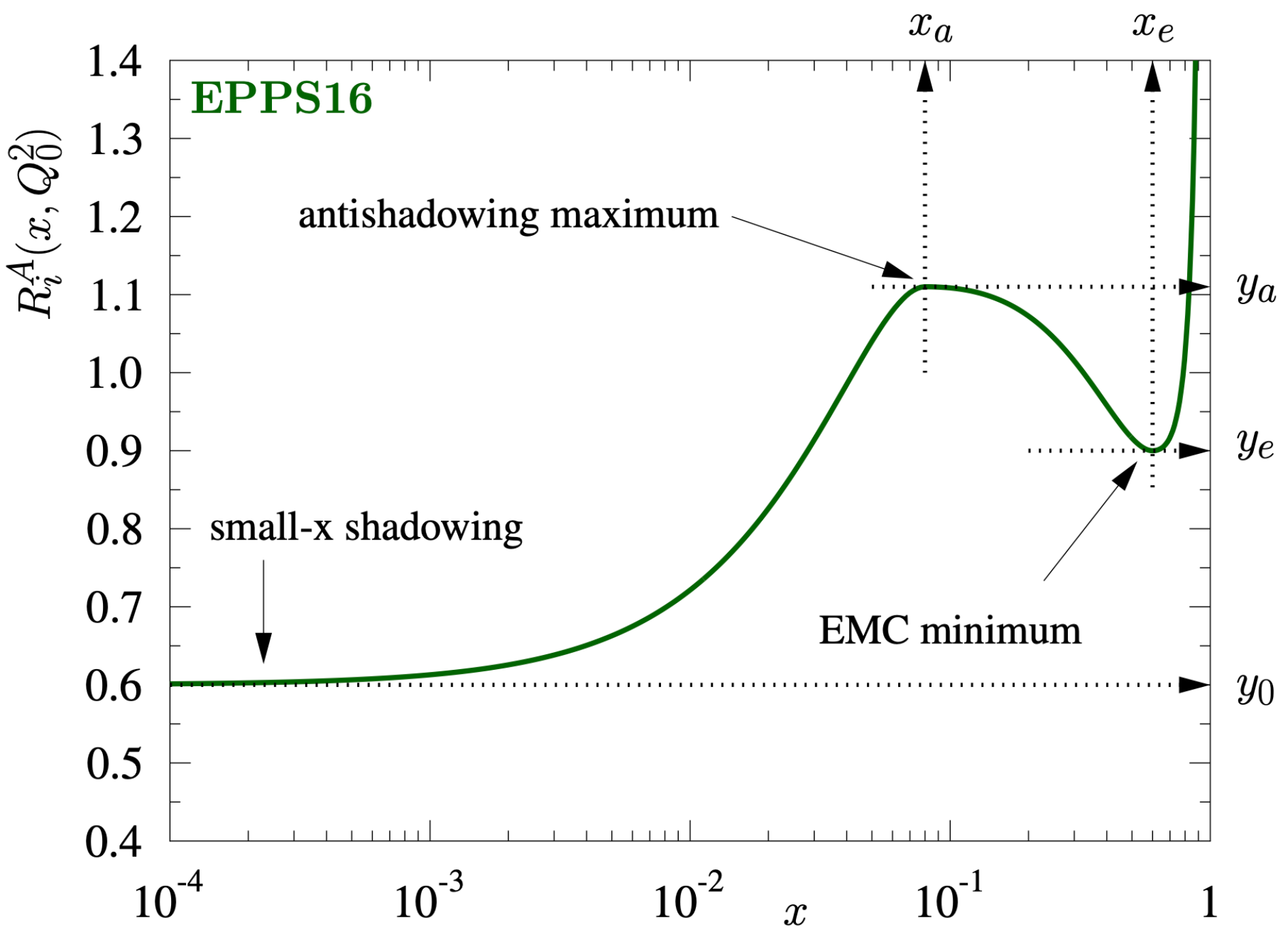}
    \caption{Schematic form of the nuclear modification factor $R_i^A(x,Q_0^2)$ used in the EPPS16 parametrisation~\cite{Eskola:2016oht}, illustrating the characteristic small-$x$ shadowing, antishadowing, EMC-suppression and Fermi-motion regions. Adapted from~\cite{Eskola:2016oht}.}
    \label{fig:nPDF_illustration}
\end{figure}

The nPDFs generalise the proton PDFs introduced previously to nucleons bound inside nuclei. In general, the partonic structure of a nucleus is not equal to a simple incoherent sum of free proton and neutron PDFs, but is modified by the binding nuclear environment. These modifications depend on Bjorken-$x$, parton flavour, and the resolution scale $Q$, and are commonly organised into shadowing (suppression at small $x$), anti-shadowing (moderate enhancement at intermediate $x$), EMC suppression in the region $x\sim0.3$--$0.7$, and Fermi-motion broadening as $x\to1$~\cite{Eskola:2016oht}. Shadowing, for example, reflects the coherent overlap of small-$x$ partons in neighbouring nucleons. This set of modifications is illustrated in Fig.~\ref{fig:nPDF_illustration}. Since final-state kinematics depends on the momentum fractions sampled in the hard scattering, these modifications translate into characteristic changes of production rates and differential spectra relative to the proton-proton baseline. Thus, understanding how hard-probe spectra are modified by nPDFs and other cold nuclear matter (CNM) effects, such as isospin differences (proton/neutron content) and initial-state multiple scattering, is essential for isolating genuine final-state interactions with the hot medium. Proton-nucleus collisions are particularly valuable in this regard, since they retain nuclear initial-state structure while largely avoiding extended medium effects. Comparisons between pp, pA, and AA systems therefore provide a systematic framework for isolating initial-state nuclear effects and inferring modifications generated by the hot medium.

Now, note that Eq.~\eqref{eq:factorisation_hard_probe} should be understood as describing hard probe production, rather than as a universal expression for observables. In general, the factorisation formula for a given observable depends on its sensitivity to additional scales or to the details of the final state. For instance, predictions for identified hadron spectra generally require additional \emph{fragmentation functions}, while observables based on reconstructed jets may involve further \emph{jet} or \emph{soft functions}. The predictive power of such factorised descriptions relies on the universality of the long-distance functions within a given class of processes or observables, so that once constrained in one case they can be used in multiple related calculations. This has been realised for a broad class of sufficiently inclusive hard observables in pp and pA collisions, and for electroweak boson production in AA collisions, where nuclear dependence is encoded in the nPDFs as in Eq.~\eqref{eq:factorisation_hard_probe}. However, for observables that are sensitive to the propagation of the hard probe through the evolving matter, no comparably general factorisation theorem has been established. While the initial hard production is well described by Eq.~\eqref{eq:factorisation_hard_probe}, the subsequent in-medium evolution introduces additional scales associated with multiple scattering, energy exchange, and medium dynamics. Thus, the description of such observables typically combines a perturbative production baseline with effective or model-based descriptions of in-medium propagation, adapted to the process and kinematic regime under consideration.

\subsection{Classes of hard probes}
\label{subsec:probe_classes}

Different hard probes couple to different aspects of the medium, with the relevant hard scales spanning from a few GeV to the TeV regime. Electroweak bosons and prompt photons are insensitive to final-state colour interactions, thus constraining the initial hard production rate. Heavy quarks provide a distinct probe through the additional scale set by their masses, modifying their production and subsequent dynamics. Quarkonium states probe the thermal and colour-screening properties of the medium through their in-medium dissociation and regeneration. High-$p_T$ hadrons and reconstructed jets arise from highly energetic coloured partons produced in early hard scatterings and probe how such partons radiate energy and redistribute momentum while traversing the medium. Taken together, these observables provide complementary constraints on the properties of the matter.

\underline{\emph{Electroweak bosons and prompt photons}}: Once produced in a short-distance scattering, isolated photons and the weak bosons $W^\pm$ and $Z$ do not carry colour charge and therefore traverse the medium without final-state interaction. Their yields in nucleus--nucleus collisions, after accounting for isospin and nuclear PDF effects~\cite{Eskola:2016oht}, are found to be consistent with binary-collision scaling, $R_{AA}^{\gamma,Z,W}\simeq1$ within uncertainties~\cite{ATLAS:2019xlu,ALICE:2024vbe}. These channels therefore provide important baselines for hard-process normalisation. In addition, photon--jet and $Z$--jet events offer calibrated recoil configurations in which the boson momentum constrains the initial hard scale, while the recoiling coloured parton remains sensitive to subsequent medium effects.

\underline{\emph{High-$p_T$ hadrons}}: They predominantly originate from leading fragments of energetic partons and therefore provide an indirect measure of partonic energy loss, integrated over much of the accompanying final-state structure. The strong suppression of inclusive high-$p_T$ hadron yields in nucleus--nucleus collisions, first established at RHIC and later extended to LHC energies, demonstrated that energetic colour charges lose substantial energy while traversing the hot medium~\cite{PHENIX:2004vcz,CMS:2016xef}. Owing to their experimental simplicity and broad kinematic reach, inclusive hadron spectra remain an important constraint on the path-length and partonic flavour dependence of quenching. In the former, hard partons are produced at different positions in the nuclear overlap region and leave the medium along different directions, so the in-medium distance travelled varies event by event and influences the observed suppression. The latter arises due to gluons, light quarks and heavy quarks, whose fragmentation will contribute to different hadronic species, coupling differently to the medium.

\underline{\emph{Open heavy flavour}}: Charm and bottom quarks are produced predominantly in the earliest hard scatterings ($m_Q \gg \Lambda_{\rm QCD}$), while thermal production is not significant ($m_Q \gg T$). Their production and fragmentation at high momentum can be described within perturbative QCD, while their subsequent propagation through the medium is sensitive to both collisional transport and radiative energy loss. At moderate momentum, transport approaches based on drag and diffusion are often employed (see e.g.~\cite{Svetitsky:1987gq,Dong:2019byy}). At higher energies, the dynamics increasingly resembles that of energetic partons undergoing quenching, modified by finite-mass effects (see e.g.~\cite{Armesto:2003jh,Cao:2018ews}). In particular, gluon emission at angles $\theta \lesssim m_Q/E$ is suppressed relative to the massless case, the so-called \emph{dead-cone effect}~\cite{Dokshitzer:2001zm}. Mass-dependent quenching effects are broadly consistent with the tendency for beauty observables to be less suppressed than charm or light-flavour probes in comparable kinematic regimes, although no universal ordering applies across all observables. The dead-cone effect itself has been directly observed in charm-tagged jets, i.e. jets containing a reconstructed charm hadron such as a $D^0$ meson ($c\bar u$) among their constituents, in proton--proton collisions by ALICE~\cite{ALICE:2021aqk}, while analogous measurements in heavy-ion collisions remain an active experimental effort.

\underline{\emph{Quarkonia}}: These are heavy $Q\bar Q$ bound states and they interact with the underlying matter in such a way that they probe deconfinement. Colour screening weakens the confining potential in the plasma and can suppress bound-state formation or survival, while interactions with the medium may induce dissociation. Since more weakly bound states are expected to melt at lower temperatures, sequential suppression patterns provide sensitivity to the thermal scales of the medium, consistent for example with the observed hierarchy of bottomonium suppression in PbPb collisions~\cite{CMS:2018zza}. At the same time, recombination of independently produced heavy quarks can partially regenerate quarkonia, particularly for charmonium at LHC energies where charm production is abundant. Quarkonium phenomenology therefore reflects the interplay of screening, dissociation, and regeneration rather than a single mechanism~\cite{Rothkopf:2019ipj}.

\underline{\emph{Hadronic jets}}: Jets are collimated sprays of hadrons produced from hard scattered quarks or gluons and constitute the most differential hard probes currently available. Following their production, these partons undergo successive QCD branchings ($q\to qg$, $g\to q\bar q$, $g\to gg$) and eventually hadronise, yielding correlated final states that are reconstructed experimentally by suitable jet clustering algorithms. Unlike single inclusive hadrons, which predominantly reflect the leading fragment of this evolution, jets retain partial information about the angular structure and momentum flow associated with the underlying branching cascade. Their modification in heavy-ion collisions can therefore be studied not only through inclusive suppression, but also through momentum imbalance, angular broadening, fragmentation patterns, and internal substructure. These modifications are collectively termed \emph{jet quenching}. Because they combine high production rates with multi-scale sensitivity to in-medium dynamics, jets have become a central tool for precision studies of QCD matter. Since jet quenching is a core theme of this thesis, its phenomenology is developed in the following section, while its theoretical description is discussed at length in Chapter~\ref{chap:jet_quenching}.

\subsection{Heavy-ion jet phenomenology}
\label{subsec:jets_intro}

Jets provided the first final-state manifestation of energetic quarks and gluons in high-energy collisions. In $e^+e^-$ annihilation, the observation of hadrons emerging in two approximately back-to-back collimated sprays~\cite{Hanson:1975fe} showed that the primary quark and antiquark produced in the hard process ($e^+e^-\to q\bar q$) materialise experimentally as clustered hadronic energy flow after hadronisation. The subsequent discovery of three-jet topologies at PETRA, interpreted as $e^+e^- \to q\bar q g$, furnished the first direct evidence for hard gluon radiation and for the gluon as a propagating QCD degree of freedom~\cite{Brandelik:1979bd}. In hadronic collisions at the Tevatron and LHC, inclusive jet and multijet production evolved into precision tests of QCD, constraining parton distributions, the running coupling and higher-order perturbative predictions over a wide kinematic range.

From the theory side, Sterman and Weinberg provided the first perturbative QCD treatment of jets by defining a two-jet cross section in which most of the event energy, a fraction $1-\epsilon$, is required to lie within two back-to-back cones of half-angle $\delta$~\cite{Sterman:1977wj}. The need for such a definition arises because, beyond leading order, the probability for additional gluon emission becomes arbitrarily large when the gluon is emitted with vanishing energy or at vanishing angle relative to an energetic parton, e.g., $e^+ e^- \to q\bar q g$. By allowing unresolved soft radiation and collinear splittings without changing the event classification, the Sterman-Weinberg construction yields a finite perturbative two-jet cross section after the cancellation of singularities from real and virtual (loop) diagrams. This provided one of the first \emph{infrared and collinear safe} (IRC safe) observables in perturbative QCD. 

In modern collider phenomenology, jets are operationally defined through a specific reconstruction procedure on the measured final-state particles, and their theoretical description is rooted in the enhancement of QCD radiation in soft and collinear regions of phase space. Concretely, after the hard production of an energetic quark or gluon, successive branchings generate a cascade that gradually lowers the characteristic virtuality from the initial hard scale to non-perturbative scales associated with hadronisation. The resulting correlated hadrons are then clustered through a \emph{jet algorithm} into composite objects, identified as jets, which serve as proxies for the underlying hard-scattered parton and its fragmentation pattern. As mentioned for the Sterman-Weinberg construction, for meaningful comparison with perturbative QCD, this jet algorithm should be IRC safe, meaning that the identified jets should be unchanged by the addition of arbitrarily soft or collinear particles. Technically, IRC safety is required to ensure that fixed-order predictions remain finite and that soft/collinear enhanced higher-order corrections can be systematically organised. We shall address this in detail in Section~\ref{sec:jet_evol_vac}. Modern collider analyses predominantly employ sequential recombination algorithms, the most standard being the generalized-$k_t$ family~\cite{Catani:1993hr,Dokshitzer:1997in,Wobisch:1998wt,Cacciari:2008gp}. These algorithms cluster a given set of particles according to the distance metrics
\begin{align}\label{eq:jet_distance_metric}
d_{ij} = \min\!\left(p_{T,i}^{2p},p_{T,j}^{2p}\right)\frac{\Delta R_{ij}^2}{R^2},
\qquad
d_{iB}=p_{T,i}^{-2},
\end{align}
where $\Delta R_{ij}^2=(\Delta\eta_{ij})^2+(\Delta\phi_{ij})^2$ is the rapidity-azimuth collider distance and $R$ is the \emph{jet radius} parameter. The azimuthal angle is defined in the plane transverse to the beamline and rapidity is related to the angle to the beamline in the case of massless particles as $\eta = -\ln\tan\left(\frac{\theta}{2}\right)$. At each step of clustering, the smallest value among all pairwise distances $d_{ij}$ and beam distances $d_{iB}$ is identified. If a $d_{ij}$ is smallest, particles $i$ and $j$ are recombined, otherwise if a $d_{iB}$ is smallest, particle $i$ is declared a jet and removed from the list. The procedure is repeated until no particles remain. In HICs, the standard choice is the anti-$k_t$ algorithm~\cite{Cacciari:2008gp}, which fixes $p=-1$. Inspecting Eq.~\eqref{eq:jet_distance_metric} for this value of $p$, this implies that soft particles are preferentially absorbed into nearby hard cores, producing stable, approximately conical boundaries that are particularly useful in the high-multiplicity environment of heavy-ion collisions, where the uncorrelated soft background can be subtracted in a controlled way. The $k_t$ ($p=1$) and Cambridge/Aachen ($p=0$) algorithms are also widely used, particularly in studies of the fragmentation pattern of jets, i.e., \emph{jet substructure}, where their clustering histories are informative~\cite{Catani:1993hr,Dokshitzer:1997in,Wobisch:1998wt}. The jet radius controls the angular reach of the jet definition. A larger $R$ captures more perturbative radiation and, in the case of HICs, more of the energy flow correlated with the jet and transported to wider angles, while a smaller $R$ reduces contamination from the underlying event at the cost of increased loss of information on wide-angle jet structure.

~\\

In HICs, jet reconstruction is complicated by the soft background generated by the underlying bulk medium. A large multiplicity of low-momentum particles unrelated to the primary hard scattering overlaps with the jet signal, requiring dedicated background subtraction to isolate genuine correlations with the jet, i.e., \emph{medium response}. At the same time, it is often useful to engineer observables which focus on specific phase space regions of a jet by suppressing others, e.g., so called \emph{grooming} observables which suppress soft, typically wide-angle radiation. Thus, heavy-ion jet observables are sensitive not only to the microscopic modifications of the energetic parton cascade and how energy is redistributed, but also to the reconstruction choices such as constituent thresholds, subtraction procedures, grooming parameters, and the jet radius $R$. For instance, energy transported to large angles or degraded to sufficiently soft momenta may fall outside the jet definition, while part of the medium response can be clustered back into the measured object. Consequently, quantitative interpretation requires disentangling genuine modifications of the partonic cascade, the jet-correlated part of the background and reconstruction effects. These complications are compensated by the exceptional information content of jets. Because they retain differential information on energy flow across multiple angular and momentum scales, jet measurements probe several qualitatively distinct aspects of the interaction between the energetic QCD cascade and the produced matter in HICs. The interested reader is pointed to, e.g.,~\cite{Connors:2017ptx, Cao:2020wlm, Apolinario:2022vzg} for recent, detailed reviews on heavy-ion jet phenomenology. Below we summarize the main classes of jet quenching observables, noting that the boundary between them is not sharp and that many observables are sensitive to several aspects of medium-induced modifications on jets.

\paragraph{Inclusive suppression and energy loss.}

The most inclusive measurements are jet spectra and the corresponding nuclear modification factor $R_{AA}$, introduced in Eq.~\eqref{eq:intro_RAA}. They compare reconstructed jet yields in nucleus-nucleus collisions with binary-scaled proton-proton baselines and therefore quantify the net depletion of jets passing a given kinematic selection. Physically, this suppression is sensitive to the amount of energy transported outside the chosen jet reconstructed boundary, shifted below constituent thresholds, or redistributed into very soft particles that thermalise with the medium. Since the production spectrum falls steeply with transverse momentum, even moderate energy shifts can induce sizeable suppressions observable in $R_{AA}$. Inclusive observables are also sensitive to geometry and fragmentation -- jets created near the surface of the medium, propagating along shorter in-medium paths, or with a narrow fragmentation pattern are more likely to satisfy a fixed final-state selection than jets traversing longer distances or fragmenting more broadly.

Although inclusive suppression does not isolate a unique microscopic mechanism, it provides an important integral constraint on the overall strength of jet-medium interactions. In practice, comparisons across centrality, rapidity, initiating flavour, and jet radius help disentangle path-length dependence, colour-charge effects, and the angular structure of the lost energy. This is a key aspect explored in Chapter~\ref{chap:jet_energy_loss}.

\paragraph{Momentum balance and calibrated probes.}

More information is obtained by comparing a quenched jet to a correlated counterpart. Examples include dijet momentum imbalance, as well as photon($\gamma$)-jet and $Z$-jet recoil measurements. In dijet events, the two hard partons are produced approximately back-to-back, so an imbalance in reconstructed transverse momentum reflects asymmetric energy loss and fluctuations in quenching. In electroweak-boson tagged events, since the boson is not sensitive to the underlying background it provides a clean reference for the initial hard scale.

Thus, such observables are valuable because they sharpen sensitivity to fluctuations in quenching, path-length dependence associated with the production geometry, and selection biases, while quantifying how much of the original energy remains inside the measured jet cone after the jet traverses the medium. They can additionally probe how the momentum imbalance is compensated by soft particles from the medium response on the recoil side of the event.

\paragraph{Angular broadening and large-angle transport.}

The medium can also modify the angular distribution of jet energy. Multiple soft momentum transfers induce transverse broadening, while spatial gradients, collective flow, and the anisotropic overlap geometry of non-central collisions can generate directional deflections and/or azimuthal modulations of reconstructed jet yields and recoil distributions. Induced radiation may additionally transport energy to large angles. Additionally, correlated soft response from the medium can likewise contribute to large-angle energy flow around the jet axis or recoil direction. These effects are probed through: dijet acoplanarity and recoil-jet decorrelations, which quantify deviations from an exactly back-to-back azimuthal configuration; jet-shape observables that measure how the jet energy is distributed as a function of radial distance from the jet axis; measurements of energy flow outside jet cone that track transport to large angles; event-plane dependent jet yields, often quantified by Fourier coefficients such as $v_2^{\rm jet}$, which probe the path-length dependence of jet quenching in non-central collisions.

\paragraph{Jet substructure.}

Jets can also be studied beyond their total momentum and direction through measurements of their internal structure, see for instance~\cite{Larkoski:2017jix, Andrews:2018jcm} for detailed reviews. Such observables are sensitive to how energy is shared among the constituents of the reconstructed jet, how radiation is distributed across angular scales, and how the hardest splittings of the shower are modified in the medium. They can also be influenced by correlated soft response and by local properties of the surrounding matter, such as flow, inhomogeneities, or momentum-space anisotropies, which may distort the angular pattern of energy flow inside the jet. The latter type of anisotropies and its effect on jets is of central importance in Chapter~\ref{chap:anisotropic_early_time}.

This internal structure can be characterised in several complementary ways. Fragmentation functions quantify how the jet momentum is distributed among its constituents, while subjet observables identify resolved hard prongs and their relative momentum fractions or angular separations. Grooming techniques suppress soft radiation, typically at large angles, in order to expose harder branchings, making observables such as groomed momentum sharing fractions, groomed opening angles, and groomed masses particularly sensitive to the resolved jet core. A complementary and especially powerful class of probes is provided by energy correlators, which measure weighted correlations of energy flow at fixed angular scales. Because they do not rely on an explicit subjet assignment and can be systematically computed to higher orders in perturbative QCD, they offer a particularly clean, multi-scale characterisation of the angular structure within jets. For this reason, they are computed in different contexts in Chapters~\ref{chap:anisotropic_early_time} and~\ref{chap:towards_multi_point_structure}.

~\\

The effects discussed above generally reflect a combination of hard production, vacuum shower evolution, in-medium branching and scattering, medium response and reconstruction effects. Thus, a quantitative interpretation requires a dynamical framework in which the reconstructed jet is understood as a final-state manifestation of an energetic, multi-scale QCD cascade propagating through distinct QCD states of matter. Formulating this picture, and the theoretical tools used to describe it, is the central topic of this thesis. The basic building blocks for this theoretical formulation are introduced in the following chapter.

 % file "Thesis_Introduction.tex"
\cleardoublepage

%%%%%%%%%%%%%%%%%%%%%%%%%%%%%%%%%%%%%%%%%%%%%%%%%%%%%%%%%%%%%%%%%%%%%%%%
%  Jet_quenching.tex  --  Chapter 2: Theoretical background: jet quenching
%  Tex Master: Thesis.tex
%%%%%%%%%%%%%%%%%%%%%%%%%%%%%%%%%%%%%%%%%%%%%%%%%%%%%%%%%%%%%%%%%%%%%%%%

\chapter{Theoretical background: jet quenching}\label{chap:jet_quenching}

\section{Vacuum jet evolution in perturbative QCD}\label{sec:jet_evol_vac}

As discussed in Section~\ref{sec:hard_probes}, jets provide experimentally accessible proxies of energetic quarks and gluons produced in hard scatterings. We now shift from that phenomenological viewpoint to a dynamical one, in which a jet is understood as a perturbative cascade developing over a hierarchy of momentum scales before hadronisation. Since medium-induced modifications act on this pre-existing structure, it is first necessary to establish the QCD fundamentals that govern jet formation: the infrared enhancement of emissions, the resummation of multiple branchings, and the coherence effects that organise their space-time evolution.

\subsection{Soft and collinear enhancements}

A jet may be understood, at the perturbative level, as a multi-particle final state generated by radiative corrections to an underlying hard scattering. While the lowest-order process (Born level) produces only a small number of energetic partons, higher-order QCD emissions populate the final state with additional quarks and gluons. Crucially, these corrections are not distributed uniformly across phase space. Instead, they are enhanced when an emitted parton is unresolved, namely soft in energy or collinear to an energetic parent. In these infrared limits, the $(n+1)$-parton squared matrix element factorises schematically as~\cite{Catani:1996vz, Catani:1999ss}
\begin{align}
|{\cal M}_{n+1}|^2 \xrightarrow{\rm IR} |{\cal M}_n|^2 \otimes {\cal K}_{\rm IR},
\end{align}
where ${\cal K}_{\rm IR}$ is a process-independent kernel whose explicit form depends on the kinematics of the singular limit. This relation should be understood as a leading-power factorisation formula in
the corresponding infrared limit and $\otimes$ denotes the
appropriate convolution, which includes the spin and colour structures of both objects. 
%In such limit, the iterative enhancement of emissions dominates the perturbative evolution away from the hard interaction, converting a few hard partons into the cascades later reconstructed as jets.

A representative example is provided by $e^+e^- \to \gamma^\ast/Z \to q\bar q$, where the Born-level final state consists of a colour-singlet $q\bar q$ pair. At the next order, real gluon emission produces the three-parton state $q\bar q g$, whose singular limits illustrate the general factorisation structure discussed above. When the gluon four-momentum becomes soft, $k^\mu \to 0$, the emitted momentum is parametrically small compared with the hard scale. The leading interaction with the energetic partons is then eikonal, and the gluon can couple to any coloured hard leg. The resulting singular factor is therefore sensitive to pairwise colour correlations and to the angular configuration of the hard partons and the emitted gluon. Thus, in this limit, the leading singular behaviour of the tree-level real emission matrix element is
\begin{align}\label{eq:M2_soft}
|{\cal M}_{q\bar q g}|^2
\xrightarrow{k\to 0}
\alpha_s\, C_F\, {\cal S}_{q\bar q}(k)\,
|{\cal M}_{q\bar q}|^2,
\end{align}
where we have averaged over final state spins and colours and ${\cal S}_{q\bar q}(k)$ depends on dipole combinations such as $(p_q \cdot p_{\bar q})/[(p_q \cdot k)(p_{\bar q}\!\cdot k)]$. Physically, this corresponds to long-wavelength radiation that probes the colour flow of the hard event rather than a local branching. By contrast, when the gluon is emitted at small relative angle to the quark or antiquark direction, the relevant unresolved limit is collinear. In this case, neighbouring massless momenta become nearly parallel and the invariant mass of the branching system tends to zero, driving the intermediate propagator close to the mass shell. Consequently, the singular behaviour is local to the corresponding energetic leg and factorises into a universal splitting kernel. For instance for a gluon collinear with the quark, the leading collinear behaviour of the tree-level real emission matrix reads
\begin{align}\label{eq:M2_collinear}
|{\cal M}_{q\bar q g}|^2
\xrightarrow{k \parallel p_q}
\alpha_s\, C_F\,
\frac{P_{q\to qg}(z)}{k_\perp^2}\,
|{\cal M}_{q\bar q}|^2,
\end{align}
where we have averaged over final state spins and colours, $P_{q\to qg}(z)$ is the Altarelli-Parisi splitting kernel~\cite{Altarelli:1977zs} governing real-emission in the collinear limit, $z$ is the emitted gluon's momentum fraction with respect to the parent quark, and $k_\perp$ is the relative transverse momentum of the quark-gluon splitting. For small opening angle $\theta$, one has $k_\perp \simeq z(1-z)E\theta$, with $E$ the total energy of the $qg$ branching system. Importantly, Eq.~\eqref{eq:M2_collinear} is only valid after azimuthal averaging of the splitting kernel, which effectively removes the spin correlated contributions. 

Thus, the soft and collinear limits reflect complementary unresolved configurations: soft radiation is controlled by colour correlations across the hard event, whereas collinear radiation is controlled by the local kinematics of an individual branching and generates the internal collimated structure of jets. Importantly, the colour factors appearing in Eqs.~\eqref{eq:M2_soft} and~\eqref{eq:M2_collinear} also encode the colour representation of the participating hard partons, although the colour structure is in general more involved in the soft case, where radiation couples coherently to pairs of hard coloured legs. In the collinear limit, this dependence reduces to the colour charge associated with the corresponding splitting: emissions off quark are weighted by the fundamental charge $C_F=(N_c^2-1)/(2N_c)=4/3$, while emissions off gluons carry the adjoint charge $C_A=N_c=3$. This difference reflects the stronger unresolved-radiation enhancement associated with gluons compared with quarks. Taken together, the factorisation properties outlined above identify the universal infrared structure of QCD amplitudes, providing the starting point for understanding how singular contributions enter perturbative jet cross-sections.

\subsection{Resummation and parton showers}\label{subsec:resummation_showers}

Although the soft and collinear limits generate singular contributions at the level of individual real-emission corrections, these divergences cancel in sufficiently inclusive physical cross sections once the corresponding virtual loop contributions are included. Following the example given in the previous section, the next-to-leading order (NLO) cross-section for $e^+e^- \to \gamma^\ast/Z \to$ hadrons may be written as
\begin{align}
\sigma^{\rm NLO}
=
\int_{\Phi_2}\left(\mathrm{d}\sigma_B+\mathrm{d}\sigma_V\right)
+
\int_{\Phi_3}\!\mathrm{d}\sigma_R ,
\end{align}
where $\mathrm{d}\sigma_V$ denotes the one-loop correction to the Born process $e^+e^- \to q\bar q$ in $d\sigma_B$, and $\mathrm{d}\sigma_R$ the real-emission channel $e^+e^- \to q\bar q g$. The corresponding phase spaces are the two-particle $\Phi_2$ and the three-particle $\Phi_3$ phase spaces. When the observable is sufficiently inclusive, the singularities arising from soft or collinear real emissions in $d\sigma_R$ cancel against the corresponding infrared poles of the virtual terms in $d\sigma_V$, as guaranteed by the Kinoshita-Lee-Nauenberg theorem~\cite{Kinoshita:1962ur,Lee:1964is}. Infrared- and collinear-safe jet observables, introduced in Section~\ref{subsec:jets_intro}, are constructed so that this cancellation remains valid under the addition of soft particles or collinear splittings. 

However, measurements that constrain radiation into selected regions of phase space, or introduce strong hierarchies between physical scales, leave \emph{logarithmically enhanced} corrections at each order in $\alpha_s$. This means that fixed-order perturbative coefficients acquire logarithms of ratios of physical scales that can compensate the smallness of $\alpha_s$. To be precise, for an observable characterised by a small dimensionless parameter $v\ll1$, one typically encounters terms of the form
\begin{align}
    \alpha_s^n L^m, \qquad L=\ln(1/v),
\end{align}
so that sufficiently large $L$ renders contributions from higher perturbative orders numerically comparable. Reliable predictions then require a reorganisation of the fixed-order perturbative expansion, such that these enhanced terms can be summed to all orders -- so called \emph{resummation}~\cite{Sterman:1986aj, Catani:1989ne}\footnote{The interested reader is pointed to~\cite{Becher:2014oda} for a review on modern methods for resummation and to~\cite{Larkoski:2017jix,Dasgupta:2013ihk} for analytic resummation of jet substructure observables.}. This becomes possible because radiation in the singular limits factorises recursively into universal contributions associated with distinct momentum regions. For sufficiently global, infrared-safe observables, namely those sensitive to radiation across the full event phase space, one can then write the differential distribution of $v$ schematically as~\cite{Catani:1992ua, Becher:2014oda}
\begin{align}\label{eq:general_HJS}
    \frac{\mathrm{d}\sigma}{\mathrm{d}v}\sim H \otimes J \otimes S ,
\end{align}
where $H$ describes the short-distance hard process, the jet functions $J$ encode radiation collinear to the energetic parton directions, and the soft function $S$ encodes low-momentum wide-angle emissions coupling eikonally to their net colour charges. These functions are renormalised separately and obey distinct evolution equations governed by their respective anomalous dimensions. They are naturally computed at their respective characteristic momentum scale, where no large logarithms appear in its perturbative coefficients, and subsequent RG evolution evolves them to a common scale $\mu$, where they are consistently combined. Through this process, termed \emph{analytic resummation}, one resums the logarithms associated with the hierarchy $v\ll1$. In modern treatments, such a process is often formulated within Soft-Collinear Effective Theory, see, e.g.,~\cite{Becher:2014oda}, although equivalent resummations were developed earlier using diagrammatic QCD methods.

A representative example of the applicability of resummation is the thrust distribution in $e^+e^- \to$ hadrons, which quantifies how closely the final state resembles the Born-level back-to-back dijet configuration. This observable is defined as
\begin{align}
    T=\max_{\hat n}\frac{\sum_i |\vec p_i\cdot\hat n|}{\sum_i |\vec p_i|}, \qquad \tau \equiv 1-T ,
\end{align}
and its differential rate reads
\begin{align}\label{eq:thrust_def}
    \frac{\mathrm{d}\sigma}{\mathrm{d}\tau}
    \propto
    \sum_X \int \mathrm{d}\Phi_X\, |{\cal M}_{e^+e^-\to X}|^2 \,
    \delta\!\big(\tau-\tau(X)\big)\,,
\end{align}
which measures the cross section for events of given thrust, where the sum runs over all hadronic final states $X$. The two-jet limit corresponds to $\tau\ll1$, for which one identifies $v=\tau$. In this regime, the leading singular part of the distribution receives contributions both from energetic radiation collinear to the quark or antiquark jet directions and from low-energy soft radiation, yielding the more specific factorised form (see, e.g.,~\cite{Catani:1992ua, Becher:2008cf} for further details)
\begin{align}
    \frac{\mathrm{d}\sigma}{\mathrm{d}\tau}\sim H(Q)\otimes J_q \otimes J_{\bar q}\otimes S .
\end{align}
The incomplete cancellation between real and virtual corrections in the constrained phase space in Eq.~\eqref{eq:thrust_def} is then reorganised by the evolution of the functions $J_q$, $J_{\bar q}$ and $S$ into exponentials of large logarithms, termed \emph{Sudakov factors}, that suppress additional emissions which would significantly increase $\tau$. 

The same logarithmically enhanced radiation can be reorganised in a probabilistic language, forming the basis of Monte Carlo descriptions of jet cascade evolution, known as \emph{parton showers}~\cite{Sjostrand:1985xi,Webber:1986mc, Marchesini:1987cf}. Rather than evolving observable-dependent functions such as $J$ and $S$, these algorithms aim to generate exclusive partonic final states $\Phi_n$, specifying the momenta, flavours, and colour connections of $n$ resolved partons, from which arbitrary infrared-safe observables may subsequently be computed. This construction introduces an evolution, or shower, variable $t$, which orders branchings sequentially between the initial hard scale $Q$ and lower resolution scales. Depending on the shower formalism, $t$ can be virtuality, opening angle or transverse momentum. The interested reader is referred to, e.g.,~\cite{Buckley:2011ms,Hoeche:2014a} for pedagogical and detailed reviews on parton showers. Starting from a configuration $\Phi_n$ at shower scale $t_1$, parton shower algorithms then formulate the probability density that the next branching occurs at a lower scale $t_2<t_1$ schematically as
\begin{align}
    d\mathcal P_n(t_2)
    =
    \Delta_n(t_1,t_2)\,
    dt_2\, d\zeta\,
    \mathcal K_n(\Phi_n,\zeta;t_2).
\end{align}
where $d\zeta$ denotes the branching phase space, including variables such as momentum fractions and azimuthal angles, while $\mathcal K_n$ is the shower approximation to the singular real-emission matrix element for producing an $(n+1)$-parton configuration from the current $n$-parton state. The dependence on $\Phi_n$ reflects that the available emitters, colour connections, recoil partners, and kinematic boundaries are determined by the evolving event configuration. The Sudakov factor
\begin{align}
\Delta_n(t_1,t_2)
=
\exp\left[
-\int_{t_2}^{t_1} dt'
\int d\zeta\,
\mathcal K_n(\Phi_n,\zeta;t')
\right]
\end{align}
is the probability of evolving from $t_1$ to $t_2$ without an additional resolvable emission. The Monte Carlo algorithm samples the next branching scale $t_2$ together with the variables $\zeta$, updates the event through a map $\Phi_n\to\Phi_{n+1}$, and iterates the procedure until an infrared cutoff is reached. This cutoff $t_{\rm cut}$ is typically of order $1$ GeV, below which non-perturbative hadronisation models are invoked. In the collinear limit, the LO kernel reduces locally for a particular emitter $a$ to~\cite{Ellis:1991qj, Dokshitzer:1991wu}
\begin{align}
    dt_2\,d\zeta\,\mathcal K_n
    \sim
    \frac{dt_2}{t_2}\,dz\,\frac{d\phi}{2\pi}\,
    \frac{\alpha_s}{2\pi}P_{a\to bc}(z),
\end{align}
up to recoil prescriptions, phase-space constraints, and sums over allowed splittings. This is simply the AP splitting kernel introduced in Eq.~\eqref{eq:M2_collinear}, with the appropriate phase space singularities. In contrast, in dipole or antenna showers, see, e.g,~\cite{Gustafson:1987rq, Giele:2007di}, $\mathcal K_n$ is organised around colour-connected pairs while reproducing the same soft and collinear singular limits. Inclusive averaging over the cascade then reproduces timelike DGLAP evolution~\cite{Gribov:1972ri,Altarelli:1977zs,Dokshitzer:1977sg}, introduced in~\ref{subsec:DIS_DGLAP} in the context of PDF spacelike evolution, whereas the fully differential construction retains the exclusive branching history and event kinematics. In the collinear form above, it is clear that the logarithms discussed previously arise directly from integrating the singular measure $dt_2/t_2$ together with the endpoint structure of the splitting kernels, making explicit how sucessive unresolved emissions generate the enhanced perturbative contributions that the shower resums.

This probabilistic framework forms the basis for modern parton shower algorithms, which then differ by the choice of ordering variable, radiation kernel, recoil prescription, and treatment of colour flow\footnote{For precision phenomenology, shower evolution is further combined with fixed-order matrix elements through \emph{matching and merging} procedures, which restore exact predictions for hard wide-angle emissions while retaining resummation for subsequent unresolved radiation.}. Further, this probabilistic formulation is not equivalent to a picture of independent emissions. In this construction, interference between different emitters is not treated globally at the full squared amplitude level, but is instead encoded locally, such that the full pattern of multi-parton interference is approximated by an ordered sequence of emissions subject to phase-space constraints\footnote{This is unlike analytic resummation discussed previously, which encodes correlations globally through factorised functions and their evolution for a given observable and logarithmic accuracy.}. Thus, parton showers generate correlated radiation as an ordered sequence of emissions $t_1 > t_2 > \cdots$, with the strongly ordered limit $t_1 \gg t_2 \gg \cdots$ corresponding to the region that controls the leading logarithmic (LL) structure, i.e., the highest power of large logarithms at each order in $\alpha_s$. At LL accuracy, the correct treatment must include \emph{colour coherence} effects, i.e., the suppression of large angle soft radiation due to destructive interference between coloured emitters~\cite{Dokshitzer:1991wu}, which can be correctly reproduced within the sequential shower picture. This is encoded differently depending on the shower formalism: angular-ordered showers implement it directly through the ordering variable, while virtuality- or transverse momentum-ordered showers enforce it through additional phase-space contraints, such as angular vetoes. Beyond LL\footnote{The precise counting depends on the observable, e.g., sufficiently differential observables can deem these effects leading~\cite{Dasgupta:2020fwr}.}, emissions need not be widely separated in scale, and more general correlations, such as spin correlations or finite-$N_c$ colour structure, become relevant (see, e.g.,~\cite{Nagy:2007ty, Dasgupta:2020fwr}).

\subsection{Formation time in vacuum showers}\label{subsec:formation_time}

The discussion in the previous subsection naturally raises the question of how the full pattern of multi-parton interference can be approximated, at LL accuracy, by an ordered sequence of emissions. As we saw, this replacement is not trivial: even at this accuracy, emissions in a parton cascade are not independent, since colour interference between different emitters is required to reproduce the correct radiation pattern. More generally, a probabilistic description in terms of sequential partonic configurations $\Phi_n$ does not follow directly from the squared QCD amplitudes, where different emission amplitudes contribute with non-trivial relative phases. The point is that, in the unresolved regions responsible for the leading logarithms, the interference pattern simplifies: collinear radiation becomes local to a parent parton, while leading soft interference is reorganised into coherent radiation from colour-connected charges.

For the calculations presented in the following sections, it is useful to give this ordering a time-domain interpretation. The relevant hierarchy is fundamentally a hierarchy of propagator denominators, or equivalently of the phases generated by off-shell propagation, rather than a separate dynamical assumption about the lifetime of partons. In covariant perturbation theory, the propagation of an off-shell particle with momentum $p^\mu$ is governed by the denominator $1/(p^2-m^2+i\epsilon)$. Expressing this propagator in a representation mixed with respect to the light-cone time $x^+=(x^0+x^3)/\sqrt{2}$, see Appendix~\ref{app:light_cone_coord}, makes explicit the phase associated with off-shell propagation. In these variables, the relevant quantity is the off-shellness mismatch in the momentum component conjugate to $x^+$,
\begin{align}
    \delta p^- \equiv p^- - p^-_{\rm on} = \frac{p^2-m^2}{2p^+},
\end{align}
so that propagation over a light-cone time interval generates phases of the form $e^{-i\delta p^- x^+}$, up to convention-dependent signs. At the level of amplitudes, this gives schematically
\begin{align}
    \mathcal{M} \sim \int dx^+\,F(x^+)\,e^{-i\delta p^- x^+}.
\end{align}
Here the phase arises from the propagator, while $F(x^+)$ collects the remaining matrix-element structure, including vertices, colour factors, spin dependence, and kinematic prefactors. For a single splitting in vacuum, this non-phase factor is not intrinsically time dependent, reflecting the time-locality of the interaction. In more involved settings, however, an effective time dependence can arise because multiple emissions, propagators, ordering constraints, or finite integration domains introduce further phase structure. More generally, $F$ may also contain external time scales associated with, e.g., a finite coloured medium, which is particularly relevant for this thesis and will be addressed in the subsequent section.

The physical role of these phases becomes apparent in the squared amplitude, where amplitudes and conjugate amplitudes with different emission histories are compared. Denoting by $x^+$ and $\bar x^+$ the light-cone times associated with the same virtual fluctuation in the amplitude and its conjugate, the phase depends on their difference,
\begin{align}
    |\mathcal{M}|^2 \sim \int dx^+\,d\bar x^+F(x^+) F^\ast(\bar x^+)\exp\left(-i\,\delta p^- (x^+-\bar x^+)\right).
\end{align}
It is therefore natural to introduce the average and relative coordinates $X^+=(x^++\bar x^+)/2$ and $r^+=x^+-\bar x^+$. The phase depends only on $r^+$, while translation invariance in vacuum makes the remaining integrand independent of $X^+$. If this integrand varies slowly over the interval probed by the phase, the coherent contribution is restricted to relative separations for which $|\delta p^- r^+|\lesssim 1$. This identifies the \emph{formation time} as
\begin{align}
    x_f^+ \sim \frac{2p^+}{p^2-m^2}.
\end{align}
In this sense, the formation time is not a property of a single amplitude, but the characteristic light-cone time interval over which a virtual fluctuation contributes coherently to the squared amplitude. The same denominator that defines the $x_f^+$ also organises the strongly ordered regions of the cascade. When successive unresolved emissions have parametrically separated denominators, their formation times are separated as well, and the cascade admits a sequential space-time interpretation. If two emission amplitudes have comparable denominators, they are formed over the same parametric interval and must be treated coherently whenever they contribute at LL. Vacuum soft colour coherence is the basic example: a leading soft gluon is radiated from the coherent colour structure of the event, not from a uniquely selected emitter.

For a single collinear splitting, the denominator can be written directly in terms of the branching kinematics. If a parent parton of light-cone energy $P^+$ emits a gluon carrying an energy fraction $z$ and with relative transverse momentum $k_\perp$, the formation time is parametrically given by
\begin{align}
    x_f^+ \sim \frac{2z(1-z)P^+}{k_\perp^2}.
\end{align}
In the soft limit, where one daughter is a gluon of energy $zP^+ = \omega\ll P^+$, this reduces to
\begin{align}
    x_f^+ \sim \frac{2}{\omega\theta^2},
\end{align}
with $k_\perp\simeq \omega\theta$. This expression makes explicit the hierarchy underlying an ordered cascade: softer or more collinear emissions can remain coherent over longer light-cone times, while larger transverse-momentum fluctuations are formed over shorter intervals. In vacuum, this is mainly a space-time interpretation of the same propagator denominators that organise the soft and collinear limits. In a medium, the interpretation becomes operational, because $x_f^+$ can be compared with external scales associated with an underlying coloured medium. 

\section{Jet quenching: a multi-scale problem}\label{sec:jet_quenching}

\subsection{Medium scales and transport properties}\label{subsec:scales}

The previous section provided the baseline for vacuum jet evolution: a jet is a perturbative QCD cascade whose structure is governed by soft and collinear dynamics, colour coherence, and scale-dependent resolution. In heavy-ion collisions, this cascade propagates through a coloured medium evolving in space and time, so its evolution is further controlled by additional scales that have no vacuum counterpart. An important example is the formation time introduced in the previous section, which ceases to be only an internal scale of vacuum emissions and becomes meaningful through comparison with the space-time support and dependence of the background matter along a given parton's trajectory. Thus, jet quenching arises from the interplay between the scales intrinsic to the perturbative cascade and the spatio-temporal structure of the produced QCD matter. Throughout this thesis, as in typical theoretical descriptions, we work in the high-energy limit, such that a parton's momentum component along the jet direction (longitudinal) is much larger than the characteristic microscopic scales of the medium. At leading power, this implies the longitudinal momentum is conserved, while the non-trivial medium-induced dynamics is concentrated in transverse motion, radiation and colour evolution. We shall make this approximation concrete in Section~\ref{sec:BDMPS_Z}.

The medium itself is characterised by its space-time profile and by the microscopic structure of its degrees of freedom. The appropriate variables, and thus the relevant medium scales, depend on the state of HIC matter which the jet interacts with (see Section~\ref{subsec:st_evolution_HIC}). In early-time descriptions, such as Glasma, the natural degrees of freedom are coherent, highly anisotropic colour fields, described by gauge fields $A^\mu(x)$, field strengths $F^{\mu\nu}(x)$ and their correlation functions. In the pre-equilibrium phase, the system is often described in terms of phase-space distributions $f_a(x,p)$, whose evolution captures the redistribution of energy and momentum and the relaxation of potentially large pressure anisotropies towards hydrodynamisation. For a near-equilibrium QGP, the relevant information is encoded in local thermodynamic quantities and flow fields such as $T(x)$, $\epsilon(x)$, and $u^\mu(x)$, whose space-time dependence captures the matter profile and its gradients, together with dissipative corrections to the stress-energy tensor, such as shear stress and bulk pressure, that describe local momentum-space anisotropies. Across these descriptions, the medium is characterized by physical scales associated both with its microscopic degrees of freedom and with the macroscopic evolution of the collision system. Depending on the regime, examples of microscopic scales include the saturation momentum $Q_s$ in the Glasma, characteristic momenta and relaxation times in kinetic pre-equilibrium, and screening scales such as $m_D^{-1}$, which in a weakly coupled thermalised QGP scales as $m_D \sim g_s T$. These coexist with larger scales associated with the system's space-time extent, expansion and spatial gradients, which together influence how partonic cascades are modified.

Once a high-energy trajectory is specified, the medium information relevant for jet quenching is projected onto the path of the jet. In a static idealisation this projection is often represented by a path length $L$, while in an evolving collision it is more generally a line integral through local fields, densities, correlation functions, or transport coefficients. The interaction with the medium then introduces quantities that are not properties of the vacuum cascade alone, nor of the bulk medium alone, but of the coupling between the energetic probe and the degrees of freedom it resolves. Depending on the description, this scale information may be organised at different levels of resolution. In weakly coupled descriptions where the interaction is resolved as scattering off microscopic constituents or collective excitations of the medium, it is often reduced to an effective, differential collision kernel. This kernel depends on the phase-space distributions characterising the medium constituents and on the propagator of the exchanged gauge field, including medium-induced screening and possible anisotropic corrections. If the relevant degrees of freedom are coherent colour fields instead, the interaction is more naturally expressed through gauge-field or field-strength correlators along the trajectory. These formulations are different ways of encoding how an energetic colour charge samples the background, through transverse momentum exchange and colour rotation. Their moments or limiting forms define transport coefficients, while induced-radiation spectra computed from them, including interference effects, capture medium-modified splitting rates. A particularly relevant example is transverse-momentum broadening. In an isotropic medium, this is commonly characterised by a scalar \emph{jet quenching parameter}, operationally defined as the transverse momentum squared accumulated per unit path length,
\begin{align}
    \hat q \equiv \frac{d\langle k_\perp^2\rangle}{d\ell}.
\end{align}
where $\ell$ is the distance travelled along the parton's trajectory. In a realistic heavy-ion background, $\hat q$ can vary along this trajectory. This can be due to, for instance, longitudinal expansion affecting the accumulation of broadening through the cooling of the medium, or due to transverse gradients which make the result dependent on the production point and propagation direction of the jet. More generally, spatial gradients, collective flow, or microscopic momentum-space anisotropies can distinguish different directions in the plane transverse to the jet. The broadening is then described by a tensor,
\begin{align}
    \hat q^{ij}
    =
    \frac{d\langle k_\perp^i k_\perp^j\rangle}{d\ell},
\end{align}
whose trace gives the usual scalar broadening coefficient, while its anisotropic components encode directional broadening. Thus, $\hat q$ should be understood as a coarse-grained moment of the interaction between a given high-energy colour charge and the medium. More differential descriptions may retain the full momentum-transfer kernel rather than only its second moment. Further, in other kinematic limits where longitudinal momentum exchanges are not suppressed, collisional energy loss, heavy-flavour diffusion, and observables sensitive to the full momentum exchange may also involve drag and longitudinal momentum-diffusion coefficients.

\subsection{Jet quenching mechanisms and effective descriptions}\label{subsec:quenching_mechanisms}

The medium information introduced above becomes relevant for jet quenching through its coupling to a developing perturbative cascade. A collision kernel, a field-strength correlator, or a coefficient such as $\hat q$ specifies how an energetic colour charge samples the medium: how transverse momentum is transferred, how colour is rotated, and how these effects are distributed along the parton's trajectory. In a jet, however, the propagating object is not a single fixed charge, but a set of colour-correlated partons whose splittings have finite formation times and transverse separations that evolve during propagation. Thus, medium interactions can affect the cascade in several distinct ways: they can transfer energy and momentum, with transverse momentum broadening as the leading high-energy effect; they can modify radiation that would already occur in vacuum or stimulate additional radiation; or they can change whether the medium resolves a multi-parton system as one coherent colour charge or as several independent charges. The central theoretical problem is to determine which of these effects is resolved at the scales relevant for a given observable, and which medium information may be consistently reduced to transport coefficients.

Elastic exchange provides the most direct way in which the medium modifies the kinematics of an energetic parton. A momentum transfer $q^\mu$ from the medium changes both the colour state and the four-momentum of the propagating charge. In the high-energy limit, the largest component of the parton momentum is only weakly perturbed by a single exchange, while the accumulated transverse momentum can have leading observable consequences. The latter effect smears back-to-back angular correlations and modifies the amount of transverse momentum available for subsequent radiation. In realistic collisions this broadening need not be azimuthally symmetric: geometry, gradients, and flow make the accumulated broadening depend on the jet production point and propagation direction, while local anisotropies of the momentum-transfer distribution can further make the broadening tensor itself directional, as encoded by $\hat q^{ij}$. The level of microscopic detail required to describe these momentum exchanges depends on the scales resolved by the dynamics and the observable under consideration. If individual exchanges are resolved, the relevant information is contained in a differential momentum-transfer kernel, whose finite-$\boldsymbol  q_\perp$ structure also captures rare harder kicks. If instead the dynamics is controlled by many unresolved soft exchanges, the interaction can be reduced to a broadening coefficient such as $\hat q$, or into $\hat q^{ij}$ when directional broadening is retained. Finally, energy transfer to the medium, often called collisional energy loss, is part of the same elastic dynamics, but for highly energetic partons it is usually subleading per scattering compared with the transverse momentum exchange and colour rotation that control radiation and angular structure.

Radiative modifications are controlled by the same formation-time logic introduced in the vacuum cascade. A transverse momentum exchange can change the virtuality and colour state of an energetic parton, i.e., the medium can modify whether quantum fluctuations in the parton's wave function decohere into resolved radiation. Further, exchanges occurring during the formation of an otherwise vacuum-like emission can also alter its phase and transverse momentum. Thus, the relevant question is not only whether the medium interacts with a propagating parton, but also how formation time coherence is modified by the longitudinal structure of the jet-medium interaction. During an emission with formation time $t_f$, the splitting samples not only the local strength of the medium, but also microscopic correlation scales, the density profile, gradients, flow, and macroscopic evolution of the background over the corresponding interval. The relevant comparison is between $t_f$ and the longitudinal or temporal scales over which these structures remain correlated, vary appreciably, or cease to contribute to the interaction. This is the physical interpretation of what we alluded to in Section~\ref{subsec:formation_time} when introducing formation time in vacuum -- the medium not only changes the phase information that gives $t_f$, but also adds further scales in the non-phase integrand which compete with $t_f$. At the microscopic level, an example is an effective mean free path or a field-correlation length scale, while at macroscopic level one can think of the in-medium path length. Therefore, medium-induced radiation should not be understood simply as being produced after a given interaction, but rather as a process that is controlled by the interplay between propagator phases and the space-time structure of the medium.

The mechanisms discussed above are typically not described within a single theoretical framework. Different descriptions of jet quenching are distinguished both by how they represent the medium and by how they organise the hierarchy between intrinsic medium scales and the dynamical scales generated by the propagating jet. The high-energy limit fixes the leading kinematics of the projectile, but it does not determine whether the medium interaction is best organised through resolved scatterings, background fields, thermal propagators, operator matrix elements, or effective modes. In a quasiparticle or scattering-centre picture, one may introduce a mean free path $\lambda$, characterising the typical distance between effective scatterings, with $\lambda^{-1}\sim\rho\sigma$ given in terms of an effective density $\rho$ and the parton-medium cross section $\sigma$. Together with the in-medium path length $L$, this defines the \emph{opacity} $L/\lambda$, which estimates the number of effective scatterings sampled by the energetic parton. Small opacity corresponds to a regime where only a few momentum exchanges are resolved, while large opacity corresponds to multiple scatterings along the path. In the \emph{GLV} (Gyulassy-Levai-Vitev) formalism~\cite{Gyulassy:1999zd,Gyulassy:2000er,Gyulassy:2000fs}, the jet-medium interaction is organised as an opacity expansion, retaining the momentum transfer associated with each resolved scattering centre. The \emph{BDMPS-Z} (Baier-Dokshitzer-Mueller-Peigné-Schiff-Zakharov) formalism~\cite{Baier:1994bd,Baier:1996kr,Baier:1996sk,Zakharov:1996fv,Zakharov:1997uu,Baier:1998yf}, which underlies most results in Chapters~\ref{chap:anisotropic_early_time} and~\ref{chap:towards_multi_point_structure}, instead gives a background-field description of high-energy propagation, in which partonic Green's functions resum arbitrarily many insertions of a stochastic colour field, with an average over medium configurations performed at the level of the squared amplitude. This is naturally suited to the regime in which many medium interactions contribute during a formation time: the individual momentum transfers are no longer kept as resolved scattering centres, and their coherent cumulative effect is encoded in the transverse Green's function. Further approximations, such as the \emph{harmonic-oscillator limit}, reduce this to a diffusive picture controlled by $\hat q$. We shall make this explicit in the following section. \emph{Higher-twist} approaches~\cite{Guo:2000nz,Wang:2001ifa} also express medium effects through correlation functions, but organise them as operator matrix elements appearing in power-suppressed corrections to a hard process. Then, \emph{AMY} (Arnold-Moore-Yaffe) approaches~\cite{Arnold:2001ba,Arnold:2002ja,Arnold:2002zm} formulate medium-induced radiation in finite-temperature field theory as splitting rates for hard partons propagating through a weakly coupled, near-equilibrium plasma, including coherence effects usually associated with the Landau-Pomeranchuk-Migdal (LPM) effect~\cite{Landau:1953um, Migdal:1956tc}. The corresponding medium input is provided by \emph{hard-thermal-loop} (HTL) resummation~\cite{Braaten:1989mz,CaronHuot:2008ni} , which describes screened soft thermal interactions and determines the relevant propagators and scattering rates. Finally, \emph{soft-collinear effective theory with Glauber modes} (SCET$_{\rm G}$)~\cite{Ovanesyan:2011kn,Ovanesyan:2011xy,Kang:2014xsa} organises collinear jet modes interacting with the medium through Glauber momentum exchanges within an effective-field-theory power counting. This list is not exhaustive. For instance, strong-coupling approaches based on gauge-gravity duality provide a different class of descriptions, in which transport properties can be computed holographically~\cite{Liu:2006ug,Herzog:2006gh,Gubser:2006bz,CasalderreySolana:2006rq}. Thus, these descriptions provide complementary field-theoretic or semi-classical organisations of the same underlying problem: how medium structure is resolved by a partonic cascade through interference effects controlled by longitudinal scales, and through mechanisms such as transverse momentum exchange, colour rotation, and stimulated radiation. The interested reader is pointed to~\cite{Armesto:2011ht} for a comparison of some of these theoretical formalisms.

The calculations presented in the following chapters of this thesis are performed in the BDMPS-Z formalism. Although we will only introduce it formally in the next section, its central scale hierarchy can already be exposed in the regime where the interaction during an emission is dominated by many unresolved soft exchanges. This regime materialises in the \emph{harmonic-oscillator or multiple soft scattering approximation} by boiling down the detailed sequence of momentum transfers to a diffusion in transverse momentum, with $\hat q$ setting the rate at which momentum is accumulated along the trajectory. Now, let us consider a gluon fluctuation of energy $\omega$, radiated from an energetic parton propagating through the medium. During its formation time $t_f^{\rm med} \sim 2\omega/k_f^2$, the fluctuation acquires a typical transverse momentum from the medium given by $k_f^2 \sim \hat q\, t_f^{\rm med}$. Combining these estimates yields
\begin{align}
    t_f^{\rm med} &\sim \sqrt{\frac{2\omega}{\hat q}}, & k_f^2 &\sim \sqrt{2\omega\hat q}.
\end{align}
The quantity $t_f^{\rm med}$ is thus the longitudinal interval over which the gluon fluctuation is resolved as an emitted quantum by the medium. In this regime it is determined self-consistently, since the transverse momentum that controls the denominator is itself accumulated from the medium during the formation process. Thus, even before deriving the full radiation spectrum, the medium-induced formation process is seen to depend on the competition between the gluon energy, transverse broadening accumulated during formation, and the longitudinal interval over which the medium interaction remains effective. Medium-induced radiation is therefore not controlled by the scale $t_f^{\rm med}$ alone, but by the interference pattern associated with that interval. If several interactions with the medium occur within the same formation interval, the corresponding radiation amplitudes do not separate into independent contributions from individual momentum transfers. They interfere coherently, so that the medium acts on the developing gluon fluctuation as a whole. This is the origin of the non-Abelian LPM effect: radiation is suppressed relative to an incoherent Bethe-Heitler sum over independent scatterings whenever the formation time is longer than the microscopic longitudinal scales of the interaction~\cite{Bethe:1934za,Baier:1996sk}. In the BDMPS-Z formalism this interference is resummed through Green's functions but the physical mechanism is more general than that. Opacity-based descriptions, for example, organise the same formation-time interference order by order in resolved scatterings. Thus, the scale estimates above simply identify the relevant competition: emissions with $t_f^{\rm med}$ parametrically smaller than the macroscopic length scale $L$ can be formed within the medium, whereas emissions with $t_f^{\rm med}\gtrsim L$ remain sensitive to the finite longitudinal support and time dependence of the jet-medium interaction. Using the medium-induced formation-time estimate $t_f^{\rm med}(\omega)\sim \sqrt{2\omega/\hat q}$, this latter condition corresponds parametrically to emissions with energy $\omega \gtrsim \omega_c$, where the characteristic gluon energy is defined as
\begin{align}
    \omega_c \sim \frac{1}{2}\hat q L^2 .
\end{align}

Once radiation from a single energetic charge is understood in terms of formation and interference, the next question is how the medium modifies the pattern of multiple emissions and interferences already present in the vacuum cascade. 
The cleanest formulation of this problem is the in-medium antenna: two colour-correlated partons with an opening angle $\theta$ propagate through the medium and test its transverse resolving power~\cite{MehtarTani:2010ebp,MehtarTani:2011tz,MehtarTani:2011gf,CasalderreySolana:2011rz,Casalderrey-Solana:2012evi,MehtarTani:2011hma,MehtarTani:2012mfa}. Their transverse separation grows parametrically as $r_\perp(t)\sim\theta t$, while the medium resolves transverse structures through the momentum broadening accumulated along the path. In the multiple soft scattering limit of BDMPS-Z, this resolution is encoded in a decoherence factor $\Delta_{\rm med}(t)$, which quantifies the rate of colour decoherence of the antenna up to time $t$. In a static homogeneous medium one obtains~\cite{MehtarTani:2012mfa},
\begin{align}\label{eq:delta_med_def}
    \Delta_{\rm med}(t) \simeq 1-\exp\left[-\frac{1}{12}\hat q\,\theta^2 t^3\right].
\end{align}
From this, one can extract a typical \emph{decoherence time} $t_d\sim(\hat q\theta^2)^{-1/3}$, and the \emph{critical angle} follows from $t_d\sim L$,
\begin{align}
    \theta_c \sim \left(\hat q L^3\right)^{-1/2}.
\end{align}
The factor in Eq.~\eqref{eq:delta_med_def} evaluated at the characteristic length scale $\Delta_{\rm med}(L)$ determines how many effective colour emitters the medium resolves over this length. When $\Delta_{\rm med}(L)\simeq0$, the antenna remains coherent and radiates approximately as a single colour charge, while when $\Delta_{\rm med}(L)\simeq1$, the interference term is suppressed and the two prongs radiate as independent colour sources~\cite{MehtarTani:2011gf,MehtarTani:2011tz,CasalderreySolana:2011rz,MehtarTani:2012mfa}. Thus, the number of effective emitters for medium-induced radiation is not fixed by the number of partons in the corresponding vacuum shower, but by the interplay between antenna geometry and medium resolution~\cite{Casalderrey-Solana:2012evi}. In terms of angular structure, antennae with $\theta\lesssim\theta_c$ remain coherent over the interaction length, while antennae with $\theta\gtrsim\theta_c$ are resolved by the medium and gradually lose memory of their original color correlation. This phenomenon interplays naturally with the vacuum color coherence we introduced at the end of Section~\ref{subsec:resummation_showers}. In vacuum, a sufficiently soft gluon emitted at angles larger than the antenna opening angle resolves only the total colour charge of the pair. For a colour-singlet antenna this suppresses large-angle radiation and confines subsequent soft emissions to the angular-ordered region. In a medium, colour exchanges rotate and decohere the antenna, weakening this cancellation and opening a medium-induced soft component at large angles. This is the origin of \emph{anti-angular ordering}: once the medium resolves the antenna, soft radiation is no longer constrained by the original vacuum colour correlation~\cite{MehtarTani:2010ebp}. In a full jet, this separates a sufficiently collimated core, which may remain protected by colour coherence and radiate as an overall colour charge, from broader substructure that is resolved by the medium and interacts as a cascade of separate colour charges. Modern descriptions of in-medium jet evolution build on this separation by combining vacuum-like branchings, medium-induced radiation, and decoherence constraints on the available phase space, thereby linking colour resolution to jet-substructure observables (see, e.g., ~\cite{Caucal:2018dla}).

This section introduced the typical perturbative effects in jet quenching: transverse broadening, medium-induced radiation, formation-time interference, finite-length effects, and medium resolution of colour substructure. The backreaction from the medium to the energy deposited by a jet, termed \emph{medium response}, is an additional mechanism which modifies reconstructed jet observables. It is typically treated with phenomenological models rather than with jet quenching formalisms such as the ones we have covered. We shall come back to this part of the problem in Chapters~\ref{chap:jet_energy_loss} and~\ref{chap:towards_multi_point_structure}. The following section provides the technical basis for treating the perturbative part of the problem within the BDMPS-Z framework.

\section{The BDMPS-Z formalism}\label{sec:BDMPS_Z}

\subsection{High-energy propagation in a background coloured field}\label{subsec:propagator_derivation}

Let us consider a highly energetic parton propagating through QCD matter, with light-cone momentum
\begin{align}
    p^\mu=(p^+,p^-,\p_\perp) = \left(\frac{p_0+p_z}{\sqrt{2}}, \frac{p_0-p_z}{\sqrt{2}}, \p_\perp\right)\,.
\end{align}
We choose the longitudinal direction such that $p^+$ is the largest scale in the problem and organize the dynamics as an expansion around the \emph{eikonal limit},
\begin{align}\label{eq:eikonal_hierarchy}
    p^+ \gg |\p_\perp|,\ q^+,\ q^-,\ |\q_\perp|\,,
\end{align}
where $q^\mu=(q^+,q^-,\q_\perp)$ denotes a typical momentum exchange with the medium. In the strict eikonal limit, the projectile propagates at fixed $p^+$ and fixed transverse position, while the medium only induces a colour rotation along its longitudinal trajectory. The BDMPS-Z formalism goes one step beyond this limit by retaining the leading correction that provides dynamics to the transverse degrees of freedom. This correction is associated to the location of the propagator pole 
\begin{align}
    p^- = \frac{\boldsymbol p^2+m^2}{2p^+}.
\end{align}
Concretely, while the dominant longitudinal motion is already encoded into the conserved large momentum $p^+$, the residual phase accumulated over light-cone time, $p^-x^+$, generates transverse evolution. Despite being subleading in eikonal accuracy, this correction is essential to describe medium-induced radiation, since it is sensitive to the finite transverse separation acquired by radiative fluctuations during propagation. In this way, it provides the minimal correction to strict eikonal propagation needed to capture finite-size effects and the LPM suppression pattern. In essence, the approximation has the structure of a paraxial expansion: $p^+$ remains a conserved label of the fast parton, while the relevant dynamics is concentrated in transverse and colour spaces\footnote{More precisely, light-front kinematics contains a transverse Galilean subgroup: $x^+$ plays the role of time, $\x_\perp$ that of space, and $p^+$ is the central charge, analogous to the nonrelativistic mass. Thus, within a sector of fixed large $p^+$, transverse boosts act as Galilean boosts and the residual dynamics can be organized as motion in the transverse plane. This is the field-theoretic content of the Galilean invariance of the infinite-momentum frame discussed in~\cite{Susskind:1967rg, Kogut:1969xa}. The interested reader is pointed to~\cite{Brodsky:1997de} for a more standard treatment of light-front dynamics.}. This expansion is natural in the context of jet propagation inside a coloured medium: the ensuing cascade is a collimated structure whose constituents all have, mostly, the same large momentum direction in common at leading eikonal power.

Let us now explicitly compute the in-medium propagation of a high-energy parton. At the level relevant for this problem, the underlying QCD medium is represented by a colour gauge field sourced by its microscopic degrees of freedom. We decompose it as
\begin{align}
    A_a^\mu(x) = A^\mu_{a,\,\rm med}(x) + \delta A_a^\mu(x),
\end{align}
where $A^\mu_{a,\,\rm med}$ denotes a background $c$-number field and $\delta A_a^\mu$ denotes quantum fluctuations around it, with $a=1,\,\cdots,\,N_c^2-1$ an adjoint colour index. In full QCD, the medium would be specified by a density matrix, and the propagating parton would couple to gauge field operators $A_a^\mu$ evaluated in that state. In the BDMPS-Z formalism, this quantum expectation value is instead approximated by an effective stochastic description: the propagation problem is first solved for a fixed classical $A_{a,\,\rm med}^\mu$ and physical quantities are then obtained by averaging over an ensemble of such background fields with a given weight functional. Diagrammatically, interactions with the medium are represented by external-field insertions on the energetic parton line. The contributions in which the fluctuations $\delta A_a^\mu$ propagate explicitly are not included in this leading fixed-background propagation problem. Rather, the unresolved structure of the medium is retained only to the extent that it is represented by the chosen ensemble of background fields. Thus, the microscopic input is supplied by the correlators of $A^\mu_{a,\,\rm med}$, which encode the medium properties resolved by the high-energy projectile and determine the transport information entering the BDMPS-Z description. In what follows, we shall denote $ \cA^\mu_a \equiv A_{a,\,\rm med}^\mu$ to simplify the notation.

The leading coupling to the background field is fixed by the high-energy hierarchy introduced in Eq.~\eqref{eq:eikonal_hierarchy}. We start by fixing the gauge of the background field to the light-cone gauge
\begin{align}
    n\cdot \cA_a = \cA_a^+ = 0\,,\qquad n^\mu = (0,1,0,0)\,,
\end{align}
which is naturally the convenient choice for a collimated jet propagating along the $(+)$ light-cone direction. Consider the $S$-matrix element for a parton with initial momentum $p$ and final momentum $p'$. We shall work with $T$-matrix elements $\cT_n(p',p)$ to represent the non-trivial part of the propagation between the two momentum states, i.e., the matrix element with $n\geq 1$ background field insertions in the propagating line. Further, for clarity of the notation, we shall omit the labels corresponding to quantum numbers, which in the case of a quark are the spin and fundamental colour indices. The leading power $T$-matrix element for a single background field insertion on a propagating quark line reads
\begin{align}\label{eq:M1}
    i\cT_1(p',p) = \bar u^r(p')\gamma^\mu u^s(p)\,\left(ig_s\left(\mathbf T_F^a\right)_{ij}\cA_{a,\mu}(p'-p)\right)
    \simeq \delta^{rs}\, 2p^+ (2\pi)\delta(p'^+-p^+)\left(ig_s t_{ij}^a\cA_a^-(\vec p'-\vec p)\right) \,,
\end{align}
where we assumed $\cA^-$ not to be parametrically suppressed with respect to $\boldsymbol{\cA}_\perp$ and used the fundamental representation of the $\mathbf T^a$ colour matrix. This result can be justified by first decomposing the spinors into their light-front $(+)$ and $(-)$ components~\cite{Kovchegov:2012mbw}
\begin{align}\label{eq:light_front_decomposition}
    u_+ = \Lambda^+ u\,,\quad u_- = \Lambda^- u\,,\qquad \Lambda^\pm = \frac{\gamma^\mp\gamma^\pm}{2}\,,
\end{align}
and observing that the non-vanishing bilinears are\footnote{The Dirac equation $(\slashed{p}-m)u(p) = 0$ constrains the $(-)$ component as~\cite{Brodsky:1997de}
\begin{align}
    u_-(p) = \gamma^+(\boldsymbol{\gamma}\cdot\p + m)/(2p^+)u_+ \,,
\end{align}
implying that it scales as $u_- \sim \lambda u_+$.}
\begin{align}\label{eq:bilinear_scaling}
    & \bar u_+\gamma^+u_+ \sim p^+\,,\nn
    & \bar u_+\boldsymbol{\gamma}^iu_-\,,\,\, \bar u_-\boldsymbol{\gamma}^iu_+ \sim \lambda p^+\,,\nn
    & \bar u_-\gamma^-u_- \sim \lambda^2 p^+\,,
\end{align}
where we defined the power counting $p^\mu = p^+(1, \lambda^2, \lambda, \lambda)$ and used the $\gamma$ matrix identities in light-cone coordinates in Appendix~\ref{app:light_cone_coord}. Thus, the $(+)$ component of the spinor current is the leading power contribution under the hierarchy in Eq.~\eqref{eq:eikonal_hierarchy}. The relation $\bar u^r(p') \gamma^+ u^s(p) \simeq \delta^{rs}\,2p^+$ then follows from the leading $p^+$ expansion of the Dirac spinors. The simple result of Eq.~\eqref{eq:M1} already shows two striking features of high-energy propagation in a medium: spin flip interactions are suppressed, i.e., the spin state of the quark remains unmodified after the interaction; at leading power the parton couples only to the $(-)$ component of the background gauge field, the field component conjugate to the large momentum component. The result is fully analogous for a propagating gluon, with conservation of its physical polarisation and colour rotation in the adjoint representation $f^{abc} \cA^{-}_c$. Finally, as we have already seen, at leading power one neglects longitudinal momentum transfers $q^+$. This implies, in particular, since $x^-$ is the conjugate variable to $q^+$, that the parton does not resolve any $x^- $ dependence of the medium, i.e., 
\begin{align}
    \cA_a^-(x) \simeq \cA_a^-(x^+,0,\x_\perp) \equiv \cA_a^-(\vec x)\,,\qquad \vec x \equiv (x^+, \x_\perp)\,.
\end{align}
In other words, a collimated jet only resolves a small range $\Delta x^- \sim 1/p^+$ near $x^- \simeq 0$, such that sensitivity to $x^-$ structures arises only in subleading corrections. In momentum-space this simplification is written as $\cA_a^-(q) \simeq (2\pi)\delta(q^+)\cA_a^-(\vec q)$ with the momentum vector defined with the $(-)$ component rather than the $(+)$ component in the coordinate vector, i.e., $\vec q \equiv (q^-, \q_\perp)$.

By inserting further momentum exchanges with the background field, one can hope to identify a pattern which can be summed exactly. The matrix element for double field insertion reads
\begin{align}
    i\cT_2(p',p) & = -g_s^2 t^b_{ik} t^a_{kj}(2\pi)\delta(p'^+-p^+)\int_{\vec l,\, \vec x_1,\, \vec x_2}e^{i(\vec p' - \vec l)\cdot \vec x_2}e^{i(\vec l - \vec p)\cdot \vec x_1}\nn 
    & \times \bar u(p')\gamma^\mu\cA_{b,\mu}(\vec x_2)\frac{i(\slashed{l}+m)}{l^2 - m^2 + i\epsilon}\gamma^\nu\cA_{a,\nu}(\vec x_1)u(p)\,,
\end{align}
where the abbreviations for momentum- and coordinate-space integration read
\begin{align}\label{eq:integration_abbreviations}
    \int_{\vec l, \vec x_i} \equiv \int_{-\infty}^{+\infty}\frac{dl^-}{2\pi}\int \frac{ d^2\l}{(2\pi)^2}\int_{-\infty}^{+\infty}dx_i^+\int d^2\x_i\,,
\end{align}
and we have dropped the $\perp$ subscript on transverse coordinates, since it is understood they are represented as bold letters. Note that in our notation one has $\vec p \cdot \vec x_1 = p^- x_1^+ - \p\cdot\x_1$. To be able to use the result in Eq.~\eqref{eq:M1} repeatedly, one starts by writing the intermediate off-shell momentum $l^\mu$ in terms of an on-shell momentum $\tilde l^\mu$ with the same large light-cone component\footnote{The definition of an on-shell momentum direction is typical in the context of computing collinear splitting functions, see, e.g.,~\cite{Catani:1999ss,Dhani:2023uxu}.}
\begin{align}\label{eq:on_shell_decomposition}
    l^\mu = \tilde l^\mu + \frac{l^2-m^2}{2l^+}n^\mu\,,
\end{align}
such that $\tilde l^\mu = (l^+, \frac{\l^2+m^2}{2l^+},\l)$ and, thus, $\tilde l^2 = m^2$. The contribution of second term to the propagator is often termed an \emph{instantaneous} contribution since it is local in light-cone time (see e.g.~\cite{Brodsky:1997de,Kovchegov:2012mbw} in the context of light-cone perturbation theory). When contracted with the remaining Dirac and Lorentz structure, this term contributes only at subleading eikonal order\footnote{It also does not contain propagator poles, since it is proportional to the off-shellness $l^2-m^2$ of the intermediate propagator. This implies it does not contribute to the light-cone time phases responsible for interference effects.}. Thus, to leading eikonal accuracy one can discard this term and use the spinor completeness relation for an on-shell momentum 
\begin{align}
    \slashed{l}+m \simeq \sum_{h=\pm 1} u^h(\tilde l)\bar u^h(\tilde l)\,,
\end{align}
such that one can use the result in Eq.~\eqref{eq:M1} for both field insertions and simplify the Dirac structure as
\begin{align}
    \bar u^r(p')\gamma^\mu (\slashed{l}+m)\gamma^\nu u^s(p)\cA_{b,\mu}(\vec x_2)\cA_{a,\nu}(\vec x_1) \simeq (2p^+)^2\cA^{-}_b(\vec x_2)\cA^{-}_a(\vec x_1)\delta^{rs}\,.
\end{align}
Thus, the matrix element for double field insertion simplifies to
\begin{align}\label{eq:M2_simplified}
    & i\cT_2(p',p) \simeq \nn
    & \delta^{rs} 2p^+(2\pi)\delta(p'^+-p^+)  \int_{\vec x_1, \vec x_2}e^ {i\vec p' \cdot \vec x_2}e^{-i\vec p \cdot \vec x_1}\left(ig_s (\mathbf T_F^b)_{ik}\cA^-_b(\vec x_2)\right)\cG_0^{km}(\vec x_2, \vec x_1|p^+)\left(ig_s(\mathbf T_F^a)_{mj}\cA^-_a(\vec x_1)\right)\,,
\end{align}
where we have defined, conveniently normalized, the vacuum scalar propagator at fixed $p^+$ as
\begin{align}\label{eq:free_propagator}
    \cG_0^{km}(\vec x_2, \vec x_1 | p^+) & = \delta^{km}\int_{l} e^{-i\vec l \cdot (\vec x_2 - \vec x_1)}\left(2l^+(2\pi)\delta(l^+-p^+)G_0(l)\right) \nn
    & = \delta^{km}\int_{\vec l} e^{-i\vec l \cdot (\vec x_2 - \vec x_1)}i\left[l^- - \left(\frac{\l^2 + m^2}{2p^+} - i\epsilon\right)\right]^{-1}\,,
\end{align}
where $G_0(l)$ is the standard scalar propagator. The picture suggested by Eq.~\eqref{eq:M2_simplified} is quite simple: in-medium high energy propagation proceeds through scalar insertions, in the Dirac and Lorentz sense, of the $\cA^-$ field component, with free scalar propagation between each insertion. For a propagating gluon, the resulting structure is exactly the same. To see that, let's start with single insertion matrix element which contains the three-gluon vertex contracted with three polarisation vectors
\begin{align}\label{eq:M1_gluon}
    i\cT_1(p',p) = g_sf^{abc}\left(-2(\epsilon_{\lambda}\cdot \cA_b)(\epsilon_{\lambda'}^{\ast}\cdot q)+2(\epsilon_{\lambda'}^{\ast}\cdot \cA_b)(\epsilon_{\lambda}\cdot q) - (\epsilon_\lambda\cdot \epsilon_{\lambda'}^{\ast})(p'+p)\cdot \cA_b\right)\,,
\end{align}
where the momentum arguments of $\cA^b(p'-p)$, $\epsilon_\lambda(p)$ and $\epsilon_{\lambda'}^{\ast}(p')$ were omitted for clarity and we used the transversality condition $p\cdot \epsilon_\lambda = 0$ of physical polarisation vectors. The latter is further constrained by $n\cdot\epsilon_\lambda = 0$ in light-cone gauge, which implies it can be written as 
\begin{align}
    \epsilon_\lambda^\mu = \left(0,\, \frac{\p\cdot\beps_\lambda}{p^+},\, \beps_\lambda\right)\,.
\end{align}
Using the same power counting as in Eq.~\eqref{eq:bilinear_scaling}, $p^\mu = p^+(1, \lambda^2, \lambda, \lambda)$, the dominant term in Eq.~\eqref{eq:M1_gluon} comes from the last term, which can be simplified at leading eikonal power as
\begin{align}\label{eq:M1_gluon_simplified}
    i\cT_1(p',p) \simeq 2p^+ g_s f^{abc}(\beps_\lambda\cdot\beps_{\lambda'}^{\ast})\cA_a^- (p'-p)\simeq \delta^{\lambda\lambda'}2p^+(2\pi)\delta(p'^+-p^+)\left(g_s f^{abc}\cA^-_a(\vec p'-\vec p)\right)\,,
\end{align}
where in light-cone gauge $\epsilon_\lambda(p)\cdot\epsilon_{\lambda'}^{\ast}(p') = -\beps_\lambda\cdot\beps_{\lambda'}^{\ast}$ and we work with orthonormal transverse polarisation vectors $\beps_\lambda\cdot\beps_{\lambda'}^{\ast} = \delta^{\lambda\lambda'}$. Thus, the single insertion amplitude on a gluon line reads the same as for a quark line in Eq.~\eqref{eq:M1}, with conservation of gluon polarisation and adjoint colour rotation, with the replacement $(\mathbf{T}_F^a)_{ij} \to (\mathbf{T}_A^a)_{bc} = -if^{abc}$. For the double insertion matrix element, there is an intermediate gluon propagator\footnote{To be exact, at the same order in $g_s$, there are further contact terms in which the propagating gluon couples to two background fields through the four-gluon vertex. However, since they contain no nearly on-shell intermediate propagation and no additional enhancement by the large momentum $p^+$, they are subleading in eikonal accuracy.}, which in light-cone gauge reads
\begin{align}\label{eq:vacuum_gluon_propagator}
    D_0^{\mu\nu}(p) = \frac{id^{\mu\nu}(p)}{p^2+i\epsilon}\,,\qquad d^{\mu\nu}(p) = -g^{\mu\nu} + \frac{p^\mu n^\nu + p^\nu n^\mu}{p^+}\,.
\end{align}
The projector $d^{\mu\nu}(p)$ can be written in terms of polarisation vectors by using the same on-shell decomposition as in Eq.~\eqref{eq:on_shell_decomposition} and the completeness relation, leading to
\begin{align}\label{eq:gluon_projector_decomposition}
    d^{\mu\nu}(p) = d^{\mu\nu}(\tilde p) + \frac{p^2}{(p^+)^2}n^\mu n^\nu = \sum_{\lambda=\pm 1}\epsilon_\lambda^{\mu}(\tilde p)\epsilon_\lambda^{\ast \nu}(\tilde p) +\frac{p^2}{(p^+)^2}n^\mu n^\nu\,,
\end{align}
where we isolate the instantaneous term, analogous to the quark case, which we shall discard as a subleading eikonal correction. Once again, this allows us to use the result in $\cT_1$ for each insertion, leading to the same result as in Eq.~\eqref{eq:M2_simplified} replacing $(\mathbf{T}_F^a)_{ij} \to (\mathbf{T}_A^a)_{bc} = -if^{abc}$ and setting $m=0$ in Eq.~\eqref{eq:free_propagator}.

Now that we have established that quark and gluon lines share the same scalar in-medium propagation, with the appropriate color structure attached to $\cA^-_a$, let us further inspect Eq.~\eqref{eq:M2_simplified}. As mentioned at the beginning of this section, in the BDMPS-Z formalism one keeps the residual minus-component during propagation, which supplies an important subleading eikonal correction that is necessary to describe essential interference effects. This statement can now be made explicit by computing the free coordinate-space propagator in Eq.~\eqref{eq:free_propagator}. The integration over $l^-$ picks up a pole in the complex plane, such that we have
\begin{align}
    \cG^{km}_0(\vec x_2, \vec x_1|p^+
    ) = \delta^{km}\,\Theta(x_2^+-x_1^+)\int_\l e^{i\l\cdot(\x_2-\x_1)}\exp{\left[-i\frac{\l^2+m^2-i\epsilon}{2p^+}(x_2^+-x_1^+)\right]}\,,
\end{align}
where forward propagation in light-cone time becomes explicit since we consider $p^+>0$. This constrains the light-cone time integrations in Eq.~\eqref{eq:integration_abbreviations}. The $i\epsilon$ prescription accounts for the damping of propagation over arbitrarily large light-cone times, such that in every cross-section computation one takes $\epsilon\to 0^+$ after all time integrations are performed. Now note that if we were to take the strict eikonal limit, then the $x_2^+-x_1^+$ phase would be discarded, freezing the transverse coordinate of the propagating parton through $\delta^{(2)}(\x_2-\x_1)$ after the integration over $\l$. Thus, to keep the leading transverse motion one must keep subleading corrections stemming from the light-cone time phases which are accumulated over a finite interaction length of the medium. It is now also clear that, upon squaring this amplitude, these phases are responsible for the interference between momentum exchanges at different field insertions, being essential to describe LPM interference and finite-size effects.

Looking at Eq.~\eqref{eq:M2_simplified}, it is clear that the structure can be straightforwardly generalized at leading power to $n$ field insertions. Schematically, this reads
\begin{align}\label{eq:Mn_simplified}
    & i\cT_n(p',p)\simeq\nn
    & \delta^{rs} 2p^+(2\pi)\delta(p'^+-p^+)  \int_{\vec x_1,\,\cdots,\, \vec x_n}e^ {i\vec p' \cdot \vec x_n}e^{-i\vec p \cdot \vec x_1}\left(ig_s \cA^-_n\right)(\cG_0)_{n,n-1}\left(ig_s \cA^-_{n-1}\right)\cdots\,(\cG_0)_{2,1}\left(ig_s\cA^-_{1}\right)\,,
\end{align}
where we used the abbreviations $\cA_n^- = (\mathbf T_R^a) \cA_a^-(\vec x_n)$, with $(\mathbf T_F^a)_{ij} = t_{ij}^a$ for a quark line and $(\mathbf T_F^a)_{bc} = -if^{abc}$ for a gluon line, and $(\cG_0)_{n,n-1} = \cG_0(\vec x_n, \vec x_{n-1}|p^+)$, $r$ and $s$ represent either quark spin projection states or gluon polarisation states and all objects carry implicit, fundamental or adjoint, color indices. The $T$-matrix elements above are amputated with respect to the external scalar propagators. To reconstruct the full two-point Green function, one attaches the external free scalar propagators in both initial and final states and then sums over all numbers of background-field insertions. Hence, the in-medium \emph{scalar} propagator reads
\begin{align}\label{eq:full_propagator_definition}
    G(p',p) = G_0(p',p) + \sum_{n=1}^\infty G_0(p')\left(i\cT_n(p',p)\right)G_0(p) \,,
\end{align}
where $G_0(p',p) =(2\pi)^4\delta^{(4)}(p'-p)G_0(p)$. Similarly to Eq.~\eqref{eq:free_propagator}, one can define a momentum-space in-medium scalar propagator at fixed $p^+$, 
\begin{align}\label{eq:fixedp+_scalar_momentum_propagator}
    G(p',p) = (2\pi)\delta(p'^+-p^+)\cG(\vec p',\vec p|p^+)/(2p^+)\,.    
\end{align}
Together with Eq.~\eqref{eq:Mn_simplified}, this definition leads to the following Dyson series in momentum-space
\begin{align}\label{eq:Dyson_series}
    \cG = \cG_0 + \cG_0 (ig_s \cA^-)\cG_0 +  \cG_0(ig_s \cA^-)\cG_0 (ig_s \cA^-)\cG_0+\cdots\,,
\end{align}
where convolutions in momentum-space were omitted for clarity and $\cA^- = (\mathbf T_R^a) \cA_a^-$ is a colour matrix in representation $R$. By Fourier transforming to the coordinate-space scalar propagator at fixed $p^+$, one can express Eq.~\eqref{eq:Dyson_series} equivalently as a Dyson equation in coordinate-space for the propagation between two fixed endpoints $\vec x$ and $\vec y$
\begin{align}\label{eq:Dyson_eq}
    \cG(\vec y, \vec x| p^+) = \cG_0(\vec y, \vec x |p^+) + \int_{\vec z}\cG_0(\vec y, \vec z|p^+)(ig_s \cA^-(\vec z))\cG(\vec z, \vec x|p^+)\,.
\end{align}
As per Eq.~\eqref{eq:free_propagator}, the fixed $p^+$ free propagator $\cG_0(\vec y, \vec x|p^+)$ is related to the standard free propagator defined in four-dimensional coordinate-space as
\begin{align}\label{eq:standard_propagator}
    G_0(y,x) = \int \frac{dp^+}{2\pi}e^{-ip^+(y^--x^-)}\frac{\cG_0(\vec y, \vec x|p^+)}{2p^+}\,,
\end{align}
which satisfies the free Klein-Gordon equation,
\begin{align}
    (\Box + m^2)G_0(y,x) = -i\delta^{(4)}(y-x)\,,
\end{align}
where the $i\epsilon$ factor from the Feynman prescription is implicit. By applying $\Box + m^2 =  2\partial_+\partial_- -\boldsymbol{\nabla}^2+ m^2$ to the $y$ coordinate in Eq.~\eqref{eq:standard_propagator}, one can show that $\cG_0(\vec y, \vec x|p^+)$ satisfies, at fixed $p^+$, the free Schr\"odinger equation with $y^+$ as the time-evolution variable, i.e.,
\begin{align}
    \left(i\frac{\partial}{\partial y^+} + \frac{\boldsymbol{\nabla_y}^2-m^2}{2p^+}\right)\cG_0(\vec y, \vec x|p^+) = i\delta(y^+-x^+)\delta^{(2)}(\y-\x)\,.
\end{align}
Thus, applying the differential operator of the free Schr\"odinger equation to the $\vec y$ coordinate in Eq.~\eqref{eq:Dyson_eq} leads to
\begin{align}
    \left(i\frac{\partial}{\partial y^+} + \frac{\boldsymbol{\nabla_y}^2-m^2}{2p^+}+g_s \cA^-(\vec y)\right)\cG(\vec y, \vec x| p^+) = i\delta(y^+-x^+)\delta^{(2)}(\y-\x)\,,
\end{align}
which is the Schr\"odinger equation in a coloured potential, with evolution variable $y^+$. The corresponding Hamiltonian is composed of the kinetic term $\left(-\boldsymbol{\nabla_y}^2 + m^2\right)/(2p^+)$, where $p^+$ plays the role analogous to a non-relativistic inertial mass, and a coloured potential $-g_s\cA^- = -g_s (\mathbf T_R^a)\cA^-_a $. The mass is set as $m=0$ for the gluon case. The solution to this equation can be written in path integral form, which is suitable for computing squared amplitudes in perturbation theory. In a given representation $R$, it reads\footnote{The normalization of $\cD\r$ is fixed by requiring the expression to reduce to the free scalar propagator in the limit $\cA^-\to 0$.}
\begin{align}\label{eq:BDMPSZ_path_integral}
    & \cG_R^{ij}(y^+, \y; x^+, \x | p^+) = \nn
    & \Theta(y^+-x^+)e^{-i\left(\frac{m^2}{2p^+}-i\epsilon\right)(y^+-x^+)}\int_{\r(x^+) = \x}^{\r(y^+) = \y}\cD \r(s^+)\exp\left\{\frac{ip^+}{2}\int_{x^+}^{y^+}ds^+ \dot\r^2\right\}U^{ij}_R(y^+,x^+, \r)\,,
\end{align}
where the object $U_R^{ij}$ is a path-ordered Wilson line in representation $R$ given by
\begin{align}\label{eq:Wilson_line_def}
    U_R^{ij}(y^+,x^+, \r(s^+)) = \cP\exp\left\{ig_s \int_{x^+}^{y^+}ds^+ \cA_a^-(s^+, \r(s^+))(\mathbf T_R^a)_{ij}\right\}\,,
\end{align}
where $(\mathbf T_F^a)_{ij} = t^a_{ij}$ in the fundamental representation ($R=F$, quark), $(\mathbf T_{\bar F}^a)_{ij} = -(t^a)_{ij}^\ast = -t^a_{ji}$ in the anti-fundamental representation ($R=\bar F$, anti-quark) and $(\mathbf T_A^a)_{bc} = -if^{abc}$, along with $m=0$, in the adjoint representation ($R=A$, gluon). We shall denote the anti-quark propagator by $\bar \cG_F\equiv \cG_{\bar F}$. In later chapters, we will identify anti-fundamental colour indices with oppositely oriented fundamental indices. In this colour-flow convention, $\bar \cG_F^{ij}$ carries $i$ as the initial colour index and $j$ as the final one, in contrast with the quark propagator $\cG_F^{ij}$. With this convention, $\bar \cG_F^{ij}$ is equal to the fundamental result of Eq.~\eqref{eq:BDMPSZ_path_integral} with the replacement $U_F \to U^\dagger_F$.  Note that we have left  the factor $i\epsilon$, inherited from the Feynman prescription on the propagator (see Eq.~\eqref{eq:free_propagator}), explicit. It is understood that one takes $\epsilon\to0^+$ after the light-cone-time integrations have been performed. The result of Eq.~\eqref{eq:BDMPSZ_path_integral} makes explicit the statement we put forward at the start of this section, based on the expansion around the eikonal limit -- the dynamics of a high-energy parton propagating in a background field is constrained to transverse space, described by the path integral over the transverse coordinate, and colour space, represented by the Wilson line which generates continuous precession of its colour state.

\subsection{Effective Feynman rules for in-medium amplitudes}\label{sec:feynman_rules}

In the previous subsection, we derived the scalar propagator for a high-energy particle propagating in a background field. Since we derived it starting from $T$-matrix elements, we effectively projected out its Dirac and Lorentz structure, ending up only with its scalar part. This was useful to explicitly see how, to leading power in the eikonal limit, the parton couples only to the $(-)$ component of the field, how the relevant object that appears in the calculation is the fixed $p^+$ propagator, and how the relevant subeikonal corrections arise in the phases of intermediate free propagators. However, one should compute the Dirac and Lorentz structure of the in-medium propagator since it will be inserted into perturbative diagrammatic calculations. In general, one can write the quark propagator as
\begin{align}\label{eq:full_fermionic_propagator_definition}
    S(p',p) = S_0(p',p) + \sum_{n=1}^\infty S_0(p')\left(i\cK_n(p',p)\right)S_0(p)\,,
\end{align}
where $S_0(p',p) = (2\pi)^4\delta^{(4)}(p'-p) S_0(p)$, $S_0(p) = (\slashed{p}+m)G_0(p)$ is the usual quark propagator and the open-index amputated kernel with $n$ insertions $\cK_n$ is related to the corresponding $T$-matrix element $\cT_n$ as
\begin{align}
    i\cT_n = \bar u(p')\left(i\cK_n(p',p)\right)u(p)\,.
\end{align}
However, as per Eq.~\eqref{eq:M1}, the eikonal limit renders the Dirac structure of the medium correction simple. In general, this correction reads
\begin{align}\label{eq:medium_correction_S}
    S_0(p')i\cK_n(p',p)S_0(p) = \int_{p_1,\cdots, p_{n-1}}S_0(p')\left(ig_s \slashed{\cA}(p'-p_{n-1})\right)S_0^{(n-1)}\cdots\, S_0^{(1)}\left(ig_s \slashed{\cA}(p_1-p)\right)S_0(p)\,, \quad n>0\,,
\end{align}
where $S_0^{(i)} = S_0(p_i)$ and $\slashed{\cA} = \mathbf T^a_F\slashed{\cA}_a$ is a matrix in fundamental colour space. Using the on-shell decomposition in Eq.~\eqref{eq:on_shell_decomposition} and the spinor completeness relation, the fermionic propagator reads
\begin{align}\label{eq:fermion_prop_PI_decomp}
    S_0(p) = \sum_{r}u^r(\tilde p)\bar u^r(\tilde p)G_0(p)  + \frac{i\gamma^+}{2p^+}  \equiv S_0^P(p)+S_0^I(p)
\end{align}
where we identified its propagating ($P$) and instantaneous ($I$) parts. Inserting this sum into Eq.~\eqref{eq:medium_correction_S}, the contribution from considering only the propagating part of each intermediate propagator is straightforward to compute at leading power since it is just an iteration of the spin-conserving result in Eq.~\eqref{eq:M1}. This leads to
\begin{align}\label{eq:propagating_correction}
    \left[S_0(p')i\cK_n(p',p)S_0(p)\right]_P \simeq \sum_{r=\pm 1} u^r(\tilde p')\bar u^r(\tilde p) G_0(p') (i\cT_n(p',p))G_0(p)\,, \quad n>0\,,
\end{align}
where we have used Eq.~\eqref{eq:fixedp+_scalar_momentum_propagator} to identify $\cT_n$ from Eq.~\eqref{eq:Mn_simplified}. As for terms containing instantaneous contributions, one can show that any chain containing consecutive $S_I$ adjacent to the same field insertion vanishes exactly in light-cone gauge and follow the identities in Appendix~\ref{app:light_cone_coord}. Either way, the non-vanishing contributions, which contribute with a non-trivial spin structure, are suppressed by at least one power of $1/p^+$ and do not contribute with propagator poles. Thus, the dominant medium correction comes from Eq.~\eqref{eq:propagating_correction}, such that the in-medium fermionic propagator at leading power reads
\begin{align}\label{eq:S_final}
    S(p',p) \simeq i\frac{\gamma^+}{2p^+}(2\pi)^4\delta^{(4)}(p'-p) + \sum_r u^r(\tilde p')\bar u^r(\tilde p)G(p',p)
\end{align}
where we identified $G(p',p)$ through its definition in terms of $\cT_n$ in Eq.~\eqref{eq:full_propagator_definition} and the first term is the instantaneous contribution to the vacuum propagator. In what follows, we will work only with the propagating near-on-shell part of the high-energy line and therefore drop this vacuum instantaneous term as well.\footnote{Effectively, we are taking the quark to be nearly on-shell, i.e., $(p^2-m^2)/(p^+)^2 \ll 1$. This implies we will not be able to properly describe hard and wide-angle radiation off it, i.e., for a parton with momentum $p$ its virtuality scales as $p^2 \sim z(1-z)(p^+)^2(1-\cos\theta)$ up to mass terms, such that in the limit $(p^2-m^2)/(p^+)^2 \ll 1$ the approximation is controlled only for collinear ($\theta \ll 1$) and/or soft radiation ($z\ll 1$ or $1-z\ll1$).}

For a propagating gluon, one writes
\begin{align}
    D^{\mu\nu}(p',p) = D_0^{\mu\nu}(p',p) + \sum_{n=1}^\infty D_0^{\mu\rho}(p') \left(i\cK_n(p',p)\right)_{\rho\sigma}D_0^{\sigma\nu}(p)
\end{align}
where $D_0^{\mu\nu}(p',p) = (2\pi)^4\delta^{(4)}(p'-p)D_0^{\mu\nu}(p)$, $D_0^{\mu\nu}(p)$ has been defined in light-cone gauge in Eq.~\eqref{eq:vacuum_gluon_propagator} and the amputated kernel relates to the corresponding $T$-matrix element as
\begin{align}
    i\cT_n(p',p) = \epsilon^{\ast,\rho}_{\lambda'}(p')\left(i\cK_n(p',p)\right)_{\rho\sigma}\epsilon_{\lambda}^\sigma(p)\,.
\end{align} 
In general, suppressing momentum and colour labels, the medium correction schematically reads
\begin{align}
    D_0^{\mu\rho}(p') \left(i\cK_n(p',p)\right)_{\rho\sigma}D_0^{\sigma\nu}(p) \sim \int D_0^{\mu\rho}\cV_{\rho\alpha_{n}\mu_{n}}\cA^{\alpha_{n}}D^{\mu_{n}\nu_{n}}_0 \cV_{\nu_{n}\alpha_{n-1}\mu_{n-1}}\cdots\,D_0^{\mu_{2}\nu_{2}}\cA^{\alpha_1}\cV_{\nu_2\alpha_1\sigma}D_0^{\sigma\nu}\,,
\end{align}
for $n>0$ and where $\cA^{\alpha_i} = \mathbf T^a_A\cA_a^{\alpha_i}$ is a matrix in adjoint colour space and $\cV_{\mu\alpha\nu}$ is the three-gluon vertex. Similarly to the fermionic vacuum propagator, by using Eq.~\eqref{eq:gluon_projector_decomposition} the gluonic vacuum propagator can be decomposed as
\begin{align}\label{eq:gluon_prop_PI_decomp}
    D_0^{\mu\nu}(p) = \sum_{\lambda=\pm 1}\epsilon_\lambda^{\mu}(\tilde p)\epsilon_\lambda^{\ast \nu}(\tilde p)D_0(p) +i\frac{n^\mu n^\nu}{(p^+)^2} \equiv D_0^P(p) + D_0^I(p)\,.
\end{align}
where $D_0(p) = i/(p^2+i\epsilon)$. As for the previous derivation, the propagating part of the medium correction is straightforwardly given at leading power by iterating Eq.~\eqref{eq:M1_gluon_simplified},
\begin{align}
    \left[D_0^{\mu\rho}(p') \left(i\cK_n(p',p)\right)_{\rho\sigma}D_0^{\sigma\nu}(p)\right]_P \simeq \sum_{\lambda}\epsilon_\lambda^\mu(\tilde p')\epsilon_\lambda^{\ast, \nu}(\tilde p)D_0(p')(i\cT_n(p',p))D_0(p)\,,\qquad n>0\,,
\end{align}
while instantaneous contributions, which have a non-trivial polarisation structure, contribute with additional suppression by $1/p^+$ and with no propagator poles. Thus, at leading power, the in-medium gluon propagator reads
\begin{align}\label{eq:D_final}
    D^{\mu\nu}(p',p) \simeq i\frac{n^\mu n^\nu}{(p^+)^2}(2\pi)^4\delta^{(4)}(p'-p) + \sum_{\lambda}\epsilon_\lambda^\mu(\tilde p')\epsilon_\lambda^{\ast, \nu}(\tilde p)G(p',p)\,,
\end{align}
The first term is the instantaneous contribution to the vacuum propagator, analogous to the one for the quark propagator in Eq.~\eqref{eq:S_final}, which we will discard.

Since the QCD vertices remain unchanged by the medium, as they are local in light-cone time, the remaining question is how in-medium propagation is included in external lines. This is answered by the standard Lehmann-Symanzik-Zimmerman (LSZ) reduction formula (see, e.g.,~\cite{Peskin:1995ev}). For an outgoing quark produced from some subamplitude $\cH$ with momentum $q$, one includes it in the diagram as
\begin{align}\label{eq:LSZ}
    \lim_{p'^2\to m^2}\bar u^r(p')S_0^{-1}(p')\int_q S(p',q)\cH(q,\cdots) & =\sum_s \lim_{p'^2\to m^2}G_0^{-1}(p')\int_q G(p',q)\bar u^r(p')\frac{\gamma^+}{2p'^+}u^s (\tilde p')\bar u^s (\tilde q)\cH(q,\cdots)\nn
    & = \lim_{p'^2\to m^2}G_0^{-1}(p')\int_q G(p',q)\bar u^r (\tilde q)\cH(q,\cdots)\,,
\end{align}
where $\tilde q^\mu = q^\mu - (q^2-m^2)/(2q^+)n^\mu$, or $\tilde q^\mu = (q^+,\frac{\q^2+m^2}{2q^+},\q)$, and the $\cdots$ in the subamplitude $\cH$ represent the dependence on the momenta and quantum numbers of the remaining outgoing and ingoing particles. The inverse vacuum propagator was expanded as in Eq.~\eqref{eq:fermion_prop_PI_decomp}, with only the second term giving a non-vanishing contribution under the limit. We have further simplified the Dirac algebra under the on-shell limit using the spinor normalization
\begin{align}\label{eq:vertex_simplification_LSZ}  
    \lim_{p'^2\to m^2} \bar u^r(p')\frac{\gamma^+}{2p'^+}u^s(\tilde p') = \delta^{rs}\,.
\end{align}
A completely analogous formula holds for an ingoing quark and for anti-quarks (replacing $u(p)$ with $v(p)$). For an outgoing gluon the formula reads~\footnote{Here we use the free inverse gluon propagator before fixing the gauge
\begin{align}
    \left(D_0^{-1}\right)_{\mu\alpha}(p') = (-i)\left(p'^2g_{\mu\alpha}-p'_\mu p'_\alpha + \text{gauge-fixing terms}\right)
\end{align}
where the gauge-fixing terms are proportional to $n^\mu$, which vanish when contracted with $D^{\mu\nu}(p')$. In light-cone gauge, one further has $p'\cdot\epsilon_\lambda (\tilde p') = 0$, so only the $g_{\mu\alpha}$ survives.}
\begin{align}\label{eq:LSZ_gluon}
    & \lim_{p'^2 \to 0}\epsilon^{\ast,\mu}_\lambda(p')(D_0^{-1})_{\mu\alpha}(p')\int_q D^{\alpha\nu}(p',q)\cH_{\nu,\,\cdots}(q,\cdots) \nn
    & = \sum_{s}\lim_{p'^2 \to 0}G_0^{-1}(p')\int_q G(p',q) \epsilon_{\alpha,\lambda}^\ast(p')\epsilon_s^\alpha(\tilde p')\epsilon^{\ast,\nu}_s(\tilde q)\cH_\nu(q,\cdots)\nn
    & = \lim_{p'^2 \to 0}G_0^{-1}(p')\int_q G(p',q) \epsilon^{\ast, \nu}_{\lambda}(\tilde q)\cH_\nu(q,\cdots)\,,
\end{align}
where in the last line we used the orthonormality of polarisation vectors in the on-shell limit
\begin{align}
    \lim_{p'^2\to 0}\epsilon_{\alpha,\lambda}^\ast(p')\epsilon_s^\alpha(\tilde p') =\delta^{s\lambda}\,.
\end{align}
In typical perturbative calculations in the BDMPS-Z formalism, it is convenient to make the light-cone time dependence explicit to connect it directly to the longitudinal structure of the medium. We shall then work with the scalar propagator at fixed $p^+$ in the mixed space of light-cone time and transverse momentum, i.e., we define
\begin{align}
    G(p',p) = (2\pi)\delta(p'^+-p^+)\int_{x^+,y^+}e^{ip'^-y^+}e^{-ip^-x^+}\cG(y^+, \p'; x^+, \p\,|\,p^+)/(2p^+)\,,
\end{align}
which is related to the coordinate-space propagator formulated as path integral in Eq.~\eqref{eq:BDMPSZ_path_integral} by a Fourier transform in transverse space. The pole residue calculation from LSZ reduction in Eqs.~\eqref{eq:LSZ} and~\eqref{eq:LSZ_gluon} is more straightforwardly defined in this space 
\begin{align}
    & \lim_{p'^2\to m^2}G_0^{-1}(p')\int_q G(p',q)(\cdots)\nn
    & \simeq (-i)\lim_{p'^-\to \Omega_{\p'}}(p'^- - (\Omega_{\p'} - i\epsilon))\int_{ q,x^+,y^+}e^{ip'^-y^+}e^{-iq^-x^+}(2\pi)\delta(p'^+-q^+)\cG(y^+,\p';x^+,\q|p'^+)(\cdots) \nn
    & = \lim_{y_\infty^+ \to \infty}\int_{\q,x^+}e^{i\Omega_{\p'}y_\infty^+}\cG(y_\infty^+,\p';x^+,\q|p'^+)\int_{q^-}e^{-iq^-x^+}(\cdots)_{q^+=p'^+}\,,
\end{align}
with $\Omega_{\p'} = (\p'^2+m^2)/(2p^+)$, with $m=0$ for gluons. In the second equality in Eq.~\eqref{eq:LSZ_time} we assumed that at sufficiently large time one recovers an asymptotic state, i.e., that medium interactions cease asymptotically\footnote{Here we also used the composition law of the propagator for $x^+<z^+<y^+$
\begin{align}\label{eq:propagator_composition_law}
    \cG(y^+,\p';x^+, \q|p^+) = \int_\l \cG(y^+,\p';z^+, \l|p^+)\cG(z^+,\l;x^+, \q|p^+)\,,
\end{align}
which can be derived from, for instance, Eq.~\eqref{eq:Dyson_eq}.},
\begin{align}\label{eq:LSZ_time}
    \cG(y^+,\p';x^+, \q|p^+)) \xrightarrow{\rm LSZ\, pole} & \lim_{y_\infty^+ \to \infty} \int_{\k}\cG_0(y^+, \p';y_\infty^+, \k)\cG(y_\infty^+, \k; x^+, \q|p^+)\nn
    = & \lim_{y_\infty^+ \to \infty} \Theta(y^+-y_\infty^+)e^{-i(\Omega_{\p'}-i\epsilon)(y^+-y_\infty^+)}\cG(y_\infty^+, \p'; x^+, \q|p^+)\,,
\end{align}
where $\xrightarrow{\rm LSZ \, pole}$ indicates that the replacement is to be understood at the level of residue calculation, i.e., corrections to this form do not generate additional poles in $p^- = \Omega_{\p'}-i\epsilon$, and we have used the Fourier transform of the free propagator at fixed $p^+$ in Eq.~\eqref{eq:free_propagator}. In any case, dependence on $y_\infty^+$ drops out, in principle, in cross-sections or, at the level of the amplitude, for finite length media. 

The results above imply that, at leading eikonal power, the medium modifies only the scalar propagation, such that the amplitude for an in-medium process  can be reduced to momentum and spin/polarisation convolutions of scalar in-medium propagators with vertex matrix elements carrying spin/polarisation information and fully contracted in Dirac and Lorentz structure\footnote{For a diagram with multiple emissions, this simplification in terms of convolutions of scalar functions corresponds to treating the emissions as an ordered sequence of lower-point splittings. In vacuum, this boils down to discarding the instantaneous term in the propagator in Eqs.~\eqref{eq:fermion_prop_PI_decomp} and~\eqref{eq:gluon_prop_PI_decomp}. This requires, beyond the leading eikonal expansion of each propagating line, an additional ordering approximation, e.g., in virtuality. We shall assume this further ahead in Section~\ref{sec:ch5_spin_correlations}. Without such an ordering, interference effects must be taken into account -- see, e.g., the full $1\to3$ collinear splitting functions and their strongly ordered, iterated $1\to2$ limits in vacuum~\cite{Somogyi:2005xz, Braun-White:2022rtg}, and similarly in medium~\cite{Fickinger:2013xwa}.}. The core reason for this is that, after dropping the instantaneous vacuum term,  the Dirac/Lorentz structure of the leading power in-medium propagator in Eqs.~\eqref{eq:S_final} and~\eqref{eq:D_final}, is factorised into external spin/polarisation projectors, 
\begin{align}
    S(p',p) \simeq \sum_r u^r(\tilde p')\bar u^r(\tilde p)G(p',p)\,, \qquad D^{\mu\nu}(p',p) \simeq \sum_{\lambda = \pm 1}\epsilon_\lambda^\mu(\tilde p')\epsilon_\lambda^{\ast, \nu}(\tilde p)G(p',p) \,,
\end{align}
which are then contracted with neighbouring QCD vertices or put on-shell according to Eqs.~\eqref{eq:LSZ} and~\eqref{eq:LSZ_gluon}. For instance for an internal quark line in between two subamplitudes one has
\begin{align}\label{eq:internal_line_rule}
    i\cT & = \int_{p',p}\cH_{\rm out}(\cdots, p')S(p',p)\cH_{\rm in}(p,\cdots) \nn
    & \simeq \int_{p'^+,\,\p',\,y^+,\,\p,\,x^+}H_{\rm out}(\cdots,p'^+,y^+,\p')\left[\sum_r u^r(\tilde p')\frac{\cG(y^+, \p';x^+,\p|p'^+)}{2p'^+}\bar u^r(\tilde q)\right]H_{\rm in}(p'^+,x^+,\p,\cdots)\,,
\end{align}
where
\begin{align}\label{eq:H_in_out_minus_tf}
    H_{\rm in}(p^+,x^+,\p,\cdots) = \int_{p^-}e^{-ip^-x^+}\cH_{\rm in}(p,\cdots)\,,\qquad H_{\rm out}(\cdots,p'^+,y^+,\p') = \int_{p'^-}e^{ip'^-y^+}\cH_{\rm out}(\cdots, p')\,.
\end{align}
Both objects will then be contracted either with further propagators with the same structure or with, respectively, ingoing and outgoing spinors and polarisation vectors following Eqs.~\eqref{eq:LSZ} and ~\eqref{eq:LSZ_gluon}. The expression for an internal gluon line is completely analogous using $D^{\mu\nu}(p',p)$ given in Eq.~\eqref{eq:internal_line_rule}. Finally, regarding the vertices, since they are local in light-cone time, most integrations over it are effectively trivialized. For instance, considering the quark-gluon vertex with endpoints $q(p_0^+,x_0^+,\p_0) \to q(p_1^+,x_1^+,\p_1)+g(p_2^+,x_2^+,\p_2)$, the integrations over light-cone times and over the $(-)$ component of momenta coming from Eq.~\eqref{eq:LSZ_time} and from the $(-)$ momentum integration in the $H_{\rm out}$ in Eq.~\eqref{eq:H_in_out_minus_tf} read
\begin{align}
    & \int_{p_1^-, p_2^-, p_0^-,x_1^+,x_2^+}(ig_st^a_{ij}\gamma^\mu (2\pi)^4\delta^{(4)}(p_1+p_2-p_0))e^{ip_0^- x_0^+}e^{-ip_1^- x_1^+}e^{-ip_2^- x_2^+} (\cdots)(x_0^+,x_1^+,x_2^+)\nn
    & = (2\pi)\delta(p_1^++p_2^+-p_0^+)(2\pi)^2\delta^{(2)}(\p_1+\p_2-\p_0)(ig_st^a_{ij}\gamma^\mu)(\cdots)(x_0^+,x_1^+=x_0^+,x_2^+=x_0^+)\,.
\end{align}
Explicitly, this means the propagator lines all connect at the same light-cone time in a given vertex. Further, the remaining $\delta^{(2)}$ function in transverse momenta removes one of the transverse integrations over the three intervening momenta, while the $\delta$ function in $(+)$ momentum removes the remaining integration over $(+)$ momentum in Eq.~\eqref{eq:internal_line_rule}. Thus, the results in Eqs.~\eqref{eq:LSZ},~\eqref{eq:LSZ_gluon},~\eqref{eq:LSZ_time} and~\eqref{eq:internal_line_rule} provide us with effective Feynman rules in mixed $(x^+,\p)$ space, with $p^+$ conserved throughout the whole process, with all lines labeled by on-shell momenta (we shall drop the overhead $\,\,\Huge\tilde{}\,\,$ on momenta in Eqs.~\eqref{eq:LSZ},~\eqref{eq:LSZ_gluon} and~\eqref{eq:internal_line_rule}) and with all propagators attached to a given vertex sharing the same light-cone time coordinate. We can summarize these rules as
\begin{align}\label{eq:effective_feynman_rules}
    & \text{\textbf{internal quark line:}}\quad \sum_r u^r( p)\frac{\cG_F^{ij}(y^+, \p;x^+,\q|p^+)}{2p^+}\bar u^r(q)\,,\nn
    & \text{\textbf{external outgoing quark line:}}\quad \lim_{y_\infty^+ \to \infty}e^{i\left(\frac{\p^2+m^2}{2p^+}\right)y_\infty^+}\cG_F^{ij}(y_\infty^+,\p;x^+,\q|p^+)\bar u^r(q)\,,\nn
    & \text{\textbf{internal gluon line:}}\quad \sum_\lambda \epsilon_\lambda^{\mu}( p)\frac{\cG_A^{ab}(y^+, \p;x^+,\q|p^+)}{2p^+}\epsilon_\lambda^{*,\nu}( q)\,,\nn
    & \text{\textbf{external outgoing gluon line:}}\quad \quad \lim_{y_\infty^+ \to \infty}e^{i\frac{\p^2}{2p^+}y_\infty^+}\cG_A^{ab}(y_\infty^+,\p;x^+,\q|p^+)\epsilon_\lambda^{*,\nu}(q)\,.
\end{align}
and similar for ingoing lines, although we will not consider them in the calculations presented in this thesis. Note that we have restored the appropriate colour indices. In terms of Eq.~\eqref{eq:BDMPSZ_path_integral}, the mixed space in-medium scalar propagator explicitly reads
\begin{align}\label{eq:mixed_space_FT}
    \cG_R^{ij}(y^+, \p;x^+,\q|p^+) = \int_{\x,\y}e^{-i\p\cdot\y}e^{i\q\cdot\x}\cG_R^{ij}(y^+, \y;x^+,\x|p^+)\,,
\end{align}
which in the vacuum limit reduces to
\begin{align}\label{eq:vacuum_mixed_propagator}
    \cG_0^{ij}(y^+, \p;x^+,\q|p^+) = \delta^{ij}\cG_0(y^+, \p;x^+,\q|p^+) = \delta^{ij}(2\pi)^2\delta^{(2)}(\p-\q)e^{-i\frac{\p^2+m^2}{2p^+}(y^+-x^+)}e^{-\epsilon(y^+-x^+)}\,.
\end{align}
As argued already, amplitudes reduce to convolutions of scalar propagators $\cG(y^+, \p;x^+,\q|p^+)$ and spin-dependent vertex matrix elements, where four-momentum is conserved, which can be explicitly calculated from spinors and polarisation vectors, for a given spin quantization axis and a polarisation basis. Consider a splitting $(p_0, \lambda_0) \rightarrow (p_1, \lambda_1) +(p_2, \lambda_2)$, where $p_i$ are the particle on-shell momenta, with $p_0 = p_1+p_2$, and $\lambda_i = \pm 1$ are their spin states, i.e., twice the spin projection along a chosen spin quantization axis for quarks and the polarisation state in a chosen basis for gluons. The spin-dependent vertex matrix elements can be summarised as
\begin{align}\label{eq:vertices_summarized}
	&V_{q\rightarrow gq}^{\lambda_0 \lambda_1 \lambda_2}(p_1,p_2) = \bar u^{\lambda_2}(p_2) \slashed{\epsilon}^{\ast, \lambda_1}(p_1)u^{\lambda_0}(p_0)\,, \nn
	&V_{g\rightarrow q\bar q}^{\lambda_0 \lambda_1 \lambda_2}(p_1,p_2) = \bar u^{\lambda_1}(p_1)\slashed{\epsilon}^{\lambda_0}(p_0)v^{\lambda_2}(p_2)\,,
    \nn
	 &V_{g\rightarrow gg}^{\lambda_0 \lambda_1 \lambda_2}(p_1,p_2) = \epsilon^{\lambda_0}_{\alpha}(p_0)\epsilon^{\ast, \lambda_1}_{\beta}(p_1)\epsilon^{\ast,\lambda_2}_{\gamma}(p_2)[g^{\alpha\beta}(p_0+p_1)^{\gamma} - g^{\beta\gamma}(p_1-p_2)^{\alpha}-g^{\gamma\alpha}(p_0+p_2)^{\beta}]\,,
    \end{align}
where we omitted the standard $ig_st^a$ for $q\to gq$ and $g\to q\bar q$ and $g_sf^{abc}$ for $g\to gg$. In the three-gluon vertex, momentum $p_0$ is ingoing, while $p_1$ and $p_2$ are outgoing. These vertex matrix elements are computed in Appendix~\ref{app:vertices}.

\subsection{Cross-section and medium average}\label{sec:xsecs_BDMPSZ}

To illustrate how the effective Feynman rules derived in the previous subsections can be applied to compute an in-medium cross-section, let us sketch the calculation of a $1\to 2$ process in the BDMPS-Z formalism, leaving detailed computations for later chapters. 

Since our aim will be to compute medium modifications to $1\to n$ processes, we start by assuming that the energetic initial parton is produced from a hard scattering process and that only afterwards it undergoes modifications due to the underlying HIC matter\footnote{It could be the case that the initial production spectrum is not given only by the hard production cross-section, but by a modified spectrum due to earlier interactions which are not captured by BDMPS-Z. Considering this effect is not trivial and it is out of the scope of this thesis.}. This implies that the subamplitude producing the initial parton, for instance $H_{\rm in}$ in Eq.~\eqref{eq:internal_line_rule}, produces it with fixed four-momentum, such that after taking into account subsequent $\delta$ functions in the outgoing subamplitude, there is an overall $\delta$ function conserving $(+)$ momentum. Therefore, one defines the amplitude for a given in-medium process by only factoring out from the $T$-matrix element the $\delta$ function conserving $(+)$ momentum
\begin{align}
    i\cT = (2\pi)\delta(p_f^+-p_i^+)(i\cM)\,,
\end{align}
such that the cross-section reads
\begin{align}
    d\sigma \sim (2\pi)\delta(p_f^+-p_i^+)|\cM|^2 d\Pi_f\,,
\end{align}
where $d\Pi_f$ is the phase-space of the generic final state considered in the process and $p_i^+$ is an externally fixed $(+)$ momentum. Now consider a generic process $p_0(\lambda_0,j_0)\to p_1(\lambda_1,j_1)+p_2(\lambda_2,j_2)$ where $\lambda_i$ are spin or polarisation states and $j_i$ are, respectively, fundamental or adjoint color indices. Let us write the amplitude for this process in general, such that we can first focus on the dependence on the initial production amplitude. Following Eq.~\eqref{eq:internal_line_rule}, the amplitude can be written as
\begin{align}\label{eq:M_gqqbar}
    i\cM^{\lambda_1,\lambda_2;j_1,j_2}(p_1^+,\p_1, p_2^+,\p_2) = & \sum_{\lambda_0} \int_{x_0^+,\,\p_0^+} \cM_{\rm med}^{\lambda_0 \lambda_1 \lambda_2;\, j_0 j_1 j_2}(p_1^+,\p_1,p_2^+,\p_2; x_0^+,\p_0)\cM_0^{\lambda_0;\,j_0}(p_0^+,x_0^+,\p_0)\,,
\end{align}
where $j_0$ is summed over, $x_0^+\in(-\infty,+\infty)$, $p_0^+ = p_1^++p_2^+$ and both $\cM_{\rm med}$ and $\cM_0$ are scalars in Dirac and Lorentz structures. In terms of the initial momentum space amplitude, $\cM_0^{\lambda_0;j_0}$ reads
\begin{align}
\cM_0^{\lambda_0;j_0}(p_0^+,x_0^+,\p_0)
=
\begin{cases}
\displaystyle
\int_{p_0^-} e^{-ip_0^-x_0^+}\,
\epsilon^{\ast,\mu}_{\lambda_0}(\tilde p_0)\,
\cM^{j_0}_{0,\mu}(p_0),
& \text{(gluon/photon)}, \\[1.2em]
\displaystyle
\int_{p_0^-} e^{-ip_0^-x_0^+}\,
\bar u^{\lambda_0}(\tilde p_0)\,
\cM^{j_0}_{0}(p_0),
& \text{(quark)},
\end{cases}
\end{align}
for, respectively, an initial gluon/photon or quark. In the calculations presented in later chapters, we will not resolve the light-cone-time structure of the hard production process described by $\cM_0$, instead representing it by an \emph{effective local source}. We shall further localize this source at the initial boundary of the medium, denoted by $l_i^+$. Thus, we approximate it as
\begin{align}\label{eq:instantaneous_source}
    \cM_0^{\lambda_0;j_0}(p_0^+,x_0^+,\p_0)
    \simeq
    \cM_0^{\lambda_0;j_0}(p_0^+,\p_0)\,
    \delta(x_0^+-l_i^+).
\end{align}
These two assumptions should be understood as follows. First, that the subsequent evolution does not resolve the finite time structure of the hard production process. This is a typical assumption in factorized descriptions and, parametrically, it is justifiable when the time scale of the hard process is much shorter than the time scales of subsequent branchings, whether vacuum-like or medium-induced. For hard processes with $Q_{\rm hard}\sim p_0^+$, such as mid-rapidity jet production, this approximation is not parametrically controlled for hard wide-angle splittings. Second, that vacuum propagation of the initial particle before the medium, as well as possible interference effects between pre-medium production or radiation and subsequent in-medium splittings, are not described explicitly. In addition, we will assume that the initial particle is produced in an \emph{unpolarised, incoherent state}. Therefore, when computing cross-sections, we take its polarisation density matrix to satisfy
\begin{align}\label{eq:diagonal_pol}
    \cM_0^{\lambda_0}\cM_0^{\dagger\bar\lambda_0}
    \longrightarrow
    \cM_0 \cM_0^\dagger\frac{\delta^{\lambda_0\bar\lambda_0}}{2}
    \,,
\end{align}
where the $1/2$ factor account for an average over the two physical polarisations and we have omitted the dependence on momenta and colour states. The assumption of incoherence discards interference between different initial spin/polarisation states, corresponding to the off-diagonal elements of the density matrix. The assumption of an unpolarised source further sets the diagonal entries to be equal, thereby discarding any spin/polarisation asymmetry inherited from the hard production subprocess. In Section~\ref{sec:ch5_spin_correlations}, we shall explore correlations of this type between a medium-induced splitting and a subsequent vacuum splitting.

With these assumptions about the initial production amplitude, namely that the initial parton is produced unpolarised and incoherently, and near the initial boundary of the medium, let us now specialize this discussion to the $\gamma\to q\bar q$ process. Since the photon does not interact with the medium, this process has a simple colour structure, making it standard in the literature (see e.g. ~\cite{Dominguez:2019ges, Isaksen:2023nlr}). Following the effective rules summarized in Eq.~\eqref{eq:effective_feynman_rules}, the in-medium amplitude for the $\gamma\to q\bar q$ process at leading eikonal accuracy reads
\begin{align}\label{eq:M_gqqbar_simplified}
    & i\cM^{\lambda_1,\lambda_2;j_1,j_2}(p_1^+,\p_1, p_2^+,\p_2) = \frac{ie_q}{2p_0^+}\int_{\p_0,\, \q_1,\,\q_2,\,x^+} e^{i\frac{\p_1^2+m^2}{2p_1^+}x_\infty^+} e^{i\frac{\p_2^2+m^2}{2p_2^+}x_\infty^+} \nn
    & \times \cG_F^{j_1 k}(x_\infty^+, \p_1; x^+, \q_1|p_1^+)\bar \cG_F^{kj_2}(x_\infty^+, \p_2; x^+, \q_2|p_2^+)V_{\gamma \to q\bar q}^{\lambda_0 \lambda_1\lambda_2}(q_1,  q_2)\cG_0(x^+,\q_0;0,\p_0|p_0^+)\cM_0(p_0^+,\p_0)\,,
\end{align}
where we have set the initial boundary of the medium to $l_i^+ \equiv 0$, $x^+\in (0,x_\infty^+)$, $\q_0 = \q_1+\q_2$, $V_{\gamma \to q\bar q}(q_1,  q_2)$ is the same as the $g\to q\bar q$ vertex given by Eq.~\eqref{eq:vertices_summarized}, with $q_i$ on-shell momenta. The coupling constant $e_q$ denotes the quark's fractional electric charge, whose exact value, dependent on quark flavour, is not relevant for the current presentation. Given that the photon is colourless, it is described by the vacuum scalar propagator in Eq.~\eqref{eq:vacuum_mixed_propagator}. Since vacuum propagation conserves transverse momentum, it sets $\p_0 = \q_0 = \q_1+\q_2$. Further, it is implicit that we take $x_\infty^+\to +\infty$ following the on-shell prescription in Eq.~\eqref{eq:LSZ_time}, followed by $\epsilon\to 0$ (see Eq.~\eqref{eq:free_propagator}). Upon squaring, summing over final quantum numbers and, as per Eq.~\eqref{eq:diagonal_pol}, averaging over the photon's polarisation ($\lambda_0$), the differential cross-section contains the following object
\begin{align}\label{eq:xsec_example}
    \frac{1}{N_c}\left \langle {\rm Tr} \left(\bar \cG_F^\dagger \cG_F^\dagger \cG_F \bar \cG_F \right)\right\rangle\,,
\end{align}
where the trace is over fundamental colour indices. Note that the daggered propagators are evaluated at different intermediate momenta $\bar \q_1$, $\bar \q_2$ and light-cone time $\bar x^+$, while the remaining kinematic quantities are fixed. The angular brackets $\langle \cdots\rangle$ represent the average over field configurations which we introduced in Section~\ref{subsec:propagator_derivation}. It is the computation of medium averages such as the four-point function in Eq.~\eqref{eq:xsec_example} that lies at the technical core of calculating matter-induced modifications to jets in the BDMPS-Z formalism.

The medium average is defined by specifiying a statistical ensemble for the background field $\cA^\mu$. In the BDMPS-Z formalism, this
ensemble is usually taken to have Gaussian statistics and to be local in
light-cone time, corresponding to a white-noise model for the colour field seen
by the energetic projectile.\footnote{This model is inspired by the McLerran-Venugopalan (MV) model~\cite{McLerran:1993ni,McLerran:1993ka,McLerran:1994vd}, which describes the distribution of colour sources in a highly boosted nucleus. In its original formulation, the stochastic variable is the colour charge density $\rho^a$ of the large-$x$ degrees of freedom, taken to satisfy
$\langle \rho^a\rangle=0$ and a local two-point correlator of the form
$\langle \rho^a(x_\perp)\rho^b(y_\perp)\rangle\propto
\delta^{ab}\delta^{(2)}(x_\perp-y_\perp)$. The corresponding classical field $\cA^\mu$ is then obtained by solving the
Yang-Mills equations with this stochastic source. In the CGC, the MV
model is commonly used as an initial condition for the rapidity evolution of
the statistical distribution, governed by the JIMWLK equation (see, e.g.,~\cite{Iancu:2001ad,Gelis:2010nm}). In jet quenching, by
contrast, the medium field correlator is usually specified directly and treated
as a phenomenological input, without dynamical evolution. Thus, it should be viewed as an effective description of the colour fields resolved by the hard projectile, rather than as a literal small-$x$ nuclear wave
function.} The only non-vanishing connected field correlator is then
\begin{align}\label{eq:A_correlator}
    \left\langle
    \cA_a^-(x^+,\x)\,
    \cA_b^-(y^+,\y)
    \right\rangle
    =
    \delta_{ab}\,
    n(x^+,\x)\,
    \delta(x^+-y^+)\,
    \gamma(\y,\x) .
\end{align}
All higher-point field correlators are generated by Wick contractions. This approximation then allows for the analytical treatment of objects such as the four-point function in Eq.~\eqref{eq:xsec_example} in terms of evolution equations or path-integral expressions (see, e.g.,~\cite{Blaizot:2012fh, Apolinario:2014csa, Isaksen:2023nlr}). The
function $\gamma(\y,\x)$ encodes the transverse momentum-transfer kernel of the medium, while $n(x^+,\x)$ is the density profile controlling the local strength of medium interactions. The symmetries of the medium are reflected in these two objects and in the colour and light-cone time locality of Eq.~\eqref{eq:A_correlator}. First, the factor $\delta_{ab}$ in the two-point correlator implies that the colour indices in each Wick pair are contracted into a \emph{colour-neutral} combination. Thus, the medium ensemble cannot inject a net colour charge: the medium can rotate the colour of the amplitude and of its complex conjugate (cc), but these rotations are correlated by the colour contractions. As a result, the medium average does not mix different total colour channels of the amplitude--conjugate-amplitude system: a singlet cannot be turned into an octet, or vice versa. Moreover, when computing a cross-section, the final colour indices of the amplitude are contracted with those of the cc amplitude, which selects the singlet channel of the amplitude--conjugate-amplitude system. Because the averaged evolution is local in $x^+$, this statement holds locally along the propagation: the system relevant for the inclusive cross-section stays in the singlet sector at all times. Since we will be assuming a \emph{transversely homogeneous} medium, i.e., $n(x^+,\x) = n(x^+)$ and $\gamma(\y,\x)=\gamma(\y-\x)$, the same reasoning applies to transverse momentum. In this case, the medium average is invariant under a common transverse translation of the amplitude--conjugate-amplitude system. Hence, the total transverse momentum in the amplitude is matched by that in the complex-conjugate amplitude. With locality in $x^+$, this matching holds at all times along the propagation.

The time locality of Eq.~\eqref{eq:A_correlator} has one further important consequence. It allows objects such as the one in Eq.~\eqref{eq:xsec_example}, and the more complex structures appearing in other processes, to be decomposed into transverse momentum convolutions of simpler medium average defined in disjoint light-cone time intervals. To see this in the present example, consider the ordering $\bar x^+>x^+$, corresponding to the contribution where the splitting happens first in the amplitude and later in the cc amplitude. Between the two splitting times, only the amplitude contains a $q\bar q$ pair, while the cc amplitude still contains the photon, which does not enter the average. After the later splitting, both amplitudes contain a $q\bar q$ pair. Introducing intermediate transverse momenta $\k_1,\k_2$ at the time $\bar x^+$, and using the composition law of the propagator in Eq.~\eqref{eq:propagator_composition_law}, the four-point function in Eq.~\eqref{eq:xsec_example} can be written schematically as
\begin{align}\label{eq:time_factorisation_photon_qqb}
    &\frac{1}{N_c}\Big \langle \Tr \Big(
        \bar \cG_F^\dagger
        \cG_F^\dagger
        \cG_F
        \bar \cG_F
    \Big)\Big\rangle\sim
    \int_{\k_1,\k_2}
    \cS_4(x_\infty^+,\p_1,\p_2;\bar x^+,\k_1,\k_2,\bar\q_1,\bar\q_2)\cS_2(\bar x^+,\k_1,\k_2;x^+,\q_1,\q_2)\,
    \, .
\end{align}
Here $\cS_2$ denotes the two-point $q\bar q$ average between the two splitting times. It evolves the pair from momenta $\q_1,\q_2$ at $x^+$ to momenta $\k_1,\k_2$ at $\bar x^+$. The object $\cS_4$ denotes the four-point average after the later splitting. It evolves the amplitude momenta $\k_1,\k_2$ and the cc-amplitude momenta $\bar\q_1,\bar\q_2$ at $\bar x^+$ to the final measured momenta at $x_\infty^+$. The opposite ordering, $\bar x^+<x^+$, is obtained analogously. Then, for each object $\cS_i$ separately, the total transverse momentum flowing through the amplitude matches that flowing through the cc amplitude, and the corresponding colour structure is projected onto its singlet component. We shall leave more detailed computations of these objects for Sections~\ref{sec:ch3_gqqbar} and~\ref{subsec:ch5_medium_vacuum_spin_correlations}.

The $n$-point functions $\cS_4$ and $\cS_2$ contain medium averages of, respectively, a trace of four Wilson lines, termed \emph{quadrupole}, and a trace of two Wilson lines, termed \emph{dipole}. Although QCD vertices contain more complex colour structures involving averages with a larger number of lines, at leading $N_c$ power the dominant colour structure can be shown to include only dipoles and quadrupoles~\cite{Dominguez:2011wm, Dominguez:2012ad, Isaksen:2020npj}. These two objects are thus central for the calculation of in-medium processes, and they are the only ones we will be computing in later chapters, since we will be working in the \emph{large-$N_c$ limit}. While the $n$-point functions involving quadrupoles are usually either solved numerically or approximated by a simpler object\footnote{We shall return to this in Chapters~\ref{chap:anisotropic_early_time} and~\ref{chap:towards_multi_point_structure}, where we choose to approximate the quadrupole (see, e.g.,~\cite{Blaizot:2012fh,Apolinario:2014csa,Isaksen:2020npj, Isaksen:2023nlr}).}, dipoles have a closed analytic solution for an average of the form of Eq.~\eqref{eq:A_correlator}. For a general representation $R$ it reads
\begin{align}\label{eq:dipole_def}
    D_R(\r;\bar x^+, x^+) \equiv \frac{1}{d_R}\Big\langle\Tr\Big(U_R(\r_1)U_R^\dagger(\r_2)\Big)\Big\rangle  = \exp\left\{-\frac{C_R}{2}\int_{x^+}^{\bar x^+}ds^+\,
  n(s^+)\,\sigma(\r_1-\r_2)\right\}\,,
\end{align}
where $\r = \r_1-\r_2$ and we have defined the \emph{dipole cross-section} $\sigma(\r) = 2g^2\bigl(\gamma(0) - \gamma(\r))$, $d_R$ is the dimension of the colour representation and we have left the time arguments of the Wilson lines, defined in Eq.~\eqref{eq:Wilson_line_def}, implicit. For fundamental Wilson lines, $d_R = N_c$ and $C_R = C_F$, which is the case for the $\gamma\to q\bar q$ splitting, while for adjoint Wilson lines, $d_R = N_c^2-1$ and $C_R = C_A$. Physically, the trace of two Wilson lines measures how a colour dipole with fixed transverse separation loses colour coherence while crossing the medium over a given light-cone time interval. To compute this object explicitly one must specify the kernel $\gamma(\q)$, with different choices corresponding to different assumptions about the distribution of transverse momentum transfer between the energetic partons and the medium. In a scattering-centre picture, this kernel is proportional to the single-scattering potential squared. The Gyulassy-Wang (GW) model corresponds to a screened Coulomb potential $V(\q)\propto \left(\q^2+\mu^2\right)^{-1}$ and, therefore, to the kernel~\cite{Gyulassy:1993hr}
\begin{align}
    \gamma_{\rm GW}(\q)
    \propto
    |V(\q)|^2
    \propto
    \frac{1}{(\q^2+\mu^2)^2}\,,
\end{align}
where $\mu$ is an infrared screening scale. Another common choice is the HTL-screened kernel~\cite{Arnold:2008vd}
\begin{align}
    \gamma_{\rm HTL}(\q)
    \propto
    \frac{1}{\q^2(\q^2+m_D^2)}\,,
\end{align}
where $m_D$ is the Debye mass, parametrically of order $g_sT$. This form encodes the electric screening structure of a weakly coupled thermal plasma. Both the GW and HTL-screened kernels have the same Coulombic large-$\q$ behaviour, $\gamma(\q)\sim 1/\q^4$, but they differ in their infrared behaviour and in the semi-hard region where $\q^2$ is comparable to the screening scale.

Although these kernels can be kept in numerical approaches~\cite{Andres:2020vxs}, analytic progress is often obtained by expanding the dipole cross-section for small transverse separations. Since colour transparency implies $\sigma(0)=0$, the leading term is quadratic. For a dipole in representation $R$, this is written as
\begin{align}\label{eq:HO_approx}
    \frac{C_R}{2}\,n(s^+)\,\sigma(\r)
    \simeq
    \frac{1}{4}\,\hat q_R(s^+)\,\r^2\,.
\end{align}
This is the \emph{harmonic-oscillator} (HO), or \emph{multiple soft scattering, approximation}. It replaces the detailed distribution of momentum transfers by Brownian transverse momentum broadening governed by the \emph{jet quenching parameter} $\hat q_R$, making the corresponding path integrals analytically tractable. For kernels with a large-$\q$ Coulomb tail, $\gamma(\q)\sim 1/\q^4$, such as the GW and HTL-screened forms, the coefficient of the quadratic term is logarithmically sensitive to the upper momentum scale at which the dipole is resolved. This logarithmic dependence is usually absorbed into an effective value of $\hat q_R$. In this approximation, the dipole becomes
\begin{align}\label{eq:HO_dipole}
    D_R(\r;\bar x^+, x^+) 
    \simeq
    \exp\left[
        -\frac{1}{4}
        \int_{x^+}^{\bar x^+}ds^+\,
        \hat q_R(s^+)\,\r^2
    \right]\,,
\end{align}
As introduced in Section~\ref{subsec:scales}, the parameter $\hat q_R \sim d\langle k_\perp^2\rangle/d\ell^+$ quantifies the mean transverse momentum squared transferred to a coloured charge in representation $R$ per unit light-cone time. The price of the harmonic approximation is that all information about the detailed shape of the scattering kernel is compressed into $\hat q_R$, so rare hard momentum transfers are not faithfully retained.

A particularly important result is the closed-form solution, in the HO approximation, of the path integral entering the two-point average $\cS_2$. This path integral, for the $\gamma\to q\bar q$ process, reads
\begin{align}\label{eq:K_def}
    \cK(\bar x^+,\u_2;x^+, \u_1) = \int_{\r(x^+) = \u_1}^{\r(\bar x^+) = \u_2}\cD \r \exp\left\{i\frac{\omega^+}{2}\int_{x^+}^{\bar x^+}ds^+ \dot{\r}^2 \right\}D_F(\r; \bar x^+, x^+)\,,
\end{align}
where $\dot{\r} = \frac{d\r}{\,ds^+}$ and $\omega^+ = z(1-z)p_0^+$, with $z$ the energy fraction of the splitting. In the HO approximation, $D_F$ is given by Eq.~\eqref{eq:HO_dipole} in the fundamental representation, so that $\cK$ is describes the real-time Schr\"odinger kernel of a non-relativistic particle of transverse mass $\omega^+$ evolving in $x^+$ in a purely imaginary quadratic potential, or equivalently a harmonic oscillator with complex frequency. This is a standard problem with a known closed form solution~\cite{Feynman:100771, Zakharov:1996fv,Kleinert:2004ev}
\begin{align}\label{eq:K_HO}
    \cK(\bar x^+,\u_2;x^+, \u_1) = \frac{\omega^+\Omega}{2\pi i \sin(\Omega \Delta x^+)}\exp\left\{i\frac{\omega^+ \Omega}{2\sin(\Omega\Delta x^+)}\left((\u_1^2+\u_2^2)\cos(\Omega\Delta x^+) - 2\u_1\cdot\u_2\right)\right\}\,,
\end{align}
where $\Delta x^+ = \bar x^+ - x^+$ and the complex frequency reads\footnote{Note that we have fixed $\hat q \equiv \hat q_F$.}
\begin{align}\label{eq:Omega_HO}
  \Omega = \frac{1-i}{\sqrt{2}}\sqrt{\frac{\hat{q}\,f(z)}{2\omega^+}}\,,
\end{align}
where $f(z) = 1$ for $\gamma\to q\bar q$, although it is generally a process dependent function. For general QCD splittings it acquires a $z$ dependence from additional dipoles in Eq.~\eqref{eq:K_def}. An essential feature of~\eqref{eq:K_HO} is that the complex nature of $\Omega$ is the path-integral manifestation of the LPM effect introduced in Section~\ref{subsec:quenching_mechanisms}. Its real part controls the oscillatory quantum phase associated with the formation of the splitting, while its imaginary part gives an exponential damping over long light-cone-time separations. As a result, splittings with
\begin{align}
    \Delta x^+ \gg t_f^+ \sim |\Omega|^{-1} \sim \sqrt{\frac{2\omega^+}{\hat q}}\,,
\end{align}
are suppressed, and the dominant contribution comes from formation times of order $t_f^+$. This gives a precise realization of the parametric estimate of formation time discussed in Section~\ref{subsec:quenching_mechanisms}.
 % file "Thesis_Introduction.tex"
\cleardoublepage

%\input{Thesis_Background} % file "Thesis_Background.tex"
%\cleardoublepage

%\input{Thesis_Implementation} % file "Thesis_Implementation.tex"
%\cleardoublepage

%\input{Thesis_new_file} % add new .tex files for new chapters
%\cleardoublepage

%\input{Thesis_new_file} % add new .tex files for new chapters
%\cleardoublepage

%\input{Thesis_new_file} % add new .tex files for new chapters
%\cleardoublepage

% \input{Thesis_Results} % file "Thesis_Results.tex"
% \cleardoublepage

%%%%%%%%%%%%%%%%%%%%%%%%%%%%%%%%%%%%%%%%%%%%%%%%%%%%%%%%%%%%%%%%%%%%%%%%
%  Anisotropies.tex
%  Chapter 3: Anisotropies...
%%%%%%%%%%%%%%%%%%%%%%%%%%%%%%%%%%%%%%%%%%%%%%%%%%%%%%%%%%%%%%%%%%%%%%%%

%%%%%%%%%%%%%%%%%%%%%%%%%%%%%%%%%%%%%%%%%%%%%%%%%%%%%%%%%%%%%%%%%%%%%%%%
%  Chapter 3: Jets as probes of anisotropic early-time QCD matter
%%%%%%%%%%%%%%%%%%%%%%%%%%%%%%%%%%%%%%%%%%%%%%%%%%%%%%%%%%%%%%%%%%%%%%%%

\chapter{Jets as probes of anisotropic QCD matter}
\label{chap:anisotropic_early_time}

\section{Directional information from early-time QCD matter}
\label{sec:ch3_motivation}

As discussed in Section~\ref{subsec:st_evolution_HIC}, the QCD matter produced in HICs starts out from a highly anisotropic, far-from-equilibrium state and evolves towards a near-equilibrium QGP. The earliest stage is characterised by strong longitudinal chromoelectric and chromomagnetic fields in the forward light cone, a state commonly referred to as the Glasma~\cite{Gelis:2010nm}. As the system expands, these coherent field configurations decohere and the medium enters a pre-equilibrium stage in which quark and gluon quasiparticles undergo elastic scatterings and inelastic splittings, reducing the initially large pressure imbalance before hydrodynamisation~\cite{Schlichting:2019abc}. Even after the onset of viscous hydrodynamics, remnants of this anisotropy can persist through shear-stress corrections to the local energy-momentum tensor. Thus, independently of how the early stage is modelled, the system evolves from a state with strong anisotropies, in particular \emph{momentum-space anisotropies} reflecting asymmetries between longitudinal and transverse dynamics, towards an isotropic, thermal QGP. 

The basic question to be addressed in this chapter is whether high-energy jets, produced concurrently with these stages, can retain perturbative imprints of such anisotropies. This question is motivated by the space-time overlap between the perturbative cascade and the evolving QCD matter. A hard parton produced with virtuality or transverse momentum scale $Q$ is created on a time scale $\tau\sim 1/Q$, so sufficiently energetic jets are, in principle, already developing while the matter is still carrying a pronounced anisotropic structure. If part of the substructure of the jet is formed while the medium still carries sizeable directional information, then the angular distribution of energy and particles inside the jet can become sensitive not only to scalar measures of jet-medium interactions, but also to the directionality of the momentum transfers induced by the medium. This information may still be diluted by the subsequent hydrodynamic evolution, or lost if the observable averages over the relevant angular orientation. This motivates the use of more differential jet substructure observables, in particular \emph{azimuthally differential observables}, as perturbative probes of early-time dynamics complementary to typical inclusive jet quenching observables. 

The strategy followed here is deliberately more modest than a first-principles calculation of jet propagation through the earliest stages of HIC matter, e.g., as in classical-statistical simulations of Glasma or in a complete effective kinetic theory (EKT) background. Along these lines, calculations of hard probe momentum broadening in Glasma fields find large, time-dependent and anisotropic broadening~\cite{Ipp:2020mjc,Ipp:2020nfu,Avramescu:2023qvv,Carrington:2021dvw,Carrington:2022bnv,Carrington:2026qlg,Avramescu:2026fgv}, while EKT provides the evolution of the jet quenching parameter, including its tensorial structure, in the pre-hydrodynamic quasiparticle regime~\cite{Boguslavski:2023alu,Boguslavski:2023waw}.\footnote{Medium-induced radiation in the Glasma stage and within EKT has also begun to be addressed directly~\cite{Barata:2024xwy,Altenburger:2025iqa,Barata:2025agq}.} We instead ask how far one can go in describing directional effects in jets using the BDMPS-Z formalism, i.e. the framework based on leading-power high-energy propagation in a stochastic background field introduced in Section~\ref{sec:BDMPS_Z}. In this sense, the anisotropies of the medium are treated as phenomenological inputs to the BDMPS-Z calculation, encoded in the directional structure and, possibly, the time dependence of the transport coefficient controlling transverse propagation. In the simplest implementation this is described by a static, homogeneous but anisotropic broadening kernel, while later it is promoted to a time-dependent input directly extracted from early-time simulations. This keeps the perturbative calculation under analytic control while retaining the feature of early-time matter most relevant for the observables considered below: a preferred direction in the momentum exchanges between the jet and the medium. The cross section computed in the BDMPS-Z formalism must retain not only the amount of radiation or broadening induced by the medium, but also the azimuthal orientation of the resulting energy flow around the initial hard parton, whose direction would define the jet axis in an eventual reconstructed jet and will often be referred to as such. Thus, the central object of the chapter is the azimuthally differential in-medium cross section for a $1\to 2$ process, from which jet substructure observables sensitive to early anisotropies can be constructed.

\section{Anisotropic broadening and azimuthally differential cross sections}
\label{sec:ch3_anisotropic_broadening}

Within the BDMPS-Z formalism, the directionality of the medium enters through the transverse structure of the medium average introduced in Eq.~\eqref{eq:A_correlator}, characteristic of a Gaussian white-noise ensemble of background fields. In the transversely homogeneous case relevant in this thesis, this reduces to $n(x^+,\x)=n(x^+)$ and $\gamma(\y,\x)=\gamma(\y-\x)$. The usual isotropic approximation further assumes that the collision kernel depends only on $|\y-\x|$. The anisotropic extension considered here keeps transverse homogeneity, and thus total transverse momentum conservation between the amplitude and its complex conjugate as discussed in Section~\ref{sec:xsecs_BDMPSZ}, but relaxes the rotational invariance of the kernel.

Equivalently, in momentum space one allows the transverse momentum-transfer distribution to depend on the orientation of the exchanged momentum. A collision kernel that is invariant under rotations in the transverse plane depends only on $\q^2$, while an anisotropic kernel distinguishes two orthogonal directions. Such tensorial or direction-dependent broadening has been studied in perturbative anisotropic plasmas~\cite{Baier:2008js,Majumder:2009cf,Hauksson:2021okc}. This motivates the introduction of two broadening coefficients in the plane transverse to the jet axis,
\begin{align}
    \hat q_x
    =
    \frac{d\langle k_x^2\rangle}{d\ell},
    \qquad
    \hat q_y
    =
    \frac{d\langle k_y^2\rangle}{d\ell},
\end{align}
which correspond to the diagonal components of the tensor $\hat q^{ij}$ introduced in Section~\ref{subsec:scales}, with $\ell$ the distance travelled by the energetic parton. The isotropic limit is recovered for $\hat q_x=\hat q_y$. This provides an effective way of incorporating direction-dependent momentum exchange in a BDMPS-Z calculation without specifying the full microscopic origin of the anisotropic kernel.\footnote{The implementation of an axis-dependent broadening in a BDMPS-Z calculation was first considered in~\cite{Hauksson:2023tze} in the context of polarisation-dependent energy rates for $g\to gg$ splittings and their inclusion in a Markovian cascade.} In weakly coupled non-equilibrium matter, such anisotropies can arise from anisotropic momentum distributions of the medium constituents generated by rapid longitudinal expansion, or from viscous corrections to a near-hydrodynamic background. A first-principles determination of the soft collision kernel in an anisotropic plasma is complicated by plasma instabilities~\cite{Mrowczynski:2000ed,Romatschke:2003ms,Arnold:2003rq}. In the present approach, the kernel is instead treated as an effective input and expanded in the anisotropic version of the harmonic oscillator approximation in Eq.~\eqref{eq:HO_approx}, i.e.,
\begin{align}\label{eq:ch3_anisotropic_HO}
    \frac{C_R}{2}n\,\sigma(\r) \simeq \frac{1}{4}\hat q_R\,\r^2
    \longrightarrow
    \frac{1}{4}\left(\hat q_{R,x} \,r_x^2+\hat q_{R,y}\,r_y^2 \right)\,,
\end{align}
where $\r$ is the transverse separation characterizing the relevant colour dipole, taken to be small in the expansion, and we have omitted the possible light-cone time dependence of $n$ and $\hat q_{R,i}$. As discussed in Section~\ref{sec:xsecs_BDMPSZ}, one relevant object entering in-medium cross-sections is the reduced splitting kernel, defined in Eq.~\eqref{eq:K_def} for the $\gamma\to q\bar q$ splitting. For a static homogeneous medium, the HO limit of this kernel generalises from its isotropic version in Eq.~\eqref{eq:K_HO} to an anisotropic version as
\begin{align}\label{eq:ch3_K_aniso}
    \cK(\bar x^+, \u_2; x^+, \u_1)
    =
    \cK_{x}(\bar x^+, \u_{2,x}; x^+, \u_{1,x})\,
    \cK_{y}(\bar x^+, \u_{2,y}; x^+, \u_{1,y}),
\end{align}
with
\begin{align}
    \cK_{i}(\bar x^+, \u_{2,i}; x^+, \u_{1,i})
    =
    \sqrt{\frac{\omega^+\Omega_{i}}{2\pi i \sin(\Omega_{i} \Delta x^+)}}\exp\left\{i\frac{\omega^+ \Omega_{i}}{2\sin(\Omega_{i}\Delta x^+)}\left((\u_{1,i}^2+\u_{2,i}^2)\cos(\Omega_{i}\Delta x^+) - 2\u_{1,i}\u_{2,i}\right)\right\}\,,
\end{align}
with $i=x,y$, $\Delta x^+=\bar x^+-x^+$, and $\omega^+$ denotes the reduced light-cone energy exactly as defined in Eq.~\eqref{eq:K_HO}. The frequencies $\Omega_{i} \sim \sqrt{\hat q_{i}}$ are fixed by the corresponding broadening coefficients analogously to Eq.~\eqref{eq:Omega_HO}. Thus, the resulting transverse dynamics remains Gaussian, but with propagation in the two directions transverse to the jet axis now controlled by two different scales.

The directional structure of the medium can be probed whenever the observable retains information about a transverse orientation. This orientation may be kinematic, as in a cross section differential in the relative transverse momentum of two final partons, or internal, as in rates resolved in the linear polarisation states of a gluon~\cite{Hauksson:2023tze}. The latter case can remain sensitive to anisotropic broadening even after integrating over transverse momenta, since the polarisation basis itself refers to the two transverse directions. In what follows, the relevant orientation will be the azimuthal angle of the final-state energy flow. For a generic in-medium $1\to2$ process, let $\p$ denote the relative transverse momentum of the two final partons. We write
\begin{align}
    \p = |\p| \left(\cos\phi,\sin\phi\right),
\end{align}
where $\phi$ is the azimuthal orientation of the relative transverse momentum measured with respect to the $x$ axis used to define $\hat q_x$ and $\hat q_y$. In a real event, the axes used to define $\hat q_i$ must ultimately be related to a physical direction of the medium, such as the direction of maximal broadening or the projection of the beam direction onto the plane transverse to the jet. This amounts to a shift of the azimuthal origin, $\phi\to\phi-\phi_0$, with $\phi_0$ fixed by the relevant physical geometry. At fixed energy fraction and in the collinear limit, $|\p|$ can equivalently be traded for the opening angle of the pair $\theta$ through $|\p| \sim z(1-z)p_0^+ \theta$, where $p_0^+$ is the energy of the initial parton. The usefulness of this parametrisation will become explicit once the in-medium cross section is computed. The dependence on $\theta$ reflects standard medium-induced modifications to partonic splittings already present in isotropic media, while the dependence on $\phi$ follows directly from the axis-dependent broadening, generating non-vanishing Fourier harmonics that will be analysed in detail in Section~\ref{subsec:harmonics_spin_sensitivity}.

In a realistic heavy-ion event, the observed azimuthal structure can receive contributions from several physical sources. The local anisotropy isolated here is axis-like: the medium broadens more efficiently along one transverse axis than the other, thereby modifying the transverse structure of the jet around its axis. Spatial gradients and collective flow instead provide local vector-like sources of angular dependence, which can induce directed distortions of the jet cascade; see~\cite{Sadofyev:2021ohn,Barata:2022krd,Andres:2022ndd,Barata:2023qds,He:2020iow,Fu:2022idl,Kuzmin:2023hko}. In non-central collisions, the global collision geometry gives another contribution, controlled by path-length differences and by the orientation of the whole jet with respect to the reaction or participant plane (see, e.g.,~\cite{Wang:2000fq}). Interference between multiple emissions (see, e.g.,~\cite{Fickinger:2013xwa}), hadronisation, jet reconstruction and medium response can also shape the measured angular correlations. These effects are in general entangled in realistic observables and require a full space-time description of jet propagation through the evolving medium. In what follows, we focus on the idealised setup introduced above, in which the only source of azimuthal dependence is the momentum-space anisotropy in the BDMPS-Z collision kernel.

\section{Spin-resolved $q\bar q$ production in anisotropic matter}
\label{sec:ch3_gqqbar}

We now turn to the calculation of the cross section for the $g(p_0,b,\lambda_0)\to q(p_1,j_1,h)\bar q(p_2,j_2,h')$ (see Fig.~\ref{fig:gqqb_diagram}) process in anisotropic QCD matter, following the setup described in the previous subsections and the high-energy BDMPS-Z formalism introduced in Section~\ref{sec:BDMPS_Z}. Following the effective Feynman rules summarized in Eq.~\eqref{eq:effective_feynman_rules} and the assumptions discussed in Section~\ref{sec:xsecs_BDMPSZ}, namely that the gluon is produced unpolarized and incoherently, near the initial boundary of the medium (fixed to $l_i^+\equiv 0$), the amplitude reads
\begin{align}\label{eq:ch3_gqqb_M}
    & i\cM^{\lambda_0hh';bj_1j_2} = \frac{ig_s}{2p_0^+}\int_{\p_0,\, \q_0,\,\q_1,\,x^+} e^{i\frac{\p_1^2+m^2}{2p_1^+}x_\infty^+} e^{i\frac{\p_2^2+m^2}{2p_2^+}x_\infty^+} \nn
    & \times \cG_F^{j_1 k}(x_\infty^+, \p_1; x^+, \q_1|p_1^+)t^a_{kl}\bar \cG_F^{lj_2}(x_\infty^+, \p_2; x^+, \q_2|p_2^+)V_{g \to q\bar q}^{\lambda_0 hh'}(q_1,  q_2)\cG_A^{ab}(x^+,\q_0;0,\p_0|p_0^+)\cM_0(p_0^+,\p_0)\,,
\end{align}
where $p_0^+ = p_1^+ + p_2^+$, $\q_2 = \q_0-\q_1$ and the transverse momentum and light-cone time integrations read
\begin{align}
    \int_{\p_0,x^+} = \int_{-\infty}^{+\infty}\frac{d^2\p_0}{(2\pi)^2}\int_0^{x_\infty^+}dx^+\,.
\end{align}
We have further assumed that the initial gluon is produced in an incoherent colour state $b$ to be averaged over at the cross-section level, exactly analogous to the discussion in the context of polarisation in Section~\ref{sec:xsecs_BDMPSZ}. Each propagator is given by the mixed-space Fourier transform (see Eq.~\eqref{eq:mixed_space_FT}) of the coordinate-space propagator in Eq.~\eqref{eq:BDMPSZ_path_integral}.The vertex matrix element $V_{g \to q\bar q}^{\lambda_0hh'}(q_1,  q_2)$ are computed in Appendix~\ref{app:vertices} for a given polarisation basis and spin quantization axis. For the particular case of circularly polarised gluons and for the spin quantization axis of quarks along the $z$-axis (i.e. along the large momentum component), at leading eikonal accuracy it reads
\begin{align}\label{eq:ch3_explicit_Vgqqb}
    V_{g\rightarrow q\bar q}^{\lambda_0 h h'}(z,\bkappa) =\frac{2h \delta^{h,h'} }{\sqrt{z(1-z)}}(z\delta^{\lambda_0,h} - (1-z)\delta^{\lambda_0, -h})(\bkappa\cdot \beps^{\lambda_0}) -\delta^{h, -h'}\delta^{\lambda_0, h} \frac{\sqrt{2}m}{\sqrt{z(1-z)}}\,,
\end{align}
where $\beps^{\lambda_0} =  (1,i\lambda_0)/\sqrt{2}$, $z$ is the energy fraction of the quark, so that $p_1^+ = zp_0^+$ and $p_2^+ = (1-z)p_0^+$, and the relative transverse momentum at the vertex reads $\bkappa = \q_1 - z\q_0$. In defining the cross section, we shall keep the final-state spin label explicit. As discussed in Section~\ref{sec:BDMPS_Z}, spin-flip interactions are power suppressed in the high-energy expansion, so the leading in-medium propagators are diagonal in spin. Thus, the spin information fixed at the vertex matrix element in Eq.~\eqref{eq:ch3_explicit_Vgqqb} is preserved by the subsequent eikonal propagation of the quark and anti-quark through the medium. This makes spin a useful additional handle on anisotropy-induced modifications, since due to the product $(\bkappa\cdot\beps^{\lambda_0})$ the spin dependence of the $q\bar q$ is tied to transverse-momentum, or equivalently transverse-derivative, structures which can distinguish $x$ and $y$ directions. Naturally, summing over $h$ recovers the spin-averaged in-medium splitting spectrum.

Let us then define the corresponding spin-resolved cross section
\begin{align}
    2z(1-z)(2\pi)^3\frac{d\sigma^{hh'}}{d^3\Omega_0\, dz\,d^2\p} = \frac{1}{2(N_c^2-1)}\sum_{\lambda_0,b}\sum_{j_1,j_2}\left\langle |\cM^{\lambda_0hh';bj_1j_2}|^2\right\rangle\,,
\end{align}
where $\langle \dots \rangle$ denotes the average over medium configurations introduced in Section~\ref{sec:xsecs_BDMPSZ} and where the Lorentz-invariant two-particle phase space 
\begin{align}
    d\Pi_2
    =
    \frac{d^3p_1}{(2\pi)^3 2E_1}
    \frac{d^3p_2}{(2\pi)^3 2E_2}
    =
    \frac{d^3\Omega_0 \,dz\,d^2\p}{2z(1-z)(2\pi)^3}\,,
\end{align}
was written in terms of total $(p_0^+, \P)$ and relative momentum $(z, \p)$ variables, with
\begin{align}
    \P \equiv \p_1+\p_2\,,\qquad \p \equiv (1-z)\p_1 - z\p_2\,,
\end{align}
which is a transformation with jacobian $p_0^+$. The total momentum phase space is contained in $d^3\Omega_0 \equiv dp_0^+ d^2\P /(2p_0^+(2\pi)^3)$. The $1/(2(N_c^2-1))$ accounts for the average over the polarisation ($\lambda_0$) and colour state ($b$) of the initial gluon, as per the assumptions of unpolarized and incoherent initial production. Upon squaring the amplitude in Eq.~\eqref{eq:ch3_gqqb_M} and using the composition law of the propagator in Eq.~\eqref{eq:propagator_composition_law} together with the time locality of the medium average in Eq.~\eqref{eq:A_correlator}, the fully differential cross-section reads\footnote{We have written the integration over the whole range of light-cone times as an ordered integration $\bar x^+ > x^+$, accounting for the opposite ordering through $2\Re$. This follows trivially from
\begin{align}
    \int_{0< \bar x^+ < x^+} f(x^+)f^\ast(\bar x^+) = \left(\int_{0< x^+ <\bar x^+}f(x^+)f^\ast(\bar x^+)\right)^\ast\,.
\end{align}}
\begin{figure}[t]
\centering
\begin{tikzpicture}[
    scale=1.0,
    >=stealth,
    gluon/.style={decorate, decoration={coil, amplitude=3pt, segment length=5pt}, thick},
    fermion/.style={thick, postaction={decorate}, decoration={markings, mark=at position 0.58 with {\arrow{>}}}},
    axis/.style={->, thick},
    vline/.style={dashed, gray},
    dot/.style={circle, fill=black, inner sep=1.4pt},
    lab/.style={font=\small}
]

% Coordinates
\coordinate (M)      at (0.0,4.0);
\coordinate (V)      at (4.0,4.0);
\coordinate (Mb)     at (0.0,1.15);
\coordinate (Vb)     at (6.2,1.15);

% Final endpoints: separated, no joining, all at x_\infty^+
\coordinate (P1a)    at (8.7,5.55);
\coordinate (P2a)    at (8.7,2.90);
\coordinate (P1b)    at (8.7,1.97);
\coordinate (P2b)    at (8.7,0.56);

% Time axis
\draw[axis] (-0.3,-0.45) -- (10,-0.45) node[right] {$x^+$};

% Vertical time markers
\draw[vline] (M) -- (0.0,-0.45);
\draw[vline] (V) -- (4.0,-0.45);
\draw[vline] (Vb) -- (6.2,-0.45);
\draw[vline] (8.7,-0.45) -- (P1a);

\node[below, lab] at (0.0,-0.45) {$0$};
\node[below, lab] at (4.0,-0.45) {$x^+$};
\node[below, lab] at (6.2,-0.45) {$\bar x^+$};
\node[below, lab] at (8.7,-0.45) {$x_\infty^+$};

% Hard amplitudes
\node[dot] at (M) {};
\node[dot] at (Mb) {};
\node[left, lab] at (M) {$\cM_0$};
\node[left, lab] at (Mb) {$\cM_0^\dagger$};

% Incoming gluons
\draw[gluon] (M) -- (V);
\draw[gluon] (Mb) -- (Vb);

% Incoming gluon labels
\node[above, lab] at (0.9,4.1) {$\lambda_0,b,p_0^+,\p_0$};
\node[above, lab] at (3.55,4.1) {$a,\q_0$};

\node[above, lab] at (0.9,1.25) {$\lambda_0,b,p_0^+,\bar\p_0$};
\node[above, lab] at (5.6,1.25) {$\bar a,\bar\q_0$};

% Vertices
\node[dot] at (V) {};
\node[dot] at (Vb) {};
% Vertex labels: momenta and colour indices only
\node[above right, lab] at ($(V)+(0.10,0.2)$) {$k,\q_1$};
\node[below right, lab] at ($(V)+(0.10,-0.2)$) {$l,\q_0-\q_1$};
\node[above right, lab] at ($(Vb)+(0.10,0.2)$) {$\bar k,\bar\q_1$};
\node[below right, lab] at ($(Vb)+(0.10,-0.2)$) {$\bar l,\bar\q_0-\bar\q_1$};

% Amplitude daughters
\draw[fermion] (V) -- (P1a);
\draw[fermion] (V) -- (P2a);

% Conjugate-amplitude daughters
\draw[fermion] (Vb) -- (P1b);
\draw[fermion] (Vb) -- (P2b);

% Amplitude daughter labels
\node[right, lab] at (P1a) {$h,j_1,p_1^+,\p_1$};
\node[right, lab] at (P2a) {$h',j_2,p_2^+,\p_2$};

% Conjugate daughter labels
\node[right, lab] at (P1b) {$h,j_1,p_1^+,\p_1$};
\node[right, lab] at (P2b) {$h',j_2,p_2^+,\p_2$};
\end{tikzpicture}
\caption{Mixed-space representation of the squared amplitude for the ordering $x^+<\bar x^+$. The splitting occurs at $x^+$ in the amplitude and at $\bar x^+$ in the complex-conjugate amplitude and the final particles propagate to $x_\infty^+$. The labels indicate the transverse momenta ($\p_0,\,\bar\p_0,\,\q_0,\,\bar\q_0,\,\q_1,\,\bar\q_1,\,\p_1,\,\p_2$), colour indices ($b,\,a,\,\bar a,\,k,\,\bar k,\,l,\,\bar l,\,j_1,\,j_2$) and spin/polarisation indices ($\lambda_0,\,h,\,h'$) appearing in the corresponding propagators and vertices.}
\label{fig:gqqb_diagram}
\end{figure}
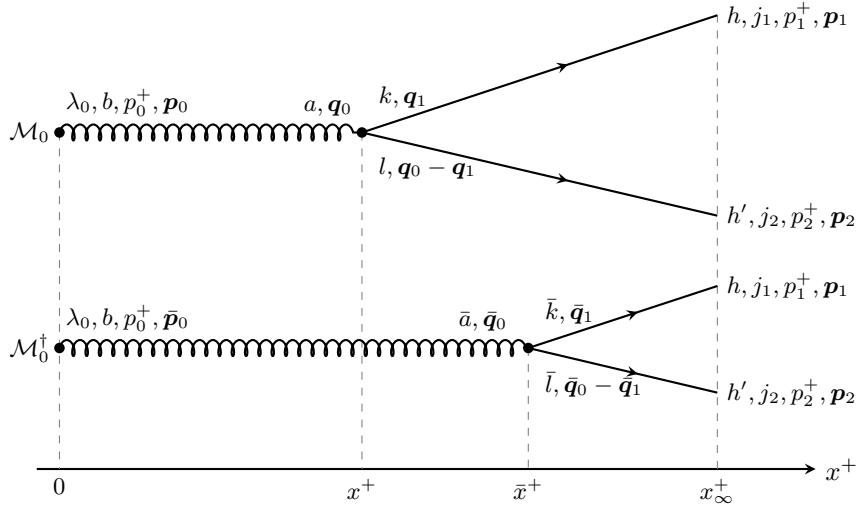
\begin{align}\label{eq:ch3_squared_M_gqqb}
    & z(1-z)(2\pi)^2\frac{d\sigma^{hh'}}{d^3\Omega_0\, dz\,d^2\p}  = \frac{\alpha_s}{(2p_0^+)^2}2\Re\int_{\p_0,\bar \p_0, \q_0,\bar \q_0, \q_1, \bar \q_1,\l_0,\l_1,\l_2, x^+<\bar x^+}\cT_{g\to q\bar q}^{hh'}(z,\bkappa,\bar\bkappa)\nn
    & \times \cM_0(p_0^+,\p_0)\cM_0^{\dagger}(p_0^+,\bar\p_0)\frac{1}{N_c^2-1}\Big\langle \cG_A^{ab}(\q_0; \p_0|p_0^+)(\cG_A^{\dagger})^{b c}(\l_0; \bar \p_0|p_0^+)\Big\rangle_{[0,x^+]}\nn
    & \times t^a_{kl}t^{\bar a}_{\bar l\bar k}\Big\langle \cG_F^{k_1 k}(\l_1;\q_1|p_1^+)\bar \cG_F^{lk_2}(\l_2;  \q_0-\q_1|p_2^+)(\cG_A^\dagger)^{c \bar a}(\bar\q_0;\l_0|p_0^+)\Big\rangle_{[x^+,\bar x^+]}\nn
    & \times \Big\langle \cG_F^{j_1 k_1}(\p_1; \l_1|p_1^+)\bar \cG_F^{k_2j_2}(\p_2; \l_2|p_2^+)(\bar \cG^\dagger_F)^{j_2\bar l}(\p_2; \bar\q_0-\bar\q_1|p_2^+)(\cG^\dagger_F)^{\bar k j_1}(\p_1; \bar\q_1|p_1^+)\Big\rangle_{[\bar x^+, x_\infty^+]}\,,
\end{align}
where $\bar\bkappa = \bar \q_1 - z\bar \q_0$, analogously to $\bkappa$, $\alpha_s = g_s^2/(4\pi)$, and we have omitted the light-cone time arguments of the propagators, since for a given medium average they share the common interval used as a subscript. The time integrations are constrained as $0<x^+<\bar x^+< x_\infty^+$. We have further defined an abbreviation for the polarisation-averaged squared vertex matrix element
\begin{align}\label{eq:ch3_T_matrix_element}
    \cT_{g\to q\bar q}^{hh'}(z,\bkappa,\bar\bkappa) \equiv 
    \frac{1}{2}\sum_{\lambda_0}V_{g\rightarrow q\bar q}^{\lambda_0 h h'}(z,\bkappa)V_{g\rightarrow q\bar q}^{\ast, \lambda_0 h h'}(z,\bar\bkappa)\,.
\end{align}
We now apply the symmetries inherent to the medium average, namely its translation invariance and colour locality, whose consequences were discussed below Eq.~\eqref{eq:A_correlator}. As a consequence, the amplitude and its complex conjugate (cc) have matching total transverse momentum and form a colour singlet at all light-cone times. The former implies the following combination of total momentum-conserving functions 
\begin{align}
    \delta^{(2)}(\l_1+\l_2 - \bar \q_0)\delta^{(2)}(\l_0-\q_0)\delta^{(2)}(\bar \p_0-\p_0)\,,
\end{align}
while in the latter the open-index colour structure of each of the three medium averages is projected onto its colour singlet contribution. This is done in detail in Appendix~\ref{app:colour_structure}, where we identify each propagator in Eq.~\eqref{eq:ch3_squared_M_gqqb} with a Wilson line dependent on a given path integration variable. As a result, the cross-section can be written as a transverse momentum convolution of objects $\cS^{(2)}\cS^{(3)}\cS^{(4)}$ containing closed-index colour structures (see Eq.~\eqref{eq:colour_regions}). The object $\mathcal{S}^{(2)}$ describes the gluon broadening in the region $[0,x^+]$, $\mathcal{S}^{(3)}$ the formation of the $q\bar q$ pair antenna in $[x^+,\bar x^+]$ and $\mathcal{S}^{(4)}$ the propagation of the resulting colour dipole in the interval $[\bar x^+,x_\infty^+]$. To further simplify the calculation, we apply the large-$N_c$ limit. In the corresponding power counting, each individual fundamental (adjoint) colour trace carries one (two) power of $N_c$, such that it is straightforward to identify the leading colour structures. Furthermore, in this limit the average of products of multiple traces reduces to products of averages of individual traces; see, e.g.,~\cite{Isaksen:2020npj} for a detailed account of Wilson line averages at all orders in $N_c$. Consequently, in comparison with the $\gamma \to q\bar q$ case discussed in Section~\ref{sec:xsecs_BDMPSZ} and to the remaining QCD vertices, the colour structure of the $g\to q\bar q$ process at large-$N_c$ is much simpler since it can be written fully in terms of dipoles (average of two Wilson lines), as shown in Eq.~\eqref{eq:colour_regions_largeNc}. Effectively, in this limit the final colour dipole propagates incoherently, i.e., the quark and the anti-quark broaden independently. Again, due to translation invariance of the average, this factorization implies total momentum is conserved individually between the quark in amplitude and cc and the same for the anti-quark. With these simplifications, the cross-section simplifies to
\begin{align}\label{eq:ch3_xsec_simplified1}
    & z(1-z)(2\pi)^2\frac{d\sigma^{hh'}}{d^3\Omega_0\, dz\,d^2\p}  = \frac{T_F\,\alpha_s}{(2p_0^+)^2}2\Re\int_{\p_0,\q_0,\q_1,\bar \q_0, \bar \q_1, x^+<\bar x^+}e^{-i\frac{m^2-i\epsilon}{2\omega^+}(\bar x^+ - x^+)}\cT_{g\to q\bar q}^{hh'}(z,\bkappa,\bar\bkappa)\nn
    & \times |\cM_0(p_0^+,\p_0)|^2\cP_A(x^+,\q_0; 0,\p_0)\cS^{(3)}(\bar x^+,\bar \q_1, \bar \q_0; x^+, \q_1, \q_0)\cP_F(x_\infty^+ ,\p_1; \bar x^+, \bar \q_1)\cP_F(x_\infty^+,\p_2;\bar x^+,\bar \q_0-\bar \q_1)
\end{align}
where $\cP$ and $\cS^{(3)}$ encode the physics of in-medium broadening and dipole formation, respectively. They are defined and discussed in detail in Appendix~\ref{app:n_point_functions}, such that after carrying out some integrations explicitly and further integrating over total final momentum variables $d^3\Omega_0$, the cross-section reads
\begin{align}\label{eq:ch3_xsec_simplified2}
    & z(1-z)(2\pi)^2\frac{d\sigma^{hh'}}{\sigma_0\, dz\,d^2\p}  =\nn
    & \frac{T_F\,\alpha_s}{(2p_0^+)^2}2\Re\int_{\bkappa,\bar\bkappa,x^+<\bar x^+}e^{-i\frac{m^2-i\epsilon}{2\omega^+}(\bar x^+ - x^+)}\cT_{g\to q\bar q}^{hh'}(z,\bkappa,\bar\bkappa)\cK(\bar x^+,\bar \bkappa; x^+, \bkappa)\cP(\p-\bar\bkappa; x_\infty^+ ,\bar x^+)\,,
\end{align}
where we assumed $\cM_0(p_0^+,\p_0)$ is a sufficiently peaked function around some fixed gluon energy, such that the integral over $p_0^+$ can be factored out. This defines the initial production cross-section, which factorises from the in-medium modification, as
\begin{align}
    \sigma_0 = \int d^3\Omega_0\, |\cM_0(p_0^+, \P)|^2\,.
\end{align}
Further, note that due to the integration over total final transverse momentum, the initial gluon broadening has been integrated out. The broadening and splitting kernels read, respectively, 
\begin{align}
    & \cP(\p-\bar \bkappa; x_\infty^+, \bar x^+) = \int_\u e^{-i(\p-\bar\bkappa)\cdot \u}D_F((1-z)\u)D_F(z\u)\,,\nn
    & \cK(\bar x^+,\bar\bkappa;x^+,\bkappa) = \int_{\r_f, \r_i} e^{i\r_f\cdot\bar\bkappa}e^{-i\r_i\cdot \bkappa }\int_{\r_i}^{\r_f}\cD \r \exp\left\{i\frac{\omega^+}{2} \int_{x^+}^{\bar x^+}ds^+\dot{\r}^2\right\}D_F((1-z)\r)D_F(z\r)\,.
\end{align}
with $\omega^+ = z(1-z)p_0^+$. Note that $\cK(\bar\bkappa, \bkappa)$ is analogous to the object we had previously briefly discussed in Eq.~\eqref{eq:K_def} for the $\gamma\to q\bar q$ process. From Appendix~\ref{app:vertices}, the polarisation-averaged vertex matrix element in Eq.~\eqref{eq:ch3_T_matrix_element}, for a spin quantization axis along $z$, reads
\begin{align}
    & \frac{1}{2}\sum_{\lambda_0}V_{g\to q\bar q}^{\lambda_0 h h'}(z,\bkappa)V_{g\to q\bar q}^{\ast,\lambda_0 h h'}(z,\bar\bkappa) =\frac{1}{z(1-z)} \left(\delta^{hh'}\left[P_{g\to q\bar q}^{\rm vac}(z)(\bkappa\cdot\bar\bkappa) + ih(1-2z)(\bkappa \times \bar\bkappa)_z\right] + \delta^{h,-h'}m^2\right) \, ,
\end{align}
where the $g\rightarrow q\bar q$ vacuum splitting function stripped off its colour factor reads
\begin{align}
    P_{g\to q\bar q}^{\rm vac}(z) = z^2+ (1-z)^2\,,
\end{align}
and the last term is the $z$ component of the cross product between the two transverse vectors, i.e., $(\bkappa \times \bar\bkappa)_z = \bkappa_x\bar\bkappa_y-\bkappa_y\bar\bkappa_x$. It is now clear what was meant by spin coupling to directional transverse structures, in particular an odd-parity structure. For an isotropic medium such spin coupling naturally vanishes. Let now write the cross-section directly in coordinate-space, such that $\bkappa$ and $\bar\bkappa$ are replaced by appropriate derivatives\footnote{With a slight abuse of notation, we denote the kernels $\cP$, $\cK$ and their Fourier transforms by the same respective symbols.}
\begin{align}\label{eq:ch3_xsec_final_before_HO}
    & (2\pi)^2\frac{d\sigma^{hh'}}{\sigma_0\, dz\,d^2\p}  = \frac{T_F\,\alpha_s}{(2\omega^+)^2}2\Re\int_{\u,x^+<\bar x^+}e^{-i\frac{m^2-i\epsilon}{2\omega^+}(\bar x^+ - x^+)}\,e^{i\p\cdot\u}\,\cP(\u; x_\infty^+ ,\bar x^+)\nn
    & \times \left\{\delta^{hh'}\left[P_{g\to q\bar q}^{\rm vac}(z)\left(\partial_{\r_i}\cdot\partial_{\r_f}\right) + ih(1-2z)\left(\partial_{\r_i}\times \partial_{\r_f}\right)_z\right] + \delta^{h,-h'} m^2\right\}\left.\cK(\bar x^+,\r_f; x^+, \r_i)\right|_{\r_i=0,\,\,\r_f = \u}\,,
\end{align}
To proceed with the calculation of the in-medium cross-section, we employ the HO approximation introduced in Section~\ref{eq:HO_approx}, extending it to the anisotropic version discussed and motivated in Section~\ref{eq:ch3_anisotropic_HO}. In this case, the splitting kernel appearing in the cross-section calculation is the one we anticipated in Eq.~\eqref{eq:ch3_K_aniso}, which is valid for a static, transversely homogeneous matter. Note, however, that we will be working with a medium with \emph{finite length}, i.e., $\hat q(s^+) = \Theta(s^+<L^+)\hat q $, with $L^+$ typically a few fm. The solution in Eq.~\eqref{eq:ch3_K_aniso}, which we shall denote by $\cK^{\rm in-in}$, is valid for the ordering $x^+ < \bar x^+ < L < x_\infty^+$, while the solution for the remaining two orderings is distinct. These read
\begin{align}
    & \cK^{\rm in-out}(\bar x^+, \r_f; x^+, \r_i) = \int_\u \cK_{0}(\bar x^+,\r_f; L, \u)\cK^{\rm in-in}(L,\u; x^+,\r_i)\,,\qquad x^+ < L < \bar x^+ < x_\infty^+\,,\nn
    & \cK^{\rm out-out}(\bar x^+, \r_f; x^+, \r_i) = \cK_{0}(\bar x^+, \r_f; x^+, \r_i)\,,\qquad L < x^+ < \bar x^+ < x_\infty^+\,,
\end{align}
where $\cK_0$ is the vacuum propagator in coordinate space
\begin{align}
    \cK_0(\bar x^+, \r_f; x^+, \r_i) = \frac{\omega^+}{2\pi i \Delta x^+}\exp\left\{\frac{i\omega^+}{2\Delta x^+}(\r_f-\r_i)^2\right\}\,.
\end{align}
with $\Delta x^+ = \bar x^+ - x^+$. The in-out term corresponds to the contribution where the splitting occurs outside the medium in the cc amplitude and out-out to the contribution where it occurs in vacuum both in the amplitude and its cc. The broadening kernel also reads differently depending on the time ordering. In the HO approximation, the solution is given by Eq.~\eqref{eq:HO_dipole}, where the exponent is only non-vanishing for the in-in term. In this case it reads explicitly
\begin{align}
    \cP^{\rm in-in}(\u; x_\infty^+,x^+) = \exp\left\{-\frac{1}{4}(L-\bar x^+)(z^2+(1-z)^2)\left(\hat q_x \u_x^2 + \hat q_y \u_y^2\right)\right\}\,,\qquad x^+ < \bar x^+ < L < x_\infty^+\,,
\end{align}
while it is $1$ for the other two orderings, where final state broadening is absent. Note that we are using the fundamental jet quenching parameter $\hat q \equiv \hat q_F$. Thus, this splits the cross-section into three corresponding terms. After carrying out the derivatives, the remaining Fourier transforms in $\u$ are Gaussian integrals. Parameterizing the relative transverse momentum as $\p = |\p|(\cos\phi,\sin\phi)$ and writing $|\p|^2 \simeq (\omega^+ \theta)^2/2$ in the collinear limit, the spin resolved, triple differential in-medium cross-section for $g\to q\bar q$ in the large-$N_c$ limit reads~\cite{Barata:2024bqp}
\begin{align}\label{eq:ch3_xsec_final}
    & (2\pi)^2\frac{\,d\sigma_{\rm in-in}^{hh'}}{\sigma_0\, dz\,d\theta\,d\phi} = \,T_F\,\alpha_s\,\theta\, \Re\int_{X^+}\,e^{-i\frac{m^2}{2\omega^+}(\bar x^+-x^+)}(-i)\frac{\sqrt{c_{1x}}\sqrt{c_{1y}}}{\sqrt{c_{3x}}\sqrt{c_{3y}}}\exp\left\{\frac{i\omega^+\theta^2}{4}\left(\frac{\cos^2\phi}{c_{3x}}+\frac{\sin^2\phi}{c_{3y}}\right)\right\}\nn
    & \times \Bigg\{\delta^{hh'}P_{g\to q\bar q}^{\rm vac}(z)\left[\omega^+(c_{1x}+c_{1y})+\omega^+\left(\frac{c_{1x}c_{2x}}{c_{3x}}+\frac{c_{1y}c_{2y}}{c_{3y}}\right)+i\frac{(\omega^+\theta)^2}{2}\left(\frac{c_{1x}c_{2x}}{c_{3x}^2}\cos^2\phi+\frac{c_{1y}c_{2y}}{c_{3y}^2}\sin^2\phi\right)\right]\nn
    &+\delta^{hh'}h(1-2z)\frac{(\omega^+\theta)^2}{2}\frac{(c_{1y}c_{2x}-c_{1x}c_{2y})}{c_{3x}c_{3y}}\cos\phi\sin\phi+\delta^{h,-h'}\frac{m^2}{2}\Bigg\}\,,\nn
    & (2\pi)^2\frac{\,d\sigma_{\rm in-out}^{hh'}}{\sigma_0\, dz\,d\theta\,d\phi} = T_F\,\alpha_s\,\frac{\omega^+\theta}{\theta^2+2(m/\omega^+)^2}\,2\Re\int_0^{L^+}dx^+\,e^{-i\frac{m^2}{2\omega^+}(L^+-x^+)}(i)\frac{\sqrt{c_{1x}}\sqrt{c_{1y}}}{\sqrt{c_{2x}}\sqrt{c_{2y}}}\nn
    & \times \exp\left\{-\frac{i\omega^+\theta^2}{4}\left(\frac{\cos^2\phi}{c_{2x}}+\frac{\sin^2\phi}{c_{2y}}\right)\right\}\Bigg\{\delta^{hh'}P_{g \to q\bar q}^{\rm vac}(z)\left[\frac{c_{1x}}{c_{2x}}\cos^2\phi+\frac{c_{1y}}{c_{2y}}\sin^2\phi\right]\theta^2\nn
    &+\delta^{hh'}h(1-2z)\left(\frac{c_{1x}}{c_{2x}}-\frac{c_{1y}}{c_{2y}}\right)\theta^2\cos\phi\sin\phi-i\delta^{h,-h'}\left(m/\omega^+\right)^2\Bigg\}\,,\nn
    & (2\pi)^2\frac{\,d\sigma_{\rm out-out}^{hh'}}{\sigma_0\, dz\,d\theta\,d\phi} = T_F\,\alpha_s\frac{\theta}{\left(\theta^2 + 2(m/\omega^+)^2\right)^2}\left(\delta^{hh'}P_{g\to q\bar q}^{\rm vac}(z)\theta^2+\delta^{h,-h'}2\left(m/\omega^+\right)^2\right)\,,
\end{align}
where the integration range for the in-in term  reads $X^+ = \{0<x^+<\bar x^+< L^+\}$ and we have taken $\epsilon\to 0^+$ everywhere since the remaining time integrations are compact. Note that the out-out term exactly corresponds to the pure vacuum term, leading to the massive $g\to q\bar q$ vacuum splitting function when summed over spins and normalized appropriately. We have introduced the notation
\begin{align}\label{eq:ch3_ci_definition_static}
    c_{1i}=\frac{\Omega_i}{2i\sin(\Omega_i\Delta x^+)}\,,\qquad c_{2i}=\frac{\Omega_i}{\tan(\Omega_i\Delta x^+)}\,,\qquad c_{3i}=-\frac{i(L^+-\bar x^+)\hat q_i(z^2+(1-z)^2)}{2\omega^+}-c_{2i}\,,
\end{align}
with complex oscillator frequency
\begin{align}
    \Omega_i=\frac{1-i}{\sqrt{2}}\sqrt{\frac{\hat q_i(z^2+(1-z)^2)}{2\omega^+}}\,,
\end{align}
and $i=x,y$. For the in-in term the time argument is $\Delta x^+=\bar x^+-x^+$, while for the in-out term it reads $\Delta x^+=L^+-x^+$.

\section{Azimuthal jet observables}\label{sec:azimuthal_observables}
In the previous section, we computed the spin-dependent cross-section to produce a $q\bar q$ pair in an anisotropic medium. Using this result, we now study the impact of the medium anisotropy, encapsulated into $\hat q_i$, on jet observables. We will first study the total distribution of $q\bar q$ pairs. This is complemented by a study of the obtained observables in terms of their Fourier decomposition, similar to the extraction of flow harmonics commonly performed in soft physics studies in HICs, see e.g.~\cite{Snellings:2011sz}. Finally we discuss how measurements of spin along different directions in the transverse plane can give information about the $q\bar q$ polarization and its relation to the direction of greater anisotropy.

\subsection{Azimuthal particle distribution}\label{subsec:azimuthal_distribution}

In the following two subsections, the numerical results shall be focused on the equal spins contribution ($\delta^{hh'}$) to the in-in term in Eq.~\eqref{eq:ch3_xsec_final} (the leading term for a sufficiently large medium), integrated over $|\p|\in(0,+\infty)$. This results in~\cite{Barata:2024bqp}\footnote{It is worth pointing out that $\text{Im}(c_{3i}) < 0$ for all $x^+$ and $\bar x^+$ guarantees the convergence of the integral over $|\p|$ and is responsible for the sign attached to the second term in Eq.~\eqref{eq:in_in_azimuth}.}
\begin{align}\label{eq:in_in_azimuth}
    & \sum_{h'}\delta^{hh'}(2\pi)\frac{\,d\sigma_{\rm in-in}^{hh'}}{\sigma_0\, dz\,d\phi} \equiv (2\pi)\frac{dN^{h}}{dz\,d\phi} =  T_F\,\frac{\alpha_s}{\pi}\,\Re\int_{X^+}\frac{\sqrt{c_{1x}}\sqrt{c_{1y}}}{\sqrt{c_{3x}}\sqrt{c_{3y}}}\,e^{-i\frac{m^2}{2\omega^+}(\bar x^+-x^+)}\nn
    & \times \Bigg\{P_{g\to q\bar q}^{\rm vac}(z)\left[\frac{(c_{1y}c_{2y}c_{3x}-c_{1x}c_{2x}c_{3y})(c_{3y}\cos^2\phi-c_{3x}\sin^2\phi)}{\left(c_{3y}\cos^2\phi+c_{3x}\sin^2\phi\right)^2}+\frac{c_{3x}c_{3y}(c_{1x}+c_{1y})}{c_{3y}\cos^2\phi+c_{3x}\sin^2\phi}\right]\nn
    &+2ih(1-2z)\frac{c_{3x}c_{3y}(c_{1y}c_{2x}-c_{1x}c_{2y})\sin\phi\cos\phi}{\left(c_{3y}\cos^2\phi+c_{3x}\sin^2\phi\right)^2}\Bigg\}\,.
\end{align}
Note that the distribution in Eq.~\eqref{eq:in_in_azimuth} has a divergence of the form $1/\Delta x^+$ as $\Delta x^+ =\bar x^+ - x^+\rightarrow 0$, as noted in~\cite{Hauksson:2023tze}, which is removed by subtracting a contribution corresponding to $\hat q_i \rightarrow 0$ to all our results. This divergence would cancel naturally had we included the in-out contribution in Eq.~\eqref{eq:ch3_xsec_final}, which we will not do for simplicity of the numerical results. The results are presented in terms of the dimensionless quantities

\begin{figure}[t]
     \centering
         \includegraphics[width=0.49\textwidth]{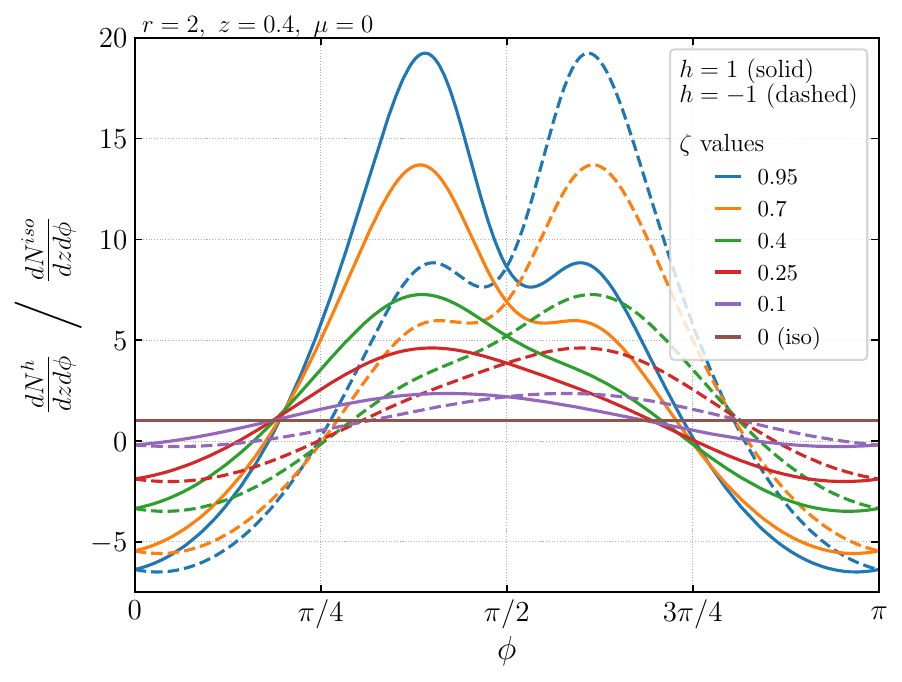}
         \centering
         \includegraphics[width=0.49\textwidth]{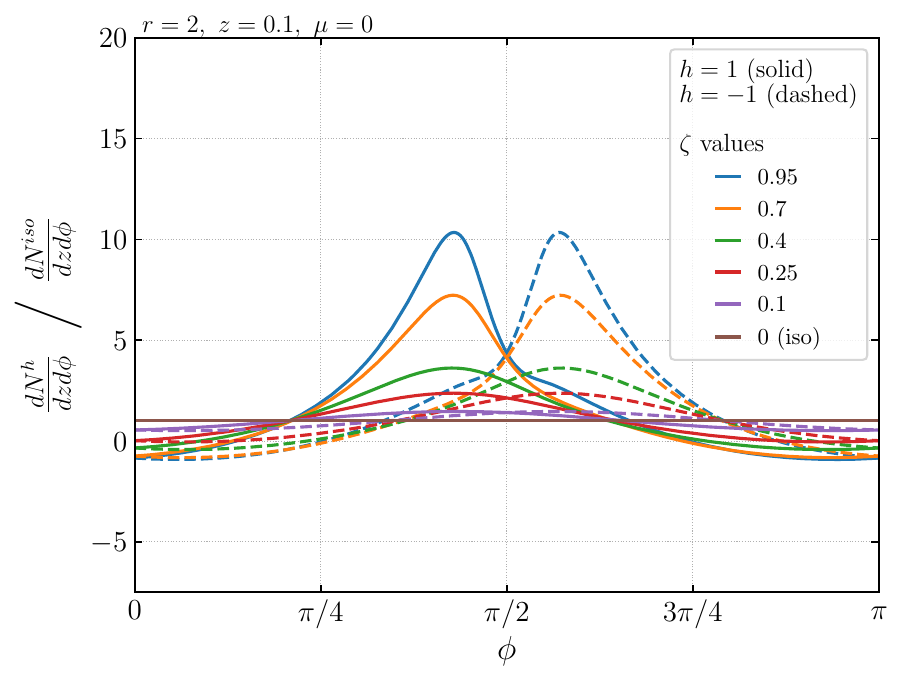}
         \centering
         \includegraphics[width=0.49\textwidth]{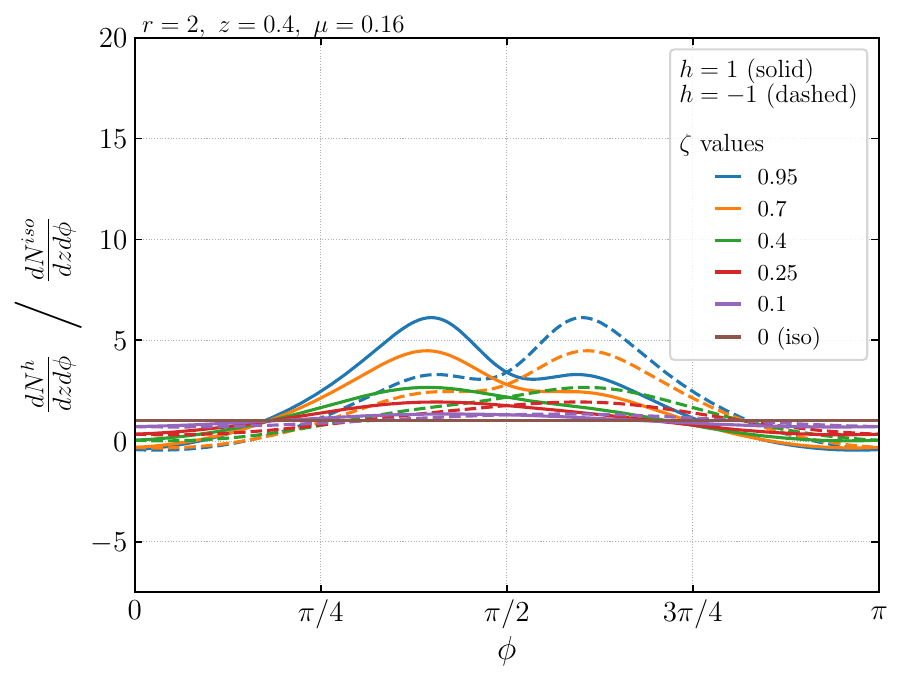}
         \centering
         \includegraphics[width=0.49\textwidth]{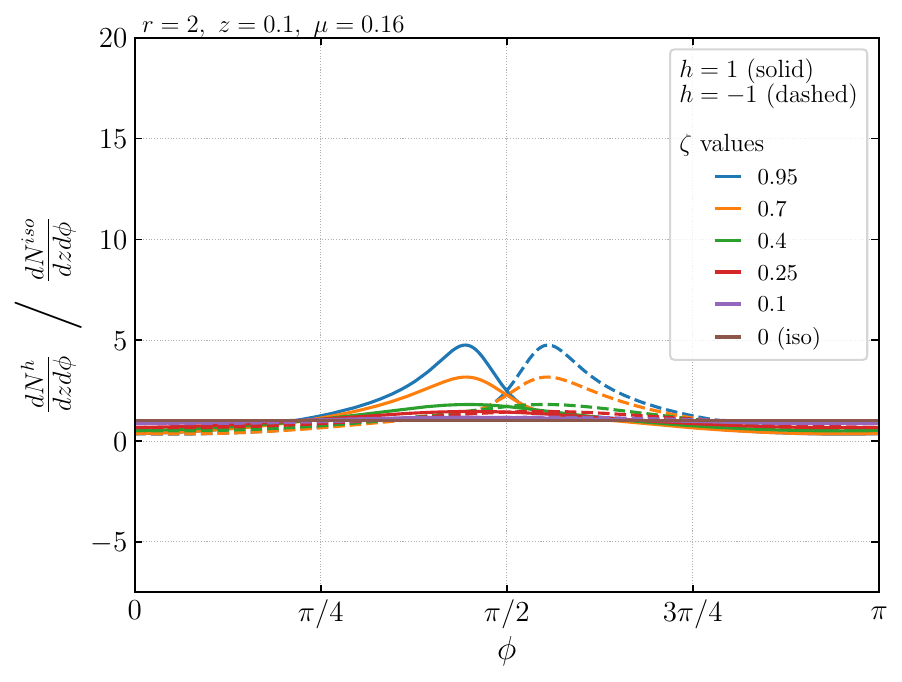}
         \label{fig:three sin x}
     \caption{Ratio of the azimuthal distribution (Eq.~\eqref{eq:in_in_azimuth}) for $h=1$ with respect to the isotropic case ($\zeta=0$) as a function of $\phi$ for different values of the anisotropy parameter $\zeta$, for fixed $r=2$, for more energetically balanced splittings ($z=0.4$, \textbf{left}) and more unbalanced splittings ($z=0.1$, \textbf{right}). The top row corresponds to the massless case ($\mu=0$) and the bottom row to the $g\rightarrow c\bar c$ case ($\mu=0.16$, $m=1.27$ GeV/$c^2$). The solid curve corresponds to spin state $h=1$ and the dashed curve to $h=-1$.}
     \label{azimuthal_plot_phi}
\end{figure}
\begin{equation}\label{dimensionless_var_def}
    \zeta = \frac{\sqrt{\hat q_y} - \sqrt{\hat q_x}}{\sqrt{\hat q_y} + \sqrt{\hat q_x}}\, ,\quad  r = L^+ \frac{\left(\sqrt{\hat q_y} + \sqrt{\hat q_x}\right)}{2\sqrt{2p_0^+}} \, , \qquad \mu = \frac{2 m^2}{\left(\sqrt{\hat q_y} + \sqrt{\hat q_x}\right)\sqrt{p_0^+}} \, ,
\end{equation}
which give a measure of the anisotropy $\zeta$, a rescaled medium length $r$ and a rescaled quark mass $\mu$. In the numerical results that follow, we shall vary the value of the anisotropy parameter over a wide range $0.10 \leq \zeta \leq 0.95$ while fixing a reasonable value for $r$ and $\mu$. In the end, we are not concerned with using exact physical values for these parameters, as our goal is not to have exact quantitative predictions, but rather to gain qualitative understanding of the effect of different degrees of anisotropy of the medium in our calculation. The values for $r$ and $\mu$ can be estimated by considering $L^+/\sqrt{2} \sim 5$ fm, $p_0^+/\sqrt{2} \sim 100$ GeV, $(\sqrt{\hat q_x} + \sqrt{\hat q_y})/2 \sim 2/\sqrt{2} $ GeV/fm$^{-1/2}$~\footnote{This corresponds to fixing the adjoint $\hat q$ as $(\sqrt{\hat q_{A,x}} + \sqrt{\hat q_{A,x}})/2 \sim 2 $ GeV/fm$^{-1/2}$, consistent with~\cite{Barata:2024bqp}.} and the charm quark mass $m = 1.27 $ GeV/$c^2$:
\begin{align}
    & r \sim 2\, ,\, \quad \mu \sim 0.16\,.
\end{align}
Note that the quenching (strength of $\hat q_i$ and size of the medium $L^+$) is purposefully exaggerated so that the anisotropy signature is clearer in the following plots.

In Fig.~\ref{azimuthal_plot_phi} we display the evolution of the azimuthal distribution for massless quarks (top panels) and massive quarks (bottom panels) as a function of $\phi$, for different values of the anisotropy parameter $\zeta$ and for two values of the energy fraction -- $z=0.4$ (left panel) and $z=0.1$ (right panel). The remaining parameters, such as the rescaled medium length, are held fixed. The plots show the ratio between the azimuthal distribution for a given spin state ($h=1$ solid, $h=-1$ dashed) and the isotropic result ($\zeta=0$), thereby measuring how sensitive the relative azimuthal orientation of the $q\bar q$ pair is to the medium anisotropy. Since the isotropic distribution is flat ($\phi$ independent), this ratio simply corresponds to a normalization change of all curves. In both the massless and massive cases, increasing the anisotropy leads to a larger ratio near $\pi/2$, where the deviation from the isotropic case appears most pronounced. The maximum is not located exactly at $\pi/2$ because of the spin-dependent contribution. In addition, one observes a mild suppression near $0$ and $\pi$, which becomes slightly stronger as the anisotropy increases. For $\hat q_x > \hat q_y$, i.e. $\zeta < 0$, the opposite pattern would be obtained -- with peaks around $0$ and $\pi$. We also find that, for smaller anisotropies, the distributions become more symmetric with respect to $\phi=\pi/2$, indicating that the spin-independent terms (symmetric under $\phi\rightarrow \pi-\phi$) dominate over the spin-dependent ones (anti-symmetric under the same transformation). Moreover, in this regime the distributions show a more regular oscillatory pattern. Together, these features point to an elliptic-type deformation of the distribution in the transverse plane. Comparing the solid and dashed curves of each color, one sees that the two spin states can be more clearly distinguished as the anisotropy increases and, seemingly, as the splitting becomes softer, owing to the $(1-2z)$ factor. This effect is again most visible near $\pi/2$. The reflection symmetry is simply a consequence of the two distributions associated with $h=\pm 1$ being mirrored with respect to $\phi=\pi/2$. The fact that the two peaks appear in clearly separated regions indicates that positive and negative spin states tend to populate different regions inside the jet. Regarding the role of mass, it appears to reduce the sensitivity to the anisotropy values, especially for larger anisotropies, as reflected by the curves moving closer together. The separation between the two spin states is also significantly reduced.

\subsection{Fourier harmonics}\label{subsec:harmonics_spin_sensitivity} 

The different symmetry properties of the spin-dependent and independent contributions to the final-state distribution suggest that one can isolate the spin terms through a harmonic Fourier decomposition. This is further motivated by the approximate symmetries observed above, in particular for small anisotropies, where the azimuthal dependence becomes increasingly regular. We can make this observation concrete by expanding the distribution in Eq.~\eqref{eq:in_in_azimuth} in powers of $\zeta$. At leading order, we find~\cite{Barata:2024bqp}
\begin{align}\label{small_anisotropy_expansion}
    & (2\pi)\frac{dN^{h}}{dzd\phi} = \frac{T_F\,\alpha_s}{\pi} \Re  \int_{X^+} e^{-i\frac{m^2}{2\omega^+}(\bar x^+-x^+)}\Bigg[2c_1^2P_{qg}^{vac}(z) - \frac{2c_1^2}{c_3}\zeta \Bigg(ih(1-2z)\Delta x^+ (2c_2^2-\bar\Omega^2)\sin(2\phi)- \nn
    & -\frac{P_{qg}^{vac}(z)}{c_3}\left(c_2^2\left((c_2+c_3)\Delta x^+-1\right)-(c_2+2c_3)(\Delta x^+\, \bar \Omega^2 + c_3)\right)\cos(2\phi)\Bigg)\Bigg]\, ,
\end{align}
where we have introduced the shorthand notation
\begin{equation}\label{expand_ci_definition}
    \begin{split}
    &c_{1} = \frac{\bar\Omega}{2i\sin{\bar\Omega\Delta x^+}} \, , \quad \qquad c_{2} = \frac{\bar\Omega}{\tan{\bar\Omega\Delta x^+}}\, , \quad  c_{3} = (L-\bar x^+)\bar\Omega^2-c_{2}\, ,
    \end{split}
\end{equation}
with 
\begin{align}
    \bar\Omega = \left(\frac{1-i}{\sqrt{2}}\right)\sqrt{\frac{z^2+(1-z)^2}{2\omega^+}}\frac{\sqrt{\hat q_x}+\sqrt{\hat q_y}}{2}\,,
\end{align}
and $\Delta x^+ =\bar x^+ - x^+$. Equation~\eqref{small_anisotropy_expansion} shows that, at leading power in $\zeta$, the spin-independent terms are proportional to $\cos(2\phi)$, corresponding to an elliptic-type modulation, while the spin dependence enters through $\sin(2\phi)$, an odd-parity contribution, as anticipated from the symmetry arguments above. This also explains why, for small anisotropies, the azimuthal distribution approaches a more regular oscillatory behaviour. It is thus useful to organize the angular dependence in terms of Fourier harmonics, in a way analogous to the harmonic analysis commonly used for flow observables in the soft sector. Performing the Fourier decomposition for finite $\zeta$ gives
\begin{align}\label{harmonic_exp}
    (2\pi)\frac{dN^{h}}{dzd\phi} = v_0\left(1+ \sum_{n=1}^\infty v^{(h)}_n \cos\left(n \phi \right) + \sum_{n=1}^\infty w^{(h)}_n \sin(n\phi)\right) \, , 
\end{align}
where the normalized Fourier coefficients read
\begin{align}
    & v_0  = \frac{1}{2\pi} \int_0^{2\pi} d\phi \,\,  (2\pi) \frac{dN^{h}}{dz d\phi} \, , \nn 
    & v_n^{(h)}  = \frac{1}{v_0}\frac{1}{\pi} \int_0^{2\pi} d\phi \,\,  (2\pi) \frac{dN^{h}}{dzd\phi} \, \cos\left(n \phi \right)\,,\qquad n\geq 1 \, ,\nn 
    & w_n^{(h)} = \frac{1}{v_0}\frac{1}{\pi} \int_0^{2\pi} d\phi \, \,    (2\pi)\frac{dN^{h}}{dzd\phi} \, \sin\left( n\phi \right)\,,\qquad n\geq 1\, .
\end{align}  
At this point, it is important to note that several of these Fourier coefficients can be determined directly from the symmetries of the spectrum in Eq.~\eqref{eq:in_in_azimuth}. In an isotropic medium, all Fourier coefficients except for $v_0$ vanish. For an anisotropic medium, both $v^{(h)}_n$ and $w_n^{(h)}$ vanish for odd $n$ due to the discrete symmetry $\phi \rightarrow \phi - \pi$ of the particle distribution. This is expected, since the anisotropic $\hat q$ cannot generate a directed vector-like contribution (as in, e.g., flow contributions). The coefficients $v_n$ for even $n$ are non-vanishing but independent of spin, $v^{(h)}_n = v_n$, as a consequence of the $\phi \rightarrow -\phi$ anti-symmetry of the spin-dependent contribution. Thus, the spin dependence can only enter through the odd-parity coefficients $w_n^{(h)}$ in the series. This is analogous to charge imbalance measurements in chiral magnetic effect studies, see e.g.~\cite{Kharzeev:2015znc}.\footnote{Although we note that in our context we do not need to consider fluctuations, as is the case in studies of the chiral magnetic effect in heavy ions.} It is also trivial that $w_n^{(+1)}=-w_n^{(-1)}$, since $w_n^{(h)} \propto h$. Thus, non-zero values of these coefficients encode information about the directional properties of the medium and about how these properties couple to the spin structure of the jet splitting. In what follows, we will focus on the leading non-vanishing harmonics, namely the elliptic coefficient $v_2$ and the odd-parity spin-sensitive coefficient $w_2$.
\begin{figure}[t]
         \centering
         \includegraphics[width=0.49\textwidth]{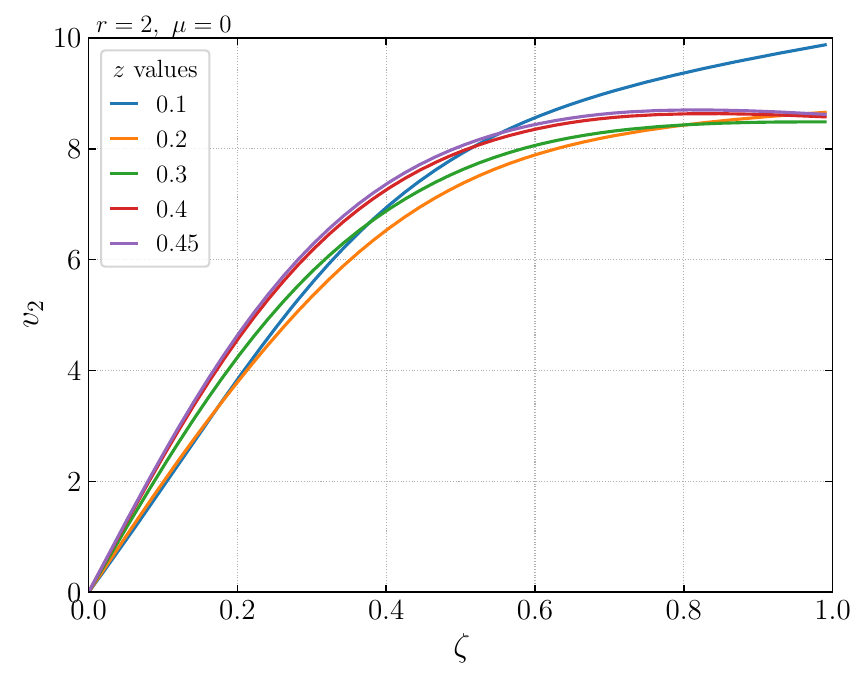}
         \centering
         \includegraphics[width=0.49\textwidth]{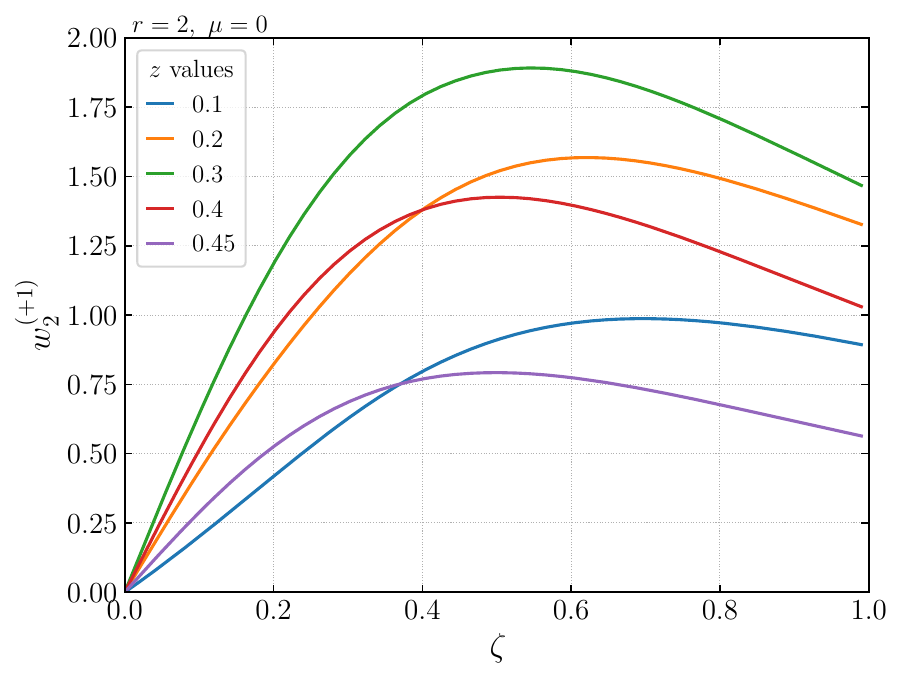}
         \centering
         \includegraphics[width=0.49\textwidth]{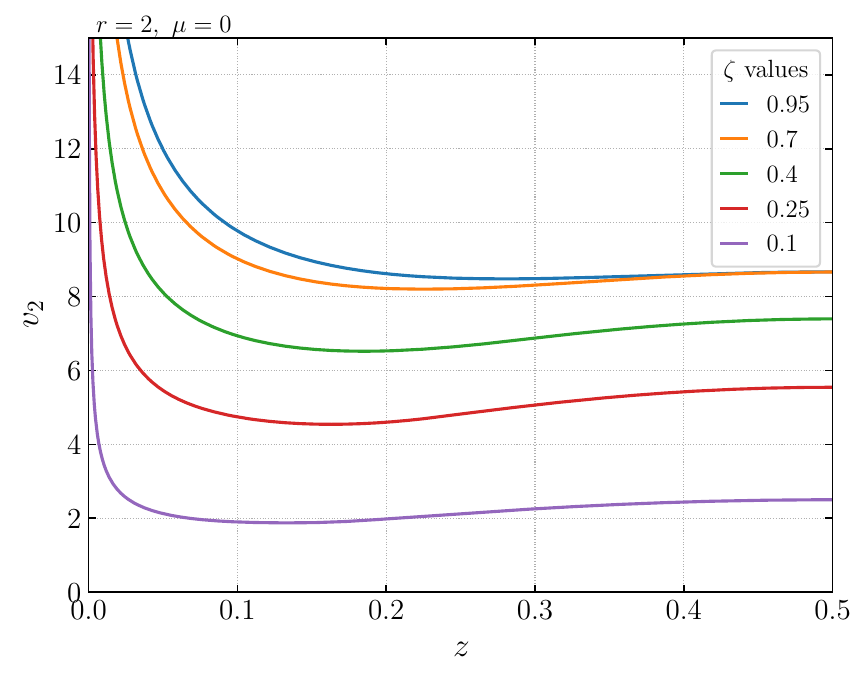}
         \centering
         \includegraphics[width=0.49\textwidth]{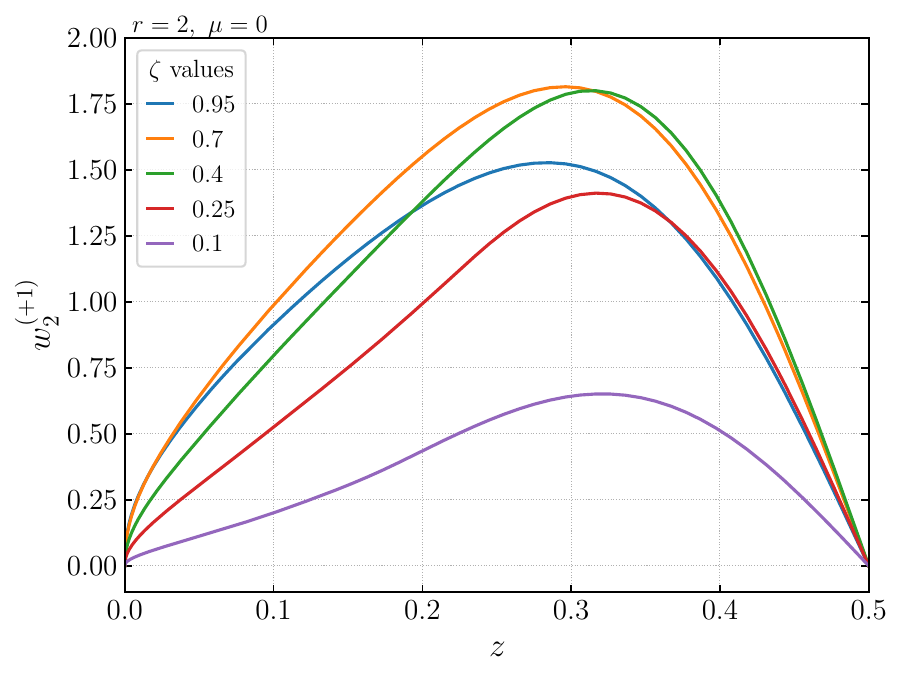}
    \caption{\textbf{Left:} Evolution of  $v_2$ in Eq.~\eqref{harmonic_exp} as a function of $\zeta$ (top line) or of $z$ (bottom line) and for different values of $z$ or of $\zeta$, respectively, with fixed $r=2$ and for massless quarks $(\mu=0)$. \textbf{Right:} equivalent plots for $w_2^{(+1)}$ in Eq.~\eqref{harmonic_exp}.}
     \label{fig:massless_harm}
\end{figure}

\begin{figure}[t]
         \centering
         \includegraphics[width=0.49\textwidth]{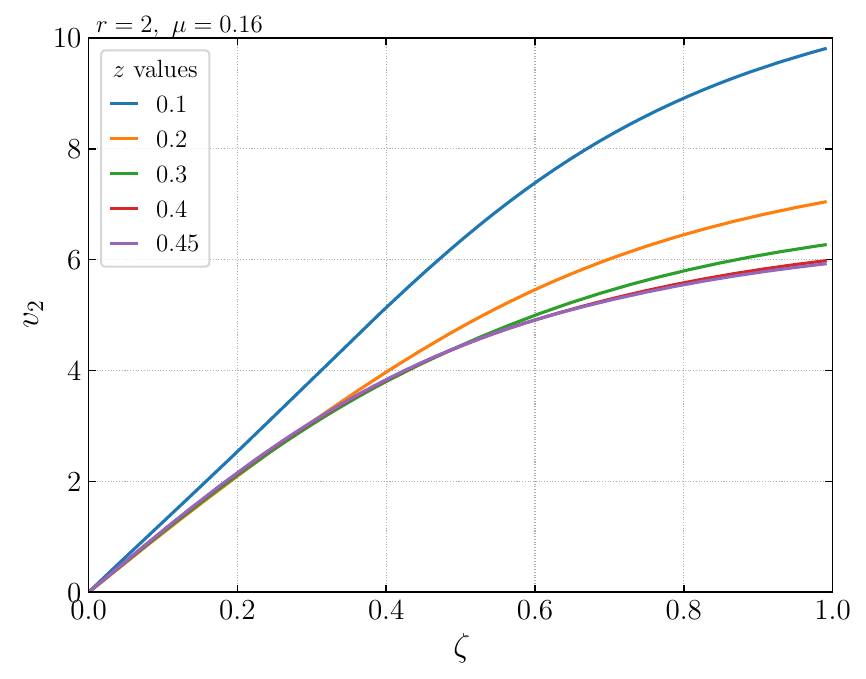}
         \centering
         \includegraphics[width=0.49\textwidth]{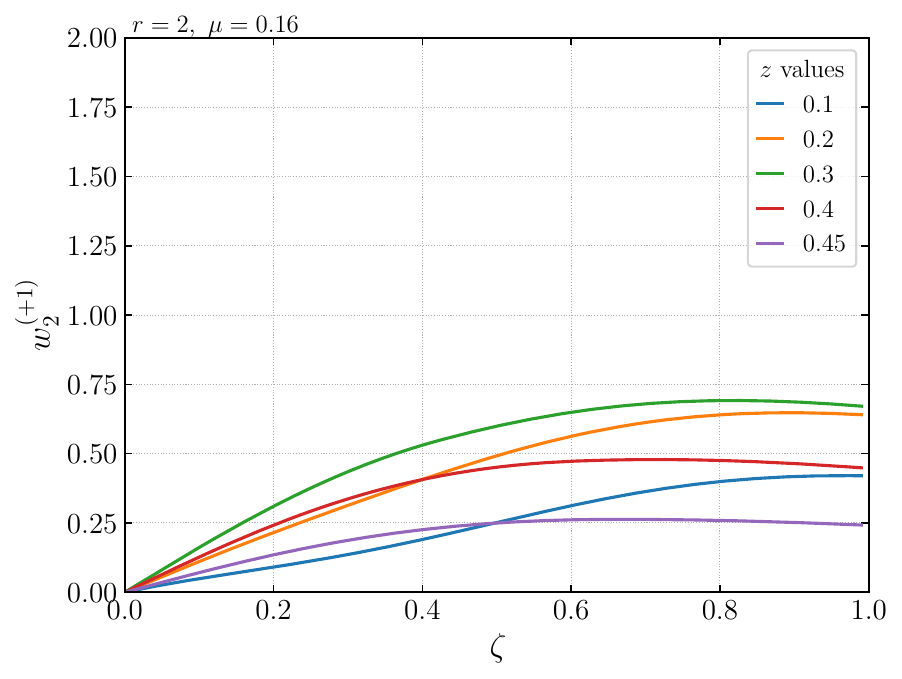}
         \centering
         \includegraphics[width=0.49\textwidth]{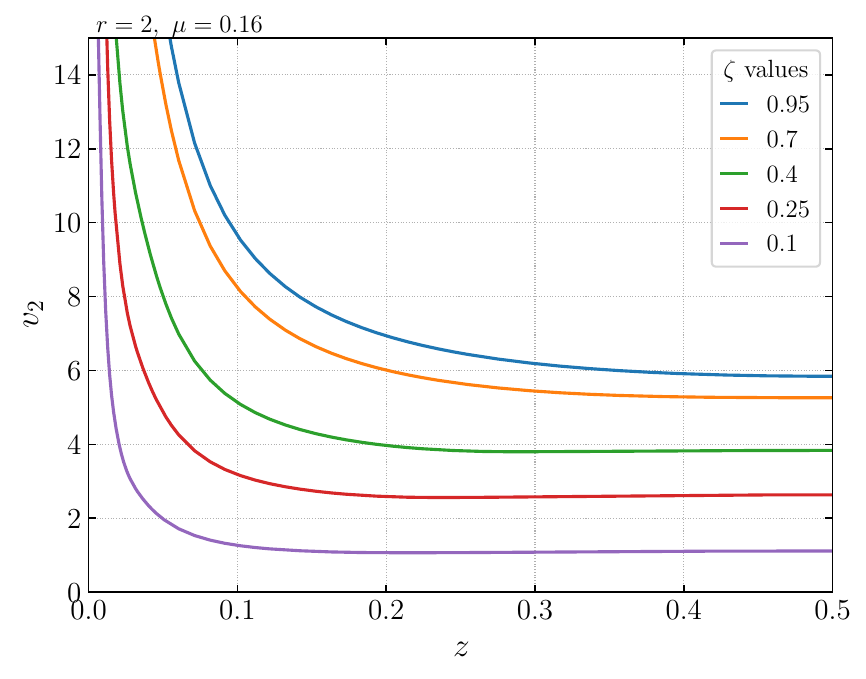}
         \centering
         \includegraphics[width=0.49\textwidth]{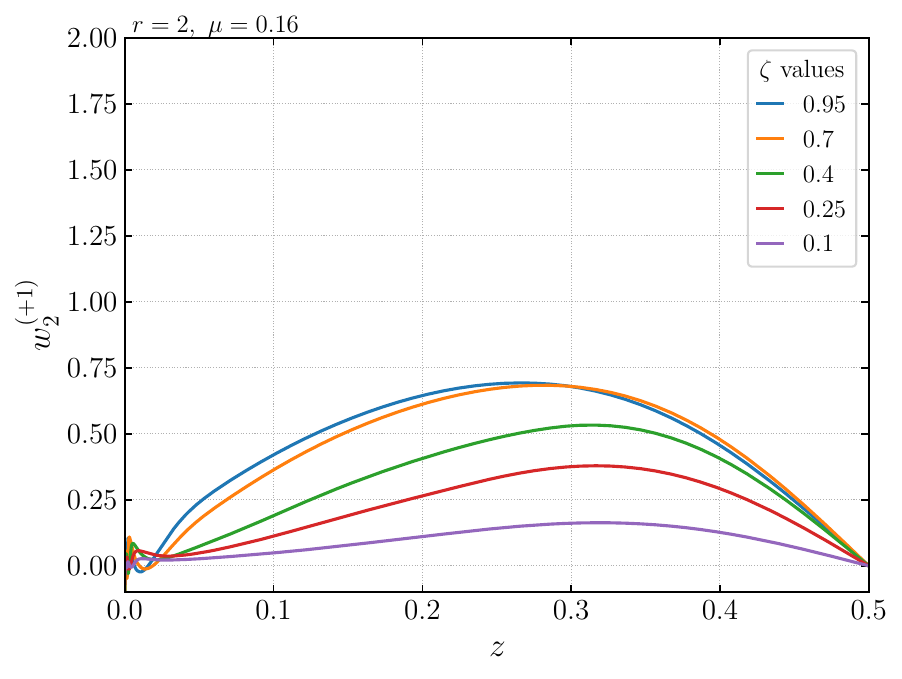}
     \caption{\textbf{Left:} Evolution of  $v_2$ in Eq.~\eqref{harmonic_exp} as a function of $\zeta$ (top line) or of $z$ (bottom line) and for different values of $z$ or of $\zeta$, respectively, with fixed $r=2$ and for $\mu = 0.16$. \textbf{Right:} equivalent plots for $w_2^{(+1)}$ in Eq.~\eqref{harmonic_exp}.}
     \label{fig:massive_harm}
\end{figure}

In Fig.~\ref{fig:massless_harm} we show the evolution of the leading harmonics, $v_2$ (left) and $w_2^{(+1)}$ (right), with $\zeta$ (top) and $z$ (bottom) for massless quarks. First, comparing the $v_2$ and $w_2^{(+1)}$ harmonics in absolute value provides a possible explanation for the near symmetry around $\pi/2$ observed in Fig.~\ref{azimuthal_plot_phi} for the smallest anisotropies. Indeed, this comparison indicates that the $\cos(2\phi)$ contribution dominates over the $\sin(2\phi)$ term, consistently with the fact that this is the leading angular modulation for small anisotropies, as shown in Eq.~\eqref{small_anisotropy_expansion}. Looking at the top left panel, we see that $v_2$ monotonically increases with the anisotropy parameter, as expected, and already reaches sizeable values for intermediate anisotropies. For sufficiently large $\zeta$, this also suggests that higher harmonics may become relevant and should eventually be included in a more complete characterization of the distribution. From the bottom left panel, we observe that $v_2$ is larger for more unbalanced splittings, although its dependence on $z$ is non-monotonic: it reaches a minimum at an intermediate value and shows only a mild dependence for sufficiently large $z$. The evolution of the $w_2$ harmonic with $\zeta$ (top right panel) is also non-monotonic. Its dependence on the energy fraction (bottom right panel) is non-monotonic as well, which follows from the fact that for $z=0$ and $z=0.5$ there is no spin dependence and therefore the harmonic must vanish, as shown. Consequently, the curves display a maximum for intermediate values of $z$, rather than at the softest splittings. The evolution in $\zeta$ is also non-trivial, since at large anisotropies the curves tend to flatten and all display a maximum.

As for the results considering massive quarks, shown in Fig.~\ref{fig:massive_harm}, we first see that $v_2$ now decreases monotonically with $z$ (bottom left panel), while reaching even larger values than in the massless case. As for $w_2$, the flattening tendency for larger anisotropies is even more evident (top right panel), and most of the curves attain a maximum for the largest anisotropy. This is supported by the behaviour of $w_2$ with $z$ in the bottom right panel, where the curves for different anisotropies are monotonically arranged for most values of the energy fraction. Some oscillations appear in the evolution of $w_2$ with $z$, which we attribute to numerical instabilities introduced by the mass phase in conjunction with small values of $z$. Finally, we see significantly smaller values of $w_2$ for the massive case in comparison with the massless case in Fig.~\ref{fig:massless_harm}, explaining the less pronounced distinction between spin states in the bottom panels of Fig.~\ref{azimuthal_plot_phi}. Overall, mass seems to obscure some of the characteristic features of the azimuthal distribution. A more interesting interplay with mass will be shown in the results of the next section.

To summarize, extracting harmonic coefficients inside jets would provide a new window into jet substructure, with a clear signal for non-trivial medium properties being separated from more traditional medium effects, such as momentum broadening and energy loss, which are mainly isotropic and therefore cannot generate any harmonic except for $v_0$.

\subsection{Transverse polarization of a massive dipole}\label{subsec:tv_polarisation}

\begin{figure}[t]
         \centering  
         \includegraphics[width=0.49\textwidth]{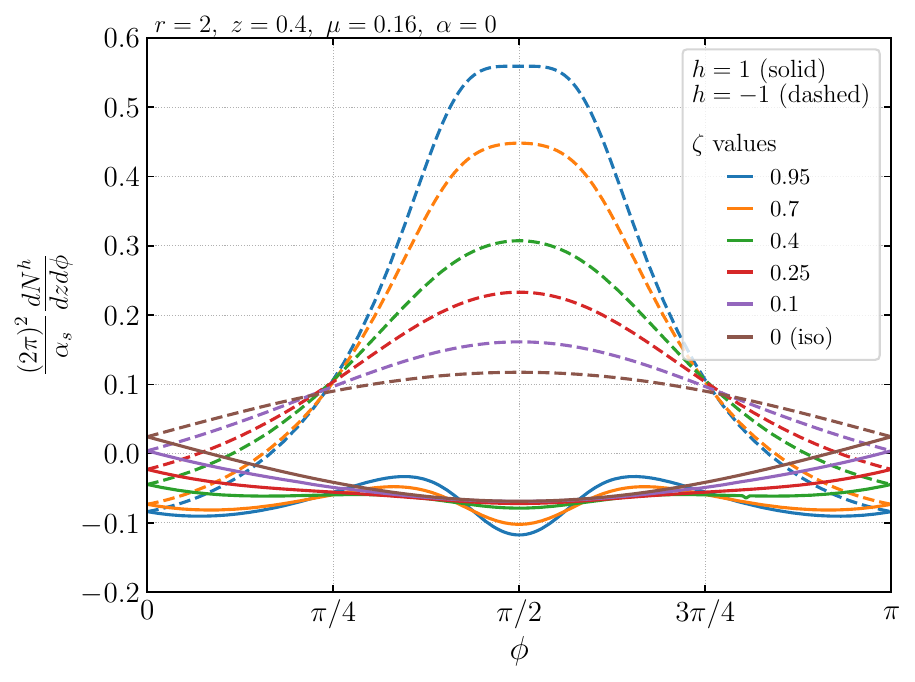}
         \centering
         \includegraphics[width=0.49\textwidth]{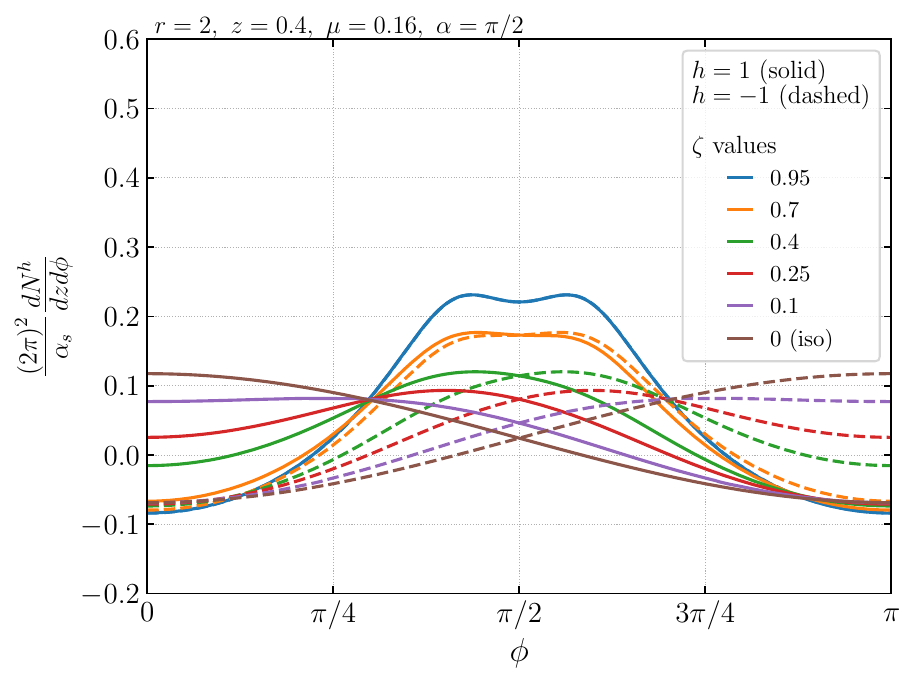}
         \centering
         \includegraphics[width=0.49\textwidth]{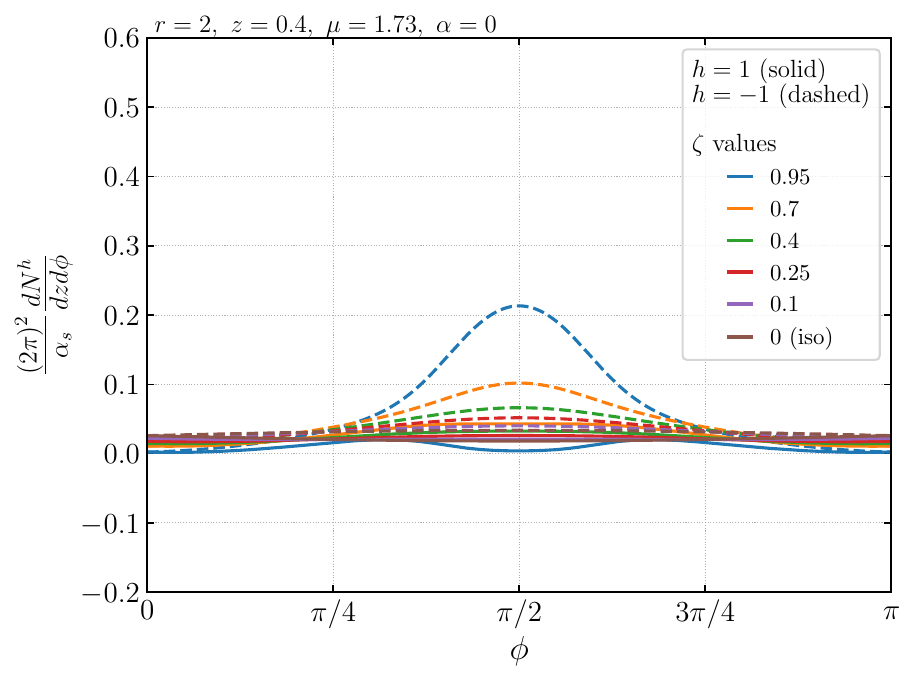}
         \centering
         \includegraphics[width=0.49\textwidth]{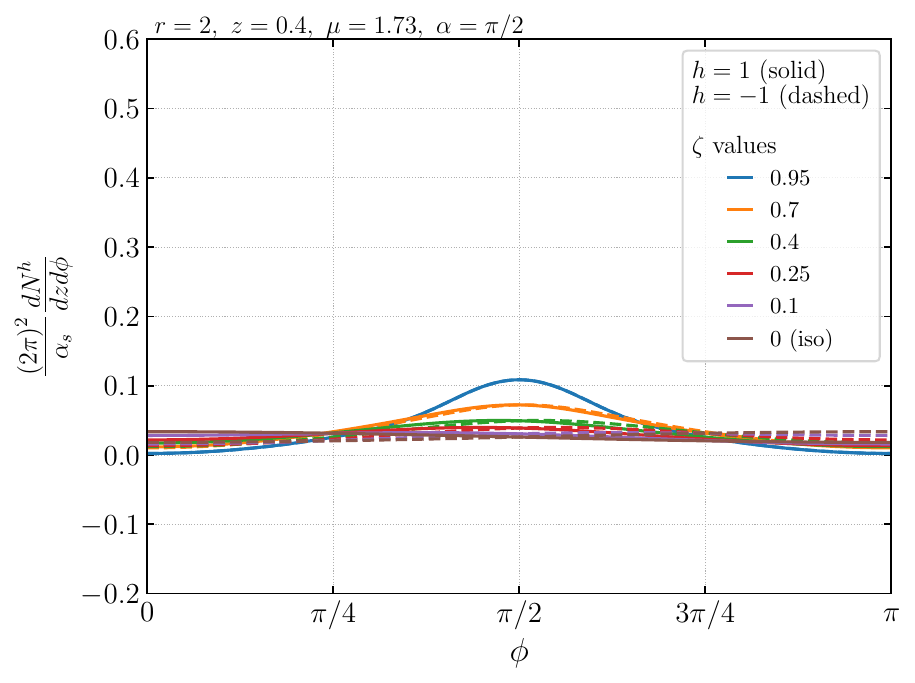}
     \caption{Plots of the distribution in Eq.~\eqref{massive_tvaligned_final} as a function of $\phi$ for different values of the anisotropy parameter $\zeta$, for $\alpha=0$ (\textbf{left}) and $\alpha = \pi/2$ (\textbf{right}), for fixed $r=2$, $z=0.4$ and for two different values of $\mu$ -- $0.16$ (upper) and $1.73$ (lower), corresponding to the charm and bottom quark masses, respectively.}
     \label{tvPol_1}
\end{figure}

\begin{figure}[t]
         \centering
         \includegraphics[width=0.49\textwidth]{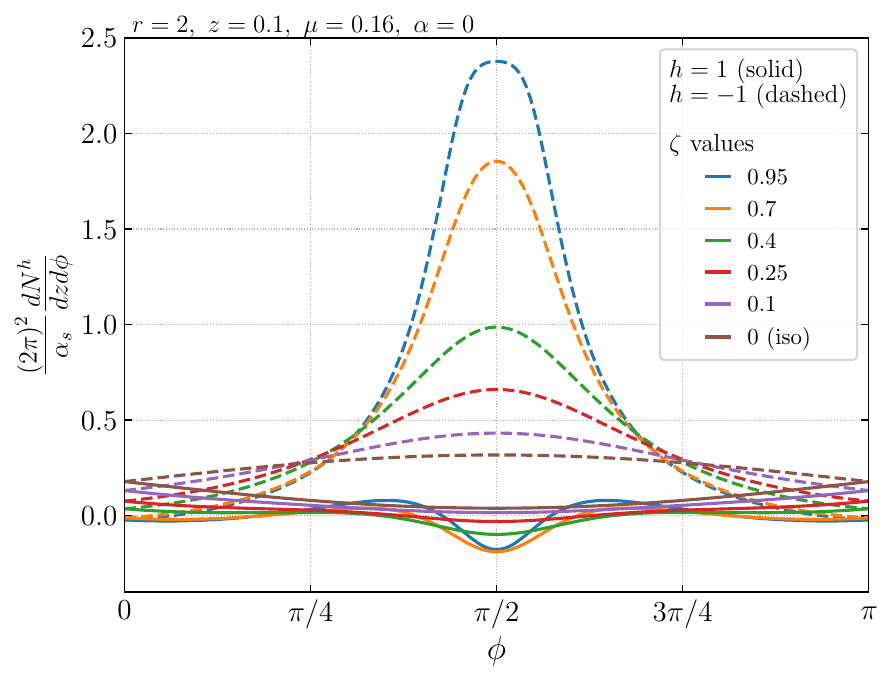}
         \centering
         \includegraphics[width=0.49\textwidth]{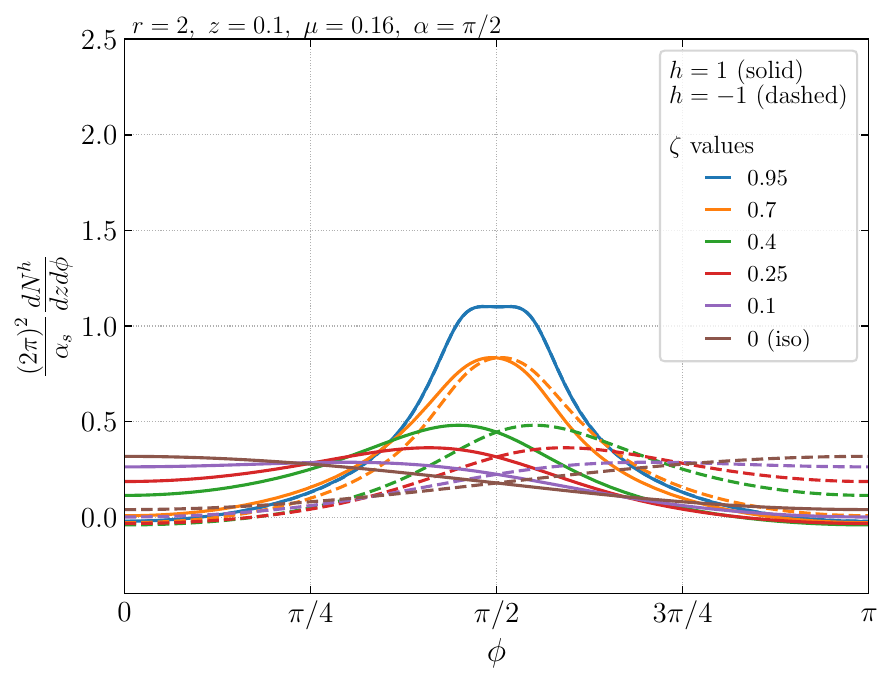}
         \centering
         \includegraphics[width=0.49\textwidth]{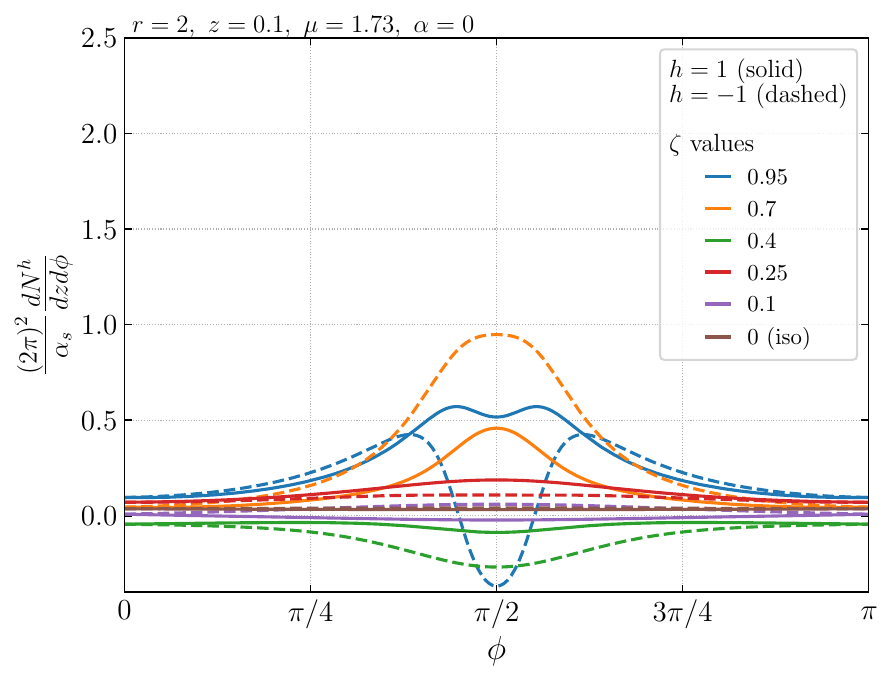}
         \centering
         \includegraphics[width=0.49\textwidth]{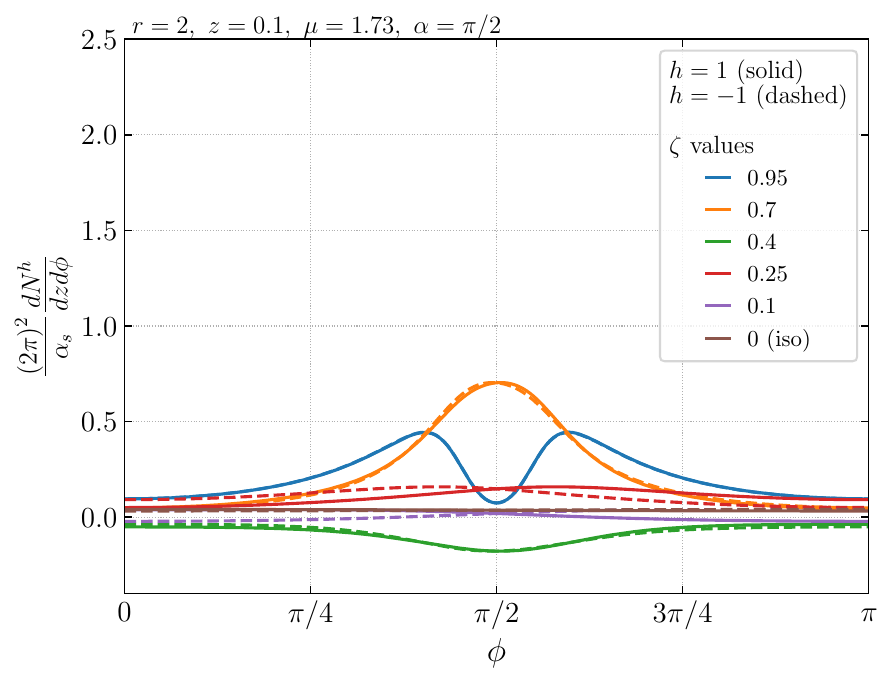}
     \caption{Plots of the distribution in Eq.~\eqref{massive_tvaligned_final} as a function of $\phi$ for different values of the anisotropy parameter $\zeta$, for $\alpha=0$ (\textbf{left}) and $\alpha = \pi/2$ (\textbf{right}), for fixed $r=2$, $z=0.1$ and for two different values of $\mu$ -- $0.16$ (upper) and $1.73$ (lower), corresponding to the charm and bottom quark masses, respectively.}
     \label{tvPol_2}
\end{figure}

So far we have defined the quark and anti-quark spin states as projections along the $z$ axis. However, since we are interested in the interplay between spin and the anisotropy experienced by the $q\bar q$ dipole in the transverse plane, it is natural to also consider spin projected along an axis in that plane. This amounts to changing the expression for the vertex matrix element, since the spinors now correspond to states with spin projected onto a transverse direction. Following the calculation in Appendix~\ref{app:vertices}, we obtain
\begin{equation}\label{eq:ch3_massive_tvpol}
    \begin{split}
    & \frac{1}{2}\sum_{\lambda_0}V_{g\to q\bar q}^{\lambda_0 hh'}(z,\bkappa)V_{g\to q\bar q}^{\ast,\lambda_0 hh'}(z,\bar\bkappa) = \frac{1}{2z(1-z)}\Big\{(\bkappa\cdot\bar\bkappa)\left(\delta_{hh'} + (1-2z)^2\delta_{h,-h'}\right) +m^2 \\
    & +imh\left[\left(\bkappa^x-\bar\bkappa^x\right)\sin\alpha - \left(\bkappa^y-\bar\bkappa^y\right)\cos\alpha\right]\left(\delta_{hh'} + (1-2z)\delta_{h,-h'}\right)\Big\} \, .
    \end{split}
\end{equation}
Here $\alpha$ is the angle that the spin quantization axis makes with the $x$-axis, so that $\alpha=0$ and $\alpha=\pi/2$ correspond to measuring spin along the $x$ and $y$ axes, respectively. Compared to the case where spin is projected along the $z$ axis, the spin-dependent term is now linear in both the transverse momenta and the quark mass. This is particularly interesting because it vanishes in vacuum, where the transverse momenta in the amplitude and complex conjugate amplitude coincide, $\bkappa^i=\bar\bkappa^i$. Therefore, a non-zero contribution for the linear terms in each direction are directly controlled by the corresponding $\hat q_i$.

The in-medium cross-section can then be computed by following the same steps that led to Eq.~\eqref{eq:ch3_xsec_final}, but replacing the squared vertex matrix element in Eq.~\eqref{eq:ch3_xsec_final_before_HO} by the one in Eq.~\eqref{eq:ch3_massive_tvpol}. Focusing on the equal-spins contribution to the in-in term and integrating over $|\p|$ leads to
\begin{align}\label{massive_tvaligned_final}
    & \sum_{h'}\delta^{hh'}(2\pi)\frac{d\sigma_{\rm in-in}^{hh'}}{\sigma_0\,dz\,d\phi}\equiv (2\pi)\frac{dN^{h}}{dzd\phi} = 
    \frac{\alpha_s}{4\pi}\Re  \int_{X^+} e^{-i\frac{m^2}{2\omega^+}(\bar x^+ - x^+)} \frac{\sqrt{c_{1x}}\sqrt{c_{1y}}}{\sqrt{c_{3x}}\sqrt{c_{3y}}}\nn
    & \left(\frac{(c_{1y}c_{2y}c_{3x}-c_{1x}c_{2x}c_{3y})(c_{3y}\cos^2{\phi}-c_{3x}\sin^2{\phi})}{\left(c_{3y}\cos^2{\phi}+c_{3x}\sin^2{\phi}\right)^2} +\left(\frac{m^2}{2\omega^+} + c_4\right)\frac{c_{3x}c_{3y}}{c_{3y}\cos^2{\phi}+c_{3x}\sin^2{\phi}} + \right.\nn
    & \left. -\frac{(1-i)m\sqrt{\pi}}{2\sqrt{\omega^+}}h\sqrt{\frac{c_{3x}c_{3y}}{c_{3y}\cos^2\phi+c_{3x}\sin^2\phi}}\frac{(2ic_{1x}-c_{2x})c_{3y}\cos\phi\sin\alpha - (2ic_{1y}-c_{2y})c_{3x}\sin\phi\cos\alpha}{\left(c_{3y}\cos^2{\phi}+c_{3x}\sin^2{\phi}\right)}\right) \, ,
\end{align}
where $X^+ = \{0<x^+ < \bar x^+ < L^+\}$. As discussed in Appendix~\ref{app:vertices}, and as is also clear from the previous two expressions, in the limit of a very large anisotropy with $\hat q_y \gg \hat q_x$ ($\zeta \rightarrow 1$), the outgoing $q\bar q$ pair becomes unpolarized along the $y$ direction ($\alpha=\pi/2$). On the other hand, when measuring spin along the $x$-axis ($\alpha=0$), the resulting distribution is polarized. This behaviour follows from the fact that, in this limit, the spin-dependent term is dominated by the component controlled by the larger transverse broadening coefficient. Notice also that this effect is entirely driven by the mass-dependent term. This observation is confirmed in Figs.~\ref{tvPol_1} and \ref{tvPol_2}, where we plot Eq.~\eqref{massive_tvaligned_final} as a function of $\phi$ for several values of $\zeta$, $z$ and $\mu$. For very large anisotropies, such as $\zeta=0.95$ and $0.7$, one clearly sees that when $\alpha=\pi/2$ (right panels) the solid and dashed curves are very close to each other, reflecting a result that is nearly independent of $h$ and therefore a $q\bar q$ pair that is only very weakly polarized along the $y$-axis. For $\alpha=0$ (left panels), the two curves are instead significantly different around $\phi=\pi/2$, giving rise to a sizeable polarization along the $x$-axis. Even for intermediate values of the anisotropy, for instance $\zeta=0.4$ (the green curve), the $h$ dependence, or equivalently the degree of polarization, is visibly larger along the $x$-axis than along the $y$-axis. If the hierarchy of broadening coefficients were reversed, $\hat q_x>\hat q_y$, corresponding to negative values of $\zeta$, the behaviour would be reversed as well, and the pair would become more polarized along the $y$ direction than along the $x$ direction. Regarding the dependence on $z$, we see that softer splittings ($z=0.1$ in Fig.~\ref{tvPol_2}) seem to lead to a more pronounced separation between distributions with different spin states, enhancing the polarization effect. The mass $\mu$, despite being the parameter that enables this effect, appears to suppress the degree of polarization when it is increased, as shown in the bottom plots of Figs.~\ref{tvPol_1} and \ref{tvPol_2}. Finally, one should note that in the isotropic case (brown curve in all plots) the $q\bar q$ pair also acquires polarization due to interactions with the medium. The crucial difference with respect to the anisotropic case is that the degree of polarization is then exactly the same in all directions $\alpha$. This becomes clear if one plots the distribution as a function of $\phi-\alpha$, since in the isotropic limit of Eq.~\eqref{massive_tvaligned_final} the spin-dependent term is simply proportional to $\cos(\phi-\alpha)$. Thus, for anisotropic matter, this direction-dependent polarization effect provides another observable feature that can be used to extract information about the medium, since it is sensitive to the direction of maximal $\hat q$.

\section{Energy correlators in jets}
\label{sec:ch3_energy_correlators}

The calculation in the previous sections gives access to the angular structure of the matter modifications to a high energy $1\to 2$ partonic cross-section. In particular, once the medium collision kernel distinguishes two transverse directions, the resulting cross section is no longer only a function of the opening angle of the final state pair, but also of its relative azimuthal orientation. To turn this cross-section level information into an experimentally meaningful jet observable, we will explore \emph{energy correlators} in the following sections. 

Energy correlators describe correlations of energy flow between asymptotic directions. They appear naturally in formal studies of quantum field theory, where correlation functions of local operators in conformal theories are constrained by symmetry and related to light-ray operator methods~\cite{Hofman:2008ar,Kravchuk:2018htv,Moult:2025nhu}. In QCD, the energy-energy correlator was first introduced in $e^+e^-$ annihilation as an IRC safe angular correlation observable to characterise perturbative radiation~\cite{Basham:1978bw,Basham:1978zq,Ellis:1980wv}. For jets, energy correlators provide access to universal collinear dynamics, anomalous scaling, hadronisation effects and precision substructure measurements~\cite{Chen:2020vvp,Lee:2022ige,Komiske:2022enw,Moult:2025nhu}. In heavy-ion collisions, the interpretation of energy correlators is more challenging than in vacuum, since the same level of analytic control is not generally available for jets evolving inside a dynamical medium. In this context, these observables receive contributions from the perturbative shower, medium-induced radiation, medium response, hadronisation and fluctuations of the background. Nevertheless, they remain useful because they retain differential angular information on how the medium modifies the radiation pattern inside jets, while being directly expressible as energy-weighted sums over final-state particles (see, e.g,~\cite{Andres:2022ovj,Yang:2023dwc,Barata:2023bhh,Singh:2024vwb,Barata:2024wsu,Bossi:2024qho,Apolinario:2025vtx}). The aim of this section is to introduce this energy-correlator language to be applied in the following sections, where heavy-flavour correlations will provide a controlled channel for studying medium-modified splittings, while azimuthally differential correlators will be used to connect the harmonic structure of energy flow to anisotropic QCD matter.

\subsection{Energy correlators as jet substructure observables}
\label{subsec:ch3_eec_observables}

Energy correlators are defined in terms of the energy-flow operator, which measures the energy deposited along an asymptotic direction $\vec n$ on the unit two-sphere, usually termed the \emph{celestial sphere}. In a quantum field theory with energy-momentum tensor $T^{\mu\nu}$, this operator can be written as~\cite{Sveshnikov:1995vi, Korchemsky:1997sy}
\begin{align}\label{eq:ch3_energy_flow_operator}
    \hat{\mathcal{E}}(\vec n)
    =
    \lim_{r\to\infty} r^2 \int_0^\infty dt\, n^i\, T^{0i}(t,r\vec n)\,,
\end{align}
i.e., the operator $\hat{\mathcal{E}}(\vec n)$ acts as an idealised calorimeter at null infinity. Correlation functions of $N$ such operators, evaluated in a given state $|\Psi\rangle$, then measure correlations of energy flow in different angular directions
\begin{align}
    \langle \Psi|
    \hat{\mathcal{E}}(\vec n_1)\hat{\mathcal{E}}(\vec n_2)\cdots\,\hat{\mathcal{E}}(\vec n_N)
    |\Psi\rangle\, .
\end{align}
In the context of collider physics, these observables are further projected onto angular variables, which we shall generally refer to as \emph{$N$-\,point energy correlators} (ENC). The simplest example is the energy-energy correlator (EEC), which measures the distribution in the relative angle between two energy-flow operators. For a state produced at a hard scale $Q$, the two-point energy correlator can be projected onto the relative angle $\chi$ between the two detectors as
\begin{align}\label{eq:ch3_eec_operator_definition}
    \frac{d\Sigma}{d\cos\chi}
    =
    \int d\Omega_{\vec n_1} d\Omega_{\vec n_2}
    \frac{
    \langle \Psi|
    \hat{\mathcal{E}}(\vec n_1)\hat{\mathcal{E}}(\vec n_2)
    |\Psi\rangle
    }{Q^2}
    \delta(\vec n_1\cdot\vec n_2-\cos\chi)\, ,
\end{align}
where $d\Omega_{\vec n_i}$ is the solid angle element along direction $\vec n_i$. In the case of jet production, the hard scale is identified with its energy. For higher-point correlators, the measurement function can retain more of the angular geometry of the energy flow, being sensitive to multi-prong structures and the detailed shape of the radiation pattern. In the context of matter induced modifications in jets, this is relevant in that they probe correlated structures across more than one angular scale. One may project these correlators onto, e.g., several independent angles or onto the largest pairwise angle as in \emph{projected $N$-\,point energy correlators} (PENC)~\cite{Chen:2020vvp,Komiske:2022enw}. The same observables have a simple particle-level interpretation in collider experiments. For a final state consisting of particles with energies $E_i$ and directions $\vec n_i$, Eq.~\eqref{eq:ch3_eec_operator_definition} becomes
\begin{align}\label{eq:ch3_eec_particle_definition}
    \frac{d\Sigma}{d\cos\chi}
    =
    \sum_{i,j\in J}
    \frac{E_iE_j}{p_t^2}
    \delta(\vec n_i\cdot\vec n_j-\cos\chi)\,,
\end{align}
where the sum runs over particles inside the jet and $p_t$ is the jet transverse momentum. Similarly, an $N$-point energy correlator is an energy-weighted sum over sets of $N$ particles. This representation extremely useful in an experimental context: energy correlators be measured directly from final-state particles, while still retaining a clean connection to the underlying energy-flow operators.

Energy correlators can be viewed as weighted cross sections. This should be contrasted with the more common situation in which a differential distribution is defined by a delta function fixing the value of a final-state measurement. Such a measurement generally depends on the full energy flow of the event or jet, and its distribution is therefore sensitive to moments of energy flow at arbitrarily high order~\cite{Chen:2020vvp}. Energy correlators instead fix the number of energy-flow insertions and the angular measurement projects these correlations onto chosen geometries. This fixed-point structure gives them a well-defined operator content and a controlled collinear expansion. In particular, in the collinear limit the EEC resolves the angular structure of the partonic splittings that produce two nearby particles or subjets. This gives direct access to QCD splitting functions and their higher-order corrections, as well as medium modified splitting functions in HICs. This is the perspective adopted below: the in-medium splitting spectra computed in the BDMPS-Z formalism provide the perturbative input for energy-correlator observables that retain the angular and, in particular, azimuthal structure generated by anisotropic matter.

\subsection{Azimuthally differential EEC and Fourier harmonics}
\label{subsec:ch3_azimuthal_eec}

The EEC in Eq.~\eqref{eq:ch3_eec_operator_definition} depends only on the opening angle $\chi$ and therefore averages over the azimuthal orientation of the pair around the jet axis. This is sufficient in an isotropic environment, but it discards precisely the directional information generated by the anisotropic broadening kernel discussed in Section~\ref{sec:ch3_anisotropic_broadening}. To retain this information, one should keep the azimuthal orientation of the pair relative to a reference, physical direction in the plane transverse to the jet, which we denote by $\b$. At the level of the analytic calculation, one can always choose $\b$ to coincide with one of the principal axes of the anisotropic broadening tensor, for instance the direction associated with $\hat q_x$. In a realistic heavy-ion event, a natural experimentally available reference direction is instead the beam direction projected onto the plane transverse to the jet axis. This choice does not introduce a new analytical ingredient, it only fixes the origin of the azimuthal angle used to analyse the pair orientation in realistic events.

\begin{figure}[t]
    \centering
    \includegraphics[width=0.45\textwidth]{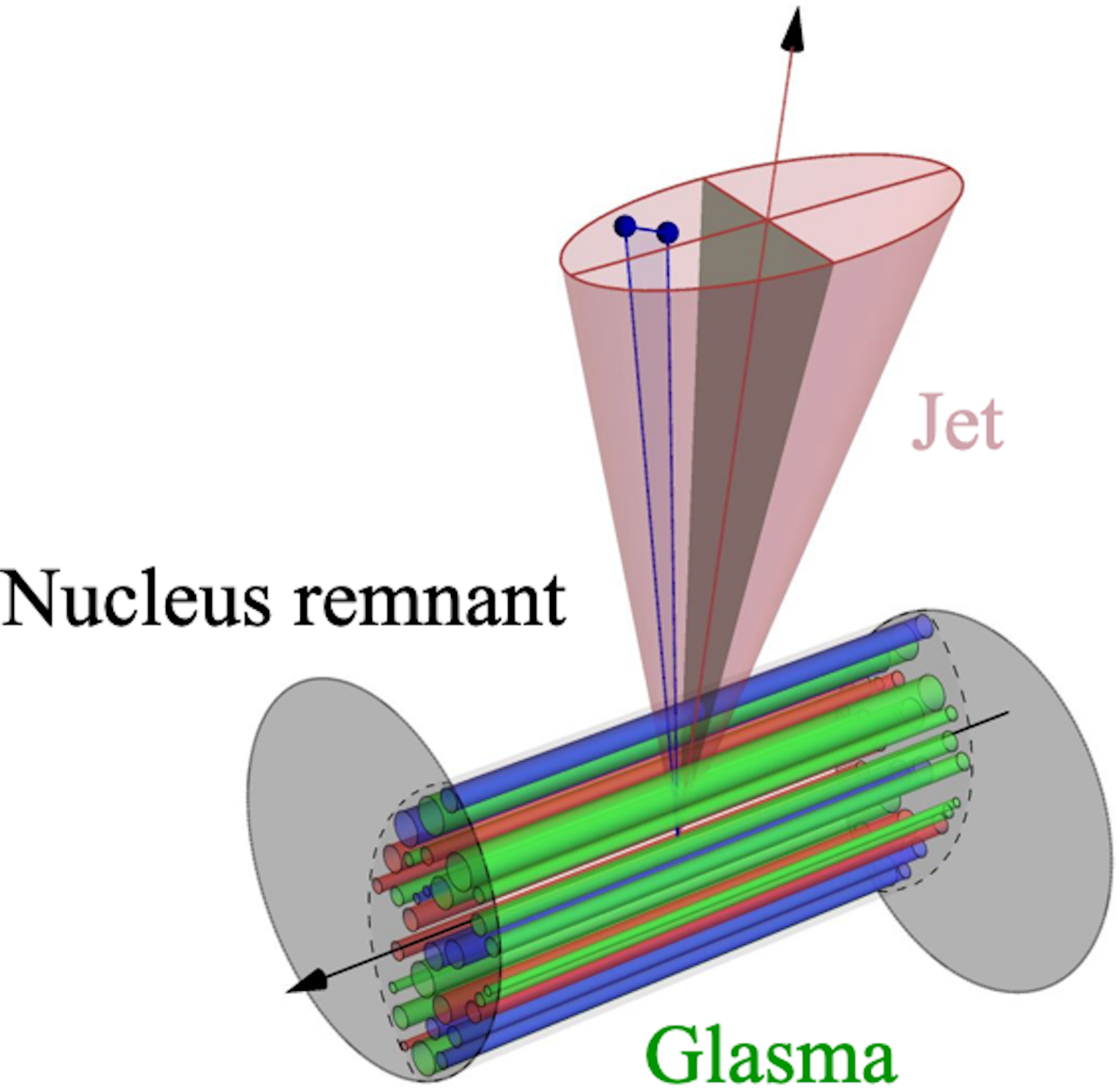}
    \centering
    \includegraphics[width=0.45\textwidth]{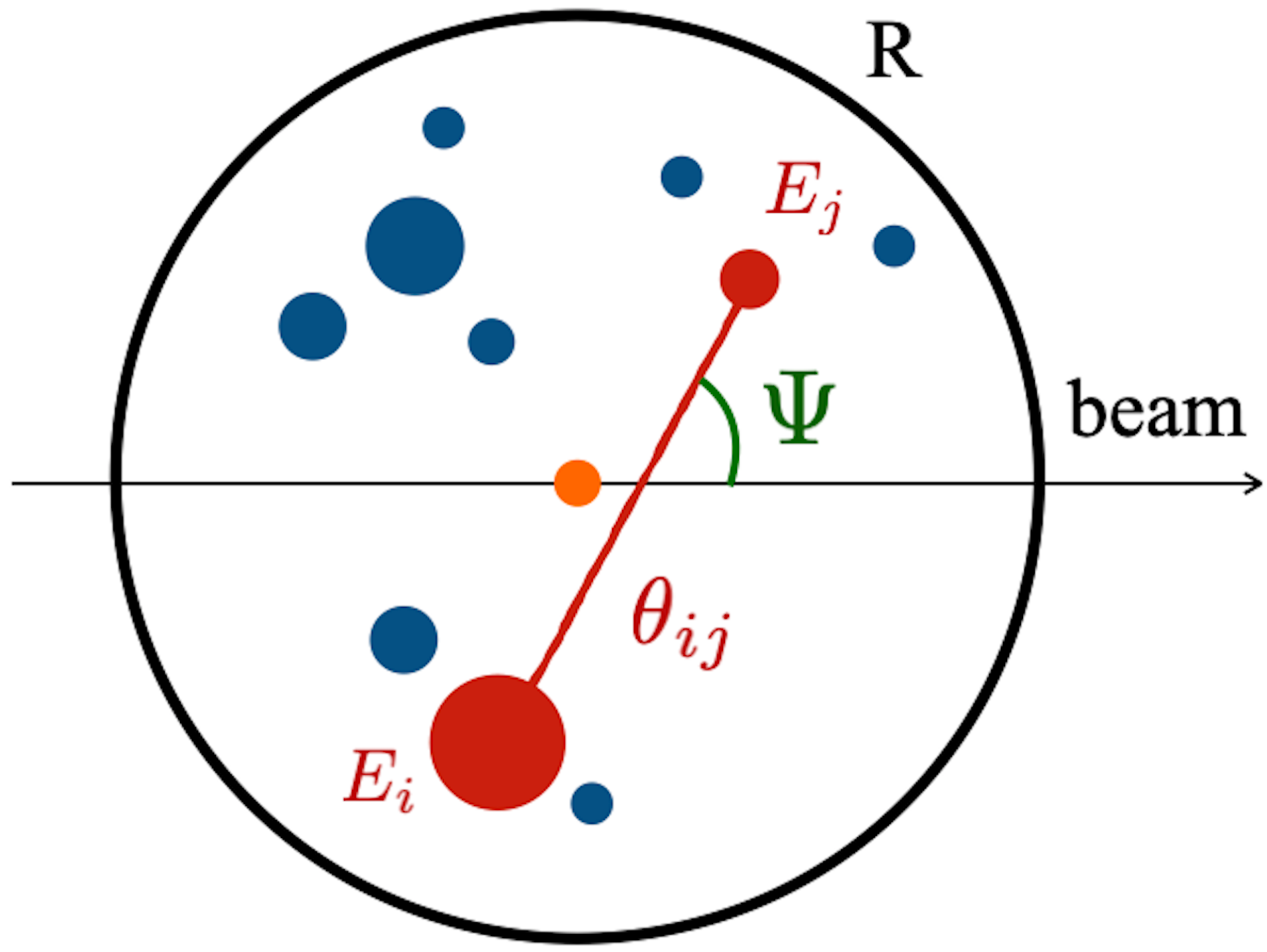}
    \caption{Schematic illustration of an azimuthally differential in-jet EEC for a typical mid-rapidity jet configuration in a heavy-ion collision. The elongated colour tubes represent the initial Glasma configuration, which dominates the earliest times after the collision. A pair of particles inside the jet contributes to the EEC with an energy weight and is classified by its opening angle, denoted by $\chi$ or equivalently $\theta_{ij}$, and by the azimuthal orientation $\Psi$ of the pair. The left panel illustrates how an anisotropic early-time configuration can imprint an elongated energy distribution inside the jet along the beam direction. The right panel shows the particle-level construction inside a jet of radius $R$: the pair carries energies $E_i$ and $E_j$, the orange point denotes the jet axis, and $\Psi$ is measured with respect to the beam direction projected onto the plane transverse to the jet axis.}
    \label{fig:ch3_eec_cartoon}
\end{figure}

The corresponding azimuthally differential EEC is defined by projecting the two-point energy correlator onto both the opening angle $\chi$ and the azimuthal angle $\Psi$~\cite{Barata:2025zku}
\begin{align}\label{eq:ch3_azimuthal_eec_operator}
    \frac{d\Sigma}{d\cos\chi d\cos\Psi}
    =
    \int d\Omega_{\vec n_1} d\Omega_{\vec n_2}
    \frac{
    \langle 
    \mathcal{E}(\vec n_1)\mathcal{E}(\vec n_2)
    \rangle
    }{p_t^2}
    \delta(\vec n_1\cdot\vec n_2-\cos\chi)
    \delta\left(
    \frac{\n_1-\n_2}{|\n_1-\n_2|}\cdot\b-\cos\Psi
    \right)\,,
\end{align}
where from now on we leave the state $|\Psi\rangle$ characterizing the heavy-ion jet implicit. The vector $\n_1-\n_2$ is proportional to the relative transverse momentum of the pair and $\n$ is the vector $\vec n$ projected onto the plane transverse to the jet axis. In the collinear limit ($\chi\ll1$), the angle $\Psi$ can be interpreted as the azimuthal angle of the pair around the jet axis, measured with respect to $\b$. This naturally coincides with the azimuthal angle used the in-medium cross-section in Section~\ref{sec:ch3_gqqbar}. It is useful to write Eq.~\eqref{eq:ch3_azimuthal_eec_operator} in a particle-level form, which is the version directly suited for experimental and Monte Carlo level analysis
\begin{align}\label{eq:ch3_azimuthal_eec_particle}
    \frac{d\Sigma}{d\cos\chi \,d\cos\Psi}
    =
    \sum_{i\neq j\in J}
    \frac{E_iE_j}{p_t^2}
    \delta(\cos\theta_{ij}-\cos\chi)
    \delta\left(
    \frac{\n_i-\n_j}{|\n_i-\n_j|}\cdot\b-\cos\Psi
    \right)\, ,
\end{align}
where $\theta_{ij}$ is the angle between particles $i$ and $j$. We will be considering mid-rapidity jets in HICs, where $\b$ is closely aligned with the beam axis, providing a particularly useful realisation of this observable. This direction is in principle aligned with the dominant initial anisotropy of the earliest Glasma-like state (see Section~\ref{subsec:st_evolution_HIC}), characterized by coherent longitudinal fields. A pair of particles inside a jet can therefore be classified not only by its opening angle, but also by whether its relative azimuthal orientation is more closely aligned with the direction of initial anisotropy or with the direction orthogonal to it\footnote{The same construction can also be used more abstractly, with $\b$ taken to be any preferred direction associated with the anisotropic medium. The beam axis, however, provides the simplest choice.}. This is illustrated in Fig.~\ref{fig:ch3_eec_cartoon}. 

Similarly to what was done for the cross-section in Section~\ref{subsec:harmonics_spin_sensitivity}, one can study the $\Psi$ dependence of the EEC by decomposing it into Fourier harmonics. For the type of axis-like anisotropy we are interested in and summing over final state spins, this decomposition reads~\cite{Barata:2024bqp, Barata:2025uxp, Barata:2025zku}
\begin{align}\label{eq:ch3_eec_harmonic_expansion}
    \frac{d\Sigma}{d\chi d\Psi}
    =
    v_0^{\rm EEC}(\chi)
    +
    \sum_{n=1}^{\infty}
    v_{2n}^{\rm EEC}(\chi)\cos(2n\Psi)\, .
\end{align}
The coefficient $v_0^{\rm EEC}$ is the azimuthally averaged EEC, while the coefficients $v_{2n}^{\rm EEC}$ measure anisotropic energy flow around the jet axis. These coefficients are defined as
\begin{align}\label{eq:ch3_eec_harmonic_coefficients}
    v_0^{\rm EEC}(\chi)
    =
    \frac{1}{2\pi}
    \int_0^{2\pi} d\Psi\,
    \frac{d\Sigma}{d\chi d\Psi}\,,
    \qquad
    v_{2n}^{\rm EEC}(\chi)
    =
    \frac{1}{\pi}
    \int_0^{2\pi} d\Psi\,
    \frac{d\Sigma}{d\chi d\Psi}\cos(2n\Psi)\, .
\end{align}
The absence of odd harmonics in Eq.~\eqref{eq:ch3_eec_harmonic_expansion} follows from the same symmetry structure discussed in Section~\ref{subsec:harmonics_spin_sensitivity}. Namely, if the medium distinguishes two axes but not an orientation along either axis, the distribution is invariant under $\Psi\to\Psi+\pi$. Further, after spin summation, the EEC is also invariant under $\Psi\to-\Psi$, i.e., there are no odd-parity contributions. Thus, these two symmetries remove, respectively, odd cosine harmonics and sine terms, leading to the even cosine series in Eq.~\eqref{eq:ch3_eec_harmonic_expansion}. This is broken if additional vector-like sources of anisotropy are present, such as strong gradients or directed flow, or if the observable keeps spin information, such as explicit spin dependence in $1\to 2$ processes or spin correlations beyond two particles in the final state (see Section~\ref{sec:ch5_spin_correlations}). 

It is useful to relate the general expansion for the azimuthally differential EEC in Eq.~\eqref{eq:ch3_eec_harmonic_expansion} to the microscopic $1\to 2$ calculation in Eq.~\eqref{eq:ch3_xsec_final}. In the anisotropic HO approximation, the in-medium cross-section depends separately on the two transverse directions through coefficients controlled by $\hat q_\parallel$, the component parallel to the direction $\b$ (the beam-axis for mid-rapidity jets), and $\hat q_\perp$, the component orthogonal to $\b$. To make the notation consistent with Section~\ref{sec:ch3_gqqbar}, we shall consider $\hat q_\parallel \equiv \hat q_x$. The relative transverse momentum of the two final partons fixes both the opening angle of the pair ($\theta\equiv\chi$) and its azimuthal orientation ($\phi\equiv\Psi$). Consequently, after weighting the splitting spectrum by the appropriate energy fractions and integrating over the energy sharing, the anisotropic dependence of the cross section becomes an anisotropic dependence of the EEC, namely
\begin{align}\label{eq:xsec_to_EEC_doublediff}
    \frac{d\Sigma}{d\chi\,d\Psi} = \int_0^1 dz\,z(1-z)\frac{d\sigma}{\sigma_0\,dz\,d\chi\,d\Psi}\,.
\end{align}
Thus, the coefficients $v_{2n}^{\rm EEC}$ can be computed from the $1\to 2$ in-medium cross-section, providing the observable-level counterpart of the azimuthal harmonics generated by the anisotropic medium collision kernel. We will characterise the anisotropy by the dimensionless parameter\footnote{Note that we used a different parameter ($\zeta$) in Section~\ref{sec:azimuthal_observables}, which used $\sqrt{\hat q_i}$ rather than $\hat q_i$. These two choices are of course equivalent in describing qualitative effects.}
\begin{align}
    \xi
    =
    \frac{\hat q_x-\hat q_y}
    {\hat q_x+\hat q_y}, .
\end{align}
The azimuthally averaged coefficient $v_0^{\rm EEC}$ is non-zero already in the isotropic limit ($\xi\to 0$), while all the remaining harmonics vanish in this limit. Importantly, some general remarks about the scaling of these harmonics with $\xi$ follow directly from symmetry. First, since exchanging the two transverse axes corresponds to the transformation $\xi\to-\xi$ or, equivalently for $\Psi$ measured from the $x$ axis, to the rotation $\Psi\to \pi/2-\Psi$, from the general even cosine harmonic expansion in Eq.~\eqref{eq:ch3_eec_harmonic_expansion} one concludes
\begin{align}
    v_{2n}^{\rm EEC}(\xi)
    =
    (-1)^n v_{2n}^{\rm EEC}(-\xi)\, .
\end{align}
Thus, $v_{2n}^{\rm EEC}$ is an odd function of $\xi$ for odd $n$ and an even function of $\xi$ for even $n$. Assuming analyticity around $\xi=0$ and using $v_{2n}^{\rm EEC}(0)=0$ for $n\geq1$, this implies $v_{2n}^{\rm EEC}=\mathcal O(\xi)$ for odd $n$ and $v_{2n}^{\rm EEC}=\mathcal O(\xi^2)$ for even $n$.

A stronger hierarchy follows from the explicit structure of the spin-summed anisotropic-HO spectrum in Eq.~\eqref{eq:ch3_xsec_final}\footnote{Although we use the spectrum for $g\to q\bar q$ in a static medium as an explicit example, the following statements do not depend on the process or on the time-independence of $\hat q$.}. Inspecting this result, one observes that differences of functions evaluated along the two principal axes, $f(c_{i x})-f(c_{i y})$, are odd under $x\leftrightarrow y$ and therefore contain only odd powers of $\xi$, while sums $f(c_{i x})+f(c_{i y})$ contain only even powers. Moreover, the odd structures multiply factors of $\cos2\Psi$, both in the exponential and in the numerator, whereas the even structures contribute to the azimuthally independent part. Thus, schematically, the azimuthal dependence of the medium-induced spectrum has the form
\begin{align}
    \exp\left[
        i\,
        \cos 2\Psi
        \sum_{i=0}^{\infty} a_i \xi^{2i+1}
    \right]
    \left[
        b(\xi)+\cos 2\Psi
        \sum_{k=0}^{\infty} c_k\xi^{2k+1}
    \right],
\end{align}
where $b(\xi)$ is an even function of $\xi$. Expanding the exponential, the $r$-th power of $\cos2\Psi$ appears with at least $\xi^r$. Since $\cos^r2\Psi$ contains harmonics up to $\cos(2r\Psi)$, we conclude that generating the $2n$-th harmonic requires at least $n$ powers of the anisotropy. The numerator has the same counting, since it is either independent of $\Psi$ or adds one extra factor of $\cos2\Psi$ together with at least one power of $\xi$. The anisotropy expansion of the harmonic coefficients reads~\cite{Barata:2024bqp, Barata:2025uxp, Barata:2025zku}
\begin{align}
    v_{2n}^{\rm EEC}
    =
    \xi^n
    \left(
        v_{2n}^{(n)}
        +
        \xi^2 v_{2n}^{(n+2)}
        +
        \cdots
    \right),
    \qquad n\geq1\, .
\end{align}
where $v^{n}_{2n}$ stands for the coefficient of the $\xi^n$ power in the expansion. Thus, $v_2^{\rm EEC}$ is the leading probe of a small axis-like anisotropy, scaling like $\cO(\xi)$, while higher harmonics are parametrically suppressed and become relevant only for stronger anisotropies or higher-order angular structures in the medium-induced spectrum.

\section{Heavy-flavour energy correlators}
\label{sec:ch3_heavy_flavour_eec}

\subsection{Heavy-flavour energy flow and the heavy-pair splitting}
\label{subsec:ch3_heavy_flavour_energy_flow}

The energy correlator observables introduced in the previous section can be made more selective by restricting the energy flow to hadrons carrying a given flavour quantum number. In this section we consider the two-point heavy-flavour energy correlator~\cite{Barata:2025uxp},\footnote{In a slight abuse of notation, we define the correlation function already normalized by $p_T^2$.}
\begin{align}\label{eq:ch3_EECheavy}
    &\langle \Psi | \mathcal{E}_{\mathcal{H}}(\vec n_1)
    \mathcal{E}_{\mathcal{H}}(\vec n_2)| \Psi\rangle 
    \equiv
    \frac{1}{\sigma_J}
    \int d\sigma_J
    \sum_{i_1,i_2 \in \mathcal{H}\subset J}
    \frac{p_{T,i_1} p_{T,i_2}}{p_T^2}
    \prod_{j=1}^2
    \delta(\eta_j-\eta_{i_j})
    \delta(\phi_j-\phi_{i_j})\, .
\end{align}
Here $|\Psi\rangle = |\Psi(\eta,p_T)\rangle$ denotes a jet state in a given transverse-momentum $p_T$ and rapidity $\eta$ bin, $\sigma_J$ is the corresponding inclusive jet cross section, and $\mathcal{E}_{\mathcal{H}}(\vec n)$ measures the energy flux carrying heavy-flavour quantum number $\mathcal{H}$ along the direction $\vec n$, here parametrised as $\vec n(\eta,\phi)$. The measure $d\sigma_J$ denotes the fully differential jet cross section restricted to the selected jet sample, with the dependence on the momenta of the particles inside the jet left implicit. The set $\mathcal{H}$ denotes heavy hadrons containing a heavy quark $Q$ or heavy antiquark $\bar Q$. The angular dependence of this correlator directly probes mass-dependent partonic splitting functions $P_{i\to Q_1 Q_2+X}$, describing the collinear splitting of a parton $i$ into a pair of heavy partons $Q_1Q_2$ plus additional collinear particles over which Eq.~\eqref{eq:ch3_EECheavy} is inclusive. At the same time, it captures the subsequent fragmentation of the heavy partons into heavy hadrons. The leading contribution to the production of two heavy flavours in the final state comes from the heavy-pair splitting
\begin{align}\label{eq:ch3_g_to_QQ_vac}
    P_{g\to Q\bar{Q}}^{(1)}(x)
    =
    T_F
    \left[
    1-\frac{2 x(1-x)}{1-\epsilon}
    +\frac{2 m_Q^2}{(1-\epsilon)s}
    \right], ,
\end{align}
where $s=(p_Q+p_{\bar Q})^2$, $x$ is the momentum fraction of the outgoing heavy quark and $\epsilon$ is the dimensional regulator, i.e., $d=4-2\epsilon$.

Thus, our proposal is to tag the heavy-quark pair splitting through correlations of heavy-flavour energy flow, rather than through a flavour-sensitive definition of the initiating jet. This is useful for two main reasons. First, the heavy-quark mass makes the angular structure of the correlator sensitive to the massive splitting kernel. Once this vacuum connection is under control, the same correlator can be used to isolate how the corresponding splitting is modified by QCD matter. Second, heavy quarks are not efficiently produced thermally in a medium or during late hadronisation, since their masses are much larger than the typical temperature of the QGP and than $\Lambda_{\rm QCD}$. As such, heavy-flavour pairs inside jets predominantly originate from perturbative production mechanisms, in particular from gluons splitting into heavy $Q\bar Q$ pairs. In this way, heavy-flavour EECs provide a controlled channel for studying heavy-quark pair production, heavy-flavour fragmentation, and medium modifications to the splitting dynamics inside jets. This approach is related to broader efforts to study heavy-flavour dynamics and parton splitting functions in jet substructure and in medium~\cite{Dhani:2025fbk,Xing:2024yrb,Dhani:2024gtx,Blok:2023ugf,Yang:2024awm,Caletti:2023spr,Andres:2023ymw,Chen:2024nfl,Cunqueiro:2022svx,Craft:2022kdo,Li:2021gjw,Cal:2021fla,Casalderrey-Solana:2019ubu,Caucal:2019uvr,Li:2017wwc,Lee:2019lge,Chang:2017gkt,Milhano:2017nzm,Mehtar-Tani:2016aco,KunnawalkamElayavalli:2017hxo,Chien:2016led,Armesto:2003jh,Sievert:2019cwq,Sievert:2018imd,Zhang:2003wk,Arleo:2012rs,Djordjevic:2003zk}. The key advantage of the energy-correlator formulation is that it can characterise the heavy-pair splitting and heavy-flavour fragmentation rate, while unambiguously tagging partonic splittings involving heavy-quark pairs and sidestepping the infrared ambiguities that challenge traditional heavy-flavour jet definitions~\cite{Banfi:2006hf,Gauld:2022lem,Caola:2023wpj,Czakon:2022wam,Caletti:2022glq,Caletti:2022hnc,Fedkevych:2022mid,Behring:2025ilo}. This precise vacuum baseline then provides the handle used in the followin subsections to probe the formation time dependence of medium-induced modifications of the splitting, and later its sensitivity to anisotropic matter.

\subsection{Vacuum baseline and heavy-flavour fragmentation}
\label{subsec:ch3_heavy_flavour_vacuum}

We first consider the vacuum baseline. The heavy-flavour EEC distribution is defined as
\begin{align}\label{eq:ch3_heavy_theta_distribution}
\frac{1}{\sigma_J}\frac{d\Sigma_{\mathcal{H}}}{d\theta}
\equiv&
\int \prod_{i=1}^2 d\phi_i d\eta_i\,\langle \Psi | \mathcal{E}_{\mathcal{H}}(\vec n_1)
\mathcal{E}_{\mathcal{H}}(\vec n_2)| \Psi\rangle
\delta\Bigl(\theta - \sqrt{(\Delta \eta_{12})^2 + (\Delta \phi_{12})^2}\Bigr)\,,
\end{align}
where here we define the angle between the two detectors in terms of the longitudinally boost-invariant collider variables. In the collinear limit, it is convenient to introduce $z=\theta^2/4$, in terms of which the cumulant of the heavy-flavour EEC factorises as~\cite{Barata:2025uxp}
\begin{align}\label{eq:ch3_heavy_EEC_fact}
    \Sigma_{\mathcal{H}}\left(z,\frac{Q^2}{\mu^2},\mu\right)
    &=
    \int_0^{z} dz^{\prime}\,
    \frac{1}{\sigma_J}
    \frac{d\Sigma_{\mathcal{H}}}{dz'}\nn\
    &=
    \int_0^1 dx\, x^2\,
    \vec J_{\mathcal{H}}
    \left(
    \ln \frac{z Q^2}{\mu^2},
    m_Q,\mu
    \right)
    \cdot
    \vec H
    \left(
    x,\frac{Q^2}{\mu^2},\mu
    \right)\, ,
\end{align}
where we emphasize that the subscript $\mathcal{H}$ indicates that the correlator is restricted to heavy hadrons carrying the heavy-quark quantum number. 

\begin{figure}[t]
    \centering
    \includegraphics[width=0.8\textwidth]{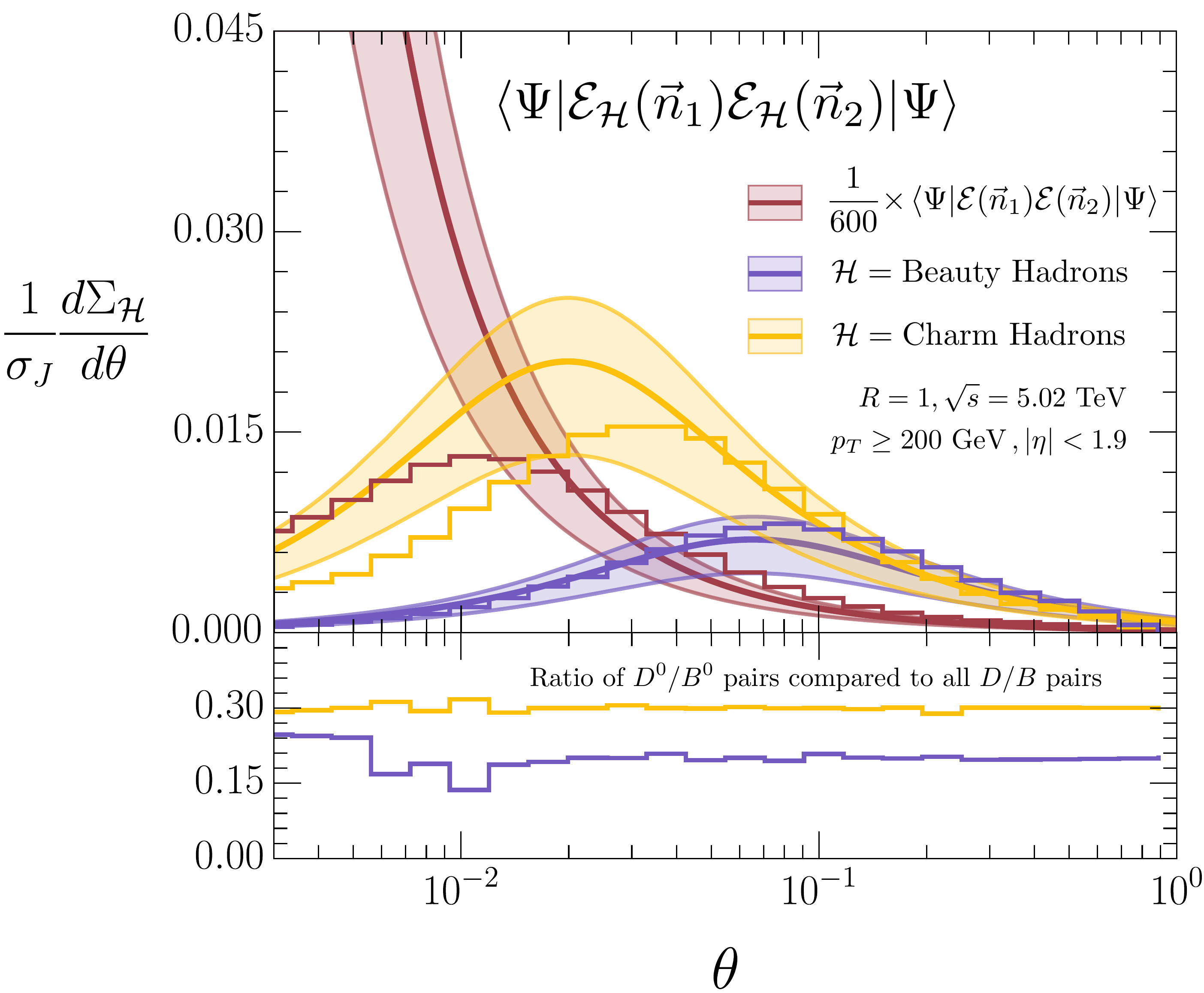}
    \caption{Heavy-flavour EEC on charm and bottom hadrons compared with the inclusive EEC, rescaled by $1/600$ for visual comparison, for jets with $p_T\geq 200$ GeV. The heavy-flavour correlator isolates the splitting function $P_{g\to Q\bar Q}^{(1)}$ inside the jet. The lower panel shows that selecting heavy-hadron subsets rescales the normalisation through HQET matrix elements, while leaving the shape essentially unchanged. The histograms are from \textsc{Pythia}~8~\cite{Sjostrand:2014zea} and the curves are NLL predictions of Eq.~\eqref{eq:ch3_heavy_EEC_fact} with scale variations.}
    \label{fig:ch3_EEC_vac}
\end{figure}

The collinear factorisation for EECs in the context of lepton colliders and jet substructure has been derived in~\cite{Dixon:2019uzg,Lee:2022ige,Lee:2024icn,Lee:2025okn}, and extended to charged-hadron measurements in~\cite{Jaarsma:2023ell,Lee:2023npz,Lee:2023tkr}. The heavy-flavour EEC factorisation follows the charged-hadron case, with the electric charge replaced by the heavy-quark quantum number. The hard and jet functions, denoted by $\vec H$ and $\vec J_{\mathcal H}$, are flavour-space vectors. The hard function $\vec H$ describes the short-distance production of the fragmenting parton that initiates the jet state $|\Psi\rangle$, including its dependence on parton distribution functions. The jet function $\vec J_{\mathcal H}$ encodes the angular correlation of two heavy hadrons inside the jet, and is therefore sensitive both to the angular scale $2p_T\sqrt{z}$ and to the heavy-quark mass $m_Q$. In the region $m_Q \ll 2p_T\sqrt{z}\ll p_T$, mass effects are power suppressed and the factorisation reduces to the massless case. The renormalisation group evolution between the jet scale $\mu_J\sim 2p_T\sqrt{z}$ and the hard scale $\mu_H\sim p_T$ resums collinear logarithms ($\ln z$), in agreement with the anomalous scaling predicted from the light-ray operator product expansion~\cite{Chang:2020qpj,Kologlu:2019mfz,Hofman:2008ar,Chen:2021gdk,Chen:2023zzh}. In the region $m_Q\sim 2p_T\sqrt{z}\ll p_T$, the resummation changes only the normalisation of the fixed-order contribution to the EEC. In Fig.~\ref{fig:ch3_EEC_vac}, we present the vacuum heavy-flavour EEC for charm and bottom hadrons, compared with the inclusive EEC for the same jet kinematics. Both the factorisation predictions from Eqs.~\eqref{eq:ch3_heavy_theta_distribution} and~\eqref{eq:ch3_heavy_EEC_fact} and the distributions obtained from \textsc{Pythia}~8~\cite{Sjostrand:2014zea} Monte-Carlo simulations, shown as functions of $\theta$, are highly sensitive to the form of the massive splitting $P_{g\to Q\bar Q}^{(1)}$, with the characteristic peak located near $2p_T\sqrt{z}\sim m_Q$. This is achieved without defining flavour-sensitive jet algorithms: the heavy-flavour EEC itself isolates the $P_{g\to Q\bar Q}^{(1)}$ splitting at leading order, while sensitivity to other splittings, such as subsequent $Q\to Qg$ emissions from the produced heavy quark or antiquark, is included through resummation and mainly modifies the normalisation. By contrast, the inclusive EEC probes an ensemble average over many different splitting functions, resulting in a markedly different angular structure.

Since the correlator is restricted to heavy-flavour hadrons, the fixed-order jet function in Eq.~\eqref{eq:ch3_heavy_EEC_fact} is sensitive to the fragmentation of a heavy quark or antiquark into the observed heavy hadrons. The perturbative nature of the heavy-quark mass allows the heavy-flavour fragmentation function $D_{Q\to H}(z,m_Q,\mu)$ to be matched, at $\mu\sim m_Q$, onto the non-perturbative matrix element $\chi_Q^H$ of boosted Heavy-Quark Effective Theory (HQET)~\cite{Eichten:1989zv,Isgur:1989vq,Isgur:1990yhj,Grinstein:1990mj,Georgi:1990um,Korner:1991kf,Mannel:1991mc,Fleming:2007qr,Fleming:2007xt}. This matrix element gives the fragmentation rate into a heavy-flavour hadron $H$ at the hadronisation scale $\Lambda_{\rm QCD}$, so that, in the region $2p_T\sqrt{z}\sim m_Q$, the heavy-flavour correlations are proportional to
\begin{align}
    \sum_{H,\bar H\in \mathcal H}
    \chi_Q^H \chi_{\bar Q}^{\bar H}\, .
\end{align}
When $\mathcal H$ contains all heavy-flavour hadrons carrying the quantum number $Q$ or $\bar Q$, heavy-quark symmetry sets this factor to unity, rendering the calculation perturbative up to power corrections in $\Lambda_{\rm QCD}/m_Q$. For a restricted set of heavy-flavour hadrons, the same factorisation formula instead predicts an overall normalisation factor determined by the production fractions of $Q$ and $\bar Q$ into the selected hadrons. This behaviour is seen in Fig.~\ref{fig:ch3_EEC_vac}. In the \textsc{Pythia}~8 simulations, we use $\mathcal H=\{B^0,\bar B^0\}$ for $Q=b$ and $\mathcal H=\{D^0,\bar D^0\}$ for $Q=c$. Relative to the selection including all $B$ and $D$ hadrons, these restricted choices leave the shape unchanged while reducing the normalisation by roughly $0.3$ and $0.2$. The corresponding production fractions, $\chi_b^{B^0}=\chi_{\bar b}^{\bar B^0}\approx\sqrt{0.2}\approx0.45$ and $\chi_c^{D^0}=\chi_{\bar c}^{\bar D^0}\approx\sqrt{0.3}\approx0.55$, are roughly consistent with $B^0$ fractions measured at $e^+e^-$ colliders~\cite{HFLAV:2016hnz,ALEPH:2001ccp} and $D^0$ fractions extracted from HERA and $e^+e^-$ data~\cite{ZEUS:2013fws,Gladilin:1999pj}. The remaining differences are expected from heavy-quark symmetry breaking, suppressed by $\Lambda_{\rm QCD}/m_Q$.

Experimentally, tagging heavy-flavour hadrons inside boosted jets is challenging, and double heavy-flavour tagging is rarer still~\cite{ATLAS:2018zhf,CDF:2004mmv}. Nevertheless, the factorisation structure shows that selecting only an experimentally accessible subset of heavy-flavour hadrons, whether due to flavour choices or decay-channel restrictions, does not change the shape of the correlator. Instead, it shifts the normalisation, which then measures the fraction of heavy-hadron pairs effectively tagged inside the jet. A tagging efficiency with a non-trivial dependence on the heavy-hadron energy, rather than only on its flavour, can also be included, since the corresponding energy distribution is perturbatively computable.

\subsection{Formation-time sensitivity of in-medium heavy EECs}
\label{subsec:ch3_heavy_flavour_formation_time}

The vacuum discussion shows that the heavy-flavour EEC effectively tags the underlying $P_{g\to Q\bar Q}^{(1)}$ splitting. In medium, this provides a direct way of studying how that splitting is modified, while reducing sensitivity to flavour mixing. The medium-modified cross-section for heavy-pair splitting, in the large-$N_c$ limit, can be obtained directly from the fully differential cross-section that we computed in Section~\ref{sec:ch3_gqqbar} in the BDMPS-Z formalism. Summing over final state spins and focusing for now on the isotropic limit of Eq.~\eqref{eq:ch3_xsec_final}, the three contributions to the cross-section read\footnote{Note that we temporarily, in this section, changed the energy fraction variable from $z$ in Eq.~\eqref{eq:ch3_xsec_final} to $x$, to avoid confusion with $z=\theta^2/4$ used in the factorisation formula in Eq.~\eqref{eq:ch3_heavy_EEC_fact}.}
\begin{align}\label{eq:ch3_xsec_iso_int}
    & \frac{d\sigma_{\rm in-in}}{\sigma_0\, dx\,d\theta} = \,
    \frac{T_F\,\alpha_s}{\pi}\,\theta\, \, 2\Re\int_{X^+}\,
    e^{-i\frac{m^2}{2\omega^+}(\bar x^+-x^+)}
    (-i)\frac{c_1}{c_3}
    e^{\frac{i\omega^+\theta^2}{4c_3}}\nn
    & \times \Bigg\{P_{g\to Q\bar Q}^{\rm vac}(x)\left[\omega^+c_1\left(1+\frac{c_2}{c_3}\right)
    +i\left(\frac{\omega^+\theta}{2}\right)^2\frac{c_1c_2}{c_3^2}\right]
    +\frac{m^2}{4}
    \Bigg\}\,,\nn
    & \frac{d\sigma_{\rm in-out}}{\sigma_0\, dx\,d\theta} =
    \frac{T_F\,\alpha_s}{\pi}\,
    \frac{\omega^+\theta}{\theta^2+2(m/\omega^+)^2}\,
    2\Re\int_0^{L^+}dx^+\,
    e^{-i\frac{m^2}{2\omega^+}(L^+-x^+)}
    (i)\frac{c_1}{c_2}e^{-\frac{i\omega^+\theta^2}{4c_2}}\nn
    & \times \Bigg\{
    P_{g \to Q\bar Q}^{\rm vac}(x)\frac{c_1}{c_2}\theta^2
    -i\left(m/\omega^+\right)^2
    \Bigg\}\,,\nn
    & \frac{d\sigma_{\rm out-out}}{\sigma_0\, dx\,d\theta} =
    \frac{T_F\,\alpha_s}{\pi}
    \frac{\theta}{\left(\theta^2 + 2(m/\omega^+)^2\right)^2}
    \left(P_{g\to Q\bar Q}^{\rm vac}(x)\theta^2
    +2\left(m/\omega^+\right)^2\right)\,.
\end{align}
where $\omega^+ = x(1-x)p_0^+$, $p_0^+$ is the energy of the initial gluon and
\begin{align}
    c_{1}=\frac{\Omega}{2i\sin(\Omega\Delta x^+)}\,,\qquad c_{2}=\frac{\Omega}{\tan(\Omega\Delta x^+)}\,,\qquad c_{3}=-\frac{i(L^+-\bar x^+)\hat q(x^2+(1-x)^2}{2\omega^+}-c_{2}\,,
\end{align}
with 
\begin{align}
    \Omega = \frac{1-i}{\sqrt{2}}\sqrt{\frac{\hat q(x^2+(1-x)^2)}{2\omega^+}}
\end{align}
This result was discussed at length in~\cite{Attems:2022otp, Attems:2022ubu}. As discussed in Section~\ref{subsec:formation_time}, the medium modification to the cross-section depends on the formation time of splitting. In particular, this scale can give a proxy for whether the heavy $Q\bar Q$ pair is formed inside or outside the medium.\footnote{Formation-time information has also been used more broadly to organise the time structure of jet quenching and to construct related jet-substructure observables; see~\cite{Apolinario:2017sob,Apolinario:2020uvt,Apolinario:2024klh,Brewer:2025wol}.} In the collinear limit, it can be estimated as
\begin{align}\label{eq:ch3_heavy_formation_time}
    \tau_f
    \sim
    \frac{
    2x(1-x)(p_0^+)^{-1}
    }{
    2(m_Q/p_0^+)^2+x^2(1-x)^2\theta^2
    }\,.
\end{align}
In the $\theta\to0$ limit, the light-parton formation time diverges, while the heavy-flavour formation time remains finite because of the mass term. Consequently, very small-angle light-parton splittings tend to occur outside the medium, whereas heavy-quark splittings can occur inside the medium at any $\theta$, provided the ratio $(m_Q/p_0^+)^2$ is sufficiently large. 

\begin{figure}[t]
    \centering
    \includegraphics[width=0.49\textwidth]{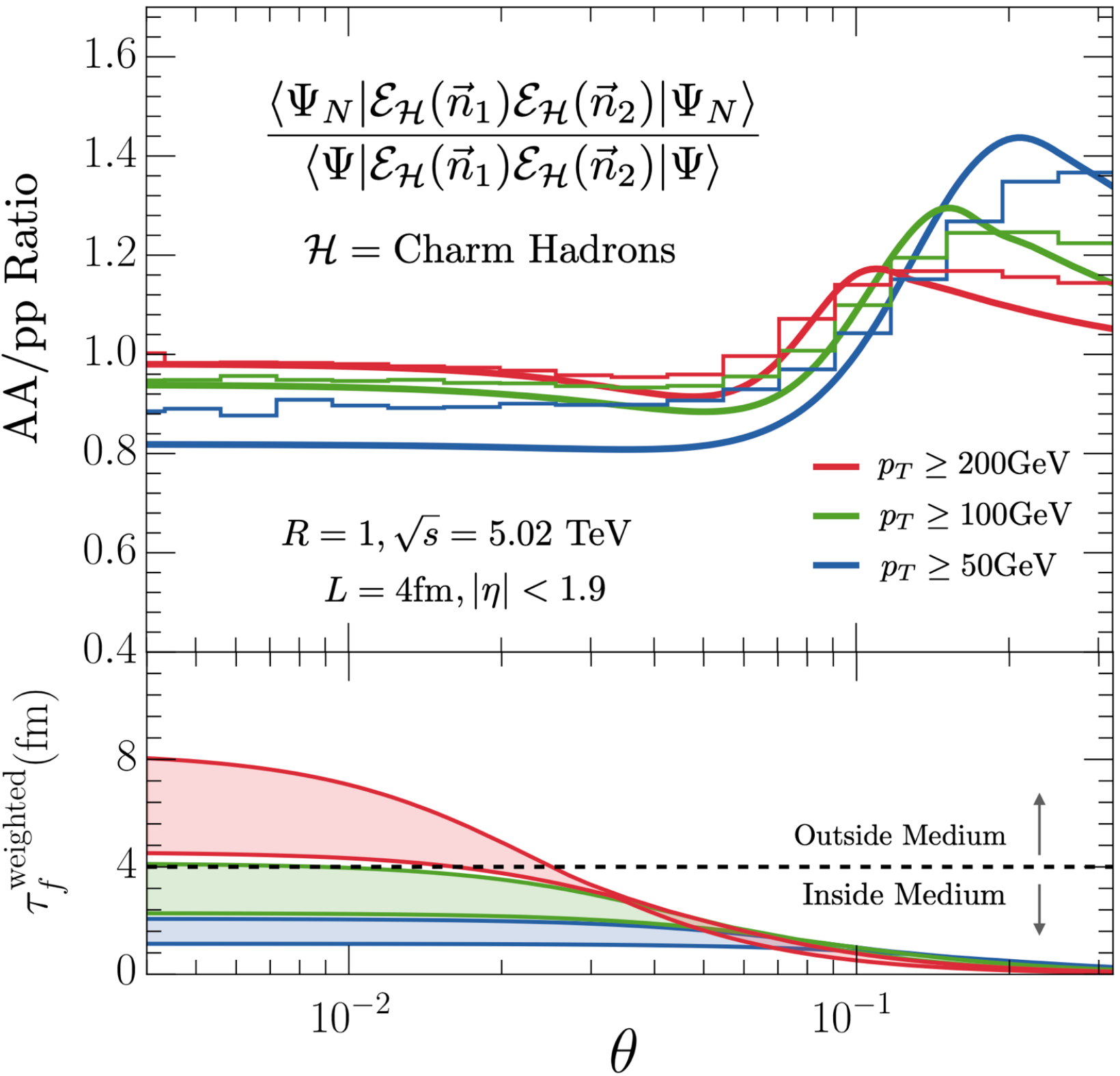}
    \includegraphics[width=0.49\textwidth]{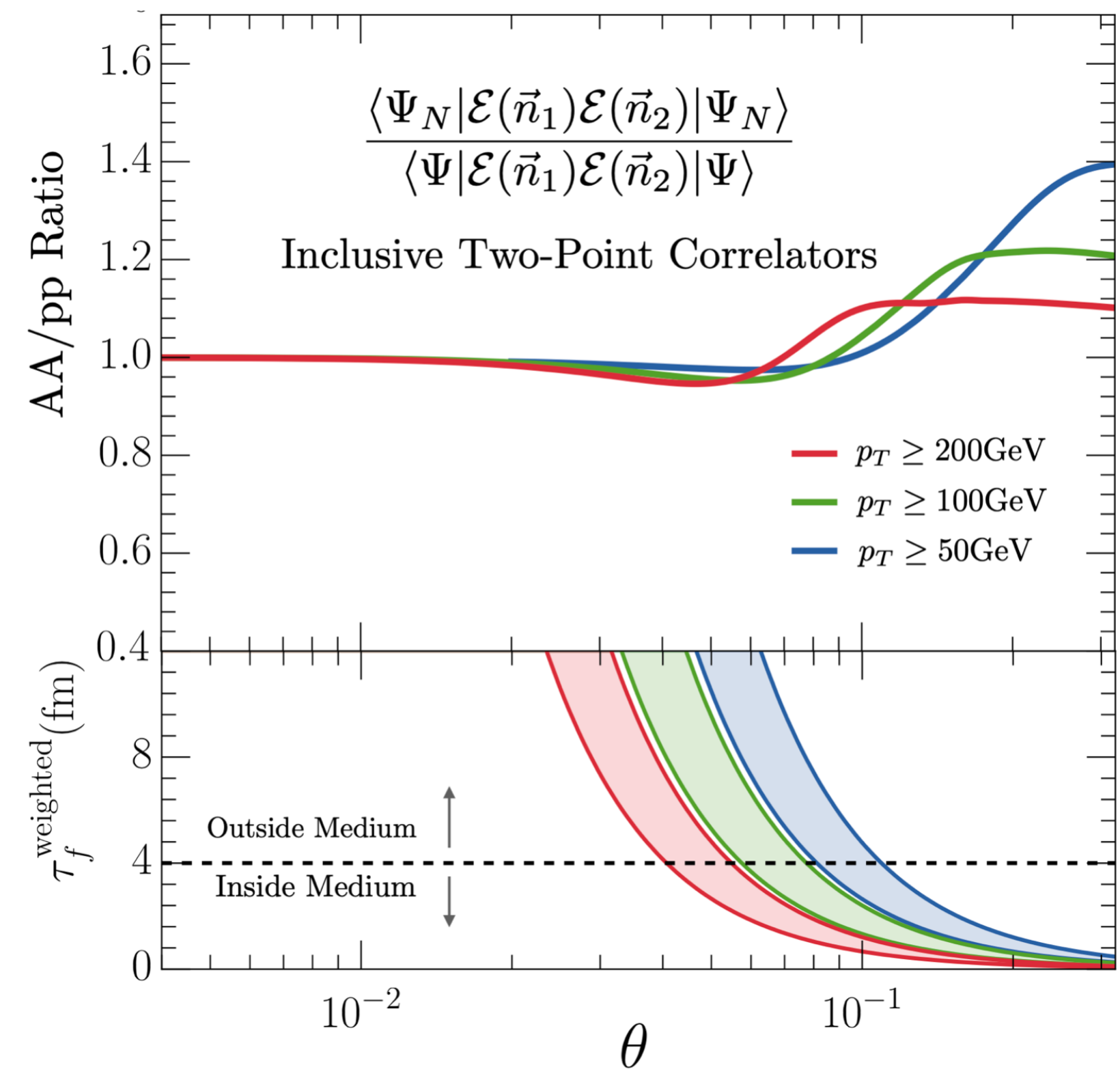}
    \caption{Medium modification of heavy-flavour and inclusive EECs in a medium of length $L=4$ fm for several jet-$p_T$ bins. Heavy-flavour histograms from \textsc{Pythia}~8~\cite{Sjostrand:2014zea} use a reweighted $g\to c\bar c$ kernel~\cite{Attems:2022otp}, while the solid curves are NLL predictions of Eq.~\eqref{eq:ch3_heavy_EEC_fact} with medium-modified splitting functions. The lower panels show $\tau_f^{\rm weighted}$ from Eq.~\eqref{eq:ch3_tauweighted} for light- and charm-quark pairs, with uncertainty given as band of initial energy from $0.5\,p_T$ to $p_T$.}
    \label{fig:ch3_EEC_med}
\end{figure}

In Fig.~\ref{fig:ch3_EEC_med}, we show the medium modification of the heavy-flavour (charm hadrons) and inclusive EECs, computed using Eq.~\eqref{eq:ch3_heavy_EEC_fact} and including medium-modifications to the jet function\footnote{Medium modifications to splittings beyond $Q\bar Q$ production were included, with their cross-section being given by a straightforward modification of the $x$ dependent factors in the $c_i$ functions in Eq.~\eqref{eq:ch3_xsec_iso_int} (see~\cite{Isaksen:2023nlr, Barata:2025fzd}). This is only valid when neglecting the "non-factorizable" piece in the quadrupole object~\cite{Blaizot:2012fh, Apolinario:2014csa,Isaksen:2023nlr}.}. We consider a static brick with length $L=4$ fm, isotropic, adjoint jet quenching parameter $\hat q_A= 2\hat q = 1.25$  GeV$^2$/fm and multiple initial parton/jet energies ($p_T = p_0^+/\sqrt{2}$). Motivated by the energy weighting in the EEC, we define the weighted formation time~\cite{Barata:2025uxp}
\begin{align}\label{eq:ch3_tauweighted}
    \tau_f^{\rm weighted}
    \equiv
    \int_0^1 dx\, 2x(1-x)\,
    \tau_f(x)\,
    P_{g\to Q\bar Q}(x), .
\end{align} 
and plot it below the corresponding EEC ratio in each panel. 
For the inclusive EEC, most of the underlying splittings are light-parton splittings, and the medium-to-vacuum ratio approaches unity at small angle while becoming enhanced at larger angles. The onset of this enhancement occurs around the angular region where the weighted formation time becomes comparable to the medium length. The charm case differs because the heavy-quark mass keeps the weighted formation time finite even at small angle. As a result, both the analytic calculation and the \textsc{Pythia}~8 simulations show small-angle modifications whose magnitude depends on the ratio $m_Q/p_0^+$. This sensitivity also depends on the jet energy: for sufficiently large-$p_T$ jets, the $c\bar c$ pair can be formed outside the medium in the smallest-angle region, whereas for lower-$p_T$ jets the weighted formation time remains below the medium length over essentially the whole angular range shown. This kinematic interplay makes the heavy-flavour EEC useful for parameterising medium modifications across different jet-energy regimes.

The angular dependence of the modification can be understood by separating the large- and small-angle regions. At large angle, one finds the enhancement characteristic of several EEC calculations~\cite{Andres:2022ovj,Barata:2023bhh,Yang:2023dwc} and measurements~\cite{CMS-PAS-HIN-23-004,talk_Jussi,talk_Anjali,talk_Ananya}. For inclusive jets, this enhancement is connected both to higher-twist contributions and to uncorrelated soft radiation~\cite{Andres:2024xvk,Barata:2025fzd,Barata:2024ukm}. Toward smaller angles, the EEC can instead be suppressed by momentum broadening: pairs that would have contributed at small angle in vacuum are broadened to larger angles in medium. Parametrically, this suppression is associated with $k_\perp^2 \sim x^2(1-x)^2\theta^2 (p_0^+)^2\sim \hat q L$, consistent with the angular regions where the medium modification appears in Fig.~\ref{fig:ch3_EEC_med}. The charm ratio shows the same broad pattern of small-angle suppression and large-angle enhancement, with the transition region shifted by the mass-dependent formation time. Although not shown here, the bottom case is qualitatively similar. Due to their larger mass, $b\bar b$ pairs are formed inside the medium over the range of $p_T$ considered for charm, but the medium effect is reduced by dead-cone-suppressed radiation and smaller momentum broadening.

\begin{figure}[t]
    \centering
    \includegraphics[width=0.62\textwidth]{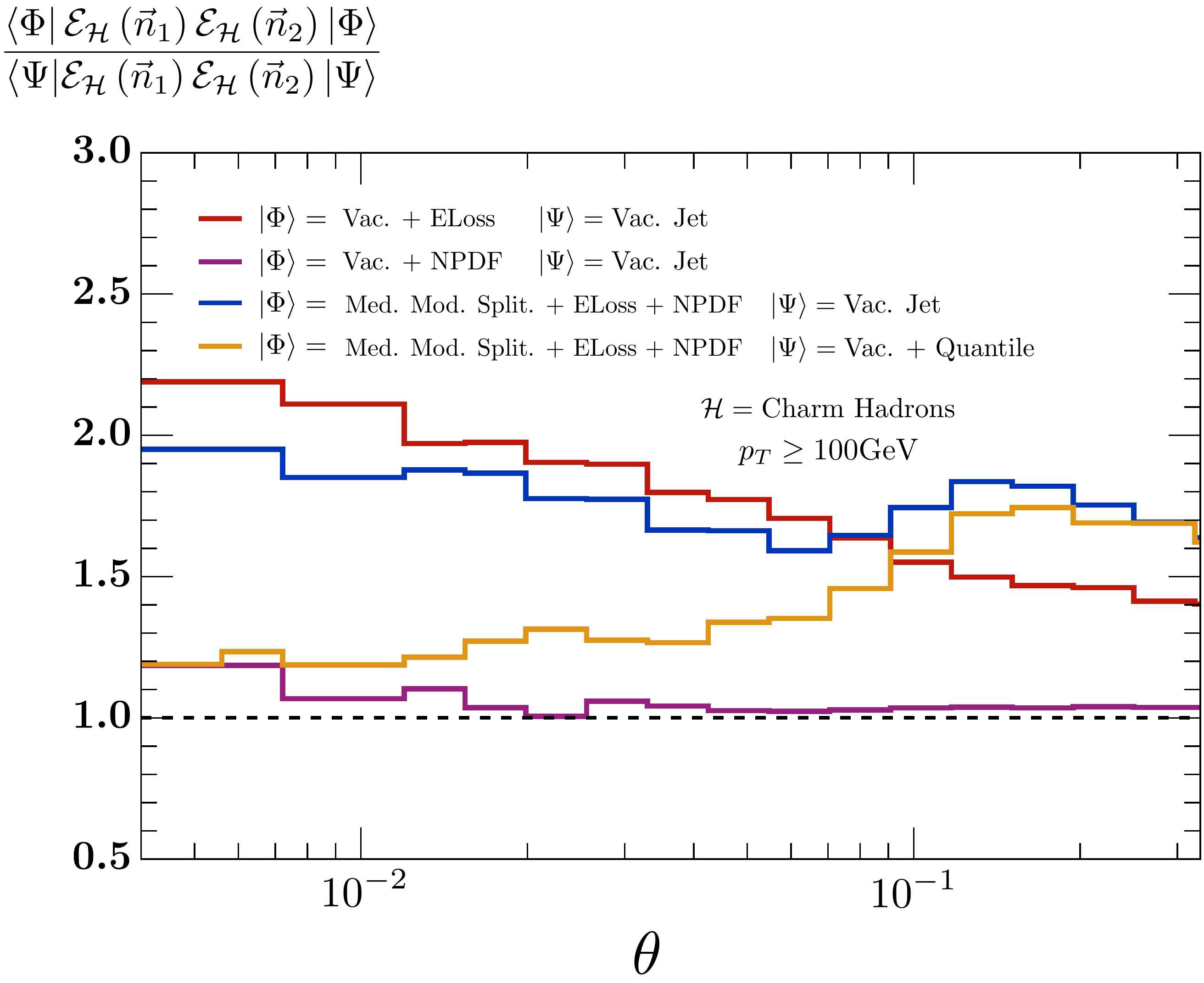}
    \caption{Impact of energy loss, nPDFs and medium-modified splittings on the heavy-flavour EEC. The quantile procedure removes the small-angle enhancement from $p_T$ migration, while the overall normalisation remains enhanced due to the reduced energy loss of heavy quarks.}
    \label{fig:ch3_EEC_eloss}
\end{figure}

The results in Fig.~\ref{fig:ch3_EEC_med} include only medium-modified splitting functions in the jet function of Eq.~\eqref{eq:ch3_heavy_EEC_fact}. In a realistic heavy-ion collision, additional effects arise from energy loss and from initial-state nuclear modifications of parton distribution functions. These effects are shown in Fig.~\ref{fig:ch3_EEC_eloss}. Initial-state nuclear effects are included through EPPS16 nuclear PDFs~\cite{Eskola:2016oht} in the hard function. The heavy-flavour EEC obtained with nPDFs is nearly identical to the result obtained with proton PDFs, reflecting the exclusive nature of the observable: it is sensitive only to jets containing at least two heavy-flavour hadrons, which are predominantly gluon jets. This differs from inclusive jet EECs~\cite{Apolinario:2025vtx,Chen:2024quk}, where changes in the quark--gluon fraction can significantly modify the EEC shape. Energy-loss effects, i.e. the final-state degradation of the energy reconstructed inside the jet cone and the associated migration between reconstructed jet-$p_T$ bins, are included using a quenching-weight approximation~\cite{Baier:2001yt}, following the prescription in~\cite{Barata:2023bhh,Brewer:2025wol}. At parton level, particles in the jet are grouped into clusters of size $\theta_c=2/\sqrt{\hat q L^3}$, the characteristic scale for radiative decoherence introduced in Section~\ref{subsec:scales}. Each cluster is assigned a net flavour according to its parton content and loses energy through the corresponding quenching factor. This produces the characteristic small-angle enhancement and large-angle suppression in the EEC. The small-angle enhancement arises because jets before energy loss had larger $p_T$, shifting the peak near $2p_T\sqrt{z}\sim m_Q$ relative to a comparison at fixed reconstructed $p_T$. The combined effect of energy loss, nPDFs and medium-modified splitting functions is shown by the blue curve in Fig.~\ref{fig:ch3_EEC_eloss}. The change in shape relative to the energy-loss-only result is similar to the shape modification found in Fig.~\ref{fig:ch3_EEC_med}, indicating that the heavy-pair splitting modification can be separated from the dominant kinematic migration effects. This separation can be further improved using the quantile procedure~\cite{Brewer:2018dfs,Apolinario:2024apr}\footnote{We shall elaborate on this method in Section~\ref{sec:qaa}.}, which compares heavy-ion jets to proton--proton jets at the corresponding original jet $p_T$ before energy loss. After this procedure, the small-angle enhancement from $p_T$ migration is removed, while the remaining angular dependence more directly reflects the medium-modified heavy-flavour splitting dynamics.

\subsection{Azimuthal EEC harmonics from anisotropic matter}
\label{subsec:ch3_heavy_flavour_anisotropy}
Having identified the heavy-pair splitting using the heavy-flavour EEC, we now use the same observable to probe directional structures of the underlying QCD matter. In this sense, we compute the heavy-flavour EEC differential in both opening angle and relative azimuthal orientation from the spin-summed medium-induced spectrum derived in Eq.~\eqref{eq:ch3_xsec_final}, as defined in Eq.~\eqref{eq:xsec_to_EEC_doublediff}. The calculation follows as discussed in previous subsections, with the medium-modified splitting functions entering the jet function in Eq.~\eqref{eq:ch3_heavy_EEC_fact} now acquiring an azimuthal dependence~\cite{Barata:2024bqp}. We emphasize that, in our model, the anisotropy is parameterised by an axis-dependent jet-quenching transport coefficient $\hat q_i$~\cite{Hauksson:2023tze,Barata:2024bqp}, and we define $\hat q=(\hat q_x+\hat q_y)/2$.
\begin{figure}[t]
    \centering
    \includegraphics[width=0.9\textwidth]{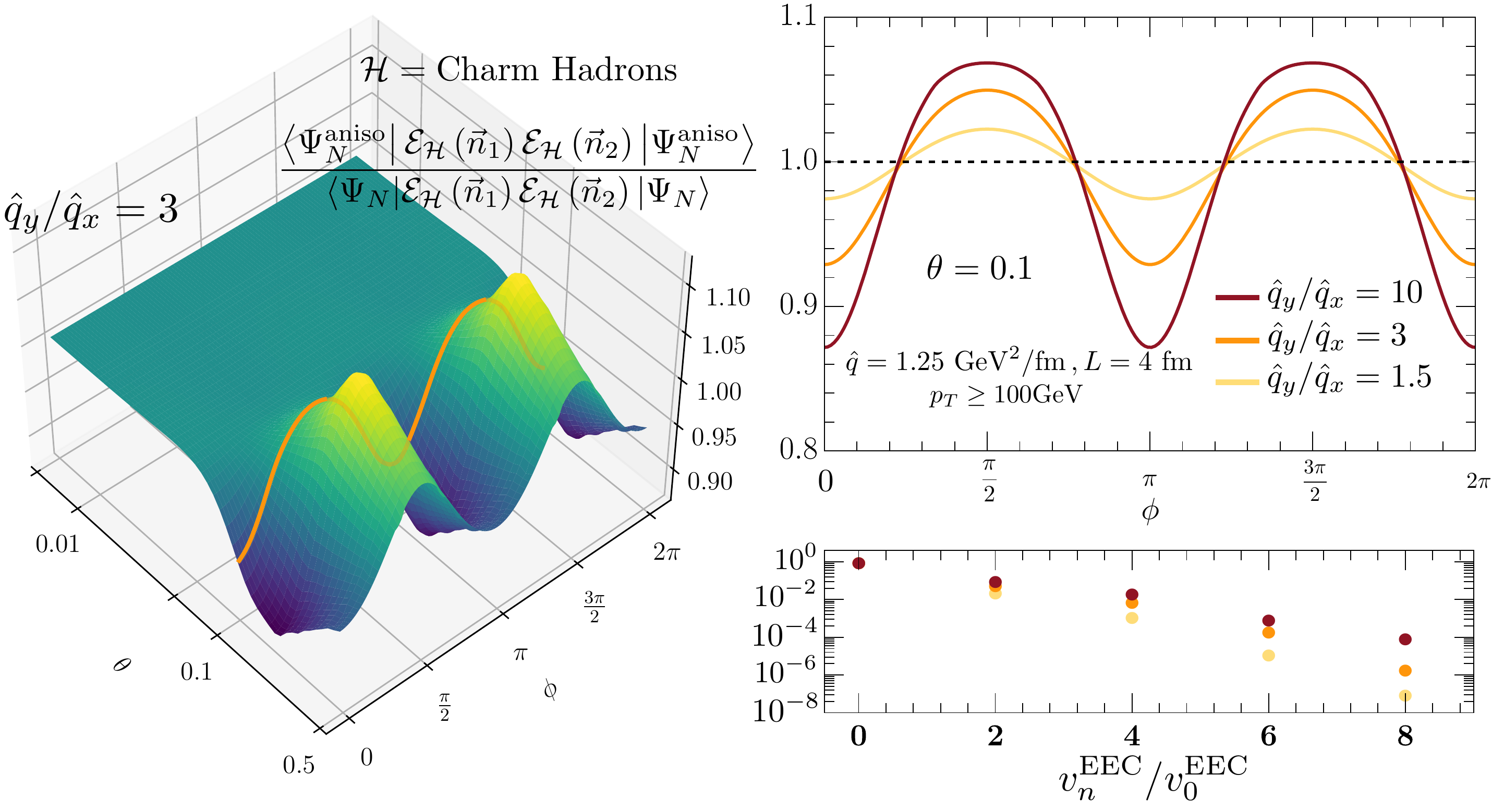}
    \caption{Ratio of the heavy-flavour EEC inside a jet produced in an anisotropic medium to that in an isotropic medium. The anisotropy induces a non-trivial azimuthal dependence in $\phi$, which can be expanded in even harmonics as in Eq.~\eqref{eq:ch3_eec_harmonic_expansion}. For $\theta=0.1$, the right panel shows the $\phi$ dependence of this ratio for several magnitudes of the anisotropy in the jet-quenching parameter, together with the harmonic coefficients $v_{2n}^{\rm EEC}$ normalized by $v_0^{\rm EEC}$. To be exact, we fixed the adjoint $\hat q$, i.e., $\hat q_A = 1.25$ GeV$^2$/fm, which, consistent with notation in the main text, fixes $\hat q = 0.625$ GeV$^2$/fm.}
    \label{fig:ch3_heavy_anisotropy}
\end{figure}

In the left panel of Fig.~\ref{fig:ch3_heavy_anisotropy} we show the double-differential EEC for an anisotropic medium with $\hat q_y/\hat q_x = 3$ divided by the same object for an isotropic medium $\hat q_x = \hat q_y$ (given by Eq.~\eqref{eq:ch3_xsec_iso_int}), as a function of $\theta$ and $\phi$. The average adjoint $\hat q_A = 2\hat q$ is set to $1.25$ GeV$^2$/fm, the medium length to $L=4$ fm and the energy of the initial parton is set to $p_T = p_0^+/\sqrt{2} = 100$ GeV. The small-$\theta$ region is relatively flat around unity, showing that the anisotropy has little effect at sufficiently small opening angle. As $\theta$ is increased, the EEC develops a non-trivial $\pi$-periodic azimuthal modulation, whose amplitude reaches a maximum at an intermediate opening angle before decreasing again at larger angles. This modulation is generated by the anisotropic structure of the medium and becomes more pronounced as the anisotropy in the jet-quenching parameter is increased ($\hat q_y/\hat q_x \in \{1.5,3,10\}$), as one can see in the top right panel, where we make a cut of the two-dimensional plot at $\theta = 0.1$. The main observation is that the modulation is close to sinusoidal in $\phi$ for small to moderate anisotropy, but it deviates increasingly from this behaviour as the anisotropy is increased. As discussed in Section~\ref{subsec:ch3_azimuthal_eec}, one can decompose this azimuthal structure into EEC harmonics, which for the current medium model and for the spin-summed spectrum is given fully in terms of even cosine harmonics. In the bottom right panel of Fig.~\ref{fig:ch3_heavy_anisotropy}, we show the resulting harmonics $\{v_2^{\rm EEC},v_4^{\rm EEC},v_6^{\rm EEC},v_8^{\rm EEC}\}$ normalized by the isotropic harmonic $v_0^{\rm EEC}$. We observe that, for fixed anisotropy ratio, higher-order harmonics are increasingly small. As the anisotropy is increased, however, higher harmonics grow rapidly and become numerically relevant. This provides numerical confirmation for the analytical statement made in Section~\ref{subsec:ch3_azimuthal_eec}, namely that the anisotropy power expansion of the harmonic $v_{2n}^{\rm EEC}$ starts at power $n$ in anisotropy.

A direct measurement of this structure would provide a new way to access the azimuthal geometry of the matter through which the jet propagates. In this sense, the heavy-flavour EEC does not only isolate the heavy-quark-pair splitting and its medium modification; once measured differentially in the pair orientation, it also turns the anisotropic broadening of the medium into a set of observable harmonic coefficients.

\section{Energy correlators in early-stage matter}
\subsection{Celestial blocks and the clover projection}
\label{subsec:ch3_celestial_clover}

The azimuthally differential EEC defined in Section~\ref{subsec:ch3_azimuthal_eec} resolves the orientation of correlated energy deposits around the jet axis. In a heavy-ion collision, this orientation can be measured with respect to the beam direction, which provides a physical reference axis for anisotropic QCD matter. The resulting angular dependence is naturally described in terms of azimuthal harmonics, whose coefficients encode both the dynamics of the jet--medium interaction and the symmetries of the underlying medium. This harmonic expansion can be formally related to the small-angle structure of energy correlators~\cite{Barata:2025zku}. The cleanest formulation of this relation is obtained in a conformal field theory (CFT), where energy correlators admit a celestial-block, or light-ray OPE, decomposition. Taking the collinear limit of that decomposition, where the two energy-flow detectors approach each other, relates the powers of the opening angle $\chi$ to twist and the azimuthal harmonics in $\Psi$ to the transverse spin of the corresponding light-ray contributions, as we will see below.

Let us introduce the ingredients of this decomposition that are needed for the present discussion. In this context, a celestial block is the symmetry-fixed contribution to an energy-flow correlator associated with a light-ray operator of definite quantum numbers. The corresponding coefficient contains the dynamical information, while the block determines the dependence on the detector geometry~\cite{Hofman:2008ar,Kologlu:2019mfz,Chang:2020qpj,Chen:2022jhb,Chang:2022ryc,Moult:2025nhu,Chen:2025rjc}. For a scalar source, rotational symmetry restricts the two-point EEC to the zero transverse spin sector. Non-zero transverse spin blocks require an additional directional structure, such as a vector-like excitation of the state or, in collider kinematics, the incoming beam axis. The latter is the relevant reference direction for the azimuthal correlations considered below. In this setting, the corresponding vacuum-CFT two-point energy correlator admits the following celestial-block decomposition~\cite{Kologlu:2019mfz,Chen:2025rjc}
\begin{align}\label{eq:ch3_celestial_block_decomp}
\langle \mathcal{E}(\vec n_1)\mathcal{E}(\vec n_2)\rangle
=
\frac{2p_t^2}{(1-\vec n_1\cdot\vec n_2)^3}
\sum_{\delta,j}
\int_\gamma
\frac{c_{\delta,j,\gamma}}{2\pi i}\,
w^\gamma
G_{\delta,j,\gamma}(z,\bar z)\, .
\end{align}
Here $p_t$ denotes the hard transverse scale of the jet, $c_{\delta,j,\gamma}$ are dynamical coefficients, and $G_{\delta,j,\gamma}$ are celestial blocks. The variable $\gamma$ is the Mellin parameter conjugate to the cross ratio $w$, and $\int_\gamma$ denotes the corresponding Mellin-Barnes contour integral. The celestial dimension is related to the twist by $\delta=\tau+2$, while $j$ is the transverse spin. The variables $z$ and $\bar z$ are collider cross ratios defined by $z\bar z=1-2e^{\Delta Y}\cos(\Delta\Phi)+e^{2\Delta Y}$ and $(1-z)(1-\bar z)=e^{2\Delta Y}$, where $\Delta Y$ and $\Delta\Phi$ denote, respectively, the rapidity and azimuthal separation of the detectors, defined with respect to the beam direction. The parity-even, exchange-symmetric celestial blocks are
\begin{align}
G_{\delta,j,\gamma}(z,\bar z)
=
z^h \bar z^{\bar h}\,
\mathcal{F}(h,\gamma,z)
+
\bar z^h z^{\bar h}\,
\mathcal{F}(h,\gamma,\bar z)\, ,
\end{align}
with $\mathcal{F}(h,\gamma,z)={}_2F_1(h,h-\gamma,2h,z)\,{}2F_1(\bar h,\bar h-\gamma,2\bar h,\bar z)$, with ${}_2F_1$ the hypergeometric function, $2h=\delta-j$, and $2\bar h=\delta+j$. We are interested in the collinear limit, $|z|\to0$, for detectors close to mid-rapidity. Parametrising $z=r e^{i\Psi}$, with $r\sim\chi$, and expanding the hypergeometric functions at small $r$, one obtains at leading power
\begin{align}
G_{\delta,j,\gamma}(z,\bar z)
\simeq
2r^{\tau+2}\cos(j\Psi)\, .
\end{align}
Importantly, this leading small-$r$ behaviour is kinematic: it follows from the form of the celestial block itself at fixed $(\tau,j)$, while the dynamics enters through the coefficients multiplying the allowed angular structures. Thus, in non-conformal theories, this kinematic expansion is expected to persist as an organizing principle, with departures from conformality encoded in the allowed spectrum and in the corresponding coefficients.

Following this leading power expansion, at mid-rapidity, the leading OPE-limit contribution is obtained from the $\gamma\to0$ pole of the Mellin integral in Eq.~\eqref{eq:ch3_celestial_block_decomp}. Doing the appropriate angular projections leads to the double-differential EEC in the OPE limit\footnote{Here the perturbation is along a defined axis rather than an oriented vector, consistent with the type of anisotropy we focus on.}
\begin{align}\label{eq:ch3_EEC_OPE}
    \frac{d\Sigma}{d\chi d\Psi}
    =
    \sum_{k=1}^{\infty}
    \left[
    \sum_{j=0,2,4\,\cdots}^{\tau-2}
    c_{\tau,j}\,
    \chi^{\tau-3}
    \cos(j\Psi)
    \right]_{\tau=2k}\, .
\end{align}
This expression makes explicit the relation between the angular Fourier structure and the light-ray OPE data: the power of $\chi$ is tied to the twist $\tau$, while the harmonic in $\Psi$ is labelled by the transverse spin $j$. The allowed vacuum contributions are constrained by the unitarity bound $\tau\geq j+2$. The first few sectors illustrate the resulting hierarchy: the azimuthally symmetric contribution corresponds to $j=0$ and starts at $\tau=2$, while the leading anisotropic harmonics correspond to $j=2$ and $j=4$, starting at $\tau=4$ and $\tau=6$, respectively. In the language of the light-ray OPE, these lowest transverse-spin structures arise from the minimal global-spin sector, with $J=3$. More generally, the global spin obeys the conformal constraint $J\geq 1+j/2$, and higher-global-spin light-ray operators can contribute with additional transverse tensor structures. In a heavy-ion environment, the coefficients multiplying these angular structures are determined by the jet-medium interaction. They can receive perturbative medium-induced corrections, as well as classical flow contributions associated with energy deposited into and carried by the bulk medium.
The latter will be discussed in Section~\ref{sec:ch5_energy_correlators}. Following the discussion in previous sections, here we focus instead on the perturbative contribution, described by the medium-induced BDMPS-Z spectrum in an anisotropic background and analysed through the azimuthal harmonic moments of the EEC. In this description, the medium-induced contribution is encoded in the coefficients $v_{2n}^{\rm EEC}(\chi)$ multiplying the $\cos(2n\Psi)$ moments, which correspond to transverse-spin sectors with $j=2n$. Thus, the celestial-block expansion classifies the allowed angular moments in terms of transverse spin and twist, with the small-$\chi$ scaling classified by twist. Any further hierarchy among the medium-induced coefficients is determined by the dynamics of the jet-medium interaction.

Having described the angular organisation of the double-differential EEC in the collinear limit, we now introduce a projection that isolates the part of the correlator distinguishing the beam-aligned and beam-orthogonal directions. For an azimuthal region $\Omega$ in the plane transverse to the jet axis, let us define
\begin{align}\label{eq:ch3_EEC_clover_region}
    \frac{d\Sigma^\Omega}{d\chi}
    =
    \int_\Omega d\Psi
    \frac{d\Sigma}{d\chi d\Psi}\, .
\end{align}
The angle $\Psi$ is measured with respect to the beam direction projected onto the plane transverse to the jet axis, as discussed in Section~\ref{subsec:ch3_azimuthal_eec}. We consider a horizontal clover region, aligned with the beam direction, and a vertical clover region, orthogonal to it,
\begin{align}
    h:\quad
    -\frac{\pi}{4}<\Psi<\frac{\pi}{4}\,,
    \qquad
    v:\quad
    \frac{\pi}{4}<\Psi<\frac{3\pi}{4}\, .
\end{align}
Thus, we define the \emph{clover EEC} as the difference between these two projections~\cite{Barata:2025zku},
\begin{align}\label{eq:ch3_EEC_clover_def}
    \frac{d\Sigma^\Delta}{d\chi}
    \equiv
    \frac{d\Sigma^h}{d\chi}
    -
    \frac{d\Sigma^v}{d\chi}\, .
\end{align}
This construction compares the energy correlated along the beam-aligned and beam-orthogonal directions, making it directly sensitive to anisotropic energy flow inside a jet, while cancelling the leading azimuthally symmetric contribution. Using the harmonic expansion defined in Section~\ref{subsec:ch3_azimuthal_eec}, valid in the absence of spin correlations/dependence and for a medium with an axis-like anisotropy, this observable reads
\begin{align}\label{eq:ch3_clover_eec_integrated}
    \frac{d\Sigma^\Delta}{d\chi}
    =
    \sum_{n=1}^{\infty}
    \frac{2}{n}
    v_{2n}^{\rm EEC}(\chi)
    \sin\left(\frac{n\pi}{2}\right)\, .
\end{align}
with the $\chi$ dependence of the harmonics now made explicit. Importantly, the clover EEC selects only odd values of $n$ in the projection, i.e., $v_2^{\rm EEC},v_6^{\rm EEC},\ldots$. Since the leading anisotropic modulation is expected to appear in the elliptic harmonic, as per Sections~\ref{subsec:harmonics_spin_sensitivity} and~\ref{subsec:ch3_heavy_flavour_anisotropy}, the clover projection provides a one-dimensional observable that directly targets the dominant directional component of the EEC. 

\subsection{Clover EEC in time-dependent anisotropic matter}
\label{subsec:ch3_clover_time_dependent}

We now evaluate the clover EEC for a jet propagating through anisotropic matter whose properties evolve in time. We consider a medium with $\hat q_i(t)=\hat q_i(t)\Theta(t<L^+)$, where $i=x,y$ labels the two principal axes in the plane transverse to the jet. The $x$ direction is identified with the reference axis used to define $\Psi$, while $y$ denotes the orthogonal direction. In the context of early stage HIC matter, this reference axis is the beam direction projected onto the plane transverse to the jet axis. The calculation follows essentially the same derivation in the BDMPS-Z formalism as in Section~\ref{sec:ch3_gqqbar}, with the final-state spins summed. The  difference stems from the static anisotropic HO coefficients now being replaced by time-dependent functions, determined by differential equations with time-dependent oscillatory frequencies. For completeness and for consistency of the notation, the medium-induced $g\to q\bar q$ spectrum in this setup, for time-dependent $\hat q$, is given below:
\begin{align}\label{eq:ch3_spectrum_aniso_time_dependent}
    & (2\pi)\frac{d\sigma^{\rm in-in}}{\sigma_0 dzd\chi d\Psi} = \frac{\alpha_s}{\pi}T_F\,\omega^+\chi\,\mathrm{Re}\int_{X^+} -i\frac{\sqrt{c_{1x}}\sqrt{c_{1y}}}{\sqrt{c_{3x}}\sqrt{c_{3y}}}e^{-i\frac{m^2}{2\omega^+}(\bar x^+-x^+)}\exp\left\{i\frac{\omega^+\chi^2}{4}\left(\frac{\cos^2\Psi}{c_{3x}}+\frac{\sin^2\Psi}{c_{3y}}\right)\right\}\nn
    & \times \left[P_{g\to q\bar q}^{\rm vac}(z)\left(c_{1x}+c_{1y}+\frac{c_{1x}c_{2x}}{c_{3x}}+\frac{c_{1y}c_{2y}}{c_{3y}}+i\frac{\omega^+\chi^2}{2}\left(\frac{c_{1x}c_{2x}}{c_{3x}^2}\cos^2\Psi+\frac{c_{1y}c_{2y}}{c_{3y}^2}\sin^2\Psi\right)\right)+\frac{m^2}{2\omega^+}\right]\,,
    \nn
    & (2\pi)\frac{d\sigma^{\rm in-out}}{\sigma_0 dzd\chi d\Psi} = \frac{\alpha_s}{\pi}T_F\frac{\omega^+\chi}{\chi^2+2(m/\omega^+)^2}2\mathrm{Re}\int_0^{L^+}dx^+\,i\frac{\sqrt{c_{1x}}\sqrt{c_{1y}}}{\sqrt{c_{2x}}\sqrt{c_{2y}}}e^{-i\frac{m^2}{2\omega^+}(L-x^+)}\nn
    & \times \exp\left\{-i\frac{\omega^+\chi^2}{4}\left(\frac{\cos^2\Psi}{c_{2x}}+\frac{\sin^2\Psi}{c_{2y}}\right)\right\}\left[P_{g\to q\bar q}^{\rm vac}(z)\left(\frac{c_{1x}}{c_{2x}}\cos^2\Psi+\frac{c_{1y}}{c_{2y}}\sin^2\Psi\right)\chi^2-i\left(\frac{m}{\omega^+}\right)^2\right]\,,
    \nn
    & (2\pi)\frac{d\sigma^{\rm out-out}}{\sigma_0 dzd\chi d\Psi} = \frac{\alpha_s}{\pi}T_F\frac{\chi}{\left(\chi^2+2(m/\omega^+)^2\right)^2}\left[P_{g\to q\bar q}^{\rm vac}(z)\chi^2+2\left(\frac{m}{\omega^+}\right)^2\right]\, ,
\end{align}
where $\omega^+=z(1-z)p_0^+$, $z$ is the energy fraction of the quark with respect to the gluon with energy $p_0^+$, $P^{\rm vac}_{g\to q\bar q}(z)=z^2+(1-z)^2$ is the vacuum $g\to q\bar q$ splitting function without its colour factor, and $\sigma_0$ is the initial production cross section. The in-in integration region is $X^+=\{0<x^+<\bar x^+ <L^+\}$. We have adopted the notation $(\chi, \Psi)$ for the opening angle and relative azimuthal orientation of the pair, rather than $(\theta,\phi)$, for consistency of notation in this section. Following the definition of the double differential EEC in Eq.~\eqref{eq:ch3_azimuthal_eec_operator}, in terms of the direction vectors these angles read
\begin{align}
    \chi \simeq |\n_1-\n_2|\,,\qquad \cos\Psi = \frac{\n_1-\n_2}{|\n_1-\n_2|}\cdot \b\,\,.
\end{align}
For beam axis $\b = (1,0)$, this definition is consistent with the notation in Section~\ref{sec:ch3_gqqbar}, since for a pair with transverse momenta $\p_1,\p_2$, one has $\n_i=\p_i/E_i$ and $\n_1-\n_2=\p/(z(1-z)E)$, with $\p=(1-z)\p_1-z\p_2$. The HO coefficients entering Eq.~\eqref{eq:ch3_spectrum_aniso_time_dependent} read
\begin{align}
    & c_{1i}(\bar x^+,x^+)=\frac{1}{2iS_{i}(\bar x^+,x^+)}\,,\qquad c_{2i}(\bar x^+,x^+)=\frac{C_{i}(x^+,\bar x^+)}{S_{i}(\bar x^+,x^+)}\,,\nn
    & c_{3i}(\bar x^+,x^+)=-i\frac{\hat k_i^2(L,\bar x^+)\left(z^2+(1-z)^2\right)}{2\omega^+}-c_{2i}(\bar x^+, x^+)\,,
\end{align}
with $\hat k_i^2(L,\bar x^+)=\int_{\bar x^+}^L dt,\hat q_i(x^+)$ and $i=x,y$. For the in-in term all $c_{ji}$ carry $(\bar x^+,x^+)$ as arguments, while for the in-out the arguments are $(L^+, x^+)$. The functions $C_i$ and $S_i$ are solutions to 
\begin{align}\label{eq:ch3_SC_diffeq}
    & \left(\partial_t^2+\Omega_i^2(x^+)\right)S_i(x^+,x^+_0)=0\,,\qquad S_i(x^+_0,x^+_0)=0\,,\qquad \left.\partial_t S_i(x^+,x^+_0)\right|_{x^+=x^+_0}=1\,,
    \nn
    & \left(\partial_t^2+\Omega_i^2(x^+)\right)C_i(x^+,x^+_0)=0\,,\qquad C_i(x^+_0,x^+_0)=1\,,\qquad \left.\partial_t C_i(x^+,x^+_0)\right|_{x^+=x^+_0}=0\,,
\end{align}
where\footnote{We use fundamental $\hat q \equiv \hat q_F$.}
\begin{align}
    \Omega_i(x^+)=\frac{1-i}{\sqrt{2}}\sqrt{\frac{\hat q_i(x^+)\left(z^2+(1-z)^2\right)}{2\omega^+}}\, .
\end{align}
In the limit of static $\hat q$ the coefficients reduce to those in Eq.~\eqref{eq:ch3_ci_definition_static}. They further satisfy~\cite{Arnold:2008iy}
\begin{align}
    C(x^+_1,x^+_2)=\partial_{x^+_2}S(x^+_2,x^+_1)\,,\qquad S(x^+_2,x^+_1)=C(x^+_1,x^+_0)S(x^+_2,x^+_0)-C(x^+_2,x^+_0)S(x^+_1,x^+_0)\,,
\end{align}
for $x^+_2>x^+_1>x^+_0$. Thus, using this result, the full light-cone time dependence of both $S_i$ and $C_i$ can be determined from the solution to the first differential equation in Eq.~\eqref{eq:ch3_SC_diffeq}.

Before computing the clover EEC numerically for a time-dependent $\hat q$, let us inspect the structure of the cross-section in Eq.~\eqref{eq:ch3_spectrum_aniso_time_dependent} and the corresponding EEC harmonic expansion, whose only non-vanishing harmonics are even cosine harmonics. Since the structure of the cross-section for the time-dependent case is exactly the same in terms of directional structure as the static case, the same scaling argument with respect to anisotropy powers derived in Section~\ref{subsec:ch3_azimuthal_eec} applies. Namely, that the $v_{2n}^{\rm EEC}$ series in powers of the anisotropy parameter $\xi \sim (\hat q_y-\hat q_x)/(\hat q_y+\hat q_x)$ starts at order $n$. Thus, since the clover EEC only involves odd $n$ (see Eq.~\eqref{eq:ch3_clover_eec_integrated}), its leading contribution comes from $v_2^{\rm EEC} \sim \xi$, while the subleading correction comes only at power $v_6^{\rm EEC} \sim \xi^3$. It remains to determine the corresponding hierarchy in the opening angle $\chi$. First, note that each factor of $\cos^2\Psi$ or $\sin^2\Psi$, equivalently $\cos2\Psi$, both in the numerator and in the exponent of the exponential, is always accompanied by a power of $\chi^2$.\footnote{This is expected since the angular dependence comes in even power combinations of relative transverse momentum, $\p_x \sim \chi\cos\Psi $ or $\p_y \sim \chi \sin\Psi$.} Thus, when expanding the exponential in powers of $\chi$, the $n$-th power of $\cos2\Psi$, and thus at max the harmonic $\cos 2n\Psi$, is accompanied by $\chi^{2n}$. Since the spectrum contains an overall factor of $\chi$, the coefficient of the $2n$-th harmonic reads~\cite{Barata:2025zku}
\begin{align}\label{eq:ch3_v2n_twist_expansion}
    v_{2n}^{\rm EEC}(\chi)=\chi^{2n+1}\sum_{k=0}^\infty v_{2n,k}^{\rm EEC}\chi^{2k}\,.
\end{align}
Thus, the leading contribution to $v_{2n}^{\rm EEC}$ scales as $\chi^{2n+1}$, so that the medium-induced constraint on angular powers and harmonic number $(j=2n)$ is $\tau \geq j+4$ in the OPE language developed in Section~\ref{subsec:ch3_celestial_clover}, which is stronger than the vacuum unitarity constraint $\tau \geq j+2$. For the isotropic harmonic $v_0^{\rm EEC}$, both the vacuum and medium-induced contributions start as $\chi$ in the present massive final-state calculation. This differs from the massless vacuum case, where the leading vacuum contribution behaves as $\chi^{-1}$. Combining the anisotropy and angular power scalings for the EEC harmonics, the clover EEC reads
\begin{align}\label{eq:ch3_clover_scaling}
    \frac{d\Sigma^\Delta}{d\chi}=2v_2^{\rm EEC}(\chi)+\mathcal{O}(\xi^3\chi^7)\, ,
\end{align}
where the leading term in $v_2^{\rm EEC}$ is of order $\mathcal{O}(\xi\chi^3)$. Thus, the clover EEC is linearly sensitive to the axis-like anisotropy at leading power, with scaling $\sim \chi^3$ in angle, while higher order contributions to this observable are suppressed by $2$ additional powers of $\xi$ and $4$ additional powers of $\chi$.

\begin{figure}[t]
    \centering
    \includegraphics[width=0.49\textwidth]{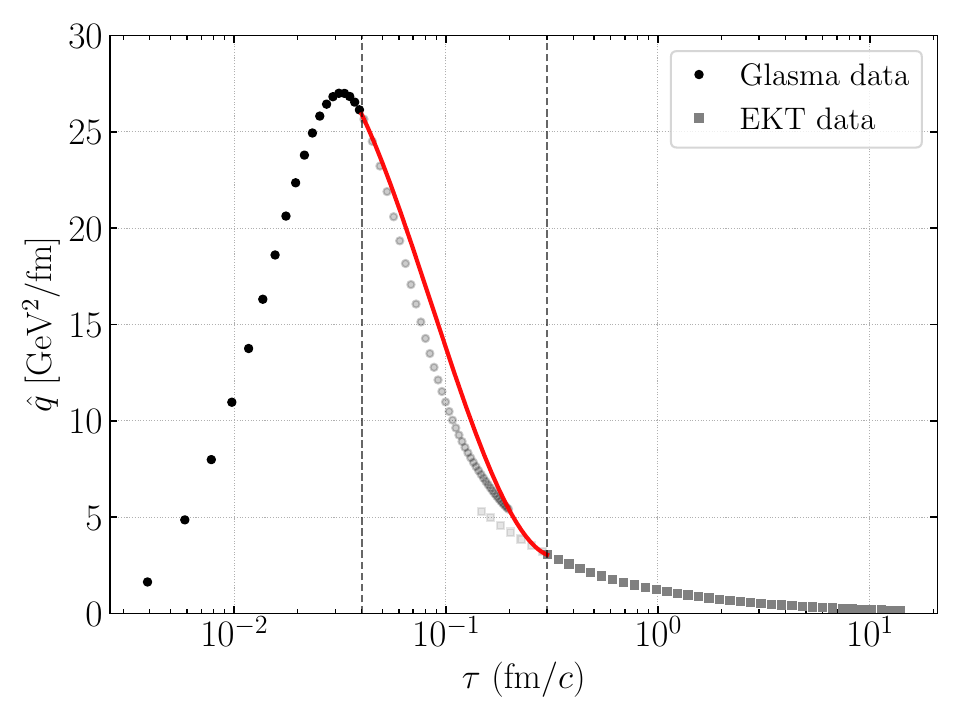}
    \includegraphics[width=0.49\textwidth]{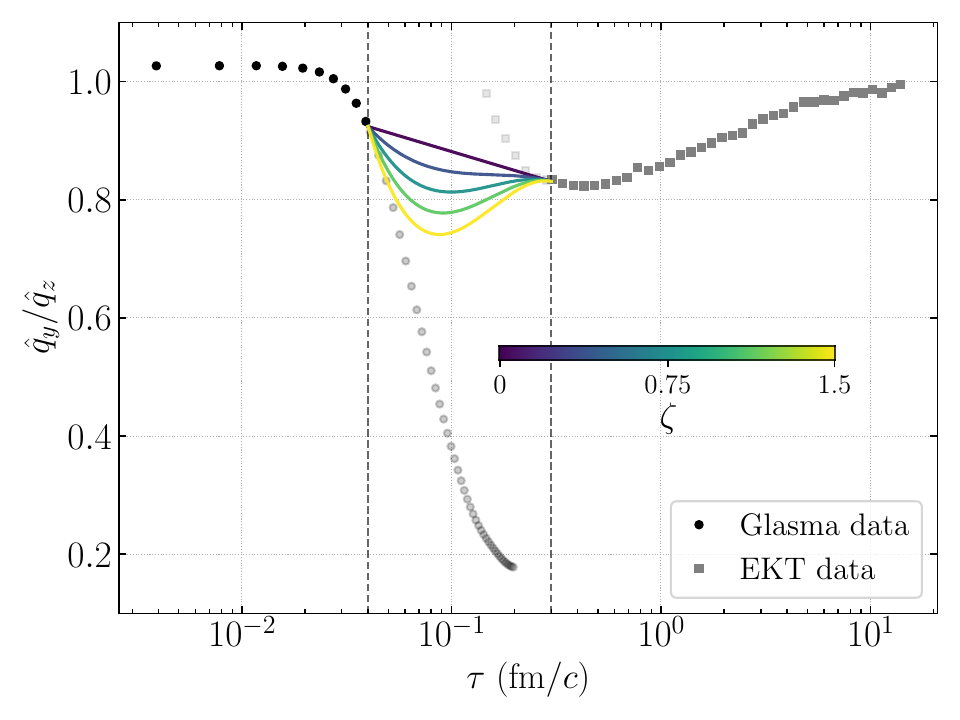}
    \caption{Scalar part of the quark jet-quenching parameter, $\hat q=\hat q_y+\hat q_z$ (\textbf{left}), and anisotropy ratio $\hat q_y/\hat q_z$ (\textbf{right}) as a function of light-cone time $\tau$. In the spectrum calculation we denote $\hat q_z$ by $\hat q_x$. The curves in the transition region $\tau\in(0.04,0.3)$ fm/c correspond to a family of Hermite interpolants depending on the parameter $\zeta\in(0,1.5)$. For $\hat q$ we use the smoothest of these curves, $\zeta=1$.}
    \label{fig:ch3_qhat_early_encs}
\end{figure}

Let us now turn to numerically computing the clover EEC, by first specifying the time dependent profile of $\hat q$ and of its associated anisotropy. To this end, we use numerical extractions of this parameter from classical-statistical simulations of the earliest Glasma stage~\cite{Avramescu:2023qvv,Ipp:2020nfu} and from the subsequent kinetic-theory epoch~\cite{Boguslavski:2024ezg,Boguslavski:2024jwr}.\footnote{We would like
to thank D. Avramescu and F. Lindenbauer for providing the Glasma and kinetic theory data for the jet
quenching parameter, respectively.} The $\hat q_i$ extracted from Glasma simulations are calculated as the $\tau$ derivative of the momentum broadening in the $i$-th direction, obtained from gauge-invariant correlators of electric and magnetic fields evaluated along the jet trajectory. In these simulations the saturation momentum is fixed to $Q_s=2$ GeV, roughly corresponding to the most central collisions at LHC energies. The EKT data are matched to the Glasma data by imposing the same energy density, which fixes $Q_s=1.4$ GeV~\cite{Boguslavski:2024ezg}. In Fig.~\ref{fig:ch3_qhat_early_encs}, we show the time evolution of this transport coefficient. In the numerical profiles, the beam direction is denoted by $z$, while in the previous analytical discussion and moving forward we identify this direction as $x$, and the orthogonal transverse direction by $y$. The left panel gives the trace $\hat q=\hat q_x+\hat q_y$, while the right panel shows the anisotropy ratio between the beam direction, $\hat q_x$, and the orthogonal direction, $\hat q_y$. The data include the region where the Glasma and EKT descriptions overlap, which is parametrically expected after $\tau\sim1/Q_s$. The precise transition between these regimes is numerically nontrivial, although the connection is conceptually understood~\cite{Epelbaum:2013waa,Berges:2014yta,Romatschke:2005ag,Romatschke:2005pm,Romatschke:2006nk,Mrowczynski:1993qm}. In practice, both sides are matched around $\tau\sim1/Q_s$ by requiring the same averaged energy density. We connect the two descriptions through a cubic Hermite interpolant in the transition region\footnote{Note that our approach here is rather practical, and we do not discuss the dependence on, e.g., the choice of saturation scale, which can change the values of $\hat q$. Although this would alter the values of the extracted observables, such modifications would not affect the conclusions of our study at a qualitative level.}. The Glasma data are truncated at $\tau_1=0.04$ fm/c and the EKT data are included after $\tau_2\equiv\tau_{\rm EKT}=0.3$ fm/c. Local least-squares polynomial fits, up to cubic order, are performed near the stitching points: to the Glasma data for $\tau<\tau_1$ and to the EKT data for $\tau>\tau_2$. These fits determine the values $y_1,y_2$ and slopes $y_1',y_2'$ at the boundaries. The transition region is parameterised by the logarithmic variable $ s=(\ln\tau-\ln\tau_1)/(\ln\tau_2-\ln\tau_1)\in[0,1]$ and the interpolation is defined as
\begin{align}\label{eq:ch3_hermite_bridge}
    p_\zeta(\tau)&=y_1h_{00}(s)+y_2h_{01}(s)+\left(\ln\tau_2-\ln\tau_1\right)\left[m_1(\zeta)h_{10}(s)+m_2(\zeta)h_{11}(s)\right], .
\end{align}
Here $h_{ij}(s)$ are the cubic Hermite basis functions and $m_i(\zeta)=(1-\zeta)m_{\rm sec}+\zeta(t_i y_i')$, with $m_{\rm sec}=(y_2-y_1)/(\ln\tau_2-\ln\tau_1)$. The parameter $\zeta$ interpolates between a straight-line connection, $\zeta=0$, and the natural slope inferred from the local fits, $\zeta=1$. The family of curves $p_\zeta(\tau)$ is shown in the lower panel of Fig.~\ref{fig:ch3_qhat_early_encs} for several values of $\zeta$, while the trace $\hat q$ in the upper panel is evaluated with $\zeta=1$.

\begin{figure}[t]
    \centering
    \includegraphics[width=0.49\textwidth]{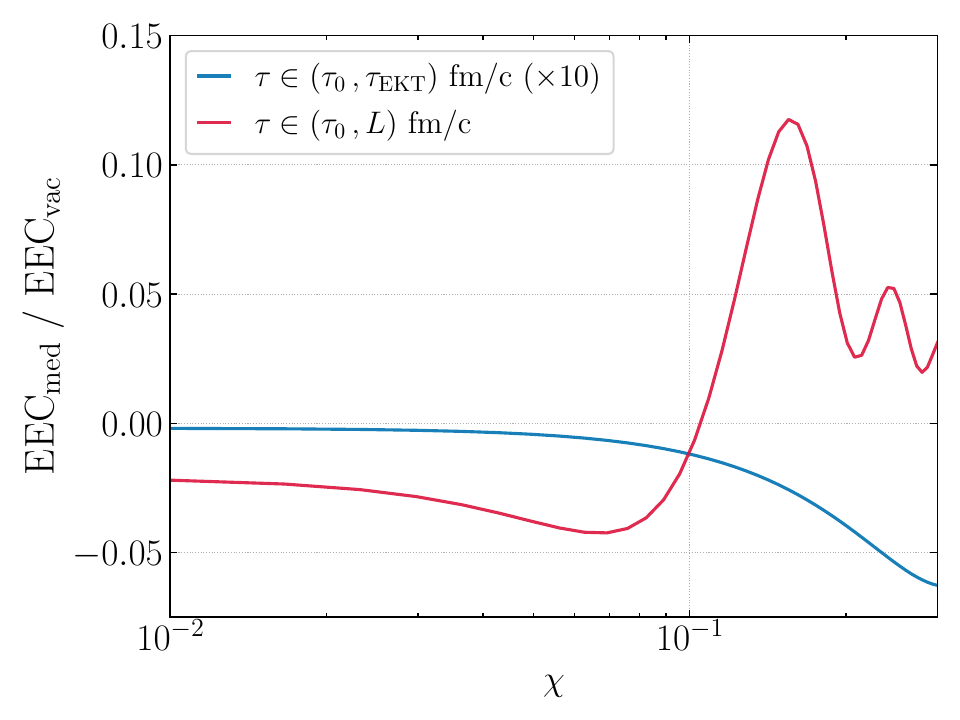}
    \includegraphics[width=0.49\textwidth]{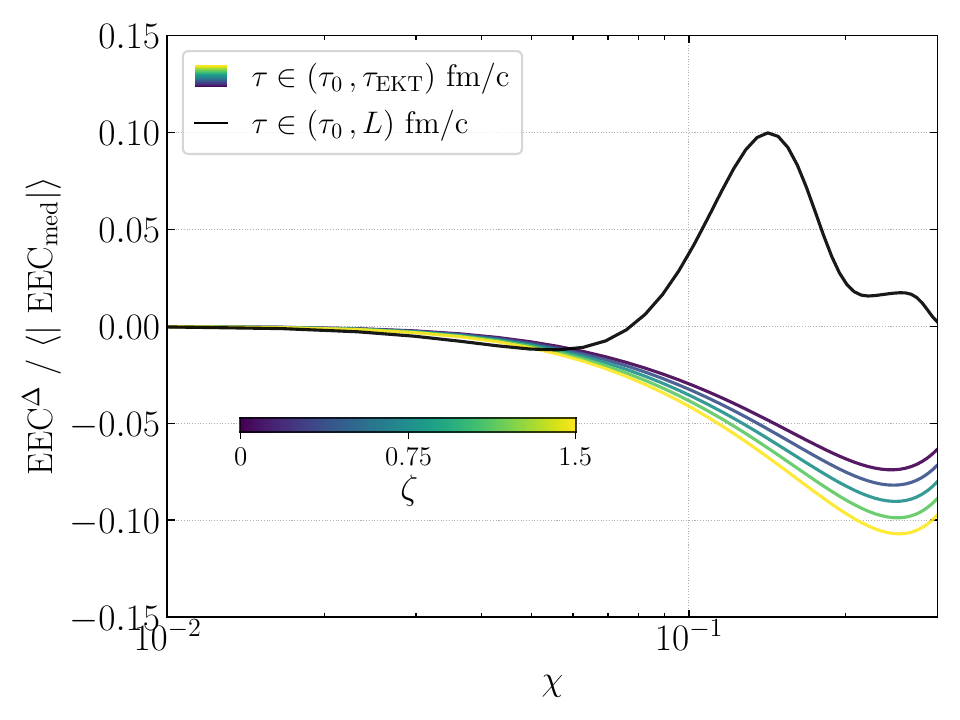}
    \caption{\textbf{Left}: Ratio between the $\Psi$-integrated medium contribution to the EEC, EEC$_{\rm med}$, and the vacuum EEC, EEC$_{\rm vac}$, for the $g\to c\bar c$ splitting. The early-time (blue) curve is multiplied by $10$ for visualisation. \textbf{Right}: Ratio between the clover EEC and the $\chi$-averaged absolute value of EEC$_{\rm med}$ in the range $\chi\in(0.01,0.3)$, for the same splitting. The denominator is divided by $4$ to normalise it to half a hemisphere. The curves are shown for initial time $\tau_0\simeq0.004$ fm/c and for two medium sizes, $\tau_{\rm EKT}=0.3$ fm/c and $L=3$ fm/c.}
    \label{fig:ch3_clover_eec}
\end{figure}
In Fig.~\ref{fig:ch3_clover_eec}, we show, in the left panel, the modification to the $g\rightarrow c\bar c$ EEC due to the medium-induced spectrum, while the right panel displays the resulting azimuthal deformation. The two time intervals correspond to disparately different medium sizes. One includes only the Glasma data and the interpolated transition region ($\tau \in (\tau_0, \tau_{\rm EKT})$, blue curve multiplied by $10$ for visualization on the left and coloured curves on the right), with $\tau_0 \simeq 0.004$ fm/c and $\tau_{\rm EKT} = 0.3$ fm/c, while the other includes all regions ($\tau \in (\tau_0,L)$, red curve on the left and black curve on the right), with the EKT data truncated at $L = 3$ fm/c. More specifically, the left panel shows the leading order result for the ratio between the $\Psi$ integrated, pure medium contribution to the EEC, EEC$_{\rm med}$, and the $\Psi$ integrated vacuum EEC, EEC$_{\rm vac}$. Note that we only plot a single curve, despite there being a family of curves for the anisotropy ratio in the right panel of Fig.~\ref{fig:ch3_qhat_early_encs}, since the $\Psi$-integrated EEC shows little sensitivity to the transition region. At the earliest stage (blue curve), the medium modification is much smaller than the red curve, which includes the EKT data, by an average factor $\sim 20$ for large $\chi$. Further, it is only non-negligible for sufficiently large $\chi$ and is always negative. For the red curve, which includes the EKT data, the medium modification has an average magnitude of a few percent. The non-vanishing small-$\chi$ behavior is expected due to an upper bound on formation time set by the mass of the heavy $c\bar c$ pair in the final state, as discussed in~\cite{Attems:2022otp,Attems:2022ubu,Barata:2025uxp}. For larger $\chi$, one sees an enhancement of the EEC, which is characteristic of medium modifications to energy-energy correlators. Note that the oscillations are not numerical in origin, but rather a consequence of the exponential term in Eq.~\eqref{eq:ch3_spectrum_aniso_time_dependent}.

In the right panel of Fig.~\ref{fig:ch3_clover_eec}, we show the ratio between the clover EEC$^\Delta$, calculated from Eq.~\eqref{eq:ch3_EEC_clover_def} for $g\rightarrow c\bar c$, and the corresponding average value of $|$EEC$_{\rm med}|$ in the range $\chi\in(0.01,0.3)$, allowing one to gauge the magnitude of the directional effect relative to the average magnitude of the overall medium modification to the EEC. First, we observe that this relative directional effect can reach $\sim 10\%$ at its maximum value for the curve containing the full medium extent (black, $\tau \in (\tau_0, L)$) and for the early-stage curve corresponding to the strongest anisotropy profile in Fig.~\ref{fig:ch3_qhat_early_encs} (yellow, $\zeta = 1.5$). Note that the opening angle at which the maximum of EEC$^\Delta$ is attained is different between the black and early-stage curves. Further, although not shown explicitly, for the interval $\tau \in (\tau_0, L)$ there is practically no sensitivity to the transition region, i.e. to the value of $\zeta$, while for the coloured curves this sensitivity is significant. This is expected since, for the early-stage time interval, the proportion of time during which the anisotropy is significant is large compared to the full time interval, which contains the EKT data (see right panel of Fig.~\ref{fig:ch3_qhat_early_encs}). Concerning the sign of EEC$^{\Delta}$, one should note that it is not necessarily indicative of the direction of maximal anisotropy. This is a consequence of ${\rm EEC}^\Delta = {\rm EEC}^h-{\rm EEC}^v = {\rm EEC}^h_{\rm med} - {\rm EEC}^v_{\rm med}$ and, thus, a larger suppression along the horizontal clover results in a negative ${\rm EEC}^\Delta$, but a larger enhancement along the vertical clover also results in the same sign. However, analysing the left panel, we see that there is only suppression for the early-stage curve, while for the curve with the whole time extent one has a suppression followed by an enhancement. Both of these behaviours are consistent with the right panel, such that we can conclude that the magnitude of medium effects is stronger along the horizontal clover, i.e. in the direction of stronger anisotropy as given by the profile in the right panel of Fig.~\ref{fig:ch3_qhat_early_encs}, where most of the time $\hat q_y < \hat q_x$. Finally, we note that both sets of curves tend to zero at sufficiently small $\chi$ and, in particular, the black curve does so much faster than the red curve in the left panel. This is a consequence of ${\rm EEC}^{\Delta} \sim v_2^{\rm EEC} \sim \cO(\chi^3)$, while EEC${\rm med} \sim (v_0^{\rm EEC}-v_{0,\,\rm vac}^{\rm EEC}) \sim \cO(\chi)$ (see Eq.~\eqref{eq:ch3_v2n_twist_expansion}).

\subsection{Toy Monte-Carlo study with JEWEL}

The previous considerations on medium-induced modifications to jet substructure in an anisotropic medium are based on a simplified analytical model for the directional effects and do not account for differences in the jet--medium interaction across the various stages of matter evolution in HICs. Moreover, it is difficult to extend this model calculation beyond leading order in the strong coupling for the computation of energy--energy correlator observables in the HIC context, or to include realistic backgrounds. A complementary approach is therefore to study jet dynamics within a dedicated Monte-Carlo framework, which incorporates partial higher-order radiative effects in the jet evolution, albeit typically at the cost of reduced analytic control and simplified modelling of the underlying jet-quenching dynamics. We emphasize that this Monte-Carlo study is intended to provide a complementary example of directional effects inside jets, not a benchmark or point of comparison for the previous analytical model, which would be highly non-trivial. To this end, we minimally extend the \jewel\,2.3.0\footnote{\jewel\ is a perturbative jet-quenching event generator in which medium-modified cascades are simulated from $2\to2$ interactions between cascade partons and scattering centres embedded in a background medium. Further details and exploration are given Section~\ref{sec:qaa}.}~\cite{Zapp:2013vla,Zapp:2013zya} medium profile to include strong directional effects through an anisotropic temperature profile. Thus, the results shown below should be regarded as a qualitative illustration of the effects expected in realistic simulations or in the phenomenological analysis of experimental data.

\begin{figure}[t]
    \centering
    \includegraphics[width=0.49\columnwidth]{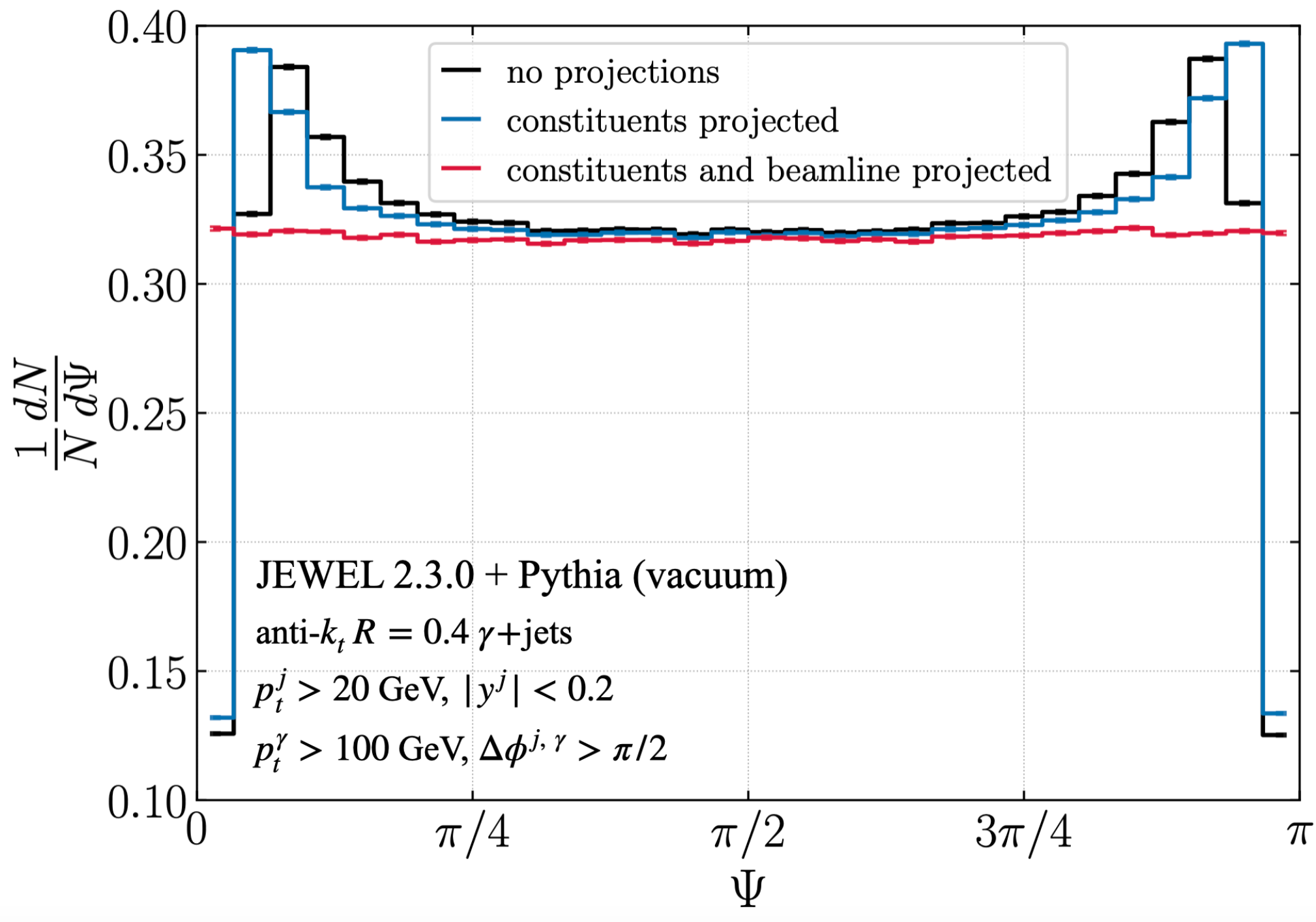}
    \includegraphics[width=0.49\columnwidth]{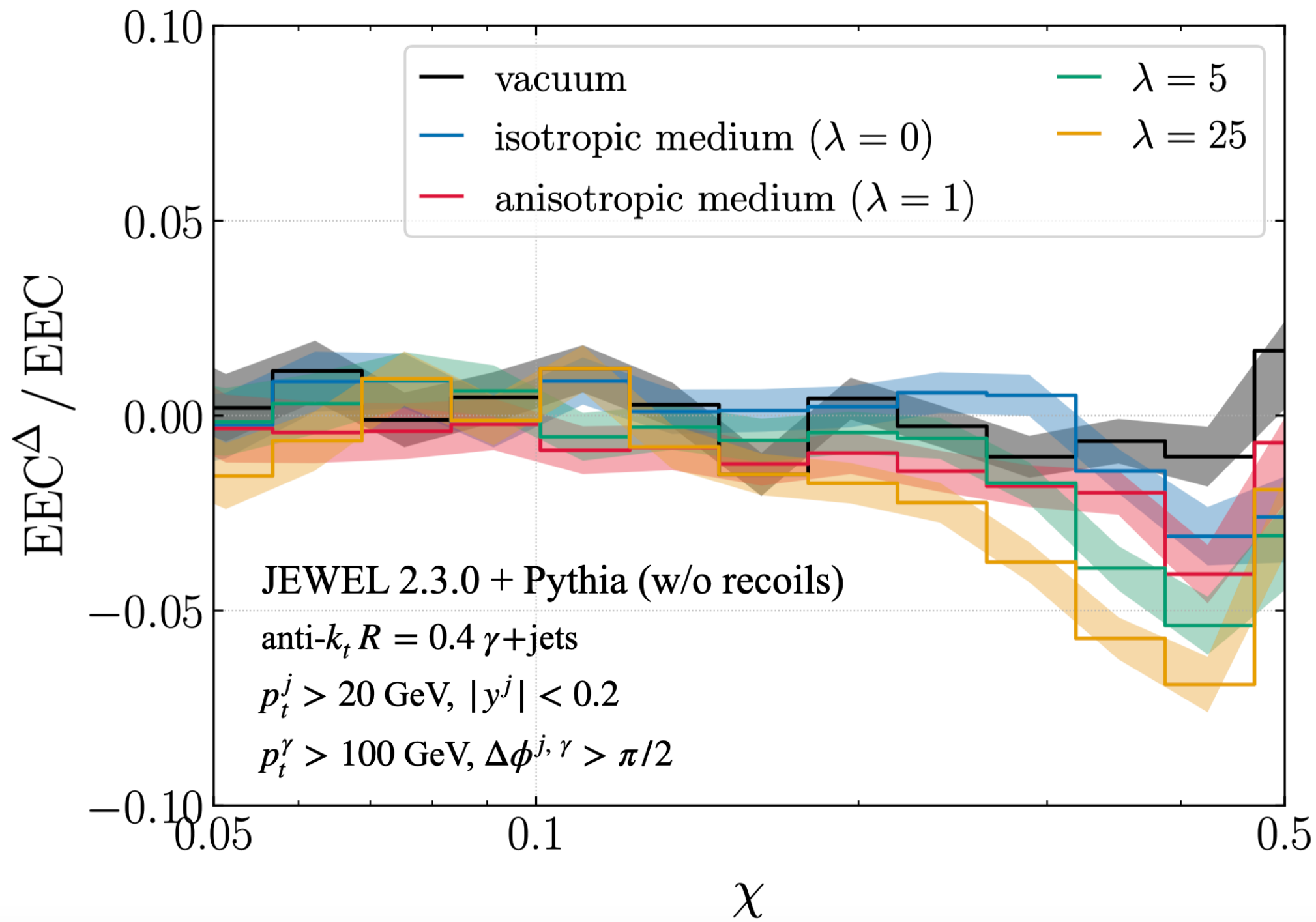}
    \caption{\textbf{Left:} Azimuthal distribution of pairs of jet constituents for vacuum \jewel\,2.3.0\ events. The three distributions correspond to different projections of the constituent momenta and of the beamline in the plane transverse to the jet axis, defining different measures of azimuthal angle $\Psi$.  \textbf{Right:} Ratio between the \textit{clover} energy-energy correlator (EEC$^{\Delta}$) and the $\Psi$ integrated energy-energy correlator (EEC) inside jets reconstructed from \jewel\,2.3.0\ events. The denominator is divided by $4$ to appropriately normalize it to half an hemisphere. The shaded band around each histogram corresponds to the statistical uncertainty.}
    \label{fig:clover_eec_jewel}
\end{figure}

We generated $10^6$ $\gamma+\rm jet$ hadronized events at $\sqrt{s_{NN}} = 5.02$ TeV, with $80$ GeV $ < \hat p_t < 120$ GeV, turning off initial-state radiation, multi-particle interactions and medium recoils (the response of the medium to the jet's passage). These events correspond to the $[0\%-10\%]$ most central collisions, and we set the medium's initial temperature to $T_i = 0.25$ GeV and the Bjorken-flow initialization time to $\tau_i = 0.6$ fm, keeping the remaining medium parameters at their default values. We reconstructed anti-$k_t$, $R=0.4$ jets at mid-rapidity ($|\eta^j| < 0.2$), with $p_T^j>20$ GeV, produced in the direction opposite to a photon with $p_T^\gamma > 100$ GeV. The remaining kinematic cuts are made explicit in Fig.~\ref{fig:clover_eec_jewel}. The modified medium profile implements a temperature gradient along the beam axis, such that the temperature increases with space-time rapidity $\eta_s$ according to a multiplicative factor
\begin{align}\label{eq:F_aniso}
    T(\eta_s) = (1 + \lambda \eta_s^2) T_{\rm default}\,,
\end{align}
where $\lambda = T^{-1}\partial T/\partial \eta_s^2$ at $\eta_s = 0$ controls the strength of the temperature gradients and $T_{\rm default}$ is the default temperature profile in \jewel\, 2.3.0.\footnote{Note that gradients in the transverse plane $(x,y)$ already exist in \jewel\ due to the initial Woods-Saxon transverse profile. Nevertheless, we have checked that, for mid-rapidity jets, these gradients have a minimal effect in comparison to the $z$-gradients coming from the anisotropy implementation in $\eta_s$.} This naturally breaks boost invariance of the underlying background medium. To place the strength of this anisotropy in the context of the medium evolution, we evaluated the temperature gradients along the beam direction near mid-rapidity, $z=0$, at a characteristic hydrodynamization time. This provides a measure of how hydrodynamic the resulting background is, assuming that in the hydrodynamic regime gradients are necessarily small compared to the temperature, while the initial-stage dynamics can exhibit strong anisotropies. The leading nonzero gradient reads
\begin{align}\label{eq:d2T}
    \left.\frac{\partial^2 T}{T^3\partial z^2}\right|_{z=0} =  \frac{1}{3}\frac{1+6\lambda}{f^2T_i^2(\tau_i\tau^2)^{2/3}} , ,
\end{align}
where $f$ characterizes the initial transverse profile of the medium. The characteristic time $t$ can be fixed, for instance, by requiring that $\hat{q}_y/\hat{q}_x$ is sufficiently close to unity in Fig.~\ref{fig:ch3_qhat_early_encs}. Choosing $\tau \sim 3$ fm/c for $\lambda \in (0,1,5,25)$, we find that the characteristic second-order gradient values are of order $\sim (0.09,0.6,3,13)$. Thus, for moderately small values of $\lambda$, the modified background relaxes sufficiently close to the hydrodynamic regime, while larger values lead to more pronounced anisotropies that go beyond the validity of a hydrodynamic description.

In Fig.~\ref{fig:clover_eec_jewel}, in the left panel, we show the self-normalized azimuthal distribution of pairs of jet constituents in vacuum events for several measures of the azimuthal angle. The red histogram corresponds to the azimuthal angle $\Psi$ defined as the angle between the projections of the momentum difference of two jet constituents and of the beamline onto the plane transverse to the jet axis. For the blue histogram, the azimuthal angle is defined as the angle between the momentum difference of two jet constituents projected onto the plane transverse to the jet axis and the unprojected beam direction. Finally, for the black histogram, the azimuthal angle is defined as the angle between the unprojected momentum difference of the jet constituents and the unprojected beam direction. Naturally, these three definitions of azimuthal angle should coincide at vanishing rapidity and in the strict collinear limit. We attribute the peaks in the black and blue curves to effects arising from the underlying spherical geometry at finite rapidity and jet radius. Note that vacuum effects correlated with the direction of the beamline could also be expected, e.g., initial color connections with beam remnants, as well as geometric distortion due to jets being reconstructed in the $(\eta, \phi)$ space rather than $(\theta, \phi)$. However, we expect these effects to be minimal for mid-rapidity jets, as supported by the red histogram, where we observe only a minimally varying azimuthal distribution.

In the right panel of Fig.~\ref{fig:clover_eec_jewel}, we show the EEC$^{\Delta}$ calculated for all pairs of particles inside a jet, with $\Psi$ defined as for the red curve in the left panel. This is further divided by the $\Psi$-integrated EEC, which includes both vacuum and medium-induced contributions. We first notice that the vacuum (black) band is mostly consistent with EEC$^\Delta \sim 0$, reinforcing that the expected vacuum effects previously alluded to are suppressed for sufficiently small rapidity jets. The increasing trend for $\chi \geq R=0.4$ is due to the finite boundary of the reconstructed jets, and it is common to all bands. Then, for $\lambda=0$ we have default \jewel\ with no added anisotropy (blue band, \textit{isotropic}) and for the remaining bands we show increasingly larger values of $\lambda$, corresponding to larger anisotropies. All bands only significantly deviate from zero for sufficiently large opening angles $\chi$. They exhibit, for most of the large-$\chi$ range, a monotonic behaviour with $\lambda$ --- the larger the anisotropy, the stronger the signal in EEC$^\Delta$. Note that for the $\lambda = 0$ case one has EEC$^\Delta \neq 0$, indicating there is some directional effect even in the default \jewel\ configuration. One can verify this by considering Eq.~\eqref{eq:d2T} for $\lambda = 0$, where the second derivative at $z=0$ is non-vanishing. Nevertheless, the points at larger $\lambda$ generate larger EEC$^\Delta$. Note that extracting information about the direction of anisotropy from EEC$^{\Delta}$ is, in this case, more involved than in the analytical example shown in Fig.~\ref{fig:ch3_clover_eec}. The reason is that, due to the temperature gradient, jet constituents at sufficiently large rapidity are not only more strongly broadened along the beamline, but also subject to enhanced suppression. This can lead to a larger fraction of the jet's energy being distributed along the vertical clover, despite the stronger longitudinal broadening. Because of this non-trivial interplay, as well as significant model differences, we refrain from making comparative statements between these curves and the colored curves in Fig.~\ref{fig:ch3_clover_eec}. % file "Anisotropies.tex"
\cleardoublepage

%%%%%%%%%%%%%%%%%%%%%%%%%%%%%%%%%%%%%%%%%%%%%%%%%%%%%%%%%%%%%%%%%%%%%%%%
%  Jet_Energy_Loss.tex
%  Chapter 4: Jet Energy Loss
%%%%%%%%%%%%%%%%%%%%%%%%%%%%%%%%%%%%%%%%%%%%%%%%%%%%%%%%%%%%%%%%%%%%%%%%
\chapter{Jet energy loss beyond suppression}
\label{chap:jet_energy_loss}

\section{Defining energy loss}
\label{sec:jel_intro}

Jet suppression is one of the main experimental manifestations of jet quenching in HICs, and its interpretation is directly tied to the question of how much energy a jet loses while traversing the medium. As discussed in Chapter~\ref{chap:jet_quenching}, energetic partons propagating through the medium exchange energy and momentum with it, and radiate medium-induced gluons whose angular distribution differs from that of accompanying vacuum-like emissions. These mechanisms reduce the energy carried by the perturbative jet, while the medium response induced by its passage can contribute with additional soft particles that can later be reconstructed inside the jet cone. Thus, the energy assigned to a jet depends both on the physical redistribution of energy between the shower and the medium and on the definition of the measured object. This point becomes especially important for jet substructure observables, which go beyond inclusive high transverse momentum ($p_T$) yields and probe increasingly differential correlations in the energy flow of the shower. Interpreting such observables requires quantifying how energy loss modifies both the measured energy flow and the ensemble of jets selected at fixed reconstructed momentum. However, energy loss is not an unambiguous quantity, and the most useful definition depends on the application.

For phenomenological studies based on reconstructed jets, it is natural to organize energy loss at the level of the measured jet object. A jet is defined by an algorithm, a radius parameter, constituent cuts and a background-subtraction prescription. Even in vacuum, a finite-radius jet captures only part of the radiation associated with the hard scattering. In a heavy-ion collision, the relevant energy loss is then the additional energy missing from the in-medium reconstructed jet relative to an appropriate vacuum reference. This includes energy transported outside the jet cone by medium-induced broadening and radiation, energy shifted below the constituent threshold, and the partial recovery of the deposited energy in the form of jet-correlated soft particles from the medium response inside the reconstructed jet. The inferred energy loss is thus an analysis-dependent reconstructed quantity, sensitive to the jet radius, the particle-$p_T$ threshold and the treatment of medium response. This is the notion of jet energy loss used in Section~\ref{sec:qaa}. There, it is the shift in reconstructed jet transverse momentum between an in-medium jet sample and a vacuum sample selected to correspond as closely as possible to the same underlying hard process. This is not an event-by-event microscopic energy loss, but rather a population-level shift in reconstructed jet $p_T$ induced by the medium. Such a measure is needed because comparisons at fixed final reconstructed $p_T$ can otherwise mix genuine modifications of jet substructure with the selection bias induced by energy-loss migration.

For field-theoretic descriptions of observables that probe the energy flow of the full cascade, it is useful to define energy loss in terms of the total energy transferred from the jet cascade to the medium. In this case, energy loss is identified as the part of the initial cascade energy that is transferred into degrees of freedom that eventually thermalize with the background. The object of interest is then not primarily a finite-radius reconstructed jet, but the fraction of the original parton energy that remains in cascade-originated final-state particles. Section~\ref{sec:elf} develops this second perspective using energy-loss functions. These functions describe the probability distribution for the fraction of the initial parton energy retained by jet-originated hadrons, in analogy with charged track functions in vacuum jet physics. The non-perturbative information is encoded in the energy-loss function itself, while its scale dependence can still be computed within perturbative QCD. This provides a way to account for energy loss at the level of the full partonic cascade in observables sensitive to correlated energy flow, including multi-point energy correlators.

\section{Jet energy loss bias and quantile matching}
\label{sec:qaa}

\subsection{Quantile matching}
\label{subsec:qaa_quantile_matching}

The standard procedure to compare vacuum and in-medium jet samples consists of selecting a window in reconstructed jet transverse momentum and calculating a given observable for both populations of jets. If the observable depends, either explicitly or implicitly, on jet $p_T$, this comparison is biased. The key problem is that it ignores that medium jets migrate to lower reconstructed $p_T$ as a result of energy loss. Thus, comparisons performed at fixed final reconstructed momentum can confound medium-induced modifications with changes that arise from comparing jets that were originated by hard partons with different transverse momenta. Thus, an estimate of jet energy loss is needed to reduce the resulting bias in the comparison of jet populations.

Experimentally, the nuclear modification factor $R_{AA}$ is useful to quantify jet suppression and qualitative discuss jet energy loss, but it cannot directly quantify the latter. The main reason is that it takes the ratio of AA over pp cross sections at the same reconstructed $p_T$, not taking $p_T$ migration into account. The resulting bias is exacerbated by the steeply falling nature of the jet spectrum, to which $R_{AA}$ is sensitive. Thus, if two jet spectra have significantly different values for steepness, for instance inclusive jet and boson+jet samples, and a fixed energy loss is imposed for all jets, then the less steep spectrum presents a smaller suppression, i.e. an $R_{AA}$ closer to unity. Hence, the same energy loss can lead to different values of $R_{AA}$ purely due to differences in the spectrum inclination. Another problem caused by $p_T$ migration is selection bias~\cite{Connors:2017ptx}: samples of in-medium jets selected in a given reconstructed-$p_T$ window are biased towards jets that were less modified, due to the steepness of the spectrum. Both problems can be partially evaded in boson+jet events~\cite{Brewer:2021hmh}, because the boson provides a reference for the transverse momentum of the hard parton that originated the recoiling jet. One can then compare pp and AA jets produced with the same hard scale. However, the statistical sample of boson+jet events is reduced in comparison to the inclusive jet sample, which motivates a procedure directly applicable to inclusive jet spectra.\footnote{The problem of $p_T$-migration bias has also been addressed in the context of energy-correlator-based observables for inclusive jet events~\cite{Andres:2024hdd}.}

Quantile matching, as introduced in~\cite{Brewer:2018dfs} and further explored in~\cite{Apolinario:2024apr}, provides such a procedure by establishing a correspondence between momenta in the vacuum and quenched spectra. Let us first define the probability of a jet having transverse momentum within a given interval $(p_{T,1},p_{T,2})$ through the cumulative jet cross section
\begin{align}
    \Sigma(p_{T,1},p_{T,2})
    =
    \int_{p_{T,1}}^{p_{T,2}} dp_T\,
    \frac{d\sigma}{dp_T}\, .
\end{align}
The matching procedure establishes a correspondence between a given transverse momentum $p_T^v$ in the pp jet spectrum and a transverse momentum $p_T^q$ in the AA jet spectrum by equating appropriately normalized upper cumulative cross sections,
\begin{align}
    \Sigma^{pp}(p_T^v,+\infty)
    =
    \Sigma^{AA}(p_T^q(p_T^v),+\infty)\, .
    \label{eq:quantile_matching}
\end{align}
Equivalently, one obtains a quantile function $p_T^q(p_T^v)$ such that AA jets with $p_T>p_T^q(p_T^v)$ occur with the same probability as pp jets with $p_T>p_T^v$. This construction admits a direct interpretation in an idealized limit. If a jet with a given transverse momentum always loses the same energy $\epsilon(p_T)$, and if this energy loss either decreases or grows sufficiently slowly with $p_T$, namely $d\epsilon/dp_T<1$, then the difference $p_T^v-p_T^q(p_T^v)$ is equal to $\epsilon(p_T^v)$~\cite{Brewer:2018dfs}. This follows from the conservation of the number of jets when jet transverse momenta are exactly and identically ordered in pp and AA. In realistic samples, the energy lost by jets originating from hard partons with the same $p_T$ can vary considerably~\cite{Milhano:2015mng,Rajagopal:2016uip,Brewer:2017fqy,Casalderrey-Solana:2016jvj}. The probabilistic nature of QCD branching and of the interaction of partons with the medium, together with the varying amount of medium traversed, all contribute to disperse the energy loss around some mean value~\cite{He:2018gks} and away from the ideally ordered scenario. Thus, in \cite{Brewer:2018dfs} the authors studied the accuracy of
\begin{align}
    1-Q_{AA}(p_T^v) = 1-\frac{p_T^q(p_T^v)}{p_T^v} \, ,
\end{align}
as a proxy for the \emph{average} fractional jet energy loss. This quantity should be interpreted as a population-level proxy for reconstructed jet energy loss, not as an event-by-event microscopic energy loss.

A crude analytic estimate of $Q_{AA}$ can be obtained under simplifying assumptions.\footnote{Detailed analytical calculations of $Q_{AA}$ are not required for the present phenomenological discussion. Energy-loss estimates using specific models were obtained in~\cite{Takacs:2021bpv}.} Assume that the jet spectrum in pp is described by a single power law,
\begin{align}
    \frac{d\sigma^{pp}}{dp_T}(p_T)
    =
    A\,p_T^{-n}\, ,
    \label{eq:assumption1}
\end{align}
where $A$ is a normalization constant, and that the quenched jet spectrum can be obtained by assigning an energy loss $\epsilon$ to each jet in the pp spectrum with probability $P(\epsilon,p_T)$~\cite{Baier:2001yt},
\begin{align}
    \frac{d\sigma^{AA}}{dp_T}(p_T)
    &=
    \int_0^{+\infty} d\epsilon\,
    P(\epsilon,p_T)
    \frac{d\sigma^{pp}}{dp_T}(p_T+\epsilon)
    \nn
    &\approx
    A\left(p_T+\langle\epsilon\rangle(p_T)\right)^{-n}\, .
    \label{eq:assumption2}
\end{align}
In the last step, the fractional energy loss $\epsilon/p_T$ is taken to be small, so that $P(\epsilon,p_T)$ is negligible unless $\epsilon/p_T\ll 1$, and $n$ is assumed to be sufficiently large, $n\sim 5$. The result is then expanded to first order in $\langle\epsilon\rangle(p_T)/p_T$, with
\begin{align}
    \langle\epsilon\rangle(p_T)
    =
    \int_0^{+\infty} d\epsilon\,
    P(\epsilon,p_T)\,\epsilon\, .
\end{align}
Therefore, the upper cumulative jet cross sections entering Eq.~\eqref{eq:quantile_matching} are
\begin{align}
    \Sigma^{pp}(p_T^v,+\infty)
    &=
    \int_{p_T^v}^{+\infty} dp_T\,
    \frac{d\sigma^{pp}}{dp_T}
    =
    \frac{A}{n-1}(p_T^v)^{1-n}\, ,
    \label{cumulative_xs_pp}
    \\
    \Sigma^{AA}(p_T^q,+\infty)
    &=
    \int_{p_T^q}^{+\infty} dp_T\,
    \frac{d\sigma^{AA}}{dp_T}
    \approx
    \frac{A}{n-1}
    \left(p_T^q+\langle\epsilon\rangle(p_T^q)\right)^{1-n}\, ,
    \label{cumulative_xs_AA}
\end{align}
where $\langle\epsilon\rangle(p_T)$ is further assumed to vary slowly with $p_T$ compared with $p_T^{-n}$, while $n$ is taken to be constant. Inserting Eqs.~\eqref{cumulative_xs_pp} and~\eqref{cumulative_xs_AA} into Eq.~\eqref{eq:quantile_matching} gives
\begin{align}
    1-\frac{p_T^q(p_T^v)}{p_T^v}
    =
    1-Q_{AA}(p_T^v)
    \approx
    \frac{\langle\epsilon\rangle(p_T^v)}{p_T^v}\, ,
    \label{eq:qaa_estimate}
\end{align}
where $\langle\epsilon\rangle(p_T^q)\approx \langle\epsilon\rangle(p_T^v)$ was used, corresponding to the zeroth-order term in the expansion of $\langle\epsilon\rangle$ in powers of $p_T^v-p_T^q$. This confirms that $1-Q_{AA}$ is a proxy for the average fractional energy loss as a function of $p_T$. In this simplified limit, the proxy is independent of the spectral slope for a $p_T^{-n}$ spectrum with constant $n$, and assumes small relative energy loss with mild $p_T$ dependence.

A local form of the integral equation in Eq.~\eqref{eq:quantile_matching} can be obtained by differentiating with respect to $p_T^v$:
\begin{align}
    \frac{dp_T^q}{dp_T^v}
    \frac{d\sigma^{AA}}{dp_T^q}
    \left(p_T^q(p_T^v)\right)
    =
    \frac{d\sigma^{pp}}{dp_T^v}(p_T^v)\, .
    \label{eq:diff_qaa}
\end{align}
This gives a first-order differential equation for $p_T^q(p_T^v)$. A determination of $p_T^q(p_T^v)$ based on Eq.~\eqref{eq:diff_qaa} is potentially more sensitive to statistical uncertainties in measured spectra. In addition, although Eq.~\eqref{eq:diff_qaa} does not require integration up to arbitrarily large $p_T$, unlike Eq.~\eqref{eq:quantile_matching}, it remains sensitive to the choice of $p_T^q$ at a given momentum. This is because Eq.~\eqref{eq:diff_qaa} is a first-order differential equation for which an initial condition must be specified. This issue will be relevant in Section~\ref{subsec:qaa_measurement}, where the finite experimental cutoff of the spectrum is discussed.

The left-hand side of Eq.~\eqref{eq:diff_qaa} was first introduced in~\cite{PHENIX:2004vcz} in the context of obtaining a simple estimate of energy loss from the suppression of the $\pi^0$ spectrum.\footnote{Note that, in Eq.~(29) of \cite{PHENIX:2004vcz}, the AA spectrum differential in vacuum momentum $p_T'$ and evaluated at a medium momentum $p_T$ ($dn/p_T'$ in that work) is taken as equal to the pp spectrum differential in $p_T'$ evaluated at $p_T'$ (i.e. $\propto 1 \, / \, (p_T')^{n-1}$ in that work).} It was used to quantify the number of particles in a bin of $p_T$ after suppression. The Jacobian factor $dp_T^q/dp_T^v$ accounts for the change in the relative particle density per $p_T$ bin as a consequence of induced energy loss. Determining $p_T^q$, either through Eq.~\eqref{eq:diff_qaa} or through Eq.~\eqref{eq:quantile_matching}, extends this logic by imposing that the number of in-medium jets in a given $p_T^q$ bin, scaled by the Jacobian, is the same as the number of vacuum jets in a given $p_T^v$ bin, under the ideal ordering assumption described above. A relation analogous to Eq.~\eqref{eq:diff_qaa} was used in~\cite{ATLAS:2023iad} as an experimental proxy for jet energy loss. The two coincide in the limit $dp_T^q/dp_T^v\to 1$, corresponding to slowly varying energy loss with $p_T$. In that limit, the two equations determine similar quantities, up to the assignment
\begin{align}
    p_T^q(p_T^v)
    =
    p_T^v-\Delta p_T(p_T^v)
    =
    p_T^v\left(1-S_{\rm loss}(p_T^v)\right)\, .
\end{align}
This comparison will be discussed in more detail in Section~\ref{subsec:qaa_measurement}. The equation used in~\cite{ATLAS:2023iad} is also a differential equation and therefore, like Eq.~\eqref{eq:diff_qaa}, requires an initial condition to be specified.\footnote{We were unable to identify the initial condition chosen in \cite{ATLAS:2023iad}.}

\subsection{Monte-Carlo setup and baseline nuclear effects}
\label{subsec:qaa_setup_nuclear}

Let us now apply the previously introduced quantile function in the context of Monte Carlo generation of heavy-ion events. We will use samples generated with \jewel\ 2.3~\cite{Zapp:2012ak,Zapp:2013vla}, and with the Hybrid Model\footnote{We thank the authors of~\cite{Bossi:2024qho} for kindly providing us with these samples.}~\cite{Casalderrey-Solana:2014bpa,Casalderrey-Solana:2015vaa,Casalderrey-Solana:2016jvj,Hulcher:2017cpt,Casalderrey-Solana:2018wrw} for model comparisons in part of the results. In both models, the medium mainly describes the QGP phase (see Section~\ref{subsec:st_evolution_HIC}), so we will use medium and QGP interchangeably in the next subsections.

\jewel\ is a Monte Carlo event generator that describes jet evolution in vacuum and in a medium, including the QCD evolution of hard partons and their scatterings with the medium in a perturbative framework. \jewel\ interleaves the final-state parton shower with scatterings in a QCD medium, with leading-logarithmic correct relative contributions for elastic and inelastic scattering, so that vacuum-like and medium-induced emissions are treated within the same perturbative language. The interplay of these two types of radiation is controlled by formation times: all emissions are allowed to occur, but the one with the longest formation time is discarded. With this prescription, and using virtuality as the ordering variable, radiative energy loss is included together with the usual vacuum parton shower. Additionally, \jewel\ implements destructive interference between emissions induced by successive medium scatterings through a probabilistic formulation of the QCD Landau--Pomeranchuk--Migdal (LPM) effect.

\jewel\ models the medium as the boost-invariant longitudinal expansion of an ideal quark--gluon gas. It can store the four-momenta of the partonic constituents of the medium that were scattered, the recoils, together with their initial-stage thermal counterparts in the event record. This information is used to perform a background subtraction that improves the quantitative treatment of medium response and allows one to reconstruct a heavy-ion jet including medium response while discarding uncorrelated background. Here, the event-wise algorithmic procedure introduced in~\cite{Milhano:2022kzx} is used to subtract the background before jet reconstruction. The cutoff distance used for subtraction is set to $\Delta R_{ij}=0.5$. All other \jewel\ parameters related to recoiling partons are kept at their default values.

In the Hybrid Model~\cite{Casalderrey-Solana:2014bpa,Casalderrey-Solana:2015vaa,Casalderrey-Solana:2016jvj,Hulcher:2017cpt,Casalderrey-Solana:2018wrw}, jet evolution inside a droplet of QGP is described by separating the strongly coupled dynamics of the QGP and of the jet--QGP interactions, both governed by the medium temperature scale $T$, from the weakly coupled dynamics that controls the production and showering of high-energy partons. Since the initial virtuality of the parton that showers and forms a jet is much larger than any scale associated with the medium, such as its temperature, the model assumes that its evolution proceeds as in vacuum, i.e. that the parton-shower branching is not modified by the presence of the plasma. More specifically, the evolution is given by the parton shower implemented in PYTHIA 8.244~\cite{Sjostrand:2014zea} and includes initial-state radiation. Thus, the modification of jets is entirely due to the interaction of each parton in the jet with the strongly coupled medium. The rate of parton energy loss is modeled using the energy loss of light-quark jets in the strongly coupled plasma of $N=4$ supersymmetric Yang--Mills theory (SYM), which has been computed holographically~\cite{Chesler:2014jva,Chesler:2015nqz}. This rate is applied to each parton of the already developed shower, using a formation-time estimate for the path length travelled by each parton. Moreover, it is assumed that all differences between the strongly coupled limit of $N=4$ SYM theory and QCD can be encoded into a single parameter, which is fitted to jet and hadron $R_{AA}$ data~\cite{Casalderrey-Solana:2018wrw}. The QGP resolution length, i.e. the angular scale below which two jet partons lose energy as a single parton, is set to zero.

Each event is embedded within a droplet of QGP that expands and cools according to relativistic viscous hydrodynamics, providing an evolving space-time background averaged over collisions within a given centrality class. The holographic energy-loss rate is evaluated in the local rest frame of this fluid and at its local temperature. The jet propagates through the expanding and cooling medium until it reaches a region where the temperature has dropped below a critical temperature $T_c$, after which no further energy loss occurs. The value of this temperature is varied within the range $145<T_c<170$ MeV, thereby gauging theory uncertainties of the model.

The energy and momentum deposited by the energetic shower partons in the medium excite a wake, which then evolves hydrodynamically. When the QGP and the jet-induced wake reach the freeze-out temperature, they are hadronized through the Cooper--Frye prescription~\cite{PhysRevD.10.186}. This prescription ensures that, by boosting the fluid in the direction of the wake, the jet enhances the production of soft particles along its direction and depletes it in the opposite direction, relative to what would be obtained without a jet wake. In the Hybrid Model, when the resulting Cooper--Frye distribution is negative, it is implemented by adding hadrons with this distribution but with negative energy. These ``negative particles'' are subtracted with the same procedure used for the thermal momenta in \jewel.

\begin{figure}[t]
     \centering
         \includegraphics[width=0.49\textwidth]{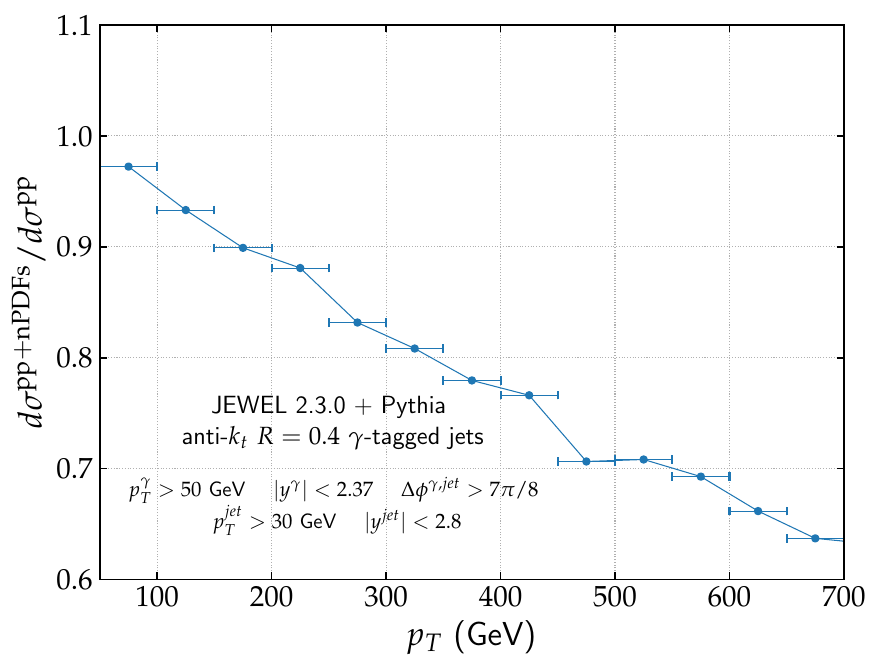}
         \label{fig:nucleareffects_spec_yj}
         \centering
         \includegraphics[width=0.49\textwidth]{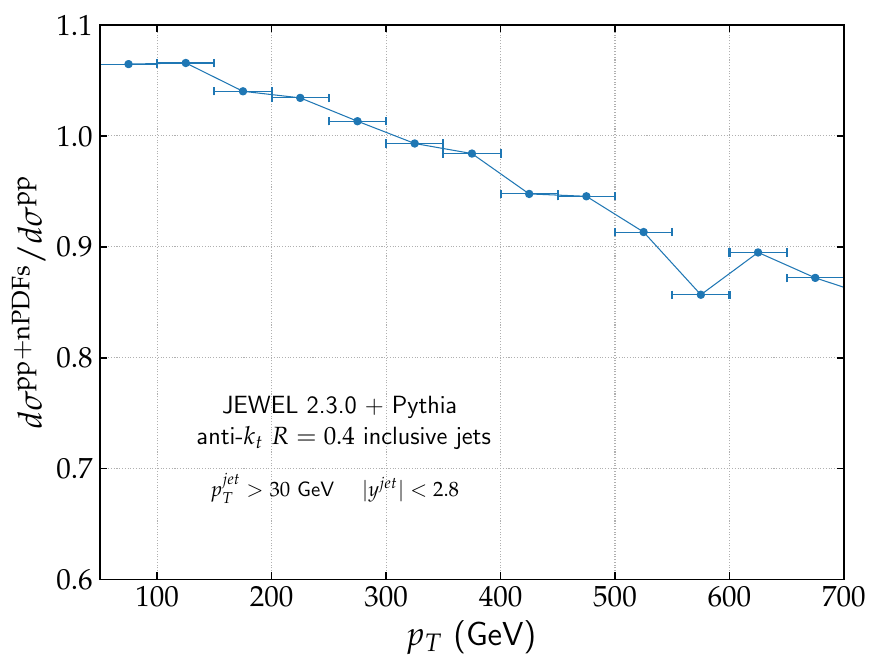}
         \label{fig:nucleareffects_spec_jj}
         \centering
         \includegraphics[width=0.49\textwidth]{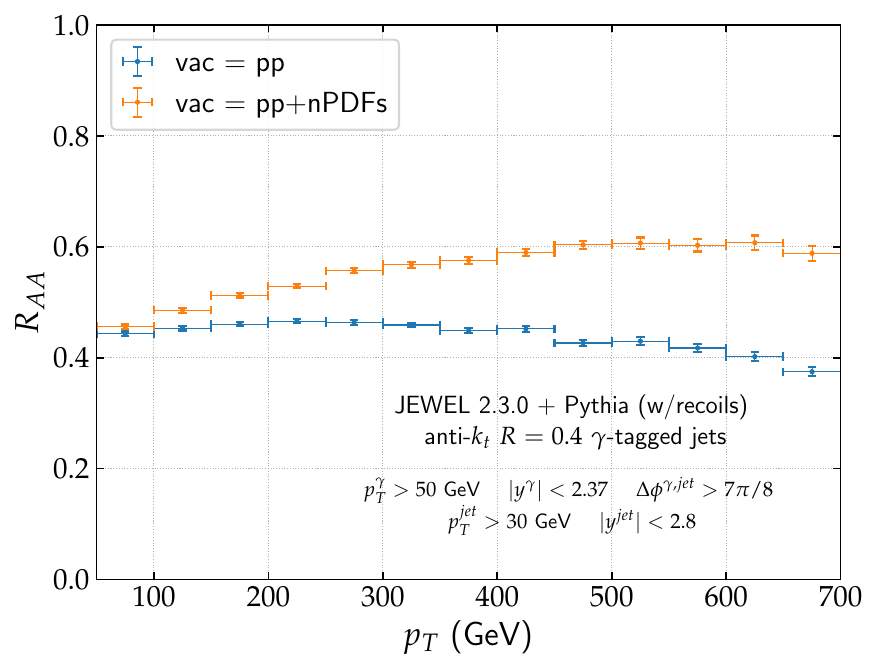}
         \label{fig:nucleareffects_raa_yj}
         \centering
         \includegraphics[width=0.49\textwidth]{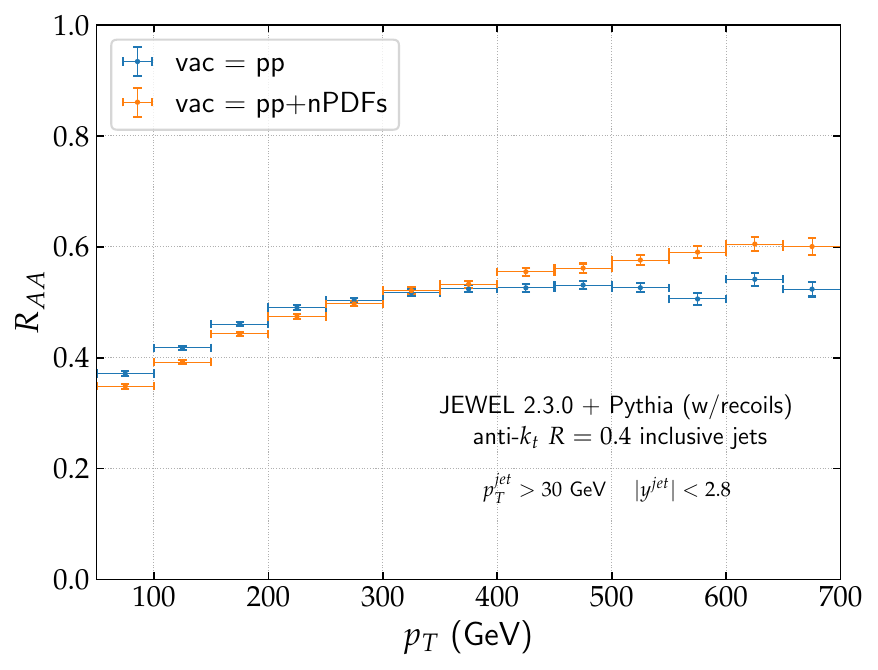}
         \label{fig:nucleareffects_raa_jj}
     \caption{\textbf{Top:} Sensitivity of the vacuum $\gamma$-jet (\textbf{left}) and inclusive-jet (\textbf{right}) spectra to nuclear effects. \textbf{Bottom:} Sensitivity of $R_{AA}$ to nuclear effects. The spectra labeled with pp+nPDFs result from average-nucleon--average-nucleon collisions generated with isospin-averaged nuclear PDFs.}
     \label{fig:nucleareffects}
\end{figure}

The \jewel\ samples were generated for $\sqrt{s}=5.02$ TeV, for both pp and PbPb collisions in the $0$--$10\%$ centrality class. These samples consist of $10^6$ dijet events and $10^6$ $\gamma+$jet events. For both dijet and $\gamma+$jet events, the EPPS16NLO nuclear PDFs~\cite{nPDFs} are used for PbPb collisions. For dijet events, the CT14NLO proton PDFs~\cite{PDFs} are used for pp collisions in most of the analysis. All PDF sets are provided by LHAPDF~\cite{lhapdf}. The particles produced in the hard scattering have a minimum transverse-momentum cutoff of $20$ GeV, and the medium is generated with $\tau_i=0.6$ fm, $T_i=0.55$ GeV and $T_c=0.17$ GeV. Initial-state radiation and hadronization are handled by the PYTHIA 6~\cite{Sjostrand:2006za} framework underlying \jewel.

Importantly, pp samples for $\gamma+$jet events are generated with the EPPS16NLO nPDFs, which are isospin averaged, as initial conditions. Thus, instead of the usual pp collisions, vacuum samples are obtained from the unphysical scenario of average-nucleon--average-nucleon collisions with no QGP. This choice attempts to isolate the role of quenching effects in the differences between vacuum and medium samples by minimizing the role of nuclear effects, including isospin~\cite{Li:2019dre} and nPDFs, which is significant for $\gamma+$jet events~\cite{ATLAS:2023iad}. Note that, for PbPb events, both \jewel\ and the Hybrid Model generate only a single average-nucleon--average-nucleon collision per event. The cross-section estimate per event already includes this, since it is weighted by the isospin-averaged nuclear PDFs. To illustrate the role of nuclear effects on the inclusive-jet and $\gamma+$jet spectra, the top panels of Fig.~\ref{fig:nucleareffects} show the ratio between vacuum jet spectra obtained from events generated with isospin-averaged nPDFs and events generated with proton PDFs. Nuclear effects are non-negligible both in the absolute magnitude of the cross section and in the shape of the spectrum. This effect is especially pronounced for $\gamma+$jets, where the isospin effect is expected to be larger since, at leading order, the hard matrix elements always include an initial-state quark or anti-quark. The bottom panels of Fig.~\ref{fig:nucleareffects} show the impact of nuclear effects on an observable such as the jet $R_{AA}$. At large $p_T$ and for $\gamma+$jets, this leads to a factor $\sim 1.5$ difference. Since nuclear effects are not as significant for dijet events, dijet vacuum samples obtained from simple pp collisions are used for most of the analysis. The exception is when comparisons are drawn with $\gamma+$jet results in Section~\ref{subsec:qaa_radius_threshold_colour}, where the same nuclear baseline is used for consistency.

The Hybrid Model samples were generated for $\sqrt{s}=5.02$ TeV, for both pp and PbPb collisions in the $0$--$5\%$ centrality class. These samples contain approximately $10^6$ dijet events. The EPS09LO nuclear PDFs~\cite{Eskola:2009uj} are used for PbPb collisions, while the NNPDF2.3 PDFs are used for pp collisions. Importantly, the particles produced in the hard scattering have a minimum transverse-momentum cutoff of $100$ GeV. Further details on the generated samples can be found in~\cite{Bossi:2024qho}.

The analysis is performed as follows. In $\gamma+$jet events, the hard-scattered photon is identified solely based on its $p_T$: the highest-$p_T$ photon from each event passing $p_T\geq 50$ GeV and $|y|\leq 2.37$ is chosen. Since \jewel\ does not implement QED radiation off either initial- or final-state quarks, most photons not coming from the hard scattering originate from $\pi^0$ decay. The photon tagging is performed before background subtraction in PbPb events, keeping the photon out of the subtraction procedure. After the subtraction procedure, for both types of events, the final-state particles are filtered according to $p_T\geq 0.5$ GeV, unless otherwise stated as in Section~\ref{subsec:qaa_radius_threshold_colour}, and $|y|\leq 3.3$, and are then given as input to the anti-$k_t$ algorithm~\cite{Cacciari:2008gp} for jet reconstruction with varying jet radius, as implemented in \texttt{FASTJET}~\cite{fastjet}. All jets have a minimum cut of $p_T^{jet}\geq 30$ GeV in \jewel\ and $p_T^{jet}\geq 130$ GeV in the Hybrid Model, and are required to have $|y^{jet}|\leq 2.8$. The $\gamma$-tagged jets are required to be azimuthally separated by $\Delta\phi\geq 7\pi/8$ from the hard photon. For results using dijet events, the jet sample is inclusive.

\subsection{Validation with photon-tagged jets}
\label{subsec:qaa_validation}

We first validate the quantile procedure as an adequate proxy for jet energy loss away from the zero-dispersion, ideally ordered case described in Section~\ref{subsec:qaa_quantile_matching}. The validation uses $\gamma+$jet events, where the transverse momentum of the photon provides a proxy for the transverse momentum of the parton that initiated the jet found back-to-back with it. As discussed in Section~\ref{subsec:qaa_setup_nuclear}, nuclear effects, namely isospin and nuclear PDFs, significantly alter the shape of the PbPb spectrum for $\gamma+$jet events. The corresponding vacuum samples are thus generated with isospin-averaged nuclear PDFs as initial conditions.

\begin{figure}[t]
     \centering
         \includegraphics[width=0.49\textwidth]{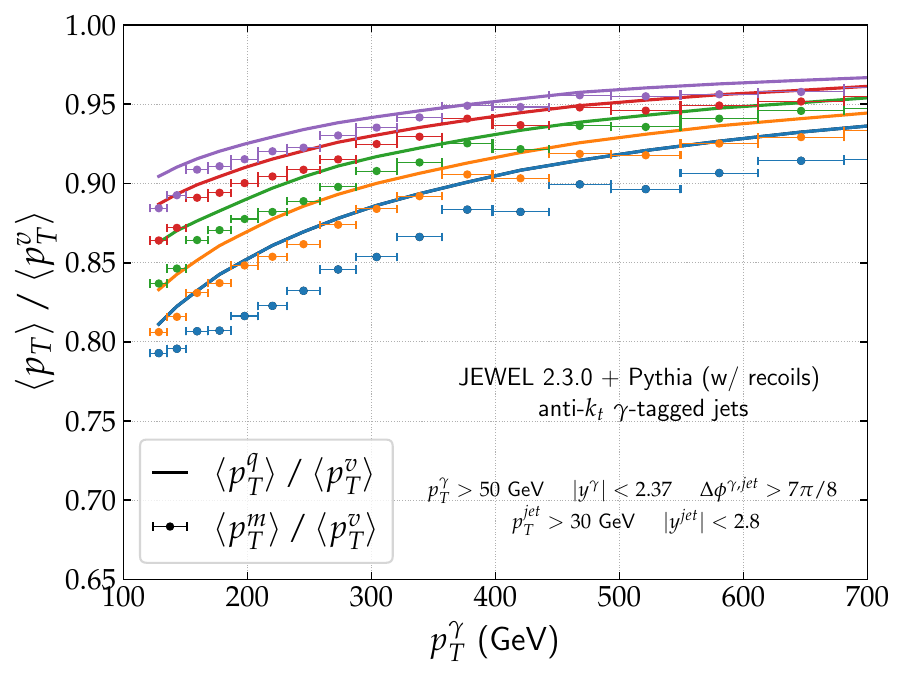}
         \label{fig:qaa_validation_r_left}
         \centering
         \includegraphics[width=0.49\textwidth]{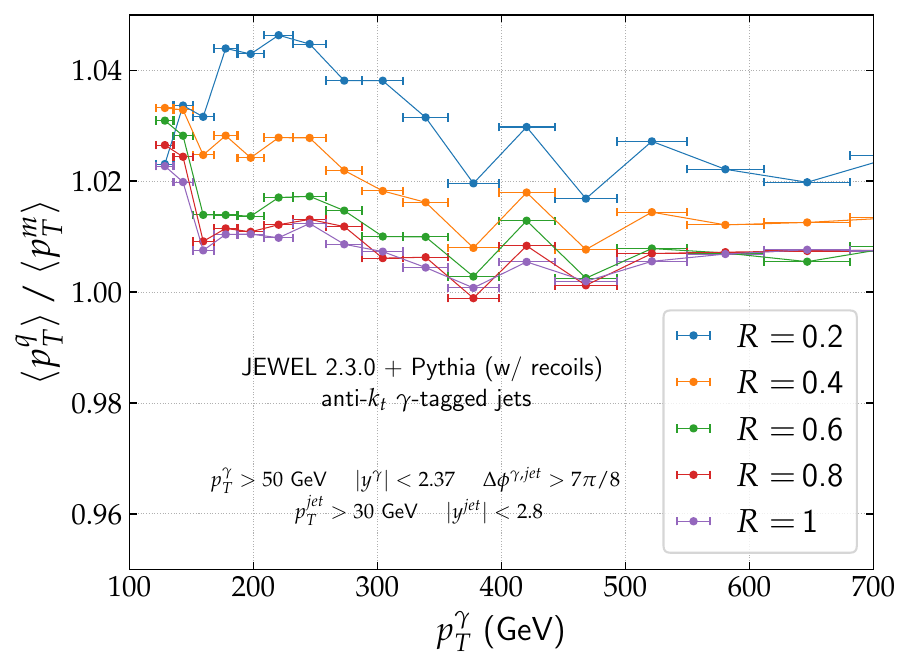}
         \label{fig:qaa_validation_r_right}
     \caption{\textbf{Left}: Dependence of average jet-energy-loss proxies on jet radius $R$: ratio of mean values of reconstructed $p_T$ (dots) and ratio of average quantile $p_T$ to mean vacuum $p_T$ (solid). \textbf{Right}: Ratio between mean $p_T^q$ and mean $p_T^m$. Vacuum samples include nuclear effects, implemented through isospin-averaged nuclear PDFs.}
     \label{fig:qaa_validation_r}
\end{figure}

The validity of the quantile procedure was previously confirmed for $R=0.4$ jets produced back-to-back with a $Z$ boson in the absence of medium-response contributions to the jet~\cite{Brewer:2018dfs}. Here, the validation is extended to the more realistic case in which medium response is accounted for, as described in Section~\ref{subsec:qaa_setup_nuclear}, and to multiple jet radii. The reasoning is the following. For a given photon transverse momentum $p_T^{\gamma}$ produced in the hard scattering, a distribution of jet transverse momenta is obtained, denoted by $p_T^v$ for vacuum samples and $p_T^m$ for medium samples. The ratio between the mean values of these distributions, $\langle p_T^m\rangle/\langle p_T^v\rangle$ (dots in the left panel of Fig.~\ref{fig:qaa_validation_r}), gives a proxy for average jet energy loss, since one is comparing jets originated by hard partons with the same transverse momentum but developed either in vacuum or in medium. Note that this is not the same as the average of the ratio of these two quantities, which is inaccessible both in \jewel\ and in experimental data. The mean $p_T$ values are defined as
\begin{align}
    \left.\langle p_T^{v/m}\rangle\right|_{p_T^{\gamma}}
    =
    \int dp_T\,
    \left.\frac{dN^{v/m}}{dp_T}\right|_{p_T^{\gamma}}
    p_T\, ,
    \label{eq:av_ptv}
\end{align}
where $\left.dN^X/dp_T\right|_{p_T^{\gamma}}$ is the normalized transverse-momentum distribution of jets for a given bin of photon transverse momentum $p_T^{\gamma}$.

On the other hand, the quantile procedure determines $p_T^q=p_T^q(p_T^v)$ for a given $p_T^v$, which should relate pp and AA jets that, on average, originated from hard partons with the same transverse momentum. One can then take the distribution of vacuum jets for a given $p_T^{\gamma}$ and construct a hypothetical medium-$p_T$ distribution by applying $p_T^q(p_T^v)$. The average of this distribution is
\begin{align}
    \left.\langle p_T^{q}\rangle\right|_{p_T^{\gamma}}
    =
    \int dp_T\,
    \left.\frac{dN^{v}}{dp_T}\right|_{p_T^{\gamma}}
    p_T^{q}(p_T)\, .
    \label{eq:av_ptq}
\end{align}
The ratio $\langle p_T^q\rangle/\langle p_T^v\rangle$ (solid curves in the left panel of Fig.~\ref{fig:qaa_validation_r}) is a proxy for the average jet energy loss for a given photon transverse momentum $p_T^{\gamma}$, and should be similar to $\langle p_T^m\rangle/\langle p_T^v\rangle$ (dots in the same panel). Again, this is not the same as the mean ratio of $p_T$ values. In this case, that would correspond to an average value of $Q_{AA}$, which, although very similar to the quantile proxy shown here, is not the correct quantity for this validation. In the left panel of Fig.~\ref{fig:qaa_validation_r}, the agreement between the two average jet-energy-loss proxies is reasonable, although the quantile procedure systematically gives a smaller average energy loss than the estimate based on the photon $p_T$.

In the right panel of Fig.~\ref{fig:qaa_validation_r}, the comparison between $\langle p_T^{q}\rangle$ and $\langle p_T^{m}\rangle$ is shown directly. A quantile procedure that perfectly quantified the correspondence between mean medium and vacuum jet transverse momenta would give a ratio exactly equal to unity. This ratio approaches unity for larger jet radii and larger jet $p_T$, saturating at about a $1\%$ difference. The largest deviation, which does not exceed $\sim 5\%$, occurs for $R=0.2$ jets. Medium response is a large source of fluctuations, pushing the system away from the ideally ordered scenario. Looking at the left panel of Fig.~\ref{fig:qaa_validation_r}, one can also see how jet energy loss varies with jet radius, for radii from $0.2$ to $1$. Using the information from the photon $p_T$, the ratio $\langle p_T^m\rangle/\langle p_T^v\rangle$ already shows that the reconstructed fractional energy loss decreases as $R$ is increased. Although this is a model-dependent statement, as discussed in Section~\ref{subsec:qaa_radius_threshold_colour}, it implies that larger jets lose a smaller fraction of their energy on average in this setup. This can be interpreted as medium response and the recapture of medium-induced radiation compensating for the fact that increasing the jet radius increases the number of jet components that can lose energy. Both proxies show the same evolution with jet radius. The same study was performed without medium response, and the agreement between proxies was maintained. Therefore, only the more interesting physical case with medium response is shown here.\footnote{In fact, this result led us to further conclude that the size of fluctuations introduced by medium response is reduced compared with that of the fluctuations intrinsic to the energy loss of a partonic cascade.}

\subsection{Dependence on radius, constituent threshold and colour charge}
\label{subsec:qaa_radius_threshold_colour}

In the following, we use $Q_{AA}$ as a proxy for average fractional jet energy loss and investigate its evolution with jet radius, its sensitivity to medium response and to the minimum particle $p_T$, and, finally, its dependence on the colour charge of the initiating parton. Only in the colour-charge study are $\gamma+$jet samples used. The remaining two studies are based on inclusive jet samples.

\subsubsection{Jet radius}
\label{subsubsec:qaa_jet_radius}

\begin{figure}[!t]
     \centering
     \includegraphics[width=0.49\textwidth]{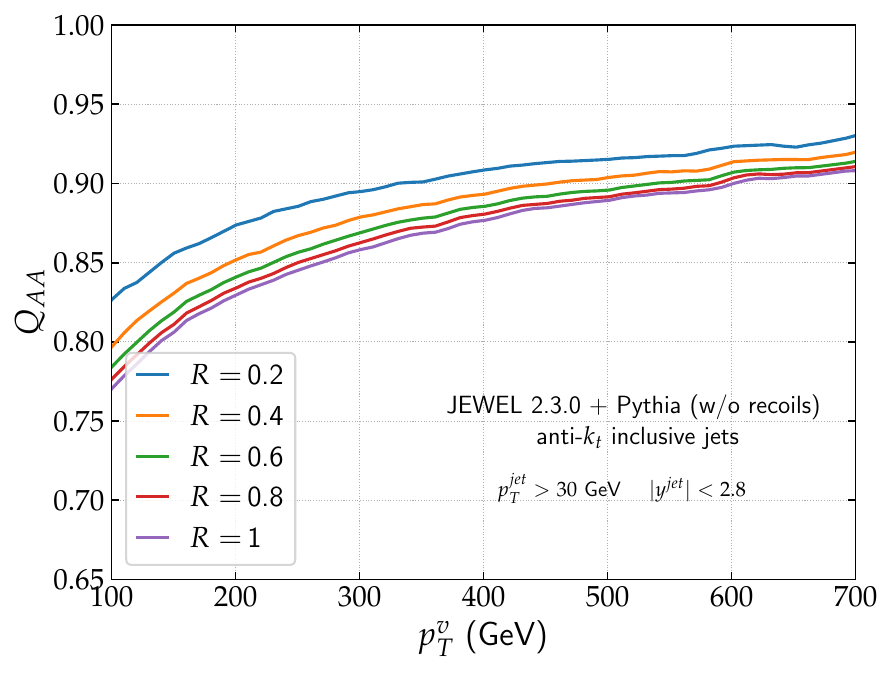}
     \includegraphics[width=0.49\textwidth]{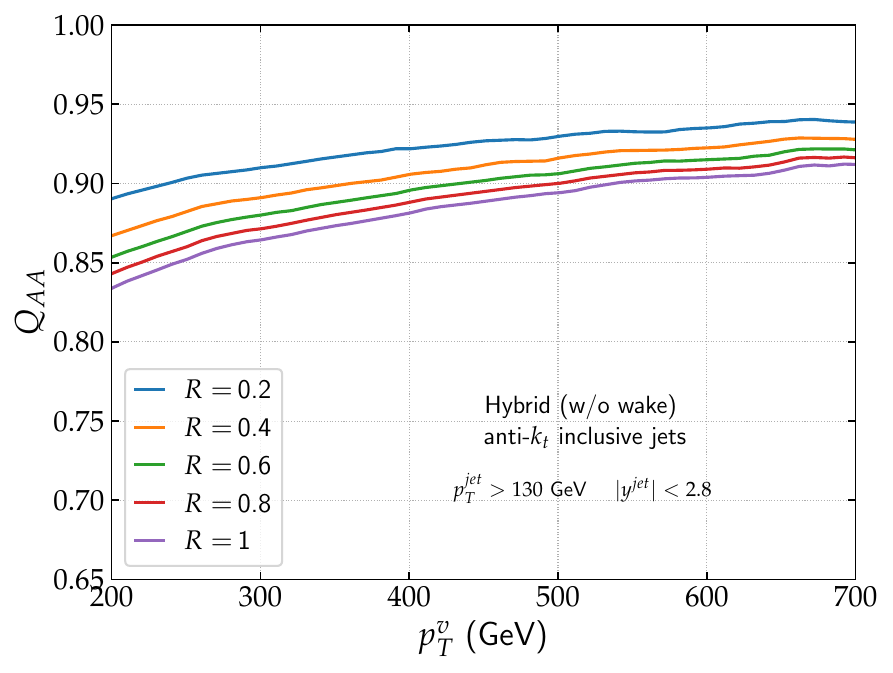}

     \includegraphics[width=0.49\textwidth]{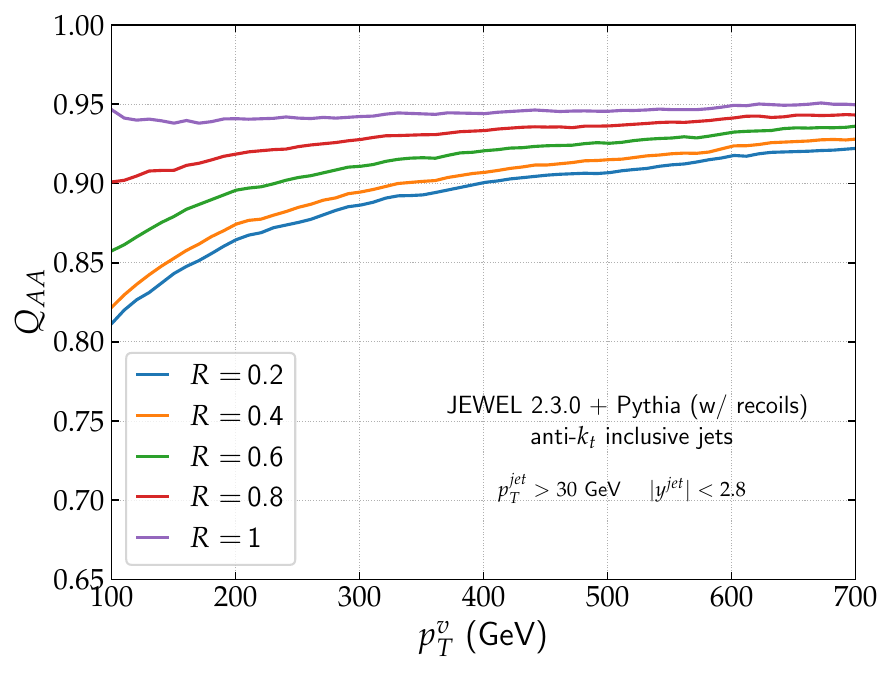}
     \includegraphics[width=0.49\textwidth]{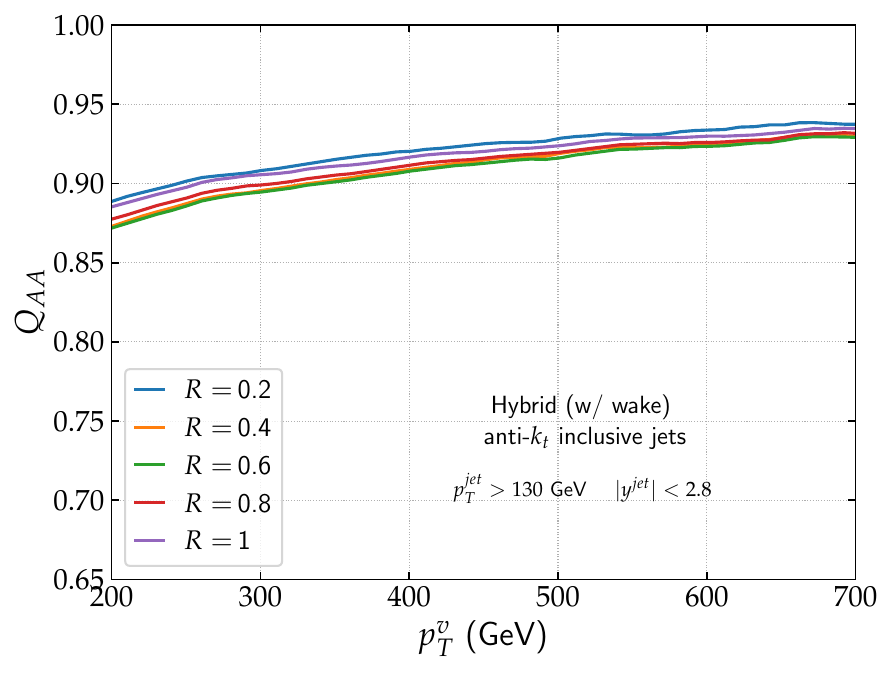}

     \caption{Dependence of $Q_{AA}$ on jet radius for \jewel\ 2.3 (\textbf{left}) and for the Hybrid Model (\textbf{right}), with (\textbf{bottom}) and without (\textbf{top}) medium response. Vacuum samples come from pp collisions.}
     \label{fig:qaa_model}
\end{figure}
In Fig.~\ref{fig:qaa_model}, we show the evolution of $Q_{AA}$ with jet radius for inclusive jet samples with (bottom panels) and without (top panels) medium response. To evaluate model dependence, we compare two radically different implementations of jet physics in the presence of a QGP: \jewel\ 2.3~\cite{Zapp:2012ak,Zapp:2013vla} (left) and the Hybrid Model~\cite{Casalderrey-Solana:2014bpa,Casalderrey-Solana:2015vaa,Casalderrey-Solana:2016jvj,Hulcher:2017cpt,Casalderrey-Solana:2018wrw} (right). Note that the curves for the Hybrid Model start at $200$ GeV. Despite the different implementations of the parton--QGP interaction in these models, the trend of $Q_{AA}$ with jet radius and $p_T$ is the same when events without medium response are considered, as shown in the top panels of Fig.~\ref{fig:qaa_model}. Even in absolute value, the two models predict similar results. The ordering of $Q_{AA}$ with jet radius is such that larger jets lose a larger fraction of their energy, on average. This implies, in particular for \jewel, that the increase in jet components that can lose energy dominates over the recapture of medium-induced radiation. When medium response is included (bottom panels in Fig.~\ref{fig:qaa_model}), however, the models predict different orderings with jet radius and only agree approximately in absolute value for large enough $p_T$ and moderate radii, $R\sim0.4,0.6$. For \jewel, the behaviour with jet radius is exactly inverted with respect to the case where medium response is absent, which is naturally attributed to the reconstruction of part of the energy lost to the medium inside the jet. This had already been observed for jet suppression, as quantified by $R_{AA}$, in~\cite{KunnawalkamElayavalli:2017hxo}. Nevertheless, the results shown for $Q_{AA}$ allow this statement to be made quantitatively with regard to average jet energy loss. In the Hybrid Model, the rate of energy recovery from reconstructing part of the wake inside the jet is not very pronounced. Thus, for small to moderate jet radii, the energy recovered from the wake does not compensate the increase in lost energy due to the larger number of jet components. This only happens for sufficiently large radii, $R\gtrsim0.6$, where the ordering with jet radius is inverted as in \jewel.

Regarding the sensitivity of $Q_{AA}$ to the jet radius, in the absence of medium response, \jewel\ predicts a radius dependence similar to that of the Hybrid Model for jet radii $R\lesssim0.6$. For larger jet radii, the dependence is milder in \jewel. When medium response is included, the radius dependence in \jewel\ is significantly stronger than in the Hybrid Model. This difference in behaviour can be explained by different rates of energy recapture. Without medium response, jets in the Hybrid Model have no way of recovering lost energy, so increasing the jet radius only increases the number of components of the vacuum parton shower that experience energy loss. On the other hand, in \jewel, the radius dependence starts saturating at intermediate jet radii, $R\sim0.6$, because of the recapture of medium-induced emissions. When medium response is included, \jewel\ jets recover energy at a significantly faster rate with increasing radius than jets in the Hybrid Model.\footnote{Note that \jewel's rate of energy recovery may be exaggerated due to the value of the subtraction parameter $\Delta R_{ij}$ being set to $0.5$. For jet radii sufficiently larger than this value, one may be undersubtracting. In fact, we have checked that this rate is slowed down when increasing $\Delta R_{ij}$ to $1$. Despite this, all statements regarding \jewel\ and Hybrid Model comparisons remain unaltered.} In~\cite{Milhano:2022kzx}, it was shown that \jewel's medium response is overpopulated in the semi-hard portion of the particle spectrum, possibly as a consequence of the recoiling medium particles not undergoing any shower evolution or re-scattering. In contrast, in the Hybrid Model, medium response is composed mostly of very soft particles and spreads to large distances~\cite{Casalderrey-Solana:2016jvj}. This means that the portion of lost energy that the jet can recover from the wake is limited, and that one must significantly increase the jet radius to recapture a substantial amount of that energy.

\subsubsection{Minimum particle \texorpdfstring{$p_T$}{pT}}
\label{subsubsec:qaa_ptmin}

\begin{figure}[!t]
     \centering
     \includegraphics[width=0.49\textwidth]{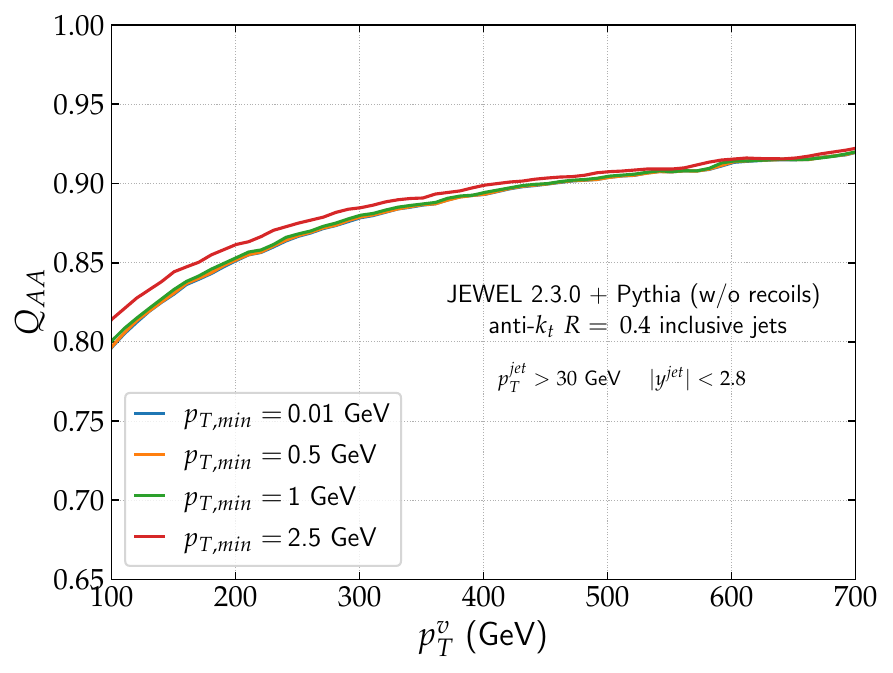}
     \includegraphics[width=0.49\textwidth]{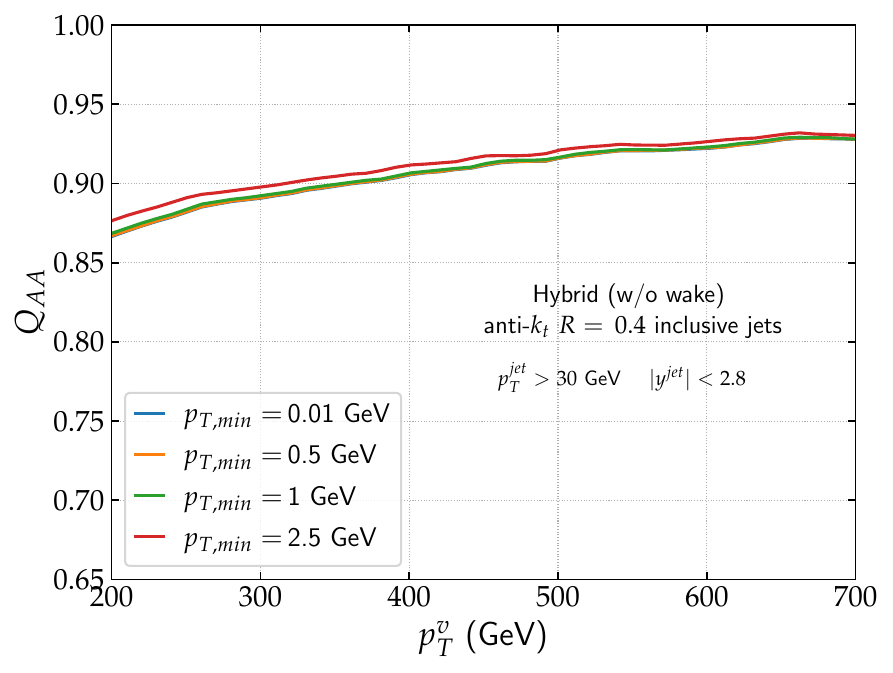}

     \includegraphics[width=0.49\textwidth]{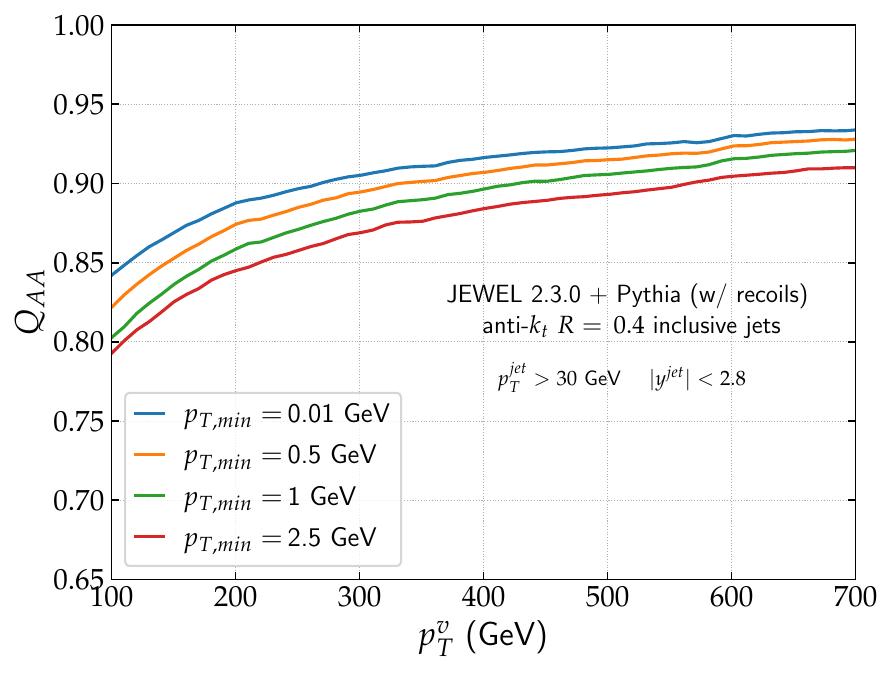}
     \includegraphics[width=0.49\textwidth]{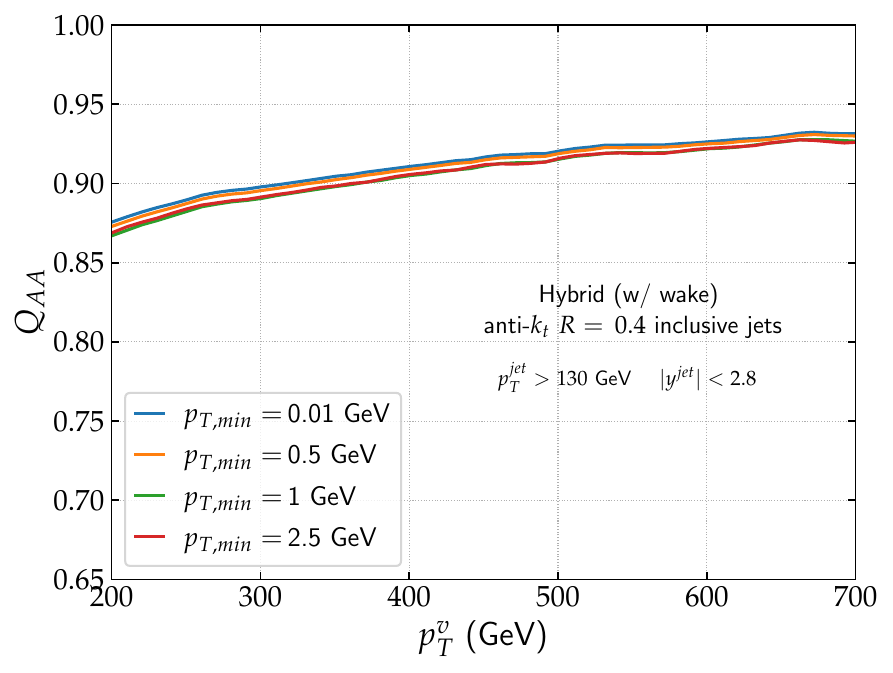}

     \includegraphics[width=0.49\textwidth]{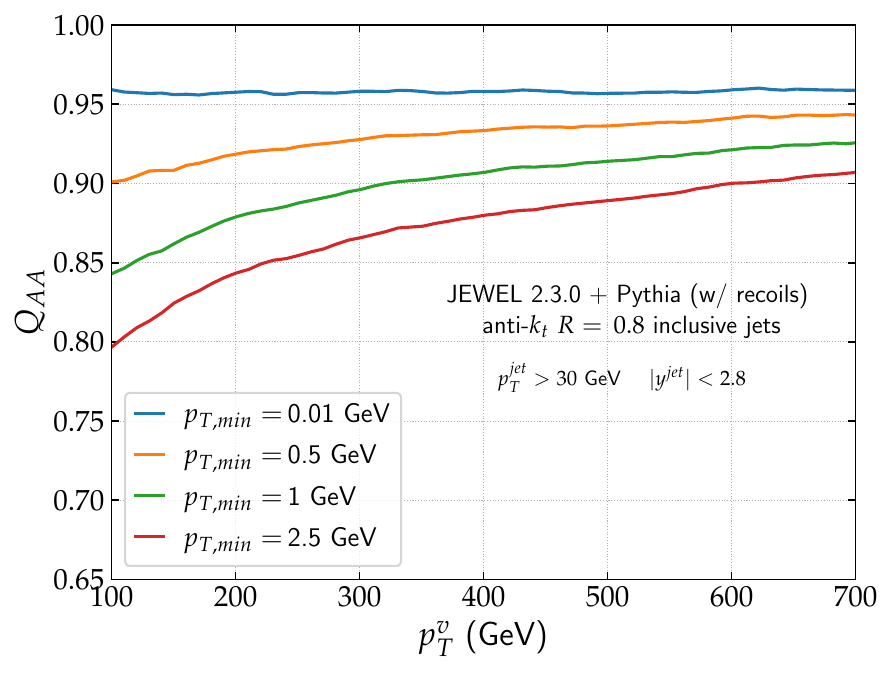}
     \includegraphics[width=0.49\textwidth]{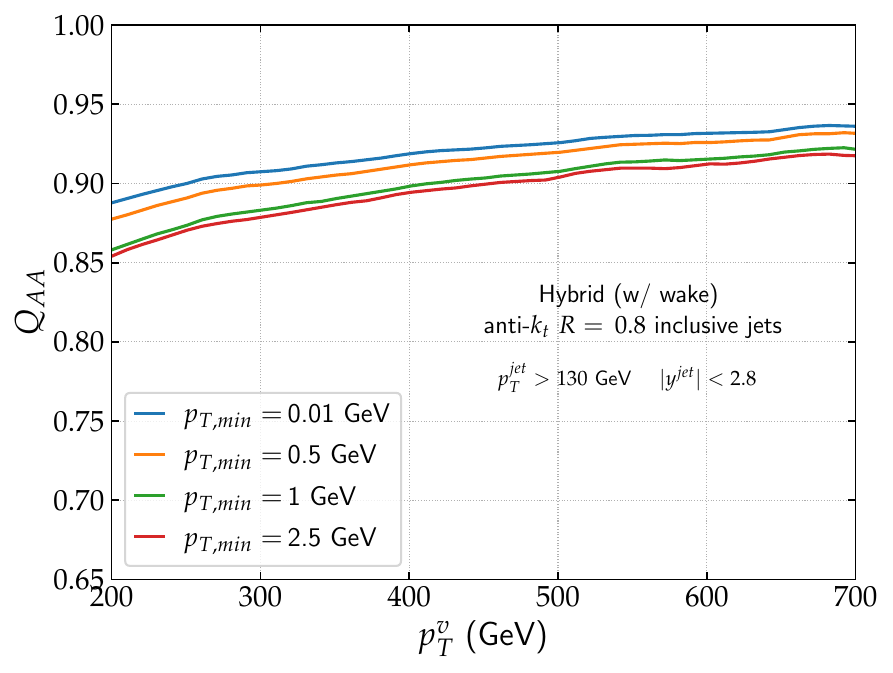}

     \caption{$Q_{AA}$ sensitivity to different minimum required $p_T$ values of final-state particles for \jewel\ 2.3 (\textbf{left}) and for the Hybrid Model (\textbf{right}). Vacuum samples are pp collisions.}
     \label{fig:ptmin_sens}
\end{figure}

Heavy-ion jets have a larger soft component than their pp counterparts, since both medium response and parton energy loss contribute to increasing the number of soft particles within the reconstruction cone. As a result, as shown in Fig.~\ref{fig:ptmin_sens}, when medium response is included, increasing the minimum particle-$p_T$ cut increases the energy lost by jets. This effect is very small for $R=0.4$ jets in the Hybrid Model, owing to the large angular spread of the wake. Naturally, the increase is greater for larger jet radii. This is because a larger jet is in principle more populated in the softest part of the particle spectrum, and therefore suffers a larger relative energy cut. This radius dependence is much more pronounced in \jewel\ than in the Hybrid Model, consistently with what was observed in Section~\ref{subsubsec:qaa_jet_radius}. As seen in~\cite{Casalderrey-Solana:2016jvj}, the implementation of medium response in the Hybrid Model underestimates the yield of semi-hard particles, leading to an excess of soft particles. This explains why, especially for $R=0.8$, there is a jump from the curve corresponding to a minimum $p_T$ of $0.5$ GeV to the one corresponding to $1$ GeV, which appears to be the range where the $p_T$ of most medium-response particles lies.

\begin{figure}[!t]
     \centering
     \includegraphics[width=0.49\textwidth]{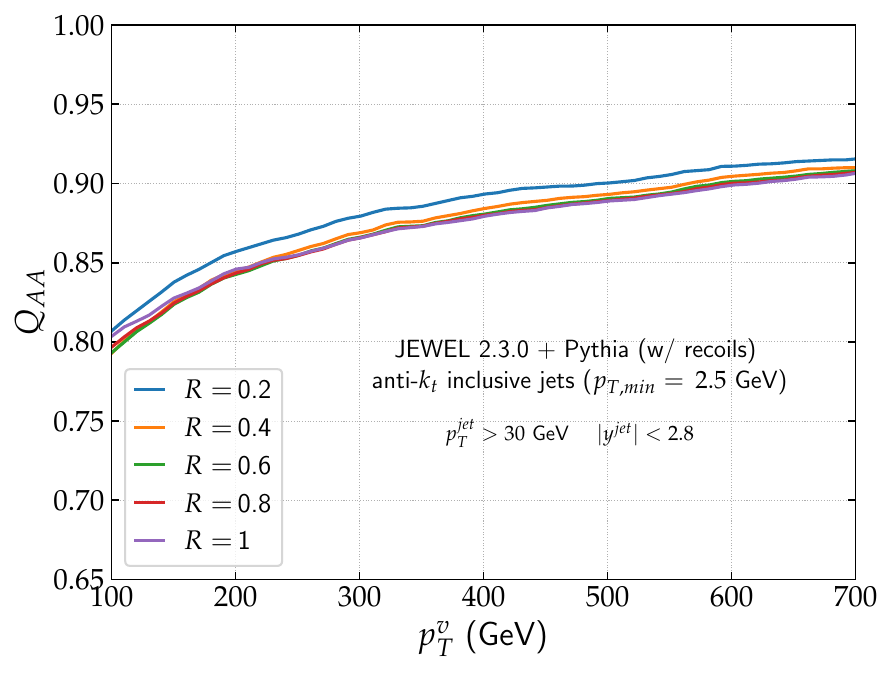}
     \includegraphics[width=0.49\textwidth]{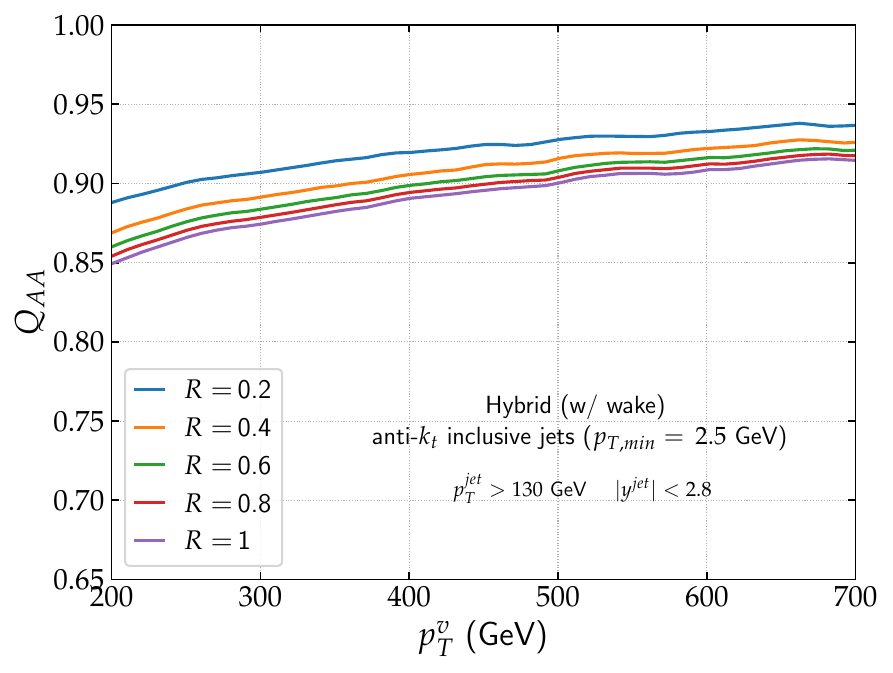}

     \caption{Evolution of $Q_{AA}$ with jet radius for \jewel\ 2.3 (\textbf{left}) and for the Hybrid Model (\textbf{right}), for an extreme value of the minimum required $p_T$ of final-state particles, $p_{T,\min}=2.5$ GeV. Medium response is included in both models. Vacuum samples are pp collisions.}
     \label{fig:radius_ptmin_sens}
\end{figure}

In the absence of medium response, the sensitivity to the minimum particle $p_T$ is almost nonexistent unless one goes to very large values of this cut. This can be seen from the curve corresponding to $p_T=2.5$ GeV, which corresponds to a smaller energy loss than the remaining values of the minimum $p_T$. Naturally, this arises from the particle constituents of a jet having a harder spectrum in vacuum than in medium. Interestingly, at such a large value of the minimum particle $p_T$, and when medium response is included, the jet-radius dependence of $Q_{AA}$, shown in Fig.~\ref{fig:radius_ptmin_sens}, is significantly reduced for both models. It is almost washed out in \jewel\ (left), except for the smaller jet radii $R\lesssim0.4$. In this case, the Hybrid Model (right) predicts a $Q_{AA}$ with a radius dependence resembling what was observed for jets in the absence of medium response in the top-right panel of Fig.~\ref{fig:qaa_model}. This is expected, since at such a large cut most of the contribution from medium response is discarded.

\subsubsection{Colour charge}
\label{subsubsec:qaa_colour_charge}

We now address the dependence of jet energy loss on the colour charge of the parton, quark or gluon, that initiates a jet. Both perturbative radiative~\cite{Baier:1996kr,Baier:1998yf,Baier:1998kq,Zakharov:1996fv,Zakharov:1997uu,Gyulassy:2000fs,Gyulassy:2000er,Wang:2001ifa,Majumder:2009zu,Arnold:2001ms,Arnold:2002ja} and collisional~\cite{Wicks:2005gt} partonic energy losses are proportional to the Casimir colour factor of the propagating parton. This implies that the energy loss of a gluon is larger, by a factor $C_A/C_F=9/4$, than that of a quark. In the absence of QGP, the amount of energy beyond a given jet reconstruction radius has been shown not to follow Casimir scaling~\cite{Dasgupta:2007wa,Dasgupta:2014yra}, since the dependence on the Casimir colour factor of the initiating parton is diluted as the parton cascade, dominated by gluon radiation, develops. When QGP is present, gluon-initiated jets lose more energy than quark-initiated jets by a factor substantially smaller than the $C_A/C_F$ Casimir scaling~\cite{Apolinario:2020nyw,Brewer:2020och}.

To establish whether the large observed difference between the $\gamma+$jet $R_{AA}$, where quark jets are dominant, and the more gluon-rich inclusive-jet $R_{AA}$ can be explained by the colour charge of the parton initiating the jet, one cannot ignore the existence of $p_T$-bin migration, exacerbated by the sensitivity of $R_{AA}$ to the steeply falling nature of the jet-$p_T$ spectrum. The $\gamma+$jet and inclusive-jet $R_{AA}$ could be different not because of different energy loss of quark and gluon jets, but simply because the steepnesses of the jet-$p_T$ spectra are different.
\begin{figure}[!t]
     \centering
     \includegraphics[width=0.49\textwidth]{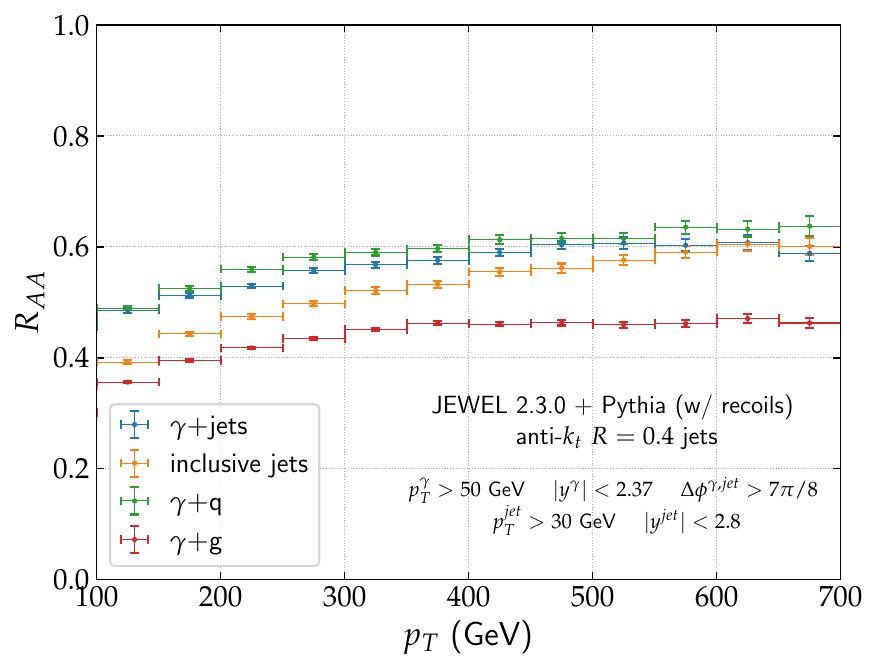}
     \includegraphics[width=0.49\textwidth]{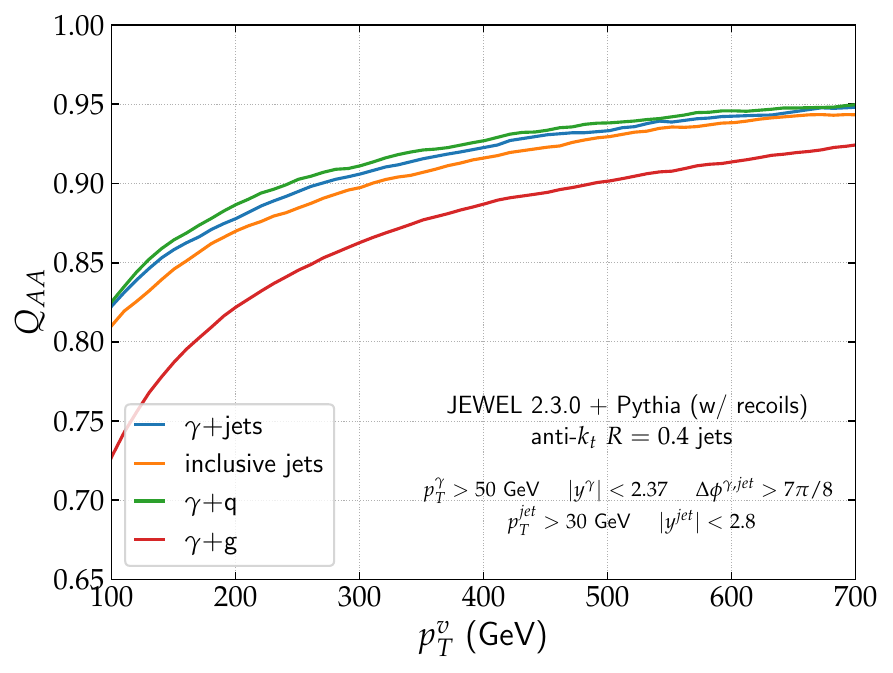}

     \includegraphics[width=0.49\textwidth]{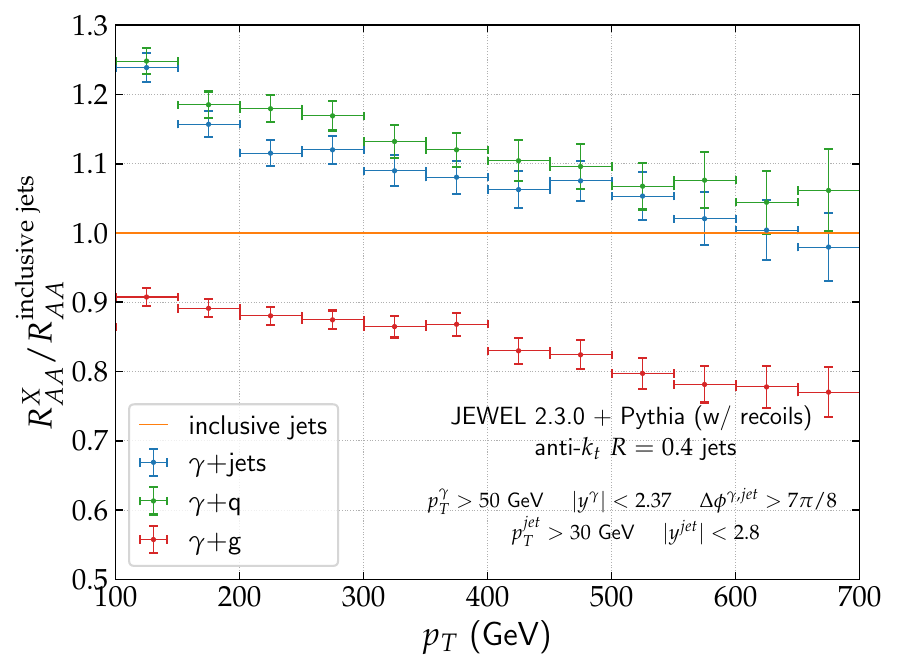}
     \includegraphics[width=0.49\textwidth]{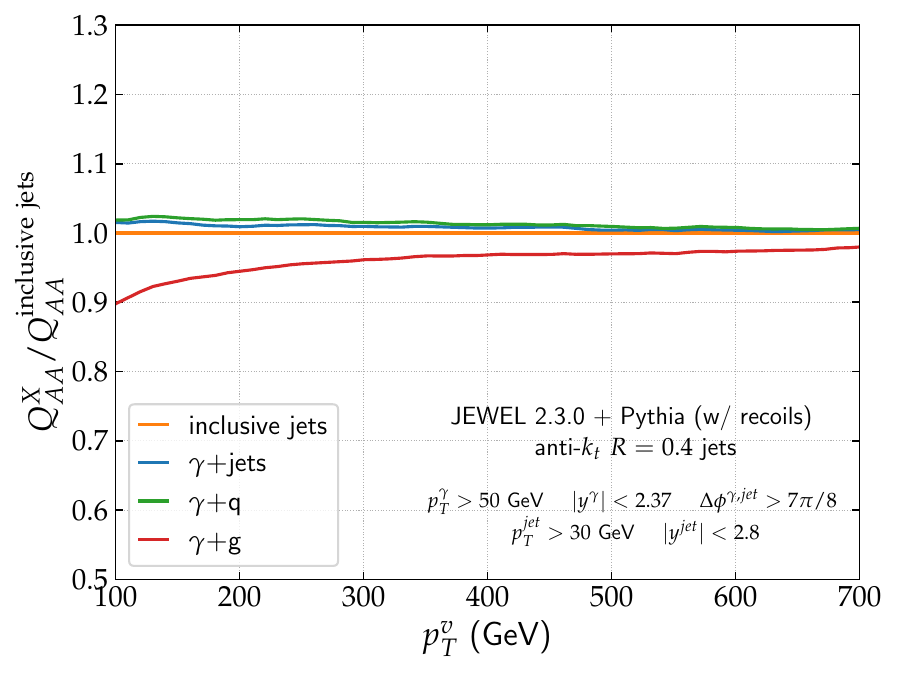}

     \caption{Dependence of $R_{AA}$ (\textbf{left}) and $Q_{AA}$ (\textbf{right}) on the hard process originating the jets. Vacuum samples include nuclear effects, implemented through isospin-averaged nuclear PDFs. The bottom panels show ratios of $R_{AA}$ and $Q_{AA}$ with respect to inclusive-jet results.}
     \label{fig:process_comp}
\end{figure}$Q_{AA}$ deconvolutes the effects of $p_T$ migration and spectral-shape dependence from jet energy loss. In Fig.~\ref{fig:process_comp}, we show both $R_{AA}$ and $Q_{AA}$ for $\gamma+$jets, inclusive jets, and pure samples of quark jets ($\gamma+q$) and gluon jets ($\gamma+g$). These two last samples were obtained by selecting, at generation level, only the relevant quark or gluon channel. The difference between $\gamma+$jets and inclusive jets in $Q_{AA}$ is significantly smaller than for $R_{AA}$, indicating that a significant part of the observed effect in $R_{AA}$ is due to $p_T$-bin migration. Interestingly, the difference in energy loss between pure samples of quark- and gluon-initiated jets, $\gamma+q$ and $\gamma+g$, is partly washed away when comparing $\gamma+$jets and inclusive jets. In fact, the inclusive-jet $Q_{AA}$ is much closer to the quark-initiated jet $Q_{AA}$ than to the gluon-initiated jet $Q_{AA}$ for all $p_T$. While inclusive jets reconstructed in pp collisions are, for $p_T\lesssim200$ GeV, a mixture of quark- and gluon-initiated jets with fractions of roughly $40/60\,\%$, the larger energy loss of gluon jets in AA leads to a survivor bias that produces a sample of reconstructed jets increasingly dominated by quark jets in each bin of $p_T$. The large steepness of the $p_T$ spectrum further exaggerates this effect. The success of simple parametric models~\cite{Spousta:2015fca,Ogrodnik:2024qug} that encode such a variation of the quark/gluon fraction supports this observation. Nevertheless, the degree to which quark/gluon fractions in AA collisions are modified with respect to pp remains to be precisely determined on both the theory~\cite{Ringer:2019rfk,Brewer:2020och,Pablos:2022mrx,Zhang:2023oid} and experimental sides~\cite{ALICE:2019ykw,CMS:2020plq}. Hence, the results in Fig.~\ref{fig:process_comp} indicate that in-medium quark/gluon jet fractions are significantly modified due to their difference in energy loss.

The large difference between inclusive jets and $\gamma+g$ $Q_{AA}$ is in apparent contradiction with what is seen for $R_{AA}$ in the left panel of Fig.~\ref{fig:process_comp}, where for $p_T\lesssim200$ GeV the two results are fairly similar. Nevertheless, the difference in spectrum steepness and the difference in energy loss affect the hierarchy of $R_{AA}$ in opposite directions. On the one hand, purely from the spectrum shape and fixing the same energy loss, one would expect the inclusive-jet $R_{AA}$ to be smaller than that of $\gamma+g$, because the former spectrum is steeper. Note that this difference in steepness depends on the initial partonic states entering the leading-order matrix element, and not only on the final-state partons that initiate the jets. This means that the shape of the inclusive spectrum does not necessarily follow a weighted sum of the $\gamma+q$ and $\gamma+g$ shapes according to the estimated quark/gluon fractions. The energy loss of $\gamma+g$, on the other hand, is greater than that of inclusive jets according to the rough estimate of a $40/60\,\%$ quark/gluon fraction. This would result in a larger $R_{AA}$ for inclusive jets than for purely gluon-initiated jets. Hence, what is seen for $p_T\lesssim200$ GeV is a cancellation of these two effects. When calculating $Q_{AA}$, energy loss is the dominant effect, so the two curves are no longer similar at $p_T\lesssim200$ GeV. At large enough $p_T$, the steepness values of both spectra are similar, so the differences in $R_{AA}$ and $Q_{AA}$ are in agreement.

\begin{figure}[!t]
     \centering
     \includegraphics[width=0.49\textwidth]{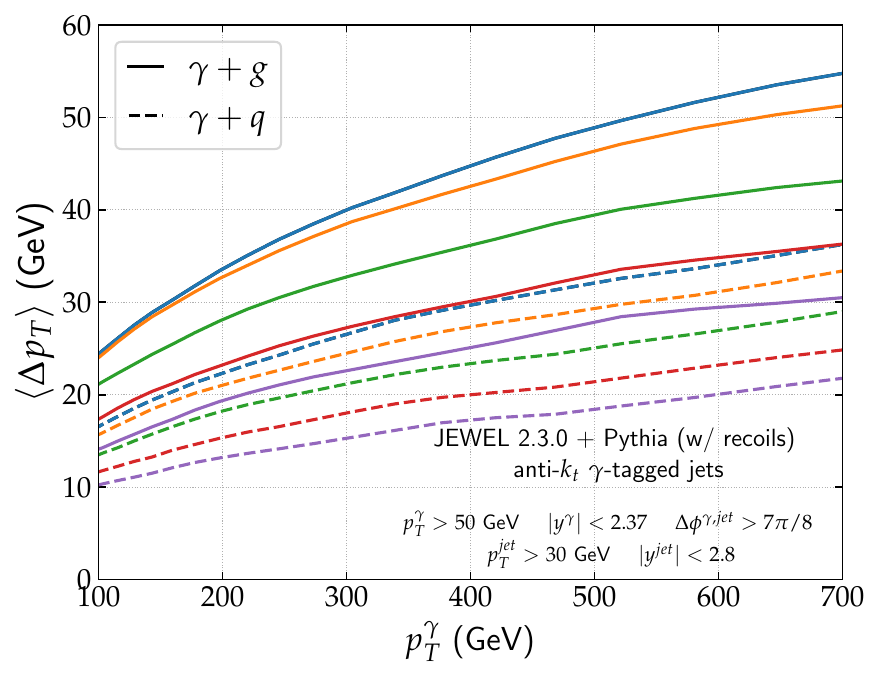}
     \includegraphics[width=0.49\textwidth]{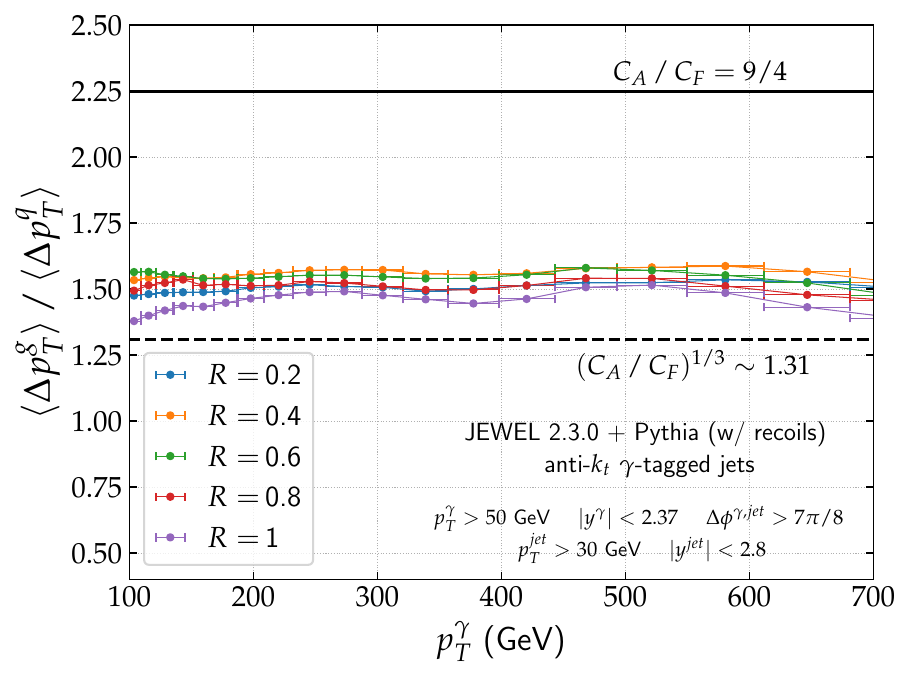}

     \caption{Evolution with jet radius of $\langle \Delta p_T\rangle=\langle p_T^v\rangle-\langle p_T^q\rangle$; see Eqs.~\eqref{eq:av_ptv} and~\eqref{eq:av_ptq} (\textbf{left}), and of the ratio between average energy losses of gluon- and quark-initiated jets (\textbf{right}), for each bin of $p_T^{\gamma}$. The black horizontal lines correspond to what is expected from single-parton Casimir scaling (solid) and from holography-based calculations (dashed)~\cite{Casalderrey-Solana:2014bpa,Chesler:2014jva,Chesler:2015nqz}. Vacuum samples include nuclear effects, implemented through isospin-averaged nuclear PDFs.}
     \label{fig:qg_comp}
\end{figure}

For a more precise statement of the dependence of jet energy loss on the colour charge of the initiating parton, the left panel of Fig.~\ref{fig:qg_comp} shows the average absolute energy loss calculated using the quantile procedure for gluon jets (solid lines) and for quark jets (dashed lines), for a given value of photon $p_T$. Making the energy loss a function of photon $p_T$ ensures that the bias coming from comparing quark and gluon jets with different initial $p_T$ in vacuum is avoided. The first observation is that the jet-radius dependence of energy loss is much more pronounced in gluon jets than in quark jets. This is expected from the previous arguments, since a gluon radiates more, implying that gluon-initiated jets have a larger angular spread. It is also clear that gluon jets lose more energy than quark jets for a wide range of jet radii. The ratio between gluon- and quark-jet energy loss for different radii is shown in the right panel of Fig.~\ref{fig:qg_comp}. It is stable around $\sim1.5$, independently of both jet radius and jet $p_T$. The Casimir scaling of single partons, represented by the black line at $9/4$, is substantially diluted in jets, with gluon jets not losing much more energy than quark jets. Interestingly, the holographic calculations of single-parton energy loss~\cite{Casalderrey-Solana:2014bpa,Chesler:2014jva,Chesler:2015nqz} that underlie the Hybrid Model predict a single-parton scaling much milder than Casimir scaling, $\sim1.31$, and much closer to the result obtained for in-medium jets. Note that this same result could be obtained by considering the energy-loss proxy in Eq.~\eqref{eq:av_ptv}, obtained by using the photon $p_T$ to match mean jet $p_T$ in vacuum and medium, rather than the one obtained through quantile matching. Although not shown here, this provides another indication of the robustness of the quantile procedure, since the result obtained using Eq.~\eqref{eq:av_ptv} shows the same features as the right panel of Fig.~\ref{fig:qg_comp}.

\subsection{Towards a measurement of \texorpdfstring{$Q_{AA}$}{QAA}}
\label{subsec:qaa_measurement}

An obvious experimental obstacle to the calculation of the quantile function as defined in Eq.~\eqref{eq:quantile_matching} is the requirement to integrate jet spectra up to arbitrarily large $p_T$, which is naturally not possible in experiment. However, performing the integration with a cutoff at $+\infty$ is not actually required, since one can cast Eq.~\eqref{eq:quantile_matching} into a form that takes into account $p_T$ cutoffs in both spectra. This is done by considering Eq.~\eqref{eq:quantile_matching} at two different values of $p_T^v$ and subtracting the two resulting equations, giving
\begin{align}
    \Sigma^{pp}(p_{T,1}^v,p_{T,2}^v)
    =
    \Sigma^{AA}\left(p_T^q(p_{T,1}^v),p_T^q(p_{T,2}^v)\right)\, .
\end{align}
Hence, if the pp spectrum has a cutoff at a value $p_{T,2}^v=p_{T,c}^v$, then the upper integration limit of $\Sigma^{AA}$ should be $p_{T,c}^q=p_T^q(p_{T,c}^v)$:
\begin{align}
    \int_{p_T^v}^{p_{T,c}^v} dp_T\,
    \frac{d\sigma^{pp}}{dp_T}
    =
    \int_{p_T^q(p_T^v)}^{p_{T,c}^q} dp_T\,
    \frac{d\sigma^{AA}}{dp_T}\, .
    \label{eq:ptc_m}
\end{align}
However, because the right-hand side depends on the unknown quantile function evaluated at two different points, $p_{T,c}^v$ and $p_T^v$, the equation is underdetermined. This is expected, since Eq.~\eqref{eq:ptc_m} is an integral equation and therefore requires an initial condition that fixes the value of $p_{T,c}^q$. For values of $p_T^v$ much smaller than $p_{T,c}^v$, the sensitivity to the upper integration limit for the medium spectrum is insignificant, and one can go around this problem by setting $p_{T,c}^q=p_{T,c}^v$. For values comparable to $p_{T,c}^v$, however, the value of this limit becomes relevant. This is clearly seen in the left panel of Fig.~\ref{fig:cutoff_sensitivity}, where setting an incorrect value for $p_{T,c}^q$ leads to a significant reduction in $Q_{AA}$ accuracy for a range of about $150$ GeV below the vacuum cutoff. The blue curve overestimates $p_{T,c}^q$ by imposing the same cutoff in both spectra, while the orange curve underestimates it by setting it to $90\%$ of the correct cutoff.

\subsubsection{Cutoff dependence}
\label{subsubsec:qaa_cutoff}

\begin{figure}[!t]
    \centering
    \includegraphics[width=0.49\textwidth]{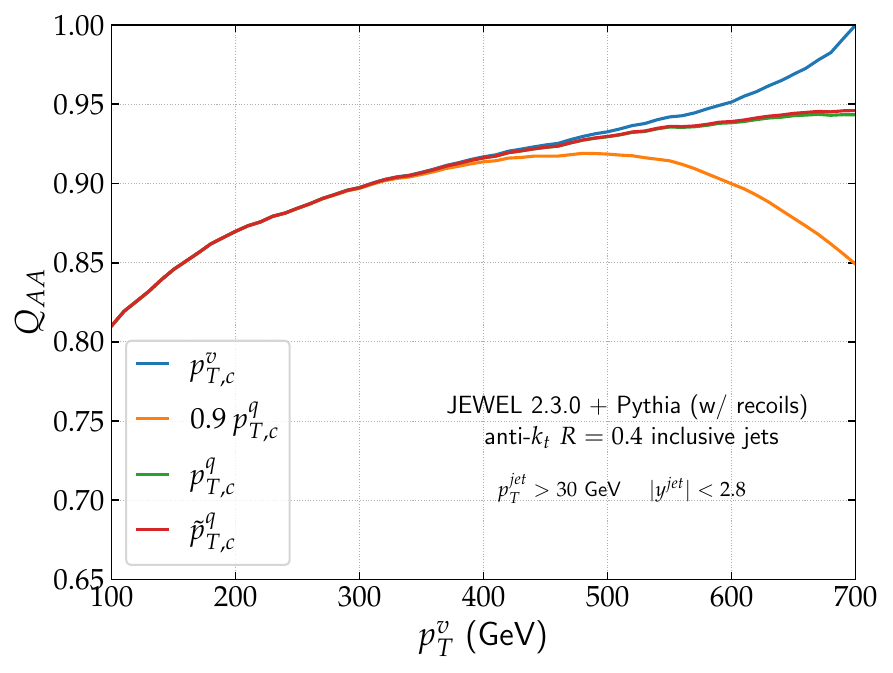}
    \includegraphics[width=0.49\textwidth]{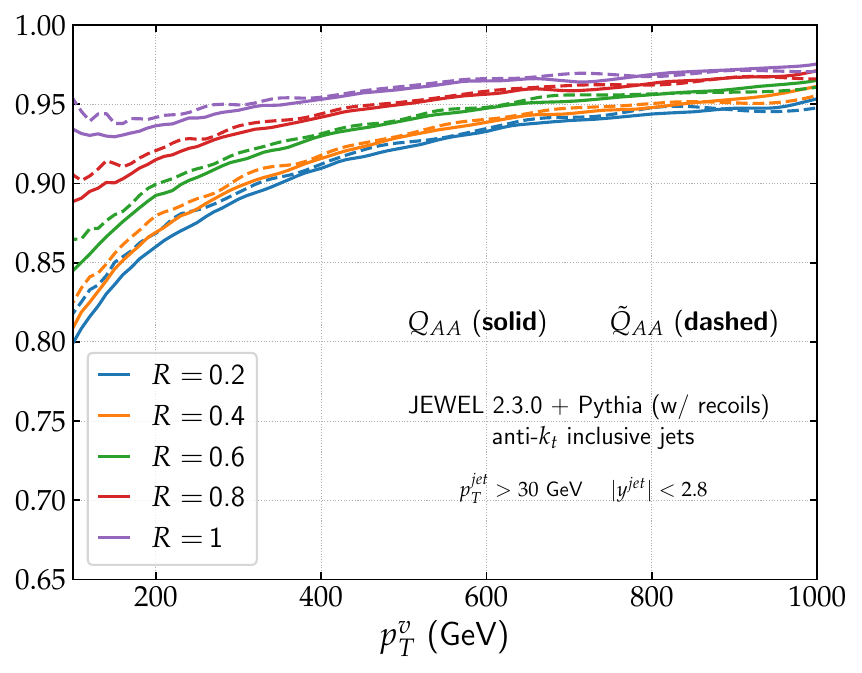}
    \caption{\textbf{Left:} $Q_{AA}$ sensitivity to different values of $p_{T,c}^m$ in Eq.~\eqref{eq:ptc_m}. \textbf{Right:} Comparison of $Q_{AA}$ and $\tilde Q_{AA}$ for multiple jet radii. Vacuum samples include nuclear effects, implemented through isospin-averaged nuclear PDFs.}
    \label{fig:cutoff_sensitivity}
\end{figure}

Possible workarounds for this problem include extracting $p_T^q(p_{T,c}^v)$ from an interpolated quantile function calculated via Monte Carlo at larger $p_T$ than what is experimentally accessible. Another option, which can be fully calculated from data, is to use a variant of the quantile function also introduced in~\cite{Brewer:2018dfs}, the pseudo-quantile $\tilde p_T^q$, to get an estimate of the initial condition $p_{T,c}^q=p_T^q(p_T^c)$. The pseudo-quantile is calculated by equating the spectra themselves instead of their cumulative versions:
\begin{align}
    \frac{d\sigma^{AA}}{d\tilde p_T^q}
    \left(\tilde p_T^q(p_T^v)\right)
    =
    \frac{d\sigma^{pp}}{dp_T^v}(p_T^v)\, .
    \label{eq:pqaa_def}
\end{align}
Interestingly, under the same assumptions as for the quantile function $p_T^q$ in Eq.~\eqref{eq:qaa_estimate}, one obtains
\begin{align}
    1-\frac{\tilde p_T^q(p_T^v)}{p_T^v}
    =
    1-\tilde Q_{AA}(p_T^v)
    \approx
    \frac{\langle\epsilon\rangle(p_T^v)}{p_T^v}\, .
    \label{eq:pqaa_estimate}
\end{align}
Hence, despite their clear conceptual difference, the quantile and pseudo-quantile functions agree well under the assumption of small and only mildly $p_T$-dependent energy loss. Indeed, by looking at Eq.~\eqref{eq:diff_qaa}, one sees that it becomes equivalent to Eq.~\eqref{eq:pqaa_def} in the limit of a constant energy-loss shift $\epsilon_0$,
\begin{align}
    p_T^q(p_T^v)
    \rightarrow
    p_T^v+\epsilon_0\, .
\end{align}
It is then expected that, when calculating $\tilde Q_{AA}$ and comparing it to $Q_{AA}$, as shown in the right panel of Fig.~\ref{fig:cutoff_sensitivity}, the two are not strikingly different for most of the $p_T$ range and that they agree very well at large enough $p_T$ and for all jet radii. By looking at the red curve of the left panel of Fig.~\ref{fig:cutoff_sensitivity}, one confirms that using $\tilde p_{T,c}^q=\tilde p_T^q(p_T^c)$ as an estimate for the initial condition $p_{T,c}^q=p_T^q(p_{T,c}^v)$ works well.

\subsubsection{Equivalence between \texorpdfstring{$1-Q_{AA}$}{1-QAA} and \texorpdfstring{$S_{\rm loss}$}{Sloss}}
\label{subsubsec:qaa_sloss}

As discussed at the end of Section~\ref{subsec:qaa_quantile_matching}, a relation similar to Eq.~\eqref{eq:diff_qaa} was used in~\cite{ATLAS:2023iad} to estimate jet energy loss. Renaming the variables, it reads
\begin{align}
    \frac{d\sigma^{AA}}{dp_T^s}(p_T^s)
    =
    \frac{d\sigma^{pp}}{dp_T^v}(p_T^v)
    \left(1+\frac{d\Delta p_T^s}{dp_T^v}\right)\, ,
    \label{eq:atlas_eq}
\end{align}
where $p_T^s(p_T^v)=p_T^v-\Delta p_T^s(p_T^v)=(1-S_{\rm loss}(p_T^v))p_T^v$. Setting $\Delta p_T^q(p_T^v)\equiv p_T^v-p_T^q(p_T^v)=(1-Q_{AA}(p_T^v))p_T^v$, the differential equation defining the quantile matching in Eq.~\eqref{eq:diff_qaa} can be rewritten as
\begin{align}
    \frac{d\sigma^{AA}}{dp_T^q}(p_T^q)
    =
    \frac{d\sigma^{pp}}{dp_T^v}(p_T^v)
    \left(1-\frac{d\Delta p_T^q}{dp_T^v}\right)^{-1}\, .
    \label{eq:diff_qaa_rewritten}
\end{align}
\begin{figure}[!t]
    \centering
    \includegraphics[width=0.49\textwidth]{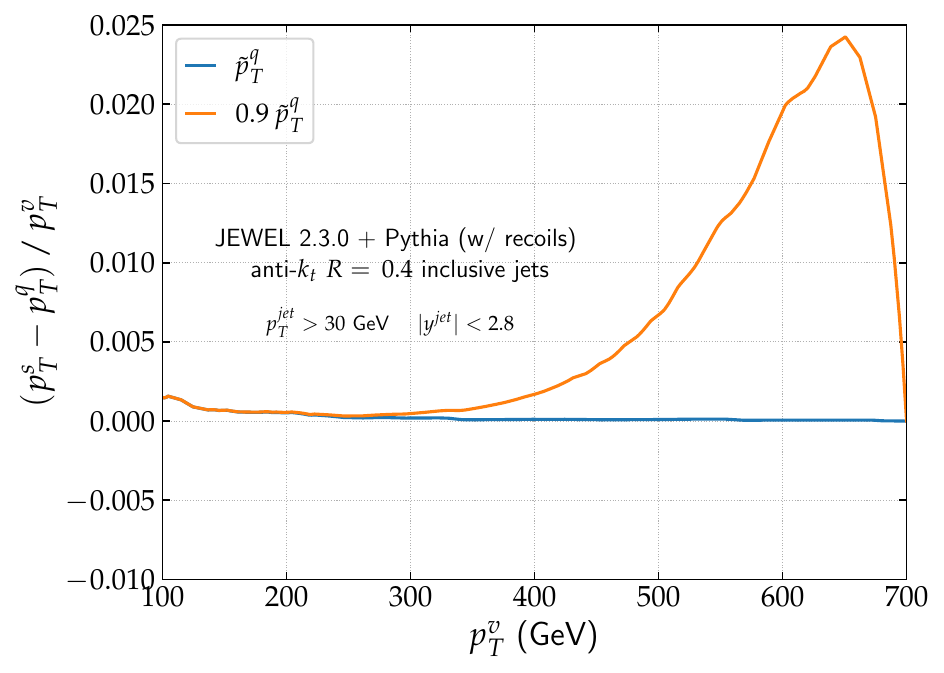}
    \includegraphics[width=0.49\textwidth]{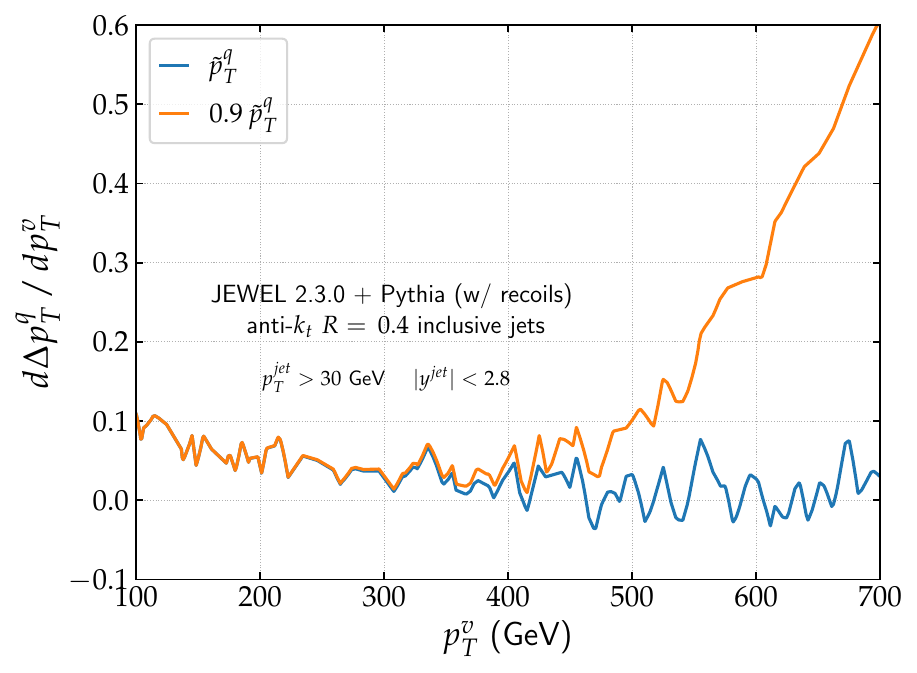}
    \caption{\textbf{Left:} Comparison of the energy-loss proxies $\Delta p_T^q=p_T^v-p_T^q$ in Eq.~\eqref{eq:diff_qaa_rewritten} and $\Delta p_T^s=p_T^v-p_T^s$ in Eq.~\eqref{eq:atlas_eq} for two different values of the initial condition, given by $\tilde p_T^q(p_T^v=700~\text{GeV})$ as a reasonable estimate and by $0.9\,\tilde p_T^q(p_T^v=700~\text{GeV})$ as an underestimate. \textbf{Right:} Derivative of the energy-loss function calculated through Eq.~\eqref{eq:diff_qaa_rewritten} for the same two values of the initial condition. Vacuum samples include nuclear effects, implemented through isospin-averaged nuclear PDFs.}
    \label{fig:proxy_comp}
\end{figure}Looking at Eqs.~\eqref{eq:atlas_eq} and~\eqref{eq:diff_qaa_rewritten}, one sees that both agree to first order in $d\Delta p_T/dp_T^v$, thus computing similar proxies, $p_T^s\approx p_T^q$, for a mildly $p_T$-dependent energy loss. Evidently, both quantities agree with the pseudo-quantile as defined in Eq.~\eqref{eq:pqaa_def} to zeroth order in $d\Delta p_T/dp_T^v$, as shown in the right panel of Fig.~\ref{fig:cutoff_sensitivity}. Nevertheless, if energy loss has a stronger $p_T$ dependence than in \jewel\ 2.3, the two equations may give significantly different results. Additionally, this dependence can be artificially introduced if one incorrectly sets the initial condition. This is clearly seen in the left panel of Fig.~\ref{fig:proxy_comp}, where the difference between the two energy-loss proxies,
\begin{align}
    \frac{\Delta p_T^q-\Delta p_T^s}{p_T^v}
    =
    \frac{p_T^s-p_T^q}{p_T^v}\, ,
\end{align}
is shown. For the blue curve, the initial condition is estimated using the pseudo-quantile as defined in Eq.~\eqref{eq:pqaa_def}, which was shown to provide an accurate approximation to the quantile at large enough $p_T$, see Fig.~\ref{fig:cutoff_sensitivity}. In this case, Eqs.~\eqref{eq:atlas_eq} and~\eqref{eq:diff_qaa_rewritten} calculate almost exactly the same quantity, with the largest difference, $\sim0.1\%$ of the lost energy, occurring at the lower end of the $p_T$ range, around $100$ GeV. This is expected since, as can be seen in the right panel of Fig.~\ref{fig:proxy_comp}, the derivative appearing on the right-hand side of both equations is quite small: it is always below $0.1$ and decreases with increasing $p_T^v$. If an inaccurate estimate is used for the initial condition, e.g. $0.9$ of the previous value, then the two quantities $\Delta p_T^s$ and $\Delta p_T^q$ separate from each other, with the largest difference reaching $\sim2.5\%$ of the lost energy. Again, this can be understood from the right panel of Fig.~\ref{fig:proxy_comp}, where the derivative increases past a value for which Eqs.~\eqref{eq:diff_qaa_rewritten} and~\eqref{eq:atlas_eq} agree to first order. Despite these differences, the quantities $1-Q_{AA}$ and $S_{\rm loss}$ should not be too different from each other when applied to experimental data and when the initial condition is reasonably estimated. Nevertheless, the quantile procedure has a much clearer physical motivation and, although not explored here, provides more solid groundwork for the study of jet energy-loss dispersion.

\subsection{Jet substructure after quantile correction}
\label{subsec:qaa_substructure}

\begin{figure}[!t]
     \centering
     \includegraphics[width=0.49\textwidth]{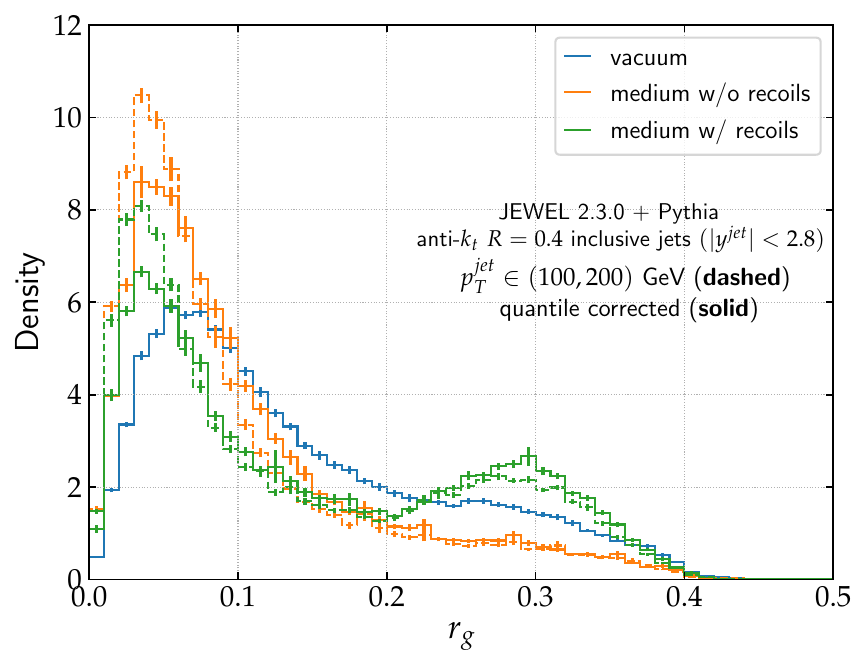}
     \includegraphics[width=0.49\textwidth]{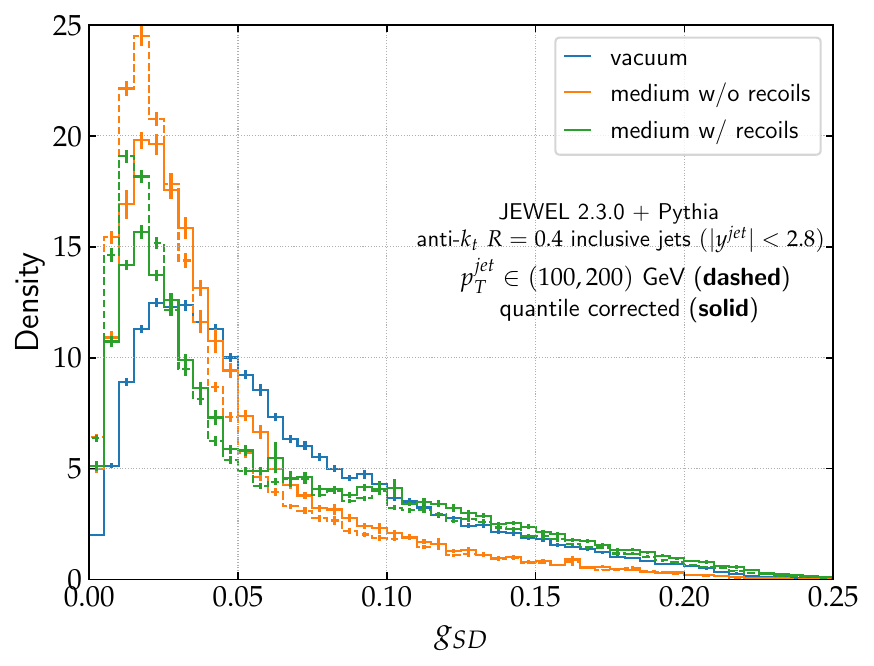}

     \includegraphics[width=0.49\textwidth]{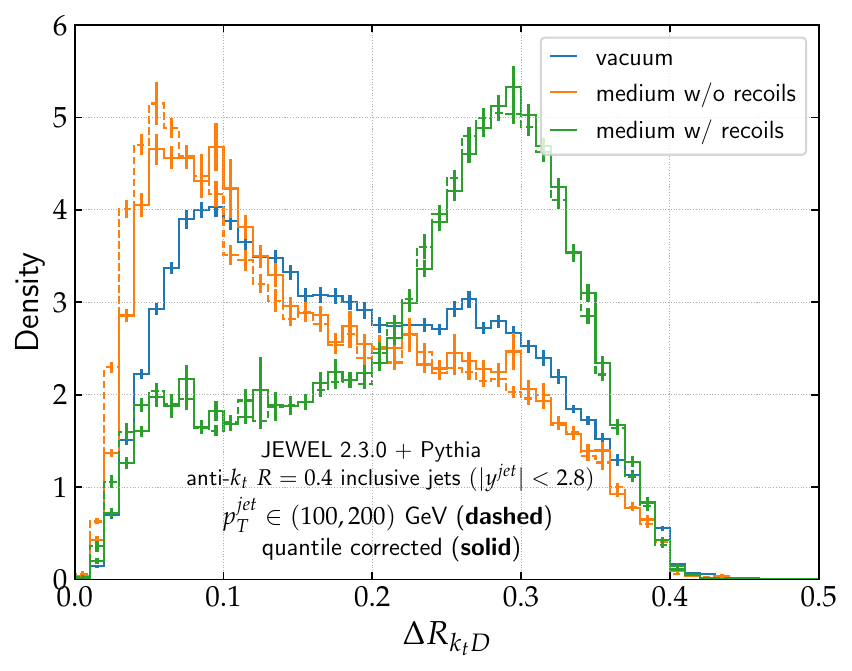}
     \includegraphics[width=0.49\textwidth]{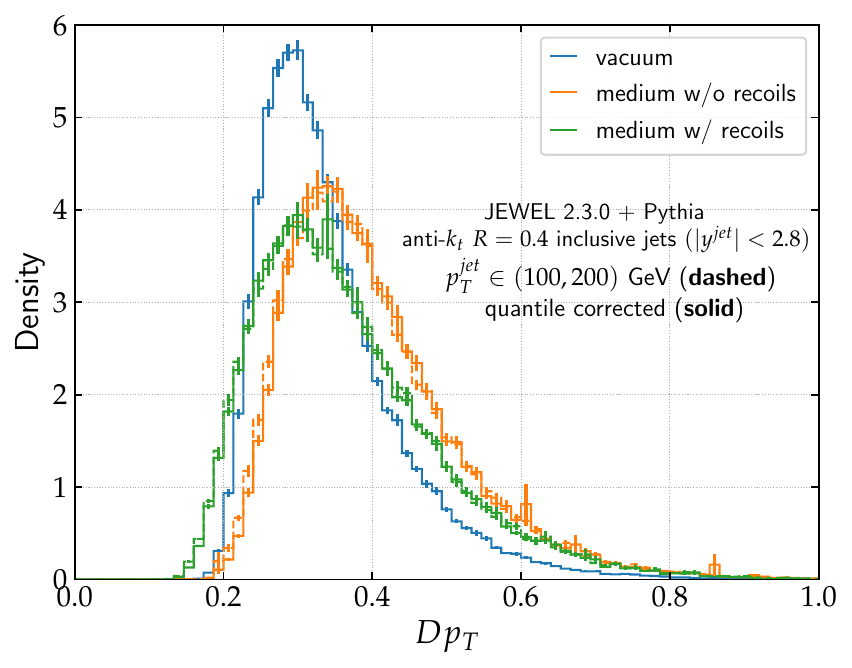}

     \caption{Distribution of four jet substructure observables for vacuum, medium without recoils and medium with recoils. The solid lines correspond to jets with $p_T^{\rm jet}\in(100,200)$ GeV and the dashed lines to $p_T^q\in(79,170)$ GeV for medium without recoils and $p_T^q\in(84,177)$ GeV for medium with recoils. Vacuum samples are pp collisions.}
     \label{fig:corrected_obs}
\end{figure}

The main purpose of the quantile matching procedure is to suppress the $p_T$-migration bias associated with comparing observables within the same reconstructed-jet momentum bin. By estimating average jet energy loss, one compares vacuum and medium jets in matched $p_T$ bins, thereby comparing jets that were, on average, initiated by hard partons with the same transverse momentum. Naturally, the sensitivity of an observable to $p_T$-migration bias depends on the correlation of that observable with jet $p_T$. This was shown in the original quantile-matching study~\cite{Brewer:2018dfs}, where the impact of $p_T$ migration was studied for the ratio between jet mass and jet transverse momentum, $m^{\rm jet}/p_T^{\rm jet}$. Here, this analysis is extended to other jet observables, including medium response as implemented in \jewel. In Fig.~\ref{fig:corrected_obs}, we show the distribution of four jet substructure observables for jets in a fixed $p_T$ bin (dashed lines) and for quantile-matched $p_T$ bins (solid lines). The list of interesting jet substructure observables is extensive~\cite{Andrews:2018jcm}; the observables shown here were chosen to illustrate both cases where the distributions are significantly corrected and cases that are practically insensitive to $p_T$ migration.

Let us first focus on the groomed jet radius $r_g$, shown in the top-left panel of Fig.~\ref{fig:corrected_obs}. To calculate $r_g$, one first reclusters a jet using the Cambridge--Aachen (C/A) algorithm and then calculates the angular distance $\Delta R_{ij}$ of the first branching or subjet that satisfies the SoftDrop condition~\cite{Larkoski:2014wba}
\begin{align}
    \frac{\min[p_{T,i},p_{T,j}]}{p_{T,i}+p_{T,j}}
    >
    z_{\rm cut}
    \left(\frac{\Delta R_{ij}}{R}\right)^{\beta}\, ,
    \label{eq:SD_condition}
\end{align}
where $z_{\rm cut}=0.1$ and $\beta=0$. For $r_g$, the contribution of $p_T$ migration is quite significant both with and without medium response. In fact, the apparent narrowing of jets from vacuum to medium is reduced when the $p_T$ bin is corrected using quantile matching.

Continuing with an observable that conveys information about the width of a jet, we now consider the jet girth $g_{SD}$ after applying SoftDrop, shown in the top-right panel of Fig.~\ref{fig:corrected_obs}:
\begin{align}
    g_{SD}
    =
    \sum_{i\in {\rm jet}_{SD}} z_i\,\Delta R_{i,{\rm jet}}\, ,
\end{align}
where $z_i$ is the fraction of transverse momentum of constituent $i$ with respect to the jet $p_T$, and $\Delta R_{i,{\rm jet}}$ is the angular distance of the constituent with respect to the jet axis. Applying SoftDrop to groom the jet amounts to discarding the softest branch at each step of the C/A sequence until the condition in Eq.~\eqref{eq:SD_condition} is met. The solid curves, corresponding to quantile-matched bins, are significantly shifted relative to the dashed curves, corresponding to unmatched bins. Thus, as for $r_g$, the apparent narrowing of jets is reduced. This is true whether or not medium recoils are included.

As examples of observables that are less sensitive to $p_T$ migration, namely observables only weakly correlated with $p_T$ in the range considered here, let us look first at another measure of the angular spread of a jet: the distance $\Delta R$ between the two subjets in the C/A reclustering sequence that satisfy the dynamical grooming condition~\cite{Mehtar-Tani:2019rrk}
\begin{align}
    \kappa^{(a)}
    =
    \frac{1}{p_T^{\rm jet}}
    \max_{i\in {\rm C/A\ seq.}}
    \left[
        z_i(1-z_i)p_{T,i}
        \left(\frac{\theta_i}{R}\right)^a
    \right]\, .
\end{align}
By setting $a=1$, corresponding to $k_T$Drop, one tags the branching with the largest relative transverse momentum $k_T\sim \kappa^{(1)}p_T^{\rm jet}$. Looking at the bottom-left panel of Fig.~\ref{fig:corrected_obs}, this observable is only slightly sensitive to $p_T$ migration in the absence of medium response, as indicated by the orange curves. However, when medium response is included, the $p_T$ correlation appears to be washed out, since the observable is very little modified after quantile matching the $p_T$ bin.

Finally, in the bottom-right panel of Fig.~\ref{fig:corrected_obs}, we show the momentum dispersion inside a jet, $Dp_T$, which quantifies how hard or soft the jet fragmentation is:
\begin{align}
    Dp_T
    =
    \frac{\sqrt{\sum_{i\in {\rm jet}} p_{T,i}^2}}{p_T^{\rm jet}}\, .
\end{align}
Whether or not medium recoils are included, this observable is almost insensitive to $p_T$ migration.

As a final remark, we point out a clear limitation of $Q_{AA}$ as currently defined. It is tempting to calculate $Q_{AA}$ in bins of a given jet substructure observable, which we denote by $X$. This would go beyond an average measure of energy loss by partially resolving its dispersion through its correlation with a single substructure variable. However, such a procedure would introduce a bias analogous to the one encountered when comparing jets in bins of equal reconstructed $p_T$. In particular, it would confound the energy loss of jets with the medium-induced modification of the distribution of $X$ relative to vacuum. One possible remedy would be to perform quantile matching in $X$, assuming that the corresponding bin migration is monotonic. However, migrations in $X$ and in $p_T$ are not, in general, independent. Treating them separately would therefore distort their correlations and yield results that do not faithfully represent the dependence of energy loss on jet substructure. The rigorously correct procedure would instead be to perform a simultaneous quantile matching in $p_T$ and $X$. This, however, is an underdetermined problem: one would have a single quantile-matching condition for two unknown quantile functions. Making such a construction possible would therefore require additional assumptions about the nature of bin migration in $X$ and its dependence on $p_T$.

\section{Energy-loss functions}
\label{sec:elf}

\subsection{From quenching weights to energy-loss functions}
\label{subsec:elf_quenching_weights}

Having discussed reconstructed jet energy loss in Section~\ref{sec:qaa}, we now turn to a more theory-driven treatment of energy loss, suited to observables that probe correlations in the energy flow of the full shower. Recent years have witnessed tremendous advances in the ability to study the differential substructure of jets in heavy-ion collisions; see~\cite{Larkoski:2017jix,Asquith:2018igt,Apolinario:2022vzg,Cao:2020wlm,Cao:2024pxc,Andrews:2018jcm,Adolfsson:2020dhm,Connors:2017ptx,Cunqueiro:2021wls} for reviews. These observables make it possible to go beyond single high-$p_T$ hadrons or jets as probes, to study correlations in energy flux, and hold the promise to improve our understanding of the microscopic structure of the HIC matter. However, the theoretical description of these new classes of observables requires techniques that account for the effect of energy loss across the entire parton shower.

The framework commonly used to describe jet energy loss is still largely based on the early seminal work of Baier, Dokshitzer, Mueller and Schiff~\cite{Baier:2001yt}. In summary, this approach relies on the observation that the medium-modified partonic cross sections can be related to the vacuum ones through the simple relation
\begin{align}
    \frac{d\sigma_i^{\rm med}}{dp_T}
    =
    \int_0^\infty d\varepsilon\,
    P_i(\varepsilon)
    \frac{d\sigma_i^{\rm vac}}{dp_T'}
    \bigg|_{p_T'=p_T+\varepsilon}\, .
    \label{eq:main_Qweights}
\end{align}
Here $P_i(\varepsilon)$ is interpreted as the probability for a single parton $i$, with initial energy $p_T$, to lose an energy $\varepsilon\ll p_T$ to the environment. Thus, the energy shift of the vacuum cross section on the right-hand side of Eq.~\eqref{eq:main_Qweights} efficiently captures the experimentally observed shift of final-particle distributions; see for example~\cite{Apolinario:2024apr,Brewer:2018dfs,Casalderrey-Solana:2007knd,Mehtar-Tani:2013pia,ATLAS:2023iad,Milhano:2015mng,Rajagopal:2016uip,Takacs:2021bpv,Spousta:2015fca}. Moreover, because of the steeply falling nature of the partonic vacuum cross sections with $p_T$, the relation in Eq.~\eqref{eq:main_Qweights} can be reduced to the form~\cite{Baier:2001yt}
\begin{align}
    \frac{d\sigma_i^{\rm med}}{dp_T}
    =
    Q_i(p_T)\,
    \frac{d\sigma_i^{\rm vac}}{dp_T}\, ,
    \label{eq:Q_weight_simplified}
\end{align}
where
\begin{align}
    Q_i(p_T)
    =
    \int d\varepsilon\,
    e^{-n\varepsilon/p_T}
    P_i(\varepsilon)\, ,
    \label{eq:Q_weight_factor}
\end{align}
with $n$ the spectral index of the partonic cross section.\footnote{The quenching factor $Q_i(p_T)$ in Eq.~\eqref{eq:Q_weight_factor} is not to be confused with the quantile ratio $Q_{AA}$ introduced in Section~\ref{sec:qaa}. The former is a cross-section-level overall suppression factor, while the latter is a ratio of matched momenta in cumulative spectra.} The relations in Eqs.~\eqref{eq:main_Qweights} and~\eqref{eq:Q_weight_simplified} give a simple prescription to capture energy-loss effects in jet quenching physics. They have seen broad phenomenological usage, with the quenching-weight factor $Q_i(p_T)$ being computed in increasingly sophisticated models~\cite{Salgado:2003gb,Caucal:2021cfb,Mehtar-Tani:2021fud,Barata:2023bhh} based on perturbative QCD considerations~\cite{Baier:2001yt,Mehtar-Tani:2017ypq}. Notwithstanding the phenomenological success of this framework, it is desirable to have a prescription for jet energy loss in which a more field-theoretic treatment of the medium modifications can be formulated in QCD, avoiding reliance on cross-section-level definitions such as Eq.~\eqref{eq:main_Qweights}.

Thus, in what follows, we introduce an alternative non-perturbative description of in-medium energy loss, which can still be studied using perturbative QCD methods. The central observation is that medium-induced energy loss shares formal similarities with the non-linear renormalization group equations developed in the study of track functions~\cite{Chang:2013rca,Chang:2013iba}. The physical reason is that both electric charge and the distribution of energy-loss sources depend on the full structure of the parton cascade. Using recent developments in the track-function formalism~\cite{Lee:2023xzv,Lee:2023tkr,Jaarsma:2022kdd,Li:2021zcf,Lee:2023npz}, this structure can be used to provide a treatment of medium-induced jet energy loss that enables calculations of energy loss in jet substructure observables, such as energy correlators, in heavy-ion collisions~\cite{Barata:2024nqo,Barata:2023zqg,Andres:2024ksi,Andres:2022ovj,Yang:2023dwc,Bossi:2024qho,Singh:2024vwb,Xing:2024yrb}.

\begin{figure}[!t]
    \centering
    \includegraphics[width=0.49\textwidth]{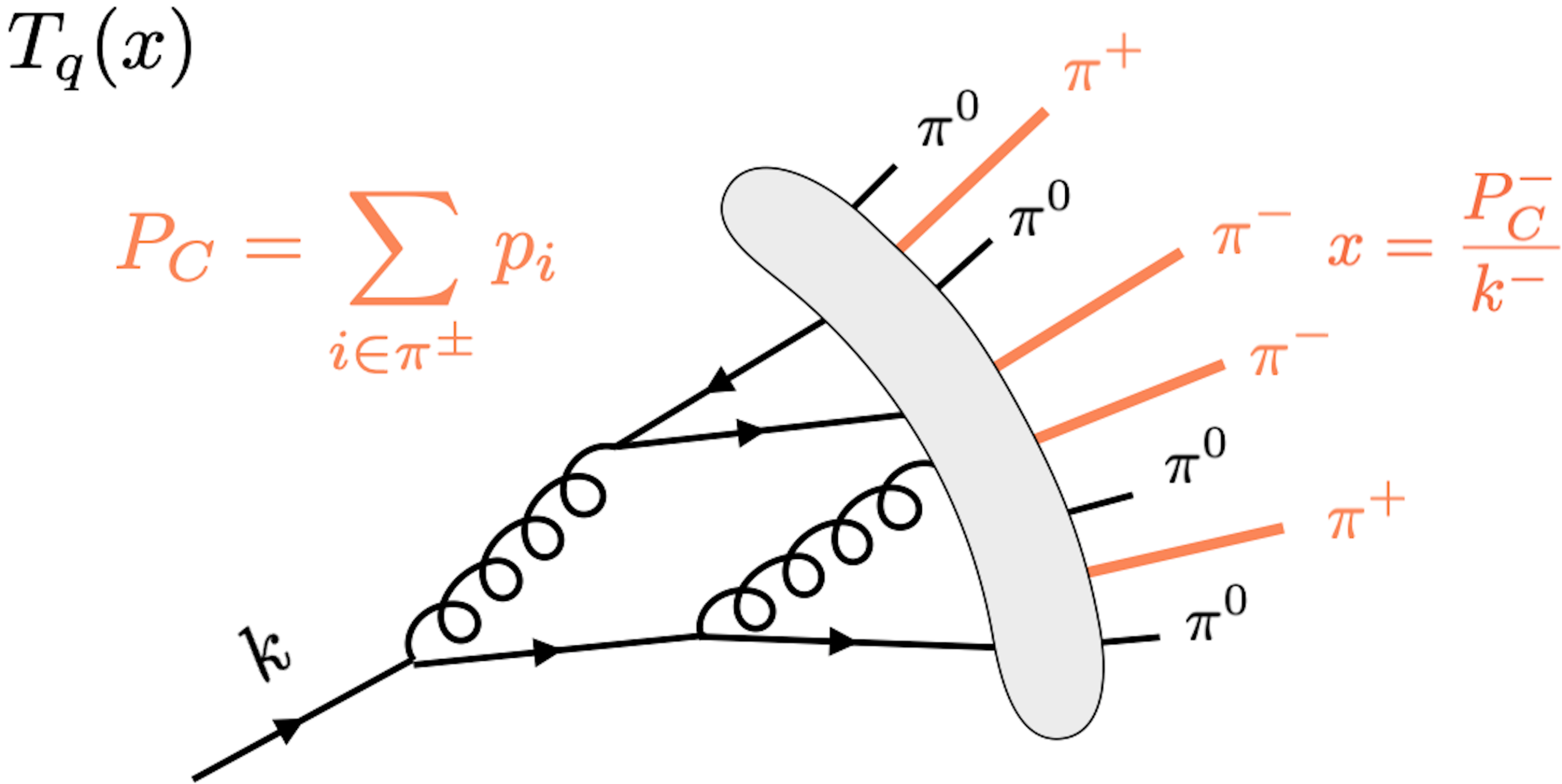}
    \includegraphics[width=0.49\textwidth]{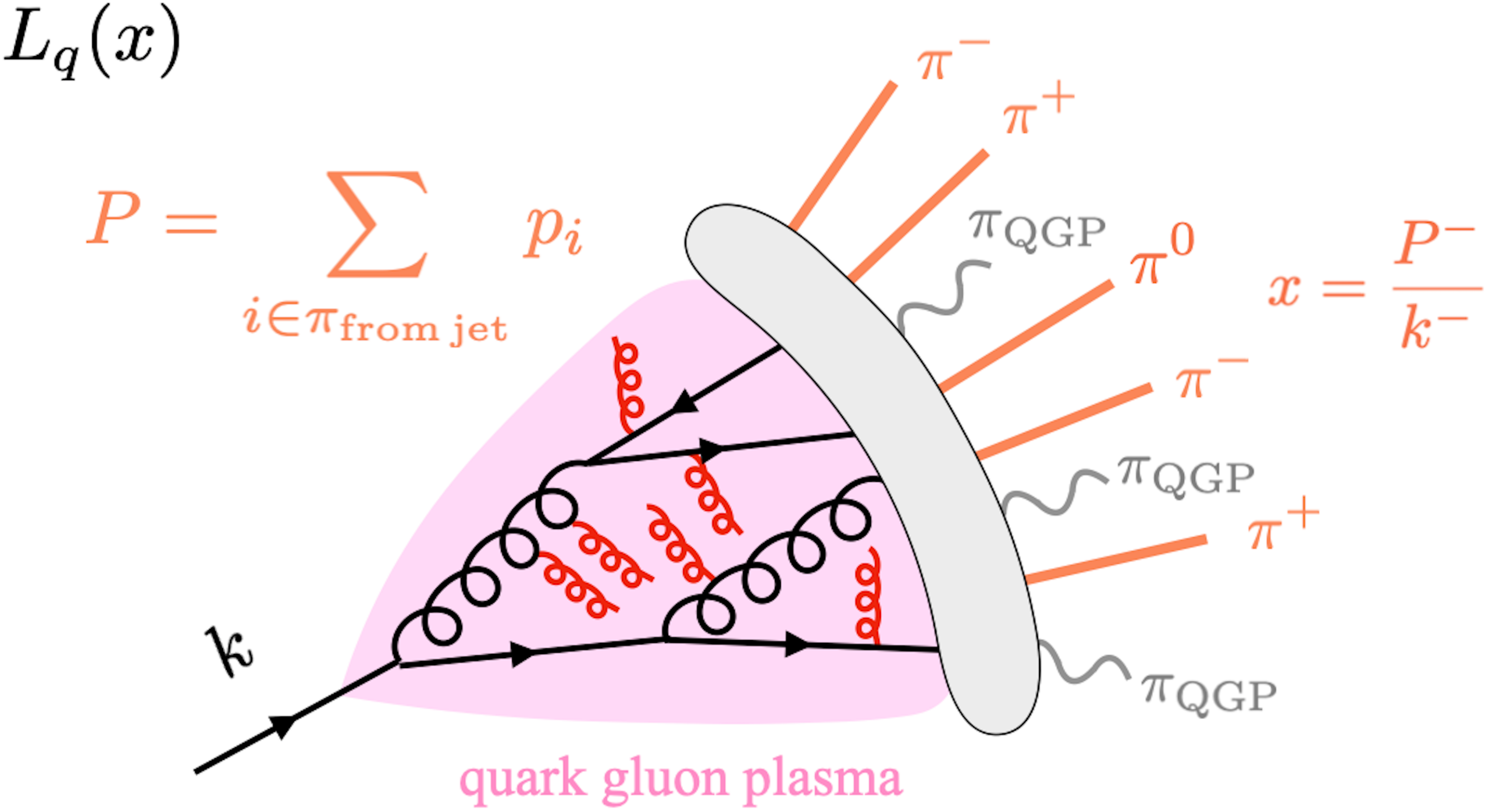}
    \caption{Illustration of the quark track (\textbf{left}) and energy-loss (\textbf{right}) functions, in the limit where all particles hadronize into pions. Assuming a large initial momentum scale $k$, the final-state pions are produced by the cascading of the initial parent quark, which can be described perturbatively (black line). For the track function, the energy fraction $x$ accounts only for the energy flow carried by charged pions ($\pi^\pm$). For the energy-loss function, all hadrons originating from the jet contribute to the energy fraction $x$, while pions originating from the thermal bath are not included (in gray). The fact that, during the evolution, part of the jet energy thermalizes into the QGP implies $x\neq1$, representing the effects of energy loss, denoted schematically by red gluons.}
    \label{fig:elf_track_schematic}
\end{figure}

To motivate the construction, let us briefly recall the track-function formalism. Track functions were introduced in~\cite{Chang:2013rca} to compute jet observables on charged particles. More generally, they provide a formalism for computing jet observables on any subset $R$ of particles specified by some particular property, achieving a rigorous factorization between perturbative and non-perturbative physics. Focusing for simplicity on track functions for charged particles, the track function $T_i(x)$ measures the energy fraction $x$ carried by charged hadrons that originated from an energetic parton $i$, as illustrated in the top panel of Fig.~\ref{fig:elf_track_schematic}. They admit a field-theoretic definition, which for quarks reads
\begin{align}
    T_q(x)
    &=
    \int dy^+ d^2\y\,
    e^{ik^-y^+/2}
    \sum_{N,C}
    \delta\left(x-\frac{P_C^-}{k^-}\right)
    \frac{1}{2N_c}
    {\rm Tr}
    \left[
        \frac{\gamma^-}{2}
        \langle \psi(y)|N,C\rangle
        \langle N,C|\bar\psi(0)\rangle
    \right]\, .
    \label{eq:def_tracks}
\end{align}
Here the state has been separated into neutral $(N)$ and charged $(C)$ components, and only the momentum of the charged component is measured. In a particle-level description, this is illustrated in Fig.~\ref{fig:elf_track_schematic} by taking the charged component to be composed of $\pi^\pm$, while the neutral states are formed by $\pi^0$. Gauge invariance is ensured by gauge links, left implicit, and an analogous definition can be given for the gluon track function. Compared to fragmentation functions, which measure the energy fraction of a single hadron, track functions exhibit a more complicated non-linear renormalization group equation because they describe the energy fraction carried by all hadrons of a particular property. Such equations have been extensively studied recently and have been calculated at next-to-leading order~\cite{Jaarsma:2023ell,Chen:2022pdu,Chen:2022muj,Jaarsma:2022kdd,Li:2021zcf}, while the track functions have also been extracted from jet data~\cite{Lee:2023xzv,ATLAS:2024jrp}. In the heavy-ion context, the modifications to the track functions induced by the medium have also been studied~\cite{Barata:2024nqo}.

While using track functions in the calculation of physical observables is in general quite complex, they have recently become a practical tool for the particular case of energy-correlator observables, introduced in Section~\ref{sec:ch3_energy_correlators}. Since energy correlators have recently been measured in heavy-ion collisions~\cite{CMS:2024ovv}, their theoretical simplicity makes them a natural target for theory development. For energy-correlator observables in vacuum, the only non-perturbative information entering the calculation is given by integer moments of the track functions~\cite{Chen:2020vvp},
\begin{align}
    T_i[n]
    =
    \int_0^1 dx\, x^n T_i(x)\, ,
    \label{eq:track_moments}
\end{align}
where $T_i[0]=1$. For example, the leading-order EEC cumulative distribution on tracks, $\Sigma_{\rm trk}$, can be expressed schematically as~\cite{Chen:2020vvp,Jaarsma:2023ell}
\begin{align}
    \frac{d\Sigma_{\rm trk}}{d\chi}
    =
    \sum_{ij}
    T_i[1]T_j[1]\,
    \frac{d\Sigma_{\rm partons}^{ij}}{d\chi}\, ,
    \label{eq:fact_theo}
\end{align}
where the sum extends over all partonic channels and $\chi>0$ denotes the separation angle between the two detectors used to define the EEC. This expression displays a clean factorization of the non-perturbative physics associated with the track measurement from the partonic cross section.

Given the advances in the description of the renormalization group equations used for the study of track functions, it is natural to transfer these developments to the study of energy loss. In particular, one would like to obtain a formula analogous to Eq.~\eqref{eq:fact_theo}, separating functions describing energy loss from a vacuum cross section. In this case, energetic jet constituents scatter off the medium and induce soft radiation, producing an energy flow from the jet to the medium. Ultimately, this effect reduces the energy fraction carried by the hadrons coming from the partonic cascade, with part of the energy now carried by the hadronic remnants of the medium. This motivates the introduction of \emph{energy-loss functions} $L_i(x)$ as the analogue of track functions for in-medium jet evolution~\cite{Barata:2024bmx}. The essential idea is to describe the fraction of the initial parton energy that remains in hadrons originating from the jet cascade, while excluding the hadronic remnants of the thermal bath. 

\subsection{Definition, evolution and application to reconstructed jets}
\label{subsec:elf_definition_jets}

We define the energy-loss function (ELF) $L_i(x)$ as the probability for an initial parton $i$, with energy $p_T$, to produce hadrons carrying a fraction $x$ of its total momentum. In vacuum, where no energy is transferred to the medium, one has $L_i(x)=\delta(1-x)$ to all orders. This reformulation of energy loss in the language of track functions is the central observation underlying the ELF formalism. The definition is sufficient for phenomenological applications, where a practical separation between bulk and jet constituents must be implemented. A formal construction of $L_i(x)$ in a framework that factorizes the collinear dynamics of particles within the jet from the dynamics of the medium is beyond the scope of the present discussion.

The ELFs, like track functions, are intrinsically non-perturbative objects. However, as for track functions, their renormalization group evolution can be studied perturbatively. From their definition, we postulate that the ELFs obey the same evolution equations as track functions. In the context of heavy-ion phenomenology, it is sufficient to consider the leading-order RG evolution, which reads~\cite{Chang:2013iba,Waalewijn:2012sv}
\begin{align}
    \mu \frac{d}{d\mu} L_i(x)
    &=
    \frac{\alpha_s}{2\pi}
    \int dx_1\,dx_2\,dz\,
    L_j(x_1)\,L_k(x_2)\,
    P_{i\to jk}(z)
    \delta\!\left(x-\left[zx_1+(1-z)x_2\right]\right)\, ,
    \label{eq:RG_D}
\end{align}
where $P_{i\to jk}$ denotes the leading-order regularized QCD splitting functions. Although track functions also satisfy Eq.~\eqref{eq:RG_D}, in the ELF context this RG evolution is only valid within the phase-space region where the vacuum anomalous dimensions are not modified by the medium, namely when $\mu$ is larger than the characteristic emergent medium scales.\footnote{The precise shape of this boundary will not be discussed here. In practice, such an effect has previously been introduced by imposing a phase-space constraint in the integration in Eq.~\eqref{eq:RG_D}; see, for example,~\cite{Mehtar-Tani:2017web,Caucal:2018dla}.} When such scales appear above the non-perturbative region, they allow the ELF to be computed directly, in analogy with what happens for track functions in the heavy-quark context. To illustrate this point without making explicit use of a medium model, one can assume that the presence of the background induces a decoherence scale $m_c$~\cite{Casalderrey-Solana:2012evi,Casalderrey-Solana:2011ule,Mehtar-Tani:2012mfa}, e.g., a dimensionful scale proportional to $\theta_c$ introduced in Section~\ref{subsec:scales}. This scale is associated with the ability of the medium to destroy quantum-mechanical coherence in the parton shower through multiple gluon exchanges with the QGP quasi-particles. It regulates the fragmentation pattern such that branchings occurring at angles smaller than $m_c/p_T$ do not generate additional colour sources that interact with the medium. In this sense, $m_c$ plays a role analogous to that of a quark mass. Thus, in analogy with boosted heavy-quark effective theory (bHQET)~\cite{Fickinger:2016rfd,Beisert:2003jj,Bigi:1993ex,Benzke:2010js,Beneke:2023nmj},\footnote{We thank R. Szafron for first pointing out this analogy and K. Lee for quantifying this observation.} and assuming the hierarchy of scales
\begin{align}
    p_T \gg m_c \gg \Lambda_{\rm QCD}\sim T\sim gT\, ,
\end{align}
with $T$ the plasma temperature, one can further factorize the track functions into a perturbative track function and a non-perturbative component.\footnote{In the bHQET context this would involve a combination of shape functions~\cite{Fickinger:2016rfd} describing the hadronization of heavy quarks~\cite{Neubert:1993um}.} An explicit realization of factorization structure is left for future work. The relevant point here is that the perturbative elements can be directly generalized from the bHQET context to jet energy loss, with the NLO partonic fragmentation function. Energy-loss effects, when hadronization effects are neglected, can then be modeled using known calculations of partonic energy loss~\cite{Mehtar-Tani:2017ypq,Baier:2001yt}, such as those entering Eq.~\eqref{eq:main_Qweights}. 

The presence of the scale $m_c$ has direct consequences for the use of ELFs in jet observables. First, taking the first moment of Eq.~\eqref{eq:RG_D},
\begin{align}
    \mu \frac{d}{d\mu}\langle x_i(\mu)\rangle
    &=
    \frac{\alpha_s}{2\pi}
    \int dx_1\,dx_2\,dz\,
    L_j(x_1)\,L_k(x_2)\,
    P_{i\to jk}(z)
    \left[zx_1+(1-z)x_2\right]\, ,
    \label{eq:elf_first_moment}
\end{align}
one obtains the evolution of the average energy loss $\langle \mathcal{E}_i\rangle
    =
    p_T-p_T\langle x_i\rangle$ as
\begin{align}
    \mu\frac{d}{d\mu}\langle\mathcal{E}_i\rangle
    =
    2\sum_{jk}
    \left(p_T-\langle\mathcal{E}_j\rangle\right)
    \gamma_2^{i\to jk}\, ,
    \label{eq:average_energy_loss_evolution}
\end{align}
where
\begin{align}
    \gamma_2^{i\to jk}
    =
    -\frac{\alpha_s}{2\pi}
    \int dz\, z\, P_{i\to jk}(z)\, .
\end{align}
This procedure can be generalized to higher moments, which give access to fluctuations of energy loss in the medium. Beyond this single-parton discussion, ELFs also enable a systematic treatment of jet energy loss. To this end, one introduces the jet ELF $\mathbb{L}_i(x,p_T,R)$, which describes the probability for a jet of radius $R$, initiated by a parton $i$ with momentum $p_T$, to carry energy $xp_T$. It can be connected to the partonic ELFs as
\begin{align}
    \mathbb{L}_i(x,p_T,R)
    &=
    \frac{1}{2}
    \sum_{jk}
    \int dz\,dx_1\,dx_2\,
    \mathcal{J}_{i\to jk}(p_T,R,z,\mu)
    L_j(x_1,\mu)\,
    L_k(x_2,\mu)\,
    \delta\!\left(x-\left[zx_1+(1-z)x_2\right]\right)\, .
    \label{eq:big_L_i}
\end{align}
The leading matching coefficients in vacuum, $\mathcal{J}_{i\to jk}$, are given in~\cite{Waalewijn:2012sv}. Medium contributions can also be obtained within a cut-off regularization scheme~\cite{Kang:2017frl,Li:2019dre}. For example, for the $q\to qg$ channel one obtains
\begin{align}
    \mathcal{J}_{qq}(p_T,R,z)
    =
    \frac{\alpha_s}{2\pi^2}
    \int_0^{p_T z(1-z)R}
    \frac{d^2\k}{\k^2}\,
    \Delta P_{q\to qg}(z,\k)\, ,
    \label{eq:medium_matching_jqq}
\end{align}
where $\Delta P_{q\to qg}(z,\k)$ denotes the power-suppressed medium correction to the leading-order regularized splitting function. This correction can be evaluated for several models of the underlying medium~\cite{Isaksen:2023nlr,Sievert:2019cwq}, which will we shall not discuss here. Further, the generalization to other flavour channels is immediate. The jet ELF $\mathbb{L}_i$ can, in principle, also be built without matching to the partonic ELFs, see for example~\cite{Lee:2023xzv}, but such a construction is considerably more involved. Using these ingredients, the probability that an energy fraction $x$ remains inside a jet initiated by a parton $i$ is given by
\begin{align}
    \frac{1}{\sigma_i}
    \frac{d\sigma_i}{dx}
    =
    \frac{\mathbb{L}_i(x,p_T,R)}
    {2(2\pi)^3 J_i(p_T R)}\, ,
    \label{eq:jet_energy_fraction_distribution}
\end{align}
where the normalization by $\sigma_i$ introduces the jet function on the right-hand side. This jet function can be obtained from the matching coefficients through the sum rule
\begin{align}
    J_i(p_T R)
    =
    \int_0^1 dz\, z
    \sum_j
    \mathcal{J}_{ij}(p_T,R,z)\, ,
    \label{eq:J_i}
\end{align}
which is valid for both the vacuum and medium components. The relations in Eqs.~\eqref{eq:big_L_i}--\eqref{eq:J_i} provide the basic ingredients required to describe energy loss in this framework, and can be systematically improved using standard perturbative techniques.

\subsection{Multipoint energy correlators and higher-twist extensions}
\label{subsec:elf_correlators_higher_twist}

The main application of the ELF formalism is to describe energy loss for observables that encode correlations between final-state fluxes. We focus on the particular case of energy correlators~\cite{Basham:1979gh,Basham:1978zq,Basham:1978bw,Basham:1977iq}. As discussed in Section~\ref{sec:ch3_energy_correlators}, these observables have recently become practical jet-substructure observables in hadron colliders~\cite{Dixon:2019uzg,Chen:2020vvp,Komiske:2022enw} and they have by now been measured in proton--proton~\cite{Komiske:2022enw,CMS:2024mlf,Tamis:2023guc,ALICE:2024dfl}, pA, and AA collisions~\cite{CMS:2024ovv}. To this end, consider the simple idealized setup of an $e^+e^-$ collision occurring in the presence of a thermal background, which mimics the presence of the QGP in heavy-ion collisions. The projected energy correlator (PENC) cumulative distribution can be written as~\cite{Dixon:2019uzg,Lee:2022uwt}
\begin{align}
    \Sigma^{[n]}\left(\chi,\log\frac{Q^2}{\mu^2},\mu\right)
    &=
    \int_0^1 dx\,x^n\,
    J^{[n]}\left(\log\frac{\chi x^2 Q^2}{\mu^2},\mu\right)
    \cdot
    H\left(x,\log\frac{Q^2}{\mu^2},\mu\right)\, ,
    \label{eq:penc_cumulative}
\end{align}
where the dot product is taken in flavour space and $Q$ denotes the large momentum scale in the process. We consider the leading-order hard function $H=2(\delta(1-x),0)$, so that the PENC depends only on the jet function. For the quark channel, this gives~\cite{Chen:2020vvp}
\begin{align}
    J_q^{[n]}
    =
    \frac{L_q[n]}{2^n}
    +
    \frac{\alpha_s}{4\pi}(-C_F)\,
    J_q^{[n],1}
    +
    \mathcal{O}(\alpha_s^2)\, .
    \label{eq:penc_jet_function}
\end{align}
The leading terms in $n$ at one loop, setting the renormalization scale to $\mu^2=\chi Q^2$ and neglecting modifications to the jet function induced by the medium, can be expressed as
\begin{align}
    J_q^{[2],1}
    &=
    \frac{37}{12}\,
    L_g[1]L_q[1]\, ,
    \nn
    J_q^{[3],1}
    &=
    \frac{611}{1200}\,
    L_g[2]L_q[1]
    +
    \frac{541}{300}\,
    L_g[1]L_q[2]\, .
    \label{eq:penc_one_loop_terms}
\end{align}
The numerical coefficients are obtained from integrals over the relevant splitting functions multiplied by the powers of energy fixed by the moment of the energy-loss function. This result can be directly compared with the one obtained by applying the quenching-weight approximation in Eq.~\eqref{eq:Q_weight_simplified}. In that case, the leading one-loop terms are
\begin{align}
    & J_q^{Q,[2],1}
    =
    \frac{37}{12}\,
    Q_g Q_q\, ,
    \qquad J_q^{Q,[3],1}
    =
    \frac{249}{48}\,
    Q_g Q_q\, .
    \label{eq:penc_quenching_weight_terms}
\end{align}
The two descriptions therefore differ for $n>2$. This mismatch between the ELF calculation and the quenching-weight approximation appears at higher order in the strong-coupling expansion. In particular, for terms proportional to $\alpha_s^k$, the two approaches do not agree for $k>n$. This illustrates the limitation of treating energy loss only through a cross-section-level quenching factor when the observable is sensitive to correlated energy flow.

Although the discussion so far has focused on leading-twist ELFs, the same logic can be systematically extended to higher twist. This extension is relevant for describing jet energy loss because, up to this point, correlations in the energy-loss and hadronization process have been neglected, in analogy with the treatment of track functions in~\cite{Waalewijn:2012sv,Chang:2013rca}. In the track-function context, charge correlations are expected to become important only at scales close to $\Lambda_{\rm QCD}$. For ELFs, this need not be the case, since multiparticle states can lose energy as a net effective colour charge~\cite{Casalderrey-Solana:2007knd}, especially at scales of order $m_c$ or smaller. In this regime, it is no longer possible to associate each outgoing parton with an independent ELF. Instead, one should use a multiparton ELF. Assuming that $m_c\gg\Lambda_{\rm QCD}$, these functions can still be studied within a perturbative picture.\footnote{Analogous studies can also be performed for track functions, but the corresponding charge-correlation corrections are power suppressed. For this reason, they have received comparatively little attention.} To illustrate the structure, consider the corresponding modification of the track-function factorization formula in Eq.~\eqref{eq:fact_theo}.\footnote{The expression is written for track functions, but the same discussion is expected apply to ELFs up to the introduction of the scale $m_c$.} One obtains
\begin{align}
    \frac{d\Sigma_{\rm trk}}{d\chi}
    =
    \sum_{ij}
    \int dx_1\,dx_2\,
    T_{ij}(x_1,x_2)\,
    x_1x_2\,
    \frac{d\Sigma_{\rm partons}^{ij}}{d\chi}\, ,
    \label{eq:higher_twist_track_factorization}
\end{align}
where $T_{ij}(x)=\int_{x_1} T_{ij}(x_1,x-x_1)$ describes the energy fraction $x$ carried by the charged hadrons that originated from the system of energetic partons $i$ and $j$ as a whole. This equation makes explicit the main structural change relative to Eq.~\eqref{eq:fact_theo}: the non-perturbative information is no longer encoded in a product of independent single-parton functions, but in a combined object associated with the multiparton system. In terms of field operators, such a double-index track function would generalize Eq.~\eqref{eq:def_tracks}. For example, for a $q\bar q$ system, the corresponding object would involve four quark field operators at two separate space-time points and would measure the energy carried by charged states relative to the total energy of the pair. At scales $\mu\gg\Lambda_{\rm QCD}$, one expects this object to be expressible as a convolution of single track functions, while at lower scales it should behave as the standard track function of an effective parton with the overall quantum numbers of the system. The renormalization group evolution of these combined track functions is correspondingly more involved than that of the single-parton functions. The exploration of the structure of higher-twist track functions and ELFs is left for future studies. The important point for the present discussion is that the ELF formalism is not restricted to independent single-parton energy loss. It provides a framework in which correlations in the energy-loss dynamics can be incorporated systematically, which is essential for a field-theoretic description of correlations in jet substructure.
 % file "Jet_Energy_Loss.tex"
\cleardoublepage

%%%%%%%%%%%%%%%%%%%%%%%%%%%%%%%%%%%%%%%%%%%%%%%%%%%%%%%%%%%%%%%%%%%%%%%%
%  Beyond_1to2.tex
%  Chapter 5: Beyond 1\to 2 splittings
%%%%%%%%%%%%%%%%%%%%%%%%%%%%%%%%%%%%%%%%%%%%%%%%%%%%%%%%%%%%%%%%%%%%%%%%
\chapter{Towards multi-point structure of jets in matter}
\label{chap:towards_multi_point_structure}

This chapter studies how QCD matter modifies jet substructure beyond the standard picture of a single medium-induced $1\to2$ splitting. In this sense, Section~\ref{sec:ch5_spin_correlations} examines spin correlations between two consecutive QCD emissions and how the resulting vacuum azimuthal modulation is altered when one emission occurs in the medium and the subsequent splitting takes place outside it. In particular, it shows how isotropic momentum broadening suppresses the spin-induced modulation, while anisotropic broadening can additionally generate a phase shift tied to the directional structure of the medium. Section~\ref{sec:ch5_energy_correlators} then turns to multi-point energy correlators, where perturbative BDMPS-Z corrections and medium-response contributions produce competing but systematically distinguishable structures in projected and shape-dependent higher-point energy correlators. The purpose of the chapter is to identify controlled calculations in which correlations among several emissions or several energy-flow insertions reveal information that may otherwise be harder to access with single-splitting observables.

\section{Collinear spin correlations in a QCD medium}\label{sec:ch5_spin_correlations}
\subsection{Overview of spin correlations in vacuum}
\label{subsec:ch5_vacuum_spin_correlations}

\begin{figure}[!t]
    \centering
    \includegraphics[width=\linewidth]{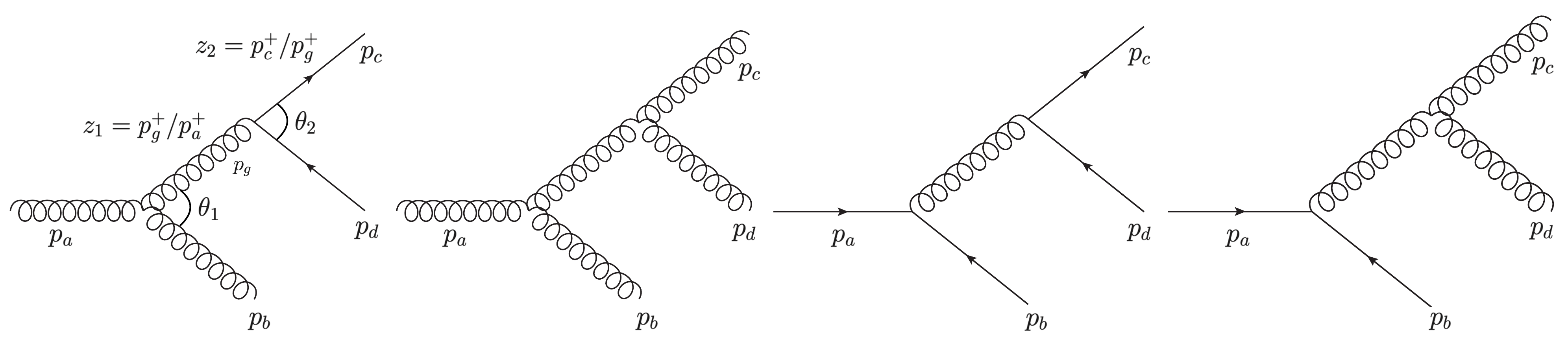}
    \caption{All relevant diagrams for the calculation of two nested collinear emissions ($\theta_2\ll\theta_1\ll 1$) with an intermediate off-shell gluon carrying energy $p_g^+ = z_1p_a^+$. The first diagram introduces the relevant splitting kinematics $(z_i,\theta_i)$ used throughout the calculation. The quark lines can be massless or massive.}
    \label{fig:diagrams}
\end{figure}

The calculation of theoretical predictions for jet observables at collider experiments involves $n$-parton matrix elements. QCD multiparton matrix elements adopt a particularly simple form in a variety of ordered limits, where kinematic properties of the emissions, such as their energy or angle, decrease from one emission to the next~\cite{Bassetto:1983mvz,Dokshitzer:1992ip,Campbell:1997hg,Catani:1999ss}. For instance, when all emissions are both soft and collinear, and widely separated in rapidity from each other, the tree-level $n$-parton matrix element reduces to the product of $n$ single-emission matrix elements. This factorisation property lies at the core of parton shower algorithms~\cite{Campbell:2022qmc}. As discussed in Section~\ref{sec:jet_evol_vac}, quantum interference effects driven by colour and spin degrees of freedom break this independent-emission picture. Colour interferences are relevant when considering multiple emissions at commensurate angles and strongly ordered in energies. They lead to a suppression of soft, wide-angle emissions, i.e., the phenomenon of angular ordering of soft radiation~\cite{Dokshitzer:1982fh,Webber:1983if,Marchesini:1987cf,Dokshitzer:1987nm,Gustafson:1987rq,Nagy:2019pjp,Hamilton:2020rcu,Forshaw:2025fif}. Spin correlations appear when considering processes with intermediate off-shell gluons and become relevant for soft, wide-angle gluons that split collinearly or in successive collinear emissions strongly ordered in angle. The latter case is the one considered in this section. These interference effects produce a non-trivial azimuthal correlation between the planes of consecutive emissions~\cite{Efremov:1981sh,Shatz:1983hv,Webber:1986mc,Webber:1987uy,Collins:1987cp,Knowles:1987cu,Knowles:1988vs,Knowles:1988hu,Richardson:2018pvo,Forshaw:2019ver,Corcella:2000bw,Karlberg:2021kwr,Hamilton:2021dyz,Bewick:2023tfi,Hoche:2025anb}, and they only arise in processes with an intermediate off-shell gluon, as a consequence of summing coherently over the spin states of this intermediate particle. Thus, in what follows, to study this type of correlations, we will focus on $1\to3$ processes of the form $a \rightarrow b\, g \rightarrow b\,c\,d$, where $b,c,d$ can be either light quarks ($q$), heavy quarks ($Q$) or gluons ($g$), i.e., $a,b,c,d\in (q, Q, g)$, with the intermediate off-shell gluon splitting as $g\to c\,d$ (see Fig.~\ref{fig:diagrams}). 

Before turning to the in-medium case, let us first discuss the kinematic regime in which the calculation is performed and review the emergence of spin-driven azimuthal correlations between subsequent splittings in vacuum. The limit of interest is the one in which all final-state particles are \textit{quasi}-collinear~\cite{Keller:1998tf,Catani:2000ef}.\footnote{This limit is a generalisation of the usual collinear limit~\cite{Catani:1999ss}, allowing for the definition of universal splitting functions that take mass effects into account~\cite{Campbell:1997hg,Catani:2000ef,Dhani:2023uxu,Craft:2023aew}. Put simply, one defines such a limit by taking all relative transverse momenta, or equivalently pair-wise angles, and on-shell masses to be small, while keeping their ratio fixed.} Further, this limit is considered in the strongly ordered regime where one enforces
\begin{equation}\label{eq:ch5_ang_ordering}
\theta_{2} \ll  \theta_{1} \ll 1\,, \quad (\theta_{1}\equiv \theta_{bg}, \theta_{2}\equiv \theta_{cd}) \,
\end{equation}
with $\theta_{ij}$ the opening angle of the $ij$ pair of particles. In this strongly ordered limit, the scattering amplitude for the $a\to b\,g \to b\,c\,d$ process factorizes into the production of a nearly on-shell gluon ($\cM_1$) and its subsequent branching ($\cM_2$).\footnote{Strictly speaking, one also needs the off-shell gluon to not take all the energy in $a\rightarrow b \,g$ for massless $a$ and $b$ and the splitting $g\rightarrow c\, d$ to not be highly asymmetric in energy for massive $c$ and $d$. See, e.g.,~\cite{Campbell:1997hg, Somogyi:2005xz, Braun-White:2022rtg,Craft:2023aew} for a more careful treatment of strongly ordered/"iterated" collinear limits.} That is, the colour-stripped amplitude reads
\begin{equation}\label{eq:ch5_general_factorised_M}
  i\cM_{a\to cbd} \propto \sum_{\lambda_g,\lambda_a} \cM_{2}^{\lambda_g \lambda_c \lambda_d}(p_c, p_d)\cM_{1}^{\lambda_{a} \lambda_{g}\lambda_b}(p_b, p_g)\cM_0^{\lambda_a}(p_a)\, ,
\end{equation}
where $\cM_0$ is some initial production amplitude, $p_a = p_b+p_c+p_d$, $p_g = p_c+p_d$ and each $\cM_i$ is fully contracted in Dirac space and involves the propagator pole of the corresponding $i$-th splitting. The colour indices have been omitted for clarity and $\lambda_{a,b,c,d}$ are polarisation indices.
The crucial point in Eq.~\eqref{eq:ch5_general_factorised_M} is that, despite $\cM_1$ and $\cM_2$ being factorised in Lorentz and Dirac indices, there is a coherent sum over physical polarisations $\lambda_g$, which upon squaring gives a double sum over polarisations, $\lambda_g$ and $\lambda_g'$. The interference terms, i.e., when $\lambda_g \neq \lambda_g'$, result in a dependence on the relative azimuthal angle between the planes of the two consecutive splittings $\psi_{12}\equiv \phi_2-\phi_1$ (see Eq.~\eqref{eq:phi_to_psi}).

To see this more clearly, the dependence on azimuthal angles can be trivially factored out from each amplitude $\cM_i$ (see Appendix~\ref{app:vertices_azimuthal_correlations}), such that the amplitude can be written as
\begin{align}\label{eq:ch5_M_decomp}
  i\cM_{a\to bg \to bcd} &\propto \sum_{\lambda_g}\tilde \cM_2^{\lambda_g\lambda_c\lambda_d}(z_2, |\q_{2}|)\tilde \cM_1^{\lambda_a\lambda_g \lambda_b}(z_1, |\q_{1}|)\cM_0(p_a)e^{i(\lambda_g-\bar \lambda_c- \bar \lambda_d) \phi_2} e^{i(\bar\lambda_a-\lambda_g - \bar\lambda_b) \phi_1}\,,
\end{align}
where the barred polarisation indices are given by $\bar\lambda=\lambda$ if the particle is a gluon, and $\bar\lambda=\pm \lambda/2$ if the particle is a quark or an anti-quark, respectively. The light-cone energy fractions\footnote{Throughout this paper we use light-cone coordinates defined as $p^\pm = (p^0+p^z)/\sqrt 2$ and $\p=(p^x,p^y)$.} are given by $z_1=p_g^+/p_a^+$ and $z_2=p_c^+/p_g^+$ and the relative transverse momenta of each splitting $\q_i$ are defined in Eq.~\eqref{eq:relative_momenta_def} and the azimuthal angle of each splitting is defined as $\phi_i=\tan^{-1}(\q^y_i/\q^x_i)$. From Eq.~\eqref{eq:ch5_M_decomp}, it is clear that, upon squaring the amplitude to obtain the cross section, all azimuthal angle phases vanish except those proportional to $\lambda_g$, the intermediate gluon's polarisation state. Thus, the amplitude squared reads\footnote{We assume that the initial parton is produced unpolarized and incoherently (see Section~\ref{sec:xsecs_BDMPSZ}). In doing so, we are discarding potential azimuthal correlations with initial-state particles.}
\begin{align}\label{eq:ch5_M2_decomp}
  |\cM_{a\to bg \to bcd}|^2 \propto |\cM_0(p_a)|^2\sum_{\lambda_g, \lambda_g'} F^{\lambda_g \lambda_g'}(z_i,|\q_i|)e^{i(\lambda_g-\lambda_g')\psi_{12}}\,,
\end{align}
where $\psi_{12}=\phi_2-\phi_1$ and
\begin{align}\label{eq:ch5_F_definition}
  F^{\lambda_g \lambda_g'}(z_i, |\q_i|) & = \frac{1}{2}\sum_{\lambda_a, \lambda_b,\lambda_c,\lambda_d}\tilde \cM_2^{\lambda_g \lambda_c \lambda_d}(z_2,|\q_{2}|)\tilde \cM_1^{\lambda_a \lambda_g \lambda_b}(z_1,|\q_{1}|)\nonumber\\
    &\times \left(\tilde \cM_2^{\lambda_g' \lambda_c \lambda_d}(z_2,|\q_{2}|)\tilde \cM_1^{\lambda_a \lambda_g' \lambda_b}(z_1,|\q_{1}|)\right)^{\ast}\,.
\end{align}
Since $|\cM|^2$ must be real, by defining $F^{\lambda_g\lambda_g'} = F_1^{\lambda_g} \delta^{{\lambda_g},{\lambda'_g}} + F_2^{\lambda_g}\delta^{\lambda_g, -\lambda'_g}$ one can write
\begin{align}\label{eq:ch5_real_imag_F_M2}
  |\cM_{a\to bcd}|^2 \propto  |\cM_0(p_a)|^2 \, \sum_{\lambda_g} \left[\Re F_1^{\lambda_g}(|\q_{i}|)+ \Re F_2^{\lambda_g}(|\q_{i}|)\cos(2\psi_{12})\lambda_g\,\Im  \,F_2^{\lambda_g}(|\q_{i}|)\sin(2\psi_{12})\right]\,.
\end{align}
At this stage, the azimuthal dependence has been isolated into two terms proportional to $\cos (2\psi_{12})$ and $\sin (2\psi_{12})$. This dependence is simply a consequence of coherently summing over gluon polarisation states, since it comes from the terms with $\lambda_g' \neq \lambda_g$. In vacuum, the functions $F_i^{\lambda_g}$ do not have an explicit polarisation dependence, i.e.,  $F_i^{\lambda_g} = F_i$. Thus, the $\sin(2\psi_{12})$ term vanishes and one is left with a $\cos(2\psi_{12})$ modulation which is well known in the literature~\cite{Shatz:1983hv, Knowles:1987cu,Collins:1987cp}. The phenomenological implications of this modulation have recently been studied in, e.g.,~\cite{Richardson:2018pvo,Chen:2020adz,Karlberg:2021kwr,Hamilton:2021dyz,Hoche:2025anb,Song:2025bdj}.

Let us now explicitly compute the azimuthal angle modulation of the vacuum cross section for the process $a\rightarrow b \, g\rightarrow b \, c \, d$ in the strongly ordered \textit{quasi}-collinear limit. As already mentioned, in this limit the amplitude can be written as a product of propagator denominators and vertex matrix elements that are fully contracted in Dirac and Lorentz indices,
\begin{equation}\label{eq:ch5_vac_M}
    i\cM_{a\to bcd} = \sum_{\lambda_g}\sum_{m,l} \left(\frac{i}{Q_{g}^2}ig\mathbf T_{ij}^a V_{g\rightarrow c d}^{\lambda_g \lambda_c \lambda_d}(z_2,\q_{2})\right)\left(\frac{i}{Q_a^2-m_1^2}  ig\mathbf T^a_{kl}V_{a\rightarrow g b}^{\lambda_a\lambda_g \lambda_b}(z_1,\q_{1})\right)\cM_0^{l}(p_a)\,,
\end{equation}
where the colour and polarisation of the intermediate gluon and the colour of the initial parton $a$ are explicitly summed over.  For $q\rightarrow gq$ the colour matrices are in the fundamental representation $(\mathbf T^a)_{ij} = t^a_{ij}$ while for $g\rightarrow gg$ one has the adjoint representation $(\mathbf T^a)_{bc} = -if^{abc}$. The vertex matrix elements $V$ are defined and calculated in Appendix~\ref{app:vertices} and the propagator denominators are written in the collinear limit as a function of the virtuality
\begin{subequations}\label{eq:ch5_ordering_vacuum}
\begin{align}
  & Q_a^2-m_1^2 = p_a^2-m_a^2= \frac{\q_1^2+z_1^2 m_1^2}{z_1(1-z_1)}+\frac{\q_2^2 + m_2^2}{z_1z_2(1-z_2)} \approx \frac{\q_1^2+z_1^2 m_1^2}{z_1(1-z_1)}\label{eq:ch5_qa_m12}\,,\\
  & Q_g^2 = (p_c + p_d)^2 = \frac{\q_2^2 + m_2^2}{z_2(1-z_2)}\,,
\end{align}
\end{subequations}
where $z_1=p_g^+/p_a^+$ and $z_2=p_c^+/p_g^+$, $\q_i$ are the relative transverse momenta of each splitting (see Eq.~\eqref{eq:phi_to_psi}), $m_i$ is the quark mass in the $i$-th splitting and in Eq.~\eqref{eq:ch5_qa_m12} the strong ordering condition $\theta_1\gg \theta_2$ was imposed. Using the explicit vertex matrix element expressions in Appendix~\ref{app:vertices}, the squared amplitude summed over quantum numbers can be straightforwardly computed. To calculate the differential cross section one then multiplies by the appropriate phase-space element (see Eq.~\eqref{eq:phase-space-final}), resulting in
\begin{equation}\label{eq:ch5_xsection_vac}
   (2\pi)^2\frac{d\sigma^{\rm vac}}{\sigma_0\,d^6\Omega_{\rm PS}} = \left[\frac{\alpha_s}{\pi} C_1 \frac{P^M_{a\rightarrow bg}(z_1,\theta_{1})}{\theta_1(1+\tilde\theta^2_{m,1})}\right]\left[\frac{\alpha_s}{\pi}C_2\frac{ P^M_{g\rightarrow cd}(z_2,\theta_{2})}{\theta_2(1+\tilde\theta^2_{m,2})}\right]\left(1 + a \cos(2\psi_{12})\right)\,,
\end{equation}
with
\begin{equation}
\label{eq:ch5_adef_vac}
a \equiv S_{g\rightarrow cd}\frac{4z_2(1-z_2)(1-z_1)}{z_1[P_{a\rightarrow b g}(z_1)+z_1\tilde\theta^2_{m,1}][P_{g\rightarrow cd}(z_2)+\tilde\theta^2_{m,2}]}\, .
\end{equation}
The cross-section in Eq.~\eqref{eq:ch5_xsection_vac} has the structure of factorized emissions, with residual correlations encoded in a spin-driven modulation in $\psi_{12} = \phi_2 - \phi_1$, the azimuthal angle between the planes of the two splittings given in terms of the azimuthal orientations of $\q_i$ (see Eq.~\eqref{eq:phi_to_psi}). The sign of $a$ indicates whether spin correlations are enhanced when the two splittings occur in the same plane ($a>0$) or in perpendicular planes ($a<0$). The factor $C_i$ is the colour factor associated to the $i$-th splitting, $S_{g\rightarrow gg} = +1$ and $S_{g\rightarrow q\bar q} = -1$. The functions $P^M_{i\rightarrow jk}$ are the spin-averaged massive splitting functions that read
\begin{subequations}\label{eq:ch5_mass_ap}
\begin{align}
    & P^{M}_{a\rightarrow bg}(z_1,\theta_{1}) = \frac{\theta_{1}^2 P_{a\rightarrow gb}(z_1)+z_1\theta_{m,1}^2}{\theta_{1}^2 + \theta_{m,1}^2 }\, ,\\
    & P^M_{g\rightarrow cd}(z_2,\theta_{2}) = \frac{\theta_{2}^2P_{g\rightarrow cd}(z_2) + \theta_{m,2}^2}{\theta_{2}^2+\theta_{m,2}^2}\,,
\end{align}
\end{subequations}
with $P_{i\rightarrow jk}$ given by Eq.~\eqref{eq:vac_splitting_functions} and where the dead-cone angles are
\begin{align}
\label{eq:ch5_deadcone_def}
  & \theta_{m,1}  = \frac{m_1}{(1-z_1)E_a}\,,\qquad \theta_{m,2} = \frac{m_2}{z_1z_2(1-z_2)E_a}\, ,
\end{align}
and their ratios with respect to the opening angle of the splitting are denoted by $\tilde\theta_{m,i} \equiv \theta_{m,i}/\theta_i$. In Eq.~\eqref{eq:ch5_deadcone_def}, $E_a = p_a^+/\sqrt{2}$ is the energy of particle $a$.
Finally, the initial production cross section is given by the following integration over total momentum $d^3\Omega_a$ (see Eq.~\eqref{eq:phase-space-final})
\begin{equation}\label{eq:ch5_sigma_0_def}
  \sigma_0 =  \sum_{l}\int d^3\Omega_a|\cM_0^{l}(p_a)|^2.
\end{equation}
Note that after integrating Eq.~\eqref{eq:ch5_xsection_vac} over $\phi_1$ and $\psi_{12}$ one simply recovers the product of the $a\rightarrow b\,g$ and $g\rightarrow c\,d$ massive splitting functions.

\begin{figure}[!t]
    \centering
         \includegraphics[width=0.49\textwidth]{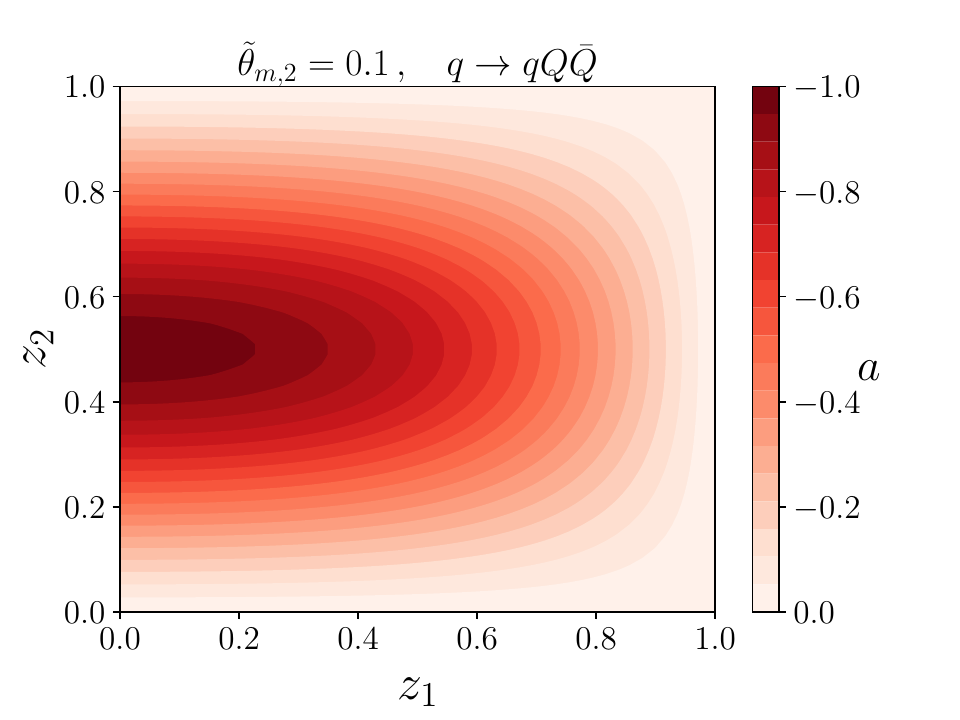}
         \includegraphics[width=0.49\textwidth]{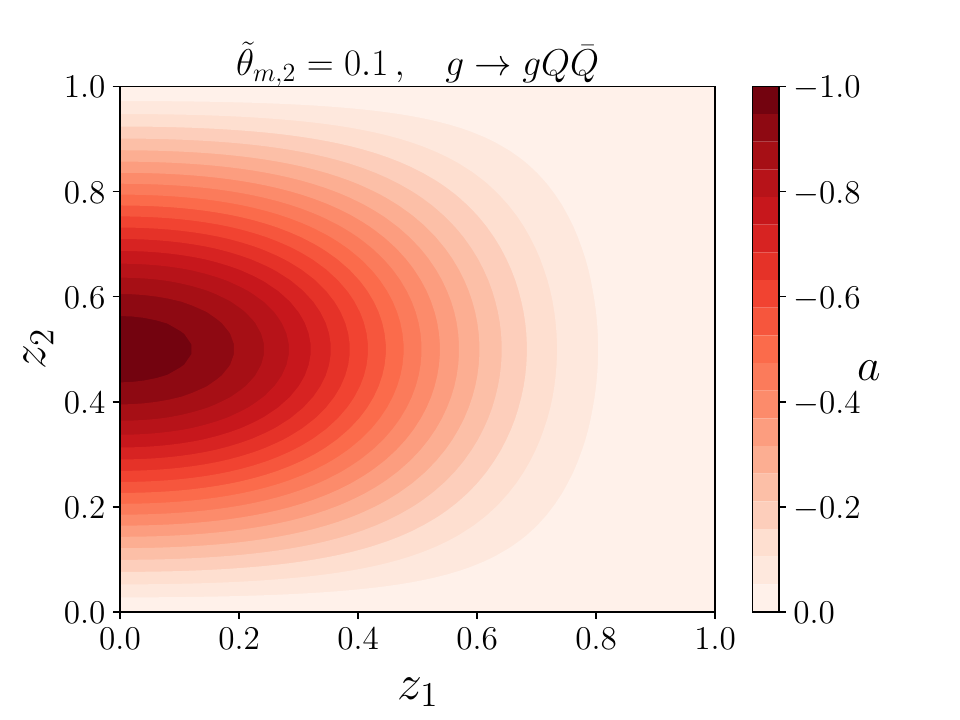} 
         \\
        \includegraphics[width=0.49\textwidth]{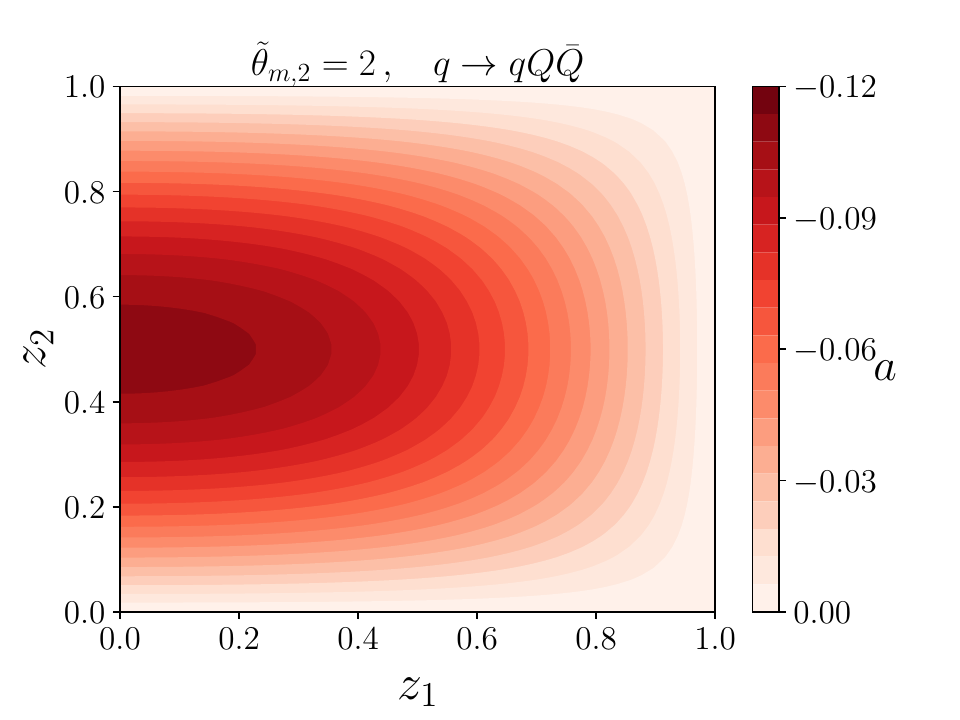}
         \includegraphics[width=0.49\textwidth]{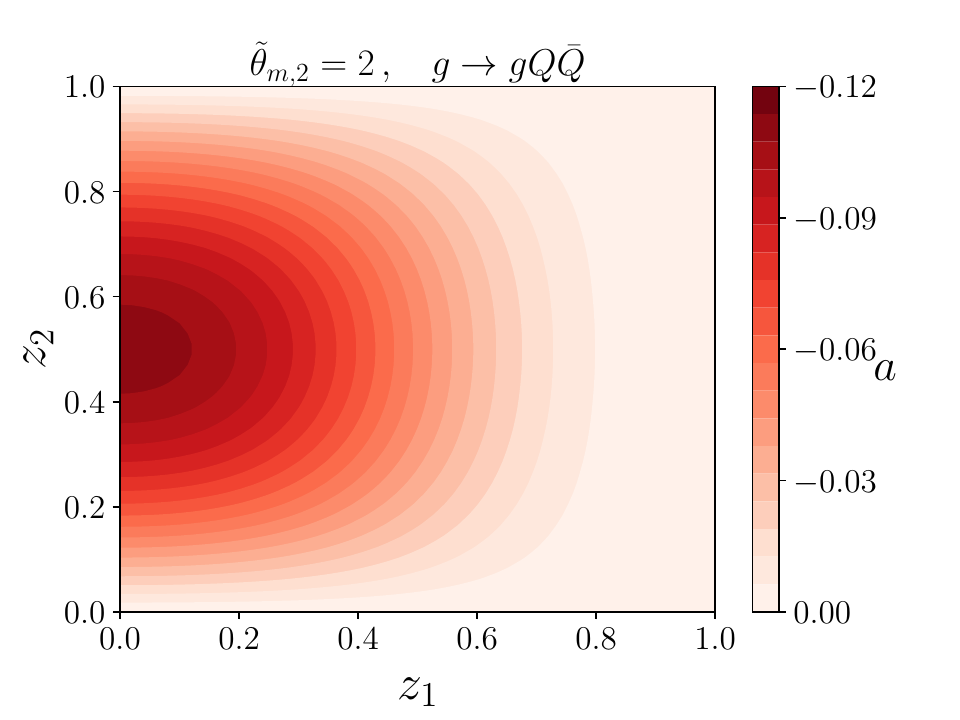} 
     \caption{Amplitude of the spin modulation signal, $a$, defined in Eq.~\eqref{eq:ch5_adef_vac}, as a function of both energy fractions $z_i$ for two different first splittings (massless $q\rightarrow qg$ on the left and $g\rightarrow gg$ on the right), fixing the second splitting to be $g\rightarrow Q\bar Q$. From top to bottom we change the opening angle of the second splitting with respect to the dead-cone angle, i.e., $\tilde\theta_{m,2}=\theta_{m,2}/\theta_2$ with $\theta_{m,2}$ defined in Eq.~\eqref{eq:ch5_deadcone_def}.} 
     \label{fig:a_vac}
\end{figure}

The amplitude of the spin-modulation signal in Eq.~\eqref{eq:ch5_adef_vac} is evaluated as a function of $(z_1,z_2)$ in Fig.~\ref{fig:a_vac}. In these results, $\tilde\theta_{m,1}=0$. The left and right panels of Fig.~\ref{fig:a_vac} contain the $q\to qg \to q q'\bar q'$ and $g\to gg \to g q\bar q$ processes, respectively. The other relevant scale in Eq.~\eqref{eq:ch5_adef_vac} is $\tilde\theta_{m,2}$. Since the massless case has been recently discussed~\cite{Richardson:2018pvo,Karlberg:2021kwr,Hoche:2025anb}, the figure focuses on a massive $g\to Q\bar Q$ second splitting and chooses $\tilde\theta_{m,2}=0.1$ in the top panels and $\tilde\theta_{m,2}=2$ in the bottom panels. For $\tilde\theta_{m,2}=0.1$, mass effects are suppressed. For both channels, the size of spin correlations reaches its maximum around $z_1\lesssim 0.1$ and $z_2\sim0.5$, i.e., when the initial parton is soft and the energy is shared democratically in the second branching. Comparing the two processes, spin correlations stay large for the $q$-initiated process in a wide region of phase space. In fact, they only vanish when the two splittings are widely asymmetric, i.e., $z_{1,2}\to 0,1$. For the $g$-initiated channel the spin modulation falls off much faster than in the quark case and becomes negligible when $z_1\gtrsim0.8$. The second row of Fig.~\ref{fig:a_vac} shows results for a second splitting whose opening angle is two times smaller than the associated dead-cone angle and thus mass effects are expected to play a role, see Eq.~\eqref{eq:ch5_adef_vac}. Indeed, for both channels there is a significant depletion of spin correlations by almost a factor 10.

\subsection{BDMPS-Z spectrum for an in-medium emission followed by a vacuum emission}
\label{subsec:ch5_medium_vacuum_spin_correlations}

The vacuum discussion of the previous subsection provides the baseline for discussing how spin-driven interference effects are modified in a QCD medium. Interference phenomena in a medium have been extensively studied from the perspective of colour coherence, most notably in antenna configurations where a first energetic splitting is followed by soft radiation emitted either inside or outside the medium~\cite{Mehtar-Tani:2010ebp,Mehtar-Tani:2011hma,Casalderrey-Solana:2011ule,Mehtar-Tani:2011vlz,Mehtar-Tani:2011lic,Armesto:2011ir,Mehtar-Tani:2012mfa,Casalderrey-Solana:2012evi,Calvo:2014cba,Barata:2021byj,Abreu:2024wka,Andres:2025prc,Kuzmin:2025fyu}. These studies have shown that medium interactions during the formation and propagation of the antenna modify the rate and phase space for a subsequent soft-gluon emission with respect to vacuum. In particular, these interactions degrade the colour correlation between the two antenna legs and eventually lead to a decoherent state in which each leg emits independently. Here we will focus on spin interference inside a QCD medium, which, to our knowledge, has not been addressed so far. Thus, the calculation below focuses on the fully differential cross-section for two emissions in the same kinematical setup as in Section~\ref{subsec:ch5_vacuum_spin_correlations}, with the first splitting occurring inside the medium and the second splitting occurring outside it. We work in this approximation for simplicity, since the full $1\rightarrow 3$ cross-section accounting for multiple scatterings between all final-state particles and the QCD medium is considerably more involved.~\cite{Casalderrey-Solana:2015bww,Arnold:2015qya,Arnold:2022mby,Arnold:2023qwi,Arnold:2024whj}.\footnote{The collinear limit of the $1\to 3$ matrix element at first order in opacity was computed in~\cite{Fickinger:2013xwa}. A systematic comparison between the dense-medium calculation considered here and the single-opacity limit is left for future work.}

We shall work within the BDMPS-Z formalism introduced in Section~\ref{sec:BDMPS_Z}, whereby the medium is treated as a stochastic coloured background field and all parton energies are parametrically larger than any medium or transverse scale. Before carrying out the calculation, let us discuss the analogue of Eq.~\eqref{eq:ch5_real_imag_F_M2} in a medium. First, the considered scale hierarchy implies that parton-medium interactions do not significantly change the parton's virtuality, such that one can consider the intermediate gluon to stay nearly on-shell during its in-medium propagation. As shown in Section~\ref{subsec:propagator_derivation}, the hierarchy further implies that spin-flip terms can be neglected. Hence, one can assume the form of the amplitude in Eq.~\eqref{eq:ch5_general_factorised_M} to hold 
\begin{equation}\label{eq:ch5_M2_decomp_medium}
  \int_{\p_a}|\cM^{\rm med}_{a\to bg\to bcd}|^2 \propto \sum_{\lambda_g, \lambda_g'} F_{\rm med}^{\lambda_g \lambda_g'}(\q_1, \q_2)e^{i(\lambda_g-\lambda_g')\psi_{12}}\,,
\end{equation}
where the dependence on the remaining relevant variables, e.g. on $z_i$, has been omitted. The fact that in Eq.~\eqref{eq:ch5_M2_decomp_medium} the total transverse momentum dependence is integrated out is valid only for the case of transversely homogeneous media with instantaneous time correlations (see Section~\ref{sec:ch3_gqqbar}). Unlike the vacuum case, $F_{\rm med}^{\lambda_g \lambda_g'}(\q_1, \q_2)$ now depends not only on the moduli of the $\q_i$ vectors but also on their orientations $\phi_i$. This means, in particular, that one should expect medium modifications to introduce a non-trivial dependence on the angle between planes $\psi_{12} = \phi_2-\phi_1$. Since the second splitting is taken to be outside the medium, the $\q_2$ dependence is given trivially by the vacuum expression for the second splitting
\begin{equation}
    F_{\rm med}^{\lambda_g \lambda_g'}(\q_1, \q_2)\underset{\substack{g\to cd \\ \text{in vacuum}}}{\rightarrow} F_{\rm med}^{\lambda_g \lambda_g'}(\q_1,|\q_2|)\,.
\end{equation}
One can now decompose $F_{\rm med}^{\lambda_g \lambda_g'}$ by projecting into the two possible helicity state combinations akin to the vacuum case, $F_{\rm med}^{\lambda_g \lambda_g'} = F_{\rm med,1}^{\lambda_g}\delta^{\lambda_g\lambda_g'}+F_{\rm med,2}^{\lambda_g}\delta^{\lambda_g, -\lambda_g'}$, and write the parametric dependence of the squared matrix element as
\begin{align}\label{eq:ch5_real_imag_F_M2_medium}
  \int_{\p_a}|\cM^{\rm med}_{a\to bg\to bcd}|^2 \propto  \sum_{\lambda_g}&\left[\Re F_{{\rm med},1}^{\lambda_g}(\q_{1},|\q_{2}|)+ \Re F_{{\rm med},2}^{\lambda_g}(\q_{1},|\q_{2}|)\cos(2\psi_{12}) \right.\nn
    &\left. - \lambda_g\,\Im  \,F_{{\rm med},2}^{\lambda_g}(\q_{1},|\q_{2}|)\sin(2\psi_{12})\right]\,.
\end{align}
Thus, one observes two major, qualitative modifications on the squared matrix-element with respect to vacuum: (i) it now depends on the orientation of $\q_1$, or equivalently on the value of $\phi_1$, and (ii) a potential non-zero $\sin(2\psi_{12})$ term if $F_{\rm med,2}^{\lambda_g}$ has a polarisation dependence. On top of these two effects, the dependence of $F_{\rm med,i}^{\lambda_g}$ on $z_1$ and $|\q_1|$ can also be modified with respect to vacuum. As argued in Appendix~\ref{app:vertices_azimuthal_correlations}, effects (i) and (ii) are only realized for a medium model that is not rotationally invariant, e.g., a medium with some preferred axis as considered in Chapter~\ref{chap:anisotropic_early_time} (see  Eq.~\eqref{eq:squared_vertex_polarisation}). Thus, if one further considers an isotropic medium, the dependence on $\phi_{1}$ and the $\sin(2\psi_{12})$-term vanish and one is left with 
\begin{equation}\label{eq:ch5_real_imag_F_M2_medium_iso}
  \int_{\p_a}|\cM^{\rm med}_{a\to bg\to bcd}|^2 \underset{\substack{\text{isotropic} \\ \text{medium}}}{\to} \propto  \sum_{\lambda_g}\left[\Re F_{{\rm med},1}^{\lambda_g}(|\q_{1}|,|\q_{2}|)+ \Re F_{{\rm med},2}^{\lambda_g}(|\q_{1}|,|\q_{2}|)\cos(2\psi_{12})\right]\,.
\end{equation}
To summarise: in an anisotropic medium, one expects a $\sin(2\psi_{12})$-term together with a non-trivial dependence of the cross-section on $\phi_1$. In turn, for an isotropic medium, the azimuthal modulation is proportional to $\cos(2\psi_{12})$ as in vacuum, with both an overall modification and a change in the relative magnitude of the $\psi_{12}$-independent and $\cos(2\psi_{12})$-terms that will depend on both the phase-space point of the first splitting and the medium properties.\footnote{The relatively simple azimuthal structure of the in-medium cross-section given in Eq.~\eqref{eq:ch5_real_imag_F_M2_medium} is not expected to hold when the second emission occurs inside the medium, as discussed in Appendix~\ref{app:vertices_azimuthal_correlations}.} These functional dependencies can now be computed explicitly.

Following the previous heuristic discussion, let us now calculate the differential cross-section for the process $a \rightarrow b\,g \rightarrow b\,c\,d$, in the strongly ordered \textit{quasi}-collinear limit, inside a QCD medium. We take the medium to be static, homogeneous and of fixed length $L^+$. As already mentioned, we work in the approximation where the $g\to c\,d$ splitting happens outside of the medium. Kinematically, this corresponds to a collinear splitting whose formation time is parametrically larger than the medium length. We work within the high energy BDMPS-Z formalism, following the effective Feynman rules summarized in Eq.~\ref{eq:effective_feynman_rules}. Further, we will keep the initial production assumptions discussed in Section~\ref{sec:xsecs_BDMPSZ}, namely that the initial parton $a$ is produced unpolarized and incoherently, at the light-cone time boundary of the medium, fixed to $x_0^+ \equiv 0$, with fixed light-cone energy $p_a^+$. Thus, in the mixed space of light-cone time and transverse momentum, the amplitude for the process $a\rightarrow b\,g\rightarrow b \,c \,d$, where the splitting $g\rightarrow c\,d$ is imposed to happen outside of the medium, can be written as
\begin{align}\label{eq:ch5_med_M}
   i\cM &= \sum_{\lambda_g}\frac{e^{i\frac{\p_b^2}{2p_b^+}x^+_{\infty}}}{4p_a^+p_g^+}[ig\mathbf T_{ij}^b V_{g\rightarrow cd}^{\lambda_g \lambda_c \lambda_d}(z_2,\q_{2})] \int_0^{x^+_{\infty}} dx_1^+\,e^{-\epsilon x_1^+}e^{i\frac{z_1^2m_1^2}{2\omega_a^+}x_1^+}\int_{x_1^+}^{x^+_{\infty}} dx_2^+\,e^{-\varepsilon x_2^+}\Theta(x_2^+ > L^+)\, \nn
    &\times e^{i\left(p_c^- + p_d^-\right)x_2^+} \int_{\k_b,\k_g}G^{ba}_{A}(x_2^+,\p_g;x_1^+,\k_g|p_g^+) G_{R_a}^{kn}(x^+_{\infty},\p_b;x_1^+,\k_b|p_b^+) [ig\mathbf T^a_{np}V_{a\rightarrow gb}^{\lambda_a \lambda_g \lambda_b}(z_1,\bkappa)]\, \nn
    &\times \int_{\p_0} G_{R_a}^{pq}(x_1^+,\k_0;0,\p_0|p_a^+)\cM_0^{q}(\vec p_0) \, ,
\end{align}
where $\mathbf T^a_{jk} = t^a_{jk}\,(-if^{ajk})$ in the fundamental (adjoint) representation and $R_a = F \,(A)$ if the initial particle is a quark (gluon). Most kinematic variables have been defined in Sec.~\ref{subsec:ch5_vacuum_spin_correlations}, while the remaining ones are  $p_c^- = (\p_c^2 + m_2^2)/(2p_c^+)$, $p_d^- = (\p_d^2 + m_2^2)/(2p_d^+)$, $\p_g = \p_c+\p_d$ and the intermediate momenta given by $\k_0 = \k_b + \k_g$, $\vec p_0 = (p_a^+,\p_0)$ and $\bkappa = (1-z_1)\k_g - z_1\k_b$. Note that light-cone energy is conserved, $p_0^+ = p_a^+$, but the total transverse momentum is not, $\p_0 \neq \p_a = \p_b+\p_c+\p_d$. Finally, as discussed in Section~\ref{sec:feynman_rules}, the $\epsilon$-prescription ($\epsilon\rightarrow 0$) assures that medium interactions are adiabatically turned off~\cite{Wiedemann:2000za}, such that one recovers the correct vacuum limit when taking $x^+_\infty \rightarrow +\infty$. 

The integration variables $x_1^+$ and $x_2^+$ in Eq.~\eqref{eq:ch5_med_M} correspond to the light-cone times of the first and second splittings, respectively. As discussed, the second splitting is taken to be outside the medium and so the condition $\Theta(x_2^+>L^+)$ is imposed in Eq.~\eqref{eq:ch5_med_M}. This restriction on the $x_2^+$ integral naturally ignores contributions to the medium modification which would only be included in the full $1\rightarrow 3$ calculation inside the medium, which is considerably more involved. Thus, this result is expected to capture the correct qualitative behaviour in the case where there is a strong hierarchy on the formation time (see Section~\ref{subsec:formation_time}) of the second splitting ($\tau^+_2$) with respect to the first splitting's ($\tau^+_1$) and the medium length, i.e.,
\begin{align}\label{eq:ch5_form_time_ordering}
    \tau^+_2 \gg L^+,\,\tau^+_1 \,, \quad \text{with}\quad \tau^+_1 \equiv 2p_a^+/(Q_a^2 -m_1^2)\,,\,\, \tau^+_2 \equiv 2p_g^+/Q_g^2 \,.
\end{align}
Additionally, this restriction on the $x_2^+$ integration introduces spurious terms when trying to recover the vacuum limit at the amplitude level. Indeed, replacing all propagators by their vacuum counterpart (see Eq.~\eqref{eq:vacuum_mixed_propagator}), leads to the following result
\begin{equation}\label{eq:ch5_M_vac_limit}
    i\cM_{a\to bcd} \underset{\substack{\text{vacuum} \\ \text{limit}}}{\to} i\cM_{a\rightarrow bcd}^{\rm vac}\left[\frac{\tau^+_2}{\tau^+_2-\tau^+_1}e^{iL^+/\tau^+_2} - \left(\frac{\tau^+_2}{\tau^+_2-\tau^+_1} - 1\right)e^{iL^+/\tau^+_1} \right]\,,
\end{equation}
where $\cM^{\rm vac}_{a\rightarrow bcd}$ is the vacuum amplitude calculated in Eq.~\eqref{eq:ch5_vac_M}. Hence, in order to recover the exact vacuum limit in Eq.~\eqref{eq:ch5_M_vac_limit}, one needs to impose the same formation time hierarchy as in Eq.~\eqref{eq:ch5_form_time_ordering}. As long as one does not have $z_1\rightarrow 1$ (intermediate gluon taking all the energy on the first branching), this strong time-ordering is equivalent to the strong angular-ordering imposed in Eq.~\eqref{eq:ch5_ang_ordering}.

Finally, the light-cone time locality of the medium average makes it convenient to separate the amplitude into two contributions: $\cM^{\rm in}$, where the first splitting is inside the medium, i.e., $x_1^+ \leq L^+$, and $\cM^{\rm out}$, where $x_1^+ > L^+$. The former contribution reads
\begin{align}\label{eq:ch5_M_in}
   i\cM^{\rm in} &= \sum_{\lambda_g}\frac{e^{i\left(Q_a^2 -m_1^2 + \p_a^2\right)L^+/(2p_a^+)}}{2p_a^+}\left[\frac{i}{Q_{g}^2} ig\mathbf T_{ij}^b V_{g\rightarrow cd}^{\lambda_g \lambda_c \lambda_d}(z_2,\q_2)\right]\int_0^{L^+} dx_1^+ e^{-i\frac{z_1^2m_1^2}{2\omega_a^+}(L^+-x_1^+)} \, \nn
  & \times \int_{\k_b,\k_g,\p_0} G^{ba}_{A}(L^+,\p_g;x_1^+,\k_g|p_g^+) G^{kn}_{R_a}(L^+,\p_b;x_1^+,\k_b|p_b^+)[ig\mathbf T^a_{np}V_{a\rightarrow gb}^{\lambda_a \lambda_g \lambda_b}(z_1,\bkappa)]\, \nn
    & \times \int_{\p_0} G^{pq}_{R_a}(x_1^+,\k_0;0,\p_0|p_0^+)\cM_0^{q}(\vec p_0)\, ,
\end{align}
while the latter is given by
\begin{align}\label{eq:ch5_M_out}
   i\cM^{\rm out}& = \sum_{\lambda_g}e^{i\left(Q_a^2 -m_1^2 + \p_a^2\right)L^+/(2p_a^+)}\left[\frac{i}{Q_{g}^2}ig\mathbf T_{ij}^a V_{g\rightarrow cd}^{\lambda_g \lambda_c \lambda_d}(z_2,\q_{2})\right]\left[\frac{i}{Q_a^2-m_1^2}  ig\mathbf T^a_{kp}V_{a\rightarrow gb}^{\lambda_a \lambda_b\lambda_g}(z_1,\q_{1})\right]\nn
    & \times \int_{\p_0} \,G^{pq}_{R_a}(L^+,\p_a;0,\p_0|p_a^+)\cM_0^{q}(\vec p_0)\, .
\end{align}
Thus, the fully differential in-medium cross-section for the process $a\rightarrow b\,g\rightarrow b\,c\,d$ can be separated into time intervals according to
\begin{equation}\label{eq:ch5_M_region_separation}
  \cM\cM^{\dagger} = \cM^{\rm in}\cM^{\rm in, \dagger} + 2\Re\,\left(\cM^{\rm in}\cM^{\rm out, \dagger}\right)+\cM^{\rm out}\cM^{\rm out, \dagger}\,.
\end{equation}
The amplitude squared then needs to be averaged over quantum numbers and over medium configurations according to Eq.~\eqref{eq:A_correlator}. Combining this amplitude squared with the corresponding phase-space element in Appendix.~\ref{app:kinematics_phase_space}, leads to the fully differential cross-section 
\begin{equation}\label{eq:medium_xsec}
  \frac{16(2\pi)^6}{z_1^3(1-z_1)z_2(1-z_2)\theta_{1}\theta_{2}}\frac{d\sigma}{d^6\Omega_{\rm PS}} = \frac{1}{2}\int\, d^3\Omega_a (p_a^+)^4\sum_{\rho_i, \lambda_i}\left\langle \cM\cM^{\dagger} \right\rangle\,,
\end{equation}
where the sum runs over colours ($\rho_i$) and spin states ($\lambda_i$), and the factor $1/2$ accounts for the average over the initial parton spin state. The term $d\sigma^{\rm out-out}$ simply reduces to the vacuum result given by Eq.~\eqref{eq:ch5_xsection_vac}, since both splittings happen outside the medium and the broadening of the initial parton is integrated out due to the integration over total transverse momentum $\p_a$.
To compute the remaining terms, we note that from the medium-averaging point of view, only the first splitting matters, since the second occurs outside of the medium. Following the discussion around Eq.~\eqref{eq:A_correlator} for an homogeneous medium with local colour and time correlations, and similarly to the calculation in Section~\ref{sec:ch3_gqqbar}, one can apply conservation of total transverse momentum between the amplitude and its cc, as well as consider them to be in a colour singlet state at all times. Finally, by integrating over total momentum information (integral in $d^3\Omega_a$), the broadening of the initial parton $a$ is integrated out and the squared initial production amplitude can be factored out. Thus, in the large-$N_c$ limit, the $d\sigma^{\rm in-in}$ and $d\sigma^{\rm in-out}$ contributions read~\cite{Silva:2025dan}
\begin{subequations}
\begin{align}
  & (2\pi)^2\frac{d\sigma^{\rm in-in}}{\sigma_0\,d^6\Omega_{\rm PS}} = \left(\frac{\alpha_s}{\pi}\right)^2\frac{C_1\theta_{1}}{8}\frac{C_2(\omega_g^+)^2 \theta_{2}P^M_{g\rightarrow cd}(z_2,\theta_{2})}{2z_2(1-z_2)Q_g^2}\nn
    & \times 2\Re \int_0^{L^+} dx_1^+ \int_{x_1^+}^{L^+} d\bar x_1^+ e^{-i\frac{z_1^2m_1^2}{2\omega_a^+}(\bar x_1^+ - x_1^+)}\int_{\bkappa,\bar \bkappa, \l} \cQ_{a\rightarrow bg}(L^+,\q_{1}, \q_{1}; \bar x_1^+, \bar \bkappa, \l)\cK_{a\rightarrow bg}(\bar x_1^+, \l; x_1^+, \bkappa) \nn
    & \times \Big[z_1^3m_1^2 + P_{a\rightarrow gb}(z_1)\left(\bkappa\cdot \bar\bkappa\right)+ S_{g\to cd}\frac{4z_2(1-z_2)(1-z_1)}{z_1(\q_{2}^2P_{g\rightarrow cd}(z_2)+m_2^2)}\Big((\q_{2}\cdot \bkappa)(\q_{2}\cdot \bar\bkappa)
    \nn
    & -(\q_{2}\times \bkappa)_z(\q_{2}\times \bar\bkappa)_z\Big)\Big] \, ,
    \label{eq:ch5_M_in_in2}
    \\
    & (2\pi)^2\frac{d\sigma^{\rm in-out}}{\sigma_0\,d^6\Omega_{\rm PS}} = -\left(\frac{\alpha_s}{\pi}\right)^2\frac{C_1(\omega_a^+)^2 \theta_{1}}{2z_1(1-z_1)(Q_a^2-m_1^2)}\frac{C_2(\omega_g^+)^2 \theta_{2}P^M_{g\rightarrow cd}(z_2,\theta_{2})}{2z_2(1-z_2)Q_g^2}\frac{1}{2\omega_a^+}\nn
    & \times 2\Re \,i\int_0^{L^+} dx_1^+ e^{-i\frac{z_1^2m_1^2}{2\omega_a^+}(L^+-x_1^+)}\int_{\bkappa} \cK_{a\rightarrow bg}(L^+, \q_1; x_1^+, \bkappa)\Big[z_1^3m_1^2 + P_{a\rightarrow gb}(z_1)\left(\bkappa\cdot \q_1\right)\nn
    & +S_{g\to cd}\frac{4z_2(1-z_2)(1-z_1)}{z_1(\q_{2}^2P_{g\rightarrow cd}(z_2)+m_2^2)} \Big((\q_{2}\cdot \bkappa)(\q_{2}\cdot \q_1)-(\q_{2}\times \bkappa)_z(\q_{2}\times \q_1)_z\Big)\Big] \, .
     \label{eq:ch5_M_in_out2}
    \end{align}
\end{subequations}
where $(\q_2\times \bkappa)_z = q_2^x\kappa^y-q_2^y\kappa^x$ and, as in the vacuum case, $S_{g\rightarrow gg} = +1$ and $S_{g\rightarrow q\bar q} = -1$. 

Let us discuss some of the ingredients that enter into Eqs.~\eqref{eq:ch5_M_in_in2} and \eqref{eq:ch5_M_in_out2}. First, we observe that massive vacuum splitting function for the second splitting ($P^M_{g\to cd}$) appears. The $C_i$ coefficient is the colour factor associated to the $i$-th splitting. The terms within brackets inside the $\bkappa$-integral correspond to the polarisation sum of the vertex product squared, see Eq.~\eqref{eq:AAdagger_all_channels}. To factor out $\sigma_0$, defined in Eq.~\eqref{eq:ch5_sigma_0_def}, $\cM_0$ has been assumed to be sharply peaked around some value of energy $p_a^+$. The objects $\cK$ and $\cQ$, defined in Eq.~\eqref{eq:K_Q_def}, involve only two-body (dipole) and four-body (quadrupole) medium averages, which is to be expected in the large-$N_c$ limit~\cite{Dominguez:2011wm,Dominguez:2012ad}. The latter can be written as a sum of a term with only dipoles, or ``factorizable'', and a term containing quadrupoles, or ``non-factorizable''~\cite{Blaizot:2012fh,Apolinario:2014csa,Dominguez:2019ges,Isaksen:2023nlr}. In what follows, the non-factorizable piece is neglected.\footnote{This approximation was tested quantitatively in the context of $\gamma \rightarrow q\bar q$~\cite{Isaksen:2023nlr}. It was concluded that the relative magnitude of the non-factorizable piece can get large when considering sufficiently large splitting angles $\theta \gtrsim 0.4$ and finite energy fractions $z \gtrsim 0.1$. Thus the large-$N_c$ result with only the factorizable piece is expected to correctly capture the soft and collinear regime of the spectrum, while reasonably describing the main qualitative features of medium modifications outside of this region. Note that although the colour structure of $\gamma \rightarrow q\bar q$ is much simpler than that of pure QCD processes, this process still provides an indication of whether neglecting the non-factorizable contribution is reasonable. This is because at large-$N_c$ the quadrupole arising from the 4-body average is common to all processes $\gamma \rightarrow q\bar q$, $q \rightarrow gq$ and $g \rightarrow gg$~\cite{Blaizot:2012fh,Apolinario:2014csa,Dominguez:2019ges}.} This amounts to taking
\begin{align}\label{eq:ch5_Q_approx}
  \cQ_{a\rightarrow bg}(L^+,\q_1, \q_1; \bar x_1^+, \bar \bkappa, \l) \approx \delta^2(\l-\bar\bkappa) \int_\v e^{-i\v\cdot(\q_1-\bar\bkappa)}\cP_{a\rightarrow bg}(\v; L^+,\bar x_1^+)\,,
\end{align}
where $\cP$ is a product of two dipole objects describing the independent broadening of the intermediate gluon and parton $b$, see Eq.~\eqref{eq:process_dep_P}. Plugging Eq.~\eqref{eq:ch5_Q_approx} into Eqs.~\eqref{eq:ch5_M_in_in2}, \eqref{eq:ch5_M_in_out2} and writing everything in coordinate space one obtains:
\begin{subequations}
\begin{align} 
  & (2\pi)^2\frac{d\sigma^{\rm in-in}}{\sigma_0\,d^6\Omega_{\rm PS}} = \left(\frac{\alpha_s}{\pi}\right)^2\frac{C_1\theta_{1}}{8}\frac{C_2(\omega_g^+)^2 \theta_{2}P^M_{g\rightarrow cd}(z_2,\theta_{2})}{2z_2(1-z_2)Q_g^2}\nn
    & \times 2\Re\int_0^{L^+} dx_1^+ \int_{x_1^+}^{L^+} d\bar x_1^+ e^{-i\frac{z_1^2m_1^2}{2\omega_a^+}(\bar x_1^+-x_1^+)}\int_{\v}\Big[z_1^3m_1^2 + P_{a\rightarrow gb}(z_1)\left(\nabla_{\u}\cdot \nabla_{\bar\u}\right)\nn
    & + S_{g\to cd}\frac{4z_2(1-z_2)(1-z_1)}{z_1(\q_{2}^2P_{g\rightarrow cd}(z_2)+m_2^2)}\left(\left(\q_{2}\cdot \nabla_{\u}\right)\left(\q_{2}\cdot \nabla_{\bar\u}\right)-\left(\q_{2}\times \nabla_{\u}\right)_z\left(\q_{2}\times \nabla_{\bar\u}\right)_z\right)\Big]\nn
  & \times e^{-i\v\cdot\q_{1}}\cP_{a\rightarrow bg}(\v; L^+,\bar x_1^+)\left.\cK_{a\rightarrow bg}(\bar x_1^+, \bar\u; x_1^+, \u)\right|_{\u = 0, \bar\u = \v} \, ,
    \label{eq:ch5_med_xsec_before_HO_inin}
    \\
  & (2\pi)^2\frac{d\sigma^{\rm in-out}}{\sigma_0\,d^6\Omega_{\rm PS}} = \left(\frac{\alpha_s}{\pi}\right)^2\frac{C_1(\omega_a^+)^2 \theta_{1}}{2z_1(1-z_1)(Q_a^2-m_1^2)}\frac{C_2(\omega_g^+)^2 \theta_{2}P^M_{g\rightarrow cd}(z_2,\theta_{2})}{2z_2(1-z_2)Q_g^2}\frac{1}{2\omega_a^+}\nn
    & \times 2\Re\int_0^{L^+} dx_1^+e^{-i\frac{z_1^2m_1^2}{2\omega_a^+}(L^+-x_1^+)}\int_\v\Big[ z_1^3m_1^2 + P_{a\rightarrow gb}(z_1)\left(\nabla_{\u}\cdot \q_{1}\right)\nn
    & + S_{g\to cd}\frac{4z_2(1-z_2)(1-z_1)}{z_1(\q_{2}^2 P_{g \rightarrow cd}(z_2) + m_2^2)}\left(\left(\q_{2}\cdot \nabla_{\u}\right)\left(\q_{2}\cdot \q_{1}\right)-\left(\q_{2}\times \nabla_{\u}\right)_z\left(\q_{2}\times \q_{1}\right)_z\right)\Big]\nn
  & \times e^{-i\v\cdot\q_{1}}\left.\cK_{a\rightarrow bg}(L^+, \v; x_1^+, \u)\right|_{\u = 0} \, .
    \label{eq:ch5_med_xsec_before_HO_inout}
    \end{align}
\end{subequations}

To explicitly compute the remaining Fourier transform, one needs to specify a form for the dipole cross-section $\sigma(\r) = 2g^2(\gamma(0)-\gamma(\r))$ entering the definition of the medium average in Eq.~\eqref{eq:A_correlator}. As discussed in Section~\ref{sec:xsecs_BDMPSZ}, here we work in the HO approximation (see Eq.~\eqref{eq:HO_approx}). Further, we consider the same type of axis-like anisotropy in the medium as in Chapter~\ref{chap:anisotropic_early_time}, breaking rotational invariance of the collision kernel. In this setup, the jet quenching parameter takes two different values along two orthogonal directions in the plane transverse to the jet axis~\cite{Hauksson:2023tze, Barata:2024bqp}, i.e., $\hat q_i$ for $i=x,y$ (see Section~\ref{sec:ch3_anisotropic_broadening} for further discussion).\footnote{Different forms of including a medium with a non-trivial structure in a BDMPS-Z calculation were considered in~\cite{Barata:2022krd,Andres:2022ndd,Barata:2023qds,Barata:2023zqg,Kuzmin:2023hko,Kuzmin:2024smy}.} In this case, the dipole cross-section is expanded as in Eq.~\eqref{eq:ch3_anisotropic_HO}, such that broadening kernel can be written as\footnote{We work with the fundamental $\hat q \equiv \hat q_F$.}
\begin{equation}\label{eq:ch5_P_after_HO}
  \cP_{a\rightarrow bg}(\v; L^+,\bar x_1^+) = \exp\left\{-\frac{1}{8}f_{a\rightarrow gb}(z_1)(L^+-\bar x_1^+)(\hat q_x v_x^2 + \hat q_y v_y^2)\right\}\, ,
\end{equation}
and the splitting kernel $\cK_{a\rightarrow bg}(\bar x_1^+, \u_2; x_1^+, \u_1)$ is given by Eq.~\eqref{eq:ch3_K_aniso} with a process-dependent complex frequency
\begin{equation}\label{eq:ch5_omega}
    \Omega_i = \left(\frac{1-i}{\sqrt{2}}\right)\sqrt{\frac{\hat{q}_i h_{a\rightarrow gb}(z_1)}{2\omega_a^+}}\,.
\end{equation}
The process dependent functions $f(z_1)$ and $h(z_1)$ read
\begin{subequations}\label{eq:ch5_process_dep_hf}
\begin{align}
    \begin{split}
    & f_{q\rightarrow g q}(z_1) = 2(1-z_1)^2 + z_1^2\,,\quad h_{q\rightarrow g q}(z_1)= 1 + (1-z_1)^2\, .
    \end{split}\\
    \begin{split}
    &f_{g\rightarrow g g}(z_1) = 2[(1-z_1)^2 + z_1^2]\,,\quad h_{g\rightarrow g g}(z_1)= 1 + z_1^2 + (1-z_1)^2\,.
    \end{split}
    \end{align}
\end{subequations}

Using the explicit form for the broadening and splitting kernels in Eqs.~\eqref{eq:ch5_P_after_HO} and~\eqref{eq:ch3_K_aniso} and parametrizing the transverse momenta according to Appendix~\ref{app:kinematics_phase_space}, Eqs.~\eqref{eq:ch5_med_xsec_before_HO_inin} and~\eqref{eq:ch5_med_xsec_before_HO_inout} can be calculated analytically modulus integrations over light-cone times. The full spectrum can then be written in the following form~\cite{Silva:2025dan}
\begin{align}\label{eq:ch5_xsec_med_full}
   (2\pi)^2\frac{d\sigma^{\rm med}}{\sigma_0\,d^6\Omega_{\rm PS}}& = \left[\frac{\alpha_s}{\pi} C_1 \frac{P^M_{a\rightarrow bg}(z_1,\theta_{1})}{\theta_1(1+\tilde\theta^2_{m,1})}\right]\left[\frac{\alpha_s}{\pi}C_2\frac{ P^M_{g\rightarrow cd}(z_2,\theta_{2})}{\theta_2(1+\tilde\theta^2_{m,2})}\right] (1+F_{\rm med}) 
     \left(1 + a_{\rm med} \cos(2\psi_{12} +\phi_{\rm med})\right) ,
\end{align}
where one observes (i) an overall modification of the spectrum encapsulated in the factor $1+F_{\rm med}$~\cite{Dominguez:2019ges,Isaksen:2023nlr}, (ii) a change in the modulation of the azimuth-dependent term ($a_{\rm med}$) and (iii) a phase-shift ($\phi_{\rm med}$). These 3 factors depend on a series of auxiliary functions that are written down explicitly as
\vspace{-0.5cm}
\begin{subequations}
\begin{align}
    & a_{\rm med} = a \frac{\sqrt{A_1^2+A_2^2}}{1+F_{\rm med}} \label{eq:ch5_amed_def}\,,\\
    & \tan\phi_{\rm med} = \frac{A_2}{A_1} 
    \label{eq:ch5_phimed_def}\,,
\end{align}
\end{subequations}
where $a$ corresponds to the amplitude of spin correlations in vacuum, see Eq.~\eqref{eq:ch5_adef_vac}. The auxiliary functions $A_1$, $A_2$ and $F_{\rm med}$ are given by~\cite{Silva:2025dan} 
\begin{subequations}\label{eq:ch5_Ai_coeff_aniso}
\begin{align}
   F_{\rm med} &= \frac{1}{\omega_a^+ P^M_{a\rightarrow bg}(z_1,\theta_{1})}\,2\Re\,i\,\Bigg\lbrace\int^{L^+}_{0} dx_1^+ e^{-i\frac{z_1^2m_1^2}{2\omega_a^+}(L^+-x_1^+)}e^{-i\frac{\q_{1}^2}{2\omega_a^+}\left(\frac{\cos^2\phi_1}{c_{2x}}+\frac{\sin^2\phi_1}{c_{2y}}\right)}\frac{\sqrt{c_{1x}}\sqrt{c_{1y}}}{\sqrt{c_{2x}}\sqrt{c_{2y}}} \nonumber \\
    &\times \Big[z_1^3m_1^2 + P_{a\rightarrow gb}(z_1)\left(\cD_+ + \cD_- \cos(2\phi_1)\right)\Big] - \left(\frac{Q_a^2-m_1^2}{2p_a^+}\right) \int_{0}^{L^+}dx_1^+ \int_{x_1^+}^{L^+}d\bar x_1^+\, e^{-i\frac{z_1^2m_1^2}{2\omega_a^+}\Delta x^+} 
    \nn
    &\times e^{i\frac{\q_{1}^2}{2\omega_a^+}\left(\frac{\cos^2\phi_1}{c_{3x}}+\frac{\sin^2\phi_1}{c_{3y}}\right)}  \frac{\sqrt{c_{1x}}\sqrt{c_{1y}}}{\sqrt{c_{3x}}\sqrt{c_{3y}}}\Big[z_1^3m_1^2 + P_{a\rightarrow gb}(z_1)\left(\cA_+ + \cB \cos(2\phi_1)\right) \Big]\Bigg\rbrace \,, \label{eq:ch5_fmed_def}\\ 
   A_1 &= 1+\frac{\q_1^2P_{a\rightarrow gb}(z_1) + z_1^3m_1^2}{\q_1^2\omega_a^+P^M_{a\rightarrow bg}(z_1,\theta_{1})}\,2\Re\,i\,\Bigg\lbrace \int_{0}^{L^+} dx_1^+ e^{-i\frac{z_1^2m_1^2}{2\omega_a^+}(L^+-x_1^+)}e^{-i\frac{\q_{1}^2}{2\omega_a^+}\left(\frac{\cos^2\phi_1}{c_{2x}}+\frac{\sin^2\phi_1}{c_{2y}}\right)}\frac{\sqrt{c_{1x}}}{\sqrt{c_{2x}}}\nn
    & \times \frac{\sqrt{c_{1y}}}{\sqrt{c_{2y}}}\left[\cD_+ + \cD_- \cos(2\phi_1)\right]-\left(\frac{Q_a^2-m_1^2}{2p_a^+}\right)\int_{0}^{L^+}  d x_1^+\int_{x_1^+}^{L^+} d\bar x_1^+\,  e^{-i\frac{z_1^2m_1^2}{2\omega_a^+}\Delta x^+}\nn
    &\times e^{i\frac{\q_{1}^2}{2\omega_a^+}\left(\frac{\cos^2\phi_1}{c_{3x}}+\frac{\sin^2\phi_1}{c_{3y}}\right)}\frac{\sqrt {c_{1x}}\sqrt{c_{1y}}}{\sqrt{ c_{3x}}\sqrt{c_{3y}}} \left[\cC_+ + \cA_- \cos(2\phi_1) + \cC_-\cos(4\phi_1) \right]\Bigg\rbrace \,, \label{eq:ch5_a1_def}\\
    A_2 &=-\frac{\q_1^2P_{a\rightarrow gb}(z_1) + z_1^3m_1^2}{\q_1^2\omega_a^+P^M_{a\rightarrow bg}(z_1,\theta_{1})}\,2\Re\,i\,\Bigg\lbrace \int_{0}^{L^+} dx_1^+\,e^{-i\frac{z_1^2m_1^2}{2\omega_a^+}(L^+-x_1^+)}e^{-i\frac{\q_{1}^2}{2\omega_a^+}\left(\frac{\cos^2\phi_1}{c_{2x}}+\frac{\sin^2\phi_1}{c_{2y}}\right)}\frac{\sqrt{c_{1x}}}{\sqrt{c_{2x}}}\,\nn
    &\times \frac{\sqrt{c_{1y}}}{\sqrt{c_{2y}}}\cD_-\sin(2\phi_1) -\left(\frac{Q_a^2-m_1^2}{2p_a^+}\right)\int_{0}^{L^+} dx_1^+\int_{x_1^+}^{L^+} d\bar x_1^+ \,e^{-i\frac{z_1^2m_1^2}{2\omega_a^+}\Delta x^+}e^{i\frac{\q_{1}^2}{2\omega_a^+}\left(\frac{\cos^2\phi_1}{c_{3x}}+\frac{\sin^2\phi_1}{c_{3y}}\right)} \nn
    &\times \frac{\sqrt{c_{1x}}\sqrt{c_{1y}}}{\sqrt{c_{3x}}\sqrt{c_{3y}}}\left[\cA_- \sin(2\phi_1) + \cC_-\sin(4\phi_1)\right]\Bigg\rbrace\,.
    \label{eq:ch5_a2_def}
    \end{align}
\end{subequations}
Finally, the process-dependent complex $c$-functions, defined analogously to Eq.~\eqref{eq:ch3_ci_definition_static},
\begin{subequations} \label{eq:ch5_ci_definition_aniso}
\begin{align}
    &c_{1i} = \frac{\Omega_i}{2i\sin{\Omega_i\Delta x^+}}\,, 
    \label{eq:ch5_c1_definition_aniso}\\
    &c_{2i} = \frac{\Omega_i}{\tan{\Omega_i\Delta x^+}}\,,
   \label{eq:ch5_c2_definition_aniso}\\
     &c_{3i} = (L^+-\bar x_1^+) \left(-i\frac{\hat q_i f_{a\rightarrow gb}(z_1)}{2\omega^+_a}\right)-c_{2i}\,,
     \label{eq:ch5_c3_definition_aniso}
     \end{align}
   \end{subequations}
encode the medium anisotropy and enter into the remaining functions 
\begin{subequations}\label{eq:ch5_abcd_functions}
\begin{align}
\begin{split}
    &\cA_{\pm} = 2\omega_a^+\left(c_{1x} \pm c_{1y} + \frac{c_{1x}c_{2x}}{c_{3x}} \pm \frac{c_{1y}c_{2y}}{c_{3y}}\right) + i\q_{1}^2\left(\frac{c_{1x}c_{2x}}{c_{3x}^2} \pm \frac{c_{1y}c_{2y}}{c_{3y}^2}\right)\,,
    \end{split}\\
    \begin{split}
  &\cB = i\q_{1}^2\left(\frac{c_{1x}c_{2x}}{c_{3x}^2}-\frac{c_{1y}c_{2y}}{c_{3y}^2}\right)\,,
    \end{split}\\
    \begin{split}
  &\cC_{\pm} = \frac{i\q_{1}^2}{2}\left(\frac{c_{1x}}{c_{3x}}\pm\frac{c_{1y}}{c_{3y}}\right)\left(\frac{c_{2x}}{c_{3x}}\pm\frac{c_{2y}}{c_{3y}}\right)\,,
    \end{split}\\
     \begin{split}
    & \cD_{\pm} = \q_{1}^2\left(\frac{c_{1x}}{c_{2x}}\pm\frac{c_{1y}}{c_{2y}}\right)\,.
       \end{split}
    \end{align}
\end{subequations}
where $\Delta x^+$ in Eq.~\eqref{eq:ch5_ci_definition_aniso} reads $\bar x_1^+-x_1^+$ for the in-in contributions (double time integral in Eq.~\eqref{eq:ch5_xsec_med_full}), while it reads $L-x_1^+$ for the in-out contributions (single time integral). Eqs.~\eqref{eq:ch5_xsec_med_full}--\eqref{eq:ch5_abcd_functions} are the main result of this subsection. A few comments are in order before presenting the numerical calculation of this cross-section. First, as a consequence of assuming only the $a\to b\,g$ splitting to be modified by the medium, the $A_1$, $A_2$ and $F_{\rm med}$ auxiliary functions solely depend on the first branching kinematics ($p_a^+,m_1,\theta_1,z_1,\phi_1$)  and on the medium parameters ($L^+$, $\hat q_x$, $\hat q_y$). Recovering the vacuum limit effectively amounts to taking $1+F_{\rm med}\to 1$, $A_1\to 1$, $A_2\to 0$. The existence of odd parity harmonics in $\psi_{12}$ and $\phi_{1}$, i.e., $\sin(2\psi_{12})$, $\sin(2\phi_{1})$ and $\sin(4\phi_{1})$, is directly related, under the assumptions of the calculation, to the medium anisotropy, as previously argued. In the context of the $1\rightarrow 2$ cross-section in anisotropic media, the odd parity harmonics in $\phi_1$ show up only when final-state spin dependence is taken into account~\cite{Barata:2024bqp}, as shown in Section~\ref{sec:ch3_gqqbar}. Regarding the direction of anisotropy, if one inverts it (i.e. $\hat q_x \leftrightarrow \hat q_y$), then the resulting $\phi_{1}$ distribution is the same as if one transforms it under $\phi_{1} \rightarrow \phi_{1}+\pi/2$. To see this, note that upon switching $\hat q_x \leftrightarrow \hat q_y$, the coefficients $\cA_-,\,\cB,\,\cD_-$ gain a minus sign, while the remaining stay unchanged. All three coefficients are attached to an even harmonic ($\cos(2\phi_1)$ and $\sin(2\phi_1)$), which gains a minus sign under $\phi_1\rightarrow \phi_1 + \pi/2$, canceling the sign from the coefficient. The $\phi_1$ dependent phase is trivially invariant under these two simultaneous transformations. 

\paragraph{Isotropic medium} Setting $\hat q_x = \hat q_y$ in Eqs.~\eqref{eq:ch5_xsec_med_full}--\eqref{eq:ch5_abcd_functions} leads to a vanishing phase-shift $\phi_{\rm med}=0$ together with $A_2=0$, i.e., $a_{\rm med}=a A_1/(1+F_{\rm med})$. The auxiliary functions reduce to
\begin{subequations}
\begin{align}
& F_{\rm med} \underset{\hat q_x=\hat q_y}{=}\frac{1}{\omega_a^+ P^M_{a\rightarrow bg}(z_1,\theta_1)}\,2\Re\,i\,\Bigg\lbrace \int_{0}^{L^+} dx_1^+  e^{-i\frac{z_1^2m_1^2}{2\omega_a^+}(L^+-x_1^+)}e^{-i\frac{\q_1^2}{2\omega_a^+c_2}}\frac{c_1}{c_2}\left[z_1^3m_1^2 + P_{a\rightarrow gb}(z_1)\cD_+\right] \nonumber \\ 
&- \left(\frac{Q_a^2-m_1^2}{2p_a^+}\right)\int_{0}^{L^+} dx_1^+ \int_{x_1^+}^{L^+} d\bar x_1^+\, e^{-i\frac{z_1^2m_1^2}{2\omega_a^+}\Delta x^+}e^{i\frac{\q_1^2}{2\omega_a^+c_3}} \frac{c_1}{c_3}\left[z_1^3m_1^2 + P_{a\rightarrow gb}(z_1)\cA_+\right]\Bigg\rbrace \, ,
\label{eq:ch5_fmed_iso}
\\
  & A_1 \underset{\hat q_x=\hat q_y}{=} 1+\frac{\q_1^2 P_{a\rightarrow gb}(z_1) + z_1^3m_1^2}{\q_1^2\omega_a^+P^M_{a\rightarrow bg}(z_1,\theta_1)}\,2\Re\,i\,\Bigg\lbrace\int_{0}^{L^+} dx_1^+ e^{-i\frac{z_1^2m_1^2}{2\omega_a^+}(L^+-x_1^+)}e^{-i\frac{\q_1^2}{2\omega_a^+c_2}}\frac{c_1}{c_2}\cD_- \nonumber\\
    & -\left(\frac{Q_a^2-m_1^2}{2p_a^+}\right)\int_{0}^{L^+} dx_1^+ \int_{x_1^+}^{L^+} d\bar x_1^+\, e^{-i\frac{z_1^2m_1^2}{2\omega_a^+}\Delta x^+}e^{i\frac{\q_1^2}{2\omega_a^+c_3}} \frac{c_1}{c_3}\cC_+\Bigg\rbrace\, .
    \label{eq:ch5_A1_iso}
\end{align}
\end{subequations}

\subsection{Isotropic medium: suppression}
\label{subsec:ch5_isotropic_spin_results}

We now numerically evaluate Eq.~\eqref{eq:ch5_xsec_med_full} and examine the impact of medium modifications on the amplitude of the spin-correlation modulation, $a_{\rm med}$, as well as on the total cross-section as a function of the azimuthal angle between splitting planes, $\psi_{12}$. This subsection considers the case of an isotropic medium, while the anisotropic case and the corresponding phase shift $\phi_{\rm med}$ are discussed in Sec.~\ref{subsec:ch5_anisotropic_spin_results}. The medium length is set to $L=L^+/\sqrt{2} = 5$ fm. As in the vacuum results of Sec.~\ref{subsec:ch5_vacuum_spin_correlations}, the first splitting is taken to be massless, $m_1=0$, and the energy of the initial parton is fixed to $E_a=100$ GeV. We stress that the kinematics of the second splitting, $(z_2,\theta_2)$, only affects the vacuum piece in Eq.~\eqref{eq:ch5_xsec_med_full}. Finally, no numerical results are shown for $F_{\rm med}$, since this contribution has already been addressed several times in the literature, see, e.g.,~\cite{Dominguez:2019ges,Isaksen:2023nlr}.

\begin{figure}[!t]
     \centering
         \includegraphics[width=0.49\textwidth]{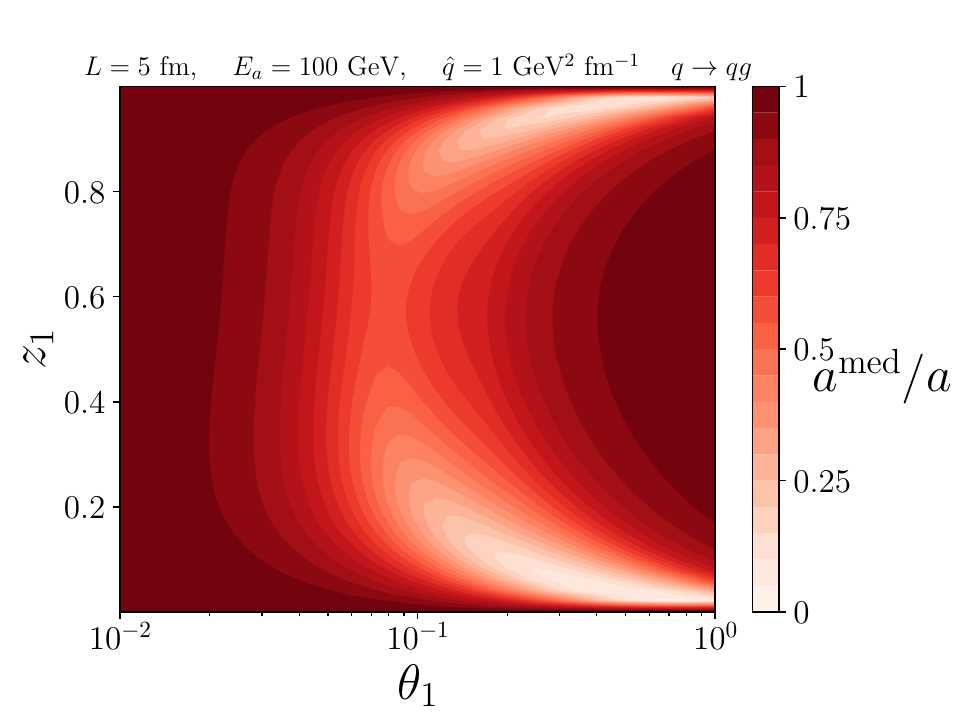}
         \includegraphics[width=0.49\textwidth]{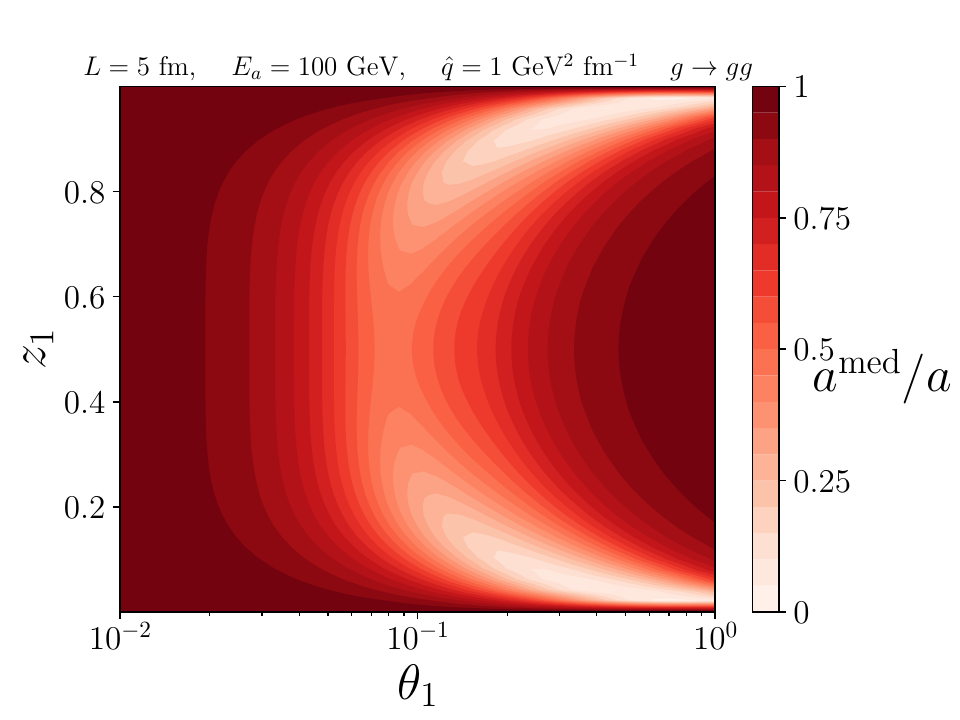}
         \\
         \includegraphics[width=0.49\textwidth]{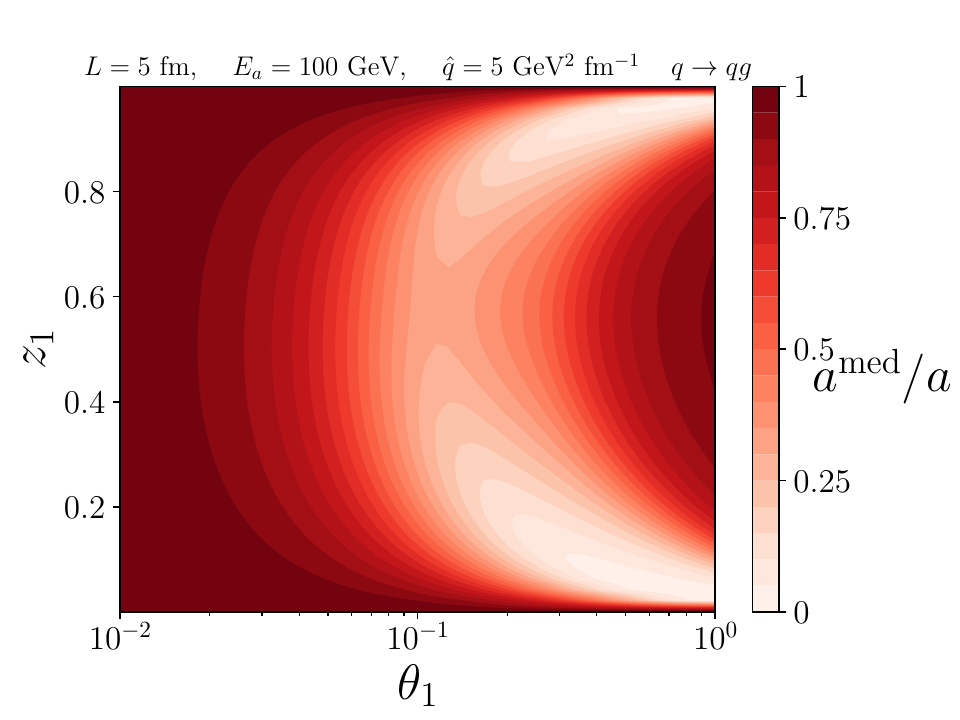}
         \includegraphics[width=0.49\textwidth]{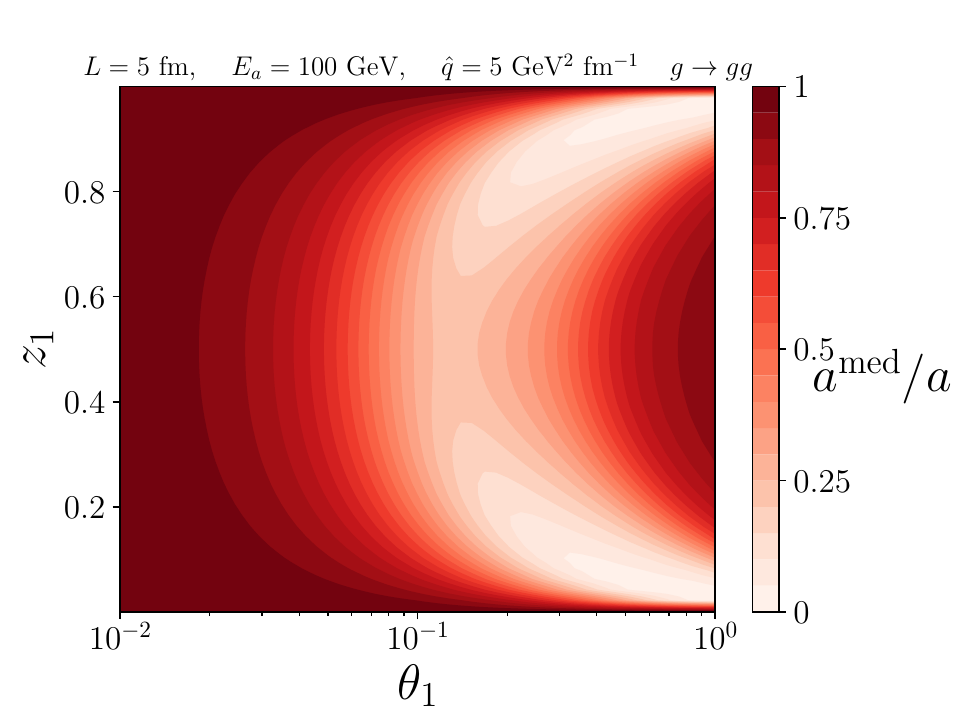}
     \caption{Ratio between the spin-correlations modulation amplitude in the medium and in vacuum $a_{\rm med}/a$ as a function of the first splitting's kinematics $z_1$ and $\theta_1$. The upper panels correspond to the first splitting being $q\rightarrow qg$ and the lower ones to $g\rightarrow gg$. The medium is isotropic and we plot $\hat q= 1$ GeV$^2$/fm (upper panels) and $5$ GeV$^2$/fm (lower panels).}
     \label{fig:A2_over_A1_iso}
\end{figure}

In Fig.~\ref{fig:A2_over_A1_iso}, we display the medium-induced modification of the spin-correlation signal, $a_{\rm med}/a$, as a function of the first splitting kinematics, $(z_1,\theta_1)$, and for two values of the adjoint jet-quenching parameter: $\hat q_A=1$ GeV$^2$/fm in the upper panels and $\hat q_A=5$ GeV$^2$/fm in the lower panels. In all plots shown in this and the following subsection, fixing $\hat q$ should be understood as fixing $\hat q_A$, while the calculation in the previous subsection uses the fundamental quantity $\hat q \equiv \hat q_F$. The left panels correspond to the first splitting being $q\rightarrow qg$, whereas the right panels correspond to $g\rightarrow gg$. The ratio $a_{\rm med}/a$ does not depend on the second splitting kinematics. The first observation is that medium interactions suppress the strength of spin correlations in the bulk of phase space. Namely, $a_{\rm med}/a$ remains $\leq1$ and, as $\hat q$ is increased, the phase-space band where this modulation suppression is significant becomes wider and, for most of phase space, stronger. In particular phase-space regions, the medium can even completely wash out the spin-correlation effect. Medium effects vanish only for very collinear splittings, $\theta_1\lesssim10^{-2}$. The $(z_1,\theta_1)$ dependence of the suppression can be traced back to the formation time of the first splitting, parametrically given by $(\tau_1^+)^{-1}\sim z_1(1-z_1)E_a\theta_1^2$. Then, for $\tau_1^>L$ or $\tau_1^+\ll L$, the effect becomes negligible, as expected. Finally, when comparing both channels, the most significant difference appears at $z_1\gtrsim0.5$. In turn, for $z_1\ll1$ no channel dependence in $a_{\rm med}/a$ is observed since, in this limit, $P_{gq}(z_1)\approx P_{gg}(z_1)\approx2/z_1$, and the same equality holds for the process-dependent functions in the medium modification in Eq.~\eqref{eq:ch5_process_dep_hf}.

\begin{figure}[!t]
     \centering
         \includegraphics[width=0.49\textwidth]{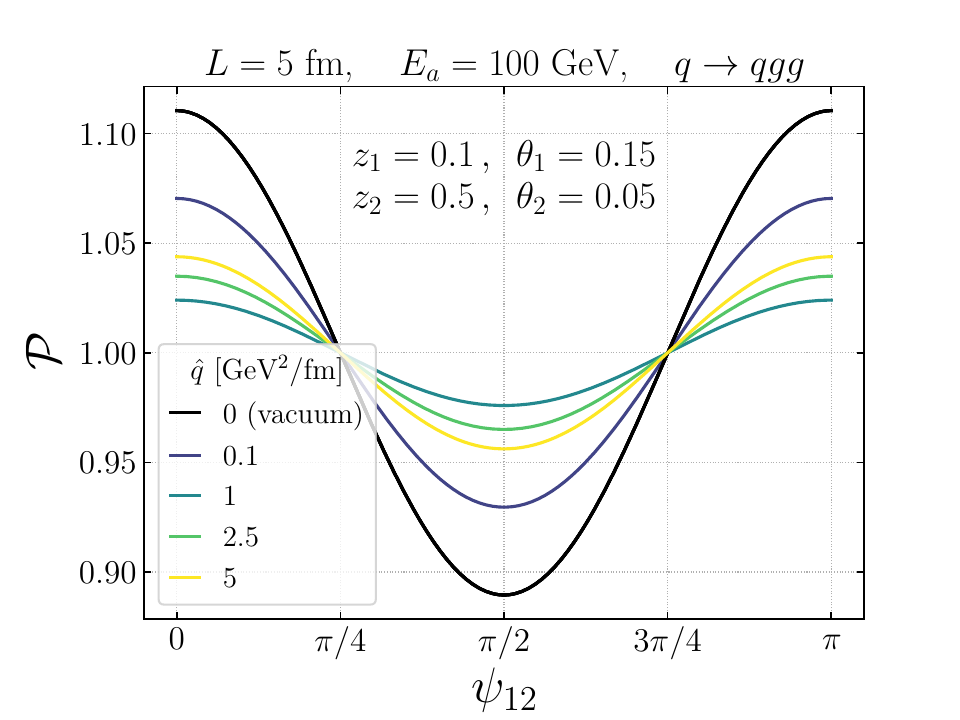}
         \includegraphics[width=0.49\textwidth]{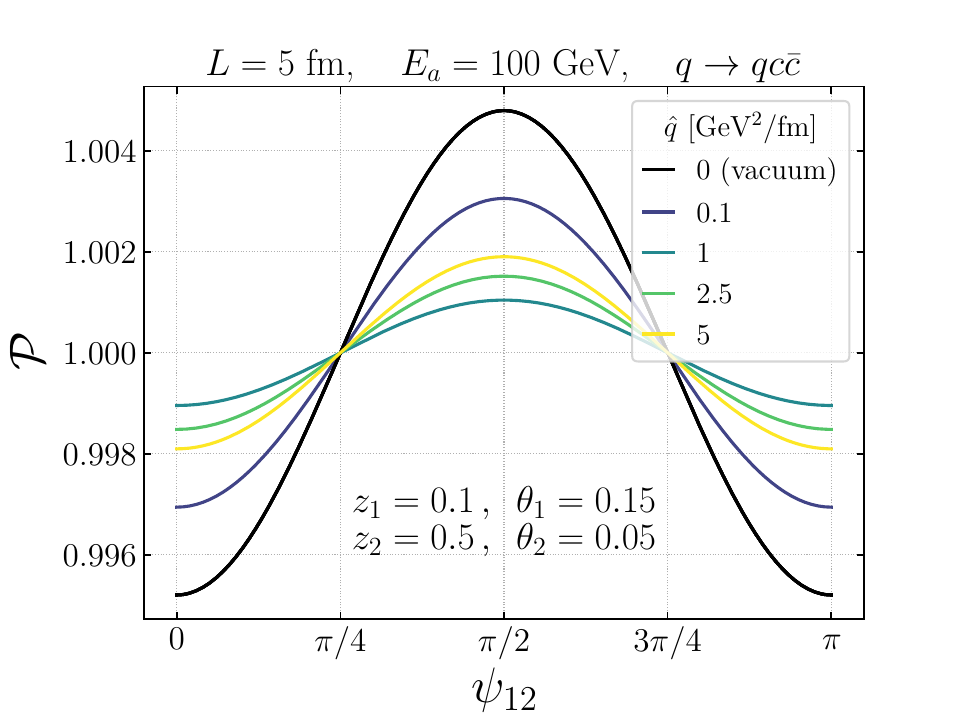}
         \\
         \includegraphics[width=0.49\textwidth]{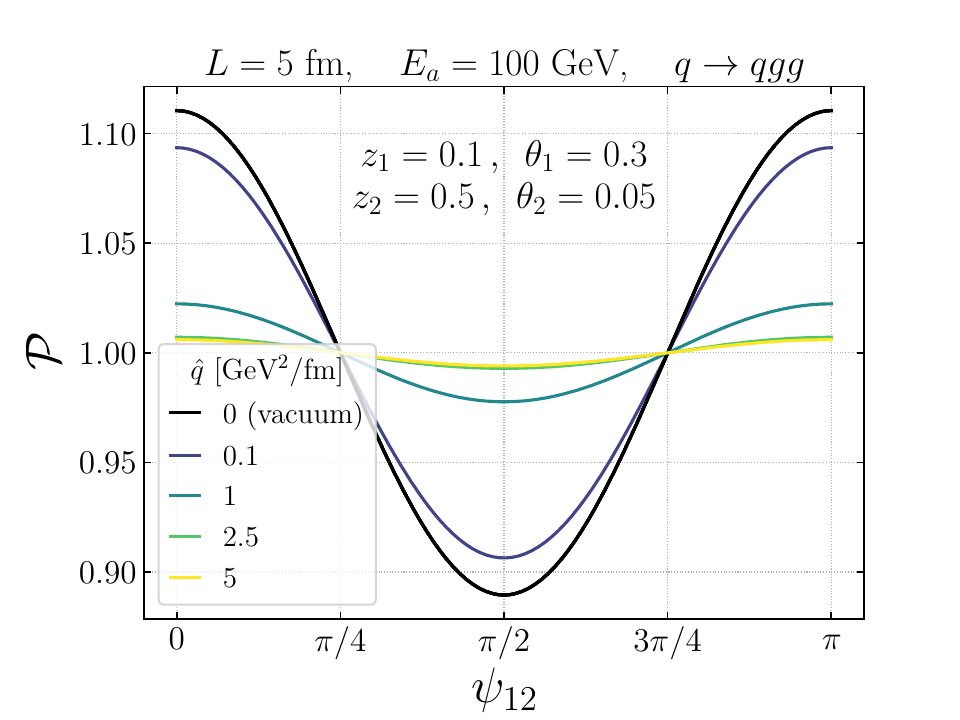}
         \includegraphics[width=0.49\textwidth]{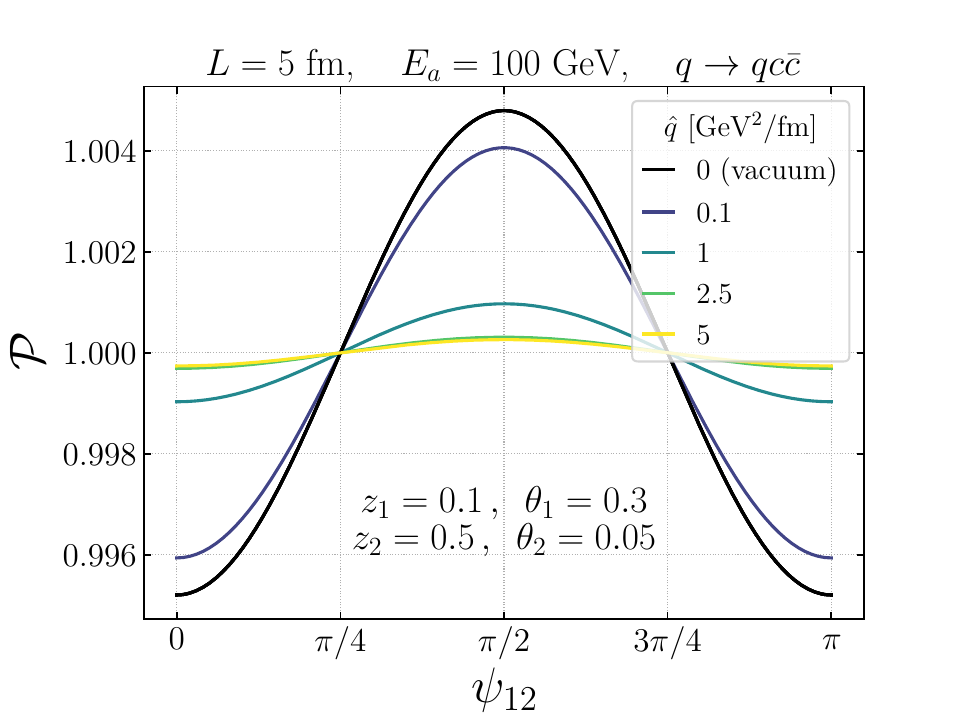}
     \caption{Evolution of the angular modulation in Eq.~\eqref{eq:ch5_p_def} with $\hat q$ as a function of $\psi_{12}$ for $\theta_1 = 0.15$ (top panels) and $\theta_1 = 0.3$ (bottom panels). The remaining kinematic variables are fixed to $z_1 = 0.1$, $z_2=0.5$ and $\theta_2 = 0.05$. The left panels correspond to $q\rightarrow qg \rightarrow qgg$ and the right panels to $q\rightarrow qg \rightarrow qc\bar c$. Note that for a massless second splitting (left panels) there is no dependence on $\theta_2$.}
     \label{fig:psi_isotropic}
\end{figure}

Next, we evaluate the in-medium cross-section stripped of both the vacuum cross-section and the $(1+F_{\rm med})$ factor, i.e., we study
\begin{equation}\label{eq:ch5_p_def}
\mathcal{P}(z_1,\theta_1,z_2,\theta_2,\psi_{12})\equiv 1 + a_{\rm med} \cos(2\psi_{12}),
\end{equation}
as a function of $\psi_{12}$ in Fig.~\ref{fig:psi_isotropic}. We choose $(z_1,\theta_1)$ values for which medium modifications are strongest in Fig.~\ref{fig:A2_over_A1_iso}. The value $z_2=0.5$ is also selected so as to maximize the azimuthal modulations, see Fig.~\ref{fig:a_vac}. Finally, we set $\theta_2=0.05<\theta_1$ to remain within the strongly ordered approximation. Let us first discuss the vacuum results, shown as black lines. As observed in the literature, $g\to gg$ and $g\to q\bar q$ splittings feature correlations with opposite signs. Namely, for $q\to qgg$ splittings, shown in the left panels, $\mathcal{P}$ is largest when the two splittings lie in the same plane, $\psi_{12}=0$, while for $g\to q\bar q$, shown in the right panels, the spin correlations are largest when the splitting planes are perpendicular. In the massless case, this modulation is much more pronounced for $g\to q\bar q$ than for $g\to gg$ branchings, see for instance~\cite{Richardson:2018pvo,Karlberg:2021kwr}. However, when considering massive quarks, in particular setting $m_2=m_c$, the modulation is almost negligible for the explored phase-space points. This is because, for the chosen kinematics, $\tilde\theta_{m,2}\approx5$, i.e., the splitting occurs deep inside the dead cone. This weak modulation of $\mathcal{P}$ is consistent with the evolution of $a$ with $\tilde\theta_{m,2}$ displayed in Fig.~\ref{fig:a_vac}. As for the medium results, shown as coloured curves, the modulation is suppressed, in agreement with Fig.~\ref{fig:A2_over_A1_iso}. Further, the evolution of $\mathcal{P}$ with $\hat q$ depends on the phase-space point under consideration. For instance, when considering moderately large opening angles such as $\theta_1=0.3$, shown in the bottom panels, there is a monotonic trend: larger $\hat q$ implies a larger suppression of the modulation. For $\theta_1=0.15$, shown in the top panels, however, the trend is non-trivial. For $\hat q\sim1$ GeV$^2$/fm, there is a turning point for which the previous trend is reversed. More concretely, for $\hat q=5$ GeV$^2$/fm the modulation is less suppressed than for $\hat q=1$ GeV$^2$/fm. This behaviour can be traced back to Fig.~\ref{fig:A2_over_A1_iso}, where an increase in $\hat q$ is not only responsible for a widening of the band where suppression is significant, but also for a slight shift of the band’s boundary. Therefore, the angular modulation in the medium is always weaker than in vacuum, and its magnitude depends both on the medium properties and on the splitting kinematics.

\subsection{Anisotropic medium: suppression and phase shift}
\label{subsec:ch5_anisotropic_spin_results}

We now discuss the case of an anisotropic medium, fixing $\hat q^{\rm av} = \frac{1}{2}(\hat q_x + \hat q_y) = 2.5$ GeV$^2$/fm and the anisotropy ratio to $\hat q_y/\hat q_x = 5$. The medium length is set to $L=L^+/\sqrt{2} = 5$ fm. Since the dependence on the first channel has been analysed in the previous subsection, see Figs.~\ref{fig:A2_over_A1_iso},~\ref{fig:psi_isotropic}, we restrict attention to a single process: $q\rightarrow qg \to q gg$. The results are shown as a function of the first splitting kinematics $(z_1,\theta_1,\phi_1)$ with fixed parent energy $E_a=100$ GeV.

\begin{figure}[!t]
\centering
\includegraphics[width=\textwidth]{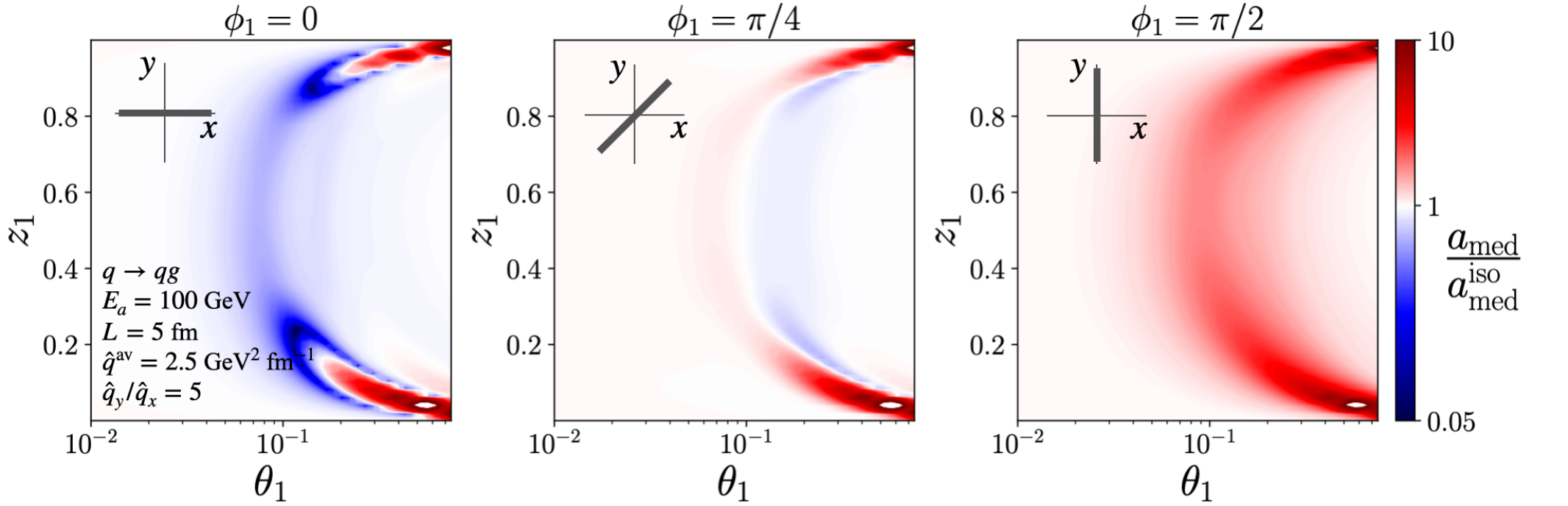}
\caption{Modulation amplitude $a_{\rm med}$ for an anisotropic medium, divided by the isotropic result with $\hat q = \hat q^{\rm av.}$ (upper panels) as a function of the first splitting kinematics $z_1$ and $\theta_1$, for the process $q\rightarrow qg$. The three columns correspond to three different azimuthal orientations of the first splitting plane: $\phi_1 = 0,,\pi/4,,\pi/2$.}
\label{fig:amed_anisotropic}
\end{figure}

\begin{figure}[!t]
\centering
\includegraphics[width=\textwidth]{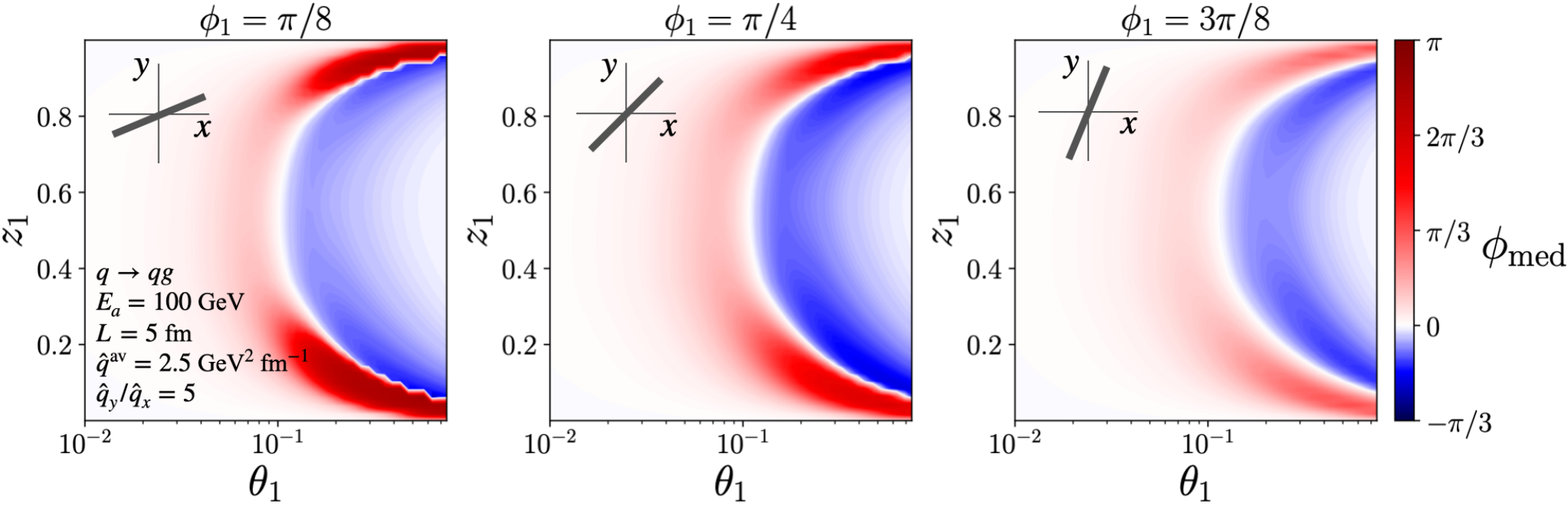}
\caption{Same as Fig.~\ref{fig:amed_anisotropic} but for the phase shift $\phi_{\rm med}$. The values of $\phi_1$ differ with respect to Fig.~\ref{fig:amed_anisotropic} since $\phi_{\rm med}=0$ for $\phi_1=0,,\pi/2$. Here we choose $\phi_1= \pi/8,,\pi/4,,3\pi/8$.}
\label{fig:phimed_anisotropic}
\end{figure}

In Fig.~\ref{fig:amed_anisotropic} we plot the ratio $a_{\rm med }/a_{\rm med}^{\rm iso}$ in the $(z_1,\theta_1)$-plane. This quantity measures the anisotropy-induced modifications of the amplitude of the spin-correlation signal. First, we notice that $a_{\rm med }/a_{\rm med}^{\rm iso}$ deviates from unity in a band of phase space which roughly coincides with the one in Fig.~\ref{fig:A2_over_A1_iso}. As discussed, the shape of this band is controlled by the hierarchy between the formation time $\tau_1$ and the medium length $L$. Depending on the azimuthal angle $\phi_1$, $a_{\rm med}$ is enhanced (red) or suppressed (blue) in the bulk of phase space with respect to the isotropic case $a_{\rm med}^{\rm iso}$. In particular, for splitting planes orthogonal to the maximal anisotropy direction ($\phi_1 = 0$), one gets mostly a suppression, while for splitting planes aligned with this direction ($\phi_1 = \pi/2$) one gets an enhancement. Interestingly, for $\phi_1 = \pi/4$ one gets a very weak anisotropy-induced modification, which becomes significant only for asymmetric, wide-angle splittings.

As discussed throughout the calculation presented in Sec.~\ref{subsec:ch5_medium_vacuum_spin_correlations}, medium anisotropies lead to a distinctive signature on the spin-correlation signal in terms of a phase shift $\cos(2\psi_{12})\to \cos(2\psi_{12}+\phi_{\rm med})$,
with $\phi_{\rm med}$ defined in Eq.~\eqref{eq:ch5_phimed_def}. We show this phase shift in Fig.~\ref{fig:phimed_anisotropic} as a function of $(z_1,\theta_1,\phi_1)$. Since $\phi_{\rm med}$ vanishes exactly for $\phi_1=0,,\pi/2$, see Eq.~\eqref{eq:ch5_Ai_coeff_aniso}, we choose a different set of $\phi_1$ values with respect to Fig.~\ref{fig:amed_anisotropic}. We find that the phase shift is more pronounced for splitting planes close to orthogonal to the maximal anisotropy direction, $\phi_1 = \pi/8$. In fact, for specific regions of phase space, e.g. $\theta_1>0.1$ and $z_1<0.2$, we obtain $\phi_{\rm med} \approx \pi$, resulting in a sign change in the azimuthal correlations. Finally, as the splitting plane becomes increasingly aligned with the maximal anisotropy direction ($\phi_1 = 3\pi/8$), $\phi_{\rm med}$ gradually takes smaller values.

\begin{figure}[!t]
\centering
\includegraphics[width=0.49\textwidth]{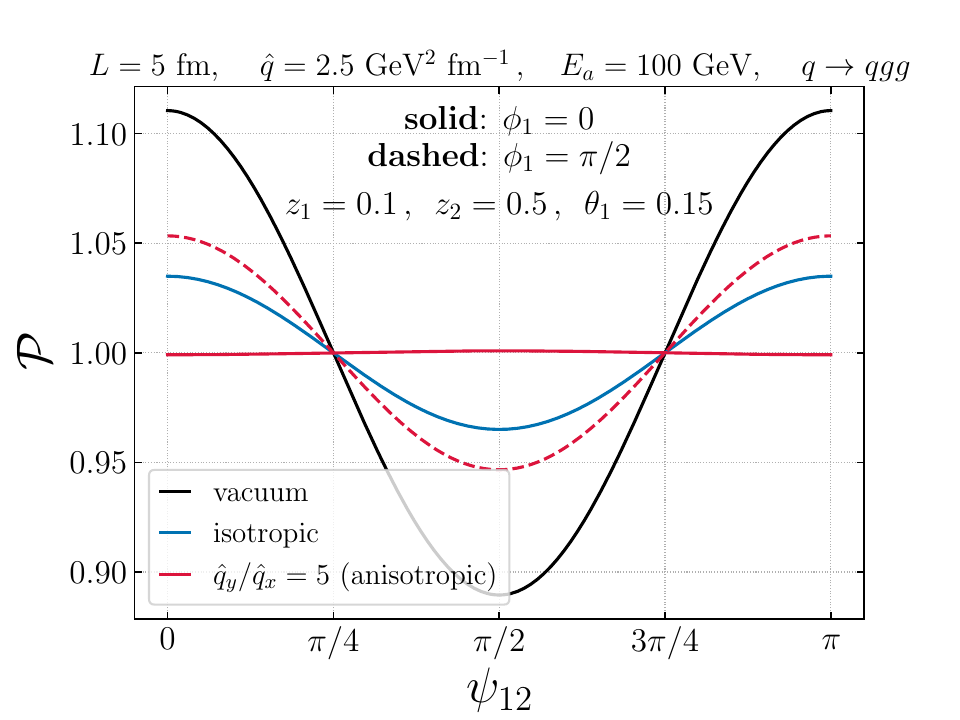}
\includegraphics[width=0.49\textwidth]{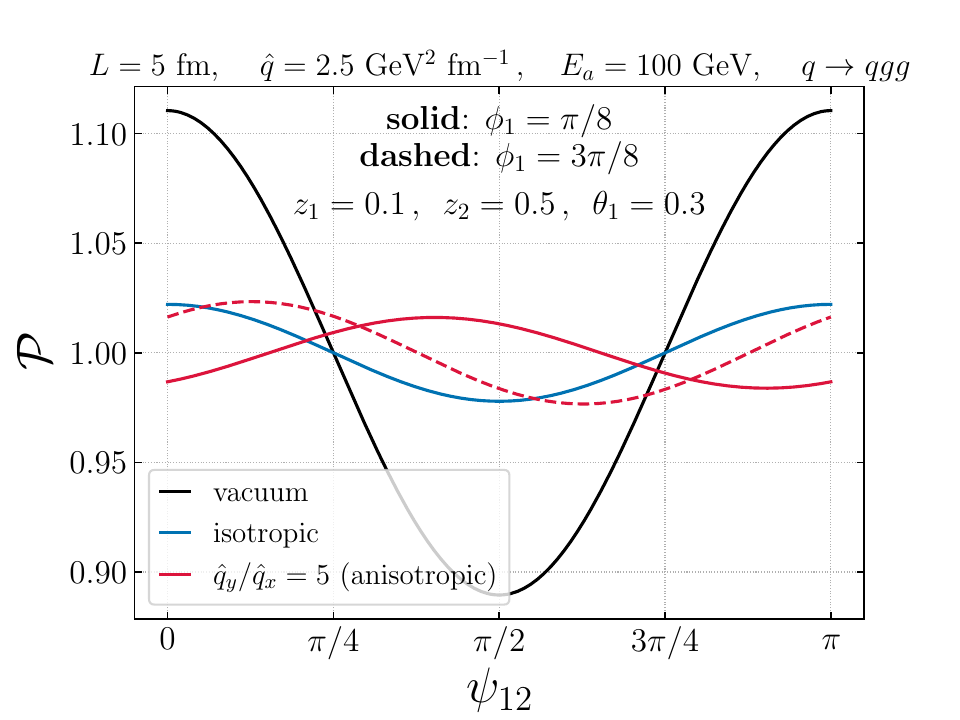}
\caption{Angular modulation in Eq.~\eqref{eq:ch5_p_anisotropic} as a function of $\psi_{12}$ for vacuum (black), isotropic medium (blue) and anisotropic medium (red). The process is fixed to be $q\rightarrow qg \rightarrow qgg$, so there is no dependence on $\theta_2$. The energy fractions are fixed to $z_1 = 0.1$ and $z_2 = 0.5$, while $\theta_1 = 0.15$ on the left panel and $\theta_1 = 0.3$ on the right panel. The solid and dashed curves correspond to the first splitting plane being more aligned with the $x$ and $y$ axes, respectively. In the left plot the medium anisotropy causes only a modulation change, while in the right one it induces only a phase shift.}
\label{fig:psi_anisotropic}
\end{figure}

To conclude, we now evaluate the in-medium modification to the cross-section, excluding the $(1+F_{\rm med})$ factor:
\begin{align}
\label{eq:ch5_p_anisotropic}
\cP(z_1,\theta_1,\phi_1,z_2,\psi_{12}) \equiv 1 + a_{\rm med}\cos(\psi_{12}+\phi_{\rm med}).
\end{align}
Similarly to Fig.~\ref{fig:psi_isotropic}, we fix the energy fractions to $z_1=0.1$ and $z_2=0.5$ and choose $\theta_1=0.15$ (left panel) and $\theta_1=0.3$ (right panel). Since we only focus on the result for a massless second splitting, $\cP$ does not depend on $\theta_2$. We plot Eq.~\eqref{eq:ch5_p_anisotropic} as a function of $\psi_{12}$ in Fig.~\ref{fig:psi_anisotropic} for vacuum (black curve), isotropic medium (blue curve) and anisotropic medium with $\hat q_y/\hat q_x = 5$ (red curves). The solid curve corresponds to the first splitting plane close to (right panel, $\phi_1=\pi/8$) or exactly (left panel, $\phi_1 = 0$) orthogonal to the maximum anisotropy direction, while the dashed curve corresponds to the first splitting plane close to (right panel, $\phi_1 = 3\pi/8$) or exactly (left panel, $\phi_1 = \pi/2$) aligned with this direction. The left panel showcases the impact of medium anisotropies on the spin-modulation signal purely at the magnitude ($a_{\rm med}$) level, since for $\phi_1 = 0,,\pi/2$ one has $\phi_{\rm med}=0$. For $\phi_1=0$ (solid) we observe $\mathcal{P}=1$, or equivalently $a_{\rm med}=1$ in Eq.~\eqref{eq:ch5_p_anisotropic}, while for $\phi_1 = \pi/2$ (dashed) there is a modulation enhancement with respect to the isotropic case. Thus, for certain regions of phase space, anisotropy can fully wash out azimuthal correlations even if they are present for the same region in the case of an isotropic medium.

The right panel highlights instead the phase shift in $\cP$, since for this choice of $(z_1,\theta_1,\phi_1)$ the value of $a_{\rm med}/a^{\rm iso}{\rm med}$ remains close to $1$, as can be seen by comparing the blue line with the red, dashed line in the right panel of Fig.~\ref{fig:psi_anisotropic}. For $\phi_1=\pi/8$ we observe that the anisotropic result for $\cP$ is shifted almost $90$ degrees with respect to the vacuum and isotropic results, $\phi{\rm med} \approx \pi$, in agreement with the findings presented in Fig.~\ref{fig:phimed_anisotropic}. This demonstrates that anisotropic broadening can modify the spin-correlation signal not only by changing its amplitude, but also through a genuine phase shift of the azimuthal modulation.

\section{Multi-point energy correlators in the QGP}
\label{sec:ch5_energy_correlators}

As introduced in Section~\ref{sec:ch3_energy_correlators}, energy correlators are defined in terms of energy-flow operators and provide a field-theoretic way of characterising the angular distribution of energy inside jets. In the heavy-ion context, most of the discussion has so far focused on the simplest two-point energy correlator, which we shall denote by E2C, equivalent to the notation used in the previous chapters (EEC). While the E2C is able to detect the presence of a scale, higher-point correlators can probe the underlying dynamics. In this sense, they are useful to access different properties of the QGP phase\footnote{In this section, we shall use the words medium and QGP interchangeably, since the present discussion is mostly focused on that phase.}, including the geometric characteristics of the medium response and the transport scales responsible for perturbative modifications to the jet cascade. In vacuum, the three-~\cite{Chen:2019bpb} and four-point~\cite{Chicherin:2024ifn} correlators have been computed, analysed~\cite{Chen:2022jhb} and measured~\cite{Chen:2022swd}. An initial exploration of the three-point correlator in the QGP was presented in~\cite{Bossi:2024qho}, highlighting its utility for mapping out the shape of the wake (medium response) of jets, while a related discussion of energy-loss effects on higher-point correlators was presented in~\cite{Barata:2024bmx}, which we addressed in Section~\ref{sec:elf}. Therfore, developing a theoretical understanding of multi-point correlators and their modification by QCD matter is a necessary step in extending energy-correlator studies beyond the two-point level.

Two recent developments make this extension tractable. First, the light-ray operator product expansion (OPE) was applied to energy correlators in nuclear environments, showing on general symmetry grounds that medium modifications can imprint themselves with well-defined scaling laws into the energy correlators~\cite{Andres:2024xvk}. Second, energy-correlator observables were analysed in hydrodynamical states, also leading to robust scaling laws~\cite{Barata:2024ieg}. Together, these two approaches provide the means to analyse both perturbative and hydrodynamic medium-response contributions to multi-point energy correlators. In what follows, we first discuss the inclusion of perturbative corrections to the jet fragmentation in projected energy correlators, before combining this contributions from the response of the medium. The full shape-dependent three-point correlator is then discussed in Sec.~\ref{subsec:ch5_shape_dependent_e3c}.

\subsection{Projected correlators and perturbative medium corrections}
\label{subsec:ch5_projected_correlators}

\begin{figure}[t!]
    \centering
    \includegraphics[width=.75\textwidth]{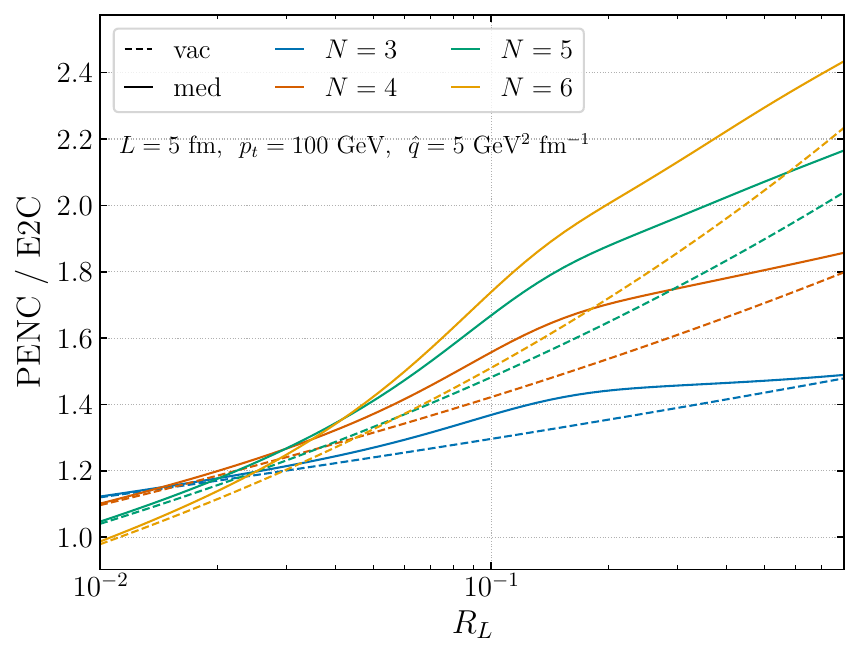} 
     \caption{PENC/E2C ratios for the vacuum (dashed) and in-medium (solid).}
     \label{fig:overall_perturbative}
\end{figure}

We begin by considering the simpler case of the projected energy correlators, which are obtained by integrating out the shape information and keeping only the information on the largest angle~\cite{Chen:2020vvp}. The projected $N$-point energy correlators (PENCs) provide a simple setting in which the collinear scaling of energy correlators is controlled by twist-two anomalous dimensions. In the collinear limit, at fixed coupling and to collinear leading-logarithmic accuracy, the quark PENC distribution, $d\Sigma_{\rm P}^{(N)}$, reads~\cite{Chen:2020vvp}
\begin{align}\label{eq:ch5_PENC_def}
 \frac{ d \Sigma^{(N)}_{\rm P}}{dR_L} = \frac{\alpha_s  R_L^{ \frac{\alpha_s}{\pi}\gamma(N+1)} }{\pi R_L}C_F \int_0^1 dz\,\bigl(1 - z^N - (1-z)^N\bigr)P_{q\to gq}(z)
\, ,
\end{align}
with $P_{q\to qg}(z)$ the vacuum splitting function striped of its colour factor (see Eq.~\eqref{eq:vac_splitting_functions})
\begin{align}
    \gamma(j)\equiv - \int_0^1 dz\, z^{j-1}\hat P_{gq}(z)
\end{align}
the twist-two anomalous dimension of spin $j=N+1$, and $\hat P_{gq}$ the regularized splitting function. The factor multiplying the splitting function is the PENC measurement function for a single collinear splitting, while the factor $R_L^{ \frac{\alpha_s}{\pi}\gamma(N+1)}$ is generated by the collinear evolution on top of the leading $dR_L/R_L$ behaviour. Thus, ratios of higher-point PENCs to the E2C suppress the common leading collinear scaling and isolate the relative perturbative anomalous scaling in QCD. We illustrate this in Fig.~\ref{fig:overall_perturbative}, in dashed lines, for $N=3-6$. 
This perturbative scaling admits a broader interpretation from the energy-correlator operator product expansion (OPE). As discussed briefly in Section~\ref{subsec:ch3_celestial_clover}, in the conformal limit the symmetries of the energy-flow operators strongly constrain the structure of energy correlators~\cite{Hofman:2008ar}. For the PENCs, this implies a collinear expansion organized by twist,
\begin{align}\label{eq:ch5_hj1}
 \frac{ d \Sigma^{(N)}_{\rm P}}{dR_L} = \sum_{k=1} a^{{\rm P}(N)}_{\tau,0} R_L^{\tau-3}\Big\vert_{\tau=2k}
\, .
 \end{align}
The leading term corresponds to $\tau=2$ and gives the characteristic $1/R_L$ collinear behaviour. Eq.~\eqref{eq:ch5_PENC_def} is the perturbative, fixed-coupling realization of this leading-twist behaviour, with the scaling exponent shifted by the corresponding twist-two anomalous dimension. 

Let us now discuss how the QGP imprints onto Eqs.~\eqref{eq:ch5_PENC_def} and~\eqref{eq:ch5_hj1}, focusing first on perturbative modifications to fragmentation, and leaving the contributions from the associated medium response to Section~\ref{subsec:ch5_projected_medium_response}. The separation in Eq.~\eqref{eq:ch5_PENC_def} between the collinear evolution factor and the splitting kernel is useful in this context. At leading order, the perturbative medium correction enters through the real $1\to2$ splitting kernel and can therefore be incorporated by replacing the vacuum splitting function with its medium modified counterpart, i.e., $P_{q\to gq}=P_{q\to gq}^{\rm vac}+P_{q\to gq}^{\rm med}$, where $P_{q\to gq}^{\rm med}$ has been widely discussed in the literature (see, e.g,~\cite{Isaksen:2023nlr,Isaksen:2020npj,Sievert:2019cwq,Apolinario:2014csa,Blaizot:2012fh}). Explicitly, these splitting functions can be computed from the $1\to 2$ cross-section
\begin{align}
    P^{\rm med}_{i\to jk}(z,\p^2) = \frac{2\pi}{\alpha_s}\frac{\p^2\,d\sigma^{\rm med}_{i\to jk}}{dz\,d\p^2}\,,
\end{align}
where $z$ is the energy fraction of particle $j$ with respect to $i$, $\p$ the relative transverse momentum of the pair $jk$, with $|\p| = z(1-z)p_t R_L$ and $p_t$ the energy of the jet/of particle $i$, and the superscript "med" implies we have already subtracted the pure vacuum contribution to the cross-section. Within the BDMPS-Z formalism introduced in Section~\ref{sec:BDMPS_Z}, and for a static, homogeneous and isotropic medium of finite length, the medium modification to the splitting function reads (see, e.g.,~\cite{Isaksen:2023nlr})
\begin{align}\label{eq:ch5_med_splitting_function}
    P^{\rm med}_{i\to jk}(z) = C_{ jk}\,P^{\rm vac}_{i\to jk}(z) \left[\frac{\p^2}{2\omega^+}2\Re\,(-i) \int_0^{L^+}dx^+\frac{c_2}{c_3}e^{\frac{i\p^2}{2\omega^+ c_3}}-2\Re\left(1 - e^{-\frac{i\p^2 \tan\Omega L^+}{2\omega^+
    \Omega}}\right)\right]\,,
\end{align}
with $\omega^+ = z(1-z)p^+$, $p^+ = p_t\sqrt{2}$ the light-cone energy of the initial particle $i$, and we have introduced the notation\footnote{We are using the fundamental jet quenching parameter $\hat q \equiv \hat q_F$.}
\begin{align}\label{eq:ch5_ci_definition_iso}
    c_2(x^+) = \frac{\Omega}{\tan \Omega x^+}\,,\qquad c_3(x^+) = x^+\left(-i\frac{\hat q Q_{jk}(z)}{2\omega^+}\right) - c_2(x^+)  \,,
\end{align}
with a process-dependent complex oscillatory frequency
\begin{align}\label{eq:ch5_Omega_definition_iso}
    \Omega = \frac{1-i}{\sqrt{2}}\sqrt{\frac{\hat q F_{jk}(z)}{2\omega^+}}\,.
\end{align}
The particular form of Eq.~\eqref{eq:ch5_med_splitting_function} was obtained in the large-$N_c$ limit and using the dipole potential in the HO approximation (see Section~\ref{sec:xsecs_BDMPSZ}). Importantly, as in Eq.~\eqref{eq:ch5_Q_approx}, we have approximated the quadrupole describing the propagation of the two particle final-state in the medium by a product of dipoles. As discussed around Eq.~\eqref{eq:ch5_Q_approx}, this approximation quantitatively describes well the kinematic region where $|\p| =z(1-z) p_t \theta \lesssim \sqrt{\hat q L^+}$, while reasonably describing the main qualitative features of medium modifications outside of this region. The process-dependent functions $Q$ and $F$ appearing in Eqs.~\eqref{eq:ch5_ci_definition_iso} and~\eqref{eq:ch5_Omega_definition_iso} read
\begin{align}
    & C_{gq} = C_F, \,\, Q_{gq}(z) = 2(1-z)^2 + z^2,\,\,F_{gq}(z) = z P_{gq}(z)\,,\nn
    & C_{gg} = C_A, \,\, Q_{gg}(z) = 2((1-z)^2 + z^2),\,\, F_{gg}(z) = 1 + z^2 + (1-z)^2\, ,
\end{align}
where, although only the $q\to qg$ is relevant for computing the medium modification to the quark PENC in this and the following subsection, we also present the factors for $g\to gg$ which are relevant for Section~\ref{subsec:ch5_shape_dependent_e3c}. 

Inserting the in-medium splitting function in Eq.~\eqref{eq:ch5_med_splitting_function} into Eq.~\eqref{eq:ch5_PENC_def} and expanding for $R_L\ll 1$, with $|\p| = z(1-z)p_t R_L$, the quark PENC distribution in the OPE limit and with the (vacuum) leading-twist contribution subtracted reads~\cite{Barata:2025fzd}
\begin{align}\label{eq:ch5_hj1_med}
 \frac{ d \Sigma^{(N)}_{\rm P, \,  med \, (no \, vac) }}{dR_L}  = \sum_{k=2} b^{{\rm P}(N)}_{\tau,0}(\omega_c/p_t,\theta_c) R_L^{\tau-3}\Big\vert_{\tau=2k} \, ,
\end{align}
where the coefficients read
\begin{align}\label{eq:perturbative_b_coeffs}
b^{{\rm P}(N)}_{\tau,0} = \frac{\alpha_s}{\pi}C_F\int_0^1 dz \, (1-z^N - (1-z)^N) P_{q\to gq}^{\rm vac}(z)    \mathcal{W}_\tau(\omega_c/p_t, \theta_c, z) \, ,
\end{align}
with
\begin{align}
    & \mathcal{W}_\tau(\omega_c/p_t, \theta_c, z) = \nn
    & \left[z(1-z)\frac{p_t}{\omega_c}\theta_c^{-2}\right]^{k-1}\frac{2}{(k-1)!}\Re \left\{-(k-1)\left[i^{k-1}\int_0^1d\xi\,\frac{\bar c_2(\xi)}{\bar c_3(\xi)^{k-1}}\right]+\left[(-i)^{k-1}\left(\frac{\tan\bar\Omega}{\bar\Omega}\right)^{k-1}\right]\right\}\Bigg\vert_{\tau=2k}\,,
\end{align}
and
\begin{align}
    \bar\Omega = (1-i)\sqrt{\frac{F_{gq}(z)}{\sqrt{2}z(1-z)}\frac{\omega_c}{p_t}}\,,\qquad \bar c_2(\xi) = \frac{\bar\Omega}{\tan(\bar\Omega \xi)}\,,\qquad \bar c_3(\xi) = -i\,\xi\,\frac{\sqrt{2}Q_{gq}(z)}{z(1-z)}\frac{\omega_c}{p_t}-\bar c_2(\xi)\,.
\end{align}
Thus, the medium-induced modification to the PENC has the same twist organization as the OPE limit expansion of the vacuum PENC in Eq.~\eqref{eq:ch5_hj1}, contributing with the coefficients $b^{{\rm P}(N)}_{\tau,0}$ which start at twist $\tau = 4$. These coefficients are controlled by two characteristic medium scales discussed in Section~\ref{subsec:quenching_mechanisms}, namely $\omega_c = (\hat q L^2)/2$, the critical gluon frequency below which the QCD LPM effect is prevalent, and $\theta_c = 2/\sqrt{\hat q L^3}$, the respective characteristic angle. This parametric dependence becomes more complex in an evolving medium, but can still be derived~\cite{Adhya:2020xcb,Baier:1998yf}. Importantly, although the explicit parametric dependence of the coefficients in Eq.~\eqref{eq:ch5_hj1_med} was realised for a specific model, the resulting collinear expansion is general since it follows from the OPE structure of energy correlators. Thus, higher-order perturbative corrections are expected to modify the coefficients, but not the symmetry-based twist organisation (also see the discussion around Eq.~\eqref{eq:ch3_EEC_OPE}). Further, all odd-twist coefficients vanish, since CP-odd interactions are neglected and the medium is structureless. Measurements yielding non-vanishing odd coefficients would then give access to non-perturbative features of the jet--medium interactions and to the spatial structure of the QGP, not yet addressed by any other jet observable. Further discussion on these aspects is left to future work. In Fig.~\ref{fig:overall_perturbative} the perturbative medium contributions to the leading order PENC ratios are shown in solid lines. Compared to the vacuum, there is an enhancement increasing with $N$ at large $R_L$, reflecting the region where $b_{\tau,0}>a_{\tau,0}$.

\subsection{Perturbative and response contributions to projected correlators}
\label{subsec:ch5_projected_medium_response}

The imprint of the jet-induced medium response on PENCs can be addressed by separating the energy flow into a fluctuating, perturbative component and a mean classical component, in line with most theoretical descriptions of the response used so far. This construction follows closely the one introduced in~\cite{Barata:2024ukm}, where the corresponding geometrical correlations were discussed in detail for the E2C. Restricting, for simplicity, the discussion to the PE3C/E2C ratio, we implement this modelling separation by writing the energy flow operator as \begin{align}
    \mathcal{E}(\bn) =\hat{\mathcal{E}}_{\rm h}(\bn)+\mathcal{E}_{\rm c}(\bn) \,. 
\end{align}
The first term represents the fluctuating part of the flow operator associated to the hard (h) jet fragmentation, discussed in the previous subsection, while the second term accounts for the classical (c) flux, including the medium response. The latter contribution is intended to capture the mean, classical contribution\footnote{Correlations of these response contributions may appear, e.g. due to hydrodynamic modes, at distinct (smaller) angular scales~\cite{Barata:2026pgh}. However, these correlations go beyond the current treatment of the medium response in phenomenological considerations and are left for future study.} of the medium response to the jet, which is usually obtained by evolving the energy and momentum deposited by the jet with linearized hydrodynamics or kinetic theory.

\begin{figure}[t!]
    \centering
    \includegraphics[width=.48\columnwidth]{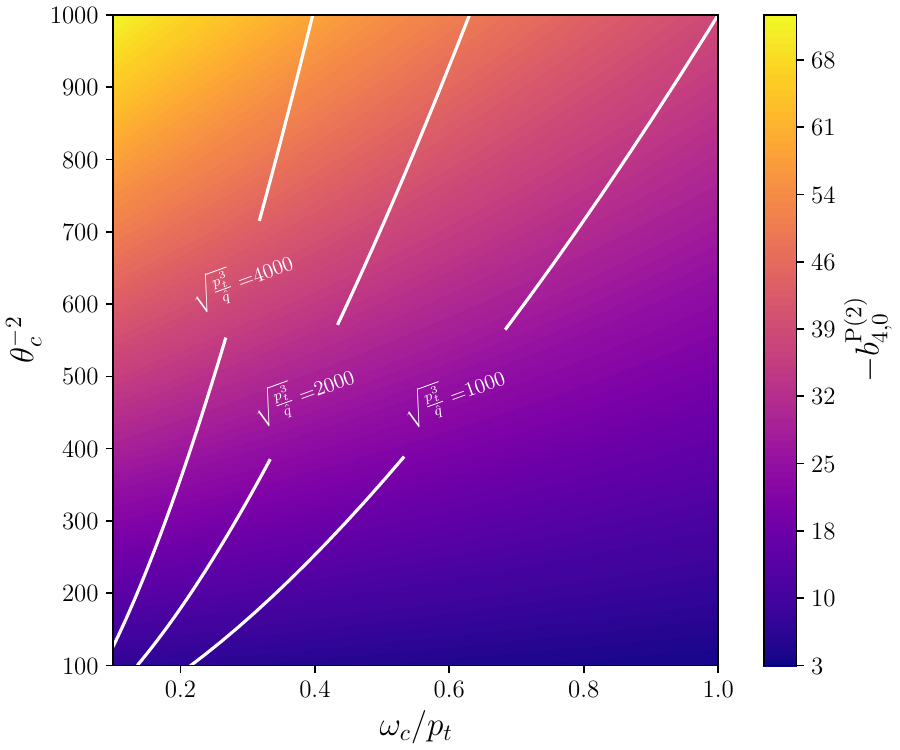}  \hspace{.0 cm}
     \includegraphics[width=.48\columnwidth]{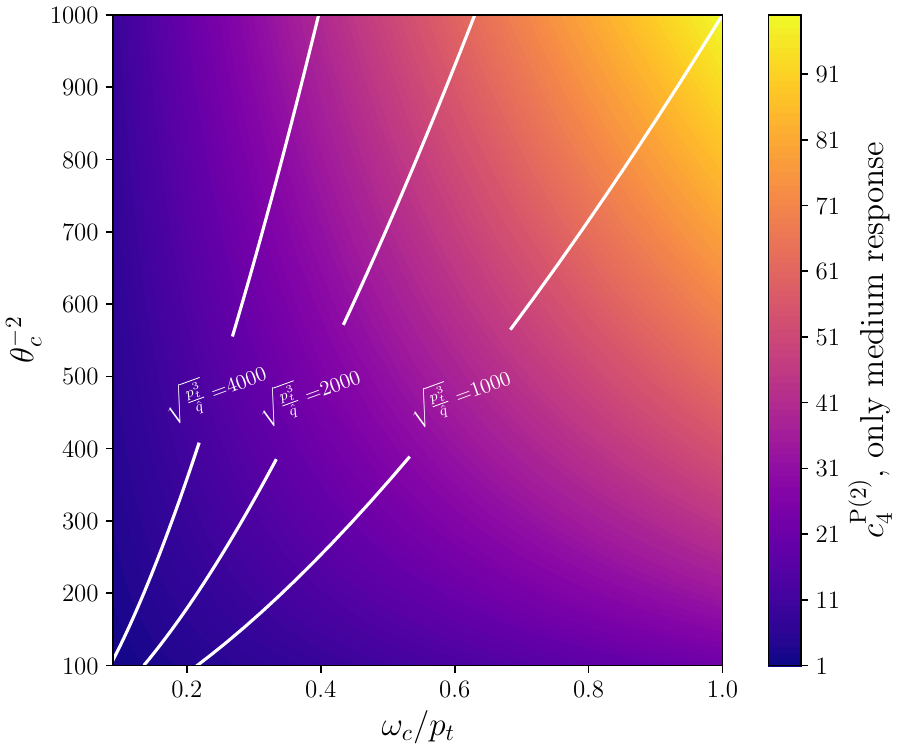}
     \caption{Coefficients of the E2C in the OPE limit, for the medium perturbative corrections, $b_{4,0}^{{\rm P}(2)}$, and the response, $c_{4}^{{\rm P}(2)}$, neglecting hard correlations. White lines have $p_t^3/\hat q$ constant.}
    \label{fig:E3C_uu1}
\end{figure}

The corresponding contributions to the PE3C, namely ccc, chh, and hhh, can be expressed through the leading perturbative correlators of the h component combined with the c piece. Note that since, by definition, $\langle \hat{\mathcal{E}}_h\rangle = 0$, terms like cch do not contribute. Here, no attempt is made to constrain the detailed event-by-event profile of the classical energy flux accompanying a jet, which would require a consistent description of both the perturbative jet--medium interaction and the subsequent medium response. Instead, one relies on a two-parameter model, $\mathcal{E}_c(\bn)=\frac{\Delta}{\pi\theta_0^2} e^{-\theta^2/\theta_0^2}$, where $\theta_0$ controls the angular size of the classical energy flux and $\Delta$ accounts for the amount of energy propagating in the medium. It should be stressed that the exact functional form of $\mathcal{E}_c$ does not qualitatively affect the leading collinear behaviour of the response contribution to energy correlators. Rather, It is rather determined by the total amount of energy carried by the response ($\Delta$) and by its characteristic angular support ($\theta_0$). The details of the medium-response contribution to ENCs and PENCs can then be made explicit without specifying a microscopic model for the classical part of the energy flux, although here we focus on the E2C and PE3C. Starting with the E2C, the fluctuating contribution (hh) is fixed by the harder particles. In the following expressions, we take vacuum scaling of this contribution, $d\Sigma^{(2)}_{\rm P, vac}/dR_L\simeq a^{{\rm P}(2)}_{2,0} R_L^{\gamma(3)-1}$ (see Eqs.~\eqref{eq:ch5_PENC_def} and~\eqref{eq:ch5_hj1}), while in the full numerical result we shall insert the appropriate perturbative medium modification. In the collinear limit ($R_L \ll 1$), one then finds
\begin{align}
&\frac{d\Sigma^{(2)}}{dR_L}\Bigg\vert_{\rm vac + response}=\frac{1}{p_t^2} \int_{n_{1,2}}\langle \mathcal{E}_{1}\mathcal{E}_{2}\rangle\delta(\cos{R_L}-n_1\cdot n_2)\sin R_L \simeq \frac{a^{{\rm P}(2)}_{2,0}}{R_L^{1-\gamma(3)}}\notag\\
&\hspace{1.5cm}+\frac{(2\pi)}{p_t^2}\hat{\sum}\left[ \int d\phi_1 d\cos\theta_1\,\mathcal{E}_c(\cos\theta_1)\mathcal{E}_c\left(\cos\theta_1\cos R_L+\sin\theta_1\sin R_L\cos\phi_1\right)\right]\sin R_L\,,
\end{align}
where $\hat{\sum}$ denotes the average over events, kept implicit in what follows, while the geometrical correlation of the classical flux in a given event is shown explicitly. Further expanding the classical energy flux in $R_L$, all even powers of $R_L$ vanish after azimuthal averaging, as long as the energy flux is symmetric around the jet axis. This gives~\cite{Barata:2025fzd}
\begin{align}
\frac{d\Sigma^{(2)}}{dR_L}\Bigg\vert_{\rm vac + response}=\frac{a^{{\rm P}(2)}_{2,0}}{R_L^{1-\gamma(3)}}+\sum_{k=2}c^{{\rm P}(2)}_\tau\left[\mathcal{E}_c\right]R_L^{\tau-3}\Big\vert_{\tau=2k}\,,
\end{align}
where for $\mathcal{E}_c(\bn)=\frac{\Delta}{\pi\theta_0^2} e^{-\theta^2/\theta_0^2}$ one has $c^{{\rm P}(2)}_4=\Delta^2/(p_t^2\theta_0^2)$. Even powers of $R_L$ can be generated for azimuthally perturbed fluxes, as can be checked by taking $\mathcal{E}_c=\delta(\phi)f(\theta)$. Turning to the PE3C, defined by
\begin{align}
  \frac{d\Sigma^{(3)}_{\rm P}}{dR_L} &=\frac{1}{p_t^3} \int_{n_{1,2,3}}\langle \mathcal{E}_1\mathcal{E}_2\mathcal{E}_3\rangle\delta(\cos{R_L}-n_1\cdot n_2)\Theta(|n_1-n_2|-|n_2-n_3|)\Theta(|n_2-n_3|-|n_1-n_3|)\sin R_L\,,
\end{align}
one finds for the purely classical contribution
\begin{align}
\frac{d\Sigma^{(3)}_{\rm P,ccc}}{dR_L} &\simeq \frac{(2\pi)^2}{p_t^3}\,\left[\int d\cos\theta_1\,\mathcal{E}_{c}^3(\theta_1)\right]\,\left(\frac{\pi}{3}-\frac{1}{4}\tan\frac{\pi}{3}\right) R_L^3 = \frac{\Delta^3}{\theta_0^2p_t^3}\,\left(\frac{2}{9}-\frac{\sqrt{3}}{6\pi}\right)R_L^3\,,
\end{align}
where in the last step the explicit form of the model classical energy flux has been used. The mixed contributions with one classical insertion and two hard insertions give
\begin{align}
\frac{d\Sigma^{(3)}_{\rm P,chh}}{dR_L}+\frac{d\Sigma^{(3)}_{\rm P,hch}}{dR_L}+\frac{d\Sigma^{(3)}_{\rm P,hhc}}{dR_L}
&\simeq \frac{ a^{{\rm P}(3)}_{2,0} \Delta}{6\pi p_t}\,\mathcal{C}\,R_L^{1+\gamma(3)}\,,
\end{align}
where, once again, for demonstration purposes, we inserted the vacuum scaling for the (hh) contribution and used $3 \, a^{P(2)}_{2,0}=2 \, a^{P(3)}_{2,0}$ at leading order. The remaining coefficient is given by the following integral
\begin{align}
\mathcal{C}=2\int_0^{\pi/3}d\phi\int_{\frac{1}{2\cos\phi}}^1dt\left(t+t^{\gamma(3)-1}+\frac{t}{(1+t^2-2t\cos\phi)^{1-\frac{\gamma(3)}{2}}}\right)\,.
\end{align}
Combining the hard (hhh) contribution, given by $d\Sigma^{(3)}_{\rm P, vac}/dR_L\simeq a^{{\rm P}(3)}_{2,0} R_L^{\gamma(4)-1}$, with the mixed and purely classical terms gives
\begin{align}\label{eq:PE3C_vac_response}
\frac{d\Sigma^{(3)}_{\rm P}}{dR_L}\Bigg\vert_{\rm vac + response}
&\simeq \frac{ a^{{\rm P}(3)}_{2,0}}{R_L^{1-\gamma(4)}} + \frac{ a^{{\rm P}(3)}_{2,0} \Delta}{6\pi p_t}\,\mathcal{C}\,R_L^{1+\gamma(3)} + \frac{\Delta^3}{\theta_0^2p_t^3}\,\left(\frac{2}{9}-\frac{\sqrt{3}}{6\pi}\right)R_L^3\,.
\end{align}
Thus, at leading order, the ratio of the leading-twist vacuum result convoluted with the classical flux of the PE3C to that of the E2C reads~\cite{Barata:2025fzd}
\begin{align}\label{eq:ch5_ratio_E3C_E2C}
&\frac{\rm PE3C}{\rm E2C}\Bigg\vert_{\rm vac + response}{\hspace{-.5 cm}} = \Bigg(\frac{ a^{P(3)}_{2,0}}{R_L^{1-\gamma(4)}} 
+ \frac{ a^{P(3)}_{2,0} \Delta}{6\pi p_t}\,\mathcal{C}\,R_L^{1+\gamma(3)} 
% \nn 
% &{\hspace{-.3 cm}}
+ \frac{\Delta^3}{\theta_0^2p_t^3}\,\left(\frac{2}{9}-\frac{\sqrt{3}}{6\pi}\right)R_L^3\Bigg)\Bigg(\frac{ a^{P(2)}_{2,0}}{R_L^{1-\gamma(3)}}+\frac{\Delta^2 R_L}{p_t^2\theta_0^2}\Bigg)^{-1} \, .
\end{align}
\begin{figure*}[t!]
    \centering
    \includegraphics[width=.49\textwidth]{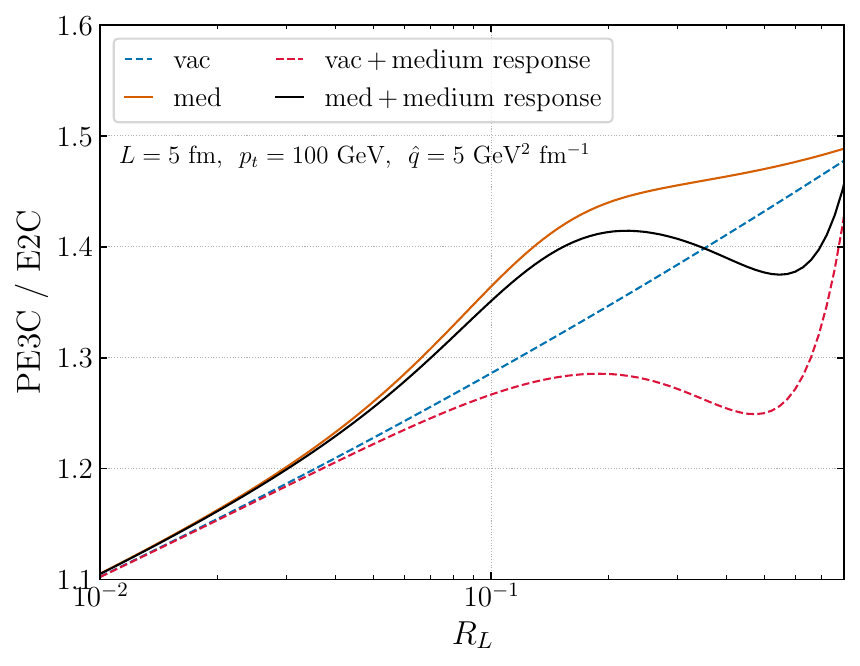}  
    \includegraphics[width=.49\textwidth]{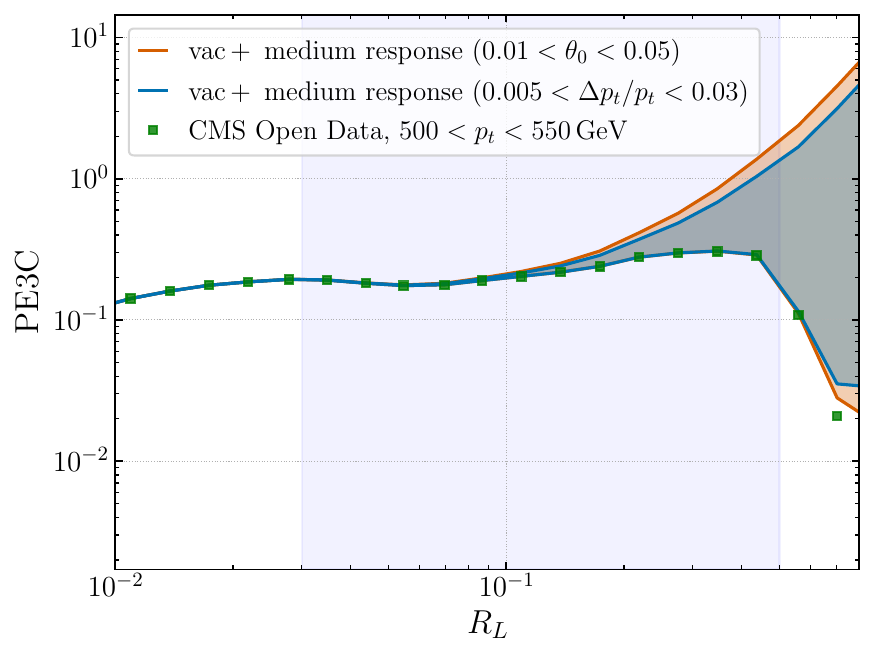}
     \caption{\textbf{Left:} PE3C/E2C including the four combinations of medium response and perturbative modifications on top of the vacuum. \textbf{Right:} PE3C extracted from CMS Open Data presented in~\cite{Komiske:2022enw}, with a theory band for the medium response, obtained by convoluting with the PE3C data. The $\theta_0$ band has $\Delta/p_t = 0.02$ while the $\Delta/p_t$ band has $\theta_0 = 0.025$.  The purple shaded area indicates the region where the perturbative calculation of the EEC is valid; the points for $R_L>0.5$ lie outside the cone of the jet where the PE3C data is extracted.}
     \label{fig:overall_response}
\end{figure*}In the left panel of Fig.~\ref{fig:overall_response}, Eq.~\eqref{eq:ch5_ratio_E3C_E2C} is evaluated for the vacuum result convoluted with the classical response (red), and compared to the purely vacuum result (blue), the result with only the perturbative medium correction from Eq.~\eqref{eq:ch5_hj1_med} added to the vacuum (orange), and the combined result with the perturbative correction also convoluted with the classical flux (black). The parameters used are $\theta_0=0.025$, $p_t=100$ GeV, $\theta_{\rm cut}=\Lambda_{\rm QCD}/p_t=0.002$, $\Delta = 0.01 , p_t$, $a^{P(3)}_{2,0} = 0.286$, and $\alpha_s = 0.3$. Including the response produces a characteristic imprint in the ratio’s shape, with a depletion at large angles, not observed when only the perturbative result is considered. This is in qualitative agreement with recent Hybrid model results~\cite{Bossi:2024qho,Casalderrey-Solana:2014bpa,Casalderrey-Solana:2015vaa,Casalderrey-Solana:2016jvj}. To better gauge the phenomenological impact of these contributions, in the right panel of Fig.~\ref{fig:overall_response} the medium response is combined with the CMS data analysis in~\cite{Komiske:2022enw}, leading to a large-angle enhancement of the PE3C. We emphasize that the shape of the PE3C in the collinear limit should depend weakly on the functional form of the classical flux, although it is sensitive to the parameters chosen. Finally, together with the perturbative coefficient derived in Eq.~\eqref{eq:perturbative_b_coeffs} (left panel), the leading response coefficient $c_{4}^{P(2)}$ to the E2C (right panel) is shown in Fig.~\ref{fig:E3C_uu1} as a function of the two characteristic medium scales $(\theta,\omega_c)$, displaying the different scaling of the response with respect to the perturbative term. Following the estimate of~\cite{Barata:2024ukm}, the leading response coefficient is assumed to scale as $c_4 \propto \hat q L^4 T^2/p_t$, or equivalently $c_4 \propto (\omega_c/p_t),\theta_c^{-4/3}$ for $T\propto \hat q^{1/3}$.

In general, the E2C associated to the medium response, in the collinear limit, takes the form
\begin{align}\label{eq:ch5_hj1_wakes}
\frac{ d \Sigma^{(2)}_{\rm P\, ,  response}}{dR_L}  = \sum_{k=2} c^{{\rm P}(2)}_{\tau}\Big[ {\mathcal{E}}_c\Big] R_L^{\tau-3}\Big\vert_{\tau=2k}\, ,
\end{align}
where the coefficients are functionals of the classical energy flux, given by angular averages of its powers and derivatives. Remarkably, the expansion in Eq.~\eqref{eq:ch5_hj1_wakes}, independent of the OPE for the energy flow operator, has the same structure as Eqs.~\eqref{eq:ch5_hj1} and~\eqref{eq:ch5_hj1_med}. Moreover, the leading coefficients in the expansion for the response’s PE3C also follow this OPE-limit structure (see Eq.~\eqref{eq:PE3C_vac_response}), with coefficients again depending only on integral moments of its powers and derivatives. These relations, i.e. Eqs.~\eqref{eq:ch5_ratio_E3C_E2C} and~\eqref{eq:ch5_hj1_wakes}, hold for any $N$ for the leading $\tau$ terms, and we conjecture they hold for all $\tau$, assuming the classical energy flux to be cylindrically symmetric and ignoring azimuthal fluctuations. However, such asymmetries may appear due to the medium evolution and anisotropy, both in the in-medium splitting function and in the medium response, and can be especially important in small HIC systems~\cite{Barata:2023zqg,Sadofyev:2021ohn,Barata:2023qds,Kuzmin:2023hko,Barata:2024bqp,Kuzmin:2024smy}. These lead to finite even contributions to Eq.~\eqref{eq:ch5_hj1_wakes}, suggesting prospective avenues to disentangle the in-medium physics from vacuum QCD, and providing a tool to access the azimuthal jet substructure.

In sum, these observations show that the perturbative terms and the medium response provide direct, competing contributions to the PENCs, requiring a systematic analysis before they can be disentangled. Eqs.~\eqref{eq:ch5_hj1_med} and~\eqref{eq:ch5_hj1_wakes} give the full parametric dependence of the coefficients in the OPE limit, indicating a path towards a theoretically driven decomposition of these effects in experiment.

\subsection{Shape dependence of the three-point energy correlator}
\label{subsec:ch5_shape_dependent_e3c}

We now consider the full shape-dependent three-point correlator, which in the collinear limit can be obtained directly by integrating the $1\to 3$ splitting function~\cite{Chen:2019bpb,Chen:2022jhb}
\begin{align}\label{eq:ch5_J3}
 J_3 = \int d\Phi_3 \, \frac{g^4 \, P_3 \, \,  z_1 z_2 z_3}{2s_{123}^2} \, , 
\end{align} 
where $(4\pi)^5d\Phi_3 =ds_{12}ds_{13} ds_{23}\prod_{i}  dz_{i} \, 4 (-D)^{-1/2} $ denotes the three-body phase space~\cite{Gehrmann-DeRidder:1997fom}, with $s_{123} = \sum_{i\neq j} s_{ij}$, $z_i \equiv E_i/p_t$ is the energy fraction carried by the final state particle $i=1,2,3$ with $\sum_i z_i = 1$, and flavour dependence is implicit. The momentum variables read $s_{ij} =  z_i z_j p_t^2 \theta_{ij}^2 $, while the determinant reads 
\begin{align}
    D =  (z_3 s_{12}-z_1 s_{23}- z_2 s_{13})^2-4 z_1 z_2 s_{13}s_{23} <0\,.
\end{align}
We restrict the discussion to the quark channel, where the vacuum splitting kernels are $P_{2, \, {\rm vac}}=P_{gq}$, $P_{3, \,{\rm vac}}  = C_F^2 P_{3, \,{\rm vac}}^{\rm Abel.} + C_F C_A P_{3, \,{\rm vac}}^{\rm non-Abel.}$. For completeness, these terms read~\cite{Catani:1998nv}
\begin{align}\label{eq:ch5_exact_vac_P3}
 & P_{3, \,{\rm vac}}  = C_F^2 P_{3, \,{\rm vac}}^{\rm Abel.} + C_F C_A P_{3, \,{\rm vac}}^{\rm non-Abel.}\, ,\nn 
 & P_{3, \,{\rm vac}}^{\rm Abel.} = \frac{s_{123}^2}{2s_{13}s_{23}} \frac{z_3(1+z_3)^2}{z_1z_2} + \frac{s_{123}}{s_{13}} \frac{z_3(1-z_1)+(1-z_3)^2}{z_1z_2} - \frac{s_{23}}{s_{13}} + (1\leftrightarrow 2) \nn 
 &P_{3, \,{\rm vac}}^{\rm non-Abel.} = \frac{(2(z_1s_{23}-z_2s_{13}) + (z_1-z_2)s_{12})^2}{4(z_1+z_2)^2s_{12}} +\frac{1}{4} + \frac{s^2_{123}}{2s_{12}s_{13}} \left(\frac{1+z_3^2}{z_2} + \frac{1+(1-z_2)^2}{1-z_3}\right)\, , \nn 
 &- \frac{s_{123}^2 z_3 (1+z_3^2)}{4s_{13}s_{23} z_1z_2} + \frac{s_{123}}{2s_{12}} \left( \frac{z_1(2-2z_1+z_1^2)-z_2(6-6z_2+z_2^2)}{z_2(1-z_3)}\right) \nn 
 &+ \frac{s_{123}}{2s_{13}} \left(\frac{(1-z_2)^3+z_3^2-z_2}{z_2(1-z_3)} -\frac{z_3(1-z_1)+(1-z_2)^3}{z_1z_2} \right) +  (1\leftrightarrow 2)\, ,
\end{align}

Since the theoretical description of $P_{3, \, {\rm med}}$ in HICs is still limited,\footnote{See, however,~\cite{Fickinger:2013xwa} for calculations of $P_{3, {\rm med}}$ in dilute matter.}, here we consider a factorized \textit{cascade} approximation~\cite{Fickinger:2013xwa}, where the $1\to 3$ process is given by a succession of $1\to 2$ branchings:
\begin{align}\label{eq:ch5_P3_cascade}
\frac{P_{3}^{\rm \, cascade}}{s_{123}} &=  \frac{P^{2,1+3}_{q\to gq}P^{1,3}_{q\to gq}  }{s_{13}} + (1\leftrightarrow 2)
+\frac{P_{q\to gq}^{1+2,3}P_{g\to gg}^{1,2} }{s_{12}}  \, ,
\end{align}
\begin{figure}[t!]
    \centering
    \includegraphics[width=.85\columnwidth]{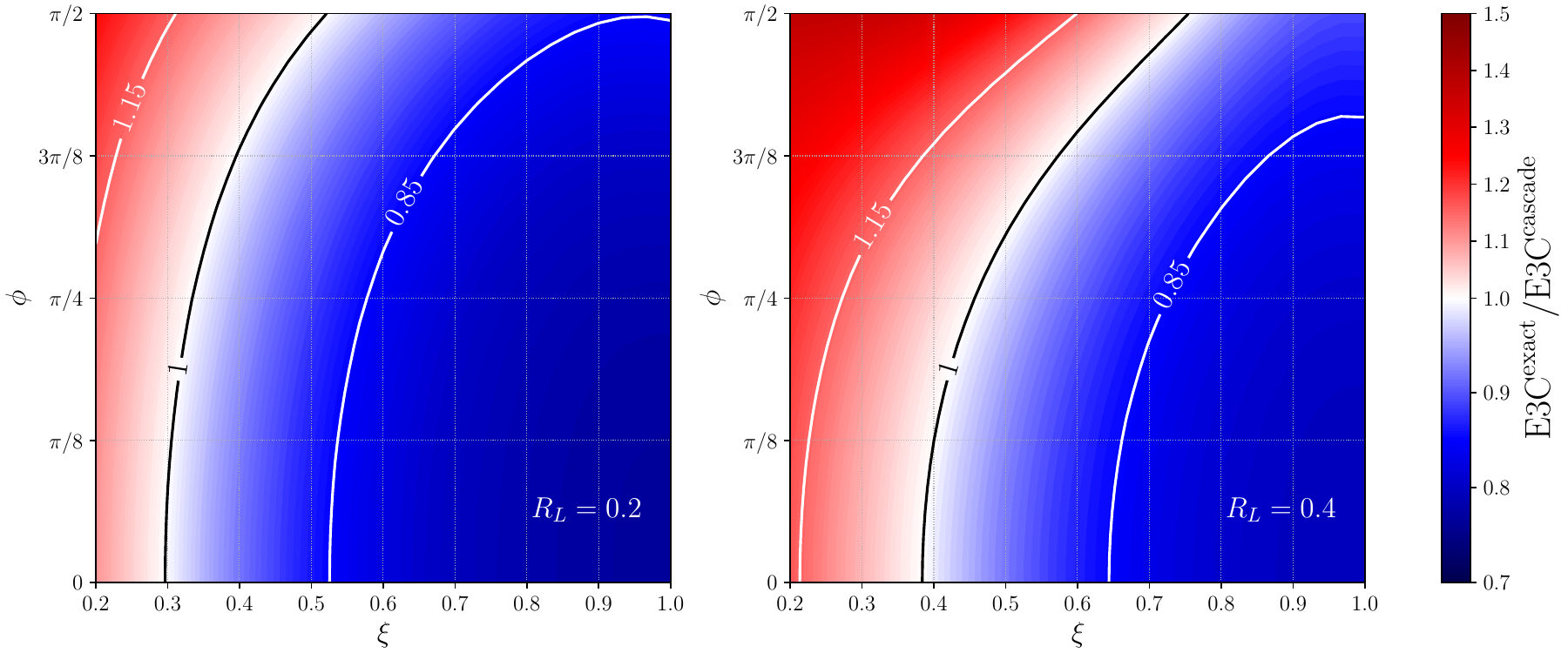}
    \caption{Leading order vacuum E3C ratio with respect to the E3C in the cascade approximation.}
    \label{fig:cascase}
\end{figure}
where the $1\to 2$ splitting functions in vacuum are given by Eq.~\eqref{eq:vacuum_splitting_functions} multiplied by the appropriate colour factors, e.g. $P_{gq}^{1,2}(x)\equiv  C_F (1+(1-x)^2)/x$, and $ P_{gg}^{1,2} =  2C_A \left(\frac{x}{1-x} + \frac{1-x}{x} + x(1-x)\right)$, with $x = \frac{E_1}{E_1+E_2} $. The medium modification to these splitting functions is given in Eq.~\eqref{eq:ch5_med_splitting_function}. For validation of this approximation, in Fig.~\ref{fig:cascase} we evaluate the ratio between the E3C obtained from $P_3$ in Eq.~\eqref{eq:ch5_exact_vac_P3} and the E3C obtained from the splitting function in the cascade approximation, following Eq.~\eqref{eq:ch5_P3_cascade}. This ratio, as well as subsequent plots, are plotted in $(\xi,\phi)$ plane, with coordinates which compactly parametrize the triangle shape~\cite{Chen:2022jhb}. They are defined in terms of the three ordered angles $R_L=\theta_{12}>\theta_{23}>\theta_{13}$ as
\begin{align}
    \xi = \frac{\theta_{13}}{\theta_{23}}\,,\qquad \sin^2 \phi = 1- \frac{(R_L-\theta_{23})^2}{\theta_{13}^2}\,,
\end{align}
with $0< 2\phi<\pi$ and $0<\xi<1$. We observe that the deviations between the cascade and exact solutions are mainly constrained to regions where the three detectors are either in flattened or squeezed configurations, with a $\mathcal{O}(30\% - 50\%)$ magnitude. Surprisingly, one would expect the cascade approximation, where successive emissions are factorized iteratively, to be applicable in the squeezed configuration after integrating out the relative azimuthal angle~\cite{Craft:2022kdo}, and similarly in a dilute medium~\cite{Fickinger:2013xwa}. Here, however, the shape-dependent E3C remains differential in the three pairwise opening angles and therefore retains this azimuthal structure, not leading to a trivial factorization. As a result, the ratio between the exact result and the cascade approximation in vacuum has a more involved behaviour in the $(\xi,\phi)$ variables~\cite{Chen:2022swd}. 

\begin{figure}[t!]
    \centering
    \includegraphics[width=0.9\columnwidth]{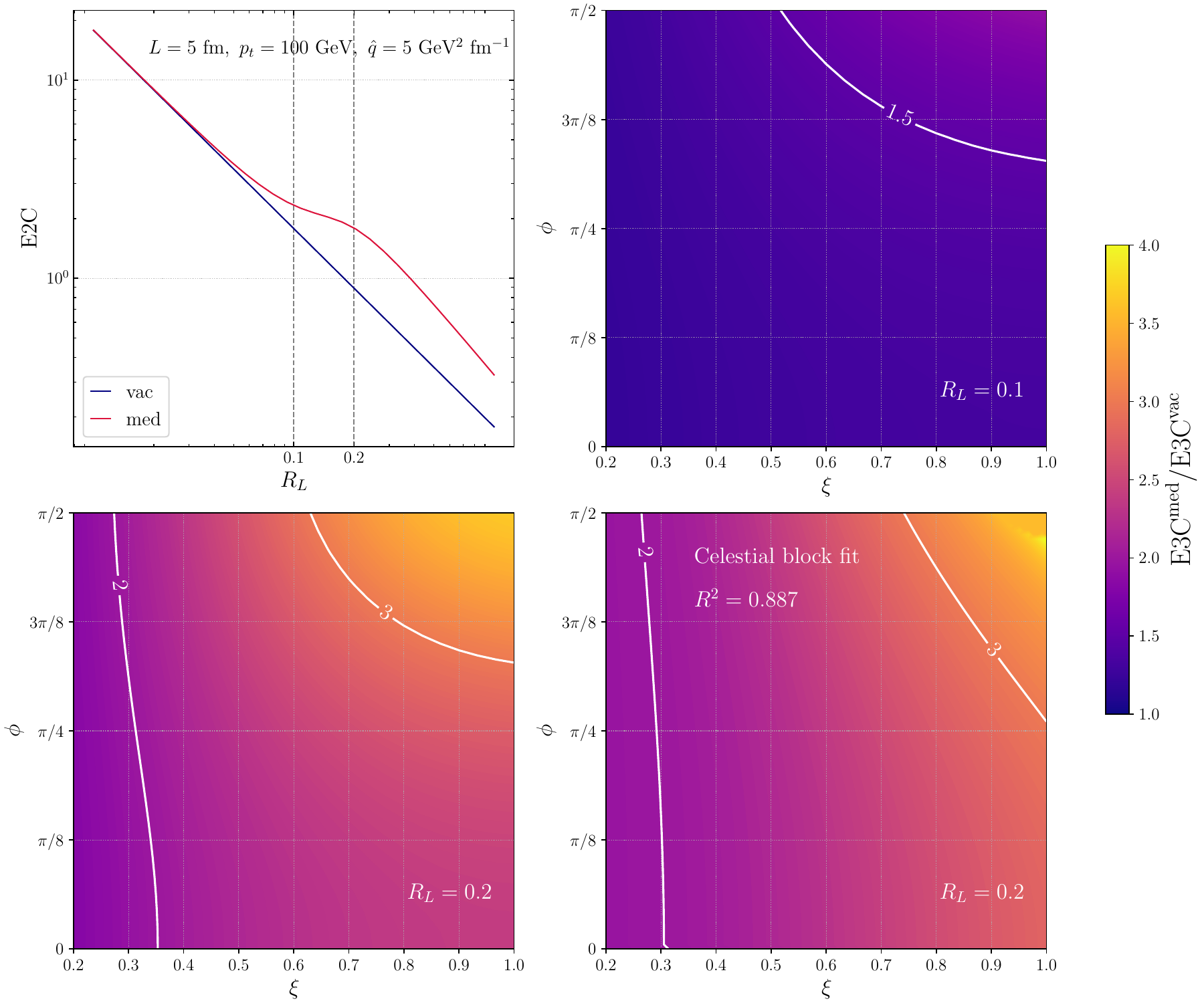}

    \caption{ 
    E2C distribution (\textbf{top left}) indicating the $R_L$ values used for the E3C/E2C ratios in-medium (\textbf{top right}, \textbf{bottom left}). Fit to the E3C/E2C ratio with $R_L=0.2$  (\textbf{bottom right}), following the celestial block structure.}

    \label{fig:2}
\end{figure}
The leading perturbative medium-induced correction to $J_3$ is obtained straightforwardly, by generalizing the cascade form in Eq.~\eqref{eq:ch5_P3_cascade} to the medium. The ratio between the in-medium and vacuum E3C distributions is shown in Fig.~\ref{fig:2}, for two values of $R_L$ chosen by examining the respective E2C, as illustrated. The medium effects lead to a characteristic enhancement in the region where $R_L\approx \theta_{13}\approx \theta_{23}$. At larger $R_L$, the enhancement extends to lower values of $\phi$ at large $\xi$. Notably, looking at Fig.~\ref{fig:cascase}, the cascade approximation in vacuum works reasonably well in the region of the $(\phi,\xi)$ plane where the medium-induced modifications are more pronounced. This supports the interpretation that the features observed in Fig.~\ref{fig:2} cannot be attributed simply to the factorized ansatz.

Similarly to the conformal-limit expansion of the E2C discussed in Section~\ref{subsec:ch3_celestial_clover}, the three-point energy correlator can also be expanded in celestial blocks~\cite{Chang:2022ryc,Chen:2019bpb,Chen:2022jhb}:
\begin{align}\label{eq:ch5_hj2}
\frac{ R_L \theta_{23}\theta_{13}  \, d\Sigma^{(3)}}{\sqrt{-D}\, dR_L d\theta_{13} d\theta_{23}} = \sum_{k=1} a^{(3)}_{\tau,0} \, G^q_{\tau,0}(z,\bar z)\Big\vert_{\tau=2k} \, ,
\end{align}
where the complex variables $z$, $\bar z$, are defined as $R_L^2 z\bar z= \theta_{13}^2$, $R_L^2(1-z)(1-\bar z) = \theta_{23}^2$, while the single-valued $G^q_{\tau,0}$ can be written in terms of hypergeometric functions, whose explicit form can be found in~\cite{Chen:2022jhb}. The data for $R_L=0.2$ is fitted to the functional form
$\left(\sum_{i=1}^3 n_i G_{2i,0}(\phi,\xi)\right/\left(\sum_{i=1}^3 d_i G_{2i,0}(\phi,\xi)\right)$, following the structure in Eq.~\eqref{eq:ch5_hj2}. This is shown in the bottom right plot of Fig.~\ref{fig:2}. A good fit is obtained, with $R^2=0.887$, while the numerator’s coefficients, for $i>1$, are approximately an order of magnitude larger than the denominator’s ones. This points to the dominance of higher-twist terms in-medium, as expected.

\begin{figure}[t!]
    \centering
    \includegraphics[width=.7\textwidth]{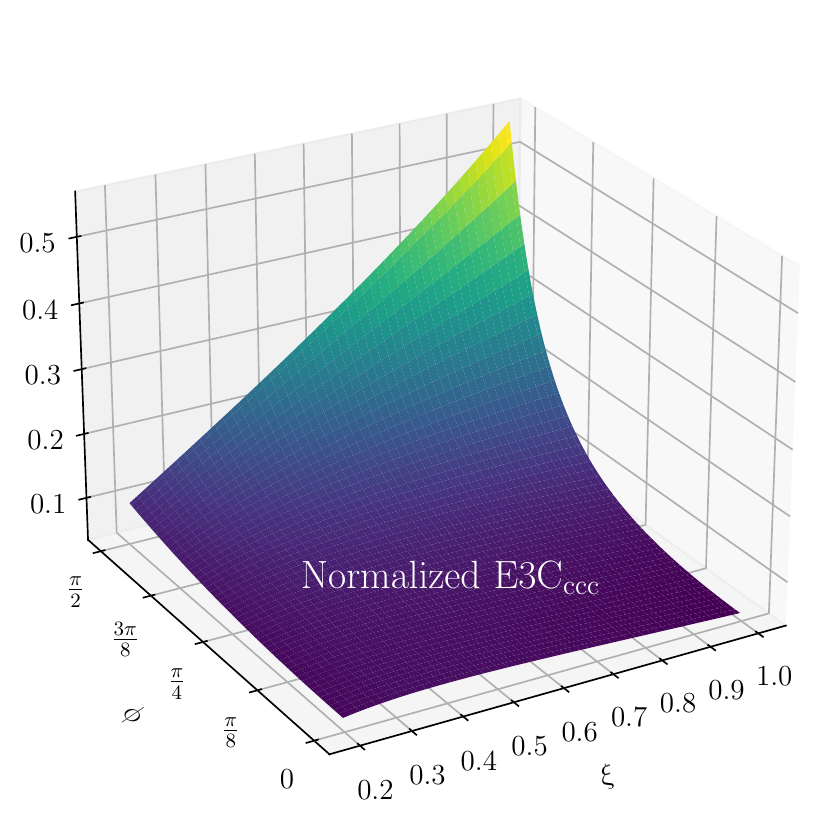} 
     \caption{Normalized E3C for the ccc term from medium response.}
   \label{fig:E3C_final_plots}
\end{figure}

The E3C extraction including the full response is far from trivial. Nonetheless, insight into the shape of the distribution can be gained from the pure classical flux term (ccc):
\begin{align}
\nonumber
\frac{d\Sigma^{(3)}_{\rm ccc}}{dR_L d\xi d\phi} = \frac{ 4 \pi^2 p_t^{-2}R_L^3\xi\,\int d\theta_1 \, \mathcal{E}^3(\cos\theta_1)\sin\theta_1}{(1+\xi\cos\phi)^3\sqrt{4+4\xi\cos\phi-\xi^2\sin^2\phi}} \, ,
\end{align}
shown in Fig.~\ref{fig:E3C_final_plots}. The maximizer of the distribution is nearly identical to the one for the perturbative terms in Fig.~\ref{fig:2}, further evidencing the overlap of these two effects, as for the PENCs. Moreover, this result closely matches the numerical result in~\cite{Bossi:2024qho}. In this case, it is determined solely by geometrical correlations. Finally, writing the ccc term in the celestial block coordinates $(z,\bar z)$,
\begin{align}
\frac{d\Sigma^{(3)}_{\rm ccc}}{dR_L d{\rm Re} z\,  d{\rm Im}z} = \frac{2\pi^2 R_L^3}{p_t^2}\left[\int d\theta_1 \, \mathcal{E}^3(\cos\theta_1)\sin\theta_1\right]\,,
\end{align}
and decomposing it in the celestial blocks, one finds in the collinear limit ($z\to 0$) at fixed $\bar z$ that the only contribution comes from the leading spin and $\tau=4$ block, demonstrating the overlap with higher terms in the twist expansion.

Thus, the shape-dependent E3C shows that perturbative modifications of the in-medium cascade and medium-response contributions populate overlapping regions of the correlator geometry. As for the projected correlators, disentangling these effects requires a systematic treatment of both contributions rather than a purely shape-based separation. % file "Beyond_1to2.tex"
\cleardoublepage

%%%%%%%%%%%%%%%%%%%%%%%%%%%%%%%%%%%%%%%%%%%%%%%%%%%%%%%%%%%%%%%%%%%%%%%%
%                                                                      %
%     File: Thesis_Conclusions.tex                                     %
%     Tex Master: Thesis.tex                                           %
%                                                                      %
%     Author: Andre C. Marta                                           %
%     Last modified :  4 Mar 2024                                      %
%                                                                      %
%%%%%%%%%%%%%%%%%%%%%%%%%%%%%%%%%%%%%%%%%%%%%%%%%%%%%%%%%%%%%%%%%%%%%%%%

\chapter{Conclusions and outlook}
\label{chap:conclusions}

The study of hot QCD matter addresses one of the central questions in strong interaction physics: how collective many-body behaviour emerges from the fundamental coloured degrees of freedom of QCD. Heavy-ion collisions (HICs) provide the experimental environment in which this medium can be explored, and the high-energy hadronic jets produced in the same collisions offer one of the richest probes of its properties. In this thesis, we explored how increasingly differential jet observables can be used to access finer structure of both the medium and the in-medium cascade. In particular, we studied directional effects in in-medium cascades associated with anisotropies of the underlying QCD matter, as well as observables that go beyond the standard single $1\to2$ medium-induced splitting functions. Taken together, these results support the idea that jet substructure can be used not only to quantify overall quenching, but also to resolve the geometry and multi-point dynamics of QCD matter.

In Chapter~\ref{chap:introduction}, we introduced the physical setting of the thesis, from QCD as the microscopic theory of the strong interaction to the formation of hot nuclear matter in HICs. The chapter established why jets are useful probes of this matter: they are produced on short time scales, propagate through and are modified by the evolving medium, and are experimentally accessible. This provided the phenomenological motivation for the theoretical developments that followed.

The next step was to formulate jet quenching in the language of perturbative QCD. In Chapter~\ref{chap:jet_quenching}, jets were treated as perturbative cascades governed by soft and collinear radiation, together with important interference effects that organize their space-time evolution. In heavy-ion collisions, this cascade propagates through an extended coloured medium, so jet quenching is framed as a multi-scale problem arising from the interplay between the intrinsic scales of the shower and the scales of the produced QCD matter. The main technical outcome of the chapter was a rigorous introduction to the BDMPS-Z formalism, where high-energy propagation through stochastic colour fields is described by Green functions encoding transverse motion and continuous colour rotation. This chapter supplied the common theoretical language used in the original calculations of the thesis.

In Chapter~\ref{chap:anisotropic_early_time}, we applied this theoretical framework to explore anisotropic QCD matter, motivated by early-stage matter in HICs. The usual scalar broadening coefficient was generalized to a tensorial structure, represented by $\hat q_x$ and $\hat q_y$, allowing directional momentum broadening to be incorporated in the BDMPS-Z calculation. We first studied spin-resolved $g\to q\bar q$ production and showed that the leading anisotropic harmonic of the spin-independent distribution is of $\cos(2\phi)$ type, while spin-dependent terms generate a corresponding $\sin(2\phi)$ structure. We then turned to energy-flow observables, where the usual angular dependence of the EEC was extended to retain the azimuthal orientation of the measured pair. Heavy-flavour EECs were used to isolate the $g\to Q\bar Q$ splitting and to expose the formation-time dependence of medium modifications, while the azimuthally differential EEC and its Fourier harmonics provided a direct language for directional broadening. In the small-angle limit, the azimuthally differential EEC was organized using the light-ray OPE/celestial-block structure of the conformal limit. There, transverse spin labels the azimuthal harmonics in $\Psi$, twist fixes the leading powers of $\chi$, and the coefficients encode the jet-medium dynamics. We then defined the clover projection, which isolates the leading $\cos(2\Psi)$-type component of the anisotropic EEC. The main lesson is that directional information from early-time matter can survive in sufficiently differential jet substructure observables.

In Chapter~\ref{chap:jet_energy_loss}, we studied jet energy loss in two complementary contexts. We first adopted a phenomenological perspective, defining it as a population-level reconstructed-$p_T$ shift between vacuum and medium-modified jet samples. Quantile matching and $Q_{AA}$ were used to estimate average fractional reconstructed energy loss and reduce fixed-$p_T$ selection bias, which otherwise mixes genuine medium-induced modifications with differences in the initial jet population. Within this framework, we studied the dependence on jet definition, medium response and initiating colour charge, and quantified its impact on selected substructure comparisons. We then introduced energy-loss functions as non-perturbative distributions of the energy retained by jet-originated hadrons, suited to observables sensitive to full-cascade energy flow. By analogy with track functions, we postulated a perturbative non-linear RG evolution for these objects. Applying this framework to projected energy correlators, we showed how energy loss enters through ELF moments, how this differs from the typical quenching-weight approximation, and how correlated energy loss can begin to be incorporated systematically.

The final chapter presented a step in pushing the discussion beyond perturbative medium modifications to a single $1\to2$ splitting. In Chapter~\ref{chap:towards_multi_point_structure}, we first studied spin-induced azimuthal correlations in two strongly ordered collinear emissions. The coherent sum over the spin states of an intermediate off-shell gluon produces a vacuum $\cos(2\psi_{12})$ modulation between the two splitting planes. Embedding only the first splitting in the medium showed that isotropic broadening suppresses this modulation, while anisotropic broadening can additionally generate a phase shift. We then turned to projected and shape-dependent higher-point energy correlators in matter. There, perturbative BDMPS-Z corrections and medium-response contributions to projected correlators were organized through OPE-like collinear power expansions, while the explicit calculation of PE3C and E3C showed that perturbative cascade modifications and response can populate overlapping regions of angular space. Together, these results demonstrated that multi-point jet structure contains information about interference, directionality and response that is absent from the simplest $1\to2$ descriptions.

The results of this thesis suggest several concrete directions for future work. One natural step is to extend the study of azimuthally differential jet observables to more realistic heavy-ion backgrounds, including pre-equilibrium and hydrodynamic evolution, in order to assess how directional broadening and clover-like projections survive in phenomenological conditions. On the perturbative side, a central open problem is the calculation of medium-induced double-emission spectra beyond the simplified ordered or factorized limits used here, which would provide a firmer basis for multi-point jet substructure in matter. It would also be useful to further explore the symmetry structure of energy correlators and its relation to more formal properties of QCD matter. In particular, their collinear expansion offers a way to organize directional structures in both perturbative cascade modifications and medium response, with the scaling of the corresponding coefficients potentially carrying information about medium parameters. Pursuing these directions would sharpen the connection between the microscopic dynamics of in-medium jets and the collective QCD matter through which they propagate. % file "Thesis_Conclusions.tex"
\cleardoublepage

% ----------------------------------------------------------------------
%  Bibliography
% ----------------------------------------------------------------------

% Add entry in the table of contents as chapter
\phantomsection
\addcontentsline{toc}{chapter}{\bibname}

\bibliographystyle{abbrvunsrtnat} % <<<<< SELECT IF USING REFERENCES BY NUMBER (CITATION ORDER)

% External bibliography database file in the BibTeX format
\bibliography{Thesis_Bibliography_DB} % file "Thesis_Bibliography_DB.bib"

\begin{thebibliography}{481}
\providecommand{\natexlab}[1]{#1}
\providecommand{\url}[1]{\texttt{#1}}
\expandafter\ifx\csname urlstyle\endcsname\relax
  \providecommand{\doi}[1]{doi: #1}\else
  \providecommand{\doi}{doi: \begingroup \urlstyle{rm}\Url}\fi

\bibitem[Apolin{\'a}rio et~al.(2025)Apolin{\'a}rio, Lu{\'\i}s, Milhano, and
  Silva]{Apolinario:2024apr}
L.~Apolin{\'a}rio, L.~Lu{\'\i}s, J.~G. Milhano, and J.~M. Silva.
\newblock {Towards an unbiased jet energy loss measurement}.
\newblock \emph{JHEP}, 01:\penalty0 164, 2025.
\newblock \doi{10.1007/JHEP01(2025)164}, arXiv:2409.12238 [hep-ph].

\bibitem[Barata et~al.(2024{\natexlab{a}})Barata, Salgado, and
  Silva]{Barata:2024bqp}
J.~Barata, C.~A. Salgado, and J.~M. Silva.
\newblock {Gluon to $ \textrm{q}\overline{\textrm{q}} $ antenna in anisotropic
  QCD matter: spin-polarized and azimuthal jet observables}.
\newblock \emph{JHEP}, 12:\penalty0 023, 2024{\natexlab{a}}.
\newblock \doi{10.1007/JHEP12(2024)023}, arXiv:2407.04774 [hep-ph].

\bibitem[Barata et~al.(2024{\natexlab{b}})Barata, Moult, and
  Silva]{Barata:2024bmx}
J.~Barata, I.~Moult, and J.~M. Silva.
\newblock {Tracking Energy Loss in Heavy Ion Collisions}.
\newblock 9 2024{\natexlab{b}}, arXiv:2409.18174 [hep-ph].

\bibitem[Barata et~al.(2025{\natexlab{a}})Barata, Moult, Sadofyev, and
  Silva]{Barata:2025fzd}
J.~Barata, I.~Moult, A.~V. Sadofyev, and J.~M. Silva.
\newblock {Dissecting Jet Modification in the QGP with Multi-Point Energy
  Correlators}.
\newblock 3 2025{\natexlab{a}}, arXiv:2503.13603 [hep-ph].

\bibitem[Kuzmin and Silva(2025)]{Kuzmin:2025fyu}
M.~V. Kuzmin and J.~M. Silva.
\newblock {QCD antenna radiative spectrum in dense media within the Improved
  Opacity Expansion}.
\newblock \emph{JHEP}, 12:\penalty0 022, 2025.
\newblock \doi{10.1007/JHEP12(2025)022}, arXiv:2507.00143 [hep-ph].

\bibitem[Barata et~al.(2025{\natexlab{b}})Barata, Brewer, Lee, and
  Silva]{Barata:2025uxp}
J.~Barata, J.~Brewer, K.~Lee, and J.~M. Silva.
\newblock {Heavy Quark Pair Energy Correlators: From Profiling Partonic
  Splittings to Probing Heavy-Flavor Fragmentation}.
\newblock 8 2025{\natexlab{b}}, arXiv:2508.19404 [hep-ph].

\bibitem[Silva and Soto-Ontoso(2025)]{Silva:2025dan}
J.~M. Silva and A.~Soto-Ontoso.
\newblock {Collinear spin correlations of final-state radiation in dense QCD
  matter}.
\newblock 11 2025, arXiv:2511.17737 [hep-ph].

\bibitem[Barata et~al.(2025{\natexlab{c}})Barata, Milhano, Sadofyev, and
  Silva]{Barata:2025zku}
J.~Barata, J.~G. Milhano, A.~V. Sadofyev, and J.~M. Silva.
\newblock {Early-Time Dynamics of Heavy-Ion Collisions through Energy
  Correlators: celestial blocks and the spacetime structure of
  out-of-equilibrium QCD matter}.
\newblock 12 2025{\natexlab{c}}, arXiv:2512.17009 [hep-ph].

\bibitem[Barata et~al.(2026{\natexlab{a}})Barata, Li, Qian, Salgado, and
  Silva]{Barata:2026icn}
J.~Barata, M.~Li, W.~Qian, C.~A. Salgado, and J.~M. Silva.
\newblock {Quantum simulating multi-particle processes in high energy nuclear
  physics: dijet production and color (de)coherence}.
\newblock 4 2026{\natexlab{a}}, arXiv:2604.11616 [hep-ph].

\bibitem[Barata et~al.(2026{\natexlab{b}})Barata, Kuzmin, Moult, Sadofyev, and
  Silva]{Barata:2026pgh}
J.~Barata, M.~V. Kuzmin, I.~Moult, A.~V. Sadofyev, and J.~M. Silva.
\newblock {Hydrodynamics and Energy Correlators}.
\newblock 4 2026{\natexlab{b}}, arXiv:2604.21971 [hep-ph].

\bibitem[Yukawa(1935)]{Yukawa:1935xg}
H.~Yukawa.
\newblock {On the Interaction of Elementary Particles I}.
\newblock \emph{Proc. Phys. Math. Soc. Jap.}, 17:\penalty0 48--57, 1935.
\newblock \doi{10.1143/PTPS.1.1}.

\bibitem[Lattes et~al.(1947)Lattes, Occhialini, and Powell]{Lattes:1947mx}
C.~M.~G. Lattes, G.~P.~S. Occhialini, and C.~F. Powell.
\newblock {Observations on the Tracks of Slow Mesons in Photographic Emulsions.
  1}.
\newblock \emph{Nature}, 160:\penalty0 453--456, 1947.
\newblock \doi{10.1038/160453a0}.

\bibitem[Gell-Mann(1961)]{GellMann:1961ky}
M.~Gell-Mann.
\newblock {The Eightfold Way: A Theory of Strong Interaction Symmetry}.
\newblock Technical report, California Institute of Technology, 1961.

\bibitem[Ne'eman(1961)]{Neeman:1961jhl}
Y.~Ne'eman.
\newblock {Derivation of strong interactions from a gauge invariance}.
\newblock \emph{Nucl. Phys.}, 26:\penalty0 222--229, 1961.
\newblock \doi{10.1016/0029-5582(61)90134-1}.

\bibitem[Gell-Mann(1964)]{GellMann:1964nj}
M.~Gell-Mann.
\newblock {A Schematic Model of Baryons and Mesons}.
\newblock \emph{Phys. Lett.}, 8:\penalty0 214--215, 1964.
\newblock \doi{10.1016/S0031-9163(64)92001-3}.

\bibitem[Zweig(1964)]{Zweig:1964jf}
G.~Zweig.
\newblock {An SU(3) model for strong interaction symmetry and its breaking}.
\newblock Technical Report CERN-TH-401, CERN-TH-412, CERN, 1964.

\bibitem[Barnes(1964)]{Barnes:1964pd}
V.~E. e.~a. Barnes.
\newblock {Observation of a Hyperon with Strangeness -3}.
\newblock \emph{Phys. Rev. Lett.}, 12:\penalty0 204--206, 1964.
\newblock \doi{10.1103/PhysRevLett.12.204}.

\bibitem[Greenberg(1964)]{Greenberg:1964pe}
O.~W. Greenberg.
\newblock {Spin and Unitary Spin Independence in a Paraquark Model of Baryons
  and Mesons}.
\newblock \emph{Phys. Rev. Lett.}, 13:\penalty0 598--602, 1964.
\newblock \doi{10.1103/PhysRevLett.13.598}.

\bibitem[Han and Nambu(1965)]{Han:1965pf}
M.-Y. Han and Y.~Nambu.
\newblock {Three Triplet Model with Double SU(3) Symmetry}.
\newblock \emph{Phys. Rev.}, 139:\penalty0 B1006--B1010, 1965.
\newblock \doi{10.1103/PhysRev.139.B1006}.

\bibitem[Bloom(1969)]{Bloom:1969kc}
E.~D. e.~a. Bloom.
\newblock {High-Energy Inelastic e p Scattering at 6-Degrees and 10-Degrees}.
\newblock \emph{Phys. Rev. Lett.}, 23:\penalty0 930--934, 1969.
\newblock \doi{10.1103/PhysRevLett.23.930}.

\bibitem[Breidenbach(1969)]{Breidenbach:1969kd}
M.~e.~a. Breidenbach.
\newblock {Observed Behavior of Highly Inelastic Electron-Proton Scattering}.
\newblock \emph{Phys. Rev. Lett.}, 23:\penalty0 935--939, 1969.
\newblock \doi{10.1103/PhysRevLett.23.935}.

\bibitem[Bjorken and Brodsky(1970)]{Bjorken:1969wi}
J.~D. Bjorken and S.~J. Brodsky.
\newblock {Statistical Model for electron-Positron Annihilation Into Hadrons}.
\newblock \emph{Phys. Rev. D}, 1:\penalty0 1416--1420, 1970.
\newblock \doi{10.1103/PhysRevD.1.1416}.

\bibitem[Feynman(1969)]{Feynman:1969ej}
R.~P. Feynman.
\newblock {Very High-Energy Collisions of Hadrons}.
\newblock \emph{Phys. Rev. Lett.}, 23:\penalty0 1415--1417, 1969.
\newblock \doi{10.1103/PhysRevLett.23.1415}.

\bibitem[Callan and Gross(1970)]{Callan:1970yg}
C.~G.~J. Callan and D.~J. Gross.
\newblock {High-Energy Electroproduction and the Constitution of the Electric
  Current}.
\newblock \emph{Phys. Rev. Lett.}, 25:\penalty0 156--159, 1970.
\newblock \doi{10.1103/PhysRevLett.25.156}.

\bibitem[Friedman et~al.(1991)Friedman, Kendall, and Taylor]{Friedman:1972sy}
J.~I. Friedman, H.~W. Kendall, and R.~E. Taylor.
\newblock {Deep Inelastic Scattering: Acknowledgements}.
\newblock \emph{Rev. Mod. Phys.}, 63:\penalty0 573--574, 1991.
\newblock \doi{10.1103/RevModPhys.63.573}.

\bibitem[Gross and Wilczek(1973)]{Gross:1973id}
D.~J. Gross and F.~Wilczek.
\newblock {Ultraviolet Behavior of Nonabelian Gauge Theories}.
\newblock \emph{Phys. Rev. Lett.}, 30:\penalty0 1343--1346, 1973.
\newblock \doi{10.1103/PhysRevLett.30.1343}.

\bibitem[Politzer(1973)]{Politzer:1973fx}
H.~D. Politzer.
\newblock {Reliable Perturbative Results for Strong Interactions?}
\newblock \emph{Phys. Rev. Lett.}, 30:\penalty0 1346--1349, 1973.
\newblock \doi{10.1103/PhysRevLett.30.1346}.

\bibitem[{ATLAS Collaboration}(2024{\natexlab{a}})]{ATLAS_SM_chart}
{ATLAS Collaboration}.
\newblock The standard model and beyond, 2024{\natexlab{a}}.
\newblock URL
  \url{https://opendata.atlas.cern/docs/documentation/introduction/SM_and_beyond}.
\newblock Accessed: 2026-04-13.

\bibitem[Peskin and Schroeder(1995)]{Peskin:1995ev}
M.~E. Peskin and D.~V. Schroeder.
\newblock \emph{{An Introduction to Quantum Field Theory}}.
\newblock Addison-Wesley, Reading, USA, 1995.

\bibitem[Kovchegov and Levin(2013)]{Kovchegov:2012mbw}
Y.~V. Kovchegov and E.~Levin.
\newblock \emph{{Quantum Chromodynamics at High Energy}}, volume~33.
\newblock Oxford University Press, 2013.
\newblock ISBN 978-1-009-29144-6, 978-1-009-29141-5, 978-1-009-29142-2,
  978-0-521-11257-4, 978-1-139-55768-9.
\newblock \doi{10.1017/9781009291446}.

\bibitem[Wilson(1974)]{Wilson:1974sk}
K.~G. Wilson.
\newblock {Confinement of Quarks}.
\newblock \emph{Phys. Rev. D}, 10:\penalty0 2445--2459, 1974.
\newblock \doi{10.1103/PhysRevD.10.2445}.

\bibitem[Roberts and Williams(1994)]{Roberts:1994dr}
C.~D. Roberts and A.~G. Williams.
\newblock {Dyson-Schwinger equations and their application to hadronic
  physics}.
\newblock \emph{Prog. Part. Nucl. Phys.}, 33:\penalty0 477--575, 1994.
\newblock \doi{10.1016/0146-6410(94)90049-3}, arXiv:hep-ph/9403224.

\bibitem[Navas et~al.(2024)]{ParticleDataGroup:2024cfk}
S.~Navas et~al.
\newblock {Review of particle physics}.
\newblock \emph{Phys. Rev. D}, 110\penalty0 (3):\penalty0 030001, 2024.
\newblock \doi{10.1103/PhysRevD.110.030001}.

\bibitem[Collins(2013)]{Collins:2011zzd}
J.~Collins.
\newblock \emph{{Foundations of perturbative QCD}}, volume~32.
\newblock Cambridge University Press, 11 2013.
\newblock ISBN 978-1-107-64525-7, 978-1-107-64525-7, 978-0-521-85533-4,
  978-1-139-09782-6.

\bibitem[Gribov and Lipatov(1972)]{Gribov:1972ri}
V.~N. Gribov and L.~N. Lipatov.
\newblock {Deep inelastic e p scattering in perturbation theory}.
\newblock \emph{Sov. J. Nucl. Phys.}, 15:\penalty0 438--450, 1972.

\bibitem[Altarelli and Parisi(1977)]{Altarelli:1977zs}
G.~Altarelli and G.~Parisi.
\newblock {Asymptotic Freedom in Parton Language}.
\newblock \emph{Nucl. Phys. B}, 126:\penalty0 298--318, 1977.
\newblock \doi{10.1016/0550-3213(77)90384-4}.

\bibitem[Dokshitzer(1977)]{Dokshitzer:1977sg}
Y.~L. Dokshitzer.
\newblock {Calculation of the Structure Functions for Deep Inelastic Scattering
  and e+ e- Annihilation by Perturbation Theory in Quantum Chromodynamics.}
\newblock \emph{Sov. Phys. JETP}, 46:\penalty0 641--653, 1977.

\bibitem[Kapusta and Gale(2006)]{Kapusta:2006pm}
J.~I. Kapusta and C.~Gale.
\newblock \emph{Finite-Temperature Field Theory: Principles and Applications}.
\newblock Cambridge University Press, 2006.

\bibitem[Stephanov(2004)]{Stephanov:2004wx}
M.~A. Stephanov.
\newblock {QCD Phase Diagram and the Critical Point}.
\newblock \emph{Prog. Theor. Phys. Suppl.}, 153:\penalty0 139--156, 2004.
\newblock \doi{10.1143/PTPS.153.139}, arXiv:hep-ph/0402115.

\bibitem[Fukushima and Hatsuda(2011)]{Fukushima:2010bq}
K.~Fukushima and T.~Hatsuda.
\newblock {The phase diagram of dense QCD}.
\newblock \emph{Rept. Prog. Phys.}, 74:\penalty0 014001, 2011.
\newblock \doi{10.1088/0034-4885/74/1/014001}, arXiv:1005.4814 [hep-ph].

\bibitem[796(2008)]{796947}
{The Frontiers of Nuclear Science, A Long Range Plan}.
\newblock 9 2008, arXiv:0809.3137 [nucl-ex].

\bibitem[Aoki et~al.(2006)Aoki, Endrodi, Fodor, Katz, and Szabo]{Aoki:2006we}
Y.~Aoki, G.~Endrodi, Z.~Fodor, S.~D. Katz, and K.~K. Szabo.
\newblock {The Order of the quantum chromodynamics transition predicted by the
  standard model of particle physics}.
\newblock \emph{Nature}, 443:\penalty0 675--678, 2006.
\newblock \doi{10.1038/nature05120}, arXiv:hep-lat/0611014.

\bibitem[Bazavov et~al.(2014)]{HotQCD:2014kol}
A.~Bazavov et~al.
\newblock {Equation of state in ( 2+1 )-flavor QCD}.
\newblock \emph{Phys. Rev. D}, 90:\penalty0 094503, 2014.
\newblock \doi{10.1103/PhysRevD.90.094503}, arXiv:1407.6387 [hep-lat].

\bibitem[Shuryak(1978)]{Shuryak:1978ij}
E.~V. Shuryak.
\newblock {Quark-Gluon Plasma and Hadronic Production of Leptons, Photons and
  Psions}.
\newblock \emph{Phys. Lett. B}, 78:\penalty0 150, 1978.
\newblock \doi{10.1016/0370-2693(78)90370-2}.

\bibitem[Bjorken(1983{\natexlab{a}})]{Bjorken:1982qr}
J.~D. Bjorken.
\newblock Highly relativistic nucleus-nucleus collisions: The central rapidity
  region.
\newblock \emph{Phys. Rev. D}, 27:\penalty0 140--151, 1983{\natexlab{a}}.

\bibitem[Alford et~al.(2008)Alford, Schmitt, Rajagopal, and
  Schafer]{Alford:2007xm}
M.~G. Alford, A.~Schmitt, K.~Rajagopal, and T.~Schafer.
\newblock Color superconductivity in dense quark matter.
\newblock \emph{Rev. Mod. Phys.}, 80:\penalty0 1455--1515, 2008.

\bibitem[Oertel et~al.(2017)Oertel, Hempel, Kl{\"a}hn, and
  Typel]{Oertel:2016bki}
M.~Oertel, M.~Hempel, T.~Kl{\"a}hn, and S.~Typel.
\newblock Equations of state for supernovae and compact stars.
\newblock \emph{Rev. Mod. Phys.}, 89:\penalty0 015007, 2017.

\bibitem[Busza et~al.(2018)Busza, Rajagopal, and van~der Schee]{Busza:2018rrf}
W.~Busza, K.~Rajagopal, and W.~van~der Schee.
\newblock {Heavy Ion Collisions: The Big Picture, and the Big Questions}.
\newblock \emph{Ann. Rev. Nucl. Part. Sci.}, 68:\penalty0 339--376, 2018.
\newblock \doi{10.1146/annurev-nucl-101917-020852}, arXiv:1802.04801 [hep-ph].

\bibitem[Heinz and Jacob(2000)]{Heinz:2000bk}
U.~W. Heinz and M.~Jacob.
\newblock Evidence for a new state of matter: An assessment of the results from
  the cern lead beam programme.
\newblock \emph{nucl-th/0002042}, 2000.

\bibitem[Collaboration(2005)]{NA50:2005}
N.~Collaboration.
\newblock Anomalous j/psi suppression in pb-pb collisions at sps.
\newblock \emph{Eur. Phys. J. C}, 39:\penalty0 335, 2005.

\bibitem[Collaboration(2002)]{WA97:2002}
W.~Collaboration.
\newblock Strangeness enhancement in pb-pb collisions at sps.
\newblock \emph{Eur. Phys. J. C}, 23:\penalty0 427, 2002.

\bibitem[Arsene et~al.(2005)]{BRAHMS:2004adc}
I.~Arsene et~al.
\newblock {Quark gluon plasma and color glass condensate at RHIC? The
  Perspective from the BRAHMS experiment}.
\newblock \emph{Nucl. Phys. A}, 757:\penalty0 1--27, 2005.
\newblock \doi{10.1016/j.nuclphysa.2005.02.130}, arXiv:nucl-ex/0410020.

\bibitem[Back et~al.(2005)]{PHOBOS:2004zne}
B.~B. Back et~al.
\newblock {The PHOBOS perspective on discoveries at RHIC}.
\newblock \emph{Nucl. Phys. A}, 757:\penalty0 28--101, 2005.
\newblock \doi{10.1016/j.nuclphysa.2005.03.084}, arXiv:nucl-ex/0410022.

\bibitem[Adcox et~al.(2005)]{PHENIX:2004vcz}
K.~Adcox et~al.
\newblock {Formation of dense partonic matter in relativistic nucleus-nucleus
  collisions at RHIC: Experimental evaluation by the PHENIX collaboration}.
\newblock \emph{Nucl. Phys. A}, 757:\penalty0 184--283, 2005.
\newblock \doi{10.1016/j.nuclphysa.2005.03.086}, arXiv:nucl-ex/0410003.

\bibitem[Adams et~al.(2005)]{STAR:2005gfr}
J.~Adams et~al.
\newblock {Experimental and theoretical challenges in the search for the quark
  gluon plasma: The STAR Collaboration's critical assessment of the evidence
  from RHIC collisions}.
\newblock \emph{Nucl. Phys. A}, 757:\penalty0 102--183, 2005.
\newblock \doi{10.1016/j.nuclphysa.2005.03.085}, arXiv:nucl-ex/0501009.

\bibitem[Bjorken(1983{\natexlab{b}})]{Bjorken:1983qr}
J.~D. Bjorken.
\newblock Highly relativistic nucleus-nucleus collisions: The central rapidity
  region.
\newblock \emph{Phys. Rev. D}, 27:\penalty0 140, 1983{\natexlab{b}}.

\bibitem[Kolb and Heinz(2003)]{Kolb:2003dz}
P.~F. Kolb and U.~W. Heinz.
\newblock {Hydrodynamic description of ultrarelativistic heavy ion collisions}.
\newblock pages 634--714, 5 2003, arXiv:nucl-th/0305084.

\bibitem[Arslandok et~al.(2023)]{Arslandok:2023utm}
M.~Arslandok et~al.
\newblock {Hot QCD White Paper}.
\newblock 3 2023, arXiv:2303.17254 [nucl-ex].

\bibitem[Gelis et~al.(2010)Gelis, Iancu, Jalilian-Marian, and
  Venugopalan]{Gelis:2010nm}
F.~Gelis, E.~Iancu, J.~Jalilian-Marian, and R.~Venugopalan.
\newblock {The Color Glass Condensate}.
\newblock \emph{Ann. Rev. Nucl. Part. Sci.}, 60:\penalty0 463--489, 2010.
\newblock \doi{10.1146/annurev.nucl.010909.083629}, arXiv:1002.0333 [hep-ph].

\bibitem[Heinz and Snellings(2013)]{Heinz:2013th}
U.~Heinz and R.~Snellings.
\newblock {Collective flow and viscosity in relativistic heavy-ion collisions}.
\newblock \emph{Ann. Rev. Nucl. Part. Sci.}, 63:\penalty0 123--151, 2013.
\newblock \doi{10.1146/annurev-nucl-102212-170540}, arXiv:1301.2826 [nucl-th].

\bibitem[Gale et~al.(2013)Gale, Jeon, and Schenke]{Gale:2013da}
C.~Gale, S.~Jeon, and B.~Schenke.
\newblock Hydrodynamic modeling of heavy-ion collisions.
\newblock \emph{Int. J. Mod. Phys. A}, 28:\penalty0 1340011, 2013.

\bibitem[Miller et~al.(2007)Miller, Reygers, Sanders, and
  Steinberg]{Miller:2007ri}
M.~L. Miller, K.~Reygers, S.~J. Sanders, and P.~Steinberg.
\newblock Glauber modeling in high-energy nuclear collisions.
\newblock \emph{Ann. Rev. Nucl. Part. Sci.}, 57:\penalty0 205--243, 2007.
\newblock \doi{10.1146/annurev.nucl.57.090506.123020}.

\bibitem[Loizides et~al.(2018)Loizides, Kamin, and
  d'Enterria]{Loizides:2017ack}
C.~Loizides, J.~Kamin, and D.~d'Enterria.
\newblock {Improved Monte Carlo Glauber predictions at present and future
  nuclear colliders}.
\newblock \emph{Phys. Rev. C}, 97\penalty0 (5):\penalty0 054910, 2018.
\newblock \doi{10.1103/PhysRevC.97.054910}, arXiv:1710.07098 [nucl-ex].

\bibitem[Glauber and Matthiae(1970)]{Glauber:1970jm}
R.~J. Glauber and G.~Matthiae.
\newblock High-energy scattering of protons by nuclei.
\newblock \emph{Nucl. Phys. B}, 21:\penalty0 135--157, 1970.

\bibitem[Woods and Saxon(1954)]{Woods:1954zz}
R.~D. Woods and D.~S. Saxon.
\newblock Diffuse surface optical model for nuclei.
\newblock \emph{Phys. Rev.}, 95:\penalty0 577--578, 1954.
\newblock \doi{10.1103/PhysRev.95.577}.

\bibitem[Kuraev et~al.(1977)Kuraev, Lipatov, and Fadin]{Kuraev:1977fs}
E.~A. Kuraev, L.~N. Lipatov, and V.~S. Fadin.
\newblock The pomeranchuk singularity in nonabelian gauge theories.
\newblock \emph{Sov. Phys. JETP}, 45:\penalty0 199--204, 1977.

\bibitem[Balitsky and Lipatov(1978)]{Balitsky:1978ic}
I.~I. Balitsky and L.~N. Lipatov.
\newblock The pomeranchuk singularity in quantum chromodynamics.
\newblock \emph{Sov. J. Nucl. Phys.}, 28:\penalty0 822--829, 1978.

\bibitem[Balitsky(1996)]{Balitsky:1995ub}
I.~Balitsky.
\newblock Operator expansion for high-energy scattering.
\newblock \emph{Nucl. Phys. B}, 463:\penalty0 99--160, 1996,
  arXiv:hep-ph/9509348.

\bibitem[Kovchegov(1999)]{Kovchegov:1999yj}
Y.~V. Kovchegov.
\newblock Small-$x$ $f_2$ structure function of a nucleus including multiple
  pomeron exchanges.
\newblock \emph{Phys. Rev. D}, 60:\penalty0 034008, 1999, arXiv:hep-ph/9901281.

\bibitem[Iancu and Venugopalan(2003)]{Iancu:2003xm}
E.~Iancu and R.~Venugopalan.
\newblock \emph{{The Color glass condensate and high-energy scattering in
  QCD}}, pages 249--3363.
\newblock 3 2003.
\newblock \doi{10.1142/9789812795533_0005}.

\bibitem[McLerran and Venugopalan(1994{\natexlab{a}})]{McLerran:1993ni}
L.~D. McLerran and R.~Venugopalan.
\newblock {Computing quark and gluon distribution functions for very large
  nuclei}.
\newblock \emph{Phys. Rev. D}, 49:\penalty0 2233--2241, 1994{\natexlab{a}}.
\newblock \doi{10.1103/PhysRevD.49.2233}, arXiv:hep-ph/9309289.

\bibitem[McLerran and Venugopalan(1994{\natexlab{b}})]{McLerran:1993ka}
L.~D. McLerran and R.~Venugopalan.
\newblock {Gluon distribution functions for very large nuclei at small
  transverse momentum}.
\newblock \emph{Phys. Rev. D}, 49:\penalty0 3352--3355, 1994{\natexlab{b}}.
\newblock \doi{10.1103/PhysRevD.49.3352}, arXiv:hep-ph/9311205.

\bibitem[H1 and Collaborations(2015)]{H1ZEUS:2015qlf}
H1 and Z.~Collaborations.
\newblock Combination of measurements of inclusive deep inelastic $e^\pm p$
  scattering cross sections and qcd analysis of hera data.
\newblock \emph{Eur. Phys. J. C}, 75:\penalty0 580, 2015, arXiv:1506.06042.

\bibitem[Golec-Biernat and Wusthoff(1998)]{GolecBiernat:1998js}
K.~J. Golec-Biernat and M.~Wusthoff.
\newblock Saturation effects in deep inelastic scattering at low $q^2$ and its
  implications on diffraction.
\newblock \emph{Phys. Rev. D}, 59:\penalty0 014017, 1998, arXiv:hep-ph/9807513.

\bibitem[Stasto et~al.(2001)Stasto, Golec-Biernat, and
  Kwiecinski]{Stasto:2000er}
A.~M. Stasto, K.~J. Golec-Biernat, and J.~Kwiecinski.
\newblock Geometric scaling for the total $\gamma^\ast p$ cross-section in the
  low $x$ region.
\newblock \emph{Phys. Rev. Lett.}, 86:\penalty0 596--599, 2001,
  arXiv:hep-ph/0007192.

\bibitem[Abdul~Khalek et~al.(2022)]{AbdulKhalek:2021gbh}
R.~Abdul~Khalek et~al.
\newblock Science requirements and detector concepts for the electron-ion
  collider: Eic yellow report.
\newblock \emph{Nucl. Phys. A}, 1026:\penalty0 122447, 2022.
\newblock \doi{10.1016/j.nuclphysa.2022.122447}, arXiv:2103.05419.

\bibitem[Kovner et~al.(1995)Kovner, McLerran, and Weigert]{Kovner:1995ja}
A.~Kovner, L.~D. McLerran, and H.~Weigert.
\newblock Gluon production from nonabelian weizsacker-williams fields in
  nucleus-nucleus collisions.
\newblock \emph{Phys. Rev. D}, 52:\penalty0 6231--6237, 1995,
  arXiv:hep-ph/9502289.

\bibitem[Krasnitz and Venugopalan(1999)]{Krasnitz:1998ns}
A.~Krasnitz and R.~Venugopalan.
\newblock {Nonperturbative computation of gluon minijet production in nuclear
  collisions at very high-energies}.
\newblock \emph{Nucl. Phys. B}, 557:\penalty0 237, 1999.
\newblock \doi{10.1016/S0550-3213(99)00366-1}, arXiv:hep-ph/9809433.

\bibitem[Lappi and McLerran(2006)]{Lappi:2006fp}
T.~Lappi and L.~McLerran.
\newblock {Some features of the glasma}.
\newblock \emph{Nucl. Phys. A}, 772:\penalty0 200--212, 2006.
\newblock \doi{10.1016/j.nuclphysa.2006.04.001}, arXiv:hep-ph/0602189.

\bibitem[Berges et~al.(2014{\natexlab{a}})Berges, Boguslavski, Schlichting, and
  Venugopalan]{Berges:2013eia}
J.~Berges, K.~Boguslavski, S.~Schlichting, and R.~Venugopalan.
\newblock {Turbulent thermalization process in heavy-ion collisions at
  ultrarelativistic energies}.
\newblock \emph{Phys. Rev. D}, 89\penalty0 (7):\penalty0 074011,
  2014{\natexlab{a}}.
\newblock \doi{10.1103/PhysRevD.89.074011}, arXiv:1303.5650 [hep-ph].

\bibitem[Lappi(2003)]{Lappi:2003bi}
T.~Lappi.
\newblock Production of gluons in the classical field model for heavy ion
  collisions.
\newblock \emph{Phys. Rev. C}, 67:\penalty0 054903, 2003, arXiv:hep-ph/0303076.

\bibitem[Schenke et~al.(2012{\natexlab{a}})Schenke, Tribedy, and
  Venugopalan]{Schenke:2012wb}
B.~Schenke, P.~Tribedy, and R.~Venugopalan.
\newblock Fluctuating glasma initial conditions and flow in heavy ion
  collisions.
\newblock \emph{Phys. Rev. Lett.}, 108:\penalty0 252301, 2012{\natexlab{a}},
  arXiv:1202.6646.

\bibitem[Schenke et~al.(2012{\natexlab{b}})Schenke, Tribedy, and
  Venugopalan]{Schenke:2012hg}
B.~Schenke, P.~Tribedy, and R.~Venugopalan.
\newblock Event-by-event gluon multiplicity, energy density, and eccentricities
  in ultrarelativistic heavy-ion collisions.
\newblock \emph{Phys. Rev. C}, 86:\penalty0 034908, 2012{\natexlab{b}},
  arXiv:1206.6805.

\bibitem[Arnold et~al.(2003{\natexlab{a}})Arnold, Moore, and
  Yaffe]{Arnold:2002zm}
P.~B. Arnold, G.~D. Moore, and L.~G. Yaffe.
\newblock {Effective kinetic theory for high temperature gauge theories}.
\newblock \emph{JHEP}, 01:\penalty0 030, 2003{\natexlab{a}}.
\newblock \doi{10.1088/1126-6708/2003/01/030}, arXiv:hep-ph/0209353.

\bibitem[Kurkela et~al.(2019{\natexlab{a}})Kurkela, Mazeliauskas, Paquet,
  Schlichting, and Teaney]{Kurkela:2018vqr}
A.~Kurkela, A.~Mazeliauskas, J.-F. Paquet, S.~Schlichting, and D.~Teaney.
\newblock {Effective kinetic description of event-by-event pre-equilibrium
  dynamics in high-energy heavy-ion collisions}.
\newblock \emph{Phys. Rev. C}, 99\penalty0 (3):\penalty0 034910,
  2019{\natexlab{a}}.
\newblock \doi{10.1103/PhysRevC.99.034910}, arXiv:1805.00961 [hep-ph].

\bibitem[Kurkela et~al.(2019{\natexlab{b}})Kurkela, Mazeliauskas, Paquet,
  Schlichting, and Teaney]{Kurkela:2018wud}
A.~Kurkela, A.~Mazeliauskas, J.-F. Paquet, S.~Schlichting, and D.~Teaney.
\newblock {Matching the Nonequilibrium Initial Stage of Heavy Ion Collisions to
  Hydrodynamics with QCD Kinetic Theory}.
\newblock \emph{Phys. Rev. Lett.}, 122\penalty0 (12):\penalty0 122302,
  2019{\natexlab{b}}.
\newblock \doi{10.1103/PhysRevLett.122.122302}, arXiv:1805.01604 [hep-ph].

\bibitem[Keegan et~al.(2016)Keegan, Kurkela, Mazeliauskas, and
  Teaney]{Keegan:2016cpi}
L.~Keegan, A.~Kurkela, A.~Mazeliauskas, and D.~Teaney.
\newblock Initial conditions for hydrodynamics from weakly coupled
  pre-equilibrium evolution.
\newblock \emph{JHEP}, 08:\penalty0 171, 2016, arXiv:1605.04287.

\bibitem[Kurkela and Zhu(2015)]{Kurkela:2015qoa}
A.~Kurkela and Y.~Zhu.
\newblock {Isotropization and hydrodynamization in weakly coupled heavy-ion
  collisions}.
\newblock \emph{Phys. Rev. Lett.}, 115\penalty0 (18):\penalty0 182301, 2015.
\newblock \doi{10.1103/PhysRevLett.115.182301}, arXiv:1506.06647 [hep-ph].

\bibitem[Romatschke(2018)]{Romatschke:2017vte}
P.~Romatschke.
\newblock Relativistic fluid dynamics far from local equilibrium.
\newblock \emph{Phys. Rev. Lett.}, 120:\penalty0 012301, 2018,
  arXiv:1704.08699.

\bibitem[Berges et~al.(2014{\natexlab{b}})Berges, Boguslavski, Schlichting, and
  Venugopalan]{Berges:2013lsa}
J.~Berges, K.~Boguslavski, S.~Schlichting, and R.~Venugopalan.
\newblock Universal attractor in a highly occupied non-abelian plasma.
\newblock \emph{Phys. Rev. D}, 89:\penalty0 074011, 2014{\natexlab{b}},
  arXiv:1303.5650.

\bibitem[Berges(2004)]{Berges:2004yj}
J.~Berges.
\newblock Introduction to nonequilibrium quantum field theory.
\newblock \emph{AIP Conf. Proc.}, 739:\penalty0 3--62, 2004,
  arXiv:hep-ph/0409233.

\bibitem[Schlichting and Teaney(2019)]{Schlichting:2019abc}
S.~Schlichting and D.~Teaney.
\newblock {The First fm/c of Heavy-Ion Collisions}.
\newblock \emph{Ann. Rev. Nucl. Part. Sci.}, 69:\penalty0 447--476, 2019.
\newblock \doi{10.1146/annurev-nucl-101918-023825}, arXiv:1908.02113 [nucl-th].

\bibitem[Borsanyi et~al.(2014)]{Borsanyi:2013bia}
S.~Borsanyi et~al.
\newblock Full result for the qcd equation of state with 2+1 flavors.
\newblock \emph{Phys. Lett. B}, 730:\penalty0 99--104, 2014.
\newblock \doi{10.1016/j.physletb.2014.01.007}, arXiv:1309.5258 [hep-lat].

\bibitem[{STAR Collaboration}(2001)]{STAR:2000ybh}
{STAR Collaboration}.
\newblock {Elliptic flow in Au+Au collisions at $\sqrt{s_{NN}} = 130$ GeV}.
\newblock \emph{Phys. Rev. Lett.}, 86:\penalty0 402--407, 2001.
\newblock \doi{10.1103/PhysRevLett.86.402}, arXiv:nucl-ex/0009011 [nucl-ex].

\bibitem[{ALICE Collaboration}(2010)]{ALICE:2010suc}
{ALICE Collaboration}.
\newblock {Elliptic flow of charged particles in Pb-Pb collisions at
  $\sqrt{s_{NN}}=2.76$ TeV}.
\newblock \emph{Phys. Rev. Lett.}, 105:\penalty0 252302, 2010.
\newblock \doi{10.1103/PhysRevLett.105.252302}, arXiv:1011.3914 [nucl-ex].

\bibitem[Romatschke and Romatschke(2019)]{Romatschke:2017ejr}
P.~Romatschke and U.~Romatschke.
\newblock \emph{{Relativistic Fluid Dynamics In and Out of Equilibrium}}.
\newblock Cambridge Monographs on Mathematical Physics. Cambridge University
  Press, 5 2019.
\newblock ISBN 978-1-108-48368-1, 978-1-108-75002-8.
\newblock \doi{10.1017/9781108651998}.

\bibitem[Bernhard et~al.(2019)Bernhard, Moreland, Bass, Liu, and
  Heinz]{Bernhard:2019bmu}
J.~E. Bernhard, J.~S. Moreland, S.~A. Bass, J.~Liu, and U.~Heinz.
\newblock Bayesian estimation of the specific shear and bulk viscosity of
  quark-gluon plasma.
\newblock \emph{Nature Phys.}, 15:\penalty0 1113--1117, 2019.
\newblock \doi{10.1038/s41567-019-0611-8}, arXiv:1905.03682 [nucl-th].

\bibitem[Parkkila et~al.(2021)]{Parkkila:2021yha}
J.~E. Parkkila et~al.
\newblock Bayesian estimation of the specific shear and bulk viscosity of the
  quark-gluon plasma with additional flow harmonic observables.
\newblock \emph{Phys. Rev. C}, 104:\penalty0 054904, 2021.
\newblock \doi{10.1103/PhysRevC.104.054904}, arXiv:2106.05019 [hep-ph].

\bibitem[Kovtun et~al.(2005)Kovtun, Son, and Starinets]{Kovtun:2004de}
P.~Kovtun, D.~T. Son, and A.~O. Starinets.
\newblock {Viscosity in strongly interacting quantum field theories from black
  hole physics}.
\newblock \emph{Phys. Rev. Lett.}, 94:\penalty0 111601, 2005.
\newblock \doi{10.1103/PhysRevLett.94.111601}, arXiv:hep-th/0405231.

\bibitem[Adam et~al.(2016)]{ALICE:2016ccg}
J.~Adam et~al.
\newblock Anisotropic flow of charged particles in pb--pb collisions at
  $\sqrt{s_{NN}}=5.02$ tev.
\newblock \emph{Phys. Rev. Lett.}, 116:\penalty0 132302, 2016.
\newblock \doi{10.1103/PhysRevLett.116.132302}, arXiv:1602.01119 [nucl-ex].

\bibitem[Koch et~al.(1986)Koch, Mueller, and Rafelski]{Koch:1986ud}
P.~Koch, B.~Mueller, and J.~Rafelski.
\newblock Strangeness in relativistic heavy ion collisions.
\newblock \emph{Phys. Rept.}, 142:\penalty0 167--262, 1986.
\newblock \doi{10.1016/0370-1573(86)90096-7}.

\bibitem[Acharya et~al.(2017)]{ALICE:2017jyt}
S.~Acharya et~al.
\newblock Enhanced production of multi-strange hadrons in high-multiplicity
  proton-proton collisions.
\newblock \emph{Nature Phys.}, 13:\penalty0 535--539, 2017.
\newblock \doi{10.1038/nphys4111}, arXiv:1606.07424 [nucl-ex].

\bibitem[Matsui and Satz(1986)]{Matsui:1986dk}
T.~Matsui and H.~Satz.
\newblock {$J/\psi$ Suppression by Quark-Gluon Plasma Formation}.
\newblock \emph{Phys. Lett. B}, 178:\penalty0 416--422, 1986.
\newblock \doi{10.1016/0370-2693(86)91404-8}.

\bibitem[Sirunyan et~al.(2019)]{CMS:2018zza}
A.~M. Sirunyan et~al.
\newblock Measurement of nuclear modification factors of $\upsilon(1s)$,
  $\upsilon(2s)$, and $\upsilon(3s)$ mesons in pbpb collisions at
  $\sqrt{s_{NN}}=5.02$ tev.
\newblock \emph{Phys. Lett. B}, 790:\penalty0 270--293, 2019.
\newblock \doi{10.1016/j.physletb.2019.01.006}, arXiv:1805.09215 [hep-ex].

\bibitem[Chatrchyan et~al.(2011)]{CMS:2011iwn}
S.~Chatrchyan et~al.
\newblock {Observation and studies of jet quenching in PbPb collisions at
  nucleon-nucleon center-of-mass energy = 2.76 TeV}.
\newblock \emph{Phys. Rev. C}, 84:\penalty0 024906, 2011.
\newblock \doi{10.1103/PhysRevC.84.024906}, arXiv:1102.1957 [nucl-ex].

\bibitem[Sirunyan et~al.(2021)]{CMS:2021vui}
A.~M. Sirunyan et~al.
\newblock {First measurement of large area jet transverse momentum spectra in
  heavy-ion collisions}.
\newblock \emph{JHEP}, 05:\penalty0 284, 2021.
\newblock \doi{10.1007/JHEP05(2021)284}, arXiv:2102.13080 [hep-ex].

\bibitem[Acharya et~al.(2024{\natexlab{a}})]{ALICE:2023waz}
S.~Acharya et~al.
\newblock {Measurement of the radius dependence of charged-particle jet
  suppression in Pb\textendash{}Pb collisions at sNN=5.02TeV}.
\newblock \emph{Phys. Lett. B}, 849:\penalty0 138412, 2024{\natexlab{a}}.
\newblock \doi{10.1016/j.physletb.2023.138412}, arXiv:2303.00592 [nucl-ex].

\bibitem[Aad et~al.(2023)]{ATLAS:2023iad}
G.~Aad et~al.
\newblock {Comparison of inclusive and photon-tagged jet suppression in 5.02
  TeV Pb+Pb collisions with ATLAS}.
\newblock \emph{Phys. Lett. B}, 846:\penalty0 138154, 2023.
\newblock \doi{10.1016/j.physletb.2023.138154}, arXiv:2303.10090 [nucl-ex].

\bibitem[Sirunyan et~al.(2017)]{CMS:2016xef}
A.~M. Sirunyan et~al.
\newblock Charged-particle nuclear modification factors in pbpb and ppb
  collisions at $\sqrt{s_{NN}}=5.02$ tev.
\newblock \emph{JHEP}, 04:\penalty0 039, 2017.
\newblock \doi{10.1007/JHEP04(2017)039}, arXiv:1611.01664 [nucl-ex].

\bibitem[Collins et~al.(1989)Collins, Soper, and Sterman]{Collins:1989gx}
J.~C. Collins, D.~E. Soper, and G.~F. Sterman.
\newblock {Factorization of Hard Processes in QCD}.
\newblock \emph{Adv. Ser. Direct. High Energy Phys.}, 5:\penalty0 1--91, 1989.
\newblock \doi{10.1142/9789814503266_0001}, arXiv:hep-ph/0409313.

\bibitem[Campbell et~al.(2018)Campbell, Huston, and Krauss]{Campbell:2017}
J.~M. Campbell, J.~Huston, and F.~Krauss.
\newblock \emph{The Black Book of Quantum Chromodynamics: A Primer for the LHC
  Era}.
\newblock Oxford University Press, Oxford, 2018.
\newblock ISBN 9780199652747.

\bibitem[Eskola et~al.(2017{\natexlab{a}})Eskola, Paakkinen, Paukkunen, and
  Salgado]{Eskola:2016oht}
K.~J. Eskola, P.~Paakkinen, H.~Paukkunen, and C.~A. Salgado.
\newblock {EPPS16: Nuclear parton distributions with LHC data}.
\newblock \emph{Eur. Phys. J. C}, 77\penalty0 (3):\penalty0 163,
  2017{\natexlab{a}}.
\newblock \doi{10.1140/epjc/s10052-017-4725-9}, arXiv:1612.05741 [hep-ph].

\bibitem[Aad et~al.(2020)]{ATLAS:2019xlu}
G.~Aad et~al.
\newblock $z$ boson production in pb+pb collisions at $\sqrt{s_{NN}}=5.02$ tev
  measured by the atlas experiment.
\newblock \emph{Phys. Lett. B}, 802:\penalty0 135262, 2020.
\newblock \doi{10.1016/j.physletb.2020.135262}, arXiv:1910.13396 [nucl-ex].

\bibitem[Acharya et~al.(2025)]{ALICE:2024vbe}
S.~Acharya et~al.
\newblock Measurement of the inclusive isolated-photon production cross section
  in pp and pb--pb collisions at $\sqrt{s_{\rm NN}}=5.02$ tev.
\newblock \emph{Eur. Phys. J. C}, 85\penalty0 (5):\penalty0 553, 2025.
\newblock \doi{10.1140/epjc/s10052-025-13971-y}, arXiv:2409.12641 [nucl-ex].

\bibitem[Svetitsky(1988)]{Svetitsky:1987gq}
B.~Svetitsky.
\newblock Diffusion of charmed quarks in the quark-gluon plasma.
\newblock \emph{Phys. Rev. D}, 37:\penalty0 2484--2491, 1988.
\newblock \doi{10.1103/PhysRevD.37.2484}.

\bibitem[Dong et~al.(2019)Dong, Lee, and Rapp]{Dong:2019byy}
X.~Dong, Y.-J. Lee, and R.~Rapp.
\newblock {Open Heavy-Flavor Production in Heavy-Ion Collisions}.
\newblock \emph{Ann. Rev. Nucl. Part. Sci.}, 69:\penalty0 417--445, 2019.
\newblock \doi{10.1146/annurev-nucl-101918-023806}, arXiv:1903.07709 [nucl-ex].

\bibitem[Armesto et~al.(2004)Armesto, Salgado, and Wiedemann]{Armesto:2003jh}
N.~Armesto, C.~A. Salgado, and U.~A. Wiedemann.
\newblock {Medium induced gluon radiation off massive quarks fills the dead
  cone}.
\newblock \emph{Phys. Rev. D}, 69:\penalty0 114003, 2004.
\newblock \doi{10.1103/PhysRevD.69.114003}, arXiv:hep-ph/0312106.

\bibitem[Cao et~al.(2019)]{Cao:2018ews}
S.~Cao et~al.
\newblock {Toward the determination of heavy-quark transport coefficients in
  quark-gluon plasma}.
\newblock \emph{Phys. Rev. C}, 99\penalty0 (5):\penalty0 054907, 2019.
\newblock \doi{10.1103/PhysRevC.99.054907}, arXiv:1809.07894 [nucl-th].

\bibitem[Dokshitzer and Kharzeev(2001)]{Dokshitzer:2001zm}
Y.~L. Dokshitzer and D.~E. Kharzeev.
\newblock {Heavy quark colorimetry of QCD matter}.
\newblock \emph{Phys. Lett. B}, 519:\penalty0 199--206, 2001.
\newblock \doi{10.1016/S0370-2693(01)01130-3}, arXiv:hep-ph/0106202.

\bibitem[Acharya et~al.(2022)]{ALICE:2021aqk}
S.~Acharya et~al.
\newblock {Direct observation of the dead-cone effect in quantum
  chromodynamics}.
\newblock \emph{Nature}, 605\penalty0 (7910):\penalty0 440--446, 2022.
\newblock \doi{10.1038/s41586-022-04572-w}, arXiv:2106.05713 [nucl-ex].
\newblock [Erratum: Nature 607, E22 (2022)].

\bibitem[Rothkopf(2020)]{Rothkopf:2019ipj}
A.~Rothkopf.
\newblock Heavy quarkonium in extreme conditions.
\newblock \emph{Phys. Rept.}, 858:\penalty0 1--117, 2020.
\newblock \doi{10.1016/j.physrep.2020.02.006}, arXiv:1912.02253 [hep-ph].

\bibitem[Hanson et~al.(1975)]{Hanson:1975fe}
G.~Hanson et~al.
\newblock Evidence for jet structure in hadron production by $e^+e^-$
  annihilation.
\newblock \emph{Phys. Rev. Lett.}, 35:\penalty0 1609--1612, 1975.
\newblock \doi{10.1103/PhysRevLett.35.1609}.

\bibitem[Brandelik et~al.(1979)]{Brandelik:1979bd}
R.~Brandelik et~al.
\newblock Evidence for planar events in $e^+e^-$ annihilation at high energies.
\newblock \emph{Phys. Lett. B}, 86:\penalty0 243--249, 1979.
\newblock \doi{10.1016/0370-2693(79)90830-X}.

\bibitem[Sterman and Weinberg(1977)]{Sterman:1977wj}
G.~F. Sterman and S.~Weinberg.
\newblock {Jets from Quantum Chromodynamics}.
\newblock \emph{Phys. Rev. Lett.}, 39:\penalty0 1436, 1977.
\newblock \doi{10.1103/PhysRevLett.39.1436}.

\bibitem[Catani et~al.(1993{\natexlab{a}})Catani, Dokshitzer, Seymour, and
  Webber]{Catani:1993hr}
S.~Catani, Y.~L. Dokshitzer, M.~H. Seymour, and B.~R. Webber.
\newblock Longitudinally invariant $k_t$ clustering algorithms for
  hadron-hadron collisions.
\newblock \emph{Nucl. Phys. B}, 406:\penalty0 187--224, 1993{\natexlab{a}}.
\newblock \doi{10.1016/0550-3213(93)90166-M}.

\bibitem[Dokshitzer et~al.(1997)Dokshitzer, Leder, Moretti, and
  Webber]{Dokshitzer:1997in}
Y.~L. Dokshitzer, G.~D. Leder, S.~Moretti, and B.~R. Webber.
\newblock Better jet clustering algorithms.
\newblock \emph{JHEP}, 08:\penalty0 001, 1997.
\newblock \doi{10.1088/1126-6708/1997/08/001}, arXiv:hep-ph/9707323.

\bibitem[Wobisch and Wengler(1999)]{Wobisch:1998wt}
M.~Wobisch and T.~Wengler.
\newblock {Hadronization Corrections to Jet Cross Sections in Deep-Inelastic
  Scattering}.
\newblock 7 1999, arXiv:hep-ph/9907280.

\bibitem[Cacciari et~al.(2008)Cacciari, Salam, and Soyez]{Cacciari:2008gp}
M.~Cacciari, G.~P. Salam, and G.~Soyez.
\newblock {The anti-$k_t$ jet clustering algorithm}.
\newblock \emph{JHEP}, 04:\penalty0 063, 2008.
\newblock \doi{10.1088/1126-6708/2008/04/063}, arXiv:0802.1189 [hep-ph].

\bibitem[Connors et~al.(2018)Connors, Nattrass, Reed, and
  Salur]{Connors:2017ptx}
M.~Connors, C.~Nattrass, R.~Reed, and S.~Salur.
\newblock {Jet measurements in heavy ion physics}.
\newblock \emph{Rev. Mod. Phys.}, 90:\penalty0 025005, 2018.
\newblock \doi{10.1103/RevModPhys.90.025005}, arXiv:1705.01974 [nucl-ex].

\bibitem[Cao and Wang(2021)]{Cao:2020wlm}
S.~Cao and X.-N. Wang.
\newblock {Jet quenching and medium response in high-energy heavy-ion
  collisions: a review}.
\newblock \emph{Rept. Prog. Phys.}, 84\penalty0 (2):\penalty0 024301, 2021.
\newblock \doi{10.1088/1361-6633/abc22b}, arXiv:2002.04028 [hep-ph].

\bibitem[Apolin\'ario et~al.(2022)Apolin\'ario, Lee, and
  Winn]{Apolinario:2022vzg}
L.~Apolin\'ario, Y.-J. Lee, and M.~Winn.
\newblock {Heavy quarks and jets as probes of the QGP}.
\newblock \emph{Prog. Part. Nucl. Phys.}, 127:\penalty0 103990, 2022.
\newblock \doi{10.1016/j.ppnp.2022.103990}, arXiv:2203.16352 [hep-ph].

\bibitem[Larkoski et~al.(2020)Larkoski, Moult, and Nachman]{Larkoski:2017jix}
A.~J. Larkoski, I.~Moult, and B.~Nachman.
\newblock {Jet Substructure at the Large Hadron Collider: A Review of Recent
  Advances in Theory and Machine Learning}.
\newblock \emph{Phys. Rept.}, 841:\penalty0 1--63, 2020.
\newblock \doi{10.1016/j.physrep.2019.11.001}, arXiv:1709.04464 [hep-ph].

\bibitem[Andrews et~al.(2020)]{Andrews:2018jcm}
H.~A. Andrews et~al.
\newblock {Novel tools and observables for jet physics in heavy-ion
  collisions}.
\newblock \emph{J. Phys. G}, 47\penalty0 (6):\penalty0 065102, 2020.
\newblock \doi{10.1088/1361-6471/ab7cbc}, arXiv:1808.03689 [hep-ph].

\bibitem[Catani and Seymour(1997)]{Catani:1996vz}
S.~Catani and M.~H. Seymour.
\newblock A general algorithm for calculating jet cross-sections in nlo qcd.
\newblock \emph{Nucl. Phys. B}, 485:\penalty0 291--419, 1997.
\newblock \doi{10.1016/0550-3213(96)00589-5}, arXiv:hep-ph/9602277.
\newblock [Erratum: Nucl.Phys.B 510, 503--504 (1998)].

\bibitem[Catani and Grazzini(2000)]{Catani:1999ss}
S.~Catani and M.~Grazzini.
\newblock {Infrared factorization of tree level QCD amplitudes at the
  next-to-next-to-leading order and beyond}.
\newblock \emph{Nucl. Phys. B}, 570:\penalty0 287--325, 2000.
\newblock \doi{10.1016/S0550-3213(99)00778-6}, arXiv:hep-ph/9908523.

\bibitem[Kinoshita(1962)]{Kinoshita:1962ur}
T.~Kinoshita.
\newblock {Mass singularities of Feynman amplitudes}.
\newblock \emph{J. Math. Phys.}, 3:\penalty0 650--677, 1962.
\newblock \doi{10.1063/1.1724268}.

\bibitem[Lee and Nauenberg(1964)]{Lee:1964is}
T.~D. Lee and M.~Nauenberg.
\newblock {Degenerate Systems and Mass Singularities}.
\newblock \emph{Phys. Rev.}, 133:\penalty0 B1549--B1562, 1964.
\newblock \doi{10.1103/PhysRev.133.B1549}.

\bibitem[Sterman(1987)]{Sterman:1986aj}
G.~F. Sterman.
\newblock Summation of large corrections to short distance hadronic
  cross-sections.
\newblock \emph{Nucl. Phys. B}, 281:\penalty0 310--364, 1987.
\newblock \doi{10.1016/0550-3213(87)90258-6}.

\bibitem[Catani and Trentadue(1989)]{Catani:1989ne}
S.~Catani and L.~Trentadue.
\newblock {Resummation of the QCD Perturbative Series for Hard Processes}.
\newblock \emph{Nucl. Phys. B}, 327:\penalty0 323--352, 1989.
\newblock \doi{10.1016/0550-3213(89)90273-3}.

\bibitem[Becher et~al.(2015)Becher, Broggio, and Ferroglia]{Becher:2014oda}
T.~Becher, A.~Broggio, and A.~Ferroglia.
\newblock \emph{Introduction to Soft-Collinear Effective Theory}, volume 896 of
  \emph{Lecture Notes in Physics}.
\newblock Springer, 2015.
\newblock \doi{10.1007/978-3-319-14848-9}.

\bibitem[Dasgupta et~al.(2013)Dasgupta, Fregoso, Marzani, and
  Salam]{Dasgupta:2013ihk}
M.~Dasgupta, A.~Fregoso, S.~Marzani, and G.~P. Salam.
\newblock {Towards an understanding of jet substructure}.
\newblock \emph{JHEP}, 09:\penalty0 029, 2013.
\newblock \doi{10.1007/JHEP09(2013)029}, arXiv:1307.0007 [hep-ph].

\bibitem[Catani et~al.(1993{\natexlab{b}})Catani, Trentadue, Turnock, and
  Webber]{Catani:1992ua}
S.~Catani, L.~Trentadue, G.~Turnock, and B.~R. Webber.
\newblock Resummation of large logarithms in $e^+e^-$ event shape
  distributions.
\newblock \emph{Nucl. Phys. B}, 407:\penalty0 3--42, 1993{\natexlab{b}}.
\newblock \doi{10.1016/0550-3213(93)90271-P}.

\bibitem[Becher and Schwartz(2008)]{Becher:2008cf}
T.~Becher and M.~D. Schwartz.
\newblock {A precise determination of $\alpha_s$ from LEP thrust data using
  effective field theory}.
\newblock \emph{JHEP}, 07:\penalty0 034, 2008.
\newblock \doi{10.1088/1126-6708/2008/07/034}, arXiv:0803.0342 [hep-ph].

\bibitem[Sjostrand(1985)]{Sjostrand:1985xi}
T.~Sjostrand.
\newblock A model for initial state parton showers.
\newblock \emph{Phys. Lett. B}, 157:\penalty0 321--325, 1985.
\newblock \doi{10.1016/0370-2693(85)90674-4}.

\bibitem[Webber(1986)]{Webber:1986mc}
B.~R. Webber.
\newblock {Monte Carlo Simulation of Hard Hadronic Processes}.
\newblock \emph{Ann. Rev. Nucl. Part. Sci.}, 36:\penalty0 253--286, 1986.
\newblock \doi{10.1146/annurev.ns.36.120186.001345}.

\bibitem[Marchesini and Webber(1988)]{Marchesini:1987cf}
G.~Marchesini and B.~R. Webber.
\newblock {Monte Carlo Simulation of General Hard Processes with Coherent QCD
  Radiation}.
\newblock \emph{Nucl. Phys. B}, 310:\penalty0 461--526, 1988.
\newblock \doi{10.1016/0550-3213(88)90089-2}.

\bibitem[Buckley et~al.(2011)Buckley, Butterworth, Lonnblad, Hoeth, Monk,
  Schulz, et~al.]{Buckley:2011ms}
A.~Buckley, J.~Butterworth, L.~Lonnblad, H.~Hoeth, J.~Monk, H.~Schulz, et~al.
\newblock General-purpose event generators for lhc physics.
\newblock \emph{Phys. Rept.}, 504:\penalty0 145--233, 2011.
\newblock \doi{10.1016/j.physrep.2011.03.005}, arXiv:1101.2599 [hep-ph].

\bibitem[H{\"o}che(2014)]{Hoeche:2014a}
S.~H{\"o}che.
\newblock Introduction to parton-shower event generators.
\newblock 2014, arXiv:1411.4085 [hep-ph].

\bibitem[Ellis et~al.(2011)Ellis, Stirling, and Webber]{Ellis:1991qj}
R.~Ellis, W.~Stirling, and B.~Webber.
\newblock \emph{{QCD and collider physics}}, volume~8.
\newblock Cambridge University Press, 2 2011.
\newblock ISBN 978-0-511-82328-2, 978-0-521-54589-1.

\bibitem[Dokshitzer et~al.(1991)Dokshitzer, Khoze, Mueller, and
  Troian]{Dokshitzer:1991wu}
Y.~L. Dokshitzer, V.~A. Khoze, A.~H. Mueller, and S.~I. Troian.
\newblock \emph{{Basics of perturbative QCD}}.
\newblock 1991.

\bibitem[Gustafson and Pettersson(1988)]{Gustafson:1987rq}
G.~Gustafson and U.~Pettersson.
\newblock {Dipole Formulation of QCD Cascades}.
\newblock \emph{Nucl. Phys. B}, 306:\penalty0 746--758, 1988.
\newblock \doi{10.1016/0550-3213(88)90441-5}.

\bibitem[Giele et~al.(2008)Giele, Kosower, and Skands]{Giele:2007di}
W.~T. Giele, D.~A. Kosower, and P.~Z. Skands.
\newblock {A simple shower and matching algorithm}.
\newblock \emph{Phys. Rev. D}, 78:\penalty0 014026, 2008.
\newblock \doi{10.1103/PhysRevD.78.014026}, arXiv:0707.3652 [hep-ph].

\bibitem[Dasgupta et~al.(2020)Dasgupta, Dreyer, Hamilton, Monni, Salam, and
  Soyez]{Dasgupta:2020fwr}
M.~Dasgupta, F.~A. Dreyer, K.~Hamilton, P.~F. Monni, G.~P. Salam, and G.~Soyez.
\newblock {Parton showers beyond leading logarithmic accuracy}.
\newblock \emph{Phys. Rev. Lett.}, 125\penalty0 (5):\penalty0 052002, 2020.
\newblock \doi{10.1103/PhysRevLett.125.052002}, arXiv:2002.11114 [hep-ph].

\bibitem[Nagy and Soper(2007)]{Nagy:2007ty}
Z.~Nagy and D.~E. Soper.
\newblock {Parton showers with quantum interference}.
\newblock \emph{JHEP}, 09:\penalty0 114, 2007.
\newblock \doi{10.1088/1126-6708/2007/09/114}, arXiv:0706.0017 [hep-ph].

\bibitem[Gyulassy et~al.(2000{\natexlab{a}})Gyulassy, Levai, and
  Vitev]{Gyulassy:1999zd}
M.~Gyulassy, P.~Levai, and I.~Vitev.
\newblock {Jet quenching in thin quark-gluon plasmas. I. Formalism}.
\newblock \emph{Nucl. Phys. B}, 571:\penalty0 197--233, 2000{\natexlab{a}}.
\newblock \doi{10.1016/S0550-3213(99)00715-7}, arXiv:hep-ph/9907461 [hep-ph].

\bibitem[Gyulassy et~al.(2001)Gyulassy, Levai, and Vitev]{Gyulassy:2000er}
M.~Gyulassy, P.~Levai, and I.~Vitev.
\newblock {Reaction operator approach to nonAbelian energy loss}.
\newblock \emph{Nucl. Phys. B}, 594:\penalty0 371--419, 2001.
\newblock \doi{10.1016/S0550-3213(00)00652-0}, arXiv:nucl-th/0006010.

\bibitem[Gyulassy et~al.(2000{\natexlab{b}})Gyulassy, Levai, and
  Vitev]{Gyulassy:2000fs}
M.~Gyulassy, P.~Levai, and I.~Vitev.
\newblock {NonAbelian energy loss at finite opacity}.
\newblock \emph{Phys. Rev. Lett.}, 85:\penalty0 5535--5538, 2000{\natexlab{b}}.
\newblock \doi{10.1103/PhysRevLett.85.5535}, arXiv:nucl-th/0005032.

\bibitem[Baier et~al.(1995)Baier, Dokshitzer, Peigne, and Schiff]{Baier:1994bd}
R.~Baier, Y.~L. Dokshitzer, S.~Peigne, and D.~Schiff.
\newblock {Induced gluon radiation in a QCD medium}.
\newblock \emph{Phys. Lett. B}, 345:\penalty0 277--286, 1995.
\newblock \doi{10.1016/0370-2693(94)01617-L}, arXiv:hep-ph/9411409.

\bibitem[Baier et~al.(1997{\natexlab{a}})Baier, Dokshitzer, Mueller, Peigne,
  and Schiff]{Baier:1996kr}
R.~Baier, Y.~L. Dokshitzer, A.~H. Mueller, S.~Peigne, and D.~Schiff.
\newblock {Radiative energy loss of high-energy quarks and gluons in a finite
  volume quark - gluon plasma}.
\newblock \emph{Nucl. Phys. B}, 483:\penalty0 291--320, 1997{\natexlab{a}}.
\newblock \doi{10.1016/S0550-3213(96)00553-6}, arXiv:hep-ph/9607355.

\bibitem[Baier et~al.(1997{\natexlab{b}})Baier, Dokshitzer, Mueller, Peigne,
  and Schiff]{Baier:1996sk}
R.~Baier, Y.~L. Dokshitzer, A.~H. Mueller, S.~Peigne, and D.~Schiff.
\newblock {Radiative energy loss and p(T) broadening of high-energy partons in
  nuclei}.
\newblock \emph{Nucl. Phys. B}, 484:\penalty0 265--282, 1997{\natexlab{b}}.
\newblock \doi{10.1016/S0550-3213(96)00581-0}, arXiv:hep-ph/9608322.

\bibitem[Zakharov(1996)]{Zakharov:1996fv}
B.~G. Zakharov.
\newblock {Fully quantum treatment of the Landau-Pomeranchuk-Migdal effect in
  QED and QCD}.
\newblock \emph{JETP Lett.}, 63:\penalty0 952--957, 1996.
\newblock \doi{10.1134/1.567126}, arXiv:hep-ph/9607440.

\bibitem[Zakharov(1997)]{Zakharov:1997uu}
B.~G. Zakharov.
\newblock {Radiative energy loss of high-energy quarks in finite size nuclear
  matter and quark - gluon plasma}.
\newblock \emph{JETP Lett.}, 65:\penalty0 615--620, 1997.
\newblock \doi{10.1134/1.567389}, arXiv:hep-ph/9704255.

\bibitem[Baier et~al.(1998{\natexlab{a}})Baier, Dokshitzer, Mueller, and
  Schiff]{Baier:1998yf}
R.~Baier, Y.~L. Dokshitzer, A.~H. Mueller, and D.~Schiff.
\newblock {Radiative energy loss of high-energy partons traversing an expanding
  QCD plasma}.
\newblock \emph{Phys. Rev. C}, 58:\penalty0 1706--1713, 1998{\natexlab{a}}.
\newblock \doi{10.1103/PhysRevC.58.1706}, arXiv:hep-ph/9803473.

\bibitem[Guo and Wang(2000)]{Guo:2000nz}
X.-f. Guo and X.-N. Wang.
\newblock {Multiple scattering, parton energy loss and modified fragmentation
  functions in deeply inelastic e A scattering}.
\newblock \emph{Phys. Rev. Lett.}, 85:\penalty0 3591--3594, 2000.
\newblock \doi{10.1103/PhysRevLett.85.3591}, arXiv:hep-ph/0005044.

\bibitem[Wang and Guo(2001)]{Wang:2001ifa}
X.-N. Wang and X.-f. Guo.
\newblock {Multiple parton scattering in nuclei: Parton energy loss}.
\newblock \emph{Nucl. Phys. A}, 696:\penalty0 788--832, 2001.
\newblock \doi{10.1016/S0375-9474(01)01130-7}, arXiv:hep-ph/0102230.

\bibitem[Arnold et~al.(2001{\natexlab{a}})Arnold, Moore, and
  Yaffe]{Arnold:2001ba}
P.~B. Arnold, G.~D. Moore, and L.~G. Yaffe.
\newblock {Photon emission from ultrarelativistic plasmas}.
\newblock \emph{JHEP}, 11:\penalty0 057, 2001{\natexlab{a}}.
\newblock \doi{10.1088/1126-6708/2001/11/057}, arXiv:hep-ph/0109064 [hep-ph].

\bibitem[Arnold et~al.(2002)Arnold, Moore, and Yaffe]{Arnold:2002ja}
P.~B. Arnold, G.~D. Moore, and L.~G. Yaffe.
\newblock {Photon and gluon emission in relativistic plasmas}.
\newblock \emph{JHEP}, 06:\penalty0 030, 2002.
\newblock \doi{10.1088/1126-6708/2002/06/030}, arXiv:hep-ph/0204343.

\bibitem[Landau and Pomeranchuk(1953)]{Landau:1953um}
L.~D. Landau and I.~Pomeranchuk.
\newblock {Limits of applicability of the theory of bremsstrahlung electrons
  and pair production at high-energies}.
\newblock \emph{Dokl. Akad. Nauk SSSR}, 92:\penalty0 535--536, 1953.

\bibitem[Migdal(1956)]{Migdal:1956tc}
A.~B. Migdal.
\newblock {Bremsstrahlung and pair production in condensed media at high
  energies}.
\newblock \emph{Phys. Rev.}, 103:\penalty0 1811--1820, 1956.
\newblock \doi{10.1103/PhysRev.103.1811}.

\bibitem[Braaten and Pisarski(1990)]{Braaten:1989mz}
E.~Braaten and R.~D. Pisarski.
\newblock {Soft amplitudes in hot gauge theories: A general analysis}.
\newblock \emph{Nucl. Phys. B}, 337:\penalty0 569--634, 1990.
\newblock \doi{10.1016/0550-3213(90)90508-B}.

\bibitem[Caron-Huot(2009)]{CaronHuot:2008ni}
S.~Caron-Huot.
\newblock {$O(g)$ plasma effects in jet quenching}.
\newblock \emph{Phys. Rev. D}, 79:\penalty0 065039, 2009.
\newblock \doi{10.1103/PhysRevD.79.065039}, arXiv:0811.1603 [hep-ph].

\bibitem[Ovanesyan and Vitev(2012)]{Ovanesyan:2011kn}
G.~Ovanesyan and I.~Vitev.
\newblock {Medium-induced parton splitting kernels from Soft Collinear
  Effective Theory with Glauber gluons}.
\newblock \emph{Phys. Lett. B}, 706:\penalty0 371--378, 2012.
\newblock \doi{10.1016/j.physletb.2011.11.040}, arXiv:1109.5619 [hep-ph].

\bibitem[Ovanesyan and Vitev(2011)]{Ovanesyan:2011xy}
G.~Ovanesyan and I.~Vitev.
\newblock {An effective theory for jet propagation in dense QCD matter: jet
  broadening and medium-induced bremsstrahlung}.
\newblock \emph{JHEP}, 06:\penalty0 080, 2011.
\newblock \doi{10.1007/JHEP06(2011)080}, arXiv:1103.1074 [hep-ph].

\bibitem[Kang et~al.(2015)Kang, Lashof-Regas, Ovanesyan, Saad, and
  Vitev]{Kang:2014xsa}
Z.-B. Kang, R.~Lashof-Regas, G.~Ovanesyan, P.~Saad, and I.~Vitev.
\newblock {Jet quenching phenomenology from soft-collinear effective theory
  with Glauber gluons}.
\newblock \emph{Phys. Rev. Lett.}, 114\penalty0 (9):\penalty0 092002, 2015.
\newblock \doi{10.1103/PhysRevLett.114.092002}, arXiv:1405.2612 [hep-ph].

\bibitem[Liu et~al.(2006)Liu, Rajagopal, and Wiedemann]{Liu:2006ug}
H.~Liu, K.~Rajagopal, and U.~A. Wiedemann.
\newblock {Calculating the jet quenching parameter from AdS/CFT}.
\newblock \emph{Phys. Rev. Lett.}, 97:\penalty0 182301, 2006.
\newblock \doi{10.1103/PhysRevLett.97.182301}, arXiv:hep-ph/0605178.

\bibitem[Herzog et~al.(2006)Herzog, Karch, Kovtun, Kozcaz, and
  Yaffe]{Herzog:2006gh}
C.~P. Herzog, A.~Karch, P.~Kovtun, C.~Kozcaz, and L.~G. Yaffe.
\newblock {Energy loss of a heavy quark moving through N=4 supersymmetric
  Yang-Mills plasma}.
\newblock \emph{JHEP}, 07:\penalty0 013, 2006.
\newblock \doi{10.1088/1126-6708/2006/07/013}, arXiv:hep-th/0605158.

\bibitem[Gubser(2006)]{Gubser:2006bz}
S.~S. Gubser.
\newblock {Drag force in AdS/CFT}.
\newblock \emph{Phys. Rev. D}, 74:\penalty0 126005, 2006.
\newblock \doi{10.1103/PhysRevD.74.126005}, arXiv:hep-th/0605182 [hep-th].

\bibitem[Casalderrey-Solana and Teaney(2006)]{CasalderreySolana:2006rq}
J.~Casalderrey-Solana and D.~Teaney.
\newblock {Heavy quark diffusion in strongly coupled N=4 Yang-Mills}.
\newblock \emph{Phys. Rev. D}, 74:\penalty0 085012, 2006.
\newblock \doi{10.1103/PhysRevD.74.085012}, arXiv:hep-ph/0605199 [hep-ph].

\bibitem[Armesto et~al.(2012{\natexlab{a}})]{Armesto:2011ht}
N.~Armesto et~al.
\newblock {Comparison of Jet Quenching Formalisms for a Quark-Gluon Plasma
  ``Brick''}.
\newblock \emph{Phys. Rev. C}, 86:\penalty0 064904, 2012{\natexlab{a}}.
\newblock \doi{10.1103/PhysRevC.86.064904}, arXiv:1106.1106 [hep-ph].

\bibitem[Bethe and Heitler(1934)]{Bethe:1934za}
H.~Bethe and W.~Heitler.
\newblock {On the Stopping of Fast Particles and on the Creation of Positive
  Electrons}.
\newblock \emph{Proc. Roy. Soc. Lond. A}, 146:\penalty0 83--112, 1934.
\newblock \doi{10.1098/rspa.1934.0140}.

\bibitem[Mehtar-Tani et~al.(2011{\natexlab{a}})Mehtar-Tani, Salgado, and
  Tywoniuk]{MehtarTani:2010ebp}
Y.~Mehtar-Tani, C.~A. Salgado, and K.~Tywoniuk.
\newblock {Anti-angular ordering of gluon radiation in QCD media}.
\newblock \emph{Phys. Rev. Lett.}, 106:\penalty0 122002, 2011{\natexlab{a}}.
\newblock \doi{10.1103/PhysRevLett.106.122002}, arXiv:1009.2965 [hep-ph].

\bibitem[Mehtar-Tani et~al.(2012{\natexlab{a}})Mehtar-Tani, Salgado, and
  Tywoniuk]{MehtarTani:2011tz}
Y.~Mehtar-Tani, C.~A. Salgado, and K.~Tywoniuk.
\newblock {Jets in QCD Media: From Color Coherence to Decoherence}.
\newblock \emph{Phys. Lett. B}, 707:\penalty0 156--159, 2012{\natexlab{a}}.
\newblock \doi{10.1016/j.physletb.2011.12.042}, arXiv:1102.4317 [hep-ph].

\bibitem[Mehtar-Tani and Tywoniuk(2013{\natexlab{a}})]{MehtarTani:2011gf}
Y.~Mehtar-Tani and K.~Tywoniuk.
\newblock {Jet coherence in QCD media: the antenna radiation spectrum}.
\newblock \emph{JHEP}, 01:\penalty0 031, 2013{\natexlab{a}}.
\newblock \doi{10.1007/JHEP01(2013)031}, arXiv:1105.1346 [hep-ph].

\bibitem[Casalderrey-Solana and
  Iancu(2011{\natexlab{a}})]{CasalderreySolana:2011rz}
J.~Casalderrey-Solana and E.~Iancu.
\newblock {Interference effects in medium-induced gluon radiation}.
\newblock \emph{JHEP}, 08:\penalty0 015, 2011{\natexlab{a}}.
\newblock \doi{10.1007/JHEP08(2011)015}, arXiv:1105.1760 [hep-ph].

\bibitem[Casalderrey-Solana et~al.(2013)Casalderrey-Solana, Mehtar-Tani,
  Salgado, and Tywoniuk]{Casalderrey-Solana:2012evi}
J.~Casalderrey-Solana, Y.~Mehtar-Tani, C.~A. Salgado, and K.~Tywoniuk.
\newblock {New picture of jet quenching dictated by color coherence}.
\newblock \emph{Phys. Lett. B}, 725:\penalty0 357--360, 2013.
\newblock \doi{10.1016/j.physletb.2013.07.046}, arXiv:1210.7765 [hep-ph].

\bibitem[Mehtar-Tani et~al.(2012{\natexlab{b}})Mehtar-Tani, Salgado, and
  Tywoniuk]{MehtarTani:2011hma}
Y.~Mehtar-Tani, C.~A. Salgado, and K.~Tywoniuk.
\newblock {The radiation pattern of a QCD antenna in a dilute medium}.
\newblock \emph{JHEP}, 04:\penalty0 064, 2012{\natexlab{b}}.
\newblock \doi{10.1007/JHEP04(2012)064}, arXiv:1112.5031 [hep-ph].

\bibitem[Mehtar-Tani et~al.(2012{\natexlab{c}})Mehtar-Tani, Salgado, and
  Tywoniuk]{MehtarTani:2012mfa}
Y.~Mehtar-Tani, C.~A. Salgado, and K.~Tywoniuk.
\newblock {The radiation pattern of a QCD antenna in a dense medium}.
\newblock \emph{JHEP}, 10:\penalty0 197, 2012{\natexlab{c}}.
\newblock \doi{10.1007/JHEP10(2012)197}, arXiv:1205.5739 [hep-ph].

\bibitem[Caucal et~al.(2018)Caucal, Iancu, Mueller, and Soyez]{Caucal:2018dla}
P.~Caucal, E.~Iancu, A.~H. Mueller, and G.~Soyez.
\newblock {Vacuum-like jet fragmentation in a dense QCD medium}.
\newblock \emph{Phys. Rev. Lett.}, 120:\penalty0 232001, 2018.
\newblock \doi{10.1103/PhysRevLett.120.232001}, arXiv:1801.09703 [hep-ph].

\bibitem[Susskind(1968)]{Susskind:1967rg}
L.~Susskind.
\newblock {Model of selfinduced strong interactions}.
\newblock \emph{Phys. Rev.}, 165:\penalty0 1535--1546, 1968.
\newblock \doi{10.1103/PhysRev.165.1535}.

\bibitem[Kogut and Soper(1970)]{Kogut:1969xa}
J.~B. Kogut and D.~E. Soper.
\newblock {Quantum Electrodynamics in the Infinite Momentum Frame}.
\newblock \emph{Phys. Rev. D}, 1:\penalty0 2901--2913, 1970.
\newblock \doi{10.1103/PhysRevD.1.2901}.

\bibitem[Brodsky et~al.(1998)Brodsky, Pauli, and Pinsky]{Brodsky:1997de}
S.~J. Brodsky, H.-C. Pauli, and S.~S. Pinsky.
\newblock {Quantum Chromodynamics and Other Field Theories on the Light Cone}.
\newblock \emph{Phys. Rept.}, 301:\penalty0 299--486, 1998.
\newblock \doi{10.1016/S0370-1573(97)00089-6}, arXiv:hep-ph/9705477.

\bibitem[Dhani et~al.(2023)Dhani, Rodrigo, and Sborlini]{Dhani:2023uxu}
P.~K. Dhani, G.~Rodrigo, and G.~F.~R. Sborlini.
\newblock {Triple-collinear splittings with massive particles}.
\newblock \emph{JHEP}, 12:\penalty0 188, 2023.
\newblock \doi{10.1007/JHEP12(2023)188}, arXiv:2310.05803 [hep-ph].

\bibitem[Somogyi et~al.(2005)Somogyi, Trocsanyi, and Del~Duca]{Somogyi:2005xz}
G.~Somogyi, Z.~Trocsanyi, and V.~Del~Duca.
\newblock {Matching of singly- and doubly-unresolved limits of tree-level QCD
  squared matrix elements}.
\newblock \emph{JHEP}, 06:\penalty0 024, 2005.
\newblock \doi{10.1088/1126-6708/2005/06/024}, arXiv:hep-ph/0502226.

\bibitem[Braun-White and Glover(2022)]{Braun-White:2022rtg}
O.~Braun-White and N.~Glover.
\newblock {Decomposition of triple collinear splitting functions}.
\newblock \emph{JHEP}, 09:\penalty0 059, 2022.
\newblock \doi{10.1007/JHEP09(2022)059}, arXiv:2204.10755 [hep-ph].

\bibitem[Fickinger et~al.(2013)Fickinger, Ovanesyan, and
  Vitev]{Fickinger:2013xwa}
M.~Fickinger, G.~Ovanesyan, and I.~Vitev.
\newblock {Angular distributions of higher order splitting functions in the
  vacuum and in dense QCD matter}.
\newblock \emph{JHEP}, 07:\penalty0 059, 2013.
\newblock \doi{10.1007/JHEP07(2013)059}, arXiv:1304.3497 [hep-ph].

\bibitem[Dom\'\i{}nguez et~al.(2020)Dom\'\i{}nguez, Milhano, Salgado, Tywoniuk,
  and Vila]{Dominguez:2019ges}
F.~Dom\'\i{}nguez, J.~G. Milhano, C.~A. Salgado, K.~Tywoniuk, and V.~Vila.
\newblock {Mapping collinear in-medium parton splittings}.
\newblock \emph{Eur. Phys. J. C}, 80\penalty0 (1):\penalty0 11, 2020.
\newblock \doi{10.1140/epjc/s10052-019-7563-0}, arXiv:1907.03653 [hep-ph].

\bibitem[Isaksen and Tywoniuk(2023)]{Isaksen:2023nlr}
J.~H. Isaksen and K.~Tywoniuk.
\newblock {Precise description of medium-induced emissions}.
\newblock \emph{JHEP}, 09:\penalty0 049, 2023.
\newblock \doi{10.1007/JHEP09(2023)049}, arXiv:2303.12119 [hep-ph].

\bibitem[McLerran and Venugopalan(1994{\natexlab{c}})]{McLerran:1994vd}
L.~D. McLerran and R.~Venugopalan.
\newblock {Green's functions in the color field of a large nucleus}.
\newblock \emph{Phys. Rev. D}, 50:\penalty0 2225--2233, 1994{\natexlab{c}}.
\newblock \doi{10.1103/PhysRevD.50.2225}, arXiv:hep-ph/9402335.

\bibitem[Iancu et~al.(2001)Iancu, Leonidov, and McLerran]{Iancu:2001ad}
E.~Iancu, A.~Leonidov, and L.~D. McLerran.
\newblock {The Renormalization group equation for the color glass condensate}.
\newblock \emph{Phys. Lett. B}, 510:\penalty0 133--144, 2001.
\newblock \doi{10.1016/S0370-2693(01)00524-X}, arXiv:hep-ph/0102009.

\bibitem[Blaizot et~al.(2013)Blaizot, Dominguez, Iancu, and
  Mehtar-Tani]{Blaizot:2012fh}
J.-P. Blaizot, F.~Dominguez, E.~Iancu, and Y.~Mehtar-Tani.
\newblock {Medium-induced gluon branching}.
\newblock \emph{JHEP}, 01:\penalty0 143, 2013.
\newblock \doi{10.1007/JHEP01(2013)143}, arXiv:1209.4585 [hep-ph].

\bibitem[Apolin\'ario et~al.(2015)Apolin\'ario, Armesto, Milhano, and
  Salgado]{Apolinario:2014csa}
L.~Apolin\'ario, N.~Armesto, J.~G. Milhano, and C.~A. Salgado.
\newblock {Medium-induced gluon radiation and colour decoherence beyond the
  soft approximation}.
\newblock \emph{JHEP}, 02:\penalty0 119, 2015.
\newblock \doi{10.1007/JHEP02(2015)119}, arXiv:1407.0599 [hep-ph].

\bibitem[Dominguez et~al.(2011)Dominguez, Marquet, Xiao, and
  Yuan]{Dominguez:2011wm}
F.~Dominguez, C.~Marquet, B.-W. Xiao, and F.~Yuan.
\newblock {Universality of Unintegrated Gluon Distributions at small x}.
\newblock \emph{Phys. Rev. D}, 83:\penalty0 105005, 2011.
\newblock \doi{10.1103/PhysRevD.83.105005}, arXiv:1101.0715 [hep-ph].

\bibitem[Dominguez et~al.(2013)Dominguez, Marquet, Stasto, and
  Xiao]{Dominguez:2012ad}
F.~Dominguez, C.~Marquet, A.~M. Stasto, and B.-W. Xiao.
\newblock {Universality of multiparticle production in QCD at high energies}.
\newblock \emph{Phys. Rev. D}, 87:\penalty0 034007, 2013.
\newblock \doi{10.1103/PhysRevD.87.034007}, arXiv:1210.1141 [hep-ph].

\bibitem[Isaksen and Tywoniuk(2020)]{Isaksen:2020npj}
J.~H. Isaksen and K.~Tywoniuk.
\newblock {Wilson line correlators beyond the large-N$_{c}$}.
\newblock \emph{JHEP}, 21:\penalty0 125, 2020.
\newblock \doi{10.1007/JHEP11(2021)125}, arXiv:2107.02542 [hep-ph].

\bibitem[Gyulassy and Wang(1994)]{Gyulassy:1993hr}
M.~Gyulassy and X.-n. Wang.
\newblock {Multiple collisions and induced gluon Bremsstrahlung in QCD}.
\newblock \emph{Nucl. Phys. B}, 420:\penalty0 583--614, 1994.
\newblock \doi{10.1016/0550-3213(94)90079-5}, arXiv:nucl-th/9306003.

\bibitem[Arnold and Xiao(2008)]{Arnold:2008vd}
P.~B. Arnold and W.~Xiao.
\newblock {High-energy jet quenching in weakly-coupled quark-gluon plasmas}.
\newblock \emph{Phys. Rev. D}, 78:\penalty0 125008, 2008.
\newblock \doi{10.1103/PhysRevD.78.125008}, arXiv:0810.1026 [hep-ph].

\bibitem[Andres et~al.(2020)Andres, Apolin{\'a}rio, and
  Dominguez]{Andres:2020vxs}
C.~Andres, L.~Apolin{\'a}rio, and F.~Dominguez.
\newblock {Medium-induced gluon radiation with full resummation of multiple
  scatterings for realistic parton-medium interactions}.
\newblock \emph{JHEP}, 07:\penalty0 114, 2020.
\newblock \doi{10.1007/JHEP07(2020)114}, arXiv:2002.01517 [hep-ph].

\bibitem[Feynman and Hibbs(1965)]{Feynman:100771}
R.~P. Feynman and A.~R. Hibbs.
\newblock \emph{{Quantum mechanics and path integrals}}.
\newblock International series in pure and applied physics. McGraw-Hill, New
  York, NY, 1965.
\newblock URL \url{https://cds.cern.ch/record/100771}.

\bibitem[Kleinert(2004)]{Kleinert:2004ev}
H.~Kleinert.
\newblock {Path Integrals in Quantum Mechanics, Statistics, Polymer Physics,
  and Financial Markets}.
\newblock 2004.

\bibitem[Ipp et~al.(2020{\natexlab{a}})Ipp, M\"uller, and Schuh]{Ipp:2020mjc}
A.~Ipp, D.~I. M\"uller, and D.~Schuh.
\newblock {Anisotropic momentum broadening in the 2+1D Glasma: analytic weak
  field approximation and lattice simulations}.
\newblock \emph{Phys. Rev. D}, 102\penalty0 (7):\penalty0 074001,
  2020{\natexlab{a}}.
\newblock \doi{10.1103/PhysRevD.102.074001}, arXiv:2001.10001 [hep-ph].

\bibitem[Ipp et~al.(2020{\natexlab{b}})Ipp, M\"uller, and Schuh]{Ipp:2020nfu}
A.~Ipp, D.~I. M\"uller, and D.~Schuh.
\newblock {Jet momentum broadening in the pre-equilibrium Glasma}.
\newblock \emph{Phys. Lett. B}, 810:\penalty0 135810, 2020{\natexlab{b}}.
\newblock \doi{10.1016/j.physletb.2020.135810}, arXiv:2009.14206 [hep-ph].

\bibitem[Avramescu et~al.(2023)Avramescu, Baran, Greco, Ipp, M\"uller, and
  Ruggieri]{Avramescu:2023qvv}
D.~Avramescu, V.~Baran, V.~Greco, A.~Ipp, D.~I. M\"uller, and M.~Ruggieri.
\newblock {Simulating jets and heavy quarks in the glasma using the colored
  particle-in-cell method}.
\newblock \emph{Phys. Rev. D}, 107\penalty0 (11):\penalty0 114021, 2023.
\newblock \doi{10.1103/PhysRevD.107.114021}, arXiv:2303.05599 [hep-ph].

\bibitem[Carrington et~al.(2022{\natexlab{a}})Carrington, Czajka, and
  Mrowczynski]{Carrington:2021dvw}
M.~E. Carrington, A.~Czajka, and S.~Mrowczynski.
\newblock {Jet quenching in glasma}.
\newblock \emph{Phys. Lett. B}, 834:\penalty0 137464, 2022{\natexlab{a}}.
\newblock \doi{10.1016/j.physletb.2022.137464}, arXiv:2112.06812 [hep-ph].

\bibitem[Carrington et~al.(2022{\natexlab{b}})Carrington, Czajka, and
  Mrowczynski]{Carrington:2022bnv}
M.~E. Carrington, A.~Czajka, and S.~Mrowczynski.
\newblock {Transport of hard probes through glasma}.
\newblock \emph{Phys. Rev. C}, 105\penalty0 (6):\penalty0 064910,
  2022{\natexlab{b}}.
\newblock \doi{10.1103/PhysRevC.105.064910}, arXiv:2202.00357 [nucl-th].

\bibitem[Carrington et~al.(2026)Carrington, Friesen, and
  Mrowczynski]{Carrington:2026qlg}
M.~E. Carrington, B.~T. Friesen, and S.~Mrowczynski.
\newblock {Gauge invariant momentum broadening of hard probes in glasma}.
\newblock 4 2026, arXiv:2604.02050 [nucl-th].

\bibitem[Avramescu et~al.(2026)Avramescu, Lamas, Lappi, Li, and
  Salgado]{Avramescu:2026fgv}
D.~Avramescu, C.~Lamas, T.~Lappi, M.~Li, and C.~A. Salgado.
\newblock {Kinetic and canonical momentum broadening in the Glasma}.
\newblock 4 2026, arXiv:2604.08520 [hep-ph].

\bibitem[Boguslavski et~al.(2024{\natexlab{a}})Boguslavski, Kurkela, Lappi,
  Lindenbauer, and Peuron]{Boguslavski:2023alu}
K.~Boguslavski, A.~Kurkela, T.~Lappi, F.~Lindenbauer, and J.~Peuron.
\newblock {Jet momentum broadening during initial stages in heavy-ion
  collisions}.
\newblock \emph{Phys. Lett. B}, 850:\penalty0 138525, 2024{\natexlab{a}}.
\newblock \doi{10.1016/j.physletb.2024.138525}, arXiv:2303.12595 [hep-ph].

\bibitem[Boguslavski et~al.(2023)Boguslavski, Kurkela, Lappi, Lindenbauer, and
  Peuron]{Boguslavski:2023waw}
K.~Boguslavski, A.~Kurkela, T.~Lappi, F.~Lindenbauer, and J.~Peuron.
\newblock {Jet quenching parameter in QCD kinetic theory}.
\newblock 12 2023, arXiv:2312.00447 [hep-ph].

\bibitem[Barata et~al.(2024{\natexlab{c}})Barata, Hauksson, Mayo~L\'opez, and
  Sadofyev]{Barata:2024xwy}
J.~Barata, S.~Hauksson, X.~Mayo~L\'opez, and A.~V. Sadofyev.
\newblock {Jet quenching in the glasma phase: medium-induced radiation}.
\newblock 6 2024{\natexlab{c}}, arXiv:2406.07615 [hep-ph].

\bibitem[Altenburger et~al.(2025)Altenburger, Boguslavski, and
  Lindenbauer]{Altenburger:2025iqa}
A.~Altenburger, K.~Boguslavski, and F.~Lindenbauer.
\newblock {Jet broadening and radiation in the early anisotropic plasma in
  heavy-ion collisions}.
\newblock 9 2025, arXiv:2509.03868 [hep-ph].

\bibitem[Barata et~al.(2025{\natexlab{d}})Barata, Boguslavski, Lindenbauer, and
  Sadofyev]{Barata:2025agq}
J.~Barata, K.~Boguslavski, F.~Lindenbauer, and A.~V. Sadofyev.
\newblock {Jet quenching in out-of-equilibrium QCD matter}.
\newblock 11 2025{\natexlab{d}}, arXiv:2511.07519 [hep-ph].

\bibitem[Baier and Mehtar-Tani(2008)]{Baier:2008js}
R.~Baier and Y.~Mehtar-Tani.
\newblock {Jet quenching and broadening: The transport coefficient $\hat{q}$ in
  an anisotropic plasma}.
\newblock \emph{Phys. Rev. C}, 78:\penalty0 064906, 2008.
\newblock \doi{10.1103/PhysRevC.78.064906}, arXiv:0806.0954 [hep-ph].

\bibitem[Majumder et~al.(2009)Majumder, Muller, and
  Mrowczynski]{Majumder:2009cf}
A.~Majumder, B.~Muller, and S.~Mrowczynski.
\newblock {Momentum Broadening of a Fast Parton in a Perturbative Quark-Gluon
  Plasma}.
\newblock \emph{Phys. Rev. D}, 80:\penalty0 125020, 2009.
\newblock \doi{10.1103/PhysRevD.80.125020}, arXiv:0903.3683 [hep-ph].

\bibitem[Hauksson et~al.(2022)Hauksson, Jeon, and Gale]{Hauksson:2021okc}
S.~Hauksson, S.~Jeon, and C.~Gale.
\newblock {Momentum broadening of energetic partons in an anisotropic plasma}.
\newblock \emph{Phys. Rev. C}, 105\penalty0 (1):\penalty0 014914, 2022.
\newblock \doi{10.1103/PhysRevC.105.014914}, arXiv:2109.04575 [hep-ph].

\bibitem[Hauksson and Iancu(2023)]{Hauksson:2023tze}
S.~Hauksson and E.~Iancu.
\newblock {Jet polarisation in an anisotropic medium}.
\newblock \emph{JHEP}, 08:\penalty0 027, 2023.
\newblock \doi{10.1007/JHEP08(2023)027}, arXiv:2303.03914 [hep-ph].

\bibitem[Mrowczynski and Thoma(2000)]{Mrowczynski:2000ed}
S.~Mrowczynski and M.~H. Thoma.
\newblock {Hard loop approach to anisotropic systems}.
\newblock \emph{Phys. Rev. D}, 62:\penalty0 036011, 2000.
\newblock \doi{10.1103/PhysRevD.62.036011}, arXiv:hep-ph/0001164.

\bibitem[Romatschke and Strickland(2003)]{Romatschke:2003ms}
P.~Romatschke and M.~Strickland.
\newblock {Collective modes of an anisotropic quark-gluon plasma}.
\newblock \emph{Phys. Rev. D}, 68:\penalty0 036004, 2003.
\newblock \doi{10.1103/PhysRevD.68.036004}, arXiv:hep-ph/0304092.

\bibitem[Arnold et~al.(2003{\natexlab{b}})Arnold, Lenaghan, and
  Moore]{Arnold:2003rq}
P.~B. Arnold, J.~Lenaghan, and G.~D. Moore.
\newblock {QCD plasma instabilities and bottom up thermalization}.
\newblock \emph{JHEP}, 08:\penalty0 002, 2003{\natexlab{b}}.
\newblock \doi{10.1088/1126-6708/2003/08/002}, arXiv:hep-ph/0307325.

\bibitem[Sadofyev et~al.(2021)Sadofyev, Sievert, and Vitev]{Sadofyev:2021ohn}
A.~V. Sadofyev, M.~D. Sievert, and I.~Vitev.
\newblock {Ab~initio coupling of jets to collective flow in the opacity
  expansion approach}.
\newblock \emph{Phys. Rev. D}, 104\penalty0 (9):\penalty0 094044, 2021.
\newblock \doi{10.1103/PhysRevD.104.094044}, arXiv:2104.09513 [hep-ph].

\bibitem[Barata et~al.(2022)Barata, Sadofyev, and Salgado]{Barata:2022krd}
J.~Barata, A.~V. Sadofyev, and C.~A. Salgado.
\newblock {Jet broadening in dense inhomogeneous matter}.
\newblock \emph{Phys. Rev. D}, 105\penalty0 (11):\penalty0 114010, 2022.
\newblock \doi{10.1103/PhysRevD.105.114010}, arXiv:2202.08847 [hep-ph].

\bibitem[Andres et~al.(2022)Andres, Dominguez, Sadofyev, and
  Salgado]{Andres:2022ndd}
C.~Andres, F.~Dominguez, A.~V. Sadofyev, and C.~A. Salgado.
\newblock {Jet broadening in flowing matter: Resummation}.
\newblock \emph{Phys. Rev. D}, 106\penalty0 (7):\penalty0 074023, 2022.
\newblock \doi{10.1103/PhysRevD.106.074023}, arXiv:2207.07141 [hep-ph].

\bibitem[Barata et~al.(2023)Barata, Mayo~L\'opez, Sadofyev, and
  Salgado]{Barata:2023qds}
J.~Barata, X.~Mayo~L\'opez, A.~V. Sadofyev, and C.~A. Salgado.
\newblock {Medium induced gluon spectrum in dense inhomogeneous matter}.
\newblock \emph{Phys. Rev. D}, 108\penalty0 (3):\penalty0 034018, 2023.
\newblock \doi{10.1103/PhysRevD.108.034018}, arXiv:2304.03712 [hep-ph].

\bibitem[He et~al.(2020)He, Pang, and Wang]{He:2020iow}
Y.~He, L.-G. Pang, and X.-N. Wang.
\newblock {Gradient Tomography of Jet Quenching in Heavy-Ion Collisions}.
\newblock \emph{Phys. Rev. Lett.}, 125\penalty0 (12):\penalty0 122301, 2020.
\newblock \doi{10.1103/PhysRevLett.125.122301}, arXiv:2001.08273 [hep-ph].

\bibitem[Fu et~al.(2023)Fu, Casalderrey-Solana, and Wang]{Fu:2022idl}
Y.~Fu, J.~Casalderrey-Solana, and X.-N. Wang.
\newblock {Asymmetric transverse momentum broadening in an inhomogeneous
  medium}.
\newblock \emph{Phys. Rev. D}, 107\penalty0 (5):\penalty0 054038, 2023.
\newblock \doi{10.1103/PhysRevD.107.054038}, arXiv:2204.05323 [hep-ph].

\bibitem[Kuzmin et~al.(2024)Kuzmin, Mayo~L\'opez, Reiten, and
  Sadofyev]{Kuzmin:2023hko}
M.~V. Kuzmin, X.~Mayo~L\'opez, J.~Reiten, and A.~V. Sadofyev.
\newblock {Jet quenching in anisotropic flowing matter}.
\newblock \emph{Phys. Rev. D}, 109\penalty0 (1):\penalty0 014036, 2024.
\newblock \doi{10.1103/PhysRevD.109.014036}, arXiv:2309.00683 [hep-ph].

\bibitem[Wang(2001)]{Wang:2000fq}
X.-N. Wang.
\newblock {Jet quenching and azimuthal anisotropy of large p(T) spectra in
  noncentral high-energy heavy ion collisions}.
\newblock \emph{Phys. Rev. C}, 63:\penalty0 054902, 2001.
\newblock \doi{10.1103/PhysRevC.63.054902}, arXiv:nucl-th/0009019.

\bibitem[Snellings(2011)]{Snellings:2011sz}
R.~Snellings.
\newblock {Elliptic Flow: A Brief Review}.
\newblock \emph{New J. Phys.}, 13:\penalty0 055008, 2011.
\newblock \doi{10.1088/1367-2630/13/5/055008}, arXiv:1102.3010 [nucl-ex].

\bibitem[Kharzeev et~al.(2016)Kharzeev, Liao, Voloshin, and
  Wang]{Kharzeev:2015znc}
D.~E. Kharzeev, J.~Liao, S.~A. Voloshin, and G.~Wang.
\newblock {Chiral magnetic and vortical effects in high-energy nuclear
  collisions\textemdash{}A status report}.
\newblock \emph{Prog. Part. Nucl. Phys.}, 88:\penalty0 1--28, 2016.
\newblock \doi{10.1016/j.ppnp.2016.01.001}, arXiv:1511.04050 [hep-ph].

\bibitem[Hofman and Maldacena(2008)]{Hofman:2008ar}
D.~M. Hofman and J.~Maldacena.
\newblock {Conformal collider physics: Energy and charge correlations}.
\newblock \emph{JHEP}, 05:\penalty0 012, 2008.
\newblock \doi{10.1088/1126-6708/2008/05/012}, arXiv:0803.1467 [hep-th].

\bibitem[Kravchuk and Simmons-Duffin(2018)]{Kravchuk:2018htv}
P.~Kravchuk and D.~Simmons-Duffin.
\newblock {Light-ray operators in conformal field theory}.
\newblock \emph{JHEP}, 11:\penalty0 102, 2018.
\newblock \doi{10.1007/JHEP11(2018)102}, arXiv:1805.00098 [hep-th].

\bibitem[Moult and Zhu(2025)]{Moult:2025nhu}
I.~Moult and H.~X. Zhu.
\newblock {Energy Correlators: A Journey From Theory to Experiment}.
\newblock 6 2025, arXiv:2506.09119 [hep-ph].

\bibitem[Basham et~al.(1978{\natexlab{a}})Basham, Brown, Ellis, and
  Love]{Basham:1978bw}
C.~Basham, L.~S. Brown, S.~D. Ellis, and S.~T. Love.
\newblock {Energy Correlations in electron - Positron Annihilation: Testing
  QCD}.
\newblock \emph{Phys. Rev. Lett.}, 41:\penalty0 1585, 1978{\natexlab{a}}.
\newblock \doi{10.1103/PhysRevLett.41.1585}.

\bibitem[Basham et~al.(1979{\natexlab{a}})Basham, Brown, Ellis, and
  Love]{Basham:1978zq}
C.~Basham, L.~Brown, S.~Ellis, and S.~Love.
\newblock {Energy Correlations in electron-Positron Annihilation in Quantum
  Chromodynamics: Asymptotically Free Perturbation Theory}.
\newblock \emph{Phys. Rev. D}, 19:\penalty0 2018, 1979{\natexlab{a}}.
\newblock \doi{10.1103/PhysRevD.19.2018}.

\bibitem[Ellis et~al.(1981)Ellis, Ross, and Terrano]{Ellis:1980wv}
R.~K. Ellis, D.~A. Ross, and A.~E. Terrano.
\newblock {The Perturbative Calculation of Jet Structure in e+ e-
  Annihilation}.
\newblock \emph{Nucl. Phys. B}, 178:\penalty0 421--456, 1981.
\newblock \doi{10.1016/0550-3213(81)90165-6}.

\bibitem[Chen et~al.(2020{\natexlab{a}})Chen, Moult, Zhang, and
  Zhu]{Chen:2020vvp}
H.~Chen, I.~Moult, X.~Zhang, and H.~X. Zhu.
\newblock {Rethinking jets with energy correlators: Tracks, resummation, and
  analytic continuation}.
\newblock \emph{Phys. Rev. D}, 102\penalty0 (5):\penalty0 054012,
  2020{\natexlab{a}}.
\newblock \doi{10.1103/PhysRevD.102.054012}, arXiv:2004.11381 [hep-ph].

\bibitem[Lee et~al.(2022)Lee, Me\c{c}aj, and Moult]{Lee:2022ige}
K.~Lee, B.~Me\c{c}aj, and I.~Moult.
\newblock {Conformal Colliders Meet the LHC}.
\newblock 5 2022, arXiv:2205.03414 [hep-ph].

\bibitem[Komiske et~al.(2023)Komiske, Moult, Thaler, and Zhu]{Komiske:2022enw}
P.~T. Komiske, I.~Moult, J.~Thaler, and H.~X. Zhu.
\newblock {Analyzing N-Point Energy Correlators inside Jets with CMS Open
  Data}.
\newblock \emph{Phys. Rev. Lett.}, 130\penalty0 (5):\penalty0 051901, 2023.
\newblock \doi{10.1103/PhysRevLett.130.051901}, arXiv:2201.07800 [hep-ph].

\bibitem[Andres et~al.(2023)Andres, Dominguez, Kunnawalkam~Elayavalli, Holguin,
  Marquet, and Moult]{Andres:2022ovj}
C.~Andres, F.~Dominguez, R.~Kunnawalkam~Elayavalli, J.~Holguin, C.~Marquet, and
  I.~Moult.
\newblock {Resolving the Scales of the Quark-Gluon Plasma with Energy
  Correlators}.
\newblock \emph{Phys. Rev. Lett.}, 130\penalty0 (26):\penalty0 262301, 2023.
\newblock \doi{10.1103/PhysRevLett.130.262301}, arXiv:2209.11236 [hep-ph].

\bibitem[Yang et~al.(2024{\natexlab{a}})Yang, He, Moult, and
  Wang]{Yang:2023dwc}
Z.~Yang, Y.~He, I.~Moult, and X.-N. Wang.
\newblock {Probing the Short-Distance Structure of the Quark-Gluon Plasma with
  Energy Correlators}.
\newblock \emph{Phys. Rev. Lett.}, 132\penalty0 (1):\penalty0 011901,
  2024{\natexlab{a}}.
\newblock \doi{10.1103/PhysRevLett.132.011901}, arXiv:2310.01500 [hep-ph].

\bibitem[Barata et~al.(2024{\natexlab{d}})Barata, Caucal, Soto-Ontoso, and
  Szafron]{Barata:2023bhh}
J.~Barata, P.~Caucal, A.~Soto-Ontoso, and R.~Szafron.
\newblock {Advancing the understanding of energy-energy correlators in
  heavy-ion collisions}.
\newblock \emph{JHEP}, 11:\penalty0 060, 2024{\natexlab{d}}.
\newblock \doi{10.1007/JHEP11(2024)060}, arXiv:2312.12527 [hep-ph].

\bibitem[Singh and Vaidya(2025)]{Singh:2024vwb}
B.~Singh and V.~Vaidya.
\newblock {Factorization for energy-energy correlator in heavy ion collision}.
\newblock \emph{JHEP}, 06:\penalty0 071, 2025.
\newblock \doi{10.1007/JHEP06(2025)071}, arXiv:2408.02753 [hep-ph].

\bibitem[Barata et~al.(2024{\natexlab{e}})Barata, Kang, Mayo~L\'opez, and
  Penttala]{Barata:2024wsu}
J.~Barata, Z.-B. Kang, X.~Mayo~L\'opez, and J.~Penttala.
\newblock {Energy-Energy Correlator for jet production in $pp$ and $pA$
  collisions}.
\newblock 11 2024{\natexlab{e}}, arXiv:2411.11782 [hep-ph].

\bibitem[Bossi et~al.(2024)Bossi, Kudinoor, Moult, Pablos, Rai, and
  Rajagopal]{Bossi:2024qho}
H.~Bossi, A.~S. Kudinoor, I.~Moult, D.~Pablos, A.~Rai, and K.~Rajagopal.
\newblock {Imaging the Wakes of Jets with Energy-Energy-Energy Correlators}.
\newblock 7 2024, arXiv:2407.13818 [hep-ph].

\bibitem[Apolin\'ario et~al.(2025)Apolin\'ario, Kunnawalkam~Elayavalli,
  Madureira, Sheng, Wang, and Yang]{Apolinario:2025vtx}
L.~Apolin\'ario, R.~Kunnawalkam~Elayavalli, N.~O. Madureira, J.-X. Sheng, X.-N.
  Wang, and Z.~Yang.
\newblock {Flavor dependence of Energy-energy correlators}.
\newblock 2 2025, arXiv:2502.11406 [hep-ph].

\bibitem[Sveshnikov and Tkachov(1996)]{Sveshnikov:1995vi}
N.~Sveshnikov and F.~Tkachov.
\newblock {Jets and quantum field theory}.
\newblock \emph{Phys. Lett. B}, 382:\penalty0 403--408, 1996.
\newblock \doi{10.1016/0370-2693(96)00558-8}, arXiv:hep-ph/9512370.

\bibitem[Korchemsky et~al.(1997)Korchemsky, Oderda, and
  Sterman]{Korchemsky:1997sy}
G.~P. Korchemsky, G.~Oderda, and G.~F. Sterman.
\newblock {Power corrections and nonlocal operators}.
\newblock \emph{AIP Conf. Proc.}, 407\penalty0 (1):\penalty0 988, 1997.
\newblock \doi{10.1063/1.53732}, arXiv:hep-ph/9708346.

\bibitem[Dhani et~al.(2025{\natexlab{a}})Dhani, Fedkevych, and
  Ghira]{Dhani:2025fbk}
P.~K. Dhani, O.~Fedkevych, and A.~Ghira.
\newblock {Heavy Flavor Jet Substructure at Lepton Colliders}.
\newblock 2 2025{\natexlab{a}}, arXiv:2502.07894 [hep-ph].

\bibitem[Xing et~al.(2025)Xing, Cao, Qin, and Wang]{Xing:2024yrb}
W.-J. Xing, S.~Cao, G.-Y. Qin, and X.-N. Wang.
\newblock {Flavor Hierarchy of Jet Energy Correlators inside the Quark-Gluon
  Plasma}.
\newblock \emph{Phys. Rev. Lett.}, 134\penalty0 (5):\penalty0 052301, 2025.
\newblock \doi{10.1103/PhysRevLett.134.052301}, arXiv:2409.12843 [hep-ph].

\bibitem[Dhani et~al.(2025{\natexlab{b}})Dhani, Fedkevych, Ghira, Marzani, and
  Soyez]{Dhani:2024gtx}
P.~K. Dhani, O.~Fedkevych, A.~Ghira, S.~Marzani, and G.~Soyez.
\newblock {Heavy flavour jet substructure}.
\newblock \emph{JHEP}, 02:\penalty0 046, 2025{\natexlab{b}}.
\newblock \doi{10.1007/JHEP02(2025)046}, arXiv:2410.05415 [hep-ph].

\bibitem[Blok and Wu(2025)]{Blok:2023ugf}
B.~Blok and C.~Wu.
\newblock {zg distribution for heavy flavor jets in dense quark-gluon plasma}.
\newblock \emph{Phys. Rev. D}, 111\penalty0 (7):\penalty0 074009, 2025.
\newblock \doi{10.1103/PhysRevD.111.074009}, arXiv:2312.15560 [hep-ph].

\bibitem[Yang et~al.(2024{\natexlab{b}})Yang, Fedkevych, and
  Esha]{Yang:2024awm}
Z.~Yang, O.~Fedkevych, and R.~Esha.
\newblock {Jet substructure of light and heavy flavor jets at RHIC}.
\newblock 12 2024{\natexlab{b}}, arXiv:2412.08682 [hep-ph].

\bibitem[Caletti et~al.(2024)Caletti, Ghira, and Marzani]{Caletti:2023spr}
S.~Caletti, A.~Ghira, and S.~Marzani.
\newblock {On heavy-flavour jets with Soft Drop}.
\newblock \emph{Eur. Phys. J. C}, 84\penalty0 (2):\penalty0 212, 2024.
\newblock \doi{10.1140/epjc/s10052-024-12562-7}, arXiv:2312.11623 [hep-ph].

\bibitem[Andres et~al.(2024{\natexlab{a}})Andres, Dominguez, Holguin, Marquet,
  and Moult]{Andres:2023ymw}
C.~Andres, F.~Dominguez, J.~Holguin, C.~Marquet, and I.~Moult.
\newblock {Seeing beauty in the quark-gluon plasma with energy correlators}.
\newblock \emph{Phys. Rev. D}, 110\penalty0 (3):\penalty0 L031503,
  2024{\natexlab{a}}.
\newblock \doi{10.1103/PhysRevD.110.L031503}, arXiv:2307.15110 [hep-ph].

\bibitem[Chen et~al.(2024{\natexlab{a}})Chen, Liu, and Ma]{Chen:2024nfl}
A.-P. Chen, X.~Liu, and Y.-Q. Ma.
\newblock {Shedding Light on Hadronization by Quarkonium Energy Correlator}.
\newblock \emph{Phys. Rev. Lett.}, 133:\penalty0 19, 2024{\natexlab{a}}.
\newblock \doi{10.1103/PhysRevLett.133.191901}, arXiv:2405.10056 [hep-ph].

\bibitem[Cunqueiro et~al.(2023)Cunqueiro, Napoletano, and
  Soto-Ontoso]{Cunqueiro:2022svx}
L.~Cunqueiro, D.~Napoletano, and A.~Soto-Ontoso.
\newblock {Dead-cone searches in heavy-ion collisions using the jet tree}.
\newblock \emph{Phys. Rev. D}, 107\penalty0 (9):\penalty0 094008, 2023.
\newblock \doi{10.1103/PhysRevD.107.094008}, arXiv:2211.11789 [hep-ph].

\bibitem[Craft et~al.(2022)Craft, Lee, Me{\c{c}}aj, and Moult]{Craft:2022kdo}
E.~Craft, K.~Lee, B.~Me{\c{c}}aj, and I.~Moult.
\newblock {Beautiful and Charming Energy Correlators}.
\newblock 10 2022, arXiv:2210.09311 [hep-ph].

\bibitem[Li et~al.(2022{\natexlab{a}})Li, Liu, and Vitev]{Li:2021gjw}
H.~T. Li, Z.~L. Liu, and I.~Vitev.
\newblock {Heavy flavor jet production and substructure in electron-nucleus
  collisions}.
\newblock \emph{Phys. Lett. B}, 827:\penalty0 137007, 2022{\natexlab{a}}.
\newblock \doi{10.1016/j.physletb.2022.137007}, arXiv:2108.07809 [hep-ph].

\bibitem[Cal et~al.(2022)Cal, Lee, Ringer, and Waalewijn]{Cal:2021fla}
P.~Cal, K.~Lee, F.~Ringer, and W.~J. Waalewijn.
\newblock {The soft drop momentum sharing fraction zg beyond
  leading-logarithmic accuracy}.
\newblock \emph{Phys. Lett. B}, 833:\penalty0 137390, 2022.
\newblock \doi{10.1016/j.physletb.2022.137390}, arXiv:2106.04589 [hep-ph].

\bibitem[Casalderrey-Solana et~al.(2020)Casalderrey-Solana, Milhano, Pablos,
  and Rajagopal]{Casalderrey-Solana:2019ubu}
J.~Casalderrey-Solana, G.~Milhano, D.~Pablos, and K.~Rajagopal.
\newblock {Modification of Jet Substructure in Heavy Ion Collisions as a Probe
  of the Resolution Length of Quark-Gluon Plasma}.
\newblock \emph{JHEP}, 01:\penalty0 044, 2020.
\newblock \doi{10.1007/JHEP01(2020)044}, arXiv:1907.11248 [hep-ph].

\bibitem[Caucal et~al.(2019)Caucal, Iancu, and Soyez]{Caucal:2019uvr}
P.~Caucal, E.~Iancu, and G.~Soyez.
\newblock {Deciphering the $z_g$ distribution in ultrarelativistic heavy ion
  collisions}.
\newblock \emph{JHEP}, 10:\penalty0 273, 2019.
\newblock \doi{10.1007/JHEP10(2019)273}, arXiv:1907.04866 [hep-ph].

\bibitem[Li and Vitev(2019)]{Li:2017wwc}
H.~T. Li and I.~Vitev.
\newblock {Inverting the mass hierarchy of jet quenching effects with prompt
  $b$-jet substructure}.
\newblock \emph{Phys. Lett. B}, 793:\penalty0 259--264, 2019.
\newblock \doi{10.1016/j.physletb.2019.04.052}, arXiv:1801.00008 [hep-ph].

\bibitem[Lee et~al.(2019)Lee, Shrivastava, and Vaidya]{Lee:2019lge}
C.~Lee, P.~Shrivastava, and V.~Vaidya.
\newblock {Predictions for energy correlators probing substructure of groomed
  heavy quark jets}.
\newblock \emph{JHEP}, 09:\penalty0 045, 2019.
\newblock \doi{10.1007/JHEP09(2019)045}, arXiv:1901.09095 [hep-ph].

\bibitem[Chang et~al.(2018)Chang, Cao, and Qin]{Chang:2017gkt}
N.-B. Chang, S.~Cao, and G.-Y. Qin.
\newblock {Probing medium-induced jet splitting and energy loss in heavy-ion
  collisions}.
\newblock \emph{Phys. Lett. B}, 781:\penalty0 423--432, 2018.
\newblock \doi{10.1016/j.physletb.2018.04.019}, arXiv:1707.03767 [hep-ph].

\bibitem[Milhano et~al.(2018)Milhano, Wiedemann, and Zapp]{Milhano:2017nzm}
G.~Milhano, U.~A. Wiedemann, and K.~C. Zapp.
\newblock {Sensitivity of jet substructure to jet-induced medium response}.
\newblock \emph{Phys. Lett. B}, 779:\penalty0 409--413, 2018.
\newblock \doi{10.1016/j.physletb.2018.01.029}, arXiv:1707.04142 [hep-ph].

\bibitem[Mehtar-Tani and Tywoniuk(2017)]{Mehtar-Tani:2016aco}
Y.~Mehtar-Tani and K.~Tywoniuk.
\newblock {Groomed jets in heavy-ion collisions: sensitivity to medium-induced
  bremsstrahlung}.
\newblock \emph{JHEP}, 04:\penalty0 125, 2017.
\newblock \doi{10.1007/JHEP04(2017)125}, arXiv:1610.08930 [hep-ph].

\bibitem[Kunnawalkam~Elayavalli and Zapp(2017)]{KunnawalkamElayavalli:2017hxo}
R.~Kunnawalkam~Elayavalli and K.~C. Zapp.
\newblock {Medium response in JEWEL and its impact on jet shape observables in
  heavy ion collisions}.
\newblock \emph{JHEP}, 07:\penalty0 141, 2017.
\newblock \doi{10.1007/JHEP07(2017)141}, arXiv:1707.01539 [hep-ph].

\bibitem[Chien and Vitev(2017)]{Chien:2016led}
Y.-T. Chien and I.~Vitev.
\newblock {Probing the Hardest Branching within Jets in Heavy-Ion Collisions}.
\newblock \emph{Phys. Rev. Lett.}, 119\penalty0 (11):\penalty0 112301, 2017.
\newblock \doi{10.1103/PhysRevLett.119.112301}, arXiv:1608.07283 [hep-ph].

\bibitem[Sievert et~al.(2019)Sievert, Vitev, and Yoon]{Sievert:2019cwq}
M.~D. Sievert, I.~Vitev, and B.~Yoon.
\newblock {A complete set of in-medium splitting functions to any order in
  opacity}.
\newblock \emph{Phys. Lett. B}, 795:\penalty0 502--510, 2019.
\newblock \doi{10.1016/j.physletb.2019.06.019}, arXiv:1903.06170 [hep-ph].

\bibitem[Sievert and Vitev(2018)]{Sievert:2018imd}
M.~D. Sievert and I.~Vitev.
\newblock {Quark branching in QCD matter to any order in opacity beyond the
  soft gluon emission limit}.
\newblock \emph{Phys. Rev. D}, 98\penalty0 (9):\penalty0 094010, 2018.
\newblock \doi{10.1103/PhysRevD.98.094010}, arXiv:1807.03799 [hep-ph].

\bibitem[Zhang et~al.(2004)Zhang, Wang, and Wang]{Zhang:2003wk}
B.-W. Zhang, E.~Wang, and X.-N. Wang.
\newblock {Heavy quark energy loss in nuclear medium}.
\newblock \emph{Phys. Rev. Lett.}, 93:\penalty0 072301, 2004.
\newblock \doi{10.1103/PhysRevLett.93.072301}, arXiv:nucl-th/0309040.

\bibitem[Arleo and Peigne(2013)]{Arleo:2012rs}
F.~Arleo and S.~Peigne.
\newblock {Heavy-quarkonium suppression in p-A collisions from parton energy
  loss in cold QCD matter}.
\newblock \emph{JHEP}, 03:\penalty0 122, 2013.
\newblock \doi{10.1007/JHEP03(2013)122}, arXiv:1212.0434 [hep-ph].

\bibitem[Djordjevic and Gyulassy(2004)]{Djordjevic:2003zk}
M.~Djordjevic and M.~Gyulassy.
\newblock {Heavy quark radiative energy loss in QCD matter}.
\newblock \emph{Nucl. Phys. A}, 733:\penalty0 265--298, 2004.
\newblock \doi{10.1016/j.nuclphysa.2003.12.020}, arXiv:nucl-th/0310076.

\bibitem[Banfi et~al.(2006)Banfi, Salam, and Zanderighi]{Banfi:2006hf}
A.~Banfi, G.~P. Salam, and G.~Zanderighi.
\newblock {Infrared safe definition of jet flavor}.
\newblock \emph{Eur. Phys. J. C}, 47:\penalty0 113--124, 2006.
\newblock \doi{10.1140/epjc/s2006-02552-4}, arXiv:hep-ph/0601139.

\bibitem[Gauld et~al.(2023)Gauld, Huss, and Stagnitto]{Gauld:2022lem}
R.~Gauld, A.~Huss, and G.~Stagnitto.
\newblock {Flavor Identification of Reconstructed Hadronic Jets}.
\newblock \emph{Phys. Rev. Lett.}, 130\penalty0 (16):\penalty0 161901, 2023.
\newblock \doi{10.1103/PhysRevLett.130.161901}, arXiv:2208.11138 [hep-ph].
\newblock [Erratum: Phys.Rev.Lett. 132, 159901 (2024)].

\bibitem[Caola et~al.(2023)Caola, Grabarczyk, Hutt, Salam, Scyboz, and
  Thaler]{Caola:2023wpj}
F.~Caola, R.~Grabarczyk, M.~L. Hutt, G.~P. Salam, L.~Scyboz, and J.~Thaler.
\newblock {Flavored jets with exact anti-kt kinematics and tests of infrared
  and collinear safety}.
\newblock \emph{Phys. Rev. D}, 108\penalty0 (9):\penalty0 094010, 2023.
\newblock \doi{10.1103/PhysRevD.108.094010}, arXiv:2306.07314 [hep-ph].

\bibitem[Czakon et~al.(2023)Czakon, Mitov, and Poncelet]{Czakon:2022wam}
M.~Czakon, A.~Mitov, and R.~Poncelet.
\newblock {Infrared-safe flavoured anti-k$_{T}$ jets}.
\newblock \emph{JHEP}, 04:\penalty0 138, 2023.
\newblock \doi{10.1007/JHEP04(2023)138}, arXiv:2205.11879 [hep-ph].

\bibitem[Caletti et~al.(2022{\natexlab{a}})Caletti, Larkoski, Marzani, and
  Reichelt]{Caletti:2022glq}
S.~Caletti, A.~J. Larkoski, S.~Marzani, and D.~Reichelt.
\newblock {A fragmentation approach to jet flavor}.
\newblock \emph{JHEP}, 10:\penalty0 158, 2022{\natexlab{a}}.
\newblock \doi{10.1007/JHEP10(2022)158}, arXiv:2205.01117 [hep-ph].

\bibitem[Caletti et~al.(2022{\natexlab{b}})Caletti, Larkoski, Marzani, and
  Reichelt]{Caletti:2022hnc}
S.~Caletti, A.~J. Larkoski, S.~Marzani, and D.~Reichelt.
\newblock {Practical jet flavour through NNLO}.
\newblock \emph{Eur. Phys. J. C}, 82\penalty0 (7):\penalty0 632,
  2022{\natexlab{b}}.
\newblock \doi{10.1140/epjc/s10052-022-10568-7}, arXiv:2205.01109 [hep-ph].

\bibitem[Fedkevych et~al.(2023)Fedkevych, Khosa, Marzani, and
  Sforza]{Fedkevych:2022mid}
O.~Fedkevych, C.~K. Khosa, S.~Marzani, and F.~Sforza.
\newblock {Identification of b jets using QCD-inspired observables}.
\newblock \emph{Phys. Rev. D}, 107\penalty0 (3):\penalty0 034032, 2023.
\newblock \doi{10.1103/PhysRevD.107.034032}, arXiv:2202.05082 [hep-ph].

\bibitem[Behring et~al.(2025)]{Behring:2025ilo}
A.~Behring et~al.
\newblock {Flavoured jet algorithms: a comparative study}.
\newblock 6 2025, arXiv:2506.13449 [hep-ph].

\bibitem[Dixon et~al.(2019)Dixon, Moult, and Zhu]{Dixon:2019uzg}
L.~J. Dixon, I.~Moult, and H.~X. Zhu.
\newblock {Collinear limit of the energy-energy correlator}.
\newblock \emph{Phys. Rev. D}, 100\penalty0 (1):\penalty0 014009, 2019.
\newblock \doi{10.1103/PhysRevD.100.014009}, arXiv:1905.01310 [hep-ph].

\bibitem[Lee et~al.(2024{\natexlab{a}})Lee, Moult, and Zhang]{Lee:2024icn}
K.~Lee, I.~Moult, and X.~Zhang.
\newblock {Revisiting Single Inclusive Jet Production: Timelike Factorization
  and Reciprocity}.
\newblock 9 2024{\natexlab{a}}, arXiv:2409.19045 [hep-ph].

\bibitem[Lee and Stewart(2025)]{Lee:2025okn}
K.~Lee and I.~Stewart.
\newblock {Dihadron Fragmentation and the Confinement Transition in Energy
  Correlators}.
\newblock 7 2025, arXiv:2507.11495 [hep-ph].

\bibitem[Jaarsma et~al.(2023)Jaarsma, Li, Moult, Waalewijn, and
  Zhu]{Jaarsma:2023ell}
M.~Jaarsma, Y.~Li, I.~Moult, W.~J. Waalewijn, and H.~X. Zhu.
\newblock {Energy correlators on tracks: resummation and non-perturbative
  effects}.
\newblock \emph{JHEP}, 12:\penalty0 087, 2023.
\newblock \doi{10.1007/JHEP12(2023)087}, arXiv:2307.15739 [hep-ph].

\bibitem[Lee and Moult(2023{\natexlab{a}})]{Lee:2023npz}
K.~Lee and I.~Moult.
\newblock {Energy Correlators Taking Charge}.
\newblock 8 2023{\natexlab{a}}, arXiv:2308.00746 [hep-ph].

\bibitem[Lee and Moult(2023{\natexlab{b}})]{Lee:2023tkr}
K.~Lee and I.~Moult.
\newblock {Joint Track Functions: Expanding the Space of Calculable
  Correlations at Colliders}.
\newblock 8 2023{\natexlab{b}}, arXiv:2308.01332 [hep-ph].

\bibitem[Chang et~al.(2022)Chang, Kologlu, Kravchuk, Simmons-Duffin, and
  Zhiboedov]{Chang:2020qpj}
C.-H. Chang, M.~Kologlu, P.~Kravchuk, D.~Simmons-Duffin, and A.~Zhiboedov.
\newblock {Transverse spin in the light-ray OPE}.
\newblock \emph{JHEP}, 05:\penalty0 059, 2022.
\newblock \doi{10.1007/JHEP05(2022)059}, arXiv:2010.04726 [hep-th].

\bibitem[Kologlu et~al.(2021)Kologlu, Kravchuk, Simmons-Duffin, and
  Zhiboedov]{Kologlu:2019mfz}
M.~Kologlu, P.~Kravchuk, D.~Simmons-Duffin, and A.~Zhiboedov.
\newblock {The light-ray OPE and conformal colliders}.
\newblock \emph{JHEP}, 01:\penalty0 128, 2021.
\newblock \doi{10.1007/JHEP01(2021)128}, arXiv:1905.01311 [hep-th].

\bibitem[Chen et~al.(2022{\natexlab{a}})Chen, Moult, and Zhu]{Chen:2021gdk}
H.~Chen, I.~Moult, and H.~X. Zhu.
\newblock {Spinning gluons from the QCD light-ray OPE}.
\newblock \emph{JHEP}, 08:\penalty0 233, 2022{\natexlab{a}}.
\newblock \doi{10.1007/JHEP08(2022)233}, arXiv:2104.00009 [hep-ph].

\bibitem[Chen(2024)]{Chen:2023zzh}
H.~Chen.
\newblock {QCD factorization from light-ray OPE}.
\newblock \emph{JHEP}, 01:\penalty0 035, 2024.
\newblock \doi{10.1007/JHEP01(2024)035}, arXiv:2311.00350 [hep-ph].

\bibitem[Sj\"ostrand et~al.(2015)Sj\"ostrand, Ask, Christiansen, Corke, Desai,
  Ilten, Mrenna, Prestel, Rasmussen, and Skands]{Sjostrand:2014zea}
T.~Sj\"ostrand, S.~Ask, J.~R. Christiansen, R.~Corke, N.~Desai, P.~Ilten,
  S.~Mrenna, S.~Prestel, C.~O. Rasmussen, and P.~Z. Skands.
\newblock {An introduction to PYTHIA 8.2}.
\newblock \emph{Comput. Phys. Commun.}, 191:\penalty0 159--177, 2015.
\newblock \doi{10.1016/j.cpc.2015.01.024}, arXiv:1410.3012 [hep-ph].

\bibitem[Eichten and Hill(1990)]{Eichten:1989zv}
E.~Eichten and B.~R. Hill.
\newblock {An Effective Field Theory for the Calculation of Matrix Elements
  Involving Heavy Quarks}.
\newblock \emph{Phys. Lett. B}, 234:\penalty0 511--516, 1990.
\newblock \doi{10.1016/0370-2693(90)92049-O}.

\bibitem[Isgur and Wise(1989)]{Isgur:1989vq}
N.~Isgur and M.~B. Wise.
\newblock {Weak Decays of Heavy Mesons in the Static Quark Approximation}.
\newblock \emph{Phys. Lett. B}, 232:\penalty0 113--117, 1989.
\newblock \doi{10.1016/0370-2693(89)90566-2}.

\bibitem[Isgur and Wise(1990)]{Isgur:1990yhj}
N.~Isgur and M.~B. Wise.
\newblock {WEAK TRANSITION FORM-FACTORS BETWEEN HEAVY MESONS}.
\newblock \emph{Phys. Lett. B}, 237:\penalty0 527--530, 1990.
\newblock \doi{10.1016/0370-2693(90)91219-2}.

\bibitem[Grinstein(1990)]{Grinstein:1990mj}
B.~Grinstein.
\newblock {The Static Quark Effective Theory}.
\newblock \emph{Nucl. Phys. B}, 339:\penalty0 253--268, 1990.
\newblock \doi{10.1016/0550-3213(90)90349-I}.

\bibitem[Georgi(1990)]{Georgi:1990um}
H.~Georgi.
\newblock {An Effective Field Theory for Heavy Quarks at Low-energies}.
\newblock \emph{Phys. Lett. B}, 240:\penalty0 447--450, 1990.
\newblock \doi{10.1016/0370-2693(90)91128-X}.

\bibitem[Korner and Thompson(1991)]{Korner:1991kf}
J.~G. Korner and G.~Thompson.
\newblock {The Heavy mass limit in field theory and the heavy quark effective
  theory}.
\newblock \emph{Phys. Lett. B}, 264:\penalty0 185--192, 1991.
\newblock \doi{10.1016/0370-2693(91)90725-6}.

\bibitem[Mannel et~al.(1992)Mannel, Roberts, and Ryzak]{Mannel:1991mc}
T.~Mannel, W.~Roberts, and Z.~Ryzak.
\newblock {A Derivation of the heavy quark effective Lagrangian from QCD}.
\newblock \emph{Nucl. Phys. B}, 368:\penalty0 204--217, 1992.
\newblock \doi{10.1016/0550-3213(92)90204-O}.

\bibitem[Fleming et~al.(2008{\natexlab{a}})Fleming, Hoang, Mantry, and
  Stewart]{Fleming:2007qr}
S.~Fleming, A.~H. Hoang, S.~Mantry, and I.~W. Stewart.
\newblock {Jets from massive unstable particles: Top-mass determination}.
\newblock \emph{Phys. Rev. D}, 77:\penalty0 074010, 2008{\natexlab{a}}.
\newblock \doi{10.1103/PhysRevD.77.074010}, arXiv:hep-ph/0703207.

\bibitem[Fleming et~al.(2008{\natexlab{b}})Fleming, Hoang, Mantry, and
  Stewart]{Fleming:2007xt}
S.~Fleming, A.~H. Hoang, S.~Mantry, and I.~W. Stewart.
\newblock {Top Jets in the Peak Region: Factorization Analysis with NLL
  Resummation}.
\newblock \emph{Phys. Rev. D}, 77:\penalty0 114003, 2008{\natexlab{b}}.
\newblock \doi{10.1103/PhysRevD.77.114003}, arXiv:0711.2079 [hep-ph].

\bibitem[Amhis et~al.(2017)]{HFLAV:2016hnz}
Y.~Amhis et~al.
\newblock {Averages of $b$-hadron, $c$-hadron, and $\tau$-lepton properties as
  of summer 2016}.
\newblock \emph{Eur. Phys. J. C}, 77\penalty0 (12):\penalty0 895, 2017.
\newblock \doi{10.1140/epjc/s10052-017-5058-4}, arXiv:1612.07233 [hep-ex].

\bibitem[Abbaneo et~al.(2001)]{ALEPH:2001ccp}
D.~Abbaneo et~al.
\newblock {Combined results on $b$ hadron production rates and decay
  properties}.
\newblock 6 2001, arXiv:hep-ex/0112028.

\bibitem[Abramowicz et~al.(2013)]{ZEUS:2013fws}
H.~Abramowicz et~al.
\newblock {Measurement of charm fragmentation fractions in photoproduction at
  HERA}.
\newblock \emph{JHEP}, 09:\penalty0 058, 2013.
\newblock \doi{10.1007/JHEP09(2013)058}, arXiv:1306.4862 [hep-ex].

\bibitem[Gladilin(1999)]{Gladilin:1999pj}
L.~Gladilin.
\newblock {Charm hadron production fractions}.
\newblock 12 1999, arXiv:hep-ex/9912064.

\bibitem[Aaboud et~al.(2019)]{ATLAS:2018zhf}
M.~Aaboud et~al.
\newblock {Properties of $g\rightarrow b\bar{b}$ at small opening angles in
  $pp$ collisions with the ATLAS detector at $\sqrt{s}=13$ TeV}.
\newblock \emph{Phys. Rev. D}, 99\penalty0 (5):\penalty0 052004, 2019.
\newblock \doi{10.1103/PhysRevD.99.052004}, arXiv:1812.09283 [hep-ex].

\bibitem[Acosta et~al.(2005)]{CDF:2004mmv}
D.~Acosta et~al.
\newblock {Measurements of $b\bar{b}$ azimuthal production correlations in
  $p\bar{p}$ collisions at $\sqrt{s} = 1.8$ TeV}.
\newblock \emph{Phys. Rev. D}, 71:\penalty0 092001, 2005.
\newblock \doi{10.1103/PhysRevD.71.092001}, arXiv:hep-ex/0412006.

\bibitem[Attems et~al.(2024)Attems, Brewer, Innocenti, Mazeliauskas, Park,
  van~der Schee, Soyez, and Wiedemann]{Attems:2022otp}
M.~Attems, J.~Brewer, G.~M. Innocenti, A.~Mazeliauskas, S.~Park, W.~van~der
  Schee, G.~Soyez, and U.~A. Wiedemann.
\newblock {Medium-Enhanced cc\textasciimacron{} Radiation}.
\newblock \emph{Phys. Rev. Lett.}, 132\penalty0 (21):\penalty0 212301, 2024.
\newblock \doi{10.1103/PhysRevLett.132.212301}, arXiv:2209.13600 [hep-ph].

\bibitem[Attems et~al.(2023)Attems, Brewer, Innocenti, Mazeliauskas, Park,
  van~der Schee, and Wiedemann]{Attems:2022ubu}
M.~Attems, J.~Brewer, G.~M. Innocenti, A.~Mazeliauskas, S.~Park, W.~van~der
  Schee, and U.~A. Wiedemann.
\newblock {The medium-modified $ g\to c\overline{c} $ splitting function in the
  BDMPS-Z formalism}.
\newblock \emph{JHEP}, 01:\penalty0 080, 2023.
\newblock \doi{10.1007/JHEP01(2023)080}, arXiv:2203.11241 [hep-ph].

\bibitem[Apolin\'ario et~al.(2018)Apolin\'ario, Milhano, Salam, and
  Salgado]{Apolinario:2017sob}
L.~Apolin\'ario, J.~G. Milhano, G.~P. Salam, and C.~A. Salgado.
\newblock {Probing the time structure of the quark-gluon plasma with top
  quarks}.
\newblock \emph{Phys. Rev. Lett.}, 120\penalty0 (23):\penalty0 232301, 2018.
\newblock \doi{10.1103/PhysRevLett.120.232301}, arXiv:1711.03105 [hep-ph].

\bibitem[Apolin\'ario et~al.(2021)Apolin\'ario, Cordeiro, and
  Zapp]{Apolinario:2020uvt}
L.~Apolin\'ario, A.~Cordeiro, and K.~Zapp.
\newblock {Time reclustering for jet quenching studies}.
\newblock \emph{Eur. Phys. J. C}, 81\penalty0 (6):\penalty0 561, 2021.
\newblock \doi{10.1140/epjc/s10052-021-09346-8}, arXiv:2012.02199 [hep-ph].

\bibitem[Apolinario et~al.(2024)Apolinario, Guerrero-Rodr{\'\i}guez, and
  Zapp]{Apolinario:2024klh}
L.~Apolinario, P.~Guerrero-Rodr{\'\i}guez, and K.~Zapp.
\newblock {Exploring the time axis within medium-modified jets}.
\newblock \emph{Eur. Phys. J. C}, 84\penalty0 (7):\penalty0 672, 2024.
\newblock \doi{10.1140/epjc/s10052-024-13048-2}, arXiv:2401.14229 [hep-ph].

\bibitem[Brewer et~al.(2025)Brewer, van~der Schee, and
  Wiedemann]{Brewer:2025wol}
J.~Brewer, W.~van~der Schee, and U.~Wiedemann.
\newblock {Testing the QCD formation time with reconstructed parton
  splittings}.
\newblock 3 2025, arXiv:2503.11764 [hep-ph].

\bibitem[CMS(2023)]{CMS-PAS-HIN-23-004}
{Energy-energy correlators from PbPb and pp collisions at 5.02 TeV,
  CMS-PAS-HIN-23-004}.
\newblock Technical report, CERN, Geneva, 2023.

\bibitem[Viinikainen()]{talk_Jussi}
J.~Viinikainen.
\newblock {Jussi Talk}.
\newblock \emph{``Energy-energy correlators from PbPb and pp collisions at 5.02
  TeV with CMS'' (July 2024), Energy Correlators at the Collider Frontier,
  MITP, Mainz (Germany)}.

\bibitem[Nambrath()]{talk_Anjali}
A.~Nambrath.
\newblock {Anjali Talk}.
\newblock \emph{``Energy-energy correlators of inclusive jets from small to
  large collision systems with the ALICE experiment'' (September 2024), Hard
  Probes 2024}.

\bibitem[Rai()]{talk_Ananya}
A.~Rai.
\newblock {Ananya Talk}.
\newblock \emph{``Probing jet modification in the QGP using N-Point Energy
  Correlators in Pb-Pb collisions with ALICE'' (April 2025), Quark Matter
  2025}.

\bibitem[Andres et~al.(2024{\natexlab{b}})Andres, Dominguez, Holguin, Marquet,
  and Moult]{Andres:2024xvk}
C.~Andres, F.~Dominguez, J.~Holguin, C.~Marquet, and I.~Moult.
\newblock {Simple Scaling Laws for Energy Correlators in Nuclear Matter}.
\newblock 11 2024{\natexlab{b}}, arXiv:2411.15298 [hep-ph].

\bibitem[Barata et~al.(2025{\natexlab{e}})Barata, Kuzmin, Milhano, and
  Sadofyev]{Barata:2024ukm}
J.~Barata, M.~V. Kuzmin, J.~G. Milhano, and A.~V. Sadofyev.
\newblock {Giving wake to energy-energy correlators: Hydrodynamic response on
  the celestial sphere}.
\newblock \emph{Phys. Rev. D}, 112\penalty0 (1):\penalty0 016005,
  2025{\natexlab{e}}.
\newblock \doi{10.1103/q8pr-djlw}, arXiv:2412.03616 [hep-ph].

\bibitem[Chen et~al.(2024{\natexlab{b}})Chen, Xu, Shen, Dai, Zhang, and
  Wang]{Chen:2024quk}
S.-Y. Chen, Z.-X. Xu, K.-M. Shen, W.~Dai, B.-W. Zhang, and E.-K. Wang.
\newblock {Study of the EEC discrimination power on quark and gluon jet
  quenching effects in heavy-ion collisions at $\sqrt{s}=5.02$ TeV}.
\newblock 9 2024{\natexlab{b}}, arXiv:2409.13996 [nucl-th].

\bibitem[Baier et~al.(2001)Baier, Dokshitzer, Mueller, and
  Schiff]{Baier:2001yt}
R.~Baier, Y.~L. Dokshitzer, A.~H. Mueller, and D.~Schiff.
\newblock {Quenching of hadron spectra in media}.
\newblock \emph{JHEP}, 09:\penalty0 033, 2001.
\newblock \doi{10.1088/1126-6708/2001/09/033}, arXiv:hep-ph/0106347.

\bibitem[Brewer et~al.(2019)Brewer, Milhano, and Thaler]{Brewer:2018dfs}
J.~Brewer, J.~G. Milhano, and J.~Thaler.
\newblock {Sorting out quenched jets}.
\newblock \emph{Phys. Rev. Lett.}, 122\penalty0 (22):\penalty0 222301, 2019.
\newblock \doi{10.1103/PhysRevLett.122.222301}, arXiv:1812.05111 [hep-ph].

\bibitem[Chen et~al.(2022{\natexlab{b}})Chen, Moult, Sandor, and
  Zhu]{Chen:2022jhb}
H.~Chen, I.~Moult, J.~Sandor, and H.~X. Zhu.
\newblock {Celestial blocks and transverse spin in the three-point energy
  correlator}.
\newblock \emph{JHEP}, 09:\penalty0 199, 2022{\natexlab{b}}.
\newblock \doi{10.1007/JHEP09(2022)199}, arXiv:2202.04085 [hep-ph].

\bibitem[Chang and Simmons-Duffin(2023)]{Chang:2022ryc}
C.-H. Chang and D.~Simmons-Duffin.
\newblock {Three-point energy correlators and the celestial block expansion}.
\newblock \emph{JHEP}, 02:\penalty0 126, 2023.
\newblock \doi{10.1007/JHEP02(2023)126}, arXiv:2202.04090 [hep-th].

\bibitem[Chen et~al.(2025{\natexlab{a}})Chen, Ruan, and Zhu]{Chen:2025rjc}
H.~Chen, H.~Ruan, and H.~X. Zhu.
\newblock {Energy-Energy Correlator at Hadron Colliders: Celestial Blocks and
  Singularities}.
\newblock 5 2025{\natexlab{a}}, arXiv:2505.16753 [hep-ph].

\bibitem[Arnold(2009)]{Arnold:2008iy}
P.~B. Arnold.
\newblock {Simple Formula for High-Energy Gluon Bremsstrahlung in a Finite,
  Expanding Medium}.
\newblock \emph{Phys. Rev. D}, 79:\penalty0 065025, 2009.
\newblock \doi{10.1103/PhysRevD.79.065025}, arXiv:0808.2767 [hep-ph].

\bibitem[Boguslavski et~al.(2024{\natexlab{b}})Boguslavski, Kurkela, Lappi,
  Lindenbauer, and Peuron]{Boguslavski:2024ezg}
K.~Boguslavski, A.~Kurkela, T.~Lappi, F.~Lindenbauer, and J.~Peuron.
\newblock {Jet momentum broadening during initial stages in heavy-ion
  collisions}.
\newblock \emph{Phys. Lett. B}, 850:\penalty0 138525, 2024{\natexlab{b}}.
\newblock \doi{10.1016/j.physletb.2024.138525}, arXiv:2303.12595 [hep-ph].

\bibitem[Boguslavski et~al.(2024{\natexlab{c}})Boguslavski, Kurkela, Lappi,
  Lindenbauer, and Peuron]{Boguslavski:2024jwr}
K.~Boguslavski, A.~Kurkela, T.~Lappi, F.~Lindenbauer, and J.~Peuron.
\newblock {Jet quenching parameter in QCD kinetic theory}.
\newblock \emph{Phys. Rev. D}, 110\penalty0 (3):\penalty0 034019,
  2024{\natexlab{c}}.
\newblock \doi{10.1103/PhysRevD.110.034019}, arXiv:2312.00447 [hep-ph].

\bibitem[Epelbaum and Gelis(2013)]{Epelbaum:2013waa}
T.~Epelbaum and F.~Gelis.
\newblock {Fluctuations of the initial color fields in high energy heavy ion
  collisions}.
\newblock \emph{Phys. Rev. D}, 88:\penalty0 085015, 2013.
\newblock \doi{10.1103/PhysRevD.88.085015}, arXiv:1307.1765 [hep-ph].

\bibitem[Berges et~al.(2014{\natexlab{c}})Berges, Schenke, Schlichting, and
  Venugopalan]{Berges:2014yta}
J.~Berges, B.~Schenke, S.~Schlichting, and R.~Venugopalan.
\newblock {Turbulent thermalization process in high-energy heavy-ion
  collisions}.
\newblock \emph{Nucl. Phys. A}, 931:\penalty0 348--353, 2014{\natexlab{c}}.
\newblock \doi{10.1016/j.nuclphysa.2014.08.103}, arXiv:1409.1638 [hep-ph].

\bibitem[Romatschke and Venugopalan(2006{\natexlab{a}})]{Romatschke:2005ag}
P.~Romatschke and R.~Venugopalan.
\newblock {A Weibel instability in the melting color glass condensate}.
\newblock \emph{Eur. Phys. J. A}, 29:\penalty0 71--75, 2006{\natexlab{a}}.
\newblock \doi{10.1140/epja/i2005-10304-y}, arXiv:hep-ph/0510292.

\bibitem[Romatschke and Venugopalan(2006{\natexlab{b}})]{Romatschke:2005pm}
P.~Romatschke and R.~Venugopalan.
\newblock {Collective non-Abelian instabilities in a melting color glass
  condensate}.
\newblock \emph{Phys. Rev. Lett.}, 96:\penalty0 062302, 2006{\natexlab{b}}.
\newblock \doi{10.1103/PhysRevLett.96.062302}, arXiv:hep-ph/0510121.

\bibitem[Romatschke and Venugopalan(2006{\natexlab{c}})]{Romatschke:2006nk}
P.~Romatschke and R.~Venugopalan.
\newblock {The Unstable Glasma}.
\newblock \emph{Phys. Rev. D}, 74:\penalty0 045011, 2006{\natexlab{c}}.
\newblock \doi{10.1103/PhysRevD.74.045011}, arXiv:hep-ph/0605045.

\bibitem[Mrowczynski(1993)]{Mrowczynski:1993qm}
S.~Mrowczynski.
\newblock {Plasma instability at the initial stage of ultrarelativistic heavy
  ion collisions}.
\newblock \emph{Phys. Lett. B}, 314:\penalty0 118--121, 1993.
\newblock \doi{10.1016/0370-2693(93)91330-P}.

\bibitem[Zapp(2014{\natexlab{a}})]{Zapp:2013vla}
K.~C. Zapp.
\newblock {JEWEL 2.0.0: directions for use}.
\newblock \emph{Eur. Phys. J. C}, 74\penalty0 (2):\penalty0 2762,
  2014{\natexlab{a}}.
\newblock \doi{10.1140/epjc/s10052-014-2762-1}, arXiv:1311.0048 [hep-ph].

\bibitem[Zapp(2014{\natexlab{b}})]{Zapp:2013zya}
K.~C. Zapp.
\newblock {Geometrical aspects of jet quenching in JEWEL}.
\newblock \emph{Phys. Lett. B}, 735:\penalty0 157--163, 2014{\natexlab{b}}.
\newblock \doi{10.1016/j.physletb.2014.06.020}, arXiv:1312.5536 [hep-ph].

\bibitem[Brewer et~al.(2022)Brewer, Brodsky, and Rajagopal]{Brewer:2021hmh}
J.~Brewer, Q.~Brodsky, and K.~Rajagopal.
\newblock {Disentangling jet modification in jet simulations and in Z+jet
  data}.
\newblock \emph{JHEP}, 02:\penalty0 175, 2022.
\newblock \doi{10.1007/JHEP02(2022)175}, arXiv:2110.13159 [hep-ph].

\bibitem[Andres et~al.(2024{\natexlab{c}})Andres, Holguin,
  Kunnawalkam~Elayavalli, and Viinikainen]{Andres:2024hdd}
C.~Andres, J.~Holguin, R.~Kunnawalkam~Elayavalli, and J.~Viinikainen.
\newblock {Minimizing Selection Bias in Inclusive Jets in Heavy-Ion Collisions
  with Energy Correlators}.
\newblock 9 2024{\natexlab{c}}, arXiv:2409.07514 [hep-ph].

\bibitem[Milhano and Zapp(2016)]{Milhano:2015mng}
J.~G. Milhano and K.~C. Zapp.
\newblock {Origins of the di-jet asymmetry in heavy ion collisions}.
\newblock \emph{Eur. Phys. J. C}, 76\penalty0 (5):\penalty0 288, 2016.
\newblock \doi{10.1140/epjc/s10052-016-4130-9}, arXiv:1512.08107 [hep-ph].

\bibitem[Rajagopal et~al.(2016)Rajagopal, Sadofyev, and van~der
  Schee]{Rajagopal:2016uip}
K.~Rajagopal, A.~V. Sadofyev, and W.~van~der Schee.
\newblock {Evolution of the jet opening angle distribution in holographic
  plasma}.
\newblock \emph{Phys. Rev. Lett.}, 116\penalty0 (21):\penalty0 211603, 2016.
\newblock \doi{10.1103/PhysRevLett.116.211603}, arXiv:1602.04187 [nucl-th].

\bibitem[Brewer et~al.(2018)Brewer, Rajagopal, Sadofyev, and Van
  Der~Schee]{Brewer:2017fqy}
J.~Brewer, K.~Rajagopal, A.~Sadofyev, and W.~Van Der~Schee.
\newblock {Evolution of the Mean Jet Shape and Dijet Asymmetry Distribution of
  an Ensemble of Holographic Jets in Strongly Coupled Plasma}.
\newblock \emph{JHEP}, 02:\penalty0 015, 2018.
\newblock \doi{10.1007/JHEP02(2018)015}, arXiv:1710.03237 [nucl-th].

\bibitem[Casalderrey-Solana et~al.(2017)Casalderrey-Solana, Gulhan, Milhano,
  Pablos, and Rajagopal]{Casalderrey-Solana:2016jvj}
J.~Casalderrey-Solana, D.~Gulhan, G.~Milhano, D.~Pablos, and K.~Rajagopal.
\newblock {Angular Structure of Jet Quenching Within a Hybrid Strong/Weak
  Coupling Model}.
\newblock \emph{JHEP}, 03:\penalty0 135, 2017.
\newblock \doi{10.1007/JHEP03(2017)135}, arXiv:1609.05842 [hep-ph].

\bibitem[He et~al.(2019)He, Pang, and Wang]{He:2018gks}
Y.~He, L.-G. Pang, and X.-N. Wang.
\newblock {Bayesian extraction of jet energy loss distributions in heavy-ion
  collisions}.
\newblock \emph{Phys. Rev. Lett.}, 122\penalty0 (25):\penalty0 252302, 2019.
\newblock \doi{10.1103/PhysRevLett.122.252302}, arXiv:1808.05310 [hep-ph].

\bibitem[Takacs and Tywoniuk(2021)]{Takacs:2021bpv}
A.~Takacs and K.~Tywoniuk.
\newblock {Quenching effects in the cumulative jet spectrum}.
\newblock \emph{JHEP}, 10:\penalty0 038, 2021.
\newblock \doi{10.1007/JHEP10(2021)038}, arXiv:2103.14676 [hep-ph].

\bibitem[Zapp et~al.(2013)Zapp, Krauss, and Wiedemann]{Zapp:2012ak}
K.~C. Zapp, F.~Krauss, and U.~A. Wiedemann.
\newblock {A perturbative framework for jet quenching}.
\newblock \emph{JHEP}, 03:\penalty0 080, 2013.
\newblock \doi{10.1007/JHEP03(2013)080}, arXiv:1212.1599 [hep-ph].

\bibitem[Casalderrey-Solana et~al.(2014)Casalderrey-Solana, Gulhan, Milhano,
  Pablos, and Rajagopal]{Casalderrey-Solana:2014bpa}
J.~Casalderrey-Solana, D.~C. Gulhan, J.~G. Milhano, D.~Pablos, and
  K.~Rajagopal.
\newblock {A Hybrid Strong/Weak Coupling Approach to Jet Quenching}.
\newblock \emph{JHEP}, 10:\penalty0 019, 2014.
\newblock \doi{10.1007/JHEP09(2015)175}, arXiv:1405.3864 [hep-ph].
\newblock [Erratum: JHEP 09, 175 (2015)].

\bibitem[Casalderrey-Solana et~al.(2016{\natexlab{a}})Casalderrey-Solana,
  Gulhan, Milhano, Pablos, and Rajagopal]{Casalderrey-Solana:2015vaa}
J.~Casalderrey-Solana, D.~C. Gulhan, J.~G. Milhano, D.~Pablos, and
  K.~Rajagopal.
\newblock {Predictions for Boson-Jet Observables and Fragmentation Function
  Ratios from a Hybrid Strong/Weak Coupling Model for Jet Quenching}.
\newblock \emph{JHEP}, 03:\penalty0 053, 2016{\natexlab{a}}.
\newblock \doi{10.1007/JHEP03(2016)053}, arXiv:1508.00815 [hep-ph].

\bibitem[Hulcher et~al.(2018)Hulcher, Pablos, and Rajagopal]{Hulcher:2017cpt}
Z.~Hulcher, D.~Pablos, and K.~Rajagopal.
\newblock {Resolution Effects in the Hybrid Strong/Weak Coupling Model}.
\newblock \emph{JHEP}, 03:\penalty0 010, 2018.
\newblock \doi{10.1007/JHEP03(2018)010}, arXiv:1707.05245 [hep-ph].

\bibitem[Casalderrey-Solana et~al.(2019)Casalderrey-Solana, Hulcher, Milhano,
  Pablos, and Rajagopal]{Casalderrey-Solana:2018wrw}
J.~Casalderrey-Solana, Z.~Hulcher, G.~Milhano, D.~Pablos, and K.~Rajagopal.
\newblock {Simultaneous description of hadron and jet suppression in heavy-ion
  collisions}.
\newblock \emph{Phys. Rev. C}, 99\penalty0 (5):\penalty0 051901, 2019.
\newblock \doi{10.1103/PhysRevC.99.051901}, arXiv:1808.07386 [hep-ph].

\bibitem[Milhano and Zapp(2022)]{Milhano:2022kzx}
J.~G. Milhano and K.~Zapp.
\newblock {Improved background subtraction and a fresh look at jet
  sub-structure in JEWEL}.
\newblock \emph{Eur. Phys. J. C}, 82\penalty0 (11):\penalty0 1010, 2022.
\newblock \doi{10.1140/epjc/s10052-022-10954-1}, arXiv:2207.14814 [hep-ph].

\bibitem[Chesler and Rajagopal(2014)]{Chesler:2014jva}
P.~M. Chesler and K.~Rajagopal.
\newblock {Jet quenching in strongly coupled plasma}.
\newblock \emph{Phys. Rev. D}, 90\penalty0 (2):\penalty0 025033, 2014.
\newblock \doi{10.1103/PhysRevD.90.025033}, arXiv:1402.6756 [hep-th].

\bibitem[Chesler and Rajagopal(2016)]{Chesler:2015nqz}
P.~M. Chesler and K.~Rajagopal.
\newblock {On the Evolution of Jet Energy and Opening Angle in Strongly Coupled
  Plasma}.
\newblock \emph{JHEP}, 05:\penalty0 098, 2016.
\newblock \doi{10.1007/JHEP05(2016)098}, arXiv:1511.07567 [hep-th].

\bibitem[Cooper and Frye(1974)]{PhysRevD.10.186}
F.~Cooper and G.~Frye.
\newblock Single-particle distribution in the hydrodynamic and statistical
  thermodynamic models of multiparticle production.
\newblock \emph{Phys. Rev. D}, 10:\penalty0 186--189, Jul 1974.
\newblock \doi{10.1103/PhysRevD.10.186}.
\newblock URL \url{https://link.aps.org/doi/10.1103/PhysRevD.10.186}.

\bibitem[Eskola et~al.(2017{\natexlab{b}})Eskola, Paakkinen, Paukkunen, and
  Salgado]{nPDFs}
K.~J. Eskola, P.~Paakkinen, H.~Paukkunen, and C.~A. Salgado.
\newblock {EPPS16: Nuclear parton distributions with LHC data}.
\newblock \emph{Eur. Phys. J. C}, 77\penalty0 (3):\penalty0 163,
  2017{\natexlab{b}}.
\newblock \doi{10.1140/epjc/s10052-017-4725-9}, arXiv:1612.05741 [hep-ph].

\bibitem[Dulat et~al.(2016)Dulat, Hou, Gao, Guzzi, Huston, Nadolsky, Pumplin,
  Schmidt, Stump, and Yuan]{PDFs}
S.~Dulat, T.-J. Hou, J.~Gao, M.~Guzzi, J.~Huston, P.~Nadolsky, J.~Pumplin,
  C.~Schmidt, D.~Stump, and C.~P. Yuan.
\newblock {New parton distribution functions from a global analysis of quantum
  chromodynamics}.
\newblock \emph{Phys. Rev. D}, 93\penalty0 (3):\penalty0 033006, 2016.
\newblock \doi{10.1103/PhysRevD.93.033006}, arXiv:1506.07443 [hep-ph].

\bibitem[Buckley et~al.(2015)Buckley, Ferrando, Lloyd, Nordstr\"om, Page,
  R\"ufenacht, Sch\"onherr, and Watt]{lhapdf}
A.~Buckley, J.~Ferrando, S.~Lloyd, K.~Nordstr\"om, B.~Page, M.~R\"ufenacht,
  M.~Sch\"onherr, and G.~Watt.
\newblock {LHAPDF6: parton density access in the LHC precision era}.
\newblock \emph{Eur. Phys. J. C}, 75:\penalty0 132, 2015.
\newblock \doi{10.1140/epjc/s10052-015-3318-8}, arXiv:1412.7420 [hep-ph].

\bibitem[Sjostrand et~al.(2006)Sjostrand, Mrenna, and Skands]{Sjostrand:2006za}
T.~Sjostrand, S.~Mrenna, and P.~Z. Skands.
\newblock {PYTHIA 6.4 Physics and Manual}.
\newblock \emph{JHEP}, 05:\penalty0 026, 2006.
\newblock \doi{10.1088/1126-6708/2006/05/026}, arXiv:hep-ph/0603175.

\bibitem[Li and Vitev(2020)]{Li:2019dre}
H.~T. Li and I.~Vitev.
\newblock {Jet charge modification in dense QCD matter}.
\newblock \emph{Phys. Rev. D}, 101:\penalty0 076020, 2020.
\newblock \doi{10.1103/PhysRevD.101.076020}, arXiv:1908.06979 [hep-ph].

\bibitem[Eskola et~al.(2009)Eskola, Paukkunen, and Salgado]{Eskola:2009uj}
K.~J. Eskola, H.~Paukkunen, and C.~A. Salgado.
\newblock {EPS09: A New Generation of NLO and LO Nuclear Parton Distribution
  Functions}.
\newblock \emph{JHEP}, 04:\penalty0 065, 2009.
\newblock \doi{10.1088/1126-6708/2009/04/065}, arXiv:0902.4154 [hep-ph].

\bibitem[Cacciari et~al.(2012)Cacciari, Salam, and Soyez]{fastjet}
M.~Cacciari, G.~P. Salam, and G.~Soyez.
\newblock {FastJet User Manual}.
\newblock \emph{Eur. Phys. J. C}, 72:\penalty0 1896, 2012.
\newblock \doi{10.1140/epjc/s10052-012-1896-2}, arXiv:1111.6097 [hep-ph].

\bibitem[Baier et~al.(1998{\natexlab{b}})Baier, Dokshitzer, Mueller, and
  Schiff]{Baier:1998kq}
R.~Baier, Y.~L. Dokshitzer, A.~H. Mueller, and D.~Schiff.
\newblock {Medium induced radiative energy loss: Equivalence between the BDMPS
  and Zakharov formalisms}.
\newblock \emph{Nucl. Phys. B}, 531:\penalty0 403--425, 1998{\natexlab{b}}.
\newblock \doi{10.1016/S0550-3213(98)00546-X}, arXiv:hep-ph/9804212.

\bibitem[Majumder(2009)]{Majumder:2009zu}
A.~Majumder.
\newblock {The In-medium scale evolution in jet modification}.
\newblock 1 2009, arXiv:0901.4516 [nucl-th].

\bibitem[Arnold et~al.(2001{\natexlab{b}})Arnold, Moore, and
  Yaffe]{Arnold:2001ms}
P.~B. Arnold, G.~D. Moore, and L.~G. Yaffe.
\newblock {Photon emission from quark gluon plasma: Complete leading order
  results}.
\newblock \emph{JHEP}, 12:\penalty0 009, 2001{\natexlab{b}}.
\newblock \doi{10.1088/1126-6708/2001/12/009}, arXiv:hep-ph/0111107.

\bibitem[Wicks et~al.(2007)Wicks, Horowitz, Djordjevic, and
  Gyulassy]{Wicks:2005gt}
S.~Wicks, W.~Horowitz, M.~Djordjevic, and M.~Gyulassy.
\newblock {Elastic, inelastic, and path length fluctuations in jet tomography}.
\newblock \emph{Nucl. Phys. A}, 784:\penalty0 426--442, 2007.
\newblock \doi{10.1016/j.nuclphysa.2006.12.048}, arXiv:nucl-th/0512076.

\bibitem[Dasgupta et~al.(2008)Dasgupta, Magnea, and Salam]{Dasgupta:2007wa}
M.~Dasgupta, L.~Magnea, and G.~P. Salam.
\newblock {Non-perturbative QCD effects in jets at hadron colliders}.
\newblock \emph{JHEP}, 02:\penalty0 055, 2008.
\newblock \doi{10.1088/1126-6708/2008/02/055}, arXiv:0712.3014 [hep-ph].

\bibitem[Dasgupta et~al.(2015)Dasgupta, Dreyer, Salam, and
  Soyez]{Dasgupta:2014yra}
M.~Dasgupta, F.~Dreyer, G.~P. Salam, and G.~Soyez.
\newblock {Small-radius jets to all orders in QCD}.
\newblock \emph{JHEP}, 04:\penalty0 039, 2015.
\newblock \doi{10.1007/JHEP04(2015)039}, arXiv:1411.5182 [hep-ph].

\bibitem[Apolin\'ario et~al.(2020)Apolin\'ario, Barata, and
  Milhano]{Apolinario:2020nyw}
L.~Apolin\'ario, J.~a. Barata, and G.~Milhano.
\newblock {On the breaking of Casimir scaling in jet quenching}.
\newblock \emph{Eur. Phys. J. C}, 80\penalty0 (6):\penalty0 586, 2020.
\newblock \doi{10.1140/epjc/s10052-020-8133-1}, arXiv:2003.02893 [hep-ph].

\bibitem[Brewer et~al.(2021)Brewer, Thaler, and Turner]{Brewer:2020och}
J.~Brewer, J.~Thaler, and A.~P. Turner.
\newblock {Data-driven quark and gluon jet modification in heavy-ion
  collisions}.
\newblock \emph{Phys. Rev. C}, 103\penalty0 (2):\penalty0 L021901, 2021.
\newblock \doi{10.1103/PhysRevC.103.L021901}, arXiv:2008.08596 [hep-ph].

\bibitem[Spousta and Cole(2016)]{Spousta:2015fca}
M.~Spousta and B.~Cole.
\newblock {Interpreting single jet measurements in Pb $+$ Pb collisions at the
  LHC}.
\newblock \emph{Eur. Phys. J. C}, 76\penalty0 (2):\penalty0 50, 2016.
\newblock \doi{10.1140/epjc/s10052-016-3896-0}, arXiv:1504.05169 [hep-ph].

\bibitem[Ogrodnik et~al.(2024)Ogrodnik, Ryb\'ar, and Spousta]{Ogrodnik:2024qug}
A.~Ogrodnik, M.~Ryb\'ar, and M.~Spousta.
\newblock {Flavor and path-length dependence of jet quenching from inclusive
  jet and \ensuremath{\gamma}-jet suppression}.
\newblock 7 2024, arXiv:2407.11234 [hep-ph].

\bibitem[Ringer et~al.(2020)Ringer, Xiao, and Yuan]{Ringer:2019rfk}
F.~Ringer, B.-W. Xiao, and F.~Yuan.
\newblock {Can we observe jet $P_T$-broadening in heavy-ion collisions at the
  LHC?}
\newblock \emph{Phys. Lett. B}, 808:\penalty0 135634, 2020.
\newblock \doi{10.1016/j.physletb.2020.135634}, arXiv:1907.12541 [hep-ph].

\bibitem[Pablos and Soto-Ontoso(2023)]{Pablos:2022mrx}
D.~Pablos and A.~Soto-Ontoso.
\newblock {Pushing forward jet substructure measurements in heavy-ion
  collisions}.
\newblock \emph{Phys. Rev. D}, 107\penalty0 (9):\penalty0 094003, 2023.
\newblock \doi{10.1103/PhysRevD.107.094003}, arXiv:2210.07901 [hep-ph].

\bibitem[Zhang et~al.(2024)Zhang, Wang, Xing, and Zhang]{Zhang:2023oid}
S.-L. Zhang, E.~Wang, H.~Xing, and B.-W. Zhang.
\newblock {Flavor dependence of jet quenching in heavy-ion collisions from a
  Bayesian analysis}.
\newblock \emph{Phys. Lett. B}, 850:\penalty0 138549, 2024.
\newblock \doi{10.1016/j.physletb.2024.138549}, arXiv:2303.14881 [hep-ph].

\bibitem[Acharya et~al.(2020)]{ALICE:2019ykw}
S.~Acharya et~al.
\newblock {Exploration of jet substructure using iterative declustering in pp
  and Pb\textendash{}Pb collisions at LHC energies}.
\newblock \emph{Phys. Lett. B}, 802:\penalty0 135227, 2020.
\newblock \doi{10.1016/j.physletb.2020.135227}, arXiv:1905.02512 [nucl-ex].

\bibitem[Sirunyan et~al.(2020)]{CMS:2020plq}
A.~M. Sirunyan et~al.
\newblock {Measurement of quark- and gluon-like jet fractions using jet charge
  in PbPb and pp collisions at 5.02 TeV}.
\newblock \emph{JHEP}, 07:\penalty0 115, 2020.
\newblock \doi{10.1007/JHEP07(2020)115}, arXiv:2004.00602 [hep-ex].

\bibitem[Larkoski et~al.(2014)Larkoski, Marzani, Soyez, and
  Thaler]{Larkoski:2014wba}
A.~J. Larkoski, S.~Marzani, G.~Soyez, and J.~Thaler.
\newblock {Soft Drop}.
\newblock \emph{JHEP}, 05:\penalty0 146, 2014.
\newblock \doi{10.1007/JHEP05(2014)146}, arXiv:1402.2657 [hep-ph].

\bibitem[Mehtar-Tani et~al.(2020)Mehtar-Tani, Soto-Ontoso, and
  Tywoniuk]{Mehtar-Tani:2019rrk}
Y.~Mehtar-Tani, A.~Soto-Ontoso, and K.~Tywoniuk.
\newblock {Dynamical grooming of QCD jets}.
\newblock \emph{Phys. Rev. D}, 101\penalty0 (3):\penalty0 034004, 2020.
\newblock \doi{10.1103/PhysRevD.101.034004}, arXiv:1911.00375 [hep-ph].

\bibitem[Kogler et~al.(2019)]{Asquith:2018igt}
R.~Kogler et~al.
\newblock {Jet Substructure at the Large Hadron Collider: Experimental Review}.
\newblock \emph{Rev. Mod. Phys.}, 91\penalty0 (4):\penalty0 045003, 2019.
\newblock \doi{10.1103/RevModPhys.91.045003}, arXiv:1803.06991 [hep-ex].

\bibitem[Cao et~al.(2024)Cao, Majumder, Modarresi-Yazdi, Soudi, and
  Tachibana]{Cao:2024pxc}
S.~Cao, A.~Majumder, R.~Modarresi-Yazdi, I.~Soudi, and Y.~Tachibana.
\newblock {Jet quenching: From theory to simulation}.
\newblock \emph{Int. J. Mod. Phys. E}, 33\penalty0 (08):\penalty0 2430002,
  2024.
\newblock \doi{10.1142/9789811294679_0001}, arXiv:2401.10026 [hep-ph].

\bibitem[Adolfsson et~al.(2020)]{Adolfsson:2020dhm}
J.~Adolfsson et~al.
\newblock {QCD challenges from pp to A{\textendash}A collisions}.
\newblock \emph{Eur. Phys. J. A}, 56\penalty0 (11):\penalty0 288, 2020.
\newblock \doi{10.1140/epja/s10050-020-00270-1}, arXiv:2003.10997 [hep-ph].

\bibitem[Cunqueiro and Sickles(2022)]{Cunqueiro:2021wls}
L.~Cunqueiro and A.~M. Sickles.
\newblock {Studying the QGP with Jets at the LHC and RHIC}.
\newblock \emph{Prog. Part. Nucl. Phys.}, 124:\penalty0 103940, 2022.
\newblock \doi{10.1016/j.ppnp.2022.103940}, arXiv:2110.14490 [nucl-ex].

\bibitem[Casalderrey-Solana and Salgado(2007)]{Casalderrey-Solana:2007knd}
J.~Casalderrey-Solana and C.~A. Salgado.
\newblock {Introductory lectures on jet quenching in heavy ion collisions}.
\newblock \emph{Acta Phys. Polon. B}, 38:\penalty0 3731--3794, 2007,
  arXiv:0712.3443 [hep-ph].

\bibitem[Mehtar-Tani et~al.(2013)Mehtar-Tani, Milhano, and
  Tywoniuk]{Mehtar-Tani:2013pia}
Y.~Mehtar-Tani, J.~G. Milhano, and K.~Tywoniuk.
\newblock {Jet physics in heavy-ion collisions}.
\newblock \emph{Int. J. Mod. Phys. A}, 28:\penalty0 1340013, 2013.
\newblock \doi{10.1142/S0217751X13400137}, arXiv:1302.2579 [hep-ph].

\bibitem[Salgado and Wiedemann(2003)]{Salgado:2003gb}
C.~A. Salgado and U.~A. Wiedemann.
\newblock {Calculating quenching weights}.
\newblock \emph{Phys. Rev. D}, 68:\penalty0 014008, 2003.
\newblock \doi{10.1103/PhysRevD.68.014008}, arXiv:hep-ph/0302184.

\bibitem[Caucal et~al.(2022)Caucal, Soto-Ontoso, and Takacs]{Caucal:2021cfb}
P.~Caucal, A.~Soto-Ontoso, and A.~Takacs.
\newblock {Dynamically groomed jet radius in heavy-ion collisions}.
\newblock \emph{Phys. Rev. D}, 105\penalty0 (11):\penalty0 114046, 2022.
\newblock \doi{10.1103/PhysRevD.105.114046}, arXiv:2111.14768 [hep-ph].

\bibitem[Mehtar-Tani et~al.(2021)Mehtar-Tani, Pablos, and
  Tywoniuk]{Mehtar-Tani:2021fud}
Y.~Mehtar-Tani, D.~Pablos, and K.~Tywoniuk.
\newblock {Cone-Size Dependence of Jet Suppression in Heavy-Ion Collisions}.
\newblock \emph{Phys. Rev. Lett.}, 127\penalty0 (25):\penalty0 252301, 2021.
\newblock \doi{10.1103/PhysRevLett.127.252301}, arXiv:2101.01742 [hep-ph].

\bibitem[Mehtar-Tani and Tywoniuk(2018{\natexlab{a}})]{Mehtar-Tani:2017ypq}
Y.~Mehtar-Tani and K.~Tywoniuk.
\newblock {Radiative energy loss of neighboring subjets}.
\newblock \emph{Nucl. Phys. A}, 979:\penalty0 165--203, 2018{\natexlab{a}}.
\newblock \doi{10.1016/j.nuclphysa.2018.09.041}, arXiv:1706.06047 [hep-ph].

\bibitem[Chang et~al.(2013{\natexlab{a}})Chang, Procura, Thaler, and
  Waalewijn]{Chang:2013rca}
H.-M. Chang, M.~Procura, J.~Thaler, and W.~J. Waalewijn.
\newblock {Calculating Track-Based Observables for the LHC}.
\newblock \emph{Phys. Rev. Lett.}, 111:\penalty0 102002, 2013{\natexlab{a}}.
\newblock \doi{10.1103/PhysRevLett.111.102002}, arXiv:1303.6637 [hep-ph].

\bibitem[Chang et~al.(2013{\natexlab{b}})Chang, Procura, Thaler, and
  Waalewijn]{Chang:2013iba}
H.-M. Chang, M.~Procura, J.~Thaler, and W.~J. Waalewijn.
\newblock {Calculating Track Thrust with Track Functions}.
\newblock \emph{Phys. Rev. D}, 88:\penalty0 034030, 2013{\natexlab{b}}.
\newblock \doi{10.1103/PhysRevD.88.034030}, arXiv:1306.6630 [hep-ph].

\bibitem[Lee et~al.(2024{\natexlab{b}})Lee, Moult, Ringer, and
  Waalewijn]{Lee:2023xzv}
K.~Lee, I.~Moult, F.~Ringer, and W.~J. Waalewijn.
\newblock {A formalism for extracting track functions from jet measurements}.
\newblock \emph{JHEP}, 01:\penalty0 194, 2024{\natexlab{b}}.
\newblock \doi{10.1007/JHEP01(2024)194}, arXiv:2308.00028 [hep-ph].

\bibitem[Jaarsma et~al.(2022)Jaarsma, Li, Moult, Waalewijn, and
  Zhu]{Jaarsma:2022kdd}
M.~Jaarsma, Y.~Li, I.~Moult, W.~Waalewijn, and H.~X. Zhu.
\newblock {Renormalization group flows for track function moments}.
\newblock \emph{JHEP}, 06:\penalty0 139, 2022.
\newblock \doi{10.1007/JHEP06(2022)139}, arXiv:2201.05166 [hep-ph].

\bibitem[Li et~al.(2022{\natexlab{b}})Li, Moult, van Velzen, Waalewijn, and
  Zhu]{Li:2021zcf}
Y.~Li, I.~Moult, S.~S. van Velzen, W.~J. Waalewijn, and H.~X. Zhu.
\newblock {Extending Precision Perturbative QCD with Track Functions}.
\newblock \emph{Phys. Rev. Lett.}, 128\penalty0 (18):\penalty0 182001,
  2022{\natexlab{b}}.
\newblock \doi{10.1103/PhysRevLett.128.182001}, arXiv:2108.01674 [hep-ph].

\bibitem[Barata and Szafron(2024)]{Barata:2024nqo}
J.~Barata and R.~Szafron.
\newblock {Leading order track functions in a hot and dense QGP}.
\newblock \emph{Phys. Rev. D}, 110\penalty0 (3):\penalty0 L031501, 2024.
\newblock \doi{10.1103/PhysRevD.110.L031501}, arXiv:2401.04164 [hep-ph].

\bibitem[Barata et~al.(2024{\natexlab{f}})Barata, Milhano, and
  Sadofyev]{Barata:2023zqg}
J.~Barata, J.~G. Milhano, and A.~V. Sadofyev.
\newblock {Picturing QCD jets in anisotropic matter: from jet shapes to energy
  energy correlators}.
\newblock \emph{Eur. Phys. J. C}, 84\penalty0 (2):\penalty0 174,
  2024{\natexlab{f}}.
\newblock \doi{10.1140/epjc/s10052-024-12514-1}, arXiv:2308.01294 [hep-ph].

\bibitem[Andres et~al.(2025{\natexlab{a}})Andres, Dominguez, Holguin, Marquet,
  and Moult]{Andres:2024ksi}
C.~Andres, F.~Dominguez, J.~Holguin, C.~Marquet, and I.~Moult.
\newblock {Towards an interpretation of the first measurements of energy
  correlators in the quark-gluon plasma}.
\newblock \emph{JHEP}, 03:\penalty0 166, 2025{\natexlab{a}}.
\newblock \doi{10.1007/JHEP03(2025)166}, arXiv:2407.07936 [hep-ph].

\bibitem[Chen et~al.(2023)Chen, Jaarsma, Li, Moult, Waalewijn, and
  Zhu]{Chen:2022pdu}
H.~Chen, M.~Jaarsma, Y.~Li, I.~Moult, W.~J. Waalewijn, and H.~X. Zhu.
\newblock {Multi-collinear splitting kernels for track function evolution}.
\newblock \emph{JHEP}, 07:\penalty0 185, 2023.
\newblock \doi{10.1007/JHEP07(2023)185}, arXiv:2210.10058 [hep-ph].

\bibitem[Chen et~al.(2025{\natexlab{b}})Chen, Jaarsma, Li, Moult, Waalewijn,
  and Zhu]{Chen:2022muj}
H.~Chen, M.~Jaarsma, Y.~Li, I.~Moult, W.~J. Waalewijn, and H.~X. Zhu.
\newblock {Collinear parton dynamics beyond
  Dokshitzer-Gribov-Lipatov-Altarelli-Parisi framework}.
\newblock \emph{Phys. Rev. D}, 111\penalty0 (7):\penalty0 076021,
  2025{\natexlab{b}}.
\newblock \doi{10.1103/PhysRevD.111.076021}, arXiv:2210.10061 [hep-ph].

\bibitem[{ATLAS Collaboration}(2024{\natexlab{b}})]{ATLAS:2024jrp}
{ATLAS Collaboration}.
\newblock {Measurement of Track Functions in ATLAS Run 2 Data},
  2024{\natexlab{b}}.
\newblock {{ATLAS-CONF-2024-012}}.

\bibitem[{CMS Collaboration}(2024)]{CMS:2024ovv}
{CMS Collaboration}.
\newblock {Energy-energy correlators from PbPb and pp collisions at 5.02 TeV},
  2024.
\newblock {CMS-PAS-HIN-23-004}.

\bibitem[Waalewijn(2012)]{Waalewijn:2012sv}
W.~J. Waalewijn.
\newblock {Calculating the Charge of a Jet}.
\newblock \emph{Phys. Rev. D}, 86:\penalty0 094030, 2012.
\newblock \doi{10.1103/PhysRevD.86.094030}, arXiv:1209.3019 [hep-ph].

\bibitem[Mehtar-Tani and Tywoniuk(2018{\natexlab{b}})]{Mehtar-Tani:2017web}
Y.~Mehtar-Tani and K.~Tywoniuk.
\newblock {Sudakov suppression of jets in QCD media}.
\newblock \emph{Phys. Rev. D}, 98\penalty0 (5):\penalty0 051501,
  2018{\natexlab{b}}.
\newblock \doi{10.1103/PhysRevD.98.051501}, arXiv:1707.07361 [hep-ph].

\bibitem[Casalderrey-Solana and
  Iancu(2011{\natexlab{b}})]{Casalderrey-Solana:2011ule}
J.~Casalderrey-Solana and E.~Iancu.
\newblock {Interference effects in medium-induced gluon radiation}.
\newblock \emph{JHEP}, 08:\penalty0 015, 2011{\natexlab{b}}.
\newblock \doi{10.1007/JHEP08(2011)015}, arXiv:1105.1760 [hep-ph].

\bibitem[Mehtar-Tani et~al.(2012{\natexlab{d}})Mehtar-Tani, Salgado, and
  Tywoniuk]{Mehtar-Tani:2012mfa}
Y.~Mehtar-Tani, C.~A. Salgado, and K.~Tywoniuk.
\newblock {The Radiation pattern of a QCD antenna in a dense medium}.
\newblock \emph{JHEP}, 10:\penalty0 197, 2012{\natexlab{d}}.
\newblock \doi{10.1007/JHEP10(2012)197}, arXiv:1205.5739 [hep-ph].

\bibitem[Fickinger et~al.(2016)Fickinger, Fleming, Kim, and
  Mereghetti]{Fickinger:2016rfd}
M.~Fickinger, S.~Fleming, C.~Kim, and E.~Mereghetti.
\newblock {Effective field theory approach to heavy quark fragmentation}.
\newblock \emph{JHEP}, 11:\penalty0 095, 2016.
\newblock \doi{10.1007/JHEP11(2016)095}, arXiv:1606.07737 [hep-ph].

\bibitem[Beisert(2004)]{Beisert:2003jj}
N.~Beisert.
\newblock {The complete one loop dilatation operator of N=4 superYang-Mills
  theory}.
\newblock \emph{Nucl. Phys. B}, 676:\penalty0 3--42, 2004.
\newblock \doi{10.1016/j.nuclphysb.2003.10.019}, arXiv:hep-th/0307015.

\bibitem[Bigi et~al.(1994)Bigi, Shifman, Uraltsev, and Vainshtein]{Bigi:1993ex}
I.~I.~Y. Bigi, M.~A. Shifman, N.~G. Uraltsev, and A.~I. Vainshtein.
\newblock {On the motion of heavy quarks inside hadrons: Universal
  distributions and inclusive decays}.
\newblock \emph{Int. J. Mod. Phys. A}, 9:\penalty0 2467--2504, 1994.
\newblock \doi{10.1142/S0217751X94000996}, arXiv:hep-ph/9312359.

\bibitem[Benzke et~al.(2010)Benzke, Lee, Neubert, and Paz]{Benzke:2010js}
M.~Benzke, S.~J. Lee, M.~Neubert, and G.~Paz.
\newblock {Factorization at Subleading Power and Irreducible Uncertainties in
  $\bar B\to X_s\gamma$ Decay}.
\newblock \emph{JHEP}, 08:\penalty0 099, 2010.
\newblock \doi{10.1007/JHEP08(2010)099}, arXiv:1003.5012 [hep-ph].

\bibitem[Beneke et~al.(2023)Beneke, Finauri, Vos, and Wei]{Beneke:2023nmj}
M.~Beneke, G.~Finauri, K.~K. Vos, and Y.~Wei.
\newblock {QCD light-cone distribution amplitudes of heavy mesons from boosted
  HQET}.
\newblock \emph{JHEP}, 09:\penalty0 066, 2023.
\newblock \doi{10.1007/JHEP09(2023)066}, arXiv:2305.06401 [hep-ph].

\bibitem[Neubert(1994)]{Neubert:1993um}
M.~Neubert.
\newblock {Analysis of the photon spectrum in inclusive B ---\ensuremath{>}
  X(s) gamma decays}.
\newblock \emph{Phys. Rev. D}, 49:\penalty0 4623--4633, 1994.
\newblock \doi{10.1103/PhysRevD.49.4623}, arXiv:hep-ph/9312311.

\bibitem[Kang et~al.(2017)Kang, Ringer, and Vitev]{Kang:2017frl}
Z.-B. Kang, F.~Ringer, and I.~Vitev.
\newblock {Inclusive production of small radius jets in heavy-ion collisions}.
\newblock \emph{Phys. Lett. B}, 769:\penalty0 242--248, 2017.
\newblock \doi{10.1016/j.physletb.2017.03.067}, arXiv:1701.05839 [hep-ph].

\bibitem[Basham et~al.(1979{\natexlab{b}})Basham, Brown, Ellis, and
  Love]{Basham:1979gh}
C.~L. Basham, L.~S. Brown, S.~D. Ellis, and S.~T. Love.
\newblock {Energy Correlations in Perturbative Quantum Chromodynamics: A
  Conjecture for All Orders}.
\newblock \emph{Phys. Lett. B}, 85:\penalty0 297--299, 1979{\natexlab{b}}.
\newblock \doi{10.1016/0370-2693(79)90601-4}.

\bibitem[Basham et~al.(1978{\natexlab{b}})Basham, Brown, Ellis, and
  Love]{Basham:1977iq}
C.~L. Basham, L.~S. Brown, S.~D. Ellis, and S.~T. Love.
\newblock {Electron - Positron Annihilation Energy Pattern in Quantum
  Chromodynamics: Asymptotically Free Perturbation Theory}.
\newblock \emph{Phys. Rev. D}, 17:\penalty0 2298, 1978{\natexlab{b}}.
\newblock \doi{10.1103/PhysRevD.17.2298}.

\bibitem[Hayrapetyan et~al.(2024)]{CMS:2024mlf}
A.~Hayrapetyan et~al.
\newblock {Measurement of Energy Correlators inside Jets and Determination of
  the Strong Coupling {\ensuremath{\alpha}}S(mZ)}.
\newblock \emph{Phys. Rev. Lett.}, 133\penalty0 (7):\penalty0 071903, 2024.
\newblock \doi{10.1103/PhysRevLett.133.071903}, arXiv:2402.13864 [hep-ex].

\bibitem[Tamis(2024)]{Tamis:2023guc}
A.~Tamis.
\newblock {Measurement of Two-Point Energy Correlators Within1 Jets in
  \ensuremath{\boldsymbol{\mathit{p}}} \ensuremath{\boldsymbol{\mathit{p}}}
  Collisions at $\sqrt{s}$ = 200 GeV at STAR}.
\newblock \emph{PoS}, HardProbes2023:\penalty0 175, 2024.
\newblock \doi{10.22323/1.438.0175}, arXiv:2309.05761 [hep-ex].

\bibitem[Acharya et~al.(2024{\natexlab{b}})]{ALICE:2024dfl}
S.~Acharya et~al.
\newblock {Exposing the parton-hadron transition within jets with energy-energy
  correlators in pp collisions at $\sqrt{\textit s}=5.02$ TeV}.
\newblock 9 2024{\natexlab{b}}, arXiv:2409.12687 [hep-ex].

\bibitem[Lee et~al.(2025)Lee, Me{\c{c}}aj, and Moult]{Lee:2022uwt}
K.~Lee, B.~Me{\c{c}}aj, and I.~Moult.
\newblock {Conformal collider physics meets LHC data}.
\newblock \emph{Phys. Rev. D}, 111\penalty0 (1):\penalty0 L011502, 2025.
\newblock \doi{10.1103/PhysRevD.111.L011502}, arXiv:2205.03414 [hep-ph].

\bibitem[Bassetto et~al.(1983)Bassetto, Ciafaloni, and
  Marchesini]{Bassetto:1983mvz}
A.~Bassetto, M.~Ciafaloni, and G.~Marchesini.
\newblock {Jet Structure and Infrared Sensitive Quantities in Perturbative
  QCD}.
\newblock \emph{Phys. Rept.}, 100:\penalty0 201--272, 1983.
\newblock \doi{10.1016/0370-1573(83)90083-2}.

\bibitem[Dokshitzer et~al.(1992)Dokshitzer, Marchesini, and
  Oriani]{Dokshitzer:1992ip}
Y.~L. Dokshitzer, G.~Marchesini, and G.~Oriani.
\newblock {Measuring color flows in hard processes: Beyond leading order}.
\newblock \emph{Nucl. Phys. B}, 387:\penalty0 675--714, 1992.
\newblock \doi{10.1016/0550-3213(92)90211-S}.

\bibitem[Campbell and Glover(1998)]{Campbell:1997hg}
J.~M. Campbell and E.~W.~N. Glover.
\newblock {Double unresolved approximations to multiparton scattering
  amplitudes}.
\newblock \emph{Nucl. Phys. B}, 527:\penalty0 264--288, 1998.
\newblock \doi{10.1016/S0550-3213(98)00295-8}, arXiv:hep-ph/9710255.

\bibitem[Campbell et~al.(2024)]{Campbell:2022qmc}
J.~M. Campbell et~al.
\newblock {Event generators for high-energy physics experiments}.
\newblock \emph{SciPost Phys.}, 16\penalty0 (5):\penalty0 130, 2024.
\newblock \doi{10.21468/SciPostPhys.16.5.130}, arXiv:2203.11110 [hep-ph].

\bibitem[Dokshitzer et~al.(1982)Dokshitzer, Fadin, and
  Khoze]{Dokshitzer:1982fh}
Y.~L. Dokshitzer, V.~S. Fadin, and V.~A. Khoze.
\newblock {Coherent Effects in the Perturbative QCD Parton Jets}.
\newblock \emph{Phys. Lett. B}, 115:\penalty0 242--246, 1982.
\newblock \doi{10.1016/0370-2693(82)90654-2}.

\bibitem[Webber(1984)]{Webber:1983if}
B.~R. Webber.
\newblock {A QCD Model for Jet Fragmentation Including Soft Gluon
  Interference}.
\newblock \emph{Nucl. Phys. B}, 238:\penalty0 492--528, 1984.
\newblock \doi{10.1016/0550-3213(84)90333-X}.

\bibitem[Dokshitzer et~al.(1988)Dokshitzer, Khoze, Troian, and
  Mueller]{Dokshitzer:1987nm}
Y.~L. Dokshitzer, V.~A. Khoze, S.~I. Troian, and A.~H. Mueller.
\newblock {QCD Coherence in High-Energy Reactions}.
\newblock \emph{Rev. Mod. Phys.}, 60:\penalty0 373, 1988.
\newblock \doi{10.1103/RevModPhys.60.373}.

\bibitem[Nagy and Soper(2019)]{Nagy:2019pjp}
Z.~Nagy and D.~E. Soper.
\newblock {Parton showers with more exact color evolution}.
\newblock \emph{Phys. Rev. D}, 99\penalty0 (5):\penalty0 054009, 2019.
\newblock \doi{10.1103/PhysRevD.99.054009}, arXiv:1902.02105 [hep-ph].

\bibitem[Hamilton et~al.(2020)Hamilton, Medves, Salam, Scyboz, and
  Soyez]{Hamilton:2020rcu}
K.~Hamilton, R.~Medves, G.~P. Salam, L.~Scyboz, and G.~Soyez.
\newblock {Colour and logarithmic accuracy in final-state parton showers}.
\newblock 11 2020.
\newblock \doi{10.1007/JHEP03(2021)041}, arXiv:2011.10054 [hep-ph].

\bibitem[Forshaw et~al.(2025)Forshaw, Pl{\"a}tzer, and
  Gonz{\'a}lez]{Forshaw:2025fif}
J.~R. Forshaw, S.~Pl{\"a}tzer, and F.~T. Gonz{\'a}lez.
\newblock {Fully differential soft gluon evolution at the amplitude level}.
\newblock \emph{JHEP}, 09:\penalty0 207, 2025.
\newblock \doi{10.1007/JHEP09(2025)207}, arXiv:2505.13183 [hep-ph].

\bibitem[Efremov and Teryaev(1982)]{Efremov:1981sh}
A.~Efremov and O.~Teryaev.
\newblock {On Spin Effects in Quantum Chromodynamics}.
\newblock \emph{Sov. J. Nucl. Phys.}, 36:\penalty0 140, 1982.

\bibitem[Shatz(1983)]{Shatz:1983hv}
M.~P. Shatz.
\newblock {Leading Log Approximation for Parton Showers}.
\newblock \emph{Nucl. Phys. B}, 224:\penalty0 218--228, 1983.
\newblock \doi{10.1016/0550-3213(83)90002-0}.

\bibitem[Webber(1987)]{Webber:1987uy}
B.~R. Webber.
\newblock {Quantum Correlations in {QCD} Jets}.
\newblock \emph{Phys. Lett. B}, 193:\penalty0 91--94, 1987.
\newblock \doi{10.1016/0370-2693(87)90461-8}.

\bibitem[Collins(1988)]{Collins:1987cp}
J.~C. Collins.
\newblock {Spin Correlations in Monte Carlo Event Generators}.
\newblock \emph{Nucl. Phys. B}, 304:\penalty0 794--804, 1988.
\newblock \doi{10.1016/0550-3213(88)90654-2}.

\bibitem[Knowles(1988{\natexlab{a}})]{Knowles:1987cu}
I.~Knowles.
\newblock {Angular Correlations in \{QCD\}}.
\newblock \emph{Nucl. Phys. B}, 304:\penalty0 767--793, 1988{\natexlab{a}}.
\newblock \doi{10.1016/0550-3213(88)90653-0}.

\bibitem[Knowles(1988{\natexlab{b}})]{Knowles:1988vs}
I.~Knowles.
\newblock {Spin Correlations in Parton - Parton Scattering}.
\newblock \emph{Nucl. Phys. B}, 310:\penalty0 571--588, 1988{\natexlab{b}}.
\newblock \doi{10.1016/0550-3213(88)90092-2}.

\bibitem[Knowles(1990)]{Knowles:1988hu}
I.~Knowles.
\newblock {A Linear Algorithm for Calculating Spin Correlations in Hadronic
  Collisions}.
\newblock \emph{Comput. Phys. Commun.}, 58:\penalty0 271--284, 1990.
\newblock \doi{10.1016/0010-4655(90)90063-7}.

\bibitem[Richardson and Webster(2020)]{Richardson:2018pvo}
P.~Richardson and S.~Webster.
\newblock {Spin Correlations in Parton Shower Simulations}.
\newblock \emph{Eur. Phys. J. C}, 80\penalty0 (2):\penalty0 83, 2020.
\newblock \doi{10.1140/epjc/s10052-019-7429-5}, arXiv:1807.01955 [hep-ph].

\bibitem[Forshaw et~al.(2019)Forshaw, Holguin, and
  Pl{\"a}tzer]{Forshaw:2019ver}
J.~R. Forshaw, J.~Holguin, and S.~Pl{\"a}tzer.
\newblock {Parton branching at amplitude level}.
\newblock \emph{JHEP}, 08:\penalty0 145, 2019.
\newblock \doi{10.1007/JHEP08(2019)145}, arXiv:1905.08686 [hep-ph].

\bibitem[Corcella et~al.(2001)Corcella, Knowles, Marchesini, Moretti, Odagiri,
  Richardson, Seymour, and Webber]{Corcella:2000bw}
G.~Corcella, I.~G. Knowles, G.~Marchesini, S.~Moretti, K.~Odagiri,
  P.~Richardson, M.~H. Seymour, and B.~R. Webber.
\newblock {HERWIG 6: An Event generator for hadron emission reactions with
  interfering gluons (including supersymmetric processes)}.
\newblock \emph{JHEP}, 01:\penalty0 010, 2001.
\newblock \doi{10.1088/1126-6708/2001/01/010}, arXiv:hep-ph/0011363.

\bibitem[Karlberg et~al.(2021)Karlberg, Salam, Scyboz, and
  Verheyen]{Karlberg:2021kwr}
A.~Karlberg, G.~P. Salam, L.~Scyboz, and R.~Verheyen.
\newblock {Spin correlations in final-state parton showers and jet
  observables}.
\newblock \emph{Eur. Phys. J. C}, 81\penalty0 (8):\penalty0 681, 2021.
\newblock \doi{10.1140/epjc/s10052-021-09378-0}, arXiv:2103.16526 [hep-ph].

\bibitem[Hamilton et~al.(2022)Hamilton, Karlberg, Salam, Scyboz, and
  Verheyen]{Hamilton:2021dyz}
K.~Hamilton, A.~Karlberg, G.~P. Salam, L.~Scyboz, and R.~Verheyen.
\newblock {Soft spin correlations in final-state parton showers}.
\newblock \emph{JHEP}, 03:\penalty0 193, 2022.
\newblock \doi{10.1007/JHEP03(2022)193}, arXiv:2111.01161 [hep-ph].

\bibitem[Bewick et~al.(2024)]{Bewick:2023tfi}
G.~Bewick et~al.
\newblock {Herwig 7.3 release note}.
\newblock \emph{Eur. Phys. J. C}, 84\penalty0 (10):\penalty0 1053, 2024.
\newblock \doi{10.1140/epjc/s10052-024-13211-9}, arXiv:2312.05175 [hep-ph].

\bibitem[H{\"o}che et~al.(2025)H{\"o}che, Hoppe, and Reichelt]{Hoche:2025anb}
S.~H{\"o}che, M.~Hoppe, and D.~Reichelt.
\newblock {A simple algorithm for polarized parton evolution}.
\newblock 8 2025, arXiv:2508.19018 [hep-ph].

\bibitem[Keller and Laenen(1999)]{Keller:1998tf}
S.~Keller and E.~Laenen.
\newblock {Next-to-leading order cross-sections for tagged reactions}.
\newblock \emph{Phys. Rev. D}, 59:\penalty0 114004, 1999.
\newblock \doi{10.1103/PhysRevD.59.114004}, arXiv:hep-ph/9812415.

\bibitem[Catani et~al.(2001)Catani, Dittmaier, and Trocsanyi]{Catani:2000ef}
S.~Catani, S.~Dittmaier, and Z.~Trocsanyi.
\newblock {One loop singular behavior of QCD and SUSY QCD amplitudes with
  massive partons}.
\newblock \emph{Phys. Lett. B}, 500:\penalty0 149--160, 2001.
\newblock \doi{10.1016/S0370-2693(01)00065-X}, arXiv:hep-ph/0011222.

\bibitem[Craft et~al.(2024)Craft, Gonzalez, Lee, Mecaj, and
  Moult]{Craft:2023aew}
E.~Craft, M.~Gonzalez, K.~Lee, B.~Mecaj, and I.~Moult.
\newblock {The 1 {\textrightarrow} 3 massive splitting functions from QCD
  factorization and SCET}.
\newblock \emph{JHEP}, 07:\penalty0 080, 2024.
\newblock \doi{10.1007/JHEP07(2024)080}, arXiv:2310.06736 [hep-ph].

\bibitem[Chen et~al.(2021)Chen, Moult, and Zhu]{Chen:2020adz}
H.~Chen, I.~Moult, and H.~X. Zhu.
\newblock {Quantum Interference in Jet Substructure from Spinning Gluons}.
\newblock \emph{Phys. Rev. Lett.}, 126\penalty0 (11):\penalty0 112003, 2021.
\newblock \doi{10.1103/PhysRevLett.126.112003}, arXiv:2011.02492 [hep-ph].

\bibitem[Song et~al.(2025)Song, Wei, Yang, and Zhou]{Song:2025bdj}
Y.-K. Song, S.-Y. Wei, L.~Yang, and J.~Zhou.
\newblock {Gluon Polarimetry with Energy-Energy Correlators}.
\newblock 9 2025, arXiv:2509.14960 [hep-ph].

\bibitem[Mehtar-Tani et~al.(2011{\natexlab{b}})Mehtar-Tani, Salgado, and
  Tywoniuk]{Mehtar-Tani:2010ebp}
Y.~Mehtar-Tani, C.~A. Salgado, and K.~Tywoniuk.
\newblock {Anti-angular ordering of gluon radiation in QCD media}.
\newblock \emph{Phys. Rev. Lett.}, 106:\penalty0 122002, 2011{\natexlab{b}}.
\newblock \doi{10.1103/PhysRevLett.106.122002}, arXiv:1009.2965 [hep-ph].

\bibitem[Mehtar-Tani et~al.(2012{\natexlab{e}})Mehtar-Tani, Salgado, and
  Tywoniuk]{Mehtar-Tani:2011hma}
Y.~Mehtar-Tani, C.~A. Salgado, and K.~Tywoniuk.
\newblock {Jets in QCD Media: From Color Coherence to Decoherence}.
\newblock \emph{Phys. Lett. B}, 707:\penalty0 156--159, 2012{\natexlab{e}}.
\newblock \doi{10.1016/j.physletb.2011.12.042}, arXiv:1102.4317 [hep-ph].

\bibitem[Mehtar-Tani and Tywoniuk(2013{\natexlab{b}})]{Mehtar-Tani:2011vlz}
Y.~Mehtar-Tani and K.~Tywoniuk.
\newblock {Jet coherence in QCD media: the antenna radiation spectrum}.
\newblock \emph{JHEP}, 01:\penalty0 031, 2013{\natexlab{b}}.
\newblock \doi{10.1007/JHEP01(2013)031}, arXiv:1105.1346 [hep-ph].

\bibitem[Mehtar-Tani et~al.(2012{\natexlab{f}})Mehtar-Tani, Salgado, and
  Tywoniuk]{Mehtar-Tani:2011lic}
Y.~Mehtar-Tani, C.~A. Salgado, and K.~Tywoniuk.
\newblock {The radiation pattern of a QCD antenna in a dilute medium}.
\newblock \emph{JHEP}, 04:\penalty0 064, 2012{\natexlab{f}}.
\newblock \doi{10.1007/JHEP04(2012)064}, arXiv:1112.5031 [hep-ph].

\bibitem[Armesto et~al.(2012{\natexlab{b}})Armesto, Ma, Mehtar-Tani, Salgado,
  and Tywoniuk]{Armesto:2011ir}
N.~Armesto, H.~Ma, Y.~Mehtar-Tani, C.~A. Salgado, and K.~Tywoniuk.
\newblock {Coherence effects and broadening in medium-induced QCD radiation off
  a massive $q {\bar q}$ antenna}.
\newblock \emph{JHEP}, 01:\penalty0 109, 2012{\natexlab{b}}.
\newblock \doi{10.1007/JHEP01(2012)109}, arXiv:1110.4343 [hep-ph].

\bibitem[Calvo et~al.(2014)Calvo, Moldes, and Salgado]{Calvo:2014cba}
M.~R. Calvo, M.~R. Moldes, and C.~A. Salgado.
\newblock {Color coherence in a heavy quark antenna radiating gluons inside a
  QCD medium}.
\newblock \emph{Phys. Lett. B}, 738:\penalty0 448--452, 2014.
\newblock \doi{10.1016/j.physletb.2014.10.010}, arXiv:1403.4892 [hep-ph].

\bibitem[Barata et~al.(2021)Barata, Dom{\'\i}nguez, Salgado, and
  Vila]{Barata:2021byj}
J.~Barata, F.~Dom{\'\i}nguez, C.~A. Salgado, and V.~Vila.
\newblock {A modified in-medium evolution equation with color coherence}.
\newblock \emph{JHEP}, 05:\penalty0 148, 2021.
\newblock \doi{10.1007/JHEP05(2021)148}, arXiv:2101.12135 [hep-ph].

\bibitem[Abreu et~al.(2025)Abreu, Mayo~L{\'o}pez, Milhano, and
  Soto-Ontoso]{Abreu:2024wka}
S.~Abreu, X.~Mayo~L{\'o}pez, G.~Milhano, and A.~Soto-Ontoso.
\newblock {A generalized picture of colour decoherence in dense QCD media}.
\newblock \emph{JHEP}, 03:\penalty0 216, 2025.
\newblock \doi{10.1007/JHEP03(2025)216}, arXiv:2410.24135 [hep-ph].

\bibitem[Andres et~al.(2025{\natexlab{b}})Andres, Apolin{\'a}rio, Armesto,
  Cordeiro, Dominguez, Guerrero-Rodr{\'\i}guez, and Milhano]{Andres:2025prc}
C.~Andres, L.~Apolin{\'a}rio, N.~Armesto, A.~Cordeiro, F.~Dominguez,
  P.~Guerrero-Rodr{\'\i}guez, and J.~G. Milhano.
\newblock {Gluon emission by a $q\bar{q}$ antenna with realistic parton-medium
  interactions}.
\newblock In \emph{{31st International Conference on Ultra-relativistic
  Nucleus-Nucleus Collisions}}, 8 2025{\natexlab{b}}.

\bibitem[Casalderrey-Solana et~al.(2016{\natexlab{b}})Casalderrey-Solana,
  Pablos, and Tywoniuk]{Casalderrey-Solana:2015bww}
J.~Casalderrey-Solana, D.~Pablos, and K.~Tywoniuk.
\newblock {Two-gluon emission and interference in a thin QCD medium: insights
  into jet formation}.
\newblock \emph{JHEP}, 11:\penalty0 174, 2016{\natexlab{b}}.
\newblock \doi{10.1007/JHEP11(2016)174}, arXiv:1512.07561 [hep-ph].

\bibitem[Arnold and Iqbal(2015)]{Arnold:2015qya}
P.~Arnold and S.~Iqbal.
\newblock {The LPM effect in sequential bremsstrahlung}.
\newblock \emph{JHEP}, 04:\penalty0 070, 2015.
\newblock \doi{10.1007/JHEP09(2016)072}, arXiv:1501.04964 [hep-ph].
\newblock [Erratum: JHEP 09, 072 (2016)].

\bibitem[Arnold et~al.(2023{\natexlab{a}})Arnold, Elgedawy, and
  Iqbal]{Arnold:2022mby}
P.~Arnold, O.~Elgedawy, and S.~Iqbal.
\newblock {Are Gluon Showers inside a Quark-Gluon Plasma Strongly Coupled? A
  Theorist{\textquoteright}s Test}.
\newblock \emph{Phys. Rev. Lett.}, 131\penalty0 (16):\penalty0 162302,
  2023{\natexlab{a}}.
\newblock \doi{10.1103/PhysRevLett.131.162302}, arXiv:2212.08086 [hep-ph].

\bibitem[Arnold et~al.(2023{\natexlab{b}})Arnold, Elgedawy, and
  Iqbal]{Arnold:2023qwi}
P.~Arnold, O.~Elgedawy, and S.~Iqbal.
\newblock {Landau-Pomeranchuk-Migdal effect in sequential bremsstrahlung: Gluon
  shower development}.
\newblock \emph{Phys. Rev. D}, 108\penalty0 (7):\penalty0 074015,
  2023{\natexlab{b}}.
\newblock \doi{10.1103/PhysRevD.108.074015}, arXiv:2302.10215 [hep-ph].

\bibitem[Arnold et~al.(2025)Arnold, Elgedawy, and Iqbal]{Arnold:2024whj}
P.~Arnold, O.~Elgedawy, and S.~Iqbal.
\newblock {Are in-medium quark-gluon showers strongly coupled? results in the
  large-N$_{f}$ limit}.
\newblock \emph{JHEP}, 01:\penalty0 193, 2025.
\newblock \doi{10.1007/JHEP01(2025)193}, arXiv:2408.07129 [hep-ph].

\bibitem[Wiedemann(2000)]{Wiedemann:2000za}
U.~A. Wiedemann.
\newblock {Gluon radiation off hard quarks in a nuclear environment: Opacity
  expansion}.
\newblock \emph{Nucl. Phys. B}, 588:\penalty0 303--344, 2000.
\newblock \doi{10.1016/S0550-3213(00)00457-0}, arXiv:hep-ph/0005129.

\bibitem[Kuzmin and Mayo~L{\'o}pez(2024)]{Kuzmin:2024smy}
M.~V. Kuzmin and X.~Mayo~L{\'o}pez.
\newblock {Gluon radiation inside a flowing medium}.
\newblock 6 2024, arXiv:2406.14628 [hep-ph].

\bibitem[Chen et~al.(2020{\natexlab{b}})Chen, Luo, Moult, Yang, Zhang, and
  Zhu]{Chen:2019bpb}
H.~Chen, M.-X. Luo, I.~Moult, T.-Z. Yang, X.~Zhang, and H.~X. Zhu.
\newblock {Three point energy correlators in the collinear limit: symmetries,
  dualities and analytic results}.
\newblock \emph{JHEP}, 08\penalty0 (08):\penalty0 028, 2020{\natexlab{b}}.
\newblock \doi{10.1007/JHEP08(2020)028}, arXiv:1912.11050 [hep-ph].

\bibitem[Chicherin et~al.(2024)Chicherin, Moult, Sokatchev, Yan, and
  Zhu]{Chicherin:2024ifn}
D.~Chicherin, I.~Moult, E.~Sokatchev, K.~Yan, and Y.~Zhu.
\newblock {Collinear limit of the four-point energy correlator in N=4
  supersymmetric Yang-Mills theory}.
\newblock \emph{Phys. Rev. D}, 110\penalty0 (9):\penalty0 L091901, 2024.
\newblock \doi{10.1103/PhysRevD.110.L091901}, arXiv:2401.06463 [hep-th].

\bibitem[Chen et~al.(2022{\natexlab{c}})Chen, Moult, Thaler, and
  Zhu]{Chen:2022swd}
H.~Chen, I.~Moult, J.~Thaler, and H.~X. Zhu.
\newblock {Non-Gaussianities in collider energy flux}.
\newblock \emph{JHEP}, 07:\penalty0 146, 2022{\natexlab{c}}.
\newblock \doi{10.1007/JHEP07(2022)146}, arXiv:2205.02857 [hep-ph].

\bibitem[Barata et~al.(2024{\natexlab{g}})Barata, Kuzmin, Milhano, and
  Sadofyev]{Barata:2024ieg}
J.~Barata, M.~V. Kuzmin, J.~G. Milhano, and A.~V. Sadofyev.
\newblock {Jet EEC aWAKEning: hydrodynamic response on the celestial sphere}.
\newblock 12 2024{\natexlab{g}}, arXiv:2412.03616 [hep-ph].

\bibitem[Adhya et~al.(2021)Adhya, Salgado, Spousta, and
  Tywoniuk]{Adhya:2020xcb}
S.~P. Adhya, C.~A. Salgado, M.~Spousta, and K.~Tywoniuk.
\newblock {Jet quenching and scaling properties of medium-evolved gluon cascade
  in expanding media}.
\newblock \emph{PoS}, HardProbes2020:\penalty0 129, 2021.
\newblock \doi{10.22323/1.387.0129}, arXiv:2009.03122 [hep-ph].

\bibitem[Gehrmann-De~Ridder and Glover(1998)]{Gehrmann-DeRidder:1997fom}
A.~Gehrmann-De~Ridder and E.~W.~N. Glover.
\newblock {A Complete O (alpha alpha-s) calculation of the photon + 1 jet rate
  in e+ e- annihilation}.
\newblock \emph{Nucl. Phys. B}, 517:\penalty0 269--323, 1998.
\newblock \doi{10.1016/S0550-3213(97)00818-3}, arXiv:hep-ph/9707224.

\bibitem[Catani and Grazzini(1999)]{Catani:1998nv}
S.~Catani and M.~Grazzini.
\newblock {Collinear factorization and splitting functions for
  next-to-next-to-leading order QCD calculations}.
\newblock \emph{Phys. Lett. B}, 446:\penalty0 143--152, 1999.
\newblock \doi{10.1016/S0370-2693(98)01513-5}, arXiv:hep-ph/9810389.

\bibitem[Pauli(2000)]{Pauli:2000gw}
H.~C. Pauli.
\newblock {A compendium of light cone quantization}.
\newblock \emph{Nucl. Phys. B Proc. Suppl.}, 90:\penalty0 259--272, 2000.
\newblock \doi{10.1016/S0920-5632(00)00904-X}, arXiv:hep-ph/0103106.

\bibitem[Lappi and Paatelainen(2017)]{Lappi:2016oup}
T.~Lappi and R.~Paatelainen.
\newblock {The one loop gluon emission light cone wave function}.
\newblock \emph{Annals Phys.}, 379:\penalty0 34--66, 2017.
\newblock \doi{10.1016/j.aop.2017.02.002}, arXiv:1611.00497 [hep-ph].

\bibitem[Fukushima and Hidaka(2007)]{Fukushima:2007dy}
K.~Fukushima and Y.~Hidaka.
\newblock {Light projectile scattering off the color glass condensate}.
\newblock \emph{JHEP}, 06:\penalty0 040, 2007.
\newblock \doi{10.1088/1126-6708/2007/06/040}, arXiv:0704.2806 [hep-ph].

\bibitem[Fukushima and Hidaka(2017)]{Fukushima:2017mko}
K.~Fukushima and Y.~Hidaka.
\newblock {General formulae for dipole Wilson line correlators with the Color
  Glass Condensate}.
\newblock \emph{JHEP}, 11:\penalty0 114, 2017.
\newblock \doi{10.1007/JHEP11(2017)114}, arXiv:1708.03051 [hep-ph].

\end{thebibliography}

\cleardoublepage

% ----------------------------------------------------------------------
%  Appendix (optional)
%
%  CAUTION: 1) the main document (up to the conclusions) shall not exceed 80 pages
%           2) the document shall not exceed a total of 100 pages (per IST regulations)
% ----------------------------------------------------------------------
\appendix

% add page number prefix according to apendix chapter (optional)
%\renewcommand{\thepage}{\thechapter.\arabic{page}}

% re-set arabic numbering (A.1,A.2,...) (optional, use only if chapter prefix is added)
%\setcounter{page}{1}

%\input{Appendix_Vertices} % file "Thesis_Appendix_A.tex"
%\cleardoublepage

% re-set arabic numbering (B.1,B.2,...) (optional, use only if chapter prefix is added)
%\setcounter{page}{1}

\chapter{Light-cone coordinates}\label{app:light_cone_coord}

Consider a four-vector defined in Minkowski coordinates $p^\mu = (p^0, p^1, p^2, p^3)$, with metric $g^{\mu\nu} = {\rm diag}(1,-1,-1,-1)$. Throughout this thesis, we use light-cone coordinates defined with respect to the direction of large momentum $p^3$, appropriate for the problem of high energy propagation. These coordinates are defined as
\begin{align}
    p^\mu = \left(p^+,\, p^-,\, \p\right)\,,\qquad p^{\pm} =  \frac{p^0\pm p^3}{\sqrt{2}}\,,
\end{align}
where the transverse components read $\p = (p^1,p^2) \equiv (\p_x, \p_y)$. The associated light-cone metric reads
\begin{align}
g^{\mu\nu}
=
\begin{pmatrix}
0 & 1 & 0 & 0 \\
1 & 0 & 0 & 0 \\
0 & 0 & -1 & 0 \\
0 & 0 & 0 & -1
\end{pmatrix}\,.
\end{align}
In this convention, the dot product between two four-vectors reads
\begin{align}
    p\cdot x = p^+x^- + p^-x^+-\p\cdot\x\,,
\end{align}
which implies that the on-shell condition for momenta determines $p^-$ as
\begin{align}
    m^2 = 2p^+p^- - \p^2 \implies p^- = \frac{\p^2+m^2}{2p^+}\,.
\end{align}

The Dirac gamma matrices are defined by their anti-commutation relations
\begin{align}
    \{\gamma^\mu, \gamma^\nu\} = 2g^{\mu\nu}
\end{align}
This implies, in particular, that they follow the useful identities used in the main text
\begin{align}
    (\gamma^+)^2 = (\gamma^-)^2 = 0\,,\qquad \gamma^\pm \gamma^\mp \gamma^\pm = 2\gamma^\pm\,,\qquad \gamma^\pm \boldsymbol{\gamma}^i\gamma^\pm = 0,.
\end{align}
Finally, the spinor projectors used in the main text read
\begin{align}
    \Lambda^\pm = \frac{\gamma^\mp\gamma^\pm}{2}\,,
\end{align}
which satisfy
\begin{align}
    \Lambda^+ + \Lambda^- = 1\,,\quad (\Lambda^\pm)^2 = \Lambda^\pm\,,\quad \Lambda^\pm\Lambda^\mp = 0\,.
\end{align} % file "App_light_cone.tex"
\cleardoublepage

\chapter{Kinematics and phase-space parametrization}\label{app:kinematics_phase_space}

In Section~\ref{sec:ch5_spin_correlations}, we consider any $1\to 3$ process such that $a\to b\,g \to b\,c\,d$, with the intermediate gluon splitting as $g\to c\,d$. We first introduce the light-cone energy fractions $z_i$ and the relative transverse momentum of each $1\to 2$ splitting 
\begin{subequations}\label{eq:relative_momenta_def}
\begin{align}
    & z_1 = p_g^+ / p_a^+\,, \quad \q_1 = (1-z_1)\p_g - z_1\p_b \\
	& z_2 = p_c^+ / p_g^+\,, \quad \q_2 = (1-z_2)\p_c - z_2\p_d
\end{align}
\end{subequations}
where we denote $\p_g = \p_c+\p_d$ and the total transverse momentum is  $\p_a = \p_b+\p_c+\p_d$. Physically, $z_1$ is the energy fraction of the intermediate gluon, and $z_2 $ is the energy fraction of the final-state particle with momentum $p_c$ (see Fig.~\ref{fig:diagrams}. It's also convenient to define
\begin{subequations} \label{eq:omega-def}
\begin{align}
\omega_a = z_1(1-z_1)p_a^+ \,,\\
\omega_g = z_2(1-z_2)p_g^+ \,.
\end{align}
\end{subequations}
We wish to make the change of coordinates $\{\p_a,\q_{1},\q_{2}\} \rightarrow \{\p_a,\theta_{1},\theta_{2},\phi_{1},\psi_{12}\}$, the definitions of which will become clear in the following text. It is useful to first invert the relations for the transverse momenta in Eq.~\eqref{eq:relative_momenta_def}
\begin{subequations}\label{eq:inverse_relative_momenta_def}
\begin{align}
	& \p_b = (1-z_1)\p_a-\q_{1}\,,\\
	& \p_c = z_1(1-z_2)\p_a+ (1-z_2) \q_{1} -\q_{2}\,, \\
	& \p_d = z_1z_2\p_a + z_2\q_{1}+ \q_{2}\,.
\end{align}
\end{subequations}
and to write the $z$-component of each momentum in terms of energy and transverse momentum i.e., $p^z_{i} = \sqrt{(p^0_i)^2- \p_i^2 - m_i^2}$, such that each cartesian three vector is defined in terms of $\p_a$, $\q_{1}$, $\q_{2}$ and $m_i$. In what follows we work in the quasi-collinear limit,  that is we scale all transverse momenta as $\p\rightarrow \lambda \p$ and masses as $m_i \rightarrow \lambda m_i$ and keep the first non-vanishing term in a power series of $\lambda\ll1$. 

Then, defining the opening angles of each splitting through the dot product of appropriate three-vectors one finds
\begin{subequations}\label{eq:opening_angles_app}
\begin{align}
	& \theta_{1}^2/2 = 1 - \frac{\vec p_b\cdot\vec p_g}{|\vec p_b||\vec p_g|} =  \q_{1}^2/\omega_a^2 + \mathcal{O}(\lambda^4) \,,\\
	& \theta_{2}^2/2 = 1 - \frac{\vec p_c\cdot\vec p_d}{|\vec p_c||\vec p_d|} = \q_{2}^2/\omega_g^2  + \mathcal{O}(\lambda^4)\,.
\end{align}
\end{subequations}
%
%where we assumed all transverse momenta and masses to be small in comparison with the energies of all partons, which is consistent with the approximations used in the main calculation. 
These expressions for the splitting opening angles are valid up to terms of order $\mathcal{O}(\lambda^4)$, where $\lambda^4$ represents any product of four transverse momenta $\p$ and/or masses $m_i$.
%is valid to leading power in $\mathcal{O}(\p^4/E_i^4)$ and $\mathcal{O}(m_i^4/E_i^4)$, where $\p^4$ stands for any product of $\p_i$ of power four. 
To relate the azimuthal orientation each relative transverse momentum $\q_i$, $\phi_{1}$ and $\phi_{2}$, to the angle between the planes spanned by the splittings, which we denote by $\psi_{12}$, we start with the definition of this quantity as the angle between the normal vectors to each plane. The normal three-vectors of unit norm are defined as the following cross product
\begin{align}
	& \vec n_{1} = \frac{\vec p_a \times \vec p_g}{|\vec p_a \times \vec p_g|}\,,\quad \vec n_{2} = \frac{\vec p_c \times \vec p_d}{|\vec p_c \times \vec p_d|}\,,
\end{align}
so that $\cos(\psi_{12})$ is defined as 
\begin{equation}
    \cos(\psi_{12}) = \vec n_{1}\cdot\vec n_{2}\,,
\end{equation}
% Using scalar and vector triple product identities, we find
% %
% \begin{align}
% 	|\vec p_a \times \vec p_g||\vec p_b \times \vec p_c|\cos(\psi_{12}) = \vec p_a \cdot [(\vec p_c-\vec p_d)(\vec p_c\cdot \vec p_d) + \vec p_c E_d^2 - \vec p_d E_c^2]
% \end{align}
%
which can now be written as a function of $\p_a$, $\q_{1}$, $\q_{2}$ and $m_i$ exactly as was done for the opening angles, using Eq.~\eqref{eq:inverse_relative_momenta_def}. By keeping only terms up to $\mathcal{O}\left(\lambda^2\right)$, where $\lambda^2$ is any quadratic product of $\p$ and $m_i$, one arrives at 
\begin{equation}\label{eq:phi_to_psi}
	\cos\psi_{12} = \frac{\q_{1}\cdot\q_{2}}{|\q_1||\q_2|} + \mathcal{O}(\lambda^2) \implies \psi_{12} = \phi_{2} - \phi_{1}\,,
\end{equation}
% \begin{align}\label{eq:phi_to_psi}
% 	\theta_{1}\theta_{2}\cos\psi = \frac{E_a}{E_bE_cE_d}\q_{1}\cdot\q_{2}+ \mathcal{O}\left(\p^2/E_i^2\right) \implies \psi_{12} = \phi_{2} - \phi_{1}\,.
% \end{align}
which we use in the main text of Section~\ref{sec:ch5_spin_correlations}. By using the relations in Eq.~\eqref{eq:opening_angles_app} together with Eq.~\eqref{eq:phi_to_psi}, we can now write the Jacobian for the change of coordinates $\{\p_a,\q_{1},\q_{2}\} \rightarrow \{\p_a,\theta_{1},\theta_{2},\phi_{1},\psi_{12}\}$
\begin{align}
\label{eq:jacobian}
	d^2\p_a d^2\q_{1}d^2\q_{2} & = \omega_a^2\omega_g^2\theta_{1}\theta_{2}/4\left(d^2\p_a d\theta_{1}d\theta_{2}d\phi_{1}d\psi_{12}\right)\nn
	& = d^2\p_a d\theta_{1}d\theta_{2}d\phi_{1}d\psi_{12} \frac{z_1^4(1-z_1)^2z_2^2(1-z_2)^2(p_a^+)^4\theta_{1}\theta_{2}}{4}.
\end{align}
This Jacobian becomes relevant when promoting splitting amplitudes $\mathcal{M}$ to differential cross-sections, by including the appropriate phase-space element, i.e.
\begin{align}
\label{eq:phase-space-element}
	2p_b^+2p_c^+2p_d^+(2\pi)^9\frac{d\sigma}{dp_b^+dp_c^+dp_d^+d^2\p_bd^2\p_cd^2\p_d} = \frac{1}{2}\sum_{\rho_i,\lambda_i} \cM\cM^{\dagger}\,.
\end{align}
where the sum runs over colours ($\rho_i$) and spin states ($\lambda_i$). The factor $1/2$ accounts for the average over the initial parton spin state that we have denoted $\lambda_a$ in the main text. Let us rewrite the phase-space element in Eq.~\eqref{eq:phase-space-element} in terms of the relative transverse momenta $\q_i$ and energy fractions $z_i$
\begin{align}
\label{eq:phase-space-two}
2z_1(1-z_1)2z_2(1-z_2)(2\pi)^6\frac{d\sigma}{d^3\Omega_a\,dz_1\,d z_2\,d^2\q_{1}d^2 \q_{2}} = \frac{1}{2}\sum_{\rho_i,\lambda_i} \cM\cM^{\dagger}\, ,
\end{align}
where
\begin{align}
d^3\Omega_a = \frac{dp_a^+}{(2\pi)2p_a^+}\frac{d^2\p_a}{(2\pi)^2}\, ,
\end{align}
corresponds to the phase-space element of the total momentum of the system.
Then, by using Eqs.~\eqref{eq:jacobian} and \eqref{eq:phase-space-two}, we can re-write Eq.~\eqref{eq:phase-space-element} as 
\begin{align}
\label{eq:phase-space-final}
	\frac{16(2\pi)^6}{z_1^3(1-z_1)z_2(1-z_2)\theta_{1}\theta_{2}}\frac{d\sigma}{dz_1\,d z_2\,d\theta_{1}d\theta_{2}d\phi_{1}d\psi_{12}} =  \frac{1}{2}\sum_{\rho_i,\lambda_i}\int\, d^3\Omega_a (p_a^+)^4 \cM\cM^{\dagger} \,.
\end{align}
To keep the notation compact, we define
\begin{equation}
\label{eq:phips-def}
d^6\Omega_{\rm PS} \equiv dz_1\,d z_2\,d\theta_{1}d\theta_{2}d\phi_{1}d\psi_{12}\,,
\end{equation}
and use it in the main text of Section~\ref{sec:ch5_spin_correlations} to express the fully differential cross-section as $d\sigma/d^6\Omega_{\rm PS}$. % file "App_kinematics.tex"
\cleardoublepage

\chapter{Vertex matrix elements}\label{app:vertex_matrix_elements}

\section{Computation at leading eikonal accuracy}\label{app:vertices}
Consider the QCD splitting $i (p_0, \lambda_0) \rightarrow j (p_1, \lambda_1) k (p_2, \lambda_2)$, where $p_i$ are the particle momenta and $\lambda_i = \pm 1$ are their spin states. The spin-dependent vertices can be directly calculated from the following expressions
\begin{subequations}
\begin{align}
	&V_{q\rightarrow gq}^{\lambda_0 \lambda_1 \lambda_2}(p_1,p_2) = \bar u^{\lambda_2}(p_2) \slashed{\epsilon}^{\ast, \lambda_1}(p_1)u^{\lambda_0}(p_0) \\
	&V_{g\rightarrow q\bar q}^{\lambda_0 \lambda_1 \lambda_2}(p_1,p_2) = \bar u^{\lambda_1}(p_1) \slashed{\epsilon}^{\lambda_0}(p_0)v^{\lambda_2}(p_2)\,,
    \\
	 &V_{g\rightarrow gg}^{\lambda_0 \lambda_1 \lambda_2}(p_1,p_2) = \epsilon^{\lambda_0}_{\alpha}(p_0)\epsilon^{\ast, \lambda_1}_{\beta}(p_1)\epsilon^{\ast,\lambda_2}_{\gamma}(p_2)[g^{\alpha\beta}(p_0+p_1)^{\gamma} - g^{\beta\gamma}(p_1-p_2)^{\alpha}-g^{\gamma\alpha}(p_0+p_2)^{\beta}]\,,
    \end{align}
\end{subequations}
where $p_0 = p_1+p_2$. The states $\lambda_i$ are defined as twice the spin projection along a given axis for quarks and as the circular polarisation state for gluons. The gluon polarisation vector is then defined in light-cone gauge as
\begin{align}\label{eq:pol_vector}
	\epsilon^{+,\lambda}(p) = 0\,,\quad  \epsilon^{-,\lambda}(p) = \frac{\beps^{\lambda}\cdot\p}{p^+}\,,\quad \beps^{\lambda} = \frac{1}{\sqrt{2}}(1,i\lambda)\,.
\end{align}
By using this polarisation vector together with the canonical Dirac spinors with spin quantized along $z$, at leading eikonal accuracy these vertices read (see e.g.~\cite{Pauli:2000gw, Kovchegov:2012mbw,Lappi:2016oup, Richardson:2018pvo})
\begin{subequations}\label{eq:vertex_all_channels}
\begin{align}
      &V_{q\rightarrow gq}^{\lambda_0 \lambda_1 \lambda_2}(z,\bkappa) =\frac{2 \delta^{\lambda_0, \lambda_2}}{z\sqrt{1-z}}(\delta^{\lambda_0, \lambda_1} + (1-z)\delta^{\lambda_0, -\lambda_1})(\bkappa\cdot \beps^{\ast,\lambda_1}) - \lambda_0\delta^{\lambda_0,-\lambda_2}\delta^{\lambda_0,\lambda_1}\frac{\sqrt{2} z m}{\sqrt{1-z}}\, , \\
      &V_{g\rightarrow q\bar q}^{\lambda_0 \lambda_1 \lambda_2}(z,\bkappa) =\frac{2\lambda_1 \delta^{\lambda_1,\lambda_2} }{\sqrt{z(1-z)}}(z\delta^{\lambda_0,\lambda_1} - (1-z)\delta^{\lambda_0, -\lambda_1})(\bkappa\cdot \beps^{\lambda_0}) -\delta^{\lambda_1, -\lambda_2}\delta^{\lambda_0, \lambda_1} \frac{\sqrt{2}m}{\sqrt{z(1-z)}}\, , \\
    &V_{g\rightarrow gg}^{\lambda_0 \lambda_1 \lambda_2}(z,\bkappa) = 2\left(\frac{\delta^{\lambda_0, \lambda_1}\delta^{\sigma,\lambda_2}}{1-z} - \delta^{\lambda_1, -\lambda_2}\delta^{\sigma, -\lambda_0} + \frac{\delta^{\lambda_0,\lambda_2}\delta^{\sigma, \lambda_1}}{z}\right)(\bkappa\cdot \beps^{\ast,\sigma})\,,
     \end{align}
\end{subequations}
where $z = p_1^+/p_0^+$ corresponds to the energy fraction of particle $j$ in the splitting and $\bkappa = (1-z)\p_1 - z\p_2$ is the relative transverse momentum of the final pair. The equal fermion spins term in both $g\to q\bar q$ and $q\to gq$ also contains mass dependent terms, but these are suppressed by additional energy denominators with respect to the remaining. 

It can be interesting to measure the spins of the final state fermions along a general spin-quantization axis $\vec n$ not necessarily aligned with $z$. We define the spin-quantization axis as
\begin{align}
    \vec{n} = \left(\sin\theta\cos\alpha, \sin\theta\sin\alpha, \cos\theta\right)\,,
\end{align}
with $\theta = 0,\, \alpha = \pi$ giving spin quantized along the $z$ axis. For the process relevant to Section~\ref{subsec:tv_polarisation}, namely $g\to q\bar q$, the vertex matrix element in this general case reads
\begin{align}\label{gen_spin_circg}
        & V_{g\to q\bar q}^{\lambda_0 hh'}(z,\bkappa) = \frac{2}{\sqrt{z(1-z)}}\nn
        & \left\{ \left. h\delta_{hh'}\left(\left[ \cos\theta\left(z\delta^{\lambda_0 h} - (1-z)\delta^{\lambda_0,-h}\right) + \sin^2(\theta/2)\left(\delta^{\lambda_0 h} - \delta^{\lambda_0, -h}\right) \right] (\boldsymbol{\epsilon}^{\lambda_0}\cdot\bkappa) - \frac{m}{2\sqrt{2}}\sin\theta e^{i\lambda_0\alpha}\right)+\right. \right.\nn
        & \left. + \delta_{h,-h'}\left(\sin\theta(z-1/2)(\boldsymbol{\epsilon}^{\lambda_0}\cdot\bkappa) + \frac{m}{\sqrt{2}}e^{i\lambda_0\alpha}\left[\delta^{\lambda_0 h}\cos^2(\theta/2) - \delta^{\lambda_0, -h} \sin^2(\theta/2)\right]\right)\right\} \, .
\end{align}
If we consider $\theta \neq 0$, i.e, a spin-quantization axis different from $z$, cross terms which are linear in the momenta (derivatives in the coordinate space amplitude) and mass appear when squaring the vertex. For instance, if we set $\theta=\pi/2$, i.e., measuring the spin of the $q\bar q$ dipole along some axis in the transverse plane, square Eq.~\eqref{gen_spin_circg} and average over polarization states we get:
\begin{equation}\label{eq:ch3_massive_tvpol}
    \begin{split}
    & \frac{1}{2}\sum_{\lambda_0}V_{g\to q\bar q}^{\lambda_0 hh'}(z,\bkappa)V_{g\to q\bar q}^{\ast,\lambda_0 hh'}(z,\bar\bkappa) = \frac{1}{2z(1-z)}\Big\{(\bkappa\cdot\bar\bkappa)\left(\delta_{hh'} + (1-2z)^2\delta_{h,-h'}\right) +m^2 \\
    & +imh\left[\left(\bkappa^x-\bar\bkappa^x\right)\sin\alpha - \left(\bkappa^y-\bar\bkappa^y\right)\cos\alpha\right]\left(\delta_{hh'} + (1-2z)\delta_{h,-h'}\right)\Big\} \, .
    \end{split}
\end{equation}
The linear terms vanish in the vacuum limit since the momenta in the direct and complex conjugate amplitudes are the same. This means that the difference $\P^i - \bar \P^i$ is controlled by $\hat q_i$ in the corresponding direction. If $\hat{q}^y \gg \hat{q}^x$, for instance, then the spin dependent term is dominated by the $\cos\alpha$ term. This implies that when measuring spin states along the $x$ axis ($\alpha=0$) the difference between the spectrum for up and down ($h=\pm 1$) is much greater than the difference obtained when measuring spin along the $y$ axis ($\alpha = \pi/2$). Put simply, in the limit where $\hat{q}^x \rightarrow 0$ and $\hat{q}^y$ is finite, then the $q\overline{q}$ pair is completely unpolarized in the $y$ direction, since there is no spin dependence, and polarized in the $x$ direction. The degree of polarization depends on the quark mass and naturally grows with $\hat{q}^y$.

\section{Azimuthal correlations}\label{app:vertices_azimuthal_correlations}
The dot product between the relative transverse momentum and the gluon polarisation vector in Eq.~\eqref{eq:vertex_all_channels} carries information about the azimuthal angle, since one can write it as
\begin{equation}
	\q\cdot\beps^{\lambda} = \frac{|\q|}{\sqrt{2}}e^{i \lambda\phi}\,.
\end{equation}
This allows to factor out the azimuthal dependence in each vertex as a single overall phase~\cite{Richardson:2018pvo, Karlberg:2021kwr}
\begin{align}\label{eq:vertex-and-phi}
	V^{\lambda_0 \lambda_1 \lambda_2 }(z,\q) = \tilde V^{\lambda_0 \lambda_1 \lambda_2}(z,|\q|)\, e^{i(\bar \lambda_0 - \bar \lambda_1 - \bar \lambda_2)\phi}\,,
\end{align}
where $\bar\lambda_i = \lambda_i$ if the particle is a gluon and $\bar \lambda_i = \pm \lambda_i / 2$ if the particle is a quark or anti-quark, respectively. In $\tilde V$ one replaces $\q\cdot\beps\rightarrow |\q|/\sqrt{2}$ in $V$. This can be readily checked by plugging this phase in Eq.~\eqref{eq:vertex_all_channels}. 

In the main text of Section~\ref{sec:ch5_spin_correlations}, we wish to compute the cross-section for the process $a\rightarrow b\,g \rightarrow b\,c\,d$, with the gluon splitting as $g\to c\,d$. At the scattering amplitude level, this involves a sum over the intermediate gluon's polarisations ($\lambda_g$) of a product of vertices,
\begin{align}\label{eq:vertex_term}
	\cA(\q_1,\q_2) \equiv \sum_{\lambda_g}V^{\lambda_g\lambda_c\lambda_d}_{g\rightarrow c d}(z_2,\q_2)V^{\lambda_a\lambda_g\lambda_b}_{a\rightarrow b g}(z_1, \q_1)\,,
\end{align}
while for the squared-matrix element we also need to sum over final state spin states $\lambda_b,\lambda_c,\lambda_d$ and average over the initial state one $\lambda_a$.
One can verify that, in vacuum, the product of any QCD vertex in Eq.~\eqref{eq:vertex_all_channels} and its complex conjugate is polarisation independent when considering a different polarisation for the off-shell gluon in the amplitude and complex conjugate amplitude, as was stated in Eq.~\eqref{eq:ch5_real_imag_F_M2}. For instance, for a $g\to q\bar q$ splitting with energy fraction $z$ and transverse momentum $\q$ we find
\begin{equation}
	\sum_{\lambda_q \lambda_{\bar q}} \tilde V^{\lambda_g \lambda_q \lambda_{\bar q}}_{g\rightarrow q\bar q}(z, \q)\tilde V^{\ast,\lambda'_g \lambda_q \lambda_{\bar q}}_{g\rightarrow q\bar q}(z, \q) = \frac{2}{z(1-z)}\left(\delta^{\lambda_g, \lambda'_g} (m_q^2 + \q^2P_{qg}(z))- \delta^{\lambda_g, -\lambda'_g}\, 2 z(1-z) \q^2 \right)\,,
\end{equation}
i.e., the functions attached to both $\delta$ tensors are independent of $\lambda_g$ and $\lambda'_g$. This implies, in particular, that there is no $\sin(2\psi_{12})$ modulation due to spin correlations (see Eq.~\eqref{eq:A_out_out}).
However, if we assume the first splitting $a\rightarrow b\,g$ to happen inside a QCD medium while the gluon splits in vacuum (see Eq.~\eqref{eq:ch5_real_imag_F_M2_medium}), one can show that polarisation dependent terms arise when there is some type of anisotropy in the transverse plane which breaks rotational invariance of the medium. To see this, consider the contribution coming from the term with $\lambda_g \neq \lambda'_g$ in the square of the $a\rightarrow b\,g$ vertex. Due to interactions with the medium, one writes this with different relative transverse momentum in the amplitude ($\bkappa$) and in the conjugate amplitude ($\bar\bkappa$),
%To see this, we make use of Eq.~\eqref{eq:vertex-and-phi} and write an in-medium ($a\to gb$) squared with transverse momentum $\kappa$ in the amplitude and $\bar\kappa$ in the conjugate amplitude as
%
\begin{align}\label{eq:squared_vertex_polarisation}
	\delta^{\lambda_g, -\lambda'_g}\sum_{\lambda_a\lambda_b}V^{\lambda_a\lambda_g \lambda_b}_{a\to gb}(z,\bkappa)V_{a\to gb}^{\ast, \lambda_a \lambda_g' \lambda_b}(z,\bar\bkappa)e^{i(\lambda_g-\lambda_g')\phi_{2}} \propto \delta^{\lambda_g, -\lambda'_g}e^{-i\lambda_g (\phi_\bkappa + \phi_{\bar\bkappa} - 2\phi_{2})}\,,
\end{align}
where $\bkappa = |\bkappa|(\cos\phi_\bkappa,\sin\phi_\bkappa)$ and the same for $\bar\bkappa$, and the angle $\phi_{2}$ is the azimuthal orientation of the final-state plane formed by $b$ and $g$. If the medium is rotationally invariant the dependence on $\phi_2$ in Eq.~\eqref{eq:squared_vertex_polarisation} vanishes. Also, the remaining function of $\bkappa$ and $\bar\bkappa$ against which Eq.~\eqref{eq:squared_vertex_polarisation} is integrated depends only on $\bkappa^2$, $\bar\bkappa^2$ and $\phi_\bkappa-\phi_{\bar\bkappa}$.\footnote{Note that this reasoning is only valid because we assumed the gluon splits in vacuum. Were this not true, then a dependence on $\psi_{12} = \phi_2-\phi_1$ would exist for the medium modification. Thus, the integrand could depend on, e.g., $(\phi_\bkappa-\psi_{12})+(\phi_{\bar\bkappa}-\psi_{12})$.} Hence, the phase $e^{i\lambda_g(\phi_\bkappa+\phi_{\bar\bkappa})}$ can be integrated out to yield a $\lambda_g$ independent result.

Finally, for all vertices in Eq.~\eqref{eq:vertex_all_channels}, one can show that the square of Eq.~\eqref{eq:vertex_term} entering the product of the amplitude and conjugate amplitude reads
\begin{align}\label{eq:AAdagger_all_channels}
    &  \frac{1}{2}\sum_{\lambda_a}\sum_{\lambda_b\lambda_c\lambda_d} \cA(\bkappa,\q_{2})\cA^{\dagger}(\bar\bkappa,\q_{2}) = \frac{4(m_2^2 + \q_{2}^2 P_{g\rightarrow cd}(z_2))}{z_2(1-z_2)z_1(1-z_1)}
    \times \Big[z_1^3 m_1^2 + P_{a\rightarrow gb}(z_1)\left(\bkappa\cdot \bar\bkappa\right)\nn
    & + S_{g\rightarrow cd}\frac{4z_2(1-z_2)(1-z_1)}{z_1(m_2^2 + \q_{2}^2 P_{g\to cd}(z_2))}\left(\left(\q_{2}\cdot \bkappa\right)\left(\q_{2}\cdot \bar\bkappa\right)-\left(\q_{2}\times \bkappa\right)_z\left(\q_{2}\times \bar\bkappa\right)_z\right)\Big]
\end{align}
where $(\q_2\times \bkappa)_z = q_2^x\kappa^y-q_2^y\kappa^x$, $\bkappa$ and $\bar\bkappa$ correspond to integration variables in the medium calculation, and all other variables are defined in Eq.~\eqref{eq:relative_momenta_def}. The channel-dependence of Eq.~\eqref{eq:AAdagger_all_channels} is encoded in the factor $S_{g\to cd}$ which takes into account if the intermediate gluon splits into a $q \bar q$ pair ($S_{g\to cd} = -1$) or a pair of gluons ($S_{g\to cd} = 1$), and in $P_{ji}(z)$ which is the massless splitting function for the process $i\rightarrow j k$ stripped of its corresponding colour factor
\begin{subequations}
\label{eq:vac_splitting_functions}
\begin{align}\label{eq:vacuum_splitting_functions}
    & P_{q\rightarrow g q}(z) = \frac{1 + (1-z)^2}{z}\,, \\
	& P_{g\rightarrow q\bar q}(z) = z^2 + (1-z)^2\, ,\\
	& P_{g\rightarrow gg}(z) = 2\left(\frac{z}{1-z}+\frac{1-z}{z}+z(1-z)\right)\,.
\end{align}
\end{subequations}
Note that $m_1=m_a=m_b$ and and $m_2=m_c=m_d$. 

In the vacuum case Eq.~\eqref{eq:AAdagger_all_channels} can be further simplified by setting $\bkappa=\bar\bkappa=\q_1$ such that the squared amplitude reads
\begin{align}
      \frac{1}{2}\sum_{\lambda_a}\sum_{\lambda_b\lambda_c\lambda_d} \cA(\q_{1},\q_{2})\cA^{\dagger}(\q_{1},\q_{2}) &\underset{\text{vacuum}}{=}  \frac{4(m_2^2 + \q_{2}^2 P_{g\rightarrow cd}(z_2))}{z_2(1-z_2)z_1(1-z_1)}\times \Big[z_1^3 m_1^2 + P_{a\rightarrow gb}(z_1)\q_{1}^2\nn
    & +S_{g\to cd}\frac{4z_2(1-z_2)(1-z_1)}{z_1(m_2^2 + \q_{2}^2 P_{g\rightarrow cd}(z_2))}\left(\left(\q_{2}\cdot \q_{1}\right)^2 -\left|\q_{2}\times \q_{1}\right|^2\right)\Big]\, .
\end{align}
By using Eq.~\eqref{eq:phi_to_psi}, we can re-write the scalar and vector products in terms of the azimuthal angle between the planes spanned by the two splittings:
\begin{align}\label{eq:A_out_out}
    & \frac{1}{2}\sum_{\lambda_a}\sum_{\lambda_b\lambda_c\lambda_d} \cA(\q_{1},\q_{2})\cA^{\dagger}(\q_{1},\q_{2}) \underset{\text{vacuum}}{=} \frac{4(m_2^2 + \q_{2}^2 P_{g\to cd}(z_2))}{z_2(1-z_2)z_1(1-z_1)}\nn
    & \times \left[z_1^3 m_1^2 + P_{a\rightarrow gb}(z_1)\q_{1}^2 + S_{g\to cd}\frac{4z_2(1-z_2)(1-z_1)\q_{1}^2\q_{2}^2}{z_1(m_2^2 + \q_{2}^2 P_{g\to cd}(z_2))}\cos(2\psi_{12})\right]\,,
\end{align}
This form for the $\cos(2\psi_{12})$ modulation in vacuum for $1\rightarrow 3$ processes with an intermediate gluon is well known in the literature (see e.g. \cite{Shatz:1983hv, Knowles:1987cu,Collins:1987cp, Richardson:2018pvo, Chen:2020adz, Karlberg:2021kwr}). Processes with no intermediate off-shell gluons (e.g. $q\rightarrow qg(q\rightarrow qg)$) do not exhibit such a dependence. % file "App_vertices.tex"
\cleardoublepage

\chapter{Colour structure}\label{app:colour_structure}
Let us explicitly write the Wilson line correlators appearing in Eq.~\eqref{eq:ch3_squared_M_gqqb} without prefactors and path integrals (each $\r_i$ is a path integration variable whose initial and final positions are conjugate to initial and final momenta appearing in Eq.~\eqref{eq:ch3_squared_M_gqqb}; the numbering is arbitrary):
\begin{align}
    & T^{\rm colour} = \frac{1}{N_c^2-1} \left\langle U_A^{ab}(\r_5)(U_A^{\dagger})^{bc}(\r_9)\right\rangle\nn
    & \times t_{kl}^a t_{\bar l\bar k}^{\bar a} \left\langle U^{k_1 k}(\r_3)(U^{\dagger})^{l k_2}(\r_4)(U_A^{\dagger})^{c\bar a}(\r_8)\right\rangle\nn
    & \times \left\langle U^{j_1 k_1}(\r_1)(U^{\dagger})^{k_2 j_2}(\r_2)(U^{\dagger})^{j_2\bar l}(\r_7)U^{\bar k j_1}(\r_6)\right\rangle\,.
\end{align}
First note that this object is in a colour singlet state. This is because we are considering the final-state quark and anti-quark to have definite colour states which are then summed over when squaring the amplitude. Since colour is conserved throughout the process (the system as a whole does not exchange colour with the medium due to the diagonal colour matrix in Eq.~\eqref{eq:A_correlator}), this means the system is in a singlet state at all times. In particular, after expanding the colour structure inside each time region as a linear combination of the state projectors in each of its irreducible representations, one can drop the colour non-singlet part (one can include the non-singlet part in the expansion, but the contraction with the final region will always only pick out the singlet parts). Finding out the colour singlet states amounts to finding out states for which the second-order Casimir operator in the appropriate representation vanishes \cite{Fukushima:2007dy, Fukushima:2017mko}. For the two-point function the only option is the normalized projector $\frac{\delta^{b\bar b_1}}{\sqrt{N_c^2-1}}$. Hence, we have
\begin{equation}
    \left\langle U_A^{ba}(\boldsymbol{r}_5)(U_A^{\dagger})^{a\bar b_1}(\boldsymbol{r}_9)\right\rangle = \frac{\delta^{b\bar b_1}}{N_c^2-1}\left\langle \text{Tr}\left[U_A(\boldsymbol{r}_5)U_A^{\dagger}(\boldsymbol{r}_9)\right]\right\rangle \, .
\end{equation}
Propagating the $\delta^{b\bar b_1}$ to the rest of the squared amplitude, writing the adjoint Wilson line in terms of fundamental ones and repeatedly making use of the Fierz identity, i.e, using 
\begin{align}\label{fierz}
     & U_A^{b\overline{b}} = 2\text{Tr}\left[t^bU^{\dagger}t^{\overline{b}}U\right]\, ,\quad  t^b_{kl}t^b_{ij} = \frac{1}{2}(\delta^{kj}\delta^{li}-\frac{1}{N_c}\delta^{kl}\delta^{ij}) \, ,
\end{align}
the other two regions read:
\begin{equation}\label{before_proj}
    \begin{split}
    & \frac{1}{2}\Bigg(\left\langle \left(U(\r_3) U^{\dagger}(\r_8)\right)^{i_1 \bar k}\left(U(\r_8)U^{\dagger}(\r_4))\right)^{\bar l j_1}\right\rangle_{(x_v^+,\bar x_v^+)} \left\langle\left( U^{\dagger}(\r_6)U(\r_1)\right)^{\bar k i_1}\left(U^{\dagger}(\r_2)U(\r_7))\right)^{j_1 \bar l}\right\rangle_{(\bar x_v^+, L^+)}  \\ 
    & - \frac{1}{N_c^2}\left\langle\text{Tr}\left[U(\r_3)U^{\dagger}(\r_4)\right]\right\rangle_{(x_v^+, \bar x_v^+)}\left\langle\text{Tr}\left[U^{\dagger}(\r_6)U(\r_1)U^{\dagger}(\r_2)U(\r_7)\right]\right\rangle_{( \bar x_v^+, L+)} \Bigg)  \, .
    \end{split}
\end{equation}
To fully contract the remaining colour structure in the first line, one needs to expand each time region's contribution in a basis of colour singlets, as we have discussed. For both cases the number of such states is two and they can be obtained through the contractions $\delta^{i_1\bar k}\delta^{\bar l j_1}$ and $\delta^{i_1 j_1}\delta^{\bar l \bar k}$. An orthonormal basis for these singlet projectors is straightforward to obtain~\cite{Fukushima:2007dy, Fukushima:2017mko}:
\begin{equation}
    \begin{split}
        & s_1^{i_1\bar k \bar l j_1} = \frac{1}{N_c}\delta^{i_1\bar k}\delta^{\bar l j_1}\\
        & s_2^{i_1\bar k \bar l j_1} = \frac{1}{\sqrt{N_c^2-1}}\left(\delta^{i_1 j_1}\delta^{\bar l \bar k} - \frac{1}{N_c}\delta^{i_1\bar k}\delta^{\bar l j_1}\right) \, .
    \end{split}
\end{equation}
After expanding each correlator in the first line of Eq.~\eqref{before_proj} as a linear combination of the projectors $s_1$ and $s_2$, the coefficients are easily obtained by taking advantage of orthonormality. For instance, we have
\begin{equation}
    \begin{split}
    & \left\langle\left( U^{\dagger}(\r_6)U(\r_1)\right)^{\bar k i_1}\left(U^{\dagger}(\r_2)U(\r_7))\right)^{j_1 \bar l}\right\rangle = \frac{1}{N_c}\left\langle\text{Tr}\left[U^{\dagger}(\r_6)U(\r_1)\right]\text{Tr}\left[U^{\dagger}(\r_2)U(\r_7)\right]\right\rangle s_1^{i_1\bar k \bar l j_1}\\
    & + \frac{1}{\sqrt{N_c^2-1}}\left(\left\langle\text{Tr}\left[U^{\dagger}(\r_6)U(\r_1)U^{\dagger}(\r_2)U(\r_7)\right]\right\rangle - \frac{1}{N_c}\text{Tr}\left[U^{\dagger}(\r_6)U(\r_1)\right]\text{Tr}\left[U^{\dagger}(\r_2)U(\r_7)\right] \right)s_2^{i_1\bar k \bar l j_1}
    \end{split}
\end{equation}
and for the remaining correlator it is completely analogous. Using these expansions, the fully contracted result of the colour part of the squared amplitude in Eq.~\eqref{eq:ch3_squared_M_gqqb} finally simplifies to
\begin{equation}
    T^{\text{colour}} = T_F\,\mathcal{C}^{(2)}\mathcal{C}^{(3)}\mathcal{C}^{(4)} \, ,
\end{equation}
where $T_F$ gives the overall colour factor for the $g\to q\bar q$ process and the colour structures read
\begin{equation}\label{eq:colour_regions}
    \begin{split}
    & \mathcal{C}^{(2)} = \frac{1}{N_c^2 -1}\left\langle \text{Tr}\left[U_A(\boldsymbol{r}_5)U_A^{\dagger}(\boldsymbol{r}_9)\right]\right\rangle\,,\\
    & \mathcal{C}^{(3)} = \frac{1}{N_c^2-1}\left\langle \text{Tr}\left[U^{\dagger}(\r_8)U(\r_3)\right]\text{Tr}\left[U^{\dagger}(\r_4)U(\r_8)\right]  - \frac{1}{N_c}\text{Tr}\left[U^{\dagger}(\r_4)U(\r_3)\right]\right\rangle\,,\\
    & \mathcal{C}^{(4)} = \frac{1}{N_c^2-1}\left\langle \text{Tr}\left[U^{\dagger}(\r_6)U(\r_1)\right]\text{Tr}\left[U^{\dagger}(\r_2)U(\r_7)\right] - \frac{1}{N_c}\text{Tr}\left[U^{\dagger}(\r_6)U(\r_1)U^{\dagger}(\r_2)U(\r_7)\right]  \right\rangle \, ,
    \end{split}
\end{equation}
which resembles results in, e.g.,~\cite{Dominguez:2012ad, Apolinario:2014csa, Isaksen:2020npj}. Both $\mathcal{C}^{(2)}$ and $\mathcal{C}^{(3)}$ have exact analytical solutions, while $\mathcal{C}^{(4)}$ is usually computed numerically~\cite{Isaksen:2020npj}. In the large-$N_c$ limit, it has been argued~\cite{Apolinario:2014csa, Blaizot:2012fh} that the quadrupole (the second term in $\mathcal{C}^{(4)}$) can be approximated by the product of two independent averages of two Wilson line traces, an approximation which is kinematically favoured when the formation time of the antenna is small compared to the medium's length. Nevertheless, in our case, this term is colour suppressed by a power $N_c^2$, so we may drop it in the large-$N_c$ limit. This simplification in the remaining $1\to 2$ processes generating a QCD final state, where the quadrupole must be dealt with it~\cite{Blaizot:2012fh,Apolinario:2014csa, Dominguez:2019ges,Isaksen:2020npj, Isaksen:2023nlr}. Thus, for the process at hand, the leading $N_c$ contributions to each colour structure can be fully written in terms of averages of traces of two Wilson lines, i.e., the dipole object as defined in Eq.~\eqref{eq:dipole_def}, since in this limit the average of products of multiple traces reduces to the product of averages of individual traces (see, e.g,~\cite{Isaksen:2020npj, Dominguez:2011wm, Dominguez:2011wm}). Explicitly, they read
\begin{align}\label{eq:colour_regions_largeNc}
    & \cC^{(2)} = D_A(\r_5-\r_9)\,,\nn
    & \cC^{(3)} = D_F(\r_8-\r_3)D_F(\r_4-\r_8) + \cO\left(\frac{1}{N_c^2}\right)\,,\nn
    & \cC^{(4)} = D_F(\r_6-\r_1)D_F(\r_2-\r_7) + \cO\left(\frac{1}{N_c^2}\right)\,.
\end{align}
 % file "App_colour_structure.tex"
\cleardoublepage

\chapter{$n$-point functions}\label{app:n_point_functions}

In Section~\ref{sec:ch3_gqqbar}, the relevant $n$-point functions appearing in the large-$N_c$ computation of the $g\to q\bar q$ in-medium cross-section in Eq.~\eqref{eq:ch3_xsec_simplified1} are the $2$-point and $3$-point functions encoding the physics of broadening and dipole formation. Explicitly, they read
\begin{align}\label{eq:app_P_K_def}
    & \cP_R(x_f^+, \p_f; x_i^+, \p_i) \equiv\nn
    & \int_{\bar \p_i}\int_{\r_i,\,\r_f,\,\bar \r_i,\,\bar \r_f}\,e^{i\p_f\cdot(\r_f-\bar \r_f)}e^{- i\p_i\cdot\r_i}e^{i\bar\p_i\cdot\bar\r_i}\int_{\r_i}^{\r_f}\cD\r\int_{\bar\r_i}^{\bar\r_f}\cD\bar\r\,\exp\left\{\frac{i p^+}{2}\int_{x_i^+}^{x_f^+}ds^+\left(\dot\r^2-\dot{\bar\r}^{2}\right)\right\}D_R(\r-\bar\r)\nn
    & \cS^{(3)}(x_f^+,\bar \q_1, \bar \q_0; x_i^+,\q_1, \q_0) \equiv\nn
    & \int_{\l_0}\int_{\r_{1,i},\,\r_{1,f},\r_{2,i},\,\r_{2,f},\,\bar\r_{0,i},\,\bar\r_{0,f}}e^{i\bar\q_1\cdot(\r_{1,f}-\r_{2,f})}e^{-i\q_1\cdot(\r_{1,i}-\r_{2,i})}e^{i\bar\q_0\cdot (\r_{2,f}-\bar\r_{0,f})}e^{-i\q_0\cdot \r_{2,i}}e^{i\l_0\cdot\bar\r_{0,i}}\nn
    & \times \int_{\r_{1,i}}^{\r_{1,f}}\cD \r_1\int_{\r_{2,i}}^{\r_{2,f}}\cD \r_2\int_{\bar\r_{0,i}}^{\bar\r_{0,f}}\cD \bar\r_0\exp\left\{\frac{i}{2}\int_{x_i^+}^{x_f^+}ds^+\left(p_1^+\dot{\r}_1^2+p_2^+\dot{\r}_2^2-p_0^+\dot{\bar{\r}}_0^2\right)\right\}D_F(\bar\r_0-\r_1)D_F(\r_2-\bar\r_0)\,.
\end{align}
with the dipole object $D_R$ defined in Eq.~\eqref{eq:dipole_def}, with time arguments kept implicit. Let us show how these objects can be simplified to give the result in Eq.~\eqref{eq:ch3_xsec_simplified2}.

To simplify $\cP_R$, we start by defining centre of mass and relative coordinates 
\begin{align}
    \u = \r-\bar \r\,,\qquad \v = \frac{\r+\bar\r}{2}\,,
\end{align}
such that the kinetic term reads
\begin{align}
    \dot{\r}^2 - \dot{\bar{\r}}^2 = 2\dot{\v}\cdot\dot{\u}
\end{align}
and the potential depends only on $\u$. Thus, both integrations can be carried out exactly, with the integration in $\v$ constraining $\u$ to its classical, straight-line path, i.e.,
\begin{align}\label{eq:app_u_classical_path}
    \u^{\rm cl.}(s^+) = \u_i + \frac{s^+ - x_i^+}{\Delta x^+}(\u_f - \u_i)\,.
\end{align}
Without the Fourier phases, $\cP$ then reads
\begin{align}
    \left(\frac{p^+}{2\pi\Delta x^+}\right)^2\exp\left\{\frac{ip^+}{\Delta x^+}(\v_f-\v_i)\cdot(\u_f-\u_i)\right\}D_R(\u^{\rm cl.})\,.
\end{align}
The phases read
\begin{align}
    e^{i\p_f\cdot(\r_f-\bar \r_f)}e^{- i\p_i\cdot\r_i}e^{i\bar\p_i\cdot\bar\r_i} = e^{i\p_f\cdot\u_f} e^{-i(\p_i-\bar\p_i)\cdot\v_i} e^{-i(\p_i+\bar\p_i)\cdot\u_i/2}\,,
\end{align}
and since they don't depend on $\v_f$, the integration over it leads to $\u_f = \u_i \equiv \u$, while integration over $\v_i$ sets $\p_i = \bar \p_i$. Thus, the broadening kernel reads
\begin{align}\label{eq:PR_simplified}
    \cP_R(\p_f-\p_i;x_f^+,x_i^+) = \int_\u e^{i(\p_f-\p_i)\cdot \u}D_R(\u;x_f^+,x_i^+)
\end{align}
where we have now made explicit the time argument of the dipole. In the main text, the integration over final total transverse momentum $\P = \p_1+\p_2$ allows for further simplification of the product of $\cP_F$ objects in Eq.~\eqref{eq:ch3_xsec_simplified1}, reading  
\begin{align}
    & \cP(\p-\bkappa; x_\infty^+,\bar x^+) \equiv \nn
    &\int_\P \cP_F(\p_1-\bar \q_1; x_\infty^+, x^+)\cP_F(\p_2-\bar \q_0+\bar\q_1; x_\infty^+, x^+) = \int_\u e^{-i(\p-\bar\bkappa)\cdot \u}D_F((1-z)\u)D_F(z\u)\,.
\end{align}
where $\p = (1-z)\p_1 - z\p_2$ and $\bar\bkappa = \bar\q_1-z\bar\q_0$. This is the object present in Eq.~\eqref{eq:ch3_xsec_simplified2} and it amounts to broadening of the relative transverse momentum of the final dipole.

To simplify $\cS^{(3)}$, we start by defining centre-of-mass and relative coordinates for the two amplitude paths ($\r_1,\r_2$), as well as the coordinate comparing that centre-of-mass coordinate in the amplitude with the coordinate in the cc amplitude ($\bar\r_0$)
\begin{align}
    & \r = \r_1-\r_2\,,\qquad \R = z\r_1 + (1-z)\r_2\nn
    & \u = \R - \bar\r_0\,,\qquad \v = \frac{\R + \bar\r_0}{2}\,,
\end{align}
such that in terms of $\r,\u,\v$ the kinetic term boils down to
\begin{align}
    p_1^+\dot{\r}_1^2+p_2^+\dot{\r}_2^2-p_0^+\dot{\bar{\r}}_0^2 = 2p_0^+ \dot{\v}\cdot\dot{\u} + \omega^+\dot{\r}^2\,,
\end{align}
the two dipole arguments read
\begin{align}
    \bar \r_0-\r_1 = \u + (1-z)\r\,,\qquad \r_2-\bar\r_0 = \u - z\r\,,
\end{align}
with $z = p_1^+/p_0^+$, and the Fourier transform phases read
\begin{align}\label{eq:app_K_phases}
    e^{i\bar\q_1\cdot\r_f}e^{-i\q_1\cdot\r_i}e^{i\bar\q_0\cdot (\u_{f}-z\r_f)}e^{-i\q_0\cdot (\v_i+\u_i/2-z\r_i)}e^{i\l_0\cdot(\v_i-\u_i/2)}\,.
\end{align}
Since the potential term does not depend on $\v$, the remaining kinetic term $\dot\v\cdot\dot\u$ can be integrated exactly, effectively constraining $\u$ to its classical, straight-line path as in Eq.~\eqref{eq:app_u_classical_path}. This results in the last line of the $\cS^{(3)}$ definition in Eq.~\eqref{eq:app_P_K_def} being written as
\begin{align}
    & \left(\frac{p_0^+}{2\pi\Delta x^+}\right)^2\exp\left(\frac{ip_0^+}{\Delta x^+}(\v_f - \v_i)\cdot(\u_f-\u_i)\right)\nn
    \times \int_{\r_i}^{\r_f}\cD \r &\exp\left\{i\frac{\omega^+}{2} \int_{x_i^+}^{x_f^+}ds^+\dot{\r}^2\right\}D_F(\u^{\rm cl.}+(1-z)\r)D_F(\u^{\rm cl.}-z\r)
\end{align}
where $\omega^+ = z(1-z)p_0^+$. Since the phases in Eq.~\eqref{eq:app_K_phases} do not depend on $\v_f$, the $\v_f$ integral sets $\u_f = \u_i \equiv \u$, while the integral over $\l_0$ sets $\v_i = \u_i/2$. Finally, the three-body average can be written as
\begin{align}
    & \cS^{(3)}(\bar \bkappa,\bkappa,\bar \q_0-\q_0; x_f^+, x_i^+) = \int_{\u,\r_f, \r_i} e^{i\r_f\cdot\bar\bkappa}e^{-i\r_i\cdot \bkappa }\,e^{i\u\cdot(\bar \q_0-\q_0)}\nn
    & \times \int_{\r_i}^{\r_f}\cD \r \exp\left\{i\frac{\omega^+}{2} \int_{x_i^+}^{x_f^+}ds^+\dot{\r}^2\right\}D_F(\u+(1-z)\r)D_F(\u-z\r)
\end{align}
where $\bkappa = \q_1-z\q_0$ and $\bar\bkappa = \bar\q_1 - z\bar\q_0$. In the main text, the integration over final total transverse momentum further allows the integration of $S^{(3)}$ over $\bar \q_0-\q_0$ to be carried out, setting $\u = 0$. This leads to the reduced splitting kernel 
\begin{align}\label{eq:K_simplified}
    \cK(x_f^+,\bar\bkappa;x_i^+,\bkappa) & = \int_\q \cS^{(3)}(\bar \bkappa,\bkappa, \q; x_f^+,x_i^+) \nn
    & = \int_{\r_f, \r_i} e^{i\r_f\cdot\bar\bkappa}e^{-i\r_i\cdot \bkappa }\int_{\r_i}^{\r_f}\cD \r \exp\left\{i\frac{\omega^+}{2} \int_{x_i^+}^{x_f^+}ds^+\dot{\r}^2\right\}D_F((1-z)\r)D_F(z\r)\,.
\end{align}
which is the object appearing in Eq.~\eqref{eq:ch3_xsec_simplified2}.

In Section~\ref{sec:ch5_spin_correlations}, we further define process-dependent objects $\cK_{a\rightarrow bg}$ and $\cQ_{a\rightarrow bg}$ which result from similar simplifications to the ones leading to Eqs.~\eqref{eq:PR_simplified} and~\eqref{eq:K_simplified} in the large-$N_c$ limit. They read
\begin{subequations}\label{eq:K_Q_def}
\begin{align}
  & \cK_{a\rightarrow bg}(\bar x_1^+, \q_{1}; x_1^+, \bkappa) = \int_{\u_1,\u_2} e^{i\u_1\cdot\bkappa}e^{-i\u_2\cdot\l}\int_{\u_1}^{\u_2}\cD\u\exp\left\{i\frac{\omega_a^+}{2}\int_{x_1^+}^{\bar x_1^+}ds\,\dot \u^2\right\}\, \cC^{(3)}_{a\rightarrow bg}(\bar x_1^+, x_1^+; \u)\,, \\
  & \cQ_{a\rightarrow bg}(L,\q_1, \q_1; \bar x_1^+, \bar \bkappa, \l) = \int_{\u_i, \u_f, \bar \u_i, \bar \u_f} e^{i\u_i\cdot\l}e^{-i\bar\u_i\cdot\bar \bkappa}e^{-i(\u_f-\bar\u_f)\cdot\q_1}\nn
  & \times  \int_{\u_i}^{\u_f}\cD\u \int_{\bar\u_i}^{\bar\u_f}\cD\bar\u\exp\left\{i\frac{\omega_a^+}{2}\int_{\bar x_1^+}^Lds\,(\dot \u^2 - \dot{\bar{\u}}^2)\right\}\cC^{(4)}_{a\rightarrow bg}(L,\bar x_1^+; \u,\bar \u)\,.
\end{align}
\end{subequations}
The $n$-point functions $\cC^{(3)}$ and $\cC^{(4)}$ are given in terms of dipole and quadrupole objects. For the relevant processes, one can write them as
\begin{subequations}
\begin{align}
    & \cC^{(3)}_{q\rightarrow qg}(\u) = D_F(\u)D_F((1-z)\u)\,,\\
    & \cC^{(3)}_{g\rightarrow gg}(\u) = D_F(\u)D_F(z\u)D_F((1-z)\u)\,,\\
    & \cC^{(4)}_{q\rightarrow qg}(\v,\bar \v) = Q(\v, \bar \v)D_F(-(1-z_1)(\v-\bar\v))\,,\\
    & \cC^{(4)}_{g\rightarrow gg}(\v,\bar \v) = Q(\v, \bar \v)D_F(-(1-z_1)(\v-\bar\v))D_F(-z_1(\v-\bar\v))\,,
\end{align}
\end{subequations}
where we have omitted the time arguments of the dipole defined in Eq.~\eqref{eq:dipole_def}.
The quadrupole is defined as
\begin{align}
    Q(\v, \bar \v) = \frac{1}{N_c}\left\langle \Tr\left(U_1^{\dagger}U_2U_3^{\dagger}U_4\right)\right\rangle\,,
\end{align} 
with Wilson lines $U_i^{jk} = U_F^{jk}(\bar x^+,x^+; \r_i)$ defined in Eq.~\eqref{eq:Wilson_line_def} and with the relevant differences of coordinates $\r_i$ being defined as 
\begin{subequations}
\begin{align}
  & \r_1-\r_2 = (1-z_1)(\v-\bar\v)\,,\\
  & \r_3-\r_4 = z_1(\v-\bar\v)\,,\\
  & \r_1-\r_4 = \v\, .
\end{align}
\end{subequations}
Keeping this object in its full form results in a quite involved calculation of the $\cQ$ function (see e.g.~\cite{Blaizot:2012fh,Apolinario:2014csa, Dominguez:2019ges, Isaksen:2023nlr}). As explained in the main text near Eq.~\eqref{eq:ch5_Q_approx}, we simplify it by approximating it as the independent broadening of the two final state partons, which amounts to writing the quadrupole as
\begin{align}\label{eq:indep_broadening}
  Q(\v,\bar\v) \approx D_F(z_1(\v-\bar \v))D_F((1-z_1)(\v-\bar \v))\,,
\end{align}
valid in a specific phase space region, as discussed in the main text. With this approximation, everything is given in terms of dipoles, such that $\cK_{a\rightarrow bg}$ and $\cQ_{a\rightarrow bg}$ are simplified to
\begin{subequations}
\begin{align}
  & \cQ_{a\rightarrow bg}(L,\q_1, \q_1; \bar x_1^+, \bar \bkappa, \l) \equiv \delta^2(\l-\bar\bkappa) \int_\v e^{-i\v\cdot(\q_1-\bar\bkappa)}\cP_{a\rightarrow bg}(\v; L,\bar x_1^+) \, ,\\
  & \int_{\l}\delta^{2}(\l-\bar\bkappa)\cK_{a\rightarrow bg}(\bar x_1^+, \l; x_1^+, \bkappa) \equiv \int_{\u_1,\u_2} e^{i\u_1\cdot\bkappa}e^{-i\u_2\cdot\bar\bkappa}\cK_{a\rightarrow bg}(\bar x_1^+, \u_2; x_1^+, \u_1)\,,
\end{align}
\end{subequations}
such that $\cP_{a\rightarrow bg}$ and $\cK_{a\rightarrow bg}$ generalize, respectively, Eqs.~\eqref{eq:PR_simplified} and~\eqref{eq:K_simplified}. For completeness, the former is given by
\begin{align}\label{eq:process_dep_P}
     & \cP_{q\to qg}(\v; L, \bar x_1^+) = D_F(z_1\v)D_F((1-z_1)\v)D_F(-(1-z_1)\v)\,,\nn
     & \cP_{g\to gg}(\v; L, \bar x_1^+) = D_F(z_1\v)D_F((1-z_1)\v)D_F(-(1-z_1)\v)D_F(-z_1\v)\,,
\end{align}
where we omitted the light-cone time coordinates of the dipoles, for clarity. This makes clear the introduction of the process dependent functions in Eq.~\eqref{eq:ch5_P_after_HO}. % file "App_npoint_functions.tex"
\cleardoublepage

% ----------------------------------------------------------------------
\end{document}